\documentstyle[11pt,epsf,psfig]{article}
\title{Exotic Kondo Effects in Metals:  Magnetic Ions in a Crystalline
Electric Field and Tunneling Centers}
\author{D.L. Cox,\\
       Department of Physics,\\
       The Ohio State University,\\
       Columbus, Ohio,  43210\\
       and\\
       Department of Physics,\\
       University of California,\\
       Davis, California, 95616 \\
       \\
       and \\
       \\
       A. Zawadowski,\\
       Institute of Physics,\\
       and\\
       Research Group of Hungarian Academy of Sciences,\\
       Technical University of Budapest, \\
       Budafoki ut 8,\\
       H1521,\\
       and\\
       Research Institute for Solid State Physics,\\
       Hungarian Academy  of Sciences,\\
       P.O.B. 49, H-1525,\\
       Budapest, Hungary}
\date{\today}
\def\cub{$O$}
\def\hex{$D_6$}
\def\tet{$D_4$}
\def\tgam{\tilde\Gamma}
\def\zar{Zar\'{a}nd$~$}

\def\thsp{\tilde H_{sp}}
\def\vld{Vlad\'{a}r$~$}
\def\noz{Nozi\`{e}res$~$}
\def\zim{Zim\'{a}nyi$~$}
\def\zow{Zawadowski$~$}
\def\ufp{U$^{4+}~$}
\def\ctp{Ce$^{3+}~$}
\def\ucon{$5f^2(J=4)~$}
\def\ccon{$4f^1(J=5/2)~$}

\def\dx{\Delta^x}
\def\dy{\Delta^y}
\def\gse{\Gamma_7}
\def\gei{\Gamma_8}
\def\gni{\Gamma_9}
\def\gon{\Gamma_1}
\def\gtw{\Gamma_2}
\def\gth{\Gamma_3}
\def\gfo{\Gamma_4}
\def\gfi{\Gamma_5}
\def\gsi{\Gamma_6}
\def\ek{\epsilon_{\vec k}}
\def\ef{\epsilon_f}
\def\tef{\tilde\epsilon_f}
\def\uff{U_{ff}}
\def\gthak{|f^2\Gamma_3,\alpha>}
\def\gthab{<f^2\Gamma_3,\alpha|}
\def\gthpk{|f^2\Gamma_3,+>}
\def\gthpb{<f^2\Gamma_3,+|}
\def\gthmk{|f^2\Gamma_3,->}
\def\gthmb{<f^2\Gamma_3,-|}
\def\gthapb{<f^2\Gamma_3,\alpha'|}
\def\gsemk{|f^1\gse,\mu>}
\def\gsemb{<f^1\gse,\mu|}
\def\gsempb{<f^1\gse,\mu'|}
\def\gsemmb{<f^1\gse,-\mu|}

\def\fnob{<f^0,\gon|}
\def\vspi{\vec S_I}
\def\vti{\vec \tau_I}
\def\vspeip{\vec S_{c8+}(0)}
\def\vspeim{\vec S_{c8-}(0)}
\def\vteiup{\vec \tau_{c8\uparrow}(0)}
\def\vteidn{\vec \tau_{c8\downarrow}(0)}
\def\vspse{\vec S_{c7}(0)}
\def\ccei{c^{\dag}_{k8\alpha\mu}}
\def\caei{c_{k8\alpha\mu}}
\def\ccse{c^{\dag}_{k7\mu}}
\def\case{c_{k7\mu}}
\def\caeip{c_{k'8\alpha\mu'}}

\def\casep{c_{k'7\mu'}}
\def\ccf{c^{\dag}_{\vec k,\sigma}}
\def\caf{c_{\vec k,\sigma}}
\def\cafp{c_{\vec k',\sigma}}
\def\yup{Y$_{1-x}$U$_x$Pd$_3$}

\def\cecs{CeCu$_2$Si$_2$}
\def\ube{UBe$_{13}$}

\def\urs{URu$_2$Si$_2$}

\def\tpi{\tau^{(0)}_I}
\def\txi{\tau^{(1)}_I}
\def\tyi{\tau^{(2)}_I}
\def\tzi{\tau^{(3)}_I}
\def\tcxi{\tau^{(1)}_c(0)}
\def\tcyi{\tau^{(2)}_c(0)}
\def\tczi{\tau^{(3)}_c(0)}
\def\scxi{S^{(1)}_c(0)}
\def\scyi{S^{(2)}_c(0)}
\def\sczi{S^{(3)}_c(0)}
\begin{document}
\maketitle
\begin{abstract}
The ordinary single channel Kondo model consists of one or more 
spin 1/2 local moments
interacting  antiferromagnetically with conduction electrons in a metal.  
This model has provided a paradigm for
understanding many phenomena of strongly correlated electronic
materials, ranging from the formation of heavy fermion fermi liquids to
the mapping to a one-band model in the cuprate superconductors.  
The simplest extension of this ordinary Kondo model 
in metals which yields exotic non-Fermi liquid physics is the
multichannel Kondo impurity model in which the conduction electrons are given an
extra quantum label known as the channel or flavor index. In the
overcompensated regime of this model non-Fermi liquid physics is
possible, in contrast to the single channel model. We overview here 
the multichannel Kondo impurity model candidates most extensively studied for
explaining real materials, specifically the two level system Kondo
model relevant for metallic glasses, nanoscale devices, and some doped
semiconductors, and the quadrupolar and magnetic two-channel Kondo
models developed for rare earth and actinide ions with crystal field
splittings in metals.   We provide an extensive justification for the
derivation of the theoretical models, noting that whenever the local
impurity degree of freedom is non-magnetic a two-channel Kondo model
must follow by virtue of the magnetic spin degeneracy of the conduction
electrons.  We carefully delineate all energetic and symmetry
restrictions on the applicability of these models.  
We describe the various methods
used to study these models along with their results  and limitations 
(multiplicative
renormalization group, numerical renormalization group, non-crossing
approximation, conformal field theory, and abelian bosonization), all of
which provide differing and useful views of the physics.  We
pay particular attention to the role that scale invariance plays in all
of these theoretical approaches.  
We point out in each case how various perturbing fields (magnetic,
crystalline electric, electric field gradients, uniaxial stress) may
destabilize the non-Fermi liquid fixed point.  
We then provide an extensive
discussion of the experimental evidence for the relevance of the
two-level system Kondo model to metallic glasses, nanoscale devices, and 
of the quadrupolar/magnetic two-channel models to a number of heavy
fermion based alloys and compounds.  We close with a discussion of the
extension of the single impurity models which comprise the main focus of
this review to other sytems (Coulomb blockade), multiple impurities, and 
lattice models.  In the latter case, we provide an overview of the
relevance  of the two-channel Kondo lattice model to non-fermi liquid
behavior and exotic superconductivity 
in heavy fermion compounds and to the theoretical possibility
of odd-frequency superconductivity, which is realized (for the first
time) in the limit of
infinite spatial dimensions for this model. 

\end{abstract}
\tableofcontents
\section{Introduction} 

Since the great success of the many body theory of metals and
superconductors
developed in the 1950's, it had been generally believed that the theory
of metals is
well understood and the only exception is in magnetic properties.  It
was accepted that
perturbation theory works well for ordinary clean and dirty metals and
no one
expected any deviation from the Landau theory at low temperature.  Then
in 1964 Kondo
[Kondo, 1964]
discovered a logarithmic divergence in the scattering of electrons by
magnetic
impurities in metals.  This invoked extensive research since that time
using very
different tools of theoretical physics including K.G. Wilson's work on
the renormalization
group [Wilson, 1975] honored by the Nobel Prize in 1982.  Indeed, the
Kondo effect
has become one of the most extensively studied many body problems in
the field of
theoretical solid state physics for the last three decades.

In the last 15 years several other problems related to the original
Kondo problem have
showed up in the literature, and the present review's main goal is to
summarize the
progress on the multi-channel Kondo model.  The possible physical
realizations of this
model generally depend on local orbital degrees of freedom as opposed
to the local
magnetic moment considered originally by Kondo.

\begin{figure}
\parindent=2in
\indent{
\epsfxsize=4.0in
\epsffile{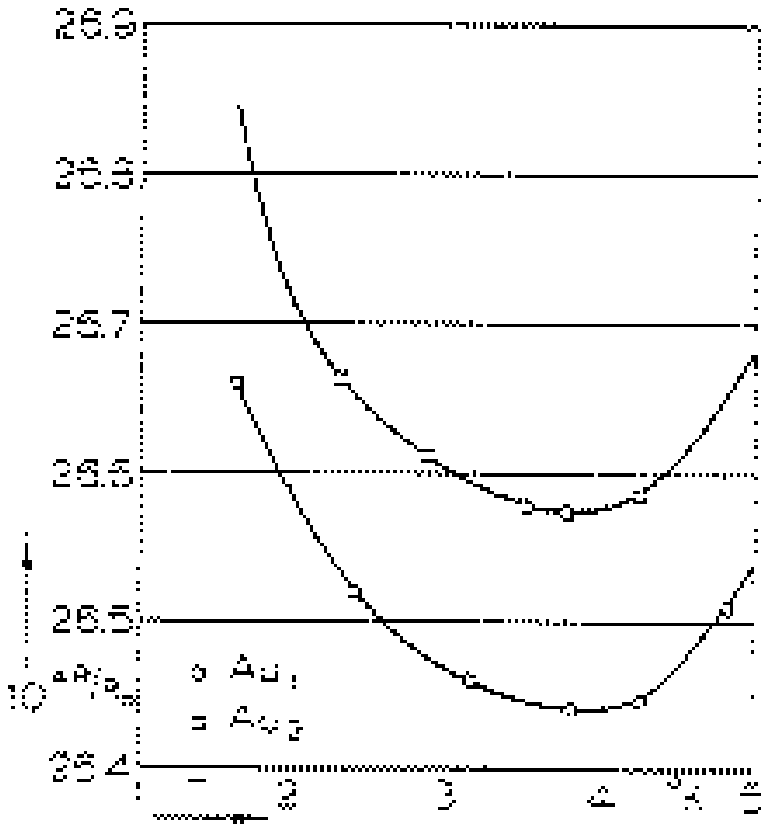}}
\parindent=0.5in
\caption{Early measurement of a resistivity minimum in metals.}
\label{fig1p1} 
\end{figure}

Before 1964, the electron-impurity interaction in metals with magnetic
impurities had been
studied by calculating the electrical resistivity in the second order
of perturbation theory
assuming simple effective Heisenberg exchange interactions between the local
moments and
the conduction electron spins.
The contribution of the magnetic impurities to the residual resistivity
was similar to
that of non-magnetic impurities except in the magnetoresistance.  This
was despite the
very early observation by W.J. de Haar, J. de Boer, and G.J. Berg in
1933 [de Haar
{\it et al.}, 1933] that in some cases the resistivity of gold shows a
minimum at around
3-4K (Fig. ~\ref{fig1p1}) and it was remarked that ``...it is of course very
interesting to investigate
the influence of the purity of the metals on this minimum.''  At that
time very little
attention was paid to the observation and only the latter extensive
studies in the 1960s
clarified that different magnetic impurities contribute to the
logarithmic temperature
dependence of the electrical restivity with minima at different
temperatures.

In the milestone work of Kondo in 1964 [Kondo, 1964], the calculation
of the electrical
resistivity was extended to third order in the local moment-conduction
electron exchange
coupling $J$ of the form
$$H_{int} = J \vec S_I\cdot \vec s_c(0) \leqno(1.1)$$
for a single impurity where $\vec S_I$ is the impurity spin and $\vec
s_c(0)$ is the conduction electron spin at the impurity site.  Note
that positive $J$ corresponds to antiferromagnetic
coupling, and negative $J$ to ferromagnetic coupling.
Kondo's result was that the resistivity took the form
$$\rho(T) = \rho_0(T) + aN(0)J^2 + bN(0)^2J^3 \ln({D \over k_BT})
\leqno(1.2)$$
where $\rho_0(T)$ is the resistivity of the pure metal, $N(0)$ is the
conduction electron
density of states at the Fermi energy, $D$ is the bandwidth of the
conduction
electron system (of the order of the Fermi energy),  $T$ is the
temperature, and $a,b$ are
constants proportional to the impurity concentration.  The origin of
this logarithmic term, as
we shall discuss in subsequent paragraphs, is the {\it non-commutative}
scattering
when the impurity has internal degrees of freedom.

 Several interesting results can be
inferred from Eq. (1.2).  First, given that in a clean metal such as
gold $\rho_0(T)\sim T^5$
at low temperatures, a resistivity minimum will appear when $J>0$
(antiferromagnetic coupling)
since the logarithmic term then grows as the temperature is
diminished.  As Kondo
noted, this minimum temperature should go as $c^{1/5}$ where, $c$ is
the impurity
concentration.    Second, given $a\simeq b$, the logarithmic term grows
to the order
of the quadratic in $J$ term at a characteristic ``Kondo'' temperature
scale
$$k_BT_K \approx De^{-1/N(0)J}~~. \leqno(1.3)$$
Below this scale systematic weak coupling perturbation expansions must
break down, and
so it represents a crossover between a high temperature, high energy
 regime which may be treated perturbatively and a low energy regime of
 complex character
which must be treated non-perturbatively.  This makes the Kondo problem
the first known
example of an asymptotically free theory in physics, and the Kondo
scale is analogous to
the so-called QCD scale $\Lambda_{QCD}$ at which perturbative quantum
chromodynamics
breaks down (see for example, Aitchison and Hey [1989], p. 298).
Finally, it is clear from this remark and the restriction of unitarity
on the scattering rate
(a finite scattering probability must exist at the Fermi energy) that
Kondo's theory is
incomplete.

In order to resolve this latter problem, Abrikosov [1965] and Suhl
[1965] introduced the
concept of the Kondo resonance.  The idea is that the appearance of the
logarithm in
Kondo's calculation signifies the development of a many body scattering
resonance at
the Fermi energy.  These treatments summed up infinite numbers of
diagrams diverging
as powers of logarithms using building blocs with only one-electron or one-hole
scattered states.  While
this led to a finite scattering rate at the Fermi energy, it did not
lead to a complete
solution of the problem.

A great boost in understanding the new phenomena came from studies on
the x-ray
absorption singularity worked out theoretically by Mahan ([1990], pp. 737-764), 
and more elaborately by \noz and de
Dominicis (\noz and de Dominicis [1969]).  
The issue for this problem is how
the conduction
electrons react to the the appearance of a deep core hole created by
x-rays and how
this response affects the x-ray absorption spectrum.  The problem can
be formulated
more generally in terms of how the conduction electrons relate to a
change in a localized
external potential.  It is well known that in the electron screening of
a localized potential
a long range Friedel oscillation is induced with wave vector $2k_F$,
where $k_F$ is the Fermi
wave vector.  This oscillation signals the presence of a singular
response for electronic states at
the Fermi energy.
Anderson reconceptualized this problem (Anderson [1967]) by pointing
out that
the ground state wave functions with two different localized potentials
are orthogonal to
one another in the thermodynamic limit.  (This change of impurity potential 
can introduce a phase shift and amplitude modification to the Friedel 
oscillations about the defect site.)  This phenomenon has been named
the ``Anderson
orthogonality catastrophe''.
The relation of the Kondo effect to this problem is easy to see:
whenever the local moment
spin is flipped by interaction with the conduction electrons, a change
in the scattering potential
for up and down spins is the result.  Hence, the orthogonality
catrastophe arises for every single
spin flip.

All these works made it obvious that the x-ray absorption problem and
Kondo problem
represent a new type of infrared divergence at low temperature and
energy in which an
infinite number of particle-hole pairs are involved on a scale
extending to large distances
away from the localized perturbation.  On the other hand, there is a
very essential difference
between these two problems, which is related to the presence of
internal degrees of
freedom in the Kondo problem.  The resulting scattering potential seen
by conduction electrons
is non-commutative in the Kondo case, while it is commutative in the
x-ray core hole
problem.  In order to demonstrate this idea, we consider a simple
scattering problem
in which conduction electrons couple to a single heavy particle (e.g.,
a localized magnetic
ion or a two-level system).  The Hamiltonian is
$$H=\sum_{k\mu} \epsilon_k c^{\dagger}_{k\mu}c_{k\mu}  +
\epsilon_0\sum_{\alpha}b^{\dagger}_{\alpha}b_{\alpha}  $$
$$~~~~~ + \sum_{k,k'}\sum_{\mu\nu\alpha\beta} V_{\mu\nu\alpha\beta}
c^{\dagger}_{k\mu}
c_{k'\nu} b^{\dagger}_{\alpha}b_{\beta} \leqno(1.4)$$
where $\epsilon_k$ is the electron kinetic energy with momentum $k$ in,
e.g., an $s$-wave
projection of plane wave states, $c^{\dagger}_{k\mu}$ creates an
electron spherical wave with radial momentum
$k$ and internal quantum numbers $\mu$, and $b^{\dagger}_{\alpha}$
creates a heavy particle
with quantum numbers $\alpha$.  Note that the internal indices of the
conduction electrons
may represent magnetic spin or orbital indices or a combination
of the two.
The interaction potential is given by $V_{\mu\nu\alpha\beta}$.
For the conduction electrons a band cutoff $D$ is applied which is
order of the Fermi energy.
In the following remarks, the limit $\epsilon_0\to \infty$ is taken to ensure
that there is only single occupancy of the heavy particle.

\begin{figure}
\parindent=2.in
\indent{
\epsfxsize=2in 
\epsffile{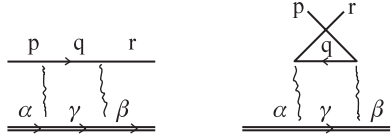}}
\parindent=0.5in
\caption{The two leading diagrams for the scattering of a light
electron on a
heavy object.  The light electron is represented by the light line and
the heavy
object by the double line.   The wavy lines indicate the interaction
matrix element
$V$.  The quantum numbers characterizing the particles are also
shown.   The expression
below gives the corresponding matrix elements of the interaction to
this order and
the minus sign arises from fermion anticommutation (the crossed lines
in the
second figure--Einstein convention applies to the indices):
$V^{(2)}_{ \beta\alpha r p} \sim V_{\beta\gamma rq}V_{\gamma\alpha qp}
-
V_{\gamma\alpha rq}V_{\beta\gamma qp} ~.$}
\label{fig1p2}
\end{figure}

In the second order of perturbation theory for the two-particle
$T$-matrix describing electron
scattering off the impurity, there are two diagrams shown in a time
ordered way in Fig. ~\ref{fig1p2}.
The direction of time corresponds to the directions of the lines on the
heavy objects.  As the
interaction is assumed independent of $k,k'$, the scattering amplitude
for an incoming
electron with energy $\omega$ is
$$V^{(2)}_{\mu\nu\alpha\beta}(\omega) = \sum_{\rho\gamma}
[V_{\mu\rho\alpha\gamma}
V_{\rho\nu\gamma\beta} -
V_{\rho\mu\alpha\gamma}V_{\nu\rho\gamma\beta}]\ln({D\over\omega})
\leqno(1.5)$$
where the quantum numbers of the internal lines are summed over and the
negative
sign arises from the fermion anticommutation relations (note the
crossed lines in the
second diagram) as the intermediate conduction state reflects a
particle or hole, respectively.
The logarithm is precisely that identified by Kondo.

Different localized perturbations can be classified by the value of the
potential matrix
``commutator'' appearing in square brackets in Eq. (1.5):  \\
{\it (i) Commutative model}:  When
$$[V_{\mu\rho\alpha\gamma}
V_{\rho\nu\gamma\beta} -
V_{\rho\mu\alpha\gamma}V_{\nu\rho\gamma\beta}] = 0 \leqno(1.6)$$
no Kondo logarithms appear.  This occurs for a point like structureless
interaction
between the heavy object and conduction electrons which is diagonal in
the internal
indices.  \\
{\it (ii) Non-Commutative model}:  When
$$[V_{\mu\rho\alpha\gamma}
V_{\rho\nu\gamma\beta} -
V_{\rho\mu\alpha\gamma}V_{\nu\rho\gamma\beta}] \ne 0 \leqno(1.7)$$
the model is non-commutative and logarithmic terms appear.

Case (i) applies to the x-ray absorption problem since the core hole has
no internal degrees
of freedom, and this also occurs in heavy particle motion of muons and
protons in metals
provided we neglect their spin degrees of freedom and focus only on the
charge degrees
of freedom.  In these cases the couplings are not renormalized in
second order and such
theories have an energy and temperature independent coupling as a
result.

Case (ii) is realized when the interaction depends, for example, on the
spin variables or
the pseudo-spin (position index) of a two-level system (TLS) describing
an atom hopping
between the lowest states of a double well anharmonic potential, or the
orbital quadrupolar
degrees of freedom of an open shell ion.  In the latter two cases, the
local orbital states of
the conduction electrons are intrinsically involved as well.  In these
examples the effective
coupling strength is strongly renormalized as a function of energy and
temperature and
can grow to infinity  (antiferromagnetic Kondo problem), shrink to
zero (ferromagnetic Kondo problem), or reach an intermediate coupling
strength independent
of the bare couplings (overcompensated multichannel Kondo problem).

The present review will be entirely devoted to non-commutative models.
In order to clarify their
solutions it is useful to look at the $M$-channel Kondo model.  The
interaction
Hamiltonian for this model describes the interaction of conduction
electrons living in
$M$ identical bands or channels with an arbitrary localized spin $\vec
S_I$, and is given
by
$$H_{int} = {J \over 2N_s} \sum_{k,k'}\sum_{\mu\nu}\sum_{\alpha=1}^M
\vec S_I\cdot
c^{\dagger}_{k\mu\alpha}\vec \sigma_{\mu\nu} c_{k'\nu\alpha}
\leqno(1.8)$$
where $N_s$ is the number of atomic sites of the host metal responsible for the 
orbitals which form the conduction band, and $\vec \sigma$ is a Pauli
matrix.   The interaction
Hamiltonian is diagonal in the channel index and an exact degeneracy of
couplings
is assumed.  The original Kondo problem specified in Eq. (1.1) has
$M=1$.  This
multi-channel extension of the Kondo problem was first introduced by
\noz and Blandin [1980]
and we follow their discussion here.

\begin{figure}
\parindent=2.in
\indent{
\epsfxsize=5in
\epsffile{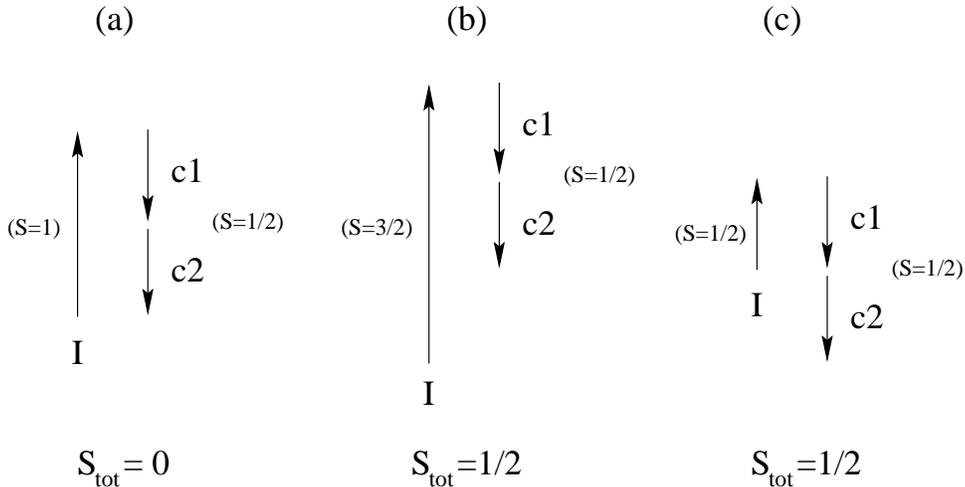}}
\parindent=0.5in
\vspace{.5in}
\caption{Compensated, Undercompensated, and Overcompensated multi-channel Kondo models.  (a) shows an example of a
compensated Kondo model.  The two spin 1/2 conduction channels exactly cancel the spin 1 impurity to form a local 
singlet ground state.  (b) shows an example of an undercompensated Kondo model.  In this case, the two spin 1/2
conduction electrons cannot fully screen out the spin 3/2 impurity moment.  The total ground state spin is determined by the 
residual, uncompensated impurity moment.  The residual coupling of this to the conduction electrons is ferromagnetic
and thus marginally irrelevant about the ground state.  Finally, (c) shows an example of the overcompensated 
situation, in which the two spin 1/2 conduction channels over-screen the spin 1/2 impurity.  In this case, the 
residual coupling of the bound spin 1/2 complex to conduction electrons at the next length scale is antiferromagnetic. 
This situation leads to a non-trivial fixed point.  }
\label{fig1p3}
\end{figure}

The ground state of the model with impurity $S_I$ but different $M$ can
fall into three  different
classes for antiferromagnetic coupling $J>0$, summarized in below and
Fig. ~\ref{fig1p3}:\\
{\it (i) Compensated ($M=2S_I$)}  In this case the ground state is a
singlet in
which the impurity spin is
completely screened by the conduction electron spins.  The entropy
tends to zero as the
temperature tends to zero. \\
{\it (ii) Undercompensated ($M<2S_I$)}.  In this case, there is not
enough conduction
spin to fully screen the impurity through the interaction of Eq.
(1.8).  $M/2$ units of spin
are compensated, leaving behind a local moment of strength $S_I - M/2$
and a corresponding
residual entropy at $T=0$ of $R\ln(2S_I-M+1)$ per mole of impurity ion
(see (b) of Fig. ~\ref{fig1p3}).
This residual entropy
would appear to violate the third law of thermodynamics, but in the
presence of more than one
impurity intersite couplings will allow the possibility of spin-spin correlations
or spin ordering to reduce the entropy.
Note that this limit cannot be achieved with $S_I=1/2$.  \\
{\it (iii) Overcompensated ($M>2S_I$)}.  In this case, there is an
excess of conduction spin
and depending on the boundary conditions for a finite size system, the
ground state will
have spin of either $M/2-S_I$ or $S_I$ with residual entropy
$R\ln(M-2S_I+1)$ or
$R\ln(2S_I+1)$ (see (c) of Fig. ~\ref{fig1p3}).    However, in the thermodynamic limit
the entropy tends to
$R\ln g(M,S_I)$ where $g$ is a universal non-integer number that we
shall discuss in
Secs. 6.3,7.2.  The case $S_I=1/2,M=2$ is the marginal one, which still 
yields the non-trivial physics and 
appears to be the most realizable model in nature as we
discuss in Sec. 2. 
This model has a residual entropy of $(R/2)\ln 2$.  This residual entropy
appears to have been observed
in several uranium based alloys recently as we shall discuss later in
the review (Sec. 8).

Let us specialize to the case $S_I=1/2$ and discuss the thermodynamic
properties at low
temperatures which reflect the low energy excitation spectrum of the
system.  A convenient
quantity to classify the behavior is the specific heat coefficient
$C_{imp}/T$ defined as the
extra heat capacity per impurity induced by the impurities.  In the
case $M=1$, the ordinary
Kondo problem, $C_{imp}/T$ tends to a constant value proportional to
$1/T_K$, with $T_K$
the Kondo temperature identified in Eq. (1.3).  This describes a local
Fermi liquid about the
impurity with effective degeneracy temperature given by $T_K$.  When
$M=2$,
$C_{imp}/T\sim \ln(T_K/T)/T_K$, which illustrates the marginal nature
of the $M=2,S_I=1/2$
overcompensated state.  Finally, when $M>2$, the specific heat
coefficient shows a
power law divergence given by $C_{imp}/T\sim T^{(2-M)/(2+M)}$.  Clearly
when $M\ge 2$ the
excitation spectrum has a non-Fermi liquid character which thus places
these models outside
the Landau paradigm discussed at the outset of this introduction.  The
logarithmic behavior
in $C_{imp}/T$ has been recently observed in numerous materials as will be
discussed in Sec. 8. 

\begin{figure}
\parindent=2in
\indent{
\epsfxsize=4in
\epsffile{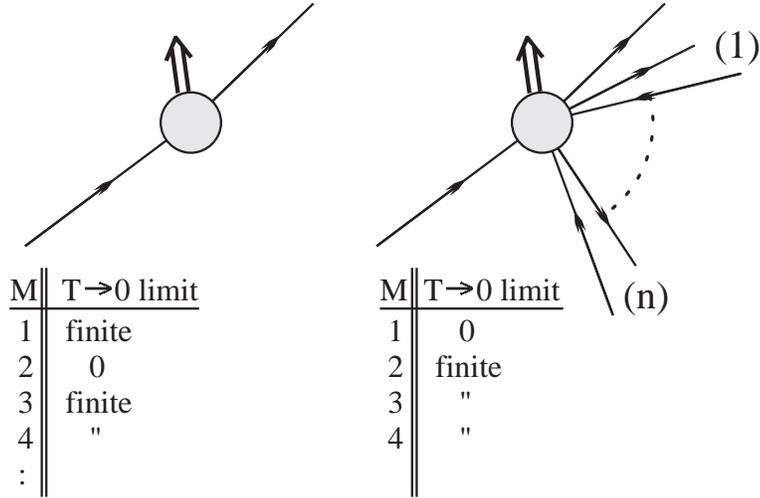}}
\parindent=0.5in
\caption{Fermi-level scattering states resulting from electrons incident on spin 1/2 Kondo impurities. 
For the $M=1$ case, all outgoing scattered states at the Fermi energy is one-body in character, and 
hence can be parameterized in terms of a phase shift.  
For $M=2$, all outgoing scattered states at the Fermi energy are {\it many-body} in character, and cannot
be parameterized by a phase shift.  For general $M$, the outgoing scattered states are a mix of
one-body and many-body in character.   }
\label{fig1p4}
\end{figure}

The scattering of an electron off the impurity in the low energy limit
further reflects the breakdown of the Landau paradigm.  We illustrate
the scattering possibilities in Fig. ~\ref{fig1p4}.
We show that for $M=1,S_I=1/2$, the $T=0$ scattering is
entirely one particle,
reflecting the complete screening of the local moment into a simple
charge scattering object.
At elevated frequency and temperature some outgoing states can be
dressed by multiple
particle hole pairs giving rise to the familiar Landau $T^2$ damping of
the excited quasiparticles.
In Fig. ~\ref{fig1p4}, we also show the case for $M=2,S_I=1/2$, in which single
particle scattering is
{\it completely} shut down on the Fermi surface. Indeed, the surprising
result is that there is no $S$-matrix element to any outgoing state 
containing {\it arbitrary} finite numbers of particle-hole pair excitations; 
rather, as phrased in the original paper, 
there is a scattering to a ``different Hilbert space hidden in 
the free field theory'' but ``opened up'' by the impurity at the non-trivial
fixed point (Maldacena and Ludwig [1996]).  These fermionic excitations
are non-local in the original variables.  
In the general case for
$M>2,S_I=1/2$, the Fermi level
scattering can be a mix of single particle and multiparticle
scattering, as illustrated in Fig. ~\ref{fig1p4}.  The outgoing scattering 
state may possibly be visualized as a ``screening cloud'' with one electron
charge which is orthogonal to any states with finite numbers of simple 
particle-hole excitations, which phenomenon must be related to Anderson's 
orthogonality catastrophe (Anderson [1967]).  

The consequence of these scattering properties on the electrical
resistivity as $T\to 0$ is
that in the $M=1,S_I=1/2$ case saturation to a value characteristic of
the unitarity limit is
accomplished with a regular $T^2$ behavior typical of a Fermi liquid.
For $M\ge 2$, the
low temperature resistivity goes as $\rho(T)/\rho(0) \simeq 1 -
AT^{2/(2+M)}$ which is clearly
non-Fermi liquid-like and related to the residual entropy and power law
divergence of the
specific heat coefficient.  In the case of $M=2$, the saturation to the
low temperature limit
follows a $T^{1/2}$ law which has been observed in recent point contact
experiments to
be discussed at length in Sec. 8.1.

A few words on notation.  In the discussion of the quadrupolar and
two-channel
magnetic Kondo effects for rare earth and actinide ions, we have chosen
to
denote all crystal field states by the convention in the tables of
[Koster {\it et al.}, 1963].
 For irreducible representations of non-Kramers (non-double group)
character, we also give the symbols used, e.g., in Tinkham's book on
group
theory[Tinkham, 1964]. Henceforth, we shall abbreviate ``irreducible
representation''
by ``irrep'' following a common practice in group theory texts.  Since
the Two-Level-System (TLS) Kondo
effects and those for rare earth and
actinide ions ``grew up'' independently, there is not surprisingly a
significant set of notational differences between papers on the two
subjects.
The TLS literature uses Pauli matrices to represent the local
pseudo-spin
variables, $\sigma$ indices to represent the real conduction spin 1/2
indices,
and $V^{x,y,z}$ to represent the couplings of the various pseudo-spin
components to the conduction electrons.  In the sections on quadrupolar
and
magnetic Kondo effects of actinide and rare earth ions, we shall use
spin 1/2
matrices throughout, with $\tau^{(1,2,3)}$ representing matrices in the
quadrupolar pseudo-spin space, and $S^{(1,2,3)}$ representing matrices
in the
magnetic pseudo-spin space. Quadrupolar indices are represented by
$\alpha=\pm$,
and magnetic indices by $\mu=\uparrow,\downarrow$.  The reason for the
abstract
$1,2,3$ labels on the pseudo-spin matrices is because they don't always
have
clear correspondence to symmetry directions of the crystal.  All
exchange
 couplings for
the rare earth and actinide ion models will be labeled by $J$, not to
be
confused with angular momentum of ground state multiplets whose usage
will be
clear in the immediate context.  Whenever an important formula is
derived, such
as a Kondo scale, we will report the results both for Pauli matrix and
spin 1/2
form.   The conduction density of states at the Fermi energy
has been denoted $\rho_0$ in the TLS literature, and $N(0)$ in the rare
earth/actinide papers.  To avoid confusion to readers looking back at
the
literature, we shall use the same custom here.  In either case, all
excitations
are measured with respect to the Fermi energy, which is therefore
chosen as the
zero of energy.   A list of what local spin and channel labels mean in 
various models appears in Table~\ref{tab1p1}.  
A ``rosetta stone'' clarifying the notational
correspondences
between the
different two-channel Kondo models to be considered in this paper
appears   in
Table~\ref{tab1p2}.  We will always use $N$ to refer to pseudo-spin degeneracy
of the
impurity or conduction
states; when it is necessary to distinguish the degeneracies, we will
use
subscripts $I$ (impurity) and $C$ (conduction).  We will always use $M$
to
refer to channel degeneracy.

\begin{table}
\parindent=2.in
\indent{
\begin{tabular}{|c|c|c|c|}\hline
Kondo & Local & Conduction & Conduction \\
Model & Pseudo-spin & Pseudo-spin & Channel\\\hline\hline
TLS & Atomic & Orbital & Magnetic \\
& Position &&  \\
& (Parity)   & (Parity) & \\\hline
Quadrupolar & Quadrupolar & Quadrupolar & Magnetic \\
& (Orbital) & (Orbital)&\\\hline
Magnetic Two-&  Magnetic & Magnetic&  Quadrupolar \\
Channel &&& (Orbital) \\\hline
\end{tabular}}
\parindent=0.5in
\caption{Meaning of local spin and channel labels in real
two-channel Kondo models.}
\label{tab1p1}
\end{table}

\begin{table}
\parindent=2.in
\indent{\begin{tabular}{|c|c|c|c|}\hline
Kondo & Local & Coupling & Density of \\
Model & Pseudo-spin & Constants & States \\\hline\hline
TLS & $\sigma^{x,y,z}$ & $V^{x,y,z}$ & $\rho_0$ \\\hline
Quadrupolar & $\tau^{1,2,3}$ & $-J^{1,2,3}$ &  $N(0)$\\\hline
Magnetic & $S^{1,2,3}$ & $-J^{1,2,3}$ & $N(0)$ \\\hline
\end{tabular} }
\parindent=0.5in
\caption{``Rosetta Stone'' of notational correspondences
between
two-channel Kondo models.  Note that the $\sigma^i$ operators for the
TLS model
are Pauli matrices, while the $\tau,S$ operators for the quadrupolar
and
magnetic two-channel models are spin 1/2 matrices.  Notice the relative
minus
sign between the couplings in the TLS models and the Quadrupolar and
Magnetic
Kondo models.  Note that while the TLS literature consistently uses
$N_s$ to
represent the number of conduction channels, we shall use this symbol
for the
number of sites in the lattice in our paper, and shall use $M$ for the
number
of channels.  }
\label{tab1p2}
\end{table}

\begin{figure}
\psfig{file=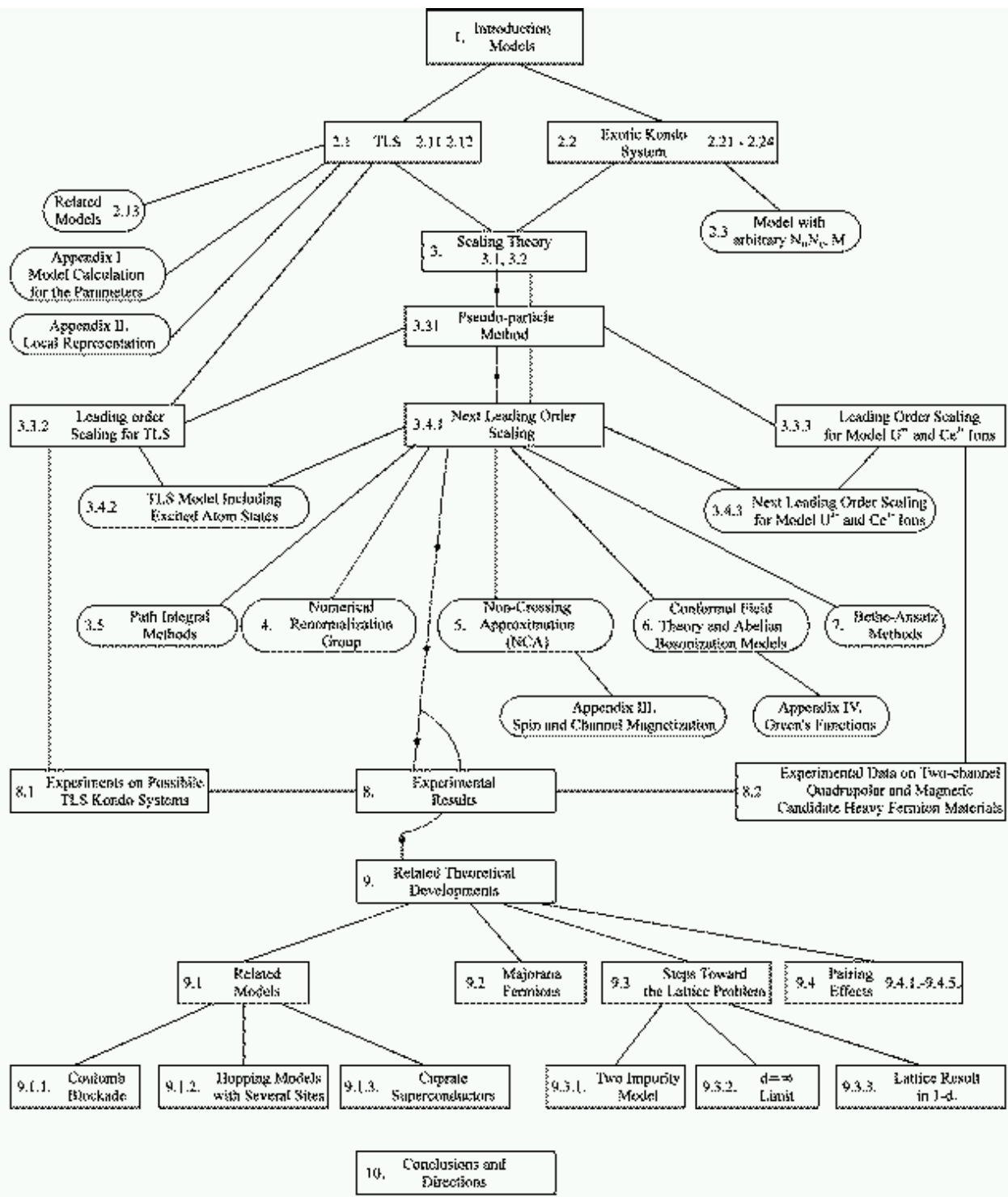,width=7in}
\caption{Guide to reading this review article}
\label{fig1p5}
\end{figure}

\section{Model Hamiltonians }

\subsection{TLS Kondo Model and Related Hamiltonians} 

\subsubsection{ TLS Kondo Model:  Physical Discussion} 

\begin{figure}
\parindent=2.in
\indent{
\epsfxsize=4in
\epsffile{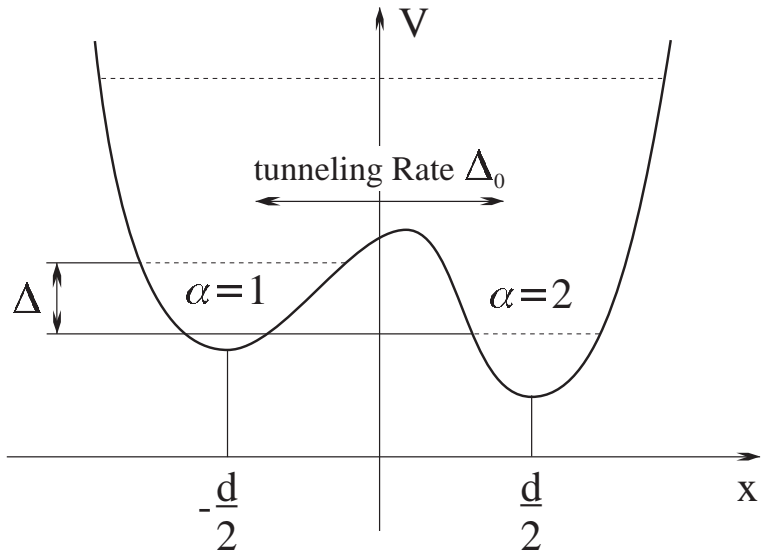}}
\parindent=0.5in
\caption{The potential for a single atom in a TLS is shown; this
has two minima.  The energies
of the states localized at the minima and of the next excited state are
shown by dashed lines.   The
energy splitting between the localized states is $\Delta$.   The
minima are labeled by
$\alpha=1,2$.}
\label{fig2p1}
\end{figure}

The simplest realization of a two-level system (TLS) is that of an atom
which
may sit in a double well potential, the two wells being localized along
a line
directed between their centers which are separated by a displacement
vector
$\vec d$.  In the absence of coupling to a bath of
excitations, the lowest two states of the atom are, approximately, the
positional eigenstates associated with harmonic oscillations within
either
well.  These are not exact eigenstates because of the overlap of their
wave
functions.  The next level usually has energy above the barrier between
the
well minima, and therefore is not localized to either well.  The basic
picture
for the atomic TLS is shown in Fig.~\ref{fig2p1}.

  To preview our discussion of the TLS
Kondo effect, one should think of the position of the atom in one well
or the
other as an Ising spin variable.  Electrons may ``flip'' the spin by
assisting
the tunneling between the wells.  The usefulness of this spin
description will
depend upon the degree of splitting of the eigenstates in the absence
of
coupling to the electrons.  If the splitting is too large, the ground
state
will be well separated from the first excited level, and the TLS will
essentially
possess a single degree of freedom leading
to only potential scattering off the TLS site and no interesting many
body
phenomena.  Hence, we shall pay close attention to the conditions which
determine the splitting of the levels.

We note that the origin of the anharmonicity which produces the double
well has
not been extensively considered.  It could derive from anharmonic
coupling to
vibrational degrees of freedom away from the TLS site [Sethna, 1981],
or
possibly from coupling to the electrons [Yu and Anderson, 1984].
The above
model is certainly not exclusive in terms of the kinds of TLS which may
occur
in real materials; for example, one may imagine a local bistable
breathing mode, a
bistable librational mode for a cage of light atoms, such as may occur
in doped
perovskite conductors, or a vibronic TLS associated with the
Jahn-Teller
effect [Gogolin, 1995], for example. For further reading see Yu and Leggett [1989],
and Burin and Kagan [1996].  

\begin{figure}
\parindent=2.in
\indent{
\epsfxsize=2in
\epsffile{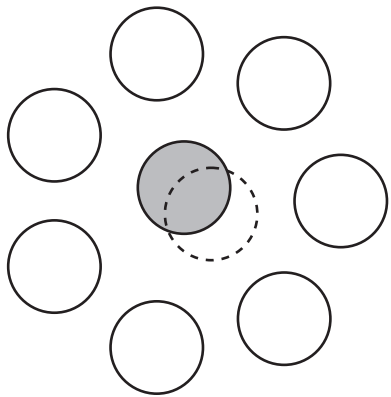}}
\parindent=0.5in
\caption{The atomic arrangement in the formation of a TLS is
shown.  The
two positions belonging to the TLS are shown by the dashed and shaded
circles.}
\label{fig2p2}
\end{figure}

An atomic tunneling TLS may also arise in a glass, due to
positional disorder as sketched in Fig. ~\ref{fig2p2} [Anderson, Halperin, and
Varma,
1971; W.A. Phillips, 1972].  Indeed, the original motivation for
studying such  a model was the observation of logarithmic anomalies in
the
resistivity of metallic glasses [Cochrane {\it et al.}, 1975; Kondo,
1976(a);
Vlad\'{a}r and
Zawadowski, 1980].  There the positional disorder of the atoms could
lead to a
TLS for individual atoms.  The complication with regard to the simple
model
presented here is that the TLS model must be concentrated (and
disordered).
 For simplicity, we shall restrict attention
to the simple model of a linear double well.

Given the TLS, in the absence of coupling to other excitations, the
atom may
move between the two positions one of two ways: \\

{\it (i) Thermal Activation}.
To make thermally activated transitions between the two minima, the
particle
must make a real excitation to one of the higher levels within the
double well
potential.  Roughly, one anticipates that the activated transition rate
will be given
by
$${1 \over \tau_{thermal}(T)} \approx \omega_0 \exp(-E_{ex}/k_BT)
\leqno(2.1.1)$$
where the ``attempt frequency'' $\omega_0$ is of order the
characteristic
vibration frequency in one of the well minima.
The excitation energies for such processes have been directly
measured in time resolved conductance experiments [Ralls and Buhrman,
1988;
Zimmerman, Golding, and Haemmerle, 1991; Golding, Zimmerman, and
Coppersmith, 1992] and are
typically in the range of tens to hundreds of kelvin.  Hence, these
must
eventually be frozen out at very low temperatures, certainly below
1-10K.\\

{\it (ii) Quantum Mechanical Tunneling}.  In this process, the atom
directly
tunnels through the potential barrier between the wells.  Given the
freezeout
of the thermally activated transitions, this process must dominate at
sufficiently low
temperatures.  In this case, neglecting the coupling to the conduction
electrons, the tunneling rate is given by
$${1\over \tau_{quantum}^s} \approx \omega_0 \exp(-w\sqrt{2MV_B}/\hbar)
= \omega_0
e^{-\lambda}  \leqno(2.1.2)$$
where the superscript $s$ stands for spontaneous tunneling,
the barrier is approximated by a square well of width $w$ and height
$V_B$, and $\lambda$ is the Gamow factor.

Based upon experimental results on mesoscopic devices, we may roughly
distinguish
between three regimes of quantum
tunneling, depending upon the observed tunneling time:\\
  (a) {\it Slow
tunneling, $(\tau_{quantum}^s)^{-1}< 10^8  s^{-1}$}.  In this case the
Gamow
factor $\lambda$ is large.  The motion of
the atom for such TLS has been measured by time resolved conductance
experiments on nanometer scale point contacts [Ralls and Buhrman, 1988]
and
thin metallic films [Zimmerman, Golding, and Haemmerle, 1991; Golding,
Zimmerman, and
Coppersmith, 1992].  The essential, and startling, idea is that each
time the
TLS hops between minimum, a measurable (order $e^2/h$) change $\delta
G$ in the
device conductance $G$ is observed, corresponding to an atomic scale
change in the
scattering cross section.  Bistable switching corresponding to isolated
TLS (of
unknown character) has been observed directly!  By producing histograms
of the
times between switching events, tunneling times can be determined.  As
our
subsequent discussion will clarify,
the spontaneous quantum mechanical tunneling rate is directly
proportional to
the electron assisted tunneling amplitude for the atom which is
responsible for
the TLS Kondo effect.
When the TLS falls into
this ``slow'' category, the TLS Kondo effect will occur at irrelevantly
small
temperature scales, if at all.  This may be modified, however, with the
consideration of excited states of the TLS, such as discussed by
\zar 
and \zow [1994(a,b)].   In this circumstance, it is possible to get an
appreciable Kondo scale {\it without}
a large splitting of the levels; see Sec. 3.4.2 for a discussion.  \\
  (b) {\it Fast tunneling, $10^8
s^{-1}>(\tau_{quantum}^s)^{-1} > 10^{12} s^{-1}$}.  In this case, the
energy
corresponding to the
tunneling rate, determined by the uncertainty principle, is in the
range of
1 mK to 10K.  As we shall explain below, this energy determines the
level
splitting which we should like to have smaller than the Kondo energy
scale
$k_BT_K$.  Given the typical estimates for electron assisted tunneling
processes and the resulting $T_K$ values, this range of tunneling times
is the
most probable for observing the
TLS Kondo effect.\\
  (c) {\it Ultrafast tunneling, $10^{12} s^{-1}<
(\tau_{quantum}^s)^{-1}$}.  In this case, the splitting of the TLS
($\sim
\hbar/\tau_{quantum}^s$) is
so
large that the ground state is uniformly spread in a bonding orbital
between
the two wells, and the Kondo effect never really happens.  In this
limit we may
practically consider only
the symmetric TLS, ($V(z)=V(-z)$).  In the absence of coupling to the
conduction electrons,
the bonding level in this case is
so well separated energetically from the anti-bonding level that
description of the TLS states with a spin
variable is essentially meaningless.  Another way of putting it is that
the
pseudospin variable of the TLS is completely polarized.

Hence, for our purposes, since we are interested in a TLS Kondo effect,
in the
absence of excited states of the TLS, our
attention will be directed to case (ii.b) of the above
paragraph:  fast quantum tunneling.  With excited states included, we
may very
well obtain a reasonable Kondo scale while possessing neglibible
splitting of
the TLS.  The reader is directed to
3.4.2 for an explanation.  \\

\subsubsection{ Hamiltonian for TLS} 

{\it (a) Non-interacting Hamiltonian}\\

The atomic degree of freedom may be described as either bosonic or
fermionic
since we are discussing a single particle in motion.  The creation
operators at
the lowest lying energy levels for the two minima are labelled by
$b^{\dag}_1$ and
$b^{\dag}_2$ and obey canonical commutation relations;
e.g., taking the operators to
be fermionic we have $\{b_i,b^{\dag}_j\}=\delta_{ij}$.  Since the
Hilbert
space is restricted, we are free to take these  to be bosonic as well.
The states $|+>=b^{\dag}_1|0>$ and
$|->=b^{\dag}_2|0>$, where $|0>$ is the particle vacuum for the TLS,
 may be regarded as pseudo-spin states, since we restrict our Hilbert
 space
to
these lowest two states.  The importance of states with higher energies
will
be discussed in Sec. 3.4.2. Hence, the most general {\it
noninteracting}
TLS Hamiltonian is given by
$$H^0_{TLS} = {1\over 2} \sum_{i,\alpha,\alpha'} \Delta^i
\sigma^i_{\alpha,\alpha'} |\alpha><\alpha'| = {1\over 2} \sum_i
\Delta^i\sigma^i\leqno(2.1.3)$$
where the $\sigma^i$ are Pauli matrices ($i=x,y,z$).
$\Delta^z$ measures the splitting between the levels in the two wells,
while
$\Delta^x$ and $\Delta^y$ are the {\it spontaneous
tunneling} matrix elements which ``flip'' the
spin with no assistance from other excitations in the system.
If the wave functions of the atom in the two wells are chosen to be
real, then the bare splitting $\Delta^y$ must vanish.  The conventional
notation in the TLS literature is
$$\Delta^z = \Delta,~~ \dx = \Delta_0, ~~\dy=0 ~~.\leqno(2.1.4)$$
[Note that the correspondence to the spin Kondo problem is that
$\Delta^z$
represents a local longitudinal magnetic field, while $\Delta^x$
represents a
local transverse magnetic field.  In the quadrupolar Kondo problem,
$\Delta^z,\Delta^x$ measure local stresses. ]

It is straightforward to diagonalize Eq. (2.1.3) by rotating to a
quantization
direction $\tilde z$ in which one has a diagonal Pauli matrix only, and
one
finds energies $\pm E/2$ with the splitting $E$ given by
$$E = \sqrt{\Delta^2+\Delta_0^2} ~~.\leqno(2.1.5)$$
The magnitude of the tunneling matrix element $\Delta_0$ is
approximately
$\hbar\omega_0 e^{-\lambda}$ as determined by applying the uncertainty
principle to the expression (2.1.2) for the spontaneous quantum
tunneling time.

The use of projection operators in Eq. (2.1.3) automatically ensures we
will
not go outside the subspace of the lowest two levels.
The price we pay is that the
projection operators no longer obey standard commutation relations.
This may
be remedied by use of the Abrikosov pseudo-fermion method [Abrikosov,
1965], in which a
fictitious chemical potential is inserted for the occupancies
$b^{\dag}_ib_i$ and
taken to infinity at the end of all calculations.  The effect of the
projection
is to remove all empty and doubly occupied states at the end of the
calculation.
This trick is particularly
convenient for the scaling analysis (Sec. 3) and non-crossing
approximation (NCA) integral equation
analysis (Sec. 5), and we
shall defer further discussion of the pseudo-particle method to that
point in
the article.  On the other hand, it is convenient when one is
implementing the
path integral approach to retain the Hamiltonian in the spin
representation,
and it is typical to then split it into longitudinal and transverse
terms. \\

{\it (b) Coupling to Electrons}\\

We model the conduction electrons as a free electron gas, with
Hamiltonian
$$H_c = \sum_{\vec k,\sigma} \ek\ccf\caf ~~,\leqno(2.1.6)$$
where $\ccf$ and $\caf$ create and destroy free electrons of wave
vector $\vec
k$ and {\it real (magnetic)}
spin $\sigma$.  We measure the excitation energies $\ek$ from the
chemical potential.  While the real spin $\sigma$ will play no direct
role in
the coupling to the TLS, it is crucial to retain it, for it plays the
role of
the channel index in the mapping of the TLS problem to a two-channel
Kondo
model.

\begin{figure}
\parindent=2.in
\indent{
\epsfxsize=3.in
\epsffile{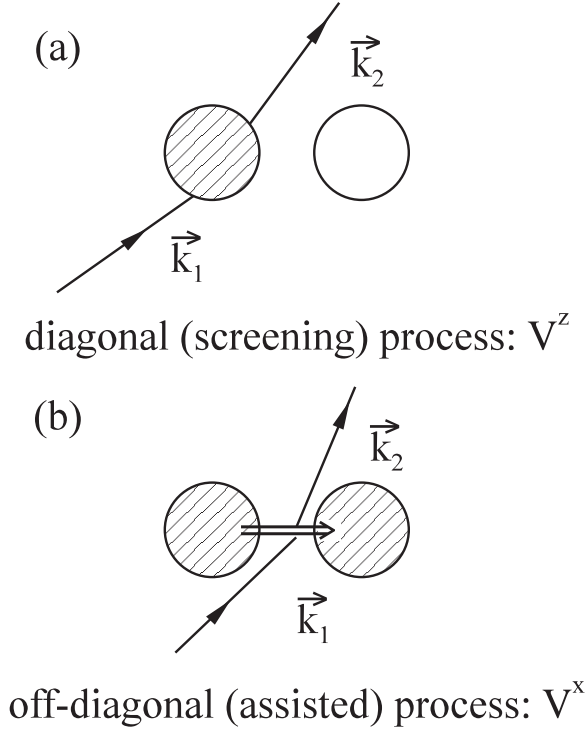}}
\parindent=.5in
\caption{Visualization of screening (a) and assisted tunneling (b) 
of TLS by conduction electrons.}
\label{fig2pbonus}
\end{figure}

\begin{figure}
\parindent=2in
\indent{
\epsfxsize=3in
\epsffile{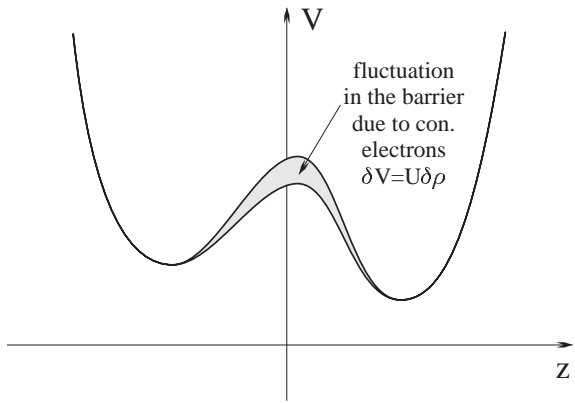}}
\parindent=0.5in
\caption{ Fluctuation in the double-well potential for the TLS is
shown.
These fluctuations are produced by the density fluctuations $\delta
\rho$ in the
conduction-electron band.  The shift of the potential is proportional
to the pseudopotential
$U$ for the electron-atom scattering.}
\label{fig2p3}
\end{figure}

The interaction Hamiltonian coupling
the TLS and conduction electrons has the form
$$H_{int} = {1\over N_s} \sum_{\vec k,\vec k',\sigma}[ V^0_{\vec k,\vec
k'}\ccf\cafp + \sum_{i=x,y,z} V^i_{\vec k,\vec k'}\sigma^i \ccf\cafp ]
\leqno(2.1.7)$$
where we have used the spin representation for the TLS.  The first term
in the
interaction describes the scattering off of the average configuration
of the
TLS, and thus is simply a potential scattering term, of little
importance
unless the total scattering strength is large [Kagan and Prokof'ev,
1989].  The
scattering form factors $V^i_{\vec k,\vec k'}$ represent the following
physical
processes (see Fig.~\ref{fig2pbonus}):\\
(i) The electrons scatter off of the atom sitting either in position
one or two
($V^z$); this process induces {\it screening} of the TLS by the
conduction
electrons.\\
(ii) The electron scattering induces a transition between the two
minima
(denoted by matrix elements $V^{x,y}$ whose physical meaning shall be 
described later--see Fig.~\ref{fig2p3}).  This is what we mean by {\it electron assisted
tunneling}, as
opposed to the matrix element $\Delta_0$ which we call {\it spontaneous
tunneling}.
(Recently, Moustakas and Fisher [1995] have noted that the $V^0$ term
not only includes ``ordinary'' potential scattering, but potential
scattering which transfers an electron from position 1 to position 2.
This term modifies the physics somewhat as we shall discuss further in
Secs. 2.1.2.c. and 3.5.) 

The general properties of the scattering
form factors $V^i_{\vec k,\vec k'}$ are as follows:\\
(i) Since the Hamiltonian is Hermitian, we must have
$$V^i_{\vec k',\vec k} = (V^i_{\vec k,\vec k'})^*  ~~.\leqno(2.1.8)$$
(ii) It is easy to determine the transformation properties under
time reversal because of the absence
of explicit dependence on the real conduction electron spin.  Assuming
we may
describe the TLS states by real wave functions, then time reversal
invariance
implies that
$$<\vec k,\sigma,\alpha|H_{int}|\vec k',\sigma,\alpha'> =
<-k',\sigma,\alpha'|H_{int}|-k,\sigma,\alpha> \leqno(2.1.9)$$
where $|k,\sigma,\alpha>$ is a direct product state of a single
electron with a
state of the TLS.  By comparison with (2.1.7), we see that
$$V^z_{\vec k,\vec k'} = V^z_{-\vec k',-\vec k}~~, \leqno(2.1.10.a)$$
$$V^x_{\vec k,\vec k'} = V^x_{-\vec k',-\vec k}~~, \leqno(2.1. 10.b)$$
and
$$V^y_{\vec k,\vec k'} = -V^y_{-\vec k',-\vec k}~~. \leqno(2.1.10.c)$$
(iii) In order to make explicit estimates of the matrix elements, we
make the
very reasonable assumption that the electrons couple to the TLS only
through
the local electronic density operator $\rho(\vec r)$ given by
$$\rho(\vec r) = {1\over N_s} \sum_{\vec k,\vec k',\sigma} \exp(i(\vec
k'-\vec
k)\cdot \vec r) \ccf\cafp~~. \leqno(2.1.11)$$
This is an obvious coupling mechanism
in the case of the diagonal matrix element $V^z$, since through
Coulombic interactions or the pseudo-potential
the electronic density is directly coupled to the
density of the atom as it sits in either well.
It is also quite reasonable for the off-diagonal (assisted tunneling)
matrix
elements, because fluctuations $\delta\rho$ in the electron density may
modulate
changes of the tunneling barrier height through the coupling of the
atomic and
electronic densities.  Hence, one may envision expanding the Gamow
factor in
powers of the density fluctuations, i.e., we regard $V^i_{\vec k,\vec
k'}$ as a
functional of $\rho(\vec r)$.  Since $\rho(\vec r)$ depends only on the
momentum transfer $\vec k-\vec k'$ for a given particle hole pair, we
have that
$$V^i_{\vec k,\vec k'} = V^i(\vec k - \vec k') ~~. \leqno(2.1. 12)$$
Property (iii) imposes a strong restriction on the TLS-electron
interaction: by
combining Eqs. (2.1.10.c) and (2.1.12) we see that
$$V^y_{\vec k,\vec k'} = 0 ~~.\leqno(2.1.13)$$
We stress that this is a property of the bare couplings; as we
renormalize the
interactions in the scaling theory and renormalization group
calculations
by integrating out virtual electronic excitations, we will generate
couplings
of $V^y$ form because the higher order terms in $\delta \rho$ are not
necessarily local in space [Kondo, 1976].

We are now in a position to explicitly estimate the matrix elements.
The
diagonal couplings $V^{0}and V^{z}$ were first estimated by Kondo
[1976] and
Black,
Gy\"{o}rffy, and J\"{a}ckle [1979].   Denote the wave function for the
atom in
positions 1 or 2 by $\phi_{1,2}(\vec r)$, and assume that the
interaction
between the electronic density and the atom at position $\vec r$ is
given by
$U(\vec r)$.  The potential scattering couples to the average of the
atomic
density over the two wells, while the $V^z$ scattering couples to the
density
difference between the two wells.  Hence
$$V^0_{\vec k,\vec k'} = U(\vec k' - \vec k) \int d^3r \exp[i(\vec
k'-\vec
k)\cdot \vec r] {1\over 2}[\phi^2_1(\vec r) + \phi^2_2(\vec r)]
~~\leqno(2.1.14.a)$$
and
$$V^z_{\vec k,\vec k'} = U(\vec k' - \vec k) \int d^3r \exp[i(\vec
k'-\vec
k)\cdot \vec r] {1\over 2}[\phi^2_1(\vec r) - \phi^2_2(\vec r)]
~~\leqno(2.1.14.b)$$
As mentioned earlier, the matrix element $V^0$ does not significantly
impinge
on the physics unless the scattering phase shift of the conduction
electrons is
near resonance; in
particular, it cannot produce any logarithmic renormalizations in the
scaling
analysis.

Derivation of the assisted tunneling matrix element is more subtle.  We
follow
Kondo [1976], who assumed the non-orthogonalized atomic wave functions
at the
two minima to be identical
apart from displacement factors, i.e.,
$$\phi_{1,2}(\vec r) = \phi(\vec r \mp {\vec d \over 2})
\leqno(2.1.15)$$
where the upper(lower) sign holds for well 1(2).  The wave functions
are
assumed separable in cylindrical coordinates ($\phi(\vec r) = \chi(\vec
r_{perp})\Phi(\theta)\zeta(z)$,
where $\vec r_{perp}$ is the displacement transverse to the TLS axis
and
$\theta$ is the azimuthal angle about the TLS axis along the $z$
direction).
The wave function dependence
along the
axis of the TLS is taken to be the harmonic oscillator form
$$\zeta(z) \approx {1\over \sqrt{\pi z_0}}\exp(-{z^2\over 2z^2_0})
\leqno(2.1.16)$$
with $z_0$ the mean square displacement of the atom about the minimum
due to
zero point motion at frequency $\omega_0$.  Explicitly, assuming the
atomic
mass to be $M$, we have
$$z_0 = \sqrt{{\hbar \over M\omega_0}}  ~~.\leqno(2.1.17)$$
If we consider atoms of intermediate mass $M\approx 50m_p$
($m_p$ the proton mass),
and with typical zero point energies of
the TLS in the regime of 100-500K, we obtain $z_0$ values in the range
of
0.03-0.1$\AA$.

Our evaluation of the coupling $V^x$ borrows from the theory of
inelastic tunneling
formulated by Scalapino and Marcus  [1967].  This derivation has the
advantage
of being physically transparent, and substantially correct
quantitatively, in
that a more formal derivation (using the
Feynman-Hellman theorem based approach of Ngai
{\it et al.} [1967] for the square barrier model and for a
potential with quartic anharmonicity) yields the
same answer to within factors of order unity.
This will be elaborated in Appendix I.
Recently \zar [1993]
has pointed out that the accurate prefactor of the wave function given
by
(2.1.16) is also affected by the barrier fluctuation and it can be
dropped only if the Gamow factor dominates, i.e., if $\exp(-\lambda) <<
1$.
The starting point of this analysis is to introduce the exact
eigenstates of the atom in the double
well potential extending over both minima but without interaction with
the electrons.  The potential
$U_{ion,el}$ describing the interaction between the ion and the
electrons induces
the matrix elements.  This method is suitable to include the higher
atomic interaction
levels in the potential in a straightforward manner (\zar  and
\zow [1994(a,b)])
and the inclusion of those levels may lead to an increase in the Kondo
temperature.
See Sec. 3.4.2 for further details. The method of \zar is also applied
to discusss
the accuracy of the concept in which the treatment starts with the
introduction of
left and right states.

The idea of Scalapino and Marcus is illustrated in Fig. ~\ref{fig2p3}.
The inelastic coupling to the conduction electron density fluctuations
modulates the barrier
height.  Hence, inside the WKB exponent, we may expand the position
dependent
inverse decay length to linear
order in the modulation.  Explicitly, we take
$$V_B(\vec r) = V_B^0(\vec r) + U\delta\rho(\vec r) \leqno(2.1.18)$$
where the coupling between electrons and atom is assumed to have a
local
pseudo-potential form
$$U_{ion,el}(\vec r_{el}-\vec r_{ion}) =v_0 U\delta(\vec
r_{el}-\vec
r_{ion}) \leqno(2.1.19)$$
$v_0$ being the unit cell volume.  The appropriate expression
for
$\delta\rho(\vec r)$ is
$$\delta\rho(\vec r) \approx \rho(\vec r) - {1\over 2} [\rho({\vec
d\over 2}) +
\rho(-{\vec d\over 2})]={1\over N_s}\sum_{\vec k,\vec
k',\sigma}[\exp(i(\vec
k'-\vec k)\cdot\vec r) - \cos((\vec k-\vec k')\cdot{\vec d\over
2})]c^{\dag}_{\vec k,\sigma}c_{\vec k',\sigma} \leqno(2.1.20)$$
which is intuitively understood as the fluctuation relative to the
averaged
density at the minimum of the wells.  This expression will be justified
in
detail in Appendix I.  Because of the assumed axial character of the
TLS, only
the $z$ dependence is relevant in Eq. (2.1.20).  We note that the local
potential approximation
can in principle be dropped at the expense of complicating the
formalism somewhat; see
\vld and \zow [1983(a)] and \zar [1993] for details.

By approximating the tunneling matrix element in the WKB
form we have, for the combined spontaneous and assisted tunneling
$\Delta^x_{total}$
$$\Delta^x_{total} = \Delta^x\exp[-{U\over 2\hbar} \int_{-d/2}^{d/2} dz
{\delta\rho(z) \over V_B^0(z) } \sqrt{2M_{ion}V_B^0(z)}]
\leqno(2.1.21)$$
which we may expand to linear order in the presumed small exponent to
obtain
$$V^x = \Delta^x_{total}-\Delta^x =- \Delta^x {\lambda U\over 2 dV_B}
\int_{-d/2}^{d/2}\delta\rho(z) dz ~~.\leqno(2.1.22)$$
In Eq. (2.1.22) we have made a square barrier approximation, with the
barrier
height $V_B$ and width $d$, so that the Gamow
factor $\lambda=d\sqrt{2MV_B}/\hbar$.  Substituting Eq. (2.1.20) for
$\delta\rho$ and carrying out the resulting integration to lowest order
in
$(k_z'-k_z)d=\Delta k_z d$, we obtain
$$V^x_{\vec k,\vec k'} = \Delta^x {\lambda U\over 24 V_B} (\Delta k_z
d)^2
~~.\leqno(2.1.23)$$
The expansion in powers of $\Delta k_z d$ is justified for the
intermediate
mass atoms given $\Delta k_z \simeq k_F \simeq 1 \AA^{-1}$, and our
presumed value $d\simeq 0.1\AA$.

If we perform the small $\Delta k_z d$ expansion on $V^z$ given by
Eq. (2.1.14.b), we obtain
$$V^z \approx {1\over 2} (\Delta k_z d) U \leqno(2.1.24)$$
so that the ratio $V^x/V^z$ is roughly estimated as
$${V^x \over V^z} \approx \Delta^x {\lambda (k_Fd) \over 12 V_B}
\approx
{e^{-\lambda} (k_F d)\over 3} \leqno(2.1.25)$$
where we have used $\Delta^x \approx \hbar\omega_0 e^{-\lambda}$,
$\lambda
\approx d^2/4z_0^2$ as justified in Appendix I, and $\hbar\omega_0 /V_B
\approx
4z^2_0/(d/2)^2$ as justified in Appendix I.  Taking $\lambda \approx 6$
so that
$e^{-\lambda} \approx 10^{-3}$, we see that the order of magnitude for
$V^x/V^z$ is given by
$${V^x\over V^z} \sim 10^{-4}-10^{-3} ~~.\leqno(2.1.26)$$
Since these will play the role of bare exchange couplings in the
mapping to the
Kondo problem, and since $V^y$=0, we see that the initial coupling is
extremely
anisotropic!  For further details, see \vld  and Zawadowski,
[1983(a)].
A numerical study of the square potential model (\zar, [1993])
provides
further justifications for the estimates presented above.  The
Hamiltonian
with excited state will be given in Sec. 3.4.2.\\

{\it (c)  Angular Momentum Representation of $H_{int}$}\\

The momentum dependence of the couplings $V^{x}~and~ V^{z}$ plays a
crucial role
in the
development of the Kondo effect.  The reason is that through the
momentum
dependence, or the form factors, the non-commutativity associated with
the Kondo scattering as discussed in the introduction will develop.

The coupling can be expressed in partial waves.  Let us first expand
the
coupling in a complete set of orthogonal functions ${f_{\alpha}(\hat
k)}$
depending only upon direction $\hat k$ so that
$$V^i_{\vec k',\vec k} \approx V^i_{\hat k',\hat k} =
\sum_{\alpha',\alpha}
f^*_{\alpha'}(\hat k')V^i_{\alpha',\alpha}f_{\alpha}(\hat k)
\leqno(2.1.27)$$
where we have assumed negligible dependence upon the magnitude of $\vec
k,\vec
k'$.  The most convenient choice for the free electron gas is
$$f_{\alpha}(\hat k) = i^l \sqrt{4\pi} Y_{l,m}(\hat k) \leqno(2.1.28)$$
where $Y_{l,m}$ is a spherical harmonic of indices $l,m$.   Hence, the
matrix
elements of the couplings are given by
$$V^i_{l'm,lm} = {i^{l-l'}\over 4\pi} \int d\hat k d\hat k'
Y^*_{l',m}(\hat k')
V^i_{\hat k',\hat k} Y_{l,m}(\hat k) \leqno(2.1.29)$$
where we have exploited the axial symmetry of the TLS in that the
azimuthal
quantum number $m$ must be conserved in all transitions provided we
choose our
quantization axis along the direction $\hat d$.
Using the previous approximation that the atomic motion is confined to
the
$z$-axis (or, more precisely, that the electronic scattering off of the
atom is $s$-wave), we may neglect all but $m$=0 in the matrix elements
of Eq.
(2.1.29).
This specifies spherical harmonics which are aligned along the
tunneling axis
$\hat d$.  We shall henceforth suppress the $m$ dependence.

Formally, the conduction ``spin'' in this model could have
an arbitrarily large number of components $N_c$.  This is seen from
equation (2.1.29): partial waves with
all orbital angular momentum values (the orbital index is the
conduction
pseudo-spin index) can be
coupled to the impurity pseudo-spin.   In practice, as we
shall discuss in the scaling theory section, such a highly anisotropic
model
will always select out just two of this large number of conduction spin
components as the
temperature is lowered.  Nevertheless, for various perturbative
approaches, it
is worth considering models in which $N_c$ is allowed to be arbitrary.

It is clear from the considerations of the previous subsection that
$V^z\sim (\vec k'-\vec k)\cdot \vec d$ is odd under inversion symmetry
while $V^x\sim [(\vec k-\vec k')\cdot \vec d]^2$ is even.  Hence,
$$V^x_{l',l} = 0 , ~~~l+l'~~odd \leqno(2.1.30.a)$$
and
$$V^z_{l',l} = 0, ~~~l+l'~~even ~~.\leqno(2.1.30.b)$$
In the approximation used in (2.1.19), $U(\vec k'-\vec k)=u_0$. We
shall use
this
and employ the small $\Delta k_z d$
approximation. In view of the discussion of the previous paragraph, we
then
 truncate the expansion at the
level of $s$ and $p$ wave harmonics, yielding
\[ (2.1.31)~~~~~~~~~~~~~~~ V^z_{l',l} \approx \tilde V^z
\left(\begin{array}{cc}
						     0 & 1\\
						    1  & 0
						   \end{array} \right)
						   \]
with $\tilde V^z = k_F du_0/ 2$,
and
\[ (2.1.32)~~~~~~~~~~~~~~~ V^x_{l',l} \approx  -\tilde V^x
					       \left(\begin{array}{cc}
						     1 & 0\\
						    0  &-1
						   \end{array} \right)
						   \]
with $\tilde V^x =- [(k_F d)^2/24] [\lambda \Delta^x u_0/V_B]$, and
where the matrix indices are $l'=0,1;l=0,1$ with $m$=0 as discussed
above (see, for further discussion, \vld and Zawadowski [1983a],
and \zar [1994]).

The matrices in Eqs. (2.1.31,32) have the form of Pauli matrices, but
they are
rotated with respect to the Pauli matrices chosen for the TLS.  We can
remove
this by performing a $\pi/2$ rotation about the $y$ axis of the space
of
conduction partial wave indices, taking the matrix for $V^z$ to
$\sigma^z$, and
the matrix for $V^x$ to $-\sigma^x$.  In so doing, one effects a
$\pi/4$
rotation in the space of orbital indices. Namely, if we denote the unit
vectors
in the restricted two-component orbital space of conduction states as
$\hat
e_0,\hat e_1$ for $s$ and $p$ wave components, respectively, then the
rotated
basis corresponds to
$$\hat e_{\pm} = {1\over \sqrt{2}} [\pm \hat e_0 + \hat e_1]
~~.\leqno(2.1.33)$$

We can now put everything together to make the mapping of the TLS model
with coupling to
conduction electrons to the $S=1/2$ two-channel Kondo model.  Within
the
restricted orbital basis,
$$H_{TLS} = {1\over 2} \sum_{i=x,z} \Delta^i\sigma^i + \sum_{i=x,z}
\tilde V^i
\sigma^i \sigma^i_c(0)  \leqno(2.1.34)$$
with
$$\sigma^i_c(0) = {1\over N_s} \sum_{k,k',\alpha,\alpha',\sigma}
\sigma^i_{\alpha,\alpha'}
c^{\dag}_{k,\alpha,\sigma}c_{k',\alpha',\sigma}~~.
\leqno(2.1.35)$$
where $N_s$ is the number of sites.
Eq. (2.1.34) has the form of a two-channel Kondo model (the real spin
of the
electrons is a spectator to the scattering off the TLS) with the local
$S=1/2$
variable measuring the position of the atom in the TLS double well, and
the
corresponding index for the conduction electrons measuring the angular
momentum.

In the introduction, where we discussed the isotropic two-channel Kondo
model,
we noted that antiferromagnetic coupling was required to produce a
growth of
coupling constants.  In the present highly anisotropic model,
independent of
the signs of $V^z,V^x$ we shall flow to strong coupling, as we will
discuss in
Section 3.3.1.

Note that without the
inclusion of $\tilde V^x$, the model would correspond to an
``Ising-Kondo''
model, which simply has screening effects and a renormalized tunneling
rate due
to the orthogonality catastrophe associated with every spontaneous
tunneling
event (see, e.g.,   Kondo [1976(a)], Black and Gy\"{o}rffy [1979],
Black,
\vld, and Zawadowski [1982], Kagan and Prokof'ev
[1986,1987,1989]).

We also note that the over-compensation of the two-channel Kondo model
discussed in the introduction is not compromised by the various
approximations
used to cast (2.1.34) into a form which has pseudo-spin 1/2 conduction
electrons.  Were we to generalize to arbitary $m$ values and not
truncate the
$l$ expansion at the $s,p$ level, we would simply have conduction
states with
larger effective spins and still have a two-channel model with the real
conduction spin.  In fact, the restriction to two-component conduction
spins is
thoroughly justified by the scaling analysis, which shows that only two
dominant spin components are selected out in the absence of additional
symmetries in the bare Hamiltonian (see Sec. 3.2 and \zar
[1994]).

Hence, on quite general grounds, the TLS undergoing assisted tunneling
maps to
a highly anisotropic two-channel Kondo Hamiltonian in which the
impurity spin
is effectively 1/2.

\subsubsection{Related lattice models with Kondo analogies} 

We will close this subsection with a discussion of two related lattice
models
in which assisted tunneling or hopping processes play a key role.  The
first is
just the generalization of the TLS model to the situation where the
atom or
heavy particle (say, a muon) may move throughout a crystalline
lattice.  The
second is the problem of electrons in two different bands, one heavy,
one
light, in which besides single particle hybridization between the bands
one includes
Coulomb assisted hopping between the bands.  \\

{\it (a) Lattice generalization of TLS with assisted hopping}\\

The TLS model can be easily generalized in the case of a heavy particle
hopping on a lattice.  This could be, for example, either a muon or
proton
diffusing in a crystal where the massive particle jumps only on a
lattice of
interstitial sites (see Kondo [1984(a,b),1985,1986], Zawadowski [1987],
\zim, \vld, and Zawadowski [1987], and Kagan [1992]).

In the metallic environment, the conduction electrons form a degenerate
gas
which couples to the heavy particle.  The non-interacting Hamiltonian
is
$$H_0 = H_h + H_c \leqno(2.1.36)$$
with $H_c$ the free electron Hamiltonian discussed previously and
$$H_h = t\sum_{<\vec R,\vec R'>} [h^{\dag}_{\vec R} h_{\vec R'} + h.c.]
\leqno(2.1.37)$$
with $\vec R,\vec R'$ nearest neighbors on the lattice through which
the heavy
particle moves, and $h^{\dag}_{\vec R}$ creating a heavy particle at
site $\vec
R$.  Eq. (2.1.37) generalizes the non-interacting TLS Hamiltonian for
the
moving heavy particle.  For convenience, we shall assume the heavy
particles
move about on a hyper-cubic lattice of dimension $d$, and we assume the
spin of
the particle has negligible coupling to the conduction electrons so we
suppress
that.

The screening interaction corresponding to $V^z$ in the TLS case is
local and
the corresponding term in the Hamiltonian is
$$H_{int,z} = {V\over N_s}
\sum_{\vec R,\sigma,}\sigma_{\vec k,\vec k'}\exp(i(\vec k'-\vec
k)\cdot\vec R)
 h^{\dag}_{\vec R}h_{\vec R}
c^{\dag}_{\vec k,\sigma} c_{\vec k',\sigma} ~~,\leqno(2.1.38)$$
where $V$ is the strength of the screening coupling.
The simplest generalization of the electron assisted hopping $V^x$ (see 
Fig.~\ref{fig2p3}) in
this
context is when the hopping along a bond depends upon the conduction
electron
density at that bond, i.e., we have an interaction term
$$H_{int,x} = t_a \sum_{<\vec R,\vec R'>}\sum_{\vec k,\vec k',\sigma}
\exp(i(\vec k'-\vec k)\cdot\vec (\vec R + \vec R')/2) h^{\dag}_{\vec
R}h_{\vec
R'} c^{\dag}_{\vec k,\sigma}c_{\vec k',\sigma}  ~~,\leqno(2.1.39)$$
where $t_a$ is the strength of the electron assisted hopping
(Zawadowski [1987]).

The problem described by the Hamiltonian $H_0 + H_{int,z}$ has been
extensively
studied in the literature for hydrogen and muon diffusion in metals.
See,
e.g., Kondo [1985,1986], Kagan [1992].
The physical relevance of the model when $t_a$ is included is not
yet clear.  The model cannot be applied to the heavy fermion systems,
for
example, because they have only extremely weak direct hopping matrix
elements
for the $f$-electrons due to the small size of the $f$-orbitals.
Moreover, the $f$-electron states hybridize with the
conduction states, which are also the electrons that produce the
screening.
Finally, while the neglect of the muon or proton spin coupling is quite
reasonable since no hybridization can occur so that only weak,
ferromagnetic
hyperfine coupling is possible and thus no Kondo effect, it is of
course
disastrous to ignore the internal degrees of freedom of the
$f$-electrons in
the heavy fermion systems.  \\

{\it (b) Occupation dependent hybridization between heavy and light
electrons}\\

The possible modification of the previous models for electronic systems
with
heavy and light electronic bands can be given in the tight binding
formalism.
The light electrons form a broad tight binding conduction band with
large
hopping rate while at certain sites there are heavy orbitals with site
energy
$\epsilon_h$ which have weak
hybridization with the conduction orbitals but no direct overlap with
heavy orbitals
on other sites.  It is assumed there is one heavy site per unit cell
but
possibly more than one light site per unit cell so that the
corresponding
creation operators are indexed by both unit cell vectors $\vec R$,
basis
vectors $\vec \delta$ which may extend into neighboring unit cells, and
that there may be an
internal orbital degeneracy to the heavy band.

\begin{figure}
\parindent=2in
\indent{
\epsfxsize=2.in
\epsffile{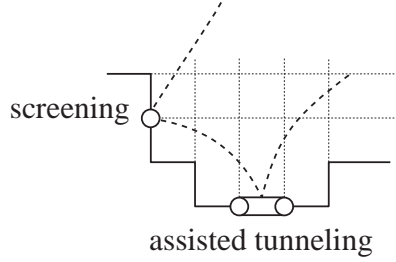}}
\parindent=0.5in
\caption{ The path of a heavy particle moving on a square lattice
is shown by the solid line.  The
electron (dashed line) is either scattered by the heavy particle
sitting on a lattice
position or just moving between two positions.  The positions are
represented by circles
and the moving particle between two positions by a double line.}
\label{fig2p4}
\end{figure}

Two physical systems to which such a model might apply are:\\
{\it (i) Cuprate superconductors}.  Here the light carrier would
be a mobile hole in the bonding oxygen bands residing in the CuO$_2$
planes,
and the heavy carrier to holes in
essentially filled non-bonding $\pi_{x,y}$ orbitals of the apex oxygen
sitting above the
copper sites. (See Fig. ~\ref{fig2p4}.)\\
{\it (i) Heavy Fermion metals}.  In this case, considering for
example UBe$_{13}$, the light carriers would derive from the Be $s-p$
bands,
and the heavy carriers from the U $5f$ orbitals.  In practice, the
model will
be more applicable to models where the physical $f$-level position (and
not the
renormalized or Kondo resonance position) is close to the Fermi level,
which
makes truly mixed valent Sm or Yb ions the most likely candidates. For
further
details we refer the reader
to Zawadowski [1989(a,b,c)], Zawadowski, Penc, and
\zim [1991], and Penc and \zow [1994].

The unperturbed Hamiltonian is given by
$$H_0 = \epsilon_h \sum_{\vec R,\gamma,\sigma} h^{\dag}_{\vec
R,\gamma,\sigma}
h_{\vec R,\gamma,\sigma} + \sum_{\vec R,\vec \delta,\delta'\sigma}
t(\delta,\delta') c^{\dag}_{\vec R,\vec \delta,\sigma}c_{\vec R,\vec
\delta',\sigma} \leqno(2.1.40)$$
$$~~~~~~~ + \sum_{\vec R,\vec \delta,\gamma,\sigma} [
t^{\gamma}(\vec R,\vec
\delta) h^{\dag}_{\vec R,\gamma,\sigma}c_{\vec R,\vec \delta,\sigma}
+h.c.]~~, $$
where $t(\delta,\delta')$ measures the direct light electron hopping
and
$t^{\gamma}_{\vec R,\vec \delta}$ the heavy electron-light electron
hybridization.

We assume that the heavy band is either completely empty or completely
full,
and thus there are only very few excited particles or holes.  We may or
may not
assume a large on-site Coulomb interaction for the heavy band
(large interaction can be dealt with
via slave boson field theory techniques), and we shall discuss each
case in
later sections.

The term which we will add to $H_0$ which is analogous to the assisted
tunneling interaction is an occupancy assisted hybridization term.
Namely,, we assume that the heavy-light hybridization
is modulated when the light orbital on the target site is
simultaneously
occupied with an electron of opposite spin.  The corresponding
interaction is
$$H_{\tilde t} = \sum_{\vec R,\vec \delta,\gamma,\sigma} [\tilde
t^{\gamma}(\vec R,\vec \delta) h^{\dag}_{\vec R,\gamma,\sigma}c_{\vec
R,\delta,\sigma} c^{\dag}_{\vec R,\vec \delta,-\sigma}
c_{\vec R,\vec \delta,-\sigma} + h.c.]~~.\leqno(2.1.41)$$

The coupling $\tilde t^{\gamma}(\vec R,\vec \delta)$ may arise in two
different
ways: \\
(i) We can assume that the radius of the light orbital depends
upon whether the orbital is singly or doubly occupied.  In this case
the
overlap and hopping rate is also modified [Zawadowski,1989].  \\
(ii) Through the off-diagonal matrix elements of the Coulomb
interaction [Hubbard, 1963; Kivelson, {\it et al.}, 1987; D. Baeriswyl,
P.
Horsch, and Maki, 1988; Gammel and Campbell, 1988; Hirsch, 1988] we may
find a
matrix element
$$\tilde t^{\gamma}(\vec R,\vec \delta) = \int d^3rd^3r'
(\phi^{\gamma}_{\vec
R,\sigma}(\vec r))^{*}\phi^*_{\vec R,\vec \delta,-\sigma}(\vec r'){e^2
\over |\vec r-\vec r'|}
\phi_{\vec R,\vec \delta,-\sigma}(\vec r')\phi_{\vec R,\vec
\delta,\sigma}(\vec
r)  \leqno(2.1.42)$$
which clearly has the required form.

As we shall briefly discuss in the next section on scaling theory, when
the
complex character of the $\tilde t$ matrix elements is taken into
account
through momentum dependent form factors which reflect the intra-cell
nature of
the hopping, a Kondo like renormalization of two particle interactions
between
light electrons occurs.  In this sense, there is a strong resemblance
to the
TLS and quadrupolar Kondo effects in which momentum dependent form
factors
drive the non-commutative algebra for interactions between the local
pseudo-spin variables and the conduction electrons.\\
{\it (c) Electric Dipole Kondo Model}\\ Emery and Kivelson [Emery and
Kivelson,
1992(a)] have considered the possibility of the electric dipole Kondo
effect
in the cuprate superconductors.  The physical origin of the dipoles is
through the formation of locally charge segragrated regions as has been
argued to occur in the $t-J$ model at finite doping.  Orbital coupling
of
itinerant electrons or holes to the extended electric dipoles maps onto
the anisotropic two-level system Kondo problem discussed in this
section.
The real carrier spin is not involved in the coupling and so again
serves as a channel index. Because of the electric dipole character
of the effective impurity spin, the current operator is modified in
this model and very different electrical conductivity results may be
obtained in comparison with other physical realizations of the
two-channel Kondo model. We defer a discussion to Sec. 9.3 where a
one-dimensional concentrated system of electric dipole Kondo
centers is discussed [Emery and Kivelson, 1993].

\subsection{Exotic Kondo Models for Rare Earth and Actinide Impurities} 

\subsubsection{  Quadrupolar Kondo Model for \ufp~ ion in cubic
symmetry} 

{\it (a) Physical Discussion and Context}\\

In this subsection, we shall derive the quadrupolar Kondo model for a
uranium
ion in a crystal field of cubic symmetry.

The idea of the quadrupolar Kondo effect was first proposed to explain
the
unusual magnetic field dependence of the heavy fermion superconductor
\ube [Cox, 1987]. Subsequently, Barnes noted that the quadrupolar
moments
of Cu$^{2+}$ ions in the cuprate superconductors could lead to such a
Kondo
effect as well [Barnes, 1988]. Later it was realized that tetravalent U
ions in
hexagonal symmetry with $\gfi$ or $\gsi$ doublets and in tetragonal
symmetry with
$\gfi$ doublets will also be subject to a quadrupolar Kondo effect of
slightly
different character, as we shall discuss in more detail in Sec. 2.2.4
[Cox, 1993].  Finally, the suggestion of Cox [1987b] for \ube~
 met some
understandable skepticism, since in this material the putative
quadrupolar
Kondo sites are distributed on a periodic lattice.  However, the
discovery of
the alloy \yup
[Seaman {\it et al.}, 1991,1992; Liu, {\it et al.}, 1992;
Andraka and Tsvelik, 1991], which for concentrations $x=0.1,0.2$
appears to display many of
the features of the quadrupolar Kondo effect, has considerably
strengthened
 the empirical basis for
this theory.

The motivation for considering the quadrupolar Kondo effect in \ube~
was that in all properties except
magnetoresistance, this material displays an extremely weak magnetic
field
dependence.  However, the heavy fermion character is presumed to derive
from a
Kondo effect.  It was noted that U$^{4+}$ ions with a nominal
configuration of
$5f^2$ have a Hund's rule angular
momentum of $J=4$.  In consequence, when placed on a site of cubic
symmetry, as
in \ube, the action of the crystal field lifting the full multiplet
degeneracy
could produce  a non-Kramers $\gth$ or $E$ doublet ground state, i.e.,
whose
degeneracy is not guaranteed by Kramers' theorem.   Kramers' theorem
states that
for an electronic configuration with an odd number of electrons, a
minimal two fold
degeneracy exists for each level due to time reversal symmetry that
cannot by lifted
by any crystalline
anisotropy (see, e.g., Lax [1974]).  No corresponding statement can be
made for an ion with
an even electron number configuration, though in some lower symmetries
symmetry distinct
singlet states may have enforced degeneracy through time reversal
symmetry (see, for example, Tinkham [1964], p. 147).

\begin{table}
\parindent=2.in
\indent{
\begin{tabular}{|c|c|c|c|c|c|c|}\hline
Configuration & State &  Parent  $J$ &Form & $<J_z>$ &
$<3J_z^2-J(J+1)>$ \\\hline\hline
$f^2$ & $\gth(+)$ & $J=4$ & $\sqrt{{5\over 24}}[|4>+|-4>]-\sqrt{{7\over
12}}|0>$ &
0 & +8 \\\hline
$f^2$ & $\gth(-)$ & $J=4$  &$\sqrt{{1\over 2}}[|2>+|-2>]$ &
0 & -8 \\\hline\hline
$f^1$ & $\gse(\uparrow)$ & $J=5/2$ & $\sqrt{{1\over 6}}
|-5/2>-\sqrt{{5\over
6}}|3/2>$ & +${5\over 6}$  &0 \\\hline
$f^1$ & $\gse(\downarrow)$ & $J=5/2$ & $\sqrt{{1\over 6}}
|5/2>-\sqrt{{5\over
6}}|-3/2>$ & -${5\over 6}$&  0 \\\hline\hline
$c^1$  &$\gei(2)$ & $J=5/2$ & $\sqrt{{5\over 6}} |5/2>+\sqrt{{1\over
6}}|-3/2>$ &
$+{11\over 6}$ & +8 \\\hline
$c^1$ & $\gei(\bar 2)$ & $J=5/2$ & $\sqrt{{5\over 6}}
|-5/2>+\sqrt{{1\over
6}}|3/2>$ &
$-{11\over 6}$&  +8 \\\hline
$c^1$ & $\gei(1)$ & $J=5/2$ & $|1/2>$
&$+{1\over 2}$&  -8 \\\hline
$c^1$ & $\gei(\bar 1)$ & $J=5/2$  &$|-1/2>$ &
$-{1\over 2}$ & -8 \\\hline
\end{tabular}}
\parindent=.5in
\caption{Angular momentum character of states for two-channel
Kondo
models for \ufp and \ctp ions in cubic symmetry. The fourth column
gives
information about the states of definite azimuthal quantum number
values which
are mixed to form the state in cubic symmetry.   The last two columns
give
information about the expectation values of magnetic and quadrupolar
moments
corresponding roughly to the $S^{(3)}$ and $\tau^{(3)}$ operators.}
\label{tab2p1}
\end{table}

 The $\gth$ doublet is not the only possible
ground state in the field of cubic symmetry for the U$^{4+}$ ion,
but will be over about half the crystal field parameter range.
[See Lea, Leask, and Wolf,
1962, for a discussion of crystal field splittings in cubic symmetry.
Some
of the relevant states are listed in Table~\ref{tab2p1}.  Note
that the sign of the $W$ parameter in Lea, Leask, and Wolf
must be negative to realize the
stable
$\gth (E)$ ground state.  While this is excluded for a point charge
model in simple
coordination environments (e.g., octahedral), it is not excluded for
complex
coordinations such as in \ube, or for splittings induced by terms
second order in
the hybridization with conduction states,
the most likely origin of the crystal field splittings in the heavy
fermion
materials.  For a simple discussion of this latter idea, see
Zhang and Levy, [1989].

Hence, since the two levels of the $\gth (E)$ doublet are not connected
by time
reversal, a magnetic field does
not split the level, at least to linear order, and the field dependent
properties will be correspondingly weaker than in a magnetic Kondo
system.  The double degeneracy of the
ground level allows one to treat it as a two-level system, i.e., as a
manifold
with a pseudo-spin of 1/2.  As we shall discuss in more
detail, such a ground state is indeed susceptible to a Kondo effect
when coupled to
conduction electrons.

Any state with internal degrees of freedom must be characterized by a
non-trivial multipole moment other than the simple occupancy or charge
operator.  The
physical meaning of the $\gth (E)$ state is that it has a non-vanishing
quadrupole
moment.  Recall from electrostatics that the quadrupole moment tensor
describes
the potential from an {\it aspherical} distribution of charge which
appears in
the energy through a coupling to the electric field gradient tensor.
Given a
charge distribution $\rho(\vec r)$, the quadrupole tensor $\hat Q_{ij}$
is
defined in cartesian form as
$$\hat Q_{ij} = \int d^3r \rho(\vec r)[3r_ir_j - r^2\delta_{ij}] ~~.
\leqno(2.2.1)$$

\begin{figure}
\parindent=1.4in
\indent{
\epsfxsize=3.in
\epsffile{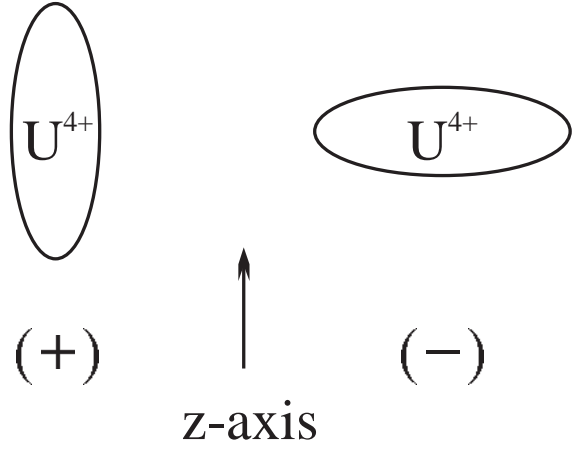}}
\parindent=0.5in
\caption{Mapping of Uranium quadrupole orientations to pseudo-spin variables 
in cubic symmetry.  
The $\gth$ quadrupole moment of the \ufp ion is quantized to two values. 
Choosing the quantization axis along the $z$ axis, the stretched or
prolate configuration maps to ``up'' or ``+'' pseudo-spin, and the 
squashed or oblate configuration maps to ``down'' or ``-'' variables.}
\label{fig2p5}
\end{figure}

Note that:\\
(i) The tensor $\hat Q_{ij}$ is symmetric and traceless, so
that it has only five independent components.  \\
(ii) For full rotational symmetry,
these five
components transform among one another like the elements of the
function
$Y_{2,m}(\hat
r)$. \\
(iii) As the point symmetry is lowered to cubic, the case at hand, the
tensor splits
into a doublet of {\it tensors} (transforming according to a
two-dimensional irreducible representation or
``irrep'' of the point group),
which corresponds to the diagonal elements of Eq.
(2.2.1) (in diad form, $\hat Q_{ii} \sim 3\hat i\hat i-1{\bf I}$, where
${\bf I}$ is the
identity tensor), and a triplet of tensors (transforming according to a
three-
dimensional irrep of the point group, which in diad form
go as $\hat Q_{ij} = \hat Q_{ji} \sim 3\hat i \hat j,~i\ne j$),
 which corresponds to the off-diagonal elements
of Eq. (2.2.1).    The doublet tensor transforms according to the so
called
$\gth (E)$ irrep of the cubic point group, and the triplet tensor
according to the so-called $\gfi (T_2)$ irrep of the cubic point
group.
The reason for only two components in the $\gth (E)$ doublet tensor
is the tracelessness of the quadrupolar tensor, {\it viz.}
$$\hat Q_{xx} +\hat Q_{yy} + \hat Q_{zz} = 0 \leqno(2.2.3)$$
so that only two components are actually independent.  It is convenient
to
write
the two components as the two traceless combinations
$$\hat Q_{+} = \sqrt{3} q_{\gth} [\hat x
\hat x - \hat y \hat y] \leqno(2.2.4.a)$$
and
$$\hat Q_{-} = q_{\gth} [2 \hat z\hat z - \hat x \hat x - \hat y
\hat y] \leqno(2.2.4.b)$$
where
$$q_{\gth} = \int d^3r \rho(\vec r) x^2 = \int d^3r\rho(\vec r)y^2=\int
d^3r
\rho(\vec r) z^2 ~~. \leqno(2.2.5)$$
The equality of the three integrals above follows from the assumed
cubic
symmetry.  Note that the diad forms to the tensors of Eqs. (2.2.4.a,b)
explicitly demonstrate the tracelessness of Eq. (2.2.3).\\
(iv) The tensor pair $\hat Q_{\pm}$ will couple linearly to electric
field
gradients of the same symmetry applied to the crystal.  The most
practical way
of effecting such gradients is through the application of external
stresses
$\hat \eta_{\pm}$ which will produce a coupling term in the energy of
the
charge distribution of the form
$$ E_{stress} = A \sum_{a=\pm}Tr[\hat Q_{a} \hat \eta_{a}].
\leqno(2.2.6)$$
The meaning of a pure stress of the form $\hat \eta_-$ is that the
crystal is
stressed either tensilely (elongated) or compressively (flattened)
along one of
the principal axes, in this case the $z$-axis.  Such a uniaxial
distortion
lowers the crystal symmetry to tetragonal.   This is clear because the
stress will couple linearly to the corresponding strain tensor
$$\hat \epsilon_{-} = 2\epsilon_{zz} - \hat \epsilon_{xx} - \hat
\epsilon_{yy}
=   2 {\partial u_z \over \partial z} - {\partial u_x \over
\partial x} - {\partial u_y \over  \partial y} = 3\epsilon_{zz}
\leqno(2.2.7)$$
where $\vec u(\vec r)$ is the atomic density displacement field at
position $\vec r$, and
where the far RHS of the above equation follows from the assumption of
a pure
stress, i.e., the uniform volume term
 $\epsilon_0=\epsilon_{xx}+\epsilon_{yy}+\epsilon_{zz}=0$.  The meaning
 of a
pure stress of the form $\eta_+$ is that the crystal will distort
orthorhombically since it will linearly couple to
$\epsilon_+=\sqrt{3}[\epsilon_{xx}-\epsilon_{yy}]$
which will elongate(shrink) the x-axis as the y-axis is
shrunk(elongated) while
the $z$-axis is unchanged.  Note that this discussion indicates the
$\gth$
doublet tensor will linearly couple to a strain or vibration field of
the same
symmetry.  A simple, useful visualization appears in Fig. ~\ref{fig2p5}

(v) Turning now to the quantum mechanical situation, we are interested
in the
charge distribution associated with a particular quantum mechanical
angular momentum
multiplet, in which case the quadrupolar
 tensor should be viewed  as a representation of
 operators that act within the multiplet.
The Wigner-Eckart
theorem may be utilized to write this tensor purely in terms of the
angular
momentum operators $\vec J$; we follow the derivation of Slichter for
nuclei
[Slichter, 1989].   Because $\vec r$
transforms as a vector (rank 1 tensor), for an open shell ion with
configuration $f^n$ and  angular momentum $J$ in
free space (full rotational symmetry) we may write
$$\hat Q_{ij} = q_{n\lambda J} [{3\over 2}(J_iJ_j+J_jJ_i) -
J(J+1)\delta_{ij}]~~,  \leqno(2.2.8)$$
where reduced matrix element $q_{n\lambda,J}$ is given by evaluation of
the above
equation in the stretched state $|n,\lambda,J,M_J=J>$; formally
$$q_{n\lambda J} = -e{<n,\lambda,J,M_J=J|\sum_{i=1}^n
[3z_i^2-r_i^2]|n,\lambda,J,M_J=J>\over
J(2J-1)} \leqno(2.2.9)$$
where $\lambda$ denotes the other quantum numbers of the system.  The
sufficiency of
using $i=j=z$ and the stretched state in evaluating the reduced Matrix
element is
guaranteed by the rotational symmetry of free space; such tricks are
customary in
applying the Wigner-Eckart theorem.  In principle,
one should proceed by computing the one-particle matrix element and
then using
angular momentum algebra to compute the effective matrix element within
the
lowest multiplet.  The idea is analogous to the calculation of the
Land\'{e}
g-factor.   $q_{n\lambda
J}$ measures the strength of the coupling to an applied field gradient,
and
thus for the quadrupolar system is analogous to the effective magnetic
moment
of a magnetic multiplet.  Note that the
above formula for $q_{n\lambda,J}$ only applies for $J>1/2$.

Under the reduction to cubic symmetry, we will split the five
dimensional
tensor operator space of  Eq. (2.2.9) into doublet ($\gth (E)$)
and triplet ($\gfi (T_2)$) spaces as in the
classical discussion above.  For the $\gth (E)$ space, the explicit
operator forms are
$$\hat Q_{+} = q_{n\lambda J}\sqrt{3}[J_x^2 - J_y^2] = q_{n\lambda J}
{\sqrt{3}\over 2} [J_+^2 + J_-^2]  \leqno(2.2.11.a)$$
and
$$\hat Q_- = q_{n\lambda J}[3J_z^2 - J(J+1)] ~~.\leqno(2.2.11.b)$$
In Eq. (2.2.11.a), $J_\pm=J_x \pm i J_y$ are the angular momentum
raising
and lowering
operators.  Note that $\hat Q_-$ is diagonal in the $J$ basis, while
$\hat Q_+$
is off diagonal in the $J$ basis.

\begin{figure}
\parindent=2in
\indent{
\epsfxsize=3.in
\epsffile{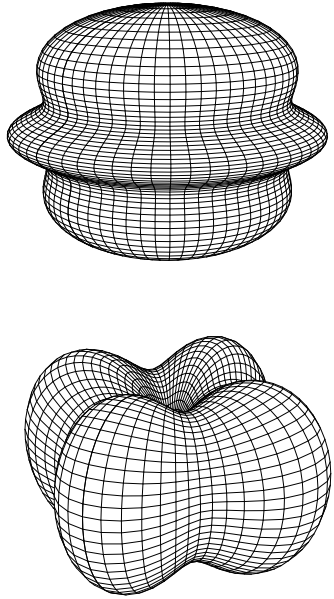}}
\parindent=0.5in
\caption{Actual orbital shapes for $4f$,$J=5/2$,$\gei$ conduction
electrons.  The upper picture is for the elongated ``+'' orbital, 
the lower for the compressed ``-'' orbital.  Each plot is a constant 
probability density contour.  Taken from Kim [1995], and Kim and Cox 
[1997] (with permission).}
\label{fig2p6}
\end{figure}

Let us now connect these ideas to the $\gth (E)$ doublet on the \ufp~
site in
cubic symmetry.  The two states of the doublet are listed in
Table~\ref{tab2p1}.  We see
that under the action of the operator $Q_-$ (apart from the
proportionality
factor of $q_{n\lambda J}$) the $+$ state corresponds to a positive
quadrupolar deformation, the $-$ state to a negative quadrupolar
deformation.
In the plus state, for which $M_J=0$ is the dominant component, we will
measure
an elongation along the chosen quantization axis in the $z$-direction.
For the
$-$ state, with dominant $M_J=\pm 2$ components, the \ufp charge
density will
be squashed along the $z$ direction.  An over-simplified picture is
shown in Fig.
~\ref{fig2p5}), where we view the ion as a ``cigar'' in the $+$ state and a
``pancake'' in
the $-$ state.  In fact, there is considerably more structure to the
charge
distribution as shown in the realistic depiction of Fig. ~\ref{fig2p6}

  At this point, the reader might be confused as to why we don't
have access instead to three  ``cigar''-shaped states along the three
principal
axes.  The reason is that the $\pm$ states transform
like a quadrupolar doublet themselves--while three cigars would be
described by
the 90$^o$ rotation of the $+$ state lined up along the $z$-axis to
instead lie along the $x$- or $y$-axes,
the $x$ and $y$
states are not
in fact orthogonal to the original $+$ state. This statement for the
states is
precisely analogous to the tracelessness condition for the tensors.
 Orthogonalization produces a
linear combination of the $x,y$ ``cigar'' orbitals which is overall
flat.

Examining the forms of the states in Table~\ref{tab2p1}, we see that the $\gth
(\pm)$
states contain states of the original $J$ manifold which differ by two
units of
angular momentum.  Hence these two states will be connected by the
operator
$\hat Q_+$ which raises or lowers angular momentum by two units.

If we now recall Eq. (2.2.6) which describes the classical energy of a
charge
distribution with cubic symmetry under the influence of an external
stress
tensor field of $\gth (E)$ symmetry, we see that the external stress
will have
two possible effects on the $\gth (E)$ ground doublet of the
\ufp ion:\\
(i) If $\hat \eta_-$ is non-zero, we will couple to $\hat Q_-$.  Since
$\hat
Q_-$ linearly splits the $\pm$ states, this is analogous the the effect
of a
longitudinal magnetic field applied to a spin-1/2 state.  This
corresponds to a
uniaxial (tetragonal) distortion.  The $\gth(+)$ state transforms as a
$\gon
(A_1)$ singlet in the tetragonal symmetry, while the $\gth(-)$ state
transforms
as a $\gth (B_1)\sim x^2-y^2$ singlet in the tetragonal symmetry.  \\
(ii) If $\hat \eta_+$ is non-zero, we will be allowed to mix the two
doublet
states.  Recall that this corresponds to an orthorhombic distortion.
Indeed,
as we lower from cubic to orthorhombic symmetry, the $\gth(\pm)$ states
both
transform as $\gon (A_1)$ irreps since now there are no symmetry
transformations of 90$^o$ rotations about the principal axes
to distinguish the two states.

Instead of using the representation of the impurity tensor operators
in terms of the angular momentum,
we may restrict our attention to the operator basis of the $\gth (E)$
doublet
of states directly.  Since the manifold is two dimensional, there are
only four
second rank tensor operators we may form.  These operators are
$$ \tpi = {1\over 2}[\gthpk\gthpb~+~\gthmk\gthmb] ~~,\leqno(2.2.12.a)$$
which describes just the charge (monopole) distribution of the \ufp
ion,
$$\txi = {1\over 2}[\gthpk\gthmb~+~\gthmk\gthpb] ~~,\leqno(2.2.12.b)$$
$$\tzi = {1\over 2}[\gthpk\gthpb~-~\gthmk\gthmb] ~~,\leqno(2.2.12.c)$$
and
$$\tyi = {1\over 2i}[\gthpk\gthmb~-~\gthmk\gthpb] ~~.\leqno(2.2.12.d)$$
The three operators $\txi,\tyi,\tzi$ clearly form a closed algebra in
the
$SU(2)$ space defined by transformations within the manifold of $\gth
(E)$
states.

Clearly, $\tzi$ which is diagonal and of opposing sign for the two
states is
proportional to $\hat Q_-$.  Also, $\txi$ which ``flips'' between the
two
members of the doublet, is clearly proportional to $\hat Q_+$. Hence,
the
$\txi,\tzi$ pair transforms as a doublet tensor of $\gth (E)$ symmetry
which
measures the corresponding $\gth (E)$ symmetry quadrupole tensor of the
ion as
restricted to the two lowest states.  The action of a uniaxial stress
$\hat \eta$
of pure
$\gth (E)$ symmetry thus adds
a term to the Hamiltonian of the quadrupolar doublet which has the form
$$H_{stress} = -\tilde A [\txi \eta_+ + \tzi\eta_-] \leqno(2.2.13)$$
where $\eta_{\pm} = \sqrt{Tr[\hat \eta_{\pm}]^2}/2$.  Thus, the doublet
external stresses $\eta_{\pm}$ for the quadrupolar Kondo model are
analogous
respectively to the spontaneous tunneling $\Delta^x$ and splitting
$\Delta^z$
of the TLS model.
This discussion
is slightly over-simplified: the $\txi,\tzi$ pair of tensors may also
have
components
from all multipole moment tensors of even rank $\le 2J$.

 The operators
$\txi,\tzi,\tpi$  clearly produce only real
matrix elements.  The operator $\tyi$, just like the $\sigma^y$
operator of the
TLS discussion, is complex.  The physical nature of this operator is
made
more clear by considering the commutation relation
$$[\tzi,\txi] = i\tyi \sim [\hat Q_-,\hat Q_+] \sim iJ_xJ_yJ_z
\leqno(2.2.14)$$
as may be readily verified by working out the commutators
$[J_i^2,J_j^2]$ using Eqs. (2.2.11a,b). Thus the operator $\tyi$
transforms as a $\gtw (A_2)$ irrep of the cubic group.
Since this tensor has three $J_i$ operators, it is clearly odd under
time
reversal, and thus represents a magnetic octupole moment tensor.

  The octupole
tensor will couple to third order polynomials in the magnetic field,
or alternatively, to the
combined action of an applied magnetic field and an applied
external stress that lowers the symmetry to rhombohedral .
We may use this latter fact to understand the meaning of
$\tyi$ more deeply.  First, apply a pure rhombic deformation (stress of
$\gfi (T_2)$
symmetry: $\hat \eta = \eta_0(\hat x\hat y+\hat y\hat z+\hat z\hat x)$)
along the body
diagonal.  This lowers the point symmetry of the uranium site
from $O$ to $D_{3d}$.  Now
apply a
magnetic field along the body diagonal. The degeneracy of the $\gth$
doublet is
now lifted.   This lowers the symmetry from $O$ to $C_3$, and breaks
time
reversal ${\cal T}$.  The $\gth$ state transforms into a pair of
singlets under
$C_3$ which are however degenerate in the absence of a ${\cal
T}$-breaking
field; the applied magnetic field then splits these states.
Thus, a term in the Hamiltonian arises of the form
$$H_{str.-field} = -\tilde B \tyi \sum_{ijk} \hat\eta_{\gfi ,ij}
H_k~~(i,j,k~cyclic)
  \leqno(2.2.15)$$
where $H_i$ is the $i$-th component of the applied magnetic field, and
$\hat\eta_{\gfi,ij}$ is the $ij$-th component of the $\gfi$ symmetry
external
stress
tensor.
A magnetic field pointing along a body diagonal will also induce this
coupling
since the field will induce the equal magnetostriction induced strains
tensors
$\epsilon_{\gfi ij} \sim H^2 \hat i\hat j,~~i\ne j$.

It should be mentioned that the subject of crystal field levels in
actinide
intermetallics has been controversial [see, e.g., Ramakrishnan,
1988].   While 5$f$
electrons have larger
spatial extent should in general lead to larger crystal field
splittings, sharp
excitations attributable to crystal fields have been clearly seen only
in a few
uranium based
compounds, most notably UPd$_3$ [Buyers {\it et al.}, 1980] and \urs
[Broholm {\it
et al.}, 1992].  In contrast, relatively sharp crystal field
excitations are
frequently seen in rare earth intermetallics, e.g., in the heavy
fermion
superconductor \cecs [Horn {\it et
al.}, 1981].  This situation will be discussed in somewhat more detail
in Sec.  5.2.3,
where the NCA equations are developed.  Application of this theory
to the
appropriate model for a crystal field split uranium ion makes it clear
that the
excited crystal field level widths should be {\it generically} broader
in the
uranium materials [Cox, 1992a)].  This material shall be discussed in
detail in
the section on experimental manifestations of the two-channel
models.\\

{(b) Coupling to Conduction Electrons}\\

{\it Analogy to TLS model}.
In this subsection we shall develop the Kondo Hamiltonian for the
coupling of
the \ufp~ ions to the conduction states.  We may think of the resulting
coupling
in a manner very similar to the TLS Hamiltonian.  Namely, if the levels
are
polarized virtually so that $\tzi$ would be non-zero, the electrons
will
attempt to relieve this polarization through a screening process
producing a
coupling $J^3$ analogous to $V^z$.
Alternatively, we may view this screening term as
representing the coupling of an electronic charge fluctuation of
uniaxial
(tetragonal) symmetry modulating the levels of the $\gth (E)$ doublet
on the
\ufp~ site.   A fluctuation of the local electron charge density with
orthorhombic symmetry coupled with strength $J^1$
 will modulate the $\gth (E)$ states by producing
transitions between them analogous to the assisted tunneling ($V^x$)
term in
the TLS model.  Unlike the generic TLS model, the couplings $J^{1,3}$
are guaranteed to
be equal due to the cubic point symmetry.   Finally, a local octupolar
distortion of the conduction electron charge density will modulate the
$\gth
(E)$ states by producing a mixing and an orbital magnetic moment.
Since the
$\tyi$ operator transforms as a $\gtw (A_2)$ irrep, nothing
generically guarantees the equality of $J^{1,3}$ with $J^2$, though we
shall
see that a good starting model in fact yields isotropic bare coupling.
In
addition, in the absence of significant disorder or external stress,
there will
be no bare splittings or spontaneous tunneling terms in the quadrupolar
Kondo
model.   Thus,
we expect the quadrupolar Kondo models to typically be nearly isotropic
in the
starting Hamiltonian, and to possess no bare splittings or
``spontaneous
tunneling'' terms.

{\it  Coupling to Conduction Electrons:  Anderson Model}  The direct
Coulombic
couplings of the conduction electrons to the \ufp~ $\gth (E)$ doublet
are likely
to be much smaller than the effective couplings mediated through the
hybridization of uranium 5$f$ states with the conduction band.  Thus
our
derivation of the Kondo model must proceed in two steps here: first, we
write
down the relevant hybridization Hamiltonian which includes the strong
atomic
correlations on the \ufp~ site, and then we transform this model to the
Kondo
form.

In order to develop a Kondo effect, we must couple this quadrupolar
moment to
the conduction electrons.  The appropriate framework for this is the
Anderson
Hamiltonian, in which atomic levels on the uranium site hybridize with
the
conduction states.  The original paper by Anderson was motivated by a
desire to
understand the formation of local moments in solids, and assumed a
single
$s=1/2$ electron present in an $s-wave$ impurity orbital [Anderson,
1961]. In order
to provide
realistical discussion of
rare earth, actinide
and transition metal impurity ions the model has been generalized to
include
orbital degeneracy by a number of workers [Coqblin and Schrieffer,
1969; Hirst,
1970, 1978;
A. Yoshimori [1976]; L. Mih\'{a}ly and A. \zow, [1978]].  A
particularly
helpful review of realistic, orbitally degenerate impurities was given
by
Nozi\`{e}res and Blandin [1980].    In this subsection we shall discuss
the simplest model
which gives rise to the quadrupolar Kondo effect.  In fact, as will be
justified in the sections of the paper discussing scaling theory
(Secs.
3.3, 3.4) and
the
non-crossing approximation (NCA) integral equations (Sec. 5.3),
this mapping is
valid so
long as the $\gth (E)$ level lies lowest in the $f^2$ configuration.

\begin{figure}
\parindent=2in
\indent{
\epsfxsize=3.in
\epsffile{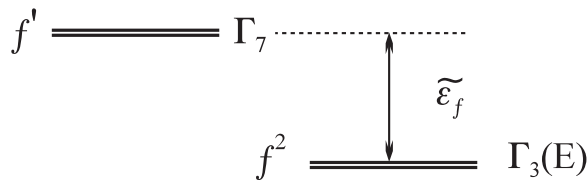}}
\parindent=0.5in
\caption{Simplified level scheme for \ufp ions in cubic symmetry
undergoing quadrupolar Kondo effect.  This simplest model involves a
ground doublet in each of the two lowest lying configurations, with the
$f^2$ having the quadrupolar or non-Kramers $\gth$ doublet, and the 
$f^2$ configuration having the magnetic or Kramers' $\gse$ doublet.  
Conduction electrons mix the two configurations through hybridization
processes.  Only a $\gei$ conduction state can couple these two 
doublets.  Similarly, if $f^3$ rather than $f^2$ is presumed to lie
lowest, the ground doublet of the excited configuration would be 
a $\gsi$ Kramers' doublet, which can mix with the $\gth$ ground 
state of the $f^2$ again only through $\gei$ conduction states. }
\label{fig2p7}
\end{figure}

The picture of the level scheme for the Anderson model relevant for
tetravalent
uranium ions in cubic symmetry appears in Fig. ~\ref{fig2p7}.  The most stable
state is a $\gth (E)$
non-magnetic doublet of
the $f^2$ configuration, and the first excited level is a $\gse$
magnetic
doublet in the $f^1$ configuration.  In fact, it is most likely that
the $f^3$
configuration which has a $\gsi$ magnetic doublet
lies lower in energy, but the qualitative features will not
depend upon this detail.  We index the $\gth (E)$ states by the label
$$\alpha=\pm$$
and the $\gse$ excited states by the label
$$\mu=\uparrow,\downarrow~~.$$

 The promotion of $f^1 \to f^2$ by removing a
conduction electron, or from $f^2 \to f^1$ by emitting a conduction
electron
specifies the symmetry of the conduction states about the impurity
which may
hybridize with the uranium $5f$ states.  Group theoretically, since
$\gth (E)\otimes\gse=\gei$, only the $\gei$ conduction quartet partial
waves may couple to
the impurity through hybridization.  These are derived from conduction
partial
waves with orbital angular momentum $l=3$ spin-orbit coupled to produce
$j=5/2,7/2$ manifolds, with each angular momentum manifold giving a
$\gei$ partial wave quartet.

The projection to the $l=3$ partial waves is implemented in terms of
conduction electron annihilation operators, through the operation
[Krishna-murthy,
Wilkins, and Wilson, 1980a)]
$$c_{k3m\sigma} = \int{d\hat k\over 4\pi} Y^*_{3m}(\hat k) c_{\vec
k,\sigma} ~~,\leqno(2.2.16)$$
and the coupling of these operators to produce $j=5/2$ states is
effected with
standard Clebsch-Gordan technology,
 {\it viz.}
$$c_{k5/2m} = \sum_{\sigma} 2\sigma\sqrt{{3-2\sigma(m-\sigma)\over  7}}
c_{k3,m-\sigma,\sigma}~~. \leqno(2.2.17)$$

The $\gei$ quartet from the $j=5/2$ manifold will be retained for
the discussion at hand; we defer a justification for this restriction
to the
later sections on scaling theory (Sec. 3) and NCA equations (Sec. 5).
By looking at
Table~\ref{tab2p1},
we see that the combinations of states which produce the conduction
$\gei$
creation and annihilation operators are written in the angular momentum
basis
as
$$c_{k8+,\uparrow} = c_{k8,2} = \sqrt{{5\over
6}}c_{k5/2,+
5/2}+\sqrt{{1\over 6}}c_{k5/2,- 3/2}~~, \leqno(2.2.18.a)$$
$$c_{k8+,\downarrow} = c_{k8,\bar 2} = \sqrt{{5\over
6}}c_{k5/2,-
5/2}+\sqrt{{1\over 6}}c_{k5/2,+ 3/2}~~, \leqno(2.2.18.b)$$
$$c_{k8-,\uparrow} = c_{k8,1} = c_{k5/2,+1/2}~~,
\leqno(2.2.18.c)$$
and
$$c_{k8-,\downarrow} = c_{k8,\bar 1} = c_{k5/2,-1/2}~~.
\leqno(2.2.18.d)$$
The $2(\bar 2)$ and $1(\bar 1)$ subscripts in the middle part of the
above equations gives the
correspondence to the notation of Table~\ref{tab2p1} and the original paper [Cox
[1987]].
The upper(lower) signs in the right hand side (RHS) of the above
equations correspond
to $\mu=\uparrow(\downarrow)$
The $\pm$ labels on the left hand side (LHS) correspond to the $\gth
(E)$ index $\alpha$.

The conduction electrons are assumed to reside in a broad band with a
flat
density of states $N(\epsilon)$ parameterized by a width $D$, which for
convenience
is typically taken to have the form
$$N(\epsilon) = {1\over 2D} \theta(D-|\epsilon|) ~~.\leqno(2.2.19) $$

In order to implement this restricted Hilbert space, some appropriate
limits of
the atomic limit parameters must be taken.  Essentially, we must take
all
crystal field, spin orbit, and exchange splittings to infinity.  This
done, we
must take the limit of the direct $f-f$ repulsion $\uff$ to $\infty$ in
a curious way.
 Denote the one-particle energy $\ef$.  To restrict to just the $f^1$
 and $f^2$
configurations, we must have the energy difference
$E(f^2)-E(f^1)=\tilde\ef=\ef+\uff$
remain finite.  Clearly this is achieved if $\ef=-\uff+\tilde\ef$ and
$\uff$ is
taken to $\infty$.  One may always restrict to just two configurations
with
similar tricks.  However, to restrict to three or more configurations
can only
be done by hand and must be viewed as an approximation to the full
Hilbert
space, rather than an exact limiting case of the full model Hamiltonian
with
all configurations included.  We remark that for 4f systems, $|\ef|$ is
of the
order of 2-4 eV, while $\uff$ is of the order of 6-10 eV [Herbst and
Wilkins, 1987; Lang {\it et al.}, 1981].
 In the actinide
systems, the greater spatial extent of the 5f orbitals leads to
correspondingly
smaller values of 1-2 eV for $\ef$ and 3-6 eV for $\uff$ [Actinide
Pars. Reference, :::].

With the above assumptions, the Anderson Hamiltonian in this restricted
Hilbert
space, which has been previously called the ``3-7-8'' model,
 may be written down as
$$H_{378} = \sum_{k\alpha\mu} \ek\ccei\caei + \tilde\ef\sum_{\alpha}
\gthak\gthab + H_{hyb} \leqno(2.2.20) $$
$$ H_{hyb} = - {V\over \sqrt{N_s}} \sum_{k\alpha\mu} sgn(\mu)
[\gthak\gsemmb \caei +
h.c.] \leqno(2.2.21)$$
where $V$ is the hybridization strength and $N_s$ is the number of
atomic unit
cells in the crystal.  We have assumed the impurity to be located at
the origin
and have taken the $f^1$ configuration at the Fermi energy--clearly,
this is
sensible since energy differences between the configurations correspond
to
electron addition and removal energies which must be measured with
respect to
the Fermi energy.  Note the presence of the phase factor $sgn(\mu)$
in the hybridization
term.  This arises because the $\gth$ doublet of the $f^2$ can mix with
two
particle states formed from a conduction electron and a single $f$
electron
which are a singlet in the channel variable $\mu$ and a doublet in the
spin
variable $\alpha$.

The use of the
``Hubbard'' operators such as $\gthak\gsemmb$ in Eq. (2.2.19) is
necessitated by
the Hilbert space
restriction. The advantage of using these operators is that strong
correlations
of atomic character are built in to the bare Hamiltonian, and the
hybridization
which is the smallest energy scale is to be viewed as the
perturbation.
 The technical nuisance of these Hubbard operators is that they don't
 obey
canonical commutation operator identities.  Neveretheless, a direct
perturbation theory may be developed [Keiter and Kimball, 1971;
Inagaki,
1979; Grewe and Keiter, 1981; Grewe, 1983,1984; Coleman, 1983;
Kuramoto, 1983; see
also
Bickers' review, 1987], or one may use pseudo-particle techniques to
map the
perturbation theory on to Feynman diagrams followed by a projection to
the
physical space [Abrikosov, 1965; Barnes, 1976; Coleman, 1983].  We
shall
implement the latter method in the NCA presentation later in the paper
(see
Secs. 3.3.1, 5).

Note that the general process to derive the Anderson Hamiltonian in
the restricted
subspace we have specified above requires explicit assumptions about
the atomic
states.  This has been discussed in general detail by Hirst
[1970,1978],
and in
detail for the problem at hand by Cox[1987b,1988a),1988b),
1991a),1992a)].
Specifically, while the single particle hybridization term may be
written down
independent of angular momentum coupling scheme (Russel-Saunders,
intermediate, or $j-j$) when it is projected to the relevant many body
states
appropriate to the restricted set of configurations, then the strength
of the
matrix element depends upon which coupling scheme is used.  However,
the
symmetry properties are unaffected by the choice of coupling scheme, as
are the
details about the ground state multiplets: $J=4$ is the ground
multiplet in
both extremes for Russell-Saunders {\it and} $j-j$ coupling.  We shall
assume
that one or the other projection scheme has been employed, and
incorporated
into the overall magnitude of the hybridization matrix element.

The hybridization matrix element is more naturally parameterized
through the
``hybridization width'' $\Gamma$, which measures the rate at which a
single
localized $f$-electron would tunnel off of the impurity site and into
the
electronic continuum.  This level width is given by
$$\Gamma = \pi N(0)V^2 ~~.\leqno(2.2.22)$$
Typically, for $f$-electron systems, this matrix element is in the
range of
0.1-0.5 eV, with 4$f$ electrons occupying the lower end of this range
and 5$f$
electrons the upper end of this range.

{\it Mapping to Kondo model:  Schrieffer-Wolff transformation}.
It has long been known that if charge fluctuations are rare, that is,
if $\Gamma/|\tilde\ef|<<1$, then for low energy scales the Anderson
Hamiltonian
may map onto an effective exchange interaction for the lowest lying
degenerate
manifold.  Physically, $\Gamma/|\tilde\ef|$ measures the deviation of
the
$f$-occupancy from 2, in this case, due to virtual fluctuations to the
excited
$f^1$ configuration.  This mapping, under the name of the
Schrieffer-Wolff [Schrieffer and Wolff, 1965; Schrieffer, 1967]
transformation,
for the non-orbitally degenerate
Anderson model, or the Coqblin-Schrieffer transformation [Coqblin and
Schrieffer, 1969], for the orbitally
degenerate case, is by now well known.  It may be implemented either
through second
order perturbation theory, or by applying a canonical transformation
which
eliminates the hybridization through second order.   The strength of
the
effective exchange interaction is proportional to the square of the
hybridization matrix element divided by the inter-configuration energy
splitting.  The most crucial result of the mapping, of course, is that
the
effective exchange interaction is {\it antiferromagnetic} as an example
of the
general ideas of superexchange formulated by Anderson [Anderson,
1950], and
hence the Kondo effect is possible.

There are no subtleties in
applying the canonical transformation to the current model, despite the
fact
that we are interested in an orbital doublet.  The resulting
exchange interaction in the space of the $\gth (E)$ orbital degrees of
freedom
is
$$H_{Quad. Kondo} =  \sum_{k\alpha\mu} \ek\ccei\caei - J\vti \cdot
(\vteiup +
\vteidn) \leqno(2.2.22)$$
where $J=2V^2/\tilde\ef<0$ and, e.g.,
$$\tau^{(i)}_I = \sum_{\alpha,\alpha'} \tau^{(i)}_{\alpha,\alpha'}
\gthak\gthapb
\leqno(2.2.23)$$
describes one of the multipole operators of the $\gth (E)$ doublet,
$\tau^{(i)}_{\alpha,\alpha'}$ being one of the spin 1/2 matrices
$(i=1,2,3)$,
and
$$\tau^{(i)}_{c8\mu}(0) = {1\over N_s} \sum_{k,k',\alpha,\alpha'}
\tau^{(i)}_{\alpha,\alpha'} c^{\dag}_{k8\alpha\mu}c_{k8\alpha'\mu}
\leqno(2.2.24)$$
describes the corresponding multipole operator formed from the
conduction
states.  Note that the exchange interaction of Eq. (2.2.22)
is isotropic, which is indeed expected
from
the 3-7-8 model. In general, the different symmetry properties of
$\tyi$ (which transforms as the  $\gtw(A_2)$ irrep) from
the $\txi,\tzi$ pair which forms a $\gth (E)$ tensor doublet will lead
to {\it
anisotropic} exchange, with $J^y\ne J^x=J^z$.
 Since this exchange anisotropy is irrelevant
about the
non-trivial fixed point, as will be discussed in
subsequent sections, and since in no case is the exchange anisotropy
expected
to be as large as for the TLS case, we are certainly justified in
writing down
isotropic exchange in Eq. (2.2.22).

Koga and Shiba [1995] have studied
a model which includes this exchange anisotropy together with
scattering
between $\gei$ and $\gse$ conduction states (cf. Sec. 4.1 of their
paper).
The latter scattering generates a coupling only to the $\txi,\tzi$
operators
since $\gse\otimes\gei=\gth\oplus\gfo\oplus\gfi$ and only the $\gth$
conduction
tensors couple to the impurity.  The resulting interaction takes the
form
$$H_{ex,78} = \bar J (\txi \tau_{c78x} + \tzi\tau_{c78z})
\leqno(2.2.25)$$
where
$$\tau_{c78x}=\sum_{k,k',\mu}
c^{\dagger}_{k'\gei,-,\mu}c_{k\gse, -\mu} + h.c. \leqno(2.2.26)$$
and
$$\tau_{c78z} = \sum_{k,k',\mu}
c^{\dagger}_{k'\gei,+,\mu}c_{k\gse, -\mu} + h.c.~~. \leqno(2.2.27)$$
They find (using the numerical renormalization group)
 that the low energy physics is still given by Eq. (2.2.22).
Their results will be discussed in more detail in Section 4 of our
paper.

We note here that the model of Eq. (2.2.22) has a counterpart for any
non-Kramers
doublet of \ufp~ ions in tetragonal or hexagonal symmetry.   The
differences are as
follows: (i) first, the bare exchange is fully anisotropic for the
tetragonal case, and
has an Ising anisotropy in the hexagonal case; (ii) in each case, since
the $c$-axis
pseudo-spin of the \ufp~ ion couples linearly to the magnetic field,
there is an unusual
conduction channel spin-impurity pseudo-spin coupling for $c$-axis
spins. This term
appears to be irrelevant in renormalization group calculations.  We
discuss the
origins of the hexagonal and tetragonal models below in Sec. 2.2.3.

\begin{figure}
\parindent=2in
\indent{
\epsfxsize=3.in
\epsffile{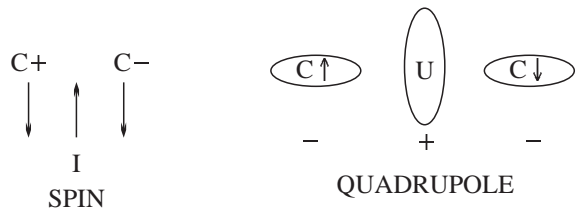}}
\parindent=0.5in
\caption{Mapping of the quadrupolar Kondo Hamiltonian to the two-channel 
Kondo model.  At left is the standard picture of the two-channel model
in spin space--two species of conduction electrons couple anti-parallel
to the impurity spin at the center of the picture.  In the quadrupolar 
Kondo case, ``spin'' is measured by quadrupolar or orbital deformations.
The two-channels arise from the real magnetic spin of the conduction
electrons (the Pauli principle allows a real-spin up electron and a
real-spin down electron in the ``negative pseudo-spin'' squashed 
orbital of the conduction electrons.}
\label{fig2p8}
\end{figure}

The most notable thing about the Hamiltonian in Eq.  (2.2.22) is that
it has the
two-channel Kondo form: two degenerate species of conduction electrons
couple
with identical exchange integrals to the local $S=1/2$ object.  In this
case,
the channel indices are the magnetic indices of
the local conduction partial wave states, and hence the degeneracy is
guaranteed by Kramers' theorem since this is a one electron state.
Fig. ~\ref{fig2p8} 
illustrates the principle
and  makes it clear why this is for very much the same
reason as in the TLS case:  the quadrupole moment, like the atomic
position in
the TLS, is invariant under time reversal.  The pseudo-spin measures
the shape
of the local orbital.  Hence, conduction electrons of opposite magnetic
index
must couple completely equivalently to the local quadrupole moment.  In
the
limit of zero spin-orbit coupling, in fact, the $\gse$ index $\mu$
becomes
the real spin index of the conduction electrons.  The only complexity
in this latter case is that it is impossible to achieve a pure orbital
doublet ground state.
We shall discuss this point in detail in one of the last sections.

{\it Jahn-Teller effect}.
There is often considerable concern about the stability of the uranium
ion
against the Jahn-Teller effect [see for example, Fulde 1978];
ordinarily, an
orbital doublet ground state
will split spontaneously upon inclusion of the linear mixing to local
nuclear
coordinates.  That will certainly happen here as well, but the Kondo
effect
provides a degree of stability against the Jahn-Teller effect in much
the same
way as the spin Kondo effect stabilizes against magnetism:  the
reduction of
the on-site susceptibility by the Kondo effect from a Curie law form
renders
the ion stable against magnetic order below a critical strength to the
intersite coupling of order $T_K$. Since the quadrupolar Kondo effect
is of
two-channel character, the susceptibility still diverges
(logarithmically) but
for a collection of such ions the collective Jahn-Teller instability is
pushed
to far lower temperature scales [Cox, 1987]. Similar conclusions have
been reached about the single site Jahn-Teller effect by Gogolin
[1995].
We shall discuss this point further
in later
sections of the paper.

{\it Use of plane wave basis}.
The reader may be concerned about the use of plane wave states here
when in
many cases a tight-binding basis would be more appropriate.  In this
latter
instance, one must only ensure that the ligand orbitals are not
forbidden by
symmetry from hybridizing with the uranium ion to write down an
Anderson
Hamiltonian.  Even if the hybridization is symmetry forbidden,
multipolar
Coulomb coupling to the uranium ion can drive the quadrupolar Kondo
effect
through highly anisotropic bare couplings in precisely analogous
fashion to the
TLS example discussed in the previous subsection.   Both possibilities
have
been discussed in detail in a tight binding basis using scaling theory
[Deisz
and Cox, 1995], and we shall outline some of these results in the next
section.
The main point is just this:  given a local non-commutative algebra
associated
with the uranium ion, the Kondo effect will ensue, and it {\it does
not}
require two bands degenerate throughout the Brillouin zone.

\subsubsection{ Two-channel Magnetic Kondo effect for a \ctp impurity} 

{\it (a) Physical Discussion and Context}\\

The original work of Nozi\`{e}res and Blandin [1980] focussed on
magnetic Kondo
effects.  The authors in fact concluded that realistic crystalline
field
anisotropy would make it unlikely to ever observe a multichannel
magnetic Kondo
effect: the anisotropies would always tend to make the system flow
eventually
to the ordinary Kondo fixed point with a singlet ground state and Fermi
liquid
excitation spectrum.

The possibility of non-trivial two-channel physics has largely been
overlooked in the
last decade with the advent of expansion techniques based upon large
orbital
degeneracy $N_I$. Noting that Ce had a single $f$ electron with total
degeneracy
of 14 neglecting spin orbit coupling, and 6 including spin orbit
coupling,
Anderson suggested that $1/N_I$ should
serve as an expansion parameter [Anderson, 1981]. Subsequently, a
number of
workers developed approaches to and applications of the $1/N_I$
expansion
[Ramakrishnan and Sur, 1982; Zhang and Lee, 1983; Read
and Newns, 1983; 
Newns and Read, 1987; Coleman, 1983,1984,1987; Rasul and Hewson,
1984a,b; Kuramoto,
1983; Gunnarsson
and Schonhammer, 1983,1984; Bickers, Cox, and Wilkins, 1985,1987;
Auerbach and
Levin, 1986; Millis and Lee, 1987].  Some
efforts were also made to extend these ideas to models for uranium and
thulium
ions [Read {\it et al.} 1986; Nunes {\it et al.} 1986].  An
extensive review of these methods appears in Bickers' article [Bickers,
1987].

In all of the above works, excepting the variational approaches of
Gunnarsson
and Schonhammer and the variational approaches to the uranium and
thulium ions,
the Coulomb repulsion $\uff$ was taken to be infinitely strong.  In all
cases,
a non-degenerate singlet ground state is obtained meaning that the
properties
will be that of a Fermi liquid.

The point of this subsection will be to derive a Hamiltonian for Ce
which
indicates that in many cases, the considerations of Nozi\`{e}res and
Blandin and
the various large $N_I$ efforts are entirely correct. Nevertheless,
the possibility
remains that Ce impurities may display a two-channel magnetic Kondo
effect
which lies outside the domain of $1/N_I$ theories.
We shall see that the conditions allowing this physics are
far more restrictive than for the quadrupolar Kondo effect.  In
particular:\\
(i)
The ground state
weight of fluctuations to the doubly occupied configuration must be
higher than
for fluctuations to the unoccupied configuration.  For this reason, it
is
unlikely that Yb ions, the hole analogue to Ce, will ever exhibit
two-channel
physics.\\
(ii)
Symmetry constrains the model to occur for only Kramers' doublets in
cubic and
hexagonal symmetry, and in the latter case only
one such doublet exists [Cox, 1991,1992a)]

Ce ions are nominally trivalent in the metallic environment, and thus
possess
a single $4f$ electron with Hund's rule ground state angular momentum
$J=5/2$.
The crystalline field will lift this degeneracy; in cubic symmetry, the
ground
state may be either a $\gse$ doublet or $\gei$ quartet [Lea, Leask, and
Wolf, 1962].
Both ground states have
been realized experimentally. For example, in La$_{1-x}$Ce$_x$Al$_2$
[see, for
example, Maple, 1984] and
  La$_{1-x}$Ce$_x$Pb$_3$ [Chen et al., 1987], the ground state is
  $\gse$, while in
La$_{1-x}$Ce$_x$B$_6$, the ground state is $\gei$ [Winzer, 1975].
Unlike the non-Kramers' $\gth (E)$ doublet discussed in the previous
subsection, the
$\gse$ doublet must remain degenerate in the absence of a magnetic
field, i.e.,
its degeneracy cannot be lifted by a Jahn-Teller effect.

{\it (b) Couplings to conduction electrons}\\

{\it Anderson Model for \ctp ion in cubic
 symmetry}.  As in the case of the \ufp ion, we must derive the Kondo
 model by first
developing the appropriate Anderson model which includes the
4$f$-conduction
hybridization and the strong electronic correlations on the $4f$ site.

\begin{figure}
\parindent=2in
\indent{
\epsfxsize=3.in
\epsffile{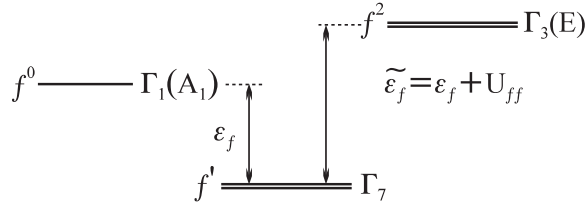}}
\parindent=0.5in
\caption{Simplest model for \ctp ions to produce the two-channel Kondo 
effect.   This requires three configurations, since the $f^0-f^1$ only
model will always produce a single channel Kondo model.  The virtual
charge fluctuations to the $\gth$ doublet of the 
$f^2$ configuration induce a two-channel
coupling of the ground $\gse$ doublet in the $f^1$ configuration to the
$\gei$ conduction electrons.  In that case, the spin index of the 
electrons is their magnetic spin, and the channel index is the
quadrupolar/orbital pseudo-spin of the conduction states.}
\label{fig2p9}
\end{figure}

We shall be concerned with the $\gse$ doublet in this subsection.
The level scheme of the simplest model which can produce the
two-channel Kondo
effect is illustrated in Fig. ~\ref{fig2p9}. In this case,
three configurations must be included, but  in the spirit of
simplicity
the $f^2$ configuration is taken to
have only the $\gth (E)$ ground doublet.  Inclusion of higher levels
can in
principle produce new physics; see the last few paragraphs of this
section
(2.2.2.(b)), Sec. 2.2.4, and Sec. 5.2 for a discussion.

In this simple
model,
in essence, the level stability
of the $f^1$ and $f^2$ have been inverted with respect to the model of
the
previous subsection.   A simplifying feature of this model is that only
$\gse$
symmetry partial waves of the conduction electrons will mix the $f^1$
doublet
to the $f^0$ singlet, and only $\gei$ symmetry partial wave states will
mix the
$f^1$ and $f^2$ doublets.\\

The Anderson Hamiltonian describing this restricted Hilbert space is
much
longer than for the uranium ion of the previous subsection, and is
given by
$$H= H_{c7} + H_{c8} + H_{ion} + H_{771,mix} + H_{378,mix}
\leqno(2.2.28)$$
where
$$H_{c7} = \sum_{k,\mu} \ek \ccse\case ~~, \leqno(2.2.29.a)$$
$$H_{c8} = \sum_{k,\alpha,\mu} \ek \ccei\caei ~~, \leqno(2.2.29.b)$$
$$H_{ion} = \ef\sum_{\mu}\gsemk\gsemb +
(2\ef+\uff)\sum_{\alpha}\gthak\gthab ~~,
 \leqno(2.2.29.c)$$
$$H_{771,mix} = {V_{17}\over \sqrt{N_s}}\sum_{k\mu} (\gsemk \fnob \case
+ h.c.)~~,
 \leqno(2.2.29.d)$$
and
$$H_{378,mix} = {-V_{37} \over
\sqrt{N_s}}\sum_{k,\mu,\alpha}sgn(\mu)(\gthak\gsemmb\caei + h.c.)
~~.\leqno(2.2.29.e)$$
Again, the phase factor which appears in the last equation derives from
the
proper admixture of $f^2$ states with states of one $f$ and one
conduction
electron.  \\

{\it  Schrieffer-Wolff transformation}.
Once again, we apply the Schrieffer-Wolff transformation [Schrieffer
and Wolff,
1965; Schrieffer, 1967; Coqblin and Schrieffer, 1969] to convert the
Anderson
Hamiltonian of Eq. (2.2.29a-e) into a Kondo Hamiltonian.  This
transformation will
produce a valid description of the low energy scale physics provided
that
$ N(0)V_{17}^2/|\ef|=w_0<<1$
and $N(0)V^2_{37}/\tilde\ef=w_2<<1$, with $\tilde\ef=\ef+\uff$ as in
the previous subsection.  The significance of the labels $w_0,w_2$ is
that these are essentially the quantum weights of $f^0,f^2$
configurations in
the ground state due to charge fluctuations.  Provided the inequalities
hold,
we have the effective Kondo Hamiltonian
$$H_{Kondo} = H_{c7} + H_{c8} +J_7\vspi\cdot\vspse +
J_8\vspi \cdot (\vspeip+\vspeim) \leqno(2.2.30)$$
with $J_7=2V^2_7/\ef>0$, $J_8 = 2V^2_8/\tilde\ef>0$ and, e.g.,
$$S^{(i)}_I = \sum_{\mu\mu'} S^{(i)}_{\mu\mu'} \gsemk\gsempb~~,
\leqno(2.2.31)$$
$$S^{(i)}_{c7}(0) = {1\over N_s}\sum_{k,k',\mu,\mu'} S^{(i)}_{\mu,\mu'}
\ccse\casep ~~,\leqno(2.2.32)$$
and
$$S^{(i)}_{c8\alpha}(0) =  {1\over N_s}\sum_{k,k',\mu,\mu'}
S^{(i)}_{\mu,\mu'}
\ccei\caeip~~.\leqno(2.2.33)$$
Here, $S^{(i)}_{\mu,\mu'}$ are spin-1/2 matrices living in the $\gse$
space, with $i=x,y,z$.
The exchange couplings in Eq. (2.2.30) are antiferromagnetic.

The first Kondo term in Eq. (2.2.30) involving $J_7$ is no surprise--it
just reflects the
effective exchange interaction mediated by virtual charge fluctuations
to the
excited empty configuration.   This coupling yields the standard
physics incorporated in the
$\uff\to\infty$ large $N_I$ theories, with $N_I=2$ here.  This term, if
$J_8$ were
set to zero, would produce a singlet ground state and Fermi liquid
excitation
spectrum by the standard methods.  The second term ($J_8$) is of course
more
interesting and has a two-channel character, with the channel indices
here
being the local $\gth (E)$ orbital labels.  (Again, two degenerate
bands throughout
the Brillouin zone are not required--only local degeneracy is necessary
to map
to the two-channel Kondo effect.)   As we shall show explicitly in
later
sections, provided $|J_8|>|J_7|$, which practically corresponds to
$w_2>w_0$,
the two-channel physics will dominate the low temperature behavior.  An
intriguing third possibility exists when $J_8=J_7$: the physics of the
three channel Kondo model will be realized.  While this requires fine
tuning of
the coupling strengths, it is in fact potentially realizable with the
application of pressure, as we shall discuss in detail in the review of
experiments (Sec. 8.2).

It is worth noting at this point that the two-channel behavior is
unlikely for
Yb$^{3+}$ ions because of the requirement $|J_8|>|J_7|$.  Yb$^{3+}$ has
a
single $4f$ hole ($f^{13}$) and excited $f^{14}$ (no hole) and $f^{12}$
(two hole) configurations.  Once the particle-hole mapping is
effected, the physics is completely analogous to the Ce case.  However,
the
details are different.  First, since for Yb$^{3+}$ the Hunds' rules
ground
state has angular momentum $J=7/2$, either $\gsi$ or $\gse$ doublets
may lie
lowest, besides the $\gei$ quartet [Lea, Leask, and Wolff, 1962].  This
is not
of concern, since
$\gsi\otimes\gth (E)=\gei$ as well as if the $\gse$ doublet lies lowest
in energy.
More significant is that previous studies have indicated that
$|\ef|<<|\ef+\uff|$ (these are now {\it hole} energies)
for Yb [Herbst and Wilkins, 1987; Lang {\it et al.},
1981], so that it will be difficult to ever realize $w_2>w_0$.

{\it Effects of excited states in the $f^2$ configuration}\\

There are several consequences which arise from including excited
states in the $f^2$ configuration.   This discussion follows primarily
the work of
Kim [Kim, 1995; Kim and Cox, 1995; Kim and Cox, 1996; Kim and Cox, 1997; Kim, 
Oliveira and Cox, 1996]
and
Koga and Shiba [1995] who applied a similar analysis to a model with
an  $f^3\gsi$ ionic ground state.   The effects are as follows:\\
{\it 1) Enhancement of two-channel coupling}.  The two-channel
antiferromagnetic
coupling
mediated by excited state $\gth$ states is enhanced when one accounts
for the presence of nine such excited states.   However, this can be
reduced
somewhat by the excited state triplets, which contribute negatively to
the
two-channel coupling.  In the work of Koga and Shiba [1995] for the
$f^3\gsi$ ground state, it was found that the net two-channel coupling
arising from the $f^2$ configuration is antiferromagnetic.  \\
{\it 2) Reduction of one-channel coupling}  While excited state
singlets
contribute to an enhancement of the antiferromagnetic one-channel
coupling that is already
present due to $f^0-f^1$ virtual charge fluctuations, the excited
triplet states
suppress this coupling strength.  In their $f^3\gsi$ model,
Koga and
Shiba [1995] find a net ferromagnetic coupling of the $\gsi$
pseudo-spin to the
conduction $\gse$ states.   \\
{\it 3) $\gse-\gei$ exchange coupling}.  The presence of excited state
$\gfo$
triplet states can give rise to exchange interactions which scatter
electrons
between the $\gse$ and $\gei$ partial wave channels.   The terms are
rather
complicated to write down.  \\
{\it 4) Channel-spin/spin coupling in the $\gei$ sector}.  Both excited
state
$\gfo$ and $\gfi$ triplets can mediate a new coupling term in which the
conduction
spin and channel spin can couple.  To understand why such a term
exists, we note
that $\Gamma_8\otimes\Gamma_8$
tensor space of the conduction states contains two $\Gamma_4$ symmetry
tensor
triplets[Kim, 1994;Kim and Cox,1995a,1995b,1996; Kim, Oliveira, and 
Cox, 1996]. The first transforms as a scalar
under
quadrupolar indices and couples to the $\Gamma_7$
doublet with the usual two-channel form.  The second transforms as an
outer product of quadrupolar
pseudo-spin and magnetic pseudo-spin operators.  These operators have
the form
$$\tilde S_{c8}^i (0) = \tilde\tau^i_{c8}(0) S_{c8}^i (0)
\leqno(2.2.34)$$
with $i=x,y,z$, the $S_c^i$ being spin 1/2 magnetic pseudo-spin
matrices
for the $\Gamma_8$
states, and the $\tau^i_{c8}(0)$ are suitable combinations of
quadrupolar
spin 1/2 matrices
given by
$$\tau^z_{c8}(0) = \tau^{(3)}_{c8}(0) \leqno(2.2.35.a)$$
$$\tau^x_{c8}(0) = -{1\over 2}\tau^{(3)}_{c8}(0) + {\sqrt{3}\over 2}
\tau^{(1)}_{c8}(0)\leqno(2.2.35.b)$$
$$\tau^y_{c8}(0) = -{1\over 2}\tau^{(3)}_{c8}(0) - {\sqrt{3}\over 2}
\tau^{(1)}_{c8}(0)~~.\leqno(2.2.35.c)$$
Note that these $\tau$ matrices obey a traceless condition, {\it viz.},
$\sum_i\tau^i_{c8}(0)=0$.
As an example of the tensor product, we write out $\tilde S^z_{c8}(0)$
as
$$\tilde S_{c8}^z(0) = c^{\dagger}_{8\uparrow +}c_{8\uparrow +} -
 c^{\dagger}_{8\uparrow -}c_{8\uparrow -}
-c^{\dagger}_{8\downarrow +}c_{8\downarrow +} +c^{\dagger}_{8\downarrow
-}
c_{8\downarrow -} ~~.
\leqno(2.2.36)$$
For further purposes in Sec. 3, we note that these operators obey the
commutation
relations
$$[\tilde S_{c8}^i(0),\tilde S_{c8}^j(0)] = {-i\over 2} \epsilon_{ijk}
S^k_{c8}(0)
= {-i\over 2} \epsilon_{ijk} \sum_{alpha} S^k_{c8\alpha}(0)
\leqno(2.2.37)$$
and
$$[\tilde S_{c8}^i(0),S_{c8}^j(0)] + [S_{c8}^i(0),\tilde S_{c8}^j(0)] =
-i\epsilon_{ijk} \tilde S^k_{c8}(0)
~~.\leqno(2.2.38)$$
The latter relation, while not immediately evident, follows with
application of
the tracelessness
condition on the $\tau_{c8}^i$ matrices.
The corresponding coupling to the $\Gamma_7$ pseudo-spin has the form
$$\tilde H_{78} =\tilde J_8 \vec S_I\cdot \vec{\tilde S}_{c8}(0)~~.
\leqno(2.2.39)$$
Koga and Shiba [1995] discuss a related model for $f^3$ states.
Indeed, in
Sec. 4.2 of their paper, the exchange Hamiltonian of Eqs. (4.19),
(4.20) has
the form of $H_{78}+ \tilde H_{78}$ provided we neglect exchange
interactions
which scatter between $\gse$ and $\gei$ conduction states.
Specifically, we
replace our $\gse$ spin operators with Koga and Shiba's $\gsi$ spin
operators,
and we can identify $J_7 = -J_0/3 - 19J_1/63$, $J_8 = 2J_0/3 +
49J_1/126$, and
$\tilde J_8 = -10J_1/63$.

 We will argue in Sec. 3.4.3 that provided $\tilde J_8$
is sufficiently small ($|\tilde J_8|<2 J_8$) it is irrelevant and the
low
temperature fixed
point will still be that of the two-channel model when $J_8$ exceeds
$J_7$.  However, interesting new
physics arises when $|\tilde J_8|>J_8$.  Kim [1995] has observed that
the special combination of operators $I_{c8}^{(i)}=-\sum_{\alpha}
S^{(i)}_{c8\alpha}(0)
\pm 2\tilde S^{(i)}_{c8}(0)$ obey the standard $SU(2)$ angular momentum
algebra [Kim, 1995; Kim, Oliveira, and Cox 1996].
Since they span a fourfold degenerate manifold, it is natural to guess
that they serve as spin 3/2 operators.  Indeed, with simple algebraic
manipulations,
one can show the $I^{(i)}$ operators are those of a spin 3/2 manifold.
A
combination of strong coupling perturbation theory,
 weak coupling scaling analysis, NRG calculations, and conformal field
 theory
confirm that for $|\tilde J_8|>J_8$, the model flows to fixed points
governed by
an exchange coupling between the spin 1/2 impurity and spin 3/2
conduction
electrons [Kim, Oliveira, and Cox, 1996].  Kim has observed that the
form of
this Hamiltonian can be obtained by keeping only an excited $\gfo$ or
$\gfi$ triplet
state in the $f^2$ configuration.
While there is only a single channel of conduction spin, the ground
state
is overcompensated because of the large conduction spin.  Hence, a
different
non-Fermi liquid fixed point is possible. We defer a  more complete 
discussion of this issue to Sec. 6.3.2.  

In summary, in this subsection we have demonstrated that when one
includes the
possibility of fluctuations to a realistic $f^2$ configuration which
includes
degenerate levels, a Ce impurity in cubic symmetry may have low
temperature
physics governed by the two-channel Kondo Hamiltonian (or,
alternatively,
a novel $S_c=3/2$ model).  However, to
ensure
this, the fluctuation weight of $f^2$ in the ground state must exceed
that of
$f^0$, so it is by no means the generic case.

\subsubsection{Excited Crystal Field States} 

For both the \ctp and \ufp models discussed above, we have excluded
excited
crystal field levels in the lowest configurations.  In fact, it is
straightforward to include these in the Hamiltonian, and we shall show
in
detail that the low energy scale physics will still map onto the
two-channel
Kondo physics in the appropriate limit when we discuss the NCA approach
to the
problem (see Secs. 6.2,6.3).  The situation is similar to the TLS with
excited
states which is discussed in detail in Sec. 3.4.2.

For the $f^2$ configuration, the additional states in the lowest $J=4$
multiplet are
$|f^2,\gfo,\eta>,|f^2,\gfi,\epsilon>,|f^2,\gon>$.  To the diagonal
$f^2$ terms
of Eqs. (2.2.19) and (2.2.29.c), we must add
$$\sum_{\Gamma_{cef},\eta_{cef}}
(E(\gth) +
\Delta_{\Gamma_{cef}})|f^2,\Gamma_{cef},\eta_{cef}><f^2,\Gamma_{cef},\eta_{cef}|
\leqno(2.2.40)$$
where $\Gamma_{cef},\eta_{cef}$ run over the excited crystal field
levels split
from the ground $\gth (E)$ doublet by amounts $\Delta_{\Gamma_{cef}}$,
and
$E(\gth)$ is the energy of the $\gth$ level ($\tilde\ef$ for the \ufp
model;
$2\ef+\uff$ for the \ctp model). $\eta_{cef}$ indexes the states of any
degenerate multiplets.  The $cef$ subscript is a reminder that these
are states in
the presence of the crystalline electric field (CEF).   For
the $f^1$ configuration, retaining only the $J=5/2$ multiplet, the only
excited
level is a $\gei$ quartet, so to Eqs. (2.2.19) and (2.2.26.c) we must
add
$$\sum_{\eta_8}
(E(\gse)+\Delta_{\gei})|f^1,\gei,\eta_8><f^1,\gei,\eta_8|
\leqno(2.2.41)$$
where $E(\gse)$  is the energy of the $\gse$ level (0 for the \ufp
model; $\ef$
for the \ctp model).  In the above two equations, the $\eta$ labels run
over
internal states of degenerate crystal field manifolds.

We must also generalize the hybridization term.  Focussing only on the
term
which mixes $f^1,f^2$ configurations, the most general form is
$$H_{hyb} = {V \over \sqrt{N_s}}
\sum_{\Gamma(f^1),\eta(f^1)}\sum_{\Gamma(f^2),\eta(f^2)}\sum_{k,\Gamma_c,\eta_c}
\Lambda(\Gamma(f^1)\eta(f^1);\Gamma(f^2)\eta(f^2);\Gamma_c\eta_c)
\times $$
$$
[|f^2,\Gamma(f^2),\eta(f^2)><f^1,\Gamma(f^1),\eta(f^1)|c_{k\Gamma_c,\eta_c}
+
h.c.] \leqno(2.2.42)$$
where the sums run over all states of the lowest $f^{1,2}$ multiplets
($\Gamma(f^{1,2}),\eta(f^{1,2})$)  and all
conduction partial waves ($\Gamma_c,\eta_c$), and
$\Lambda(\Gamma(f^1),\eta(f^1);\Gamma(f^2),\eta(f^2);\Gamma_c,\eta_c) $
contains the Clebsch-Gordan coefficient for the cubic irreps and the
reduced matrix element measuring the strength by which the $f^1$ may
attach to
the $f^2,J=4$ multiplet.  This has been worked out extensively in the
case of
Russell-Saunders ($LS$) coupling of angular momentum, and the detailed
presentation appears in Kim and Cox [1996].

Now, each crystal field level must be treated on equivalent dynamical
footing,
which means when we introduce Green's functions for the local levels,
we must
introduce a new Green's function for each excited crystal field level.
We shall discuss in Sec. 3.4.2 how
the excited states of the TLS corresponding to higher
vibrational levels play a similar role to the crystal field levels
here.

\subsubsection{ Group theory of two-channel Kondo models for Ce and U
impurities} 

The purpose of this section is to
state five selection rules which contain
minimal necessary conditions requisite to have the low energy scale
physics
for a single \ufp or \ctp ion to be
described by a two-channel Kondo model.  We shall first produce a
physical
motivation for the selection rules, and then summarize the selection
rules.  This section is
adopted from Cox [1993].

\begin{figure}
\parindent=2in
\indent{
\epsfxsize=3.in
\epsffile{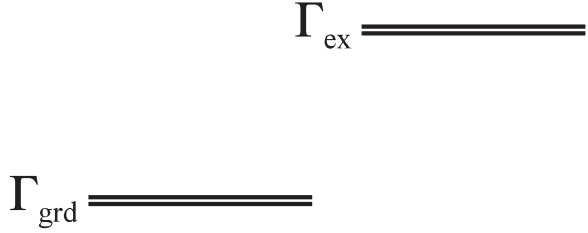}}
\parindent=0.5in
\caption{Simplest level scheme to produce a two-channel Kondo effect.  
The two-channel Kondo effect requires minimally that one have low 
lying configurations with doublets in each.  These doublets may be 
magnetic (Kramers' doublet) or non-magnetic (non-Kramers' doublet)
depending upon whether the configuration has an odd or even number of
electrons.}
\label{fig2p10}
\end{figure}

Fig. ~\ref{fig2p10}  displays the basic picture of the states of an impurity
lanthanide/actinide
ion  and conduction electrons
which are minimally required to achieve a two-channel Kondo model
description of
low energy scale physics.  This restriction to two configurations
is sufficient for U$^{4+}$(5f$^2$) because the two lowest excited
configurations
must have odd numbers of electrons and therefore have
at least doubly degenerate crystal field states. For definiteness, we
assume the
first excited configuration is $f^1$.
For Ce$^{3+}$, we need to augment this
picture to include three configurations as shown in Fig. ~\ref{fig2p7} 
because the excited $f^0$ state must be a singlet.

Regardless, we see
that two configurations of the impurity ion have doublets as the lowest
crystal
field states.  These states
span vector spaces which transform under the irreps
$\Gamma_{grd},\Gamma_{ex}$
of the group $\bar G\times {\cal T}$
where $\bar G$ is the double point group of the crystal and ${\cal T}$
is the group of time reversal (containing the identity and time
reversal operators).  The states in
the $\Gamma_{grd,ex}$ spaces have labels $\alpha_{grd,ex}$.
The extension to include ${\cal T}$ covers groups such as
$C_6$ under which certain pairs of irreducible representations are
complex singlets whose degeneracy is not assured by $\bar G$
but is assured by ${\cal T}$.  The double group is required
because one of the
hybridizing configurations will always contain an odd number of
electrons.
The subscripts
$grd,ex$ refer to ground and excited configurations.

We express the conduction operator  $c^{\dag}_{\vec k,\sigma}$ which
creates a Bloch state of momentum $\vec k$, spin $\sigma$, and energy
$\epsilon$
in a symmetry adapted basis around the impurity, i.e.,
$$c^{\dag}_{\vec k,\sigma} = \sum_{\Gamma_c,\alpha_c}
a_{\Gamma_c,\alpha_c}(\epsilon_k)c^{\dag}_{k\Gamma_c\alpha_c}
\leqno(2.2.43)$$
where $\Gamma_c,\alpha_c$ label irreps of $\bar G\times {\cal T}$
where the point group $\bar G$
is defined at the impurity site.
The local conduction states are
derived from partial waves in a plane-wave basis, or from
suitable linear combinations of ligand orbitals in a tight-binding
basis.

The Anderson Hamiltonian for U$^{4+}$ ions then takes the form
$$H = H_{cond} + H_{grd} + H_{ex} + H_{mix} \leqno(2.2.34)$$
with
$$H_{cond} = \sum_{k\Gamma_c\alpha_c}\epsilon_k
c^{\dag}_{k\Gamma_c\alpha_c}c_{k\Gamma_c\alpha_c} \leqno(2.2.45)$$
$$H_{grd} = E(f^2) \sum_{\alpha_{grd}} |\Gamma_{grd}\alpha_{grd}>
<\Gamma_{grd}\alpha_{grd}| \leqno(2.2.46)$$
$$H_{ex} =  E(f^1) \sum_{\alpha_{ex}} |\Gamma_{ex}\alpha_{ex}>
<\Gamma_{ex}\alpha_{ex}| \leqno(2.2.47)$$
and
$$H_{hyb} = {1\over
\sqrt{N_s}}\sum_{k}\sum_{\Gamma_c\alpha_c}\sum_{\alpha_{grd},\alpha_{ex}}
V[\Gamma_c\alpha_c;\alpha_{grd};\alpha_{ex}]
[|\Gamma_{grd}\alpha_{grd}>
<\Gamma_{ex}\alpha_{ex}|c_{k\Gamma_c\alpha_c} +h.c.]
\leqno(2.2.48)$$
where $V[\Gamma_c\alpha_c;\alpha_{grd};\alpha_{ex}]$ includes
the single particle matrix element, a reduced matrix element expressing
the
attachment probability for adding an $f$ electron to get this $f^2$
state
from this $f^1$ state, and a Clebsch-Gordan coefficient in the
crystal field representation basis.
Again, the restriction to $f^1$ for the excited configuration
is purely  a matter of convenience for the exposition purposes here.
In fact
it is sometimes more realistic to consider an excited $f^3$ configuration.

For the Ce$^{3+}$ ions, we interchange ground and
excited levels and (with $E(f^0)$ set at zero in this case) add
the hybridization
term
$${V_0 \over \sqrt{N_s}}\sum_{\epsilon\alpha_{grd}}
[|\Gamma_{grd}\alpha_{grd}><f^0| c_{\epsilon\Gamma_{grd}\alpha_{grd}}
+ h.c.]~~.\leqno(2.2.49)$$

In addition to discussing the symmetry properties of the states
themselves,
it is important to discuss the symmetry properties of the ground
configuration tensor operators which
live in the product space transforming according to
$\Gamma_{grd}^{ket}\otimes\Gamma_{grd}^{bra}$. The superscripts are a
reminder
that the tensors are formed from outer products of the states.  The
form of the
low energy scale interactions which will correspond to the
two-channel coupling of Eq. (2.2.22) are entirely specified by the
symmetry properties of these tensors.
The interactions arise when we integrate out
the virtual charge fluctuations to the excited configuration to
derive an effective interaction between the conduction electrons and
the
ground configuration degrees of freedom of magnitude $\sim V^2/\Delta
E$, where
$\Delta E$ is the interconfiguration energy
difference (Schrieffer and Wolff, [1966]).
(Note:  all notation for point group
representations used in
this paper follow those of Koster {\it et al.} [1963].)

We now state the {\it necessary} (and not sufficient)
selection rules which will minimally
ensure that the effective Hamiltonian
at low energy scales derived from an underlying Anderson
Hamiltonian has the two-channel $S=1/2$ Kondo form:\\
\begin{quote}{\it Selection Rule 1 (Ground Doublet Selection Rule)}:
Under the action of the crystal field, the
lowest state of the lowest angular momentum multiplet of
the ground configuration
should be a degenerate doublet which transforms as the
irrep
$\Gamma_{grd}$ of the group $\bar G\times {\cal T}$.
\end{quote}
\begin{quote}{\it Selection Rule 2 (Excited Doublet Selection Rule)}:
Under
the action of the crystal field, the lowest state of the lowest
lying angular momentum of the excited configuration must
be a degenerate doublet transforming as a representation
$\Gamma_{ex}$ of the group
$\bar G\times{\cal T}$.
\end{quote}
\begin{quote} {\it Selection Rule 3 (Hybridization Selection Rule):}
The conduction band must contain states which, when projected to the
impurity site, transform as the direct product representation
$\Gamma_c$ such that
$\Gamma_c=\Gamma_{grd}\otimes\Gamma_{ex}$.
\end{quote}
\begin{quote} {\it Corollary Selection Rule 4 (Tensor Selection
Rule):}  When
$\Gamma_c$ is a reducible representation of the group $\bar G\times
{\cal
T}$ of the form
$\Gamma_c=\Gamma_{c1}\oplus\Gamma_{c2}$ where $\Gamma_{c1,2}$ are
irreducible doublet representations of $\bar G\times{\cal T}$, then
the tensors $\Gamma_{grd}^{ket}\otimes\Gamma_{grd}^{bra}$ which are off
diagonal
in the $\Gamma_{grd}$ space must be contained in the space of the
product representation
$\Gamma_{c1}\otimes\Gamma_{c2}$ and not in $(\Gamma_{c1}\otimes
\Gamma_{c1})\oplus(\Gamma_{c2}\otimes\Gamma_{c2})$.
\end{quote}
\begin{quote}{\it Selection Rule 5 (Dynamic selection rule):}
When $\Gamma_{grd}$ is the lowest level of an odd number
electron configuration,
the exchange coupling generated by virtual
excitations to $\Gamma_{ex}$ must be larger than the coupling induced
by virtual charge fluctuations to excited singlet states.\end{quote}

Selection Rule 1 ensures that the lowest impurity states
have internal degrees of freedom so that
a Kondo effect is possible.
Selection Rules 2,3 are necessary for two-channel behavior in
conjunction with Rule 1.  The excited doublet state labels {\it are}
the channel indices, and if
the conduction band doesn't have local
$\Gamma_c=\Gamma_{grd}\otimes\Gamma_{ex}$ symmetry states,
no Kondo effect is possible.

The basis for Corollary Selection Rule 4 is an examination of the
tensor operator structure.  This rule is irrelevant for cubic
structure because the
irreducible
$\Gamma_8$ representation is the only quartet of conduction
states allowed for the $\Gamma_3$
and $\Gamma_{6,7}$ doublets of the different
configurations.  It is essential for the lower symmetry
crystal syngonies.  To see why, note that we
form the exchange term of Eq. (1) by coupling tensors
of the impurity states $\Gamma_{grd}$ to those with the
same symmetry derived from
the conduction states $\tilde\Gamma_c$.  In the lower symmetry
syngonies
of interest (hexagonal, tetragonal, rhombohedral) $\tilde\Gamma_c$ is
reducible, decomposing into $\Gamma_{c1}\oplus\Gamma_{c2}$.  The
magnitude
and antiferromagnetic sign are set by the integration
out of virtual fluctuations to the $\Gamma_{ex}$ states; hence, as in
the
conventional Anderson impurity, the exchange $J<0$ always.
Now, the tensor operators forming the basis for
$\Gamma_{grd}^{ket}\otimes\Gamma_{grd}^{bra}$ include two which
are diagonal in the
$\Gamma_{grd}$ indices and two which are
off-diagonal corresponding to $S_I^{(1,2)}$ in the pseudo-spin 1/2
space.
The identity operator in the $\Gamma_{grd}$ space gives
charge scattering, while the other diagonal term
corresponds to the $S^{(3)}_I$ component of the impurity
pseudo-spin.  This is always contained in both the
tensor spaces $\Gamma_{c1}^{ket}\otimes\Gamma_{c1}^{bra}$
and $\Gamma_{c2}^{ket}\otimes\Gamma_{c2}^{bra}$.   If the off-diagonal
impurity operators were contained in these tensor product spaces, then
no symmetry conditions would ensure the exact equality of exchange
coupling
constants between the channels (now indexed simply by the irreducible
representation labels $\Gamma_{c1,2}$). This is {\it always} the case
in
rhombohedral
symmetry, so that two-channel coupling will not be generically present
in this case for $f$-ions.  However, if the off-diagonal operators are
contained in the mixed-direct product
$\Gamma_{c1}^{ket}\otimes\Gamma_{c2}^{bra}$, then
the ``spin-flip'' conduction tensors must mix states
of the two doublets, and the channel degeneracy is automatically
ensured.

With regard to the conduction tensor operators, we note that only the
case
of the Ce$^{3+}$ ion in cubic symmetry gives rise to the additional
pseudo-spin
tensor, such as appears in Eq. (2.2.36).  For all other cases, the
relevant pseudo-spin
tensors of the conduction electrons appear only once.

Strictly speaking, Corollary Section Rule 4 is redundant given rules
1,2, and 3 in that
for the assumed ground doublets of the lowest configurations and
appropriate
conduction
hybridization, the presence of the appropriate conduction tensors is
guaranteed (Han [1995]).  However,
the physical importance of this rule leads us to state it precisely
here as a corollary
to these first three rules.

\begin{table}
\begin{tabular}{|c|c|c|c|c|c|}\hline
Ion & Ground  &Point & $\Gamma_{grd}$  &$\Gamma_{ex}$ & Conduction \\
& Config. & Group &&&   Quartet \\\hline\hline
U$^{4+}$ & $5f^2(J=4)$ & Cubic($O$) & $\gth (E)$&  $\gse$ &
$\gei=\gth\otimes\gse$\\\hline
\ufp&  \ucon & Hexag.($D_6$) & $\gfi (E_1)$ & $\gse $ &
$\gse\oplus\gni$ \\
&&&&$\gei$ & $\gei\oplus\gni$\\
&&&&$\gni$ & $\gse\oplus\gei$\\
&&&$\gsi (E_2)$&  $\gse$ & $\gei\oplus\gni$\\
&&&&$\gei$&  $\gse\oplus\gni$ \\
&&&&$\gni$&  $\gse\oplus\gei$ \\\hline
\ufp & \ucon  &Tetrag.($D_4$)&  $\gfi (E)$ & $\gsi$ or $\gse$ &
$\gse\oplus\gsi$\\\hline
\ufp & \ucon  &Tetrag.($D_4$)&  $\gon (A_1)\oplus\gth (B_1)$ & $\gse$&
$\gse\otimes(\gon\oplus\gth)$ \\\hline
\ctp &\ccon & Cubic($O$) & $\gse$ & $\gth$ & $\gei$ \\\hline
\ctp & \ccon & Hexag($D_6$) & $\gni$ & $\gfi$ or $\gsi$ &
$\gsi\oplus\gse$\\\hline
\end{tabular}
\caption{ Symmetries of ground, excited, and conduction states
for
two-channel Kondo models of \ufp and \ctp ions.  The ionic
configuration is
listed in the second column, with the Hunds' rules ground angular
momentum in
parentheses.  The crystal point symmetry is in the third column; though
we
choose the most symmetric group from each syngony, equivalent results
are found
for all smaller point groups in the given syngony.  The fourth column
lists the
ground doublet $\Gamma_{grd}$, the fifth column the doublet of the
excited
configuration in the simplest model $\Gamma_{ex}$, and the last column
the
symmetry of the conduction quartet which mixes the levels and screens
the moment
of the ground doublet.}
\label{tab2p2}
\end{table}

\begin{table}
\begin{tabular}{|c|c|c|c|c|c|}\hline
Ion &  Point Group & $\Gamma_{grd}$ &(1)&(2)&(3)  \\\hline\hline
\ufp &&& $\txi$&  $\tyi$ & $\tzi$  \\\hline\hline
& $O$ & $\gth (E)$ & $\gth (+)[Y_{2,2}+Y_{2,-2}]$ &
$\gtw[Y_{3,2}+Y_{3,-2}]$ & $\gth (-)[Y_{2,0}]$ \\
&&& $J_x^2-J_y^2$ & $J_xJ_yJ_z$ &$3J_z^2-J^2$\\\hline
& $D_6$ & $\gfi$or$\gsi$ & $\gsi(1)[Y_{2,-2}+Y_{2,2}]$ &
$\gsi(2)[Y_{2,2}-Y_{2,-2}]$ & $\gtw[\alpha Y_{3,0}+\beta Y_{1,0}]$ \\
&&& $J_x^2-J_y^2$ & $J_xJ_y+J_yJ_x$ & $\rho J_z^3 + \gamma J_z
$\\\hline
& $D_4$ & $\gfo$ & $\gth[Y_{2,-2}+Y_{2,2}]$ &
$\gfo[Y_{2,2}-Y_{2,-2}]$ & $\gtw[\alpha Y_{3,0}+\beta Y_{1,0}]$ \\
&&& $J_x^2-J_y^2$&  $J_xJ_y+J_yJ_x$ &  $\rho J_z^3+\gamma J_z$ \\\hline
& $D_4$  &$\gon\oplus\gth$&  $\gth[Y_{2,2}+Y_{2,-2}]$ &
$\gth[Y_{3,2}+Y_{3,-2}]$ & $\gon [Y_{2,0}]$\\
&&& $J_x^2-J_y^2$ & $J_xJ_yJ_z$ & $3J_z^2-J^2$\\\hline\hline
\ctp &&& $S^{(1)}$ & $S^{(2)}$ & $S^{(3)}$ \\\hline\hline
&  $O$ & $\gse$ & $\gfo(1)[Y_{1,1}-Y_{1,-1}]$ & $\gfo(2)[Y_{1,1}+Y_{1,-1}]$& $\gfo(3)[Y_{1,0}]$\\
&&&$J_x$ & $J_y$ & $J_z$ \\\hline
& $D_6$ & $\gni$ & $\gth[Y_{3,3}-Y_{3,-3}]$ & $\gfo[Y_{3,-3}+Y_{3,3}]$& $\gtw[Y_{1,0}]$\\
&&& $J_y^3-3J_yJ_x^2$ & $J_x^3 - 3J_xJ_y^3 $ & $J_z$ \\\hline
\end{tabular}
\caption{Tensors of the local pseudo-spin for model \ufp and
\ctp
impurities.  The operators indices correspond to the $1,2,3$ labels of
the
pseudo-spin operators.  Their symmetry label for the appropriate group
is the
first label, and in square braces appears the ``multipole content'',
i.e., which
set of spherical tensor operators $Y_{l,m}(\vec J)$ regarded as
polynomials in
the angular momentum corresponds to dominant term in the crystalline
tensor.
For the hexagonal and tetragonal cases involving $\gfi,\gsi$ (hex.) and
$\gfo$
(tet.) doublets, the $\tau^{(3)}$ operator is predominantly octupolar
($l=3$), with a
weak dipolar admixture, so that $\beta >> \alpha$.  Beneath each tensor
symmetry, we display the corresponding cartesian form in polynomials of
the
angular momentum operator.}
\label{tab2p3}
\end{table}

\begin{table}
\parindent=2.in
\indent{
\begin{tabular}{|c|c|c|ccc|}\hline
Ion & Pt. Group  &Conduction &&  Conduction & \\
&& Quartet&&   Tensors & \\\hline\hline
\ufp &&& $\tcxi$ & $\tcyi$ & $\tczi$ \\\hline\hline
 & $O$&  $\gei=\gse\otimes\gth$ &
  $\gon^{7\times7}\otimes\gth^{3\times3}(+)$ &
  $\gon^{7\times7}\otimes\gtw^{3\times3}$ & $\gon^{7\times
  7}\otimes\gth^{3\times
  3}(-)$ \\\hline
  &  $D_6$ &$\gse\oplus\gei$ & $\gsi^{7\times8}(1)+h.c.$ &
  $\gsi^{7\times8}(2)+h.c. $ & $\gtw^{7\times 7}+\gtw^{8\times
  8}$ \\
  &   & $\gse\oplus\gni$ & $\gsi^{7\times9}(1)+h.c.$ &
  $\gsi^{7\times9}(2)+h.c. $ & $\gtw^{7\times 7}+\gtw^{9\times
  9}$ \\
  && $\gni\oplus\gei$ & $\gsi^{9\times8}\oplus(1)+h.c.$ &
  $\gsi^{9\times8}(2)+h.c. $ &  $\gtw^{9\times 9}+\gon^{8\times
  8}$ \\\hline
  &  $D_4$ & $\gse\oplus\gsi$ & $\gfi^{7\times6}(1)+h.c.$ &
  $\gfi^{7\times6}(2)+h.c. $&  $\gtw^{7\times 7}+\gtw^{6\times
  6}$ \\\hline
  && $\gse\otimes(\gon\oplus\gth)$
  &$\gon^{7\times7}\otimes\gth^{3\times1}$ &
  $\gon^{7\times7}\otimes\gth^{3\times 1} $  &$\gon^{7\times
  7}\otimes(\gon^{3\times
  3}-\gon^{1\times 1}$) \\\hline\hline\hline
  \ctp &&&$\scxi$&  $\scyi$ & $\sczi$ \\\hline\hline
  &  $O$ & $\gei=\gth\otimes\gse$ & $\gon^{3\times3}\otimes
  \gfo^{7\times7}(x)$ & $\gon^{3\times 3}\otimes\gfo^{7\times7}(y)$
& $\gon^{3\times 3}\otimes\gfo^{7\times 7}(z)$\\\hline
&  $D_6$ & $\gse\oplus\gei$ & $\gth^{7\times8}+ h.c.$&
 $\gfo^{7\times 8}+ h.c.  $&
 $\gtw^{7\times 7}+\gon^{8\times 8}$\\\hline
 \end{tabular}}
\parindent=0.5in
\caption{Tensors of Conduction States.  This table enumerates
the
symmetries of conduction tensors which may couple to the impurity
(quadrupolar
if \ufp, magnetic if \ctp; see Table 2.3 for further information on the
impurity tensors).  The meaning of a superscript $3\times 3$ for
example, means
that this operator is formed from the tensor (outer) product of $\gth$
bra and ket
states.}
\label{tab2p4}
\end{table}

Table~\ref{tab2p2} specifies all the possible two-channel $S=1/2$
combinations of $\Gamma_{grd},\Gamma_{ex},\Gamma_c$ states
for U$^{4+}$ and Ce$^{3+}$ ions.  We note that
a split doublet could display two-channel behavior above a
crossover scale below $T_0$.
Tables~\ref{tab2p3} and ~\ref{tab2p4} 
summarize all the relevant tensor operators for
impurity doublet states and conduction states together with their
transformation properties in analogy to spherical harmonics.  These
tables
give information about their multipole content as well.

  Each of the tensors discussed in preceding paragraphs
corresponds to some multipole moment tensors of the
ion. Since the multipole formalism is very physical and familiar from
electromagnetism and elementary quantum theory, it is worth exploring
this
connection further.

The matrix elements of
multipole tensors of $l-th$ rank in first quantized notation are given
by
$$<\eta|\hat O_{l,m}|\eta'> \sim \int d^3r \psi_{\eta}^*(\vec r) r^l
Y_{l,m}(\hat r) \psi_{\eta'}(\vec r) $$
$$~~~~~~~ \sim <\eta|[ Y_{l,m}(\vec J)]|\eta'> \leqno(2.2.50)$$
where $\eta,\eta'$ label states of the ion, and the second line follows
from
the Wigner-Eckart theorem.  With regard to $[Y_{l,m}(\vec J)]$, it is
to be
understood that: (i) the square braces indicate symmetrized
combinations of the
angular momentum operator components $J^i$, (ii) each time a particular
direction cosine $\hat i$ appears in the explicit polynomial expansion
of
$Y_{l,m}$ one should insert $J^i$ and properly symmetrize, and (iii)
each time
a power of $(J^i)^{l-p}$, with $p<l$ even,
 appears in the polynomial expansion of $Y_{l,m}$ one should multiply
 that term
by a factor of
factor of $[J(J+1)]^{p/2}$.

As an example of the use of multipole terminology, consider the $\gth
(E)$
doublet of a U ion in cubic symmetry.  The $\gon (A_1)$ tensor is of
predominant monopole (charge, $l=0$) character,
but also contains components of fourth rank (hexadecapole, $l=4$) as
well as sixth rank ($l$=6) multipoles.  The $\gth (E)$ doublet is of
predominant
quadrupolar ($l=2$) character.  It is a generic fact that one of the
three
non-trivial tensors formed from the quadrupolar doublets, in this case
the $\gtw
(A_2)$ tensor, will be odd under time reversal.  The predominant
multipole character of
the $\gtw (A_2)$ singlet tensor is octupolar ($l=3$).  Hence, a
non-zero
expectation value to this operator will correspond to a combined
lattice
distortion and non-vanishing magnetic moment.  The operator will couple
to the
third power of the magnetic field, or a combined product of electrical
field
gradient (uniaxial stress) and magnetic field.

The multipole character of the relevant tensor operators is summarized
in
Table~\ref{tab2p5}.

\begin{table}
\parindent=2.in
\indent{
\begin{tabular}{|c|c|c|c|c|}\hline
Irred. Rep.&  Dim. & Labels & $J$-basis & Spherical Basis
\\\hline\hline
$\gon (A_1)$ & 1  &- & $J(J+1)$ & $Y_{0,0}$ \\\hline
$\gtw (A_2)$ & 1 & - & $J_xJ_yJ_z$  &$Y_{3,3}(\vec J)+Y_{3,-3}(\vec J)$
\\\hline
$\gth (E)$ & 2  &$\gth (+)$ & $J_x^2-J_y^2$  &$Y_{2,2}(\vec
J)+Y_{2,-2}(\vec
J)$ \\
&&$\gth (-)$ & $3J_z^2 - J(J+1)$ & $Y_{2,0}(\vec J)$ \\\hline
$\gfo (T_1)$ & 3 & $\gfo (x)$ & $J_x$  &$Y_{1,1}(\vec J) -
Y_{1,-1}(\vec J)$\\
&&$\gfo (y)$ & $J_y$ & $Y_{1,1}(\vec J)+Y_{1,-1}(\vec J)$ \\
&&$\gfo (z)$ & $J_z$ & $Y_{1,0}(\vec J)$ \\\hline
$\gfi (T_2)$ & 3  &$\gfi (xz)$ & $\{J_x,J_z\}$ &  $Y_{2,1}(\vec J) +
Y_{2,-1}(\vec J)$\\
&&$\gfi (yz)$ & $J_yJ_z+J_zJ_y$  &$Y_{2,1}(\vec J)-Y_{2,-1}(\vec J)$ \\
&&$\gfi (xy)$  &$J_xJ_y+J_yJ_x$  &$Y_{2,2}(\vec J)-Y_{2,-2}(\vec J)$
\\\hline
\end{tabular}}
\parindent=.5in
\caption{Table of second rank tensor operators for \ufp and
\ctp
ions  in cubic symmetry (point group $O$).  The first column gives the
label of the irreducible
representation, the second column its dimensionality.  The third column
gives
the label associated to the vectors spanning the space of the
representation,
and the fourth column gives a basis for the vector space in the
Cartesian
operator representation (or Qubic harmonics).  Finally, the fourth
column gives
the corresponding basis in the spherical tensor language.}
\label{tab2p5}
\end{table}

In the lower symmetry cases, one can see that the diagonal pseudo-spin
operator
for the U ions has predominant octupole character, with a weak
admixture of
magnetic dipole character.  The transverse operators, which are
essential for
the Kondo effect, have predominantly quadrupolar character.  This
justifies the
labelling of the two-channel Kondo effect in these circumstances as
quadrupolar.  For the $\gni$ Ce doublet in hexagonal symmetry, it
is interesting that the transverse operators must exchange three units
of
angular momentum and hence have predominantly octupolar character.
Hence, it
is perhaps most appropriate to label the Kondo effect in this case as
an
``octupolar'' Kondo effect!

We now return to a discussion of the selection rules.
Selection Rule 5 follows from scaling along the lines of
Nozi\`{e}res and Blandin [1980], and from
NCA analysis for the model Ce$^{3+}$ ion (Cox [1993]; 
Kim, [1995]; Kim and Cox [1995a,b]).
Specifically, we define two crossover temperatures $T^{(I,II)}_x$ where
the
superscript refers to single or two-channel crossover.
Consider the case in cubic symmetry, with $\Gamma_{grd}=\Gamma_7$,
$\Gamma_{ex}=\Gamma_3$, and $\Gamma_c=\Gamma_8$.  Define
dimensionless effective exchange coupling constants $\tilde g_7,\tilde
g_8$ by
$$\tilde g_7 = {N(0)V^2_7 \over E_0 - E(f^0)} ~~~;~~~ \tilde g_8
= {N(0)V_{37}^2\over E_0 - E(f^2)} \leqno(2.2.51)$$
where $E_0$ is the ground state energy, $N(0)$ is the
conduction electron density of states,
and $V_{17,37}$ the hybridization matrix elements with
$\Gamma_{7,8}$ partial waves.  The Kondo scale for
the single channel model is $T_0^{(2)}\simeq D\exp(1/2\tilde g_7)$,
and for the two-channel model is $T_0^{(2)}\simeq D\exp(1/2\tilde g_8)$
to leading exponential accuracy.  As we shall explain in Sec. 5.2, for
$T$ below the crossover
scale $T^x_{(1)}$
$$T_{(1)}^x \approx {T_0^{(1)}\over 3}|{1\over \tilde g_8}
- {1\over \tilde g_7}|^{3/2} \leqno(2.2.52)$$
single channel behavior will dominate for $|\tilde g_7|>|\tilde g_8|$.
For $T$ below the crossover scale $T^x_{(2)}$ given by
$$T_{(2)}^x \approx {T_0^{(2)}\over 2}|{1\over \tilde g_7}
- {1\over \tilde g_8}|^2 \leqno(2.2.53)$$
two-channel physics will dominate for $|\tilde g_8|>|\tilde g_7|$.
Practically, this is tested by examining the sign of the thermopower,
given the particle-hole asymmetry of the model.
Dominant $f^0$-induced one-channel coupling will tend to
produce positive thermopower, while dominant
$f^2$-induced two-channel coupling will
tend to produce a negative thermopower.
For hexagonal symmetry, consider the $D_6$ point
group for concreteness.
Analogous arguments go through provided
$\Gamma_{grd}=\Gamma_9 \sim |\pm3/2>$,
and $\Gamma_{ex}=\Gamma_{5,6}$, with
$\Gamma_c=\Gamma_7\oplus\Gamma_8$.

For Yb$^{3+}$ ions with
$\Gamma_{6,7}$ ground states in cubic symmetry or $\Gamma_9$
in hexagonal symmetry similar arguments go through,
with $f^n\to f^{14-n}$. However,
the very large $f^{12}-f^{13}$ splitting (order 10 eV) makes
it unlikely that Selection Rule 5 can be satisfied (where we now
require $w(f^{12})>w(f^{14})$).

U$^{4+}$ ions with $f^2$ ground configurations have Hund's rules
angular momentum $J=4$, so that non-Kramers doublets are possible.
Consider
the
ground doublets for the hexagonal and tetragonal syngonies.
All of these doublets have the
property that the non-trivial diagonal operator transforms like
the $z$-component of a real spin, while the off-diagonal elements are
quadrupolar and contained only in direct products of two distinct
irreps of local
conduction states.  The physical reason is simple, and easily
understood
by considering states with $M_J=\pm 1$ pair of states in the presence
of a
crystal field Hamiltonian of pure axial character,
viz $H_{cef} = \Delta_{cef} [3J^2_z-J(J+1)]$.  This term is even
under time reversal and
thus maintains doublet degeneracy of the $\pm 1$ states.
Hence, the non-trivial diagonal tensor is just $|1><1|-|-1><-1|$
which transforms
like $J_z$ as restricted to the doublet.  The off diagonal tensor
must change the angular momentum by two units, and hence must have
quadrupolar character.  Turning now to $\Gamma_{ex},\Gamma_{c1,2}$
in the same axial field, we must have doublets of the form
$\pm (2n+1)/2$, since the
excited configuration has an odd number of electrons and
the conduction states always transform as double valued
representations which are
descended from half integral angular momentum in the full isotropic
symmetry.   Hence, the off diagonal conduction tensors in a given
representation can only change angular momentum by an
{\it odd} number of units, and cannot `flip' the  impurity spin.
However, it is
possible to form tensors from the cross products
$\Gamma_{c1}\otimes\Gamma_{c2}$ which can change the angular
momentum index by two units.
For example, for conduction states derived from
$j=5/2$ partial waves, the operator $|3/2><-1/2|$ changes the
angular momentum by two units.
Note that the results we have discussed are properties only of
the representations,
but easily illustrated in this pure axial limit.

From the discussion of the preceding paragraph
it is apparent that any Kondo model derived from a degenerate
doublet ground state of U$^{4+}$ ions in hexagonal or
tetragonal symmetry will be of the quadrupolar form, because the only
degenerate levels in ground or excited states are doublets, time
reversal guaranteeing the channel degeneracy (indexed
by the excited state in effect).  To make the idea more explicit,
assume for
definiteness that we have hexagonal $D_6$ point symmetry and
$\Gamma_{grd}=\Gamma_5$, $\Gamma_{ex}=\Gamma_7$ in an excited $f^1$
configuration.  This
yields
$\tilde\Gamma_c=\Gamma_5\otimes\Gamma_7=\Gamma_7\oplus\Gamma_9$.
Taking the simple axial crystal
field model and using conduction plane waves
in a $j=5/2$ partial wave manifold,
representative states in $J,M_J$ form are $|\Gamma_5\pm>=
|4,\pm>$, $|\Gamma_7\pm> = |5/2,\pm1/2>$, and $|\Gamma_9\pm>=|5/2,\pm
3/2>$.
Let us reorganize the labelling of the conduction states.
Define channel 1 as labelling the states created by the pair of
operators
$c^{\dag}_{k\Gamma_7+},c^{\dag}_{k\Gamma_9-}$, and channel 2 as
labelling the pair of states
created by $c^{\dag}_{k\Gamma_9+},c^{\dag}_{k\Gamma_7-}$.  Now let
$\alpha$ be the spin index, equal to $\pm$, and $\mu=1,2$ be the
channel
index.  Denote channel spin operators by $\tau^{(i)},i=1,2,3$.
Thus, for example,
$c^{\dag}_{k,+,1} = c^{\dag}_{\Gamma_7+}$.
By performing a Schrieffer-Wolff transformation, with
the interconfiguration energy splitting given by
$\epsilon_f=E(f^2\Gamma_5)-E(f^1\Gamma_7)$, we obtain the Kondo
coupling
$$H_{Kondo} = -{1\over N_s}\sum_{i,k,k',\alpha,\alpha',\mu}
J^{(i)}S^{(i)}_I S^{(i)}_{\alpha,\alpha'}
c^{\dag}_{k\alpha\mu}c_{k'\alpha'\mu} - {K\over N_s} S^{(3)}_I
\sum_{k,k',\mu,\alpha}
\tau^{(3)}_{\mu,\mu}c^{\dag}_{k\alpha\mu}c_{k'\alpha\mu}
\leqno(2.2.54)$$
where $J^{(1,2)}= V_7V_9/\epsilon_f$, $J^{(3)}=
(V_7^2+V_9^2)/2\epsilon_f$, and $K=(V_7^2-V_9^2)/2\epsilon_f$. Here
$V_{7,9}$ are the hybridization matrix elements coupling the
$\Gamma_{7,9}$
conduction states to the impurity. Note that:
i) this exchange Hamiltonian is intrinsically anisotropic but the
diagonal ($J^{(3)}$) term is antiferromagnetic which is sufficient to
ensure the Kondo effect, and (ii) this Hamiltonian has the peculiar
term
coupling diagonal spin and channel spin operators.  These are not of
concern, since it is now well established that exchange anisotropy
is irrelevant in the $M=2,S_I=1/2$ model (Affleck, Ludwig, Pang, and
Cox [1992])
and scaling calculations about
the non-trivial fixed point indicate that the spin-channel spin
coupling is marginally irrelevant (H.-B. Pang, [1992]).  We note that
for the case of tetragonal symmetry the spin-channel spin coupling
again arises, and all bare exchange constants are
generically unequal. Again, as we shall discuss in later sections the
exchange anisotropy is irrelevant.

Koga and Shiba [1995] have studied a model related to Eq.  (2.2.43) in
which excited crystal field singlet states are retained in the $f^2$
configuration.   The excited
states yield a triplet when the crystal field splitting $\Delta$ is
taken
to zero.  The idea of the study is that in the zero splitting limit the
``triplet impurity'' spin is exactly compensated, yielding a Fermi
liquid fixed point at low temperatures, while for a range of finite
splitting
the two-channel fixed point is stable.  In particular, the two-channel
fixed point is found to be stable for all values of parameters in
tetragonal
symmetry, and for a wide range of parameters in the hexagonal crystal
field.  As the details of their model are rather technical, we refer
the reader to section
3 of their paper for a complete discussion.

We note that the two-channel Kondo coupling for a \ctp~ ion in
hexagonal
symmetry has a precisely analogous form to Eq. (2.2.54).

A final comment on symmetry concerns the particle-hole transformation.
Provided the conduction band of the host metal is symmetric about the
Fermi
energy, which it will always be for sufficiently small energy scales,
and
provided the exchange coupling is only easy axis-anisotropic
in the case of the non-magnetic
Kondo effects, the model Hamiltonians we have
discussed in this section enjoy a discrete particle hole symmetry,
which has a
different meaning in the case of two-channel magnetic Kondo effects as
opposed
to the TLS and quadrupolar Kondo effects.  No matter what the degree of
exchange anisotropy, the particle-hole symmetry will be present in
the asymptotic low energy spectrum,
since, as we shall show, the exchange anisotropy is irrelevant about
the
non-trivial two-channel fixed point (and about the ordinary strong
coupling
Kondo fixed point as well).

For the spin Kondo models, the particle hole transformation is the
usual charge
conjugation operation.  Suppressing all but magnetic labels $\mu$ on
conduction
states we map particle creation operators to hole annihilation
operators
 according to
$$ c^{\dag}_{\mu} = i\sigma^y_{\mu,\mu'} h_{\mu'} \leqno(2.2.45)$$
where summation convention has been used, and $\sigma^y$ is the Pauli
matrix
introduced in Sec. II.A.   This transformation reverses both charge and
spin,
and hence leaves the spin tensor operators unaffected in the Kondo
coupling.
This symmetry is present for magnetic Kondo effects even when one
admits
anisotropic exchange couplings.

For the non-magnetic Kondo effects, the transformation is different.
We must
first restrict our quantization axis in pseudo-spin space to the
direction
along which the lone octupolar
operator points, and perform the usual particle hole transformation
described above.
The octupolar operator is odd under time reversal ${\cal T}$ and hence
the
coupling along that direction is invariant under the usual
particle-hole
transformation.  However, the transverse operators, with this choice of
quantization axis, have quadrupolar character and are thus even under
time
reversal but odd under the reversed sign of charge.  Flipping the sign
of
transverse couplings does not remove the Kondo effect, as is well
known, so
that the spectrum will be unaffected by this transformation.  However,
to
finish the transformation, we are free to follow the particle-hole
transformation by a $\pi$ rotation about the octupolar quantization
axis.
By the end of the procedure, we have performed the transformation
$$c^{\dag}_{\alpha} = i\sigma^y_{\alpha,\alpha'}h_{\alpha'}
\leqno(2.2.46)$$
where we have suppressed all but the quadrupolar index of the
conduction
states.

To summarize the results of this subsection,
we have demonstrated that the mapping of low energy scale properties
to the
two-channel quadrupolar Kondo model is  robust for U$^{4+}$ ions
with doublet
ground states in hexagonal and tetragonal symmetries in that
{\it all} such doublets will be described by this model on coupling to
the
conduction states.  We have also shown that under more restrictive
conditions
the model will apply to Ce$^{3+}$ ions in cubic and hexagonal
symmetry, but is unlikely to apply to Yb$^{3+}$ ions.

\subsubsection{Additional ions which may display two channel Kondo
effects} 

Among the actinide ions, \ufp remains the best candidate for
the two-channel quadrupolar Kondo effect.  Np$^{2+}(5f^4)$ or
Np$^{4+}(5f^2)$ would also
have a $J=4$ ground state and possibly the quadrupolar Kondo effect
when one of
the doublet levels lies lowest in cubic, hexagonal, or tetragonal
symmetry.
Np$^{3+}(5f^3,J=9/2)$ and Pu$^{3+}(5f^5,J=5/2)$ ions could display the
two-channel magnetic
Kondo effect in cubic or hexagonal symmetry provided they have the
appropriate
ground doublets.
In the Np$^{3+}$ case, the chances would be excellent since both
the excited $5f^2$ and $5f^4$ configurations have $J=4$ ground
multiplets and
thus possibly quadrupolar doublet ground states for $\Gamma_{ex}$ as
required
by the selection rules.  For the Pu$^{3+}$ case, the ground multiplet
of the
$f^4$ excited configuration is $J=4$, but for the $f^6$ it is $J=0$, so
that a
similar competition to \ctp models between single and two-channel Kondo
effects
arises.  Heavy fermion like behavior has been reported in some
concentrated Np
and Pu based materials.

In the rare earth row, Sm$^{3+}(4f^5,J=5/2)$ and Yb$^{3+}(4f^{13})$
could in principle exhibit
two-channel magnetic Kondo effects, but unfortunately the very large
energy
splitting to  excited
configurations with degenerate ground multiplets ($4f^4(J=4)$ for
Sm$^{3+}$, $4f^{12}(J=6)$
for Yb$^{3+}$) makes it far less likely than in the Ce case.  With
regard to
the quadrupolar Kondo effect, Pr$^{3+}(4f^2)$ is the most direct
analog to
\ufp, and indeed shows valence fluctuation tendencies in some compounds
[Cox, 1988b)].  Recent work on PrInAg$_2$ has renewed this promise (Yatskar {\it et al.} [1996]).  
Tb$^{3+}(4f^8,J=6)$ and Tm$^{3+}(4f^{12},J=6)$ ions both display mixed
valence
tendencies and could exhibit the quadrupolar Kondo effect when the
appropriate
doublets lie lowest. Indeed, dilute Tb in cubic Th appears promising as
a
candidate [Sereni {\it et al.}, 1986; Cox, 1988b)]. A complication in
cubic
symmetry in this case is that for
$J=6$ the $\gth$ doublet has a much smaller window of stability, and
when it is
stable, the magnetic $\gfi$ triplet tends to be very close
energetically [Lea,
Leask and Wolf, 1962].  That means the system will always be quite
susceptible
to magnetic ordering, or at the very least the magnetic excited state
will
obfuscate attempts to prove the quadrupolar ground state exists.  For
example,
the low temperature susceptibility of Tb in Th at low concentrations
is 2-4 emu/mole-Tb [Sereni {\it et al.}, 1986].

Generically, the odds of observing the two-channel behavior are higher
for the actinide ions rather than the rare earth ions
because of the greater extent of the 5f wave functions. This larger
size of the orbitals both
increases
the hybridization and lowers the correlation energies, both of which
serve to
enhance the Kondo temperature.  What allows the Kondo
scale of the \ufp ions to be small (order 10-100K for the \ufp-based
heavy
fermion materials) in the actinide
 case is the multiplication of the hybridization by a fractional
 parentage
coefficient (to project to the lowest states) which is a number smaller
than
unity.  As a corollary, the effects of excited crystal field splittings
are
likely to be less, because the crystal field splitting in these
intermetallics is
expected to scale with the square of the hybridization  [Zhang and
Levy,
1988a,b)].  In Y$_{1-x}$U$_x$Pd$_3$, for example, the first
crystal field splitting
appears
to be 5-6 meV [Mook {\it et al.} 1993; Dai {\it et al.}
1995, McEwen {\it et al.}, 1995], while
in UBe$_{13}$ it appears
to be 15 meV [Shapiro {\it et al.}, 1985; Cox, 1987].  In contrast,
PrPb$_3$ is
a collective Jahn-Teller system with an apparent $\gth$ ground doublet
on the
Pr sites, and the {\it overall} crystal field splitting there is of
order 1 meV (Ott, [1982]).

\subsection{Models with arbitrary $N_I,N_C,M$}. 

We have now derived several model Hamiltonians which have the
properties of
spin 1/2 Kondo models possessing $M$=2 (or possibly 3) ``channels'' of
electrons.
As we have made clear in the introduction, by channel we mean those
internal
degrees of freedom of conduction electrons which are decoupled from the
impurity.  We are primarily interested in models in which the
degeneracy of the
conduction and impurity spins are $N_c=N_I=N=2$, but we have
encountered models
in which $N_I,N_c>2$ as well.  Because these different degeneracy
factors may
serve as expansion parameters for organizing terms in perturbation
theory, it
is worth considering models in which we allow these parameters to
acquire
arbitrary values.   We shall briefly list a few such models in this
subsection.

{\it $SU(N_c=2)\otimes SU(M),N_I\ge N_c$ models}.  These are the
``overscreened'' models
originally considered by Nozi\`{e}res and Blandin, and are specified in
terms of a
Kondo coupling
$$H_{Kondo} = -{J\over N_s} \vec S_I \cdot \sum_{\alpha=1}^{M} \vec
S_{c\alpha}(0) \leqno(2.3.1)$$
where $S_I$ is allowed to take any value and $S_{c\alpha}(0)$ is the
conduction
spin density in channel $\alpha$ at the impurity site, with all
conduction
electrons having $S=1/2$.

{\it $SU(N_c=N_I=N)\otimes SU(M)$ models}.  These are a straightforward
generalization of Eq. (2.3.1) with the restriction of $N_I=N_c$.  These
models
are amenable to the NCA treatment.  In the Kondo form, the interaction
Hamiltonian is given by
$$H_{Kondo} = -{J\over 2 N_s} \sum_{k,k',\mu,\mu',\alpha}
|\mu'><\mu| c^{\dag}_{k\mu\alpha}c_{k'\mu'\alpha} \leqno(2.3.2)$$
where $|\mu>$ is one of the $N_I$ states of the local ``spin''.  One
may also
write an equivalent Anderson Hamiltonian, placing the local spin states
at
energy $\ef$ and introducing an excited state field indexed by label
$\alpha$,
 leading to the
hybridization Hamiltonian
$$H_{hyb} = {V\over \sqrt{N_s}} \sum_{k\alpha\mu}
[-sgn(\alpha)|\mu><-\alpha|c_{k\mu\alpha} + h.c.] \leqno(2.3.3)$$
where the match to Eq. (2.3.2) is through the Schrieffer-Wolff
transformation
with $J=V^2/\ef$.  This simple decoupling is only possible when
$N_I=N_c$.

{\it $SU(min(N_I,N_c))\otimes SU(M)$ models}.   These models are the
most
general form and include couplings which are symmetric under unitary
transformations of the smaller of the impurity or conduction spin
degeneracy.
They are difficult to write down, and probably of marginal relevance to
real materials,
except in the case $N_I=2\le N_c$ in which case we can simply use Eq.
(2.3.1)
with $\vec S_{c\alpha}(0)$ being given by
$$S^i_{c \alpha}(0) = {1\over N_s}
\sum_{\mu,\nu}S^i_{\mu,\nu} c^{\dag}_{k,\mu,\alpha}c_{k'\nu\alpha}
\leqno(2.3.4)$$
with $S^i_{\mu,\nu}$ being the $i$-th component of the $J_c=(N_c-1)/2$
angular
momentum representation of $SU(2)$.  This model could be of possible
relevance
to the TLS theory, once the two-site problem is generalized to a multi-site 
problem, as discussed recently by \zar [1996].  As mentioned in Sec. 2.2, this model could also be
of relevance for \ctp~ impurities.

\section{Scaling Theory of Kondo Models} 

\subsection{Overview of the Physics and Interelatedness of Methods} 

\subsubsection{Concepts and Terminology of Scaling and Renormalization
Group
Theory} 

The basic philosophy of any renormalization group theory is to describe
the
physics of the problem in terms of an effective Hamiltonian at each
length
scale which is expressed in terms of a small basis set of operators
multiplied
by coupling constants which depend upon the length scale.  Following
the
notions of Kadanoff and Wilson, we imagine integrating out variables on
small
length scales to derive Hamiltonians on large length scales.  The first
to
apply such concepts to the Kondo problem in a straightforward manner
was
Anderson [1970].  Later this idea was formulated in the framework of the 
multiplicative renormalization group (\zow and Fowler [1970]; 
Fowler and \zow [1971]; Abrikosov and
Migdal [1970]).  In this
subsection, we wish to lay out some of the relevant concepts and
terminology.

{\it Equivalence of Space and Time in Kondo models}.
We may interchange
length scale with time scale in our impurity problems.  The reason is
simple:
effectively, any impurity model presents a quantum problem in one
spatial
dimension equivalent to the radial direction away from the impurity.
This is
often called a ``1+1'' dimensional problem, where the ``+1'' refers to
the time
direction.  The conversion factor between space and time or energy and
momentum
is simply the Fermi velocity $v_F$ of the conduction electrons, which
hinges on the
fact that the dispersion of the electrons near the Fermi energy may
always be
linearized. In fact, the problem may be viewed as a Lorentz invariant
model
with speed of light equal to $v_F$.   This simple idea underlies much
of the physics of the Kondo model,
and is essential for its solution using the
the renormalization group theory (Sec. 4), conformal field theory (Sec.
6),
and the Bethe-Ansatz (Sec. 7).
Hence, our approach in the scaling and numerical
renormalization
group approaches shall be to determine the effective Hamiltonians
describing
the problem at different energy scales, which are set, for example, by
the
frequency of an external probe, the temperature, the magnetic field, or
simply
by the magnitude of the conduction bandwidth.

{\it Appearance of Logarithms in Perturbation Theory}.
The methods we will employ in discussing the two-channel Kondo
problems and Kondo analogues  have features in common which derive from
the appearance of logarithms in the high energy scale {\it (ultraviolet
region)}
perturbation theory which diverge upon approach to the low energy
scales
{\it (infrared region)}.  As Wilson has noted, the appearance of
logarithms
indicates that all energy  scales are of equal importance which
strongly
supports the relevance of a renormalization group approach.
Practically, we
will imagine accessing the infrared regime through the reduction of the
temperature, frequency, magnetic field, or conduction bandwidth.

The individually logarithmically divergent terms in perturbation theory
may be treated collectively by the various techniques to generate
models
characterized by renormalized couplings appropriate to whichever
infrared energy
scale we
are sitting at in the calculation.  The spirit here is very much the
same as the
``running coupling constant'' scheme in quantum field theory.  Unlike
quantum
field theory, the lattice always provides us with an ultraviolet cutoff
(maximum
energy scale) which will be the Fermi energy or conduction bandwidth in
our
work.  Indeed, in those models where the coupling constants grow with
reduced
energy scale, we have in our hands a precise mathematical analogue of
the
asymptotically free models (such as quantum chromodynamics) studied in
high
energy theory.

{\it Fixed Point Taxonomy}.  We are said to be at the {\it fixed point}
of a
renormalization group transformation if upon further rescaling the
Hamiltonian
remains unchanged.   The fixed point may be characterized in terms of
its
{\it excitation spectrum} that may be either: (i) {\it
Fermi-liquid-like} with
a 1:1 map near the Fermi energy to
the spectrum of one (or more) one-dimensional Fermi gases
 which have uniform level spacing $v_F\pi/L$, $L$ the radial extent of
 the
metal. A trivial modification of this spectrum occurs in the presence
of an ordinary
potential scattering center which imparts a phase shift $\delta$ to the
conduction
states; this phase shift manifests in the energy levels through an
additional
shift of the amount $-v_F\delta/L$ with respect
to the free gas; (ii) {\it non-trivial} or {\it non-Fermi-liquid-like}
in which
case the excitations appear at non-uniform, fractional spacings of the
Fermi
gas, and for which application of the phase shift concept $\delta$ is
not meaningful.
The states are characterized by separated spin, charge, and channel
numbers,
relative to the free fermion gas.
(In between fixed
points, the spectrum is a typically complex many body spectrum.)

We shall generically flow to four kinds of fixed points:\\
\begin{quote} {\it (1) Zero Coupling Fixed Point}:  This
fixed point has precisely the free Fermi gas spectrum for each spin and
channel
of conduction electrons, and corresponds to an impurity uncoupled from
the
electrons, whence the phase shift $\delta=0$.  This is the stable
infrared
fixed point of the {\it ferromagnetic} isotropic Kondo model.
\end{quote}
\begin{quote}{\it (2) Strong Coupling Fixed Point}: As applied to some
of the
single channel Kondo  models of
interest here, at this  fixed point, the
effective coupling of the impurity to the conduction electrons is
infinitely
strong, and the excitation spectrum is Fermi liquid like with a phase
shift
fixed ``universally'' (i.e., independent of input parameters to the
model)
by the degeneracy of the impurity and conduction electrons.  This is
the
infrared stable fixed point of the isotropic antiferromagnetic
single-channel
Kondo problem, where $\delta=\pi/2$.
\end{quote}
\begin{quote}{\it (3) Intermediate Coupling Fermi-Liquid Fixed Point}:
These fixed points are Fermi
liquid in character but possessed of non-universal phase shift values
between 0
and $\pi/2$.
Such fixed points typically arise when some low energy field is present
in the
problem that cuts off the scaling to strong or non-trivial fixed point,
such as
the level splitting and spontaneous tunneling in the case of the TLS
model.
\end{quote}
\begin{quote}{\it (4) Non-trivial intermediate coupling fixed point}.
This is
the interesting fixed point of the multi-channel Kondo models first
identified
by Nozi\`{e}res and Blandin [1980]. At this
kind of fixed point, the internal degrees of freedom of the impurity
are never
completely compensated, and the excitation spectrum is non-trivial.  In
consequence, instead of obtaining Fermi liquid properties at low $T$,
one
obtains critical phenomena:  the incremental specific heat coefficient
and
susceptibility diverge as $T\to 0$.  This divergence may be logarithmic
($n=2$) or power-law in character ($n>2$).
Indeed, this fixed point is a true critical
point--the impurity has driven the entire metal to the edge of a phase
transition.
\end{quote}

{\it Stability of Fixed Points}.
A fixed point is also characterized by its stability properties:
namely, if you
begin with coupling constants arbitrarily close to the fixed point,
will you
flow into the fixed point with rescaling (lowering of energy)
or will you flow away?  The answer may
be mixed--some couplings or parameters may drive you away,
others towards the fixed point.  Any coupling
constant measured relative to the fixed point value which vanishes upon
rescaling is said to be {\it irrelevant}, if the coupling difference
grows, the
coupling is {\it relevant}, and if the coupling to leading order is
unchanged,
it is {\it marginal}.  Examination of next-leading order effects can
further identify if a marginal operator is ``marginally irrelevant'' or
``marginally relevant.''

{\it Universality}.
Asymptotically close to some of the above fixed points,
properties may be
{\it universal}, i.e., they can be expressed as universal functions of
thermodynamic or dynamic variables measured with respect to  a single
energy
scale that depend only
upon properties of the fixed point and not the bare couplings.  For
example, at
the strong coupling fixed point of the ordinary one-channel spin 1/2
Kondo problem, the phase shift
of the conduction electrons is $\pi/2$, a universal number independent
of the
bare exchange value, and at low temperatures the susceptibility times
the
temperature with the non-universal effective magnetic moment divided
out
($T\chi(T)/\mu^2$) is a universal function of temperature measured in
units of
the Kondo temperature $T_K$, a natural energy scale that divides high
and low
energy regimes and may be determined in the perturbative scaling
analysis.  When
it is meaningful to use it, the effective scaled
Fermi temperature $T_0$ differs from
the Kondo scale $T_K$ only by pure numerical factors of order unity.
However,
independent of the properties of the infrared fixed point, the Kondo
scale
retains meaning as the boundary between high and low energy physics.

\subsubsection{ Illustrative Example: Two-channel model in applied spin
and
channel fields} 

The concepts discussed in the previous subsection have been discussed
somewhat
in the abstract.  We would like to illustrate their meaning by
providing an
overview of the physics of the two-channel spin 1/2 Kondo model in
applied spin and
channel fields.  The derivation of these results will follow in later
sections.
For the purposes of discussion, we shall assume the exchange
interaction to be
isotropic.

\begin{figure}
\parindent=2in
\indent{
\epsfxsize=4.in
\epsffile{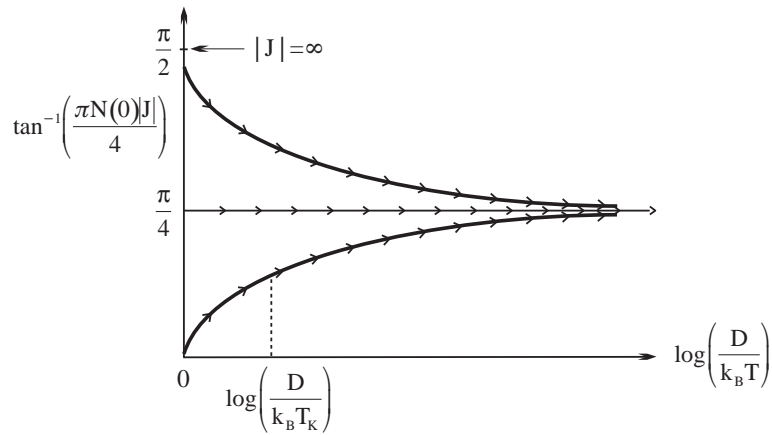}}
\parindent=0.5in
\caption{Scaling of dimensionless exchange coupling vs. logarithm of the
bandwidth for the two-channel spin 1/2 Kondo model.  
Whether starting from initially 
strong ($J\to\infty$) or weak ($J\to 0$)coupling, 
the arctangent of the dimensionless 
exchange coupling scales to the fixed point value of $\pi/4$
as the bandwidth is reduced.  The crossover scale for weak coupling
is identified as the Kondo temperature.  If one begins right at the 
fixed point, the coupling will be unchanged under renormalization.}
\label{fig3p1}
\end{figure}

{\it Zero Field Case}.  Fig.~\ref{fig3p1}  displays the behavior of the
effective
exchange coupling as a function of the logarithm of the temperature
beginning
at high $T \sim D/k_B$ and flowing to low $T$ (much smaller than the
Kondo
scale, $T_K$, in the case of initial weak coupling).  The meaning of
this plot
is that the first few levels of the excitation spectrum can be fit by a
simple
model in which conduction electrons exchange couple to an $S_I^*=1/2$
impurity.
 In fact, the ``impurity'' is a complicated bound object consisting of
 the original
impurity at its core, but surrounded by shells of alternating
conduction spin.
(An amusing metaphor is to think of the impurity spin as the seed of a
pearl,
with the layers of conduction spin accreting like the layers of the
pearl as
the temperature is lowered [S. Williams, 1992].)  At low temperatures
(large
values of $-\log(T/D)$) we ``flow'' to the fixed point coupling,
regardless of the
size of the initial, high temperature coupling.   Nothing cuts off this
scaling
process, and the spin can never be compensated away as in the single
channel
Kondo effect, so that the properties of the system as the fixed point
is
approached are those of a critical point in which the system looks the
same on
all length scales.

\begin{figure}
\parindent=2in
\indent{
\epsfxsize=4.in
\epsffile{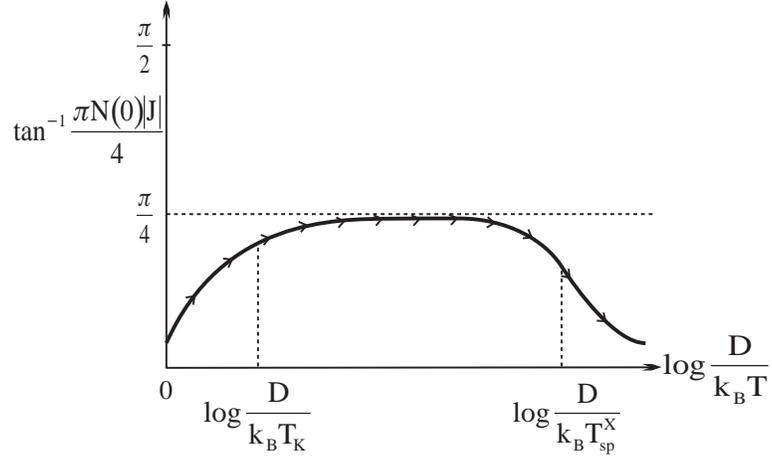}}
\parindent=0.5in
\caption{Scaling of the spin 1/2 two-channel Kondo model in the presence
of a spin-field.  In this case, where a field that linearly splits the
spin of the local moment is applied, 
then provided the applied field is small compared to the
Kondo scale, the system will first flow close to the fixed point
coupling
strength of the two-channel model, then flow to a value determined 
simply by the product of the bare exchange strength and the 
impurity spin as the system flows to the 
``polarized scatterer fermi liquid'' fixed point.  The crossover 
scale to the latter is $T_{sp}^x \sim H_{sp}^2/T_K$, where $H_{sp}$ is the 
applied spin-field. }
\label{fig3p2}
\end{figure}

{\it Applied Spin Field}.  By applied spin field, e.g. in the Kondo
problem,
we mean a
local field which couples to the impurity spin linearly, or a bulk
field which
also couples to the conduction spin.  This could correspond to the
spontaneous
tunneling and level splitting terms in the TLS model, to a uniaxial
stress in
the quadrupolar Kondo model, or to an applied magnetic field in the
magnetic
two-channel model.  The spin field
cuts off the scaling process of the previous paragraph, so that below a
``crossover temperature'' $T^x_{sp}$ determined by the field strength the
system will
behave like a Fermi liquid with energy scale set by the crossover
temperature.
If we begin from the physically relevant case of weak coupling, then
assuming
the spin field splitting $\Delta^z <<T_K$, we will have
$T_{sp}^x \approx
(\Delta^z)^2/T_K$ (see Secs. 4.2.e, Sec. 5.1.6, 6.1,7).  This crossover
effect
on the effective coupling constant is
illustrated in Fig. ~\ref{fig3p2}. 

\begin{figure}
\parindent=2in
\indent{
\epsfxsize=7.in
\epsffile{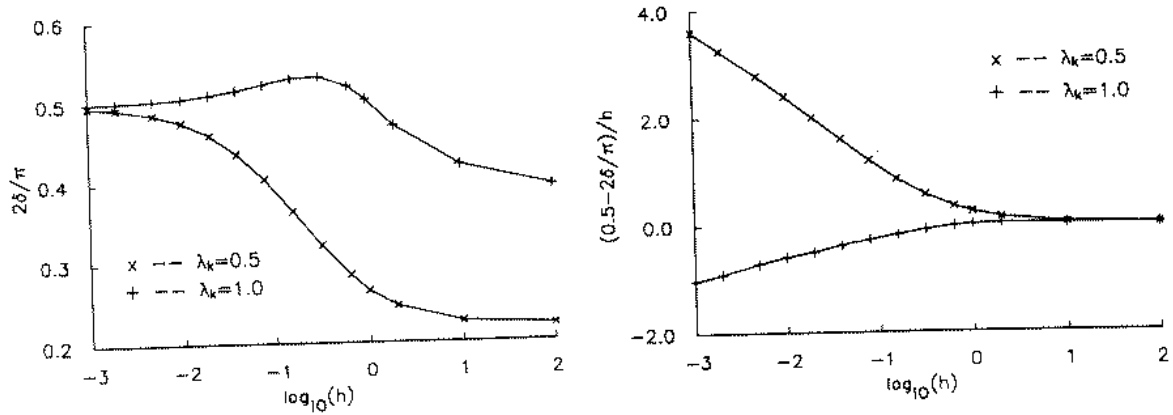}}
\parindent=0.5in
\caption{Physical effects of applied spin field on the two-channel spin 
1/2 Kondo model.  At left we  display the phase shift as a function of field 
as determined by numerical renormalization group calculations (see 
Affleck {\it et al.} [1992]).  As noted in the text, this may be used to
calculate the magnetization.  At right, we show that properly scaled
this numerically calculated phase shift displays the logarithmic
singularity expected in the differential susceptibility as a function of
field.  }
\label{fig3p3}
\end{figure}

The low temperature fixed point in this case is of variety
(3) (intermediate coupling, Fermi liquid) because the ground state
phase shift
will obtain a value determined by the magnetic field and the bare
coupling
constants.  Because up and down spin electrons experience equal and
opposite
phase shifts, we must view the excitation spectrum as deriving from two
independent Fermi gases.   Fig. ~\ref{fig3p3} illustrates how the spin field
crossover manifests
itself in the value of the conduction electron phase shift for up-spin
electrons at the Fermi energy [Affleck {\it et al.}, 1992].
As we discuss in Sec. 7.2, this effect has a rather spectacular
manifestation in the specific heat:  
above the crossover temperature, the specific heat
displays the
logarithmic behavior characteristic of the two-channel model, while
below the
ground state residual entropy is shoved out into a Schottky-like peak
which for
a range of temperatures is actually larger than the zero field specific
heat by
nearly an order of magnitude.

{\it Applied Channel Field}.  By channel field, we mean an external
probe which
couples to the channel index of the conduction electrons and acts to
lift the
degeneracy of the exchange coupling.  Practically, this is effected in
the
quadrupolar Kondo model by the application of a magnetic field, which
splits
the excited $\gse$ doublet and thus through the Schrieffer-Wolff
mapping splits
the exchange integrals for the different conduction channels by an
amount
$\Delta J \simeq (V/\tilde\ef)^2 \mu_{eff}H$ where $\mu_{eff}$ is the
effective moment
of the excited doublet and $H$ the magnetic field strength.    In the
TLS
model, magnetic field would also be a channel field in principle,
but there is no obvious mechanism by which
the couplings for up and down spin can be split.   In the magnetic
two-channel model, uniaxial stress will split the excited quadrupolar
doublet
and lift the channel degeneracy.

\begin{figure}
\parindent=2in
\indent{
\epsfxsize=3.5in
\epsffile{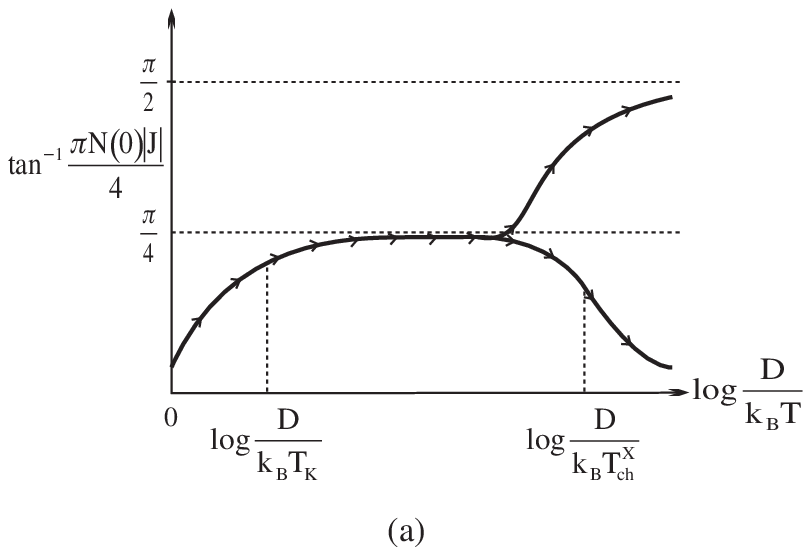}
\epsffile{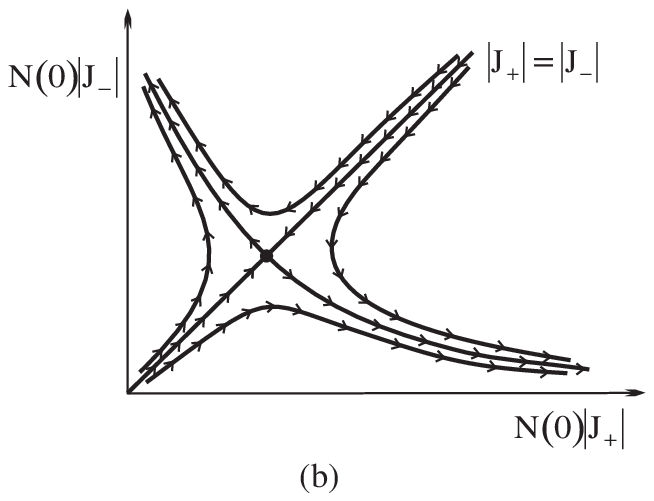}}
\parindent=0.5in
\caption{Scaling of the spin 1/2 two-channel Kondo model under
application of a channel field.  A channel field couples linearly to the
conduction electron channel spin density and thus lifts the degeneracy
of the two exchange integrals.  (a) shows that one coupling will pass 
to strong coupling and the other to weak coupling at a crossover scale 
$T^x_{ch} \sim (\delta J)^2/T_K$, where $\Delta J$ is the exchange
integral splitting.  (b) shows the corresponding flow diagram in the 
space of coupling constants.  The flows are stable along the $J_1=J_2$
separatrix, moving to the non-trivial fixed point.  They are unstable 
away from this separatrix, moving the strong coupling fixed point of
which ever exchange is initially stronger.}
\label{fig3p4}
\end{figure}

The channel field also cuts off the scaling of Fig. ~\ref{fig3p1},
but in a very different way than for the spin field, as made clear by
Nozi\`{e}res [1980].  Assume  initial weak coupling.
Below a
crossover temperature $T^x_{ch} \approx (\Delta J)^2/T_K$ (see Secs.
4.2.d,5.1.5,6.1,7),
the more strongly
coupled channel will tend towards the {\it strong coupling} ordinary
single
channel Kondo fixed point
for which the phase shift is $\pi/2$, and the weakly coupled channel
will tend
towards the {\it zero coupling}
 fixed point with zero phase shift.  Hence, as with
the applied spin field, the excitation spectrum is composed of two
fermi gas
excitation spectra with different phase shifts.  In this case  This is
illustrated in Fig. ~\ref{fig3p4}(a)  where below $T_{ch}$ the single line of high
$T$
scaling ``bifurcates'' with the upper branch corresponding to the
strongly
coupled channel, and the lower branch to the weakly coupled channel.
In Fig. ~\ref{fig3p4}(b) we show the scaling flows in the space of coupling
constants, where the non-trivial fixed point of the flows is
evident in the center.  This fixed point is stable for $J_+=J_-$,
but unstable to any small differences.
We note that in the case of the TLS Kondo model, the magnetic field
splitting
of conduction states
will not induce a splitting of exchange integrals in the same way.  The
only
possible influence on the fixed point is for physically irrelevant
fields
$\mu H \simeq D$ which cannot be realized experimentally.

A last point about the two-channel model is that it is stable with
respect to
anisotropy of the exchange integral while maintaining the channel
degeneracy (see
Secs. 3.4.1,4.2.c,6.1).  Thus, while applied spin and channel
fields destabilize the model and take it away from the non-trivial
fixed point
to one of the other three generic fixed points, the model enjoys
stability
against the realistic feature of exchange anisotropy.

\subsection{Formal Development of Scaling Theory} 

\subsubsection{Organization of Perturbation Expansion In Logarithms;
Methods of
Resummation} 

We have already pointed out in the introduction and elaborated above on
the
origin of logarithmic corrections in the perturbation expansion of
the impurity models considered here due to
the non-commutative character of the conduction electron scattering
processes
off the impurity.  The logarithmic factor which occurs is
$${\cal L}(max\{T,\omega,E\}) = \log({D\over max\{T,\omega,E\}})
\leqno(3.2.1)$$
where $D$ is the high energy cutoff which will always be the conduction
bandwidth in the problems interesting us, and $max\{T,\omega,E\}$ is
the
largest of the other external parameters in the system, such as the
temperature
$T$, the driving frequency of an external probe $\omega$, or
the splitting
of the
TLS, $E=\sqrt{(\Delta^z)^2+(\Delta^x)^2}$.  $E$ ultimately
serves as the
small energy
infrared cutoff.

In the treatment of the models of interest in this Review, we may
organize the
perturbation theory for any quantity according to powers of the
coupling $g$
between impurity and conduction electron states,
times powers of the logarithmic factor ${\cal L}$.  Assuming a given
quantity
${\cal Q}$
has a minimal power $g^p$ of the coupling (e.g., for the resistivity in
the
Kondo problem, $p=1$), we have the following hierarchy:
\[ \begin{array}{lccccc}
  (3.2.2)~~~~~~~~~~~~~ {\cal Q}&=&   g^p&& &   \\
&&                       +g^{p+1}{\cal L} & +g^{p+1} & & \\
&&                       +g^{p+2}{\cal L}^2 & +g^{p+2} {\cal L} &
+g^{p+2} & \\
&&                       +g^{p+3}{\cal L}^3 & +g^{p+3}{\cal L}^2
&+g^{p+3}{\cal
								L}
								&+g^{p+3}\\
&&                       .  .  .  . &&&\\
&&                       .  .  .  . &&&\\
&&                       .  .  .  . &&& \end{array} \]
In this array, clearly the rows represent the order in a systematic
expansion
of the coupling constants.  The columns give us the logarithmic
hierarchy:
since the logarithmic factors diverge as one tunes the infrared cutoff
to zero,
summing up the logarithms in the first column will produce a stronger
singularity
than summing up the logarithms in the second, third, fourth, etc.
columns.  We call the resummation of logarithms in the first column the
{\it
leading logarithmic approximation}, which sums
one set of diagrams to all orders in perturbation theory of the
coupling $g$.
By including also the summed up logarithms of the second column, we
include
another set of diagrams to all orders in the perturbation expansion,
and this
is called the {\it next leading logarithmic} approximation.

We could of course
proceed in this way through all the columns, which in principle would
yield an
exact calculation of the quantity ${\cal Q}$.  In practice, however, a
considerable amount of information may be obtained by considering only
the
first and second columns. (Indeed, in certain limits this approximation
may be
shown to be exact.)  For example, the Kondo scale $T_K$ is quite
accurately estimated by the next leading logarithmic approximation;
indeed, as
shown by Wilson [1975] the next to leading order scaling theory
determines the
universal parts of the $T_K$ expression.  In the
limit of large numbers of channels $M$, the non-trivial Nozi\`{e}res
and Blandin fixed
point is correctly uncovered by the next leading logarithmic
approximation.

The systematic collections of terms in the logarithmic hierarchy may be
accomplished by several methods.  We shall discuss five strongly
related
methods here:\\
{\it (i) Leading logarithmic order resummation of diagrams}.  This is
equivalent to the
direct diagrammatic summation methods of Abrikosov [1965],
as well as to the ``poor man's scaling'' method of Anderson
[1970]. In this case, the summation problem is cast in the form of a
differential or integral equation for the relevant quantity (such as
the
conduction electron $T$-matrix describing scattering off of the
impurity). This
method is known as the ``parquet'' method, and is analogous to the
lowest order
RPA approximation in a metal, where one sums the  diagrams built
from spin-conduction
particle/hole bubbles to all orders. We discuss this approach in
section 3.3.\\
{\it (ii) Multiplicative Renormalization Group in the Leading and Next
Leading
Orders of Diagrams}.  This method has been
developed by Gell-Mann and Low, Bogulubov and Shirkov for treating
infrared and
ultraviolet divergences in field theory. This approach was first
applied to the
Kondo problem by \zow and Fowler [1970] (see also 
Fowler and Zawadowski [1971]), and by Abrikosov and Migdal
[1970].
The method was later simplified by S\'{o}lyom and applied for to the
one-dimensional electron gas (S\'{o}lyom [1974,1979],  Zawadowski
[1973]).  We
shall discuss this in Sec. 3.4.   \\
{\it (iii) Path Integral Method}.  This approach used first by
Anderson, Yuval and
Hamann [1969] and
subsequently by \zim,\vld, and \zow [1988a,b].  We shall discuss this
in Sec. 3.5.  \\
{\it (iv) Numerical Renormalization Group}.   This non-perturbative
approach
was developed by Wilson [1973,1975; Krishna-murthy, Wilkins, and
Wilson, 1980a,b);
Cragg and Lloyd, 1979; Cragg, Lloyd, and Nozi\`{e}res, 1980].  In
essence, one
solves the rescaled equations exactly on a cleverly discretized mesh by
finding
the lowest energy levels of the system at each renormalization step.
It is
possible to compute properties  with this method if sufficiently many
steps are
retained.  For example, Wilson provided the first calculation of the
magnetic
susceptibility which was accurate over the entire temperature range. We
shall
discuss this method in more detail in Sec. 4.\\
{\it (v) Next-leading order resummation and NCA}.  As we shall discuss,
this procedure
sums a class of perturbation theory diagrams to all orders and includes
to leading
and next leading order the corrections to the vertex
of the conduction electron-local spin scattering, and to leading order
the
corrections to the ``self-energy'' of the local spin variable, which
shall be
precisely defined in the next subsection.  The resummation takes the
form of
coupled non-linear differential or integral equations. This
approximation was first
employed in the x-ray edge singularity problem by Moulet {\it et al.}
[1969],
and Nozi\`{e}res {\it et al.} [1969] where it was called the ``self
consistent
parquet approximation.'' This method was first applied to the Kondo
problem
by Nozi\`{e}res [1969].
A somewhat different approximation
appeared in
the work of Keiter and Kimball [1971].  Subsequent use of the
approximation as
written by Keiter and Kimball for
Anderson and Kondo models has gone under the name of the ``NCA'' or
non-crossing approximation, since in the formalism of Keiter and
Kimball, the
diagrams retained have no crossed propagator lines.  Although this
approximation may seem crude, it in fact can be justified within
a large $N$ expansion.  For all Kondo problems, it gives a
good extrappolation of the high temperature physics into the low
temperature regime, well below $T_K$
(though it eventually fails to give the correct ground state for
the single channel model).
For the multi-channel model, it has been shown exact for the
limit $N\to\infty$, $M/N$ held fixed [Cox and Ruckenstein, 1993]
and reproduces thermodynamic properties quite accurately even
for $N=2$ and $M=2,3$ [Kim, 1995; Kim and Cox, 1995,1997].
The NCA approach for
the multi-channel Kondo problem is discussed in Sec. 5.  \\
{\it (vi) Conformal Field Theory and Bosonization Techniques}.
Conformal field
theory may also be viewed in somewhat the same light as the scaling and
renormalization methods, since it exploits a scale invariant critical
point
in $1+1$ dimensions which is also conformally invariant, and
builds a description of the physics in terms of relevant and
irrelevant operators about the
conformal point.  Conformal field theory is a kind of
``non-Abelian bosonization'' scheme, since the low lying excitations
are
written in terms of current operators which satisfy the Kac-Moody
algebra
commutation
relations of non-Abelian symmetry groups.  An extensive development of this theory for the Kondo
problem
has been carried out recently [Affleck and Ludwig,
1991a,b,c;1992;1993;1994a,b;
Ludwig and Affleck, 1991; Ludwig, 1994]. We shall
discuss
this approach in Sec. 6.1.  Separately,
an Abelian bosonization scheme with ordinary harmonic oscillator
operators
has been developed by Emery and Kivelson [1992].  We shall discuss this
approach in Sec. 6.2.

We note that methods (ii) and (v) are in fact similar, except that most
formulations of (ii) have not included the imaginary part of the
dressed
exchange interaction and of the
spin self-energy, while these are included in (v),
so that all response functions have the appropriate analytic behavior.
It is
possible to apply (ii) in the energy region below the ground state
threshold at
zero temperature
where all response functions are purely real (for negative energy,
there are no
allowed decay processes) and then to analytically continue the solution
into
the positive frequency region (Zawadowski, unpublished, [1970];
S\'{o}lyom [1971]).  This
appears to be substantially equivalent to the differential form of the
NCA
developed by Kuramoto and Kojima [1984] and
M\"{u}ller-Hartmann [1984].

The common feature in methods (i)-(iii) and (v) is that they are based
on an
expansion in the coupling which presumably should not be applied when
the
renormalized dimensionless coupling grows to order unity.  Method (iv)
does not suffer from
this approach of course.  In fact, method (v) appears to cross into the
non-perturbative regime quite adequately as well, although the reasons
are not
completely clear as to why this works. What is useful about methods (i)
and
(ii) is that they correctly predict the growth of the couplings, and
they can
in some instances correctly predict other phenomena such as non-trivial
fixed
points.

In closing this subsection, we make two final observations: \\
{\it (1) Absence of Asymptotic Freedom in commutative models}.  We
still obtain
logarithmic corrections in commutative models such as the x-ray edge
singularity and the TLS model with $V^z\ne 0,V^x=0$.  However, in these
cases,
there is no renormalization of the bare coupling constant as we go to
the
infrared region.  Nevertheless, power law singularities may result in
physical
quantities with the power laws related to the phase shift associated
with the
commutative coupling constants.  As an example, in the commutative TLS
model,
for sufficiently large $V^z$, one finds that the spontaneous tunneling
rate
$\Delta^x$ is renormalized at low temperatures according to
$$\Delta^x(T) = \Delta^x(T=D)({T \over D})^K  \leqno(3.2.3)$$
where $K$ is proportional to the square of the
 phase shift $\delta(V^z) = tan^{-1}(\pi V^z)$ (Kondo [1976],
Libero and Oliveira [1991]).\\
{\it (2) Conformal Field Theory Approach and Abelian Bosonization}.  
Although we didn't include
it in
the list of methods based upon resumming logarithms, the conformal
field theory
approach first applied by Tsvelik [1990], and substantially evolved by
Affleck [1990a,1995],
Affleck
and Ludwig  [1991a,b,c;1992;1993], Ludwig and Affleck [1991,1994];
Ludwig [1994a,b], and Affleck {\it et al.},1992] has a
strong relationship to these other methods.  This method exploits
conformal invariance of the Kondo models at the fixed points to provide
exact
forms of universal functions (both thermodynamic and dynamic) and exact
values
for universal amplitudes (such as the residual ground state entropy).
The
method is very similar in character to the eigen-operator expansion of
the
renormalization group, but provides considerably more information due
to the
local character of the conformal invariance.  The method can also be
applied to
the calculation of finite size excitation spectra, where detailed
quantitative
agreement with numerical renormalization group calculations has been
found
[Affleck {\it et al.},  1992].  Unlike methods (i-v) above,
however, the conformal theory is incapable of calculating full
crossover behavior
between fixed points.  In principle, as well, the conformal
theory requires additional confirmation that models displaying
conformal invariance correspond to fixed point Hamiltonians
of the original models.  In practice, there no doubt that
this is the case for the models considered by Affleck and
Ludwig (Affleck [1990a], Affleck and Ludwig [1991a,b,c; 1992, 1993];
Ludwig and
Affleck [1991]; Ludwig [1994a,b]).  In addition, an approach based upon an 
Abelian bosonization of the anisotropic two-channel model has been introduced by 
Emery and Kivelson [1992] which reproduces many of the features of the 
conformal theory in perturbing around a special fixed point of the model.  
We shall discuss the conformal theory  approach in some detail in
Section 6.1, and the Abelian bosonization in Sec. 6.2.  
Finally we must emphasize that the methods (i) and (ii)
describe the  high temperature results, and methods (vi) give the
correct low temperature behavior in terms of non-universal amplitudes
which
can be determined by a match to other approaches (such as the numerical
renormalization group). The NCA and NRG approaches are methods which
also describe well the crossover regime between high and low energy
scales.

\subsubsection{Multiplicative Renormalization Group} 

The multiplicative renormalization group is based upon the following
observation:  let us consider an interaction between two different
kinds of
particles described by Green's functions ${\cal G},G$ which are
functions of
frequency, temperature, field, momentum, etc.  The ${\cal G}$ Green's
function
denotes a ``heavy
particle'' which shall correspond to the localized pseudo-spin
variables or
mobile heavy particles
 in our models,
and the $G$ Green's function corresponds to a ``light particle'' which
shall be
the conduction electrons or light band electrons in our models.  In the
interaction terms of interest in our model Hamiltonians, a light
particle is
scattered by a heavy particle (Fig. ~\ref{fig1p2}).  The bare scattering
interaction
between the particles (the coupling which appears in the Hamiltonian)
is given
by $V^{(i)}$, and the corresponding vertex which includes all multiple
scattering processes is given by $\Gamma^{(i)}$.

Now, consider the following transformation of the Green's functions and
couplings, where $z_1,z_2,$ and $z^{(i)}$ are arbitrary numbers:
\[ (3.2.4) ~~~~~~~ \begin{array}{c}
			G \to z_1 G \\
		 {\cal G} \to z_2 {\cal G}\\
		   \Gamma^{(i)} \to (z^{(i)})^{-1}\Gamma^{(i)}\\
		   V^{(i)} \to  (z_1 z_2)^{(-1)} z^{(i)} V^{(i)}
		   \end{array}~~.\]
Then in any internal point of the general scattering diagram {\it
beyond lowest
(bare) order}, the $z$-factors cancel out because any higher than
lowest
order diagram may
be expressed as powers of the factor
$$ V^{(i)}G{\cal G}\Gamma^{(i)} \leqno(3.2.5) $$
which is manifestly invariant under the above transformation.  So far,
this
statement of the invariance under the rescaling specified by Eq.
(3.2.4) is
totally topological and general in character, and devoid of any
specific
contact to the problems of interest.  What we shall proceed to
demonstrate is
that the rescaling of the conduction bandwidth in our models will
generate a
transformation of the form Eq. (3.2.4).

The main idea of the renormalization group, as emphasized in the
previous
subsection, is that the particle or hole excitations with large energy
values
do not directly participate in real processes at low energies.  They
only have
an effect through virtual excitations of the low energy states to high
energies
through the scattering process.   These high energy virtual processes
may be taken into
account for the low energy electrons by introducing renormalized
parameters (or
couplings $V^{(i)}$, Fermi velocities etc.) which sum up the high
energy
processes above a certain cutoff in energy.  The goal will be to
describe the
model for all energies below a new cutoff $D'$ instead of the bare
cutoff $D$
 by integrating out the virtual
excitations to states between $D'$ and $D$ and compensating by
adjusting the
couplings so that the same physics remains, i.e., Eq. (3.2.5) remains
invariant.

\begin{figure}
\parindent=2in
\indent{
\epsfxsize=4in
\epsffile{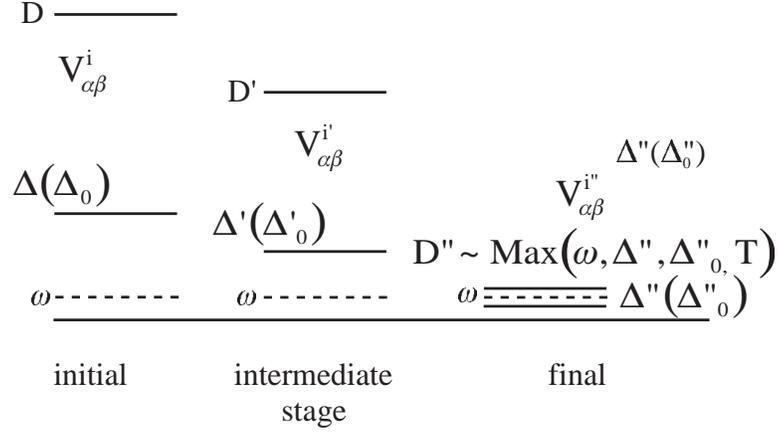}}
\parindent=0.5in
\caption{ General scheme of the renormalization group.  The
bandwidth cutoff is reduced step
by step.  The parameters like $\Delta$ and $\Delta_0$ must be changed
also with the
coupling. The procedure stops when the new cutoff reaches one of the
parameters like
$\omega,T$ or the renormalized $\Delta,\Delta_0$ values.}
\label{fig3p5}
\end{figure}

It is useful to conceptualize the smaller bandwidth as proportional
to  the
temperature $T$ if that serves as the infrared cutoff in our
perturbation expansion.
As we lower the temperature of the system from order $D$,
only those real excitations with significant Boltzmann weights, i.e.,
$E_{ex} \le
10k_BT\simeq D'(T)$, say,  will contribute significantly to the low
energy physics.  On the
other hand, through the scattering of Fig. ~\ref{fig1p2}, 
processes with energies
above
$D'(T)$ will contribute to the low energy physics. We are not free to
simply
include the real processes within $D'(T)$ to get the physics correctly,
so we
must adjust the Hamiltonian appropriate to computing properties from
purely real processes within $D'(T)$ by adjusting the couplings to
reflect the
virtual excitations to higher energies.
This general scheme is illustrated in Fig. ~\ref{fig3p5}.

We can illustrate this idea by considering what happens when we resum
only
leading logarithmic terms in the Kondo problem.  The resummation, as is
clear
from Eq. (3.2.2), will simply result in a geometric series-like formula
 times the leading power $g$ of
the dimensionless coupling.  Taking the temperature as the infrared
cutoff, the
factor obtained in the geometric series is
$F(g,D,T) =(1-g\log(D/T))^{-1}$.  We may rewrite this expression in a
trivial way that
shows the effect of the bandwidth rescaling:
$${1\over 1-g\log({D\over T})} = {1\over 1-g[\log({D'\over T}) +
\log({D\over
D'})] } \leqno(3.2.6)$$
$$~~~~~~~= z(g,{D \over D'}) {1 \over 1-g'\log({D'\over T})} $$
where
$$z(g,{D\over D'}) = {1 \over 1 - g\log({D \over D'})} \leqno(3.2.7)$$
and
$$g' = z(g,{D \over D'}) g ~~.\leqno(3.2.8)$$
Thus the original analytical form of $F(g,D,T)$ may be preserved with
the new
parameters $g',D'$ instead of $g,D$, provided we multiply by the factor
$z(g,D/D')$ which depends upon $g,D$ but not $T$.

This behavior under bandwidth rescaling makes it possible to eliminate
the
virtual high energy phase space in the framework of the renormalization
group
and to introduce new effective couplings.  Note that while the
rescaling
specified by Eqs. (3.2.4-5) is always possible, it is {\it not} always
possible
to find models which satisfy these properties in the multiplicative
form
manifest in Eqs. (3.2.6-8) characteristic of Kondo models.  Thus, the
ability
to define and implement a multiplicative renormalization group
transformation
is only realizable
for particular classes of models.
In many problems, there are further parameters which might be changed
as the
width of the
excitation spectrum is reduced (virtual excitations integrated out)
such as the
mass, Fermi velocity, etc.

Note that while the rescaling of the bandwidth corresponds to
redefining the
scale of energy on which we measure, we may also think of it as
rescaling the
length.  This fact follows from the one-dimensional character of the
impurity
models we are considering in this paper, where only the radial
direction is
important, so that energy and momentum are related through the
conversion
constant $v_F$.  Thus, when we talk about looking at problems on lower
temperature or energy scales, we are equivalently talking about looking
at
longer length scales.

The change in effective coupling strength may be cast in the form of a
differential equation by reducing the bandwidth $D\to D'$
infinitesimally.  The
general form of the differential equation is
$$ {\partial V^{(i)} \over \partial \log D} = \beta(V^{(i)})
\leqno(3.2.9)$$
which defines the $\beta$-function of the multiplicative
renormalization group.
The $\beta$ function will always vanish when the couplings vanish, and
tend to
infinity when the coupling strength tends to infinity, though it need
not
interpolate between these two points monotonically.
Eq. (3.2.9) is quite general, but we shall typically approximate the
$\beta$-function by a low order polynomial obtained from perturbation
theory.

The perturbative approach breaks down when the couplings $V^{(i)}$
expressed
in dimensionless form (measuring them in units of the
bare conduction bandwidth $D$) grow to order unity. We define  the
value of the
renormalized bandwidth $D$ or the equivalent temperature $T$ where the
couplings grow to order unity as the Kondo Temperature, $T_K$.  This is
a kind
of crossover temperature, which separates the high energy regime for
which
perturbation theory is valid, from the low energy regime where the
electrons
and heavy particles are strongly coupled.

If the $\beta$-function obtains zero for some value of couplings,
$(V^{(i)})^*$, then under further rescaling at this value, the
couplings will
remain unchanged.  That point in the space of couplings is called a
{\it fixed
point}, as we have discussed in Sec. 3.1.1.  Usually, in the models of
interest
here, that point is outside the regime of perturbative scaling theory
apart
from the trivial fixed point where all $V^{(i)}=0$ which corresponds to
decoupled heavy and light particles.  As an example, since the beta
function
diverges as $V^{(i)}$ diverge, infinite coupling strength will also
always be
a fixed point.   We may
artificially tune certain parameters, such as the channel number $M$,
to
bring the fixed points into the perturbative regime.  In that case, as
we shall
demonstrate explicitly, for isotropic couplings (all $V^{(i)}$ equal)
the $\beta$-function has the form (in terms of
dimensionless couplings $g<0$ )
$$\beta(g) = g^2 + {M\over 2} g^3 + .....\leqno(3.2.10)$$
so that a fixed point at $g^*=-2/M$ occurs in the perturbative regime
for
$M>>1$.  Wilson and Fisher introduced the $\epsilon=4-d$, $d$ the
spatial
dimensionality, to obtain a similar perturbative control in problems
where the
renormalization group is applied to critical phenomena [K.G. Wilson and
M.E.
Fisher, 1969].
Otherwise {\it
non-perturbative} methods must be employed to evaluate the
$\beta$-function or
perform equivalent calculations.

The crossover temperature scale calculated from the scaling equation
depends
upon the accuracy to which the $\beta$-function is estimated.  For
example, in
the isotropic Kondo problem with antiferromagnetic exchange coupling
$J$, we have
$$k_BT_K = D\exp[{1\over \rho_0 J} + {1\over 2}\log(|\rho_0 J|) +
P(\rho_0 J)] \leqno(3.2.11)$$
$$~~~~~ \approx D|\rho_0J|^{1/2} \exp[{1\over \rho_0J}] $$
where $P(x)$ is a polynomial, so that for $|\rho_0J|<<1$, we may
neglect
$P(\rho_0J)$ compared to the singular terms in the exponent.  These
leading two
terms will be shown to derive from the leading and next leading
logarithmic
approximations to the perturbation theory which provide terms of order
$J^2,J^3$ in the $\beta$-function.  Thus, provided we begin with a
small bare coupling strength  meaning $|J|<<D$, we need only the
leading two
orders in the perturbative expansion of $\beta(J)$ to obtain an
accurate
estimate of the Kondo scale.

\subsection{Scaling in Leading Logarithmic Order} 

In this subsection, we shall present a number of results of physical
interest
readily derivable from leading logarithmic scaling. Since there are
rather
different physics issues  to be addressed in each model case, we shall
organize
the derivations by reference to the Hamiltonian being studied.  We
shall spend
the most time, however, on the TLS model which displays the broadest
range of
coupling space to which the method can be applied.  We note that
Anderson's
``poor man's scaling'' [Anderson, 1970]
is entirely equivalent to the results presented here, and that this
approach
may be generalized beyond the leading logarithmic order (S\'{o}lyom and
Zawadowski, [1974]).  We also note that at this order, there is no
distinction
between single- and multi-channel Kondo models; the channel number $M$
enters
at next leading order.  Before we begin discussing the physics of each
model,
we will introduce the Abrikosov pseudo-particle trick (Abrikosov, [1965]),
together with the generalization by Barnes [1976], which is useful in
evaluation of perturbation theory diagrams.

\subsubsection{Pseudo-particle method} 

We wish to treat the dynamics of the local pseudo-spin variable as
manifested
in the Green's function ${\cal G}$ discussed in the previous subsection
 in a manner
which as much as possible resembles that of mobile particles with
access to the
full range of occupancies.  To do so, we follow Abrikosov [1965].
Namely, when
dealing with a local spin variable, we choose to represent the spin in
a
fermionic form, {\it viz.}
$$\sigma^i = \sum_{\mu,\nu}\sigma^i_{\mu,\nu}f^{\dag}_{\mu}f_{\nu}
\leqno(3.3.1)$$
where $f^{\dag}_{\mu}$ creates a local fermion of spin index $\mu$.
However,
the spin variable can only have two states, corresponding to the single
occupied states of the local fermion.  The correspondence of states is
$$|\mu> = f^{\dag}_{\mu}|0_{ps}> \leqno(3.3.2)$$
where $|0_{ps}>$ is the vacuum of the pseudo-fermion.
Thus, to faithfully represent the
Hilbert space of the local spin variable, we must add to our
Hamiltonian a
fictitious chemical potential term
$$H_{pseudo} = -\lambda_{ps}(\sum_{\mu} f^{\dag}_{\mu}f_{\mu} -1)
\leqno(3.3.3)$$
in which the chemical potential $\lambda_{ps}$ is taken to $-\infty$ at
the end
of calculations to project to the physical subspace where the fermion
occupancy
is unity. In so doing, we must also shift the arguments of all
pseudo-fermion
frequencies to be measured with respect to the infinitely negative
chemical
potential $\lambda_{ps}$.  The one pseudo-particle state has a
statistical
weight $(2S+1) \exp(\beta\lambda_{ps})$ with $S=1/2$, and thus physical
quantities should be normalized by this factor.  The normalization
factor is
not affected by renormalization in low orders (Black, \vld,
and Zawadowski [1982]).  Other than the minor nuisance of the
projection,
all the standard
rules of Feynman perturbation theory in the interaction may be carried
out
since the bare Hamiltonian is quadratic in the fermion operators.

 Thus, prior to projection, the pseudo-fermion propagator in
Matusubara frequency has the form
$${\cal G}^{(\lambda_{ps})}_{\mu}(\omega) = -\int_0^{\beta} d\tau
e^{i\omega\tau}
<T_{\tau}f_{\mu}(\tau)f^{\dag}_{\mu}(0)> = {1 \over i\omega-E_{\mu}
+\lambda_{ps}} \leqno(3.3.4)$$
where $\omega=2\pi k_BT(n+1/2)$ is a Fermion Matsubara frequency.  The
superscript on ${\cal G}$
is a mnemonic notation indicating that the projection and frequency
shift have not yet taken place.  We shall always drop these
superscripts after
projection.

For the purposes of discussing the rescaling of   the one particle
Green's function of the pseudo-fermion spin variable, and the
two-particle
Green's function describing the localized spin and the conduction
electrons,
this prescription alone is sufficient and has the effect of suppressing
the
unphysical states where the fermion level is empty or doubly occupied.
However, when we wish to calculate measurable properties, we must
explicitly
project to the occupancy 1 subspace with a normalization factor that
depends
upon $\lambda_{ps}$.  We defer discussion of this point to Section 4
where we
discuss the NCA method.

In the case of the Anderson Hamiltonian, discussed in conjunction with
the
models for \ufp and \ctp ions, we have a larger local Hilbert space.
For
example, in the Anderson model for the \ufp ion, we may fluctuate
between the
$f^1$ states labelled by magnetic index $\mu$ and the $f^2$ states
labelled by
quadrupolar index $\alpha$.  In this case, we follow Barnes [1976], who
pointed
out that while you may retain the pseudo-fermion label for the ground
configuration, you can generalize by describing the excited
configuration by a
bosonic variable so as to retain a net fermionic character to the
change of
configuration.  Hence, in this example, we augment the pseudo-particle
space by
a pseudo-boson operator $b^{\dag}_{\mu}$ with the state correspondence
$$b^{\dag}_{\mu}|0_{ps}> = |f^1,\gse,\mu>~~; f^{\dag}_{\alpha}|0_{ps}>
=
|f^2,\gth,\alpha> ~~.\leqno(3.3.5)$$
Correponding to the pseudo-fermion occupancy in the previous paragraph,
we
define the ``f-charge'' as
$$Q_f = \sum_{\alpha} f^{\dag}_{\alpha}f_{\alpha} + \sum_{\mu}
b^{\dag}_{\mu}b_{\mu} \leqno(3.3.6)$$
which commutes with the Hamiltonian of Eq. (2.2.19), in which we write
the
hybridization term as
$$H_{hyb} = -{V\over \sqrt{N_s}} \sum_{k\alpha\mu}sgn(\mu)[
f^{\dag}_{\alpha}b_{-\mu}c_{k,8,\alpha,\mu} +h.c.] ~~.\leqno(3.3.7)$$
Physically this commutation is obvious, since each time we create a
pseudo-boson, we destroy a pseudo-fermion, or vice versa.  We then add
a
fictitious chemical potential term of the form
$$H_{pseudo} = -\lambda_{ps}(Q_f - 1) \leqno(3.3.9)$$
and take $\lambda_{ps}\to -\infty$ at the end of all calculations.  As
alluded
to in the last part of Sec. 2, we may regard the pseudo-boson as being
the
boson mediating the exchange between the heavy particle and conduction
electrons; in the Kondo form, its dynamics is completely generated by
the
states of the electronic system, while in the Anderson form it has
independent
meaning as an excited configuration state.  The boson carries the
labels of the
channel index of the conduction electrons, the fermion the label of the
spin
index.

Thus, when working in the Anderson formulation where Eq. (3.3.7)
specifies the
perturbation term, we must also introduce the pseudo-boson Green's
function
$${\cal G}^{(\lambda_{ps})}_{\alpha}(\nu) = -\int_0^{\beta}d\tau
<T_{\tau}b_{\alpha}(\tau)b^{\dag}_{\alpha}(0)> = {1\over i\nu -
E_{\mu}+\lambda_{ps}} \leqno(3.3.10)$$
where $\nu = 2\pi k_BT n$ is a bosonic Matsubara frequency, and again
the
superscript means prior to any projection procedure.

Note that we are completely free to interchange the boson and fermion
representations for the two configurations since on carrying out the
projection
only Boltzmann statistics remains.  We must however always ensure that
one
configuration is fermionic and one bosonic to guarantee that the
electron
addition and removal processes are fermionic in character, i.e., so
that each
hybridization event conserves fermion number. We also observe that the
extension to inclusion of higher fermion or boson excited levels is
straightforward:  we must simply add the occupancy factors for the new
states to the conserved charge $Q_f$.  This is true in the case of the
excited states for the tunneling center as well (see Sec. 3.4.2).

\subsubsection{Leading Order Scaling in the TLS Model} 

The idea of the leading order scaling
is based upon the second order perturbation theory diagrams already
discussed in the introduction and displayed in Fig. ~\ref{fig1p2}.  Referring
back to
that figure, we see that the lowest order correction to the
two-particle
interaction between the local spin and conduction electrons is either
of the
form where the intermediate state includes an excited particle (Fig.
\ref{fig1p2}(a))
or hole
\ref{fig1p2}(b)).
This holds for all of the models of interest in
our review.

Turning to the TLS model with couplings specified by Eq. (2.1.7), we
see that
each of these diagrams contributes a logarithmic dependence.  To
evaluate, we
represent the dashed line of the diagram by the pseudo-fermion
propagator of
Eq. (3.3.4).  We see that the sign of the logarithmic corrections is
different,
and also the ordering of the spin matrices and momentum dependent
couplings is
different.  The total coefficient of the diagram corresponding to Fig.
\ref{fig1p2}(a) is
$$\sum_{i,j} V^i_{\vec k_2,\vec k}V^j_{\vec k,\vec k_1} \sigma^i
\sigma^j$$
where $k_{1,2}$ are the incident(outgoing) momenta, $k$ is the internal
momentum, and we see that the coefficient of the term corresponding to
Fig.
\ref{fig1p2}(b) is given by
$$\sum_{i,j} V^i_{\vec k,\vec k_1}V^j_{\vec k_2,\vec k} \sigma^i
\sigma^j $$
so that the conduction momentum indices are reversed on the second
term.  If we
now evaluate the total two-particle scattering amplitude corresponding
to Fig.
\ref{fig3p5} to second order, at zero temperature, we find
$$T^m_{\hat k_2,\hat k_1}(\omega,D,V^i) = V^m_{\hat k_2,\hat k_1} -
2i\rho_0
\sum_{i,j} \int {dS_F(\hat k) \over S_F} [V^i_{\hat k_2,\hat
k}V^j_{\hat k,\hat
k_1} \epsilon^{ijm} \log({D\over \omega})]  \leqno(3.3.11)$$
where $dS_F(\hat k)$ is an element of Fermi surface area in the
direction of
$\hat k$ of the total Fermi surface area $4\pi k_F^2$, we have
neglected the radial
dependence of the matrix elements upon
$k$, and the Levi-Civitta symbol arises from the algebra of the Pauli
matrices.
We have taken the density of states to be constant within the band of
full
width $2D$.

The change in the bandwidth $D \to D'$ discussed in the previous
subsection can
be compensated by readjusting the couplings $V^i$. That scaling
dependence may be written as
$$ {\partial T^m_{\hat k_2,\hat k_1} \over \partial D} dD +
\sum_{i,\hat k,\hat
k'} {\partial
T^m_{\hat k_2,\hat k_1}\over \partial V^i_{\hat k,\hat k'}} dV^i_{\hat
k,\hat
k'} = 0 \leqno(3.3.12)$$
which implies that the equation for the rescaled coupling constants is
given by
$${\partial V^m_{\vec k_2,\vec k_1} \over \partial lnD} = 2i\rho_0
\sum_{i,j}\epsilon^{ijm}
\int {dS_F(\vec k) \over S_F} V^i_{\hat k_2,\hat k}(D)V^j_{\hat k,\hat
k_1}(D)
~~.\leqno(3.3.13)$$
The logic of (3.3.12) is that the physics may be specified in terms of
the
amplitudes of the $T$-matrices for 1,2,3,... particle scattering.  Thus
to
ensure that the physics is unchanged, we must ensure that after
renormalization
we obtain the same T-matrix which means that in lowering the cutoff
which
explicitly affects the second term we must modify the couplings
accordingly.

Following Zawadowski [1980],
we may write this equation in a more convenient form by using the
matrix
representation $V^m_{\alpha,\alpha'}$ given in Eq. (2.1.27), and by
introducing
the dimensionless variable $x=\log(D_0/D)$ where $D_0$ is the original
bare
bandwidth and $D$ the rescaled bandwidth.  Then the leading order
scaling
equation is
$${\partial V^m_{\alpha,\alpha'} \over \partial x} = -2i\sum_{i,j}
\epsilon^{ijm} \sum_{\gamma}
V^i_{\alpha,\gamma}(x)V^j_{\gamma,\alpha'}(x) \leqno(3.3.13)$$
subject to the boundary condition that we match the bare couplings at
the
original bandwidth, so that
$$V^i_{\alpha,\alpha'}(0) = V^i_{\alpha,\alpha'} \leqno(3.3.14)$$
where the r.h.s. of the last equation just contains the couplings that
appear
in the Hamiltonian.

We shall now list several characteristic features of the equations for
the
leading order scaling:\\
{\it (i) Lack of dependence on the channel number $M$}.
In the derivation of Eq. (3.3.13), we used the diagrams of Fig.
\ref{fig1p2}.
Notice
that since
these diagrams have no closed loops,
all dependence on the channel index is through the label $\alpha$
which is conserved throughout the scattering.  At next leading order,
we will
obtain a closed electron loop inside the diagram and thus pick up a
dependence
upon the number of channels. 

{\it (ii) Results for a two-dimensional subspace:  irrelevance of
anisotropy}.  Let
us for a moment accept that the dominant space of conduction electron
orbital
indices will be two-dimensional as we scale from the band edge, a point
we
shall prove shortly.  Given this assumption, we shall illustrate that
exchange
anisotropy is irrelevant as we scale, and that if we begin scaling in a
subspace where just two couplings are non-zero, that independent of
their sign
we will ultimately flow towards the isotropic, antiferromagnetic
coupling
regime as we rescale.

As we have already seen in Sec. 2.1, even when restricted to the two
component space of orbital indices, the quantization axis of the
electron
pseudo-spin need not align with the quantization axis of the TLS
pseudo-spin.
This means that the most general form of the coupling matrices in a two
dimensional subspace is
$$V^i_{\alpha,\alpha'} = V^i_{\tilde x}\sigma^{\tilde
x}_{\alpha,\alpha'} +
V^i_{\tilde y}\sigma^{\tilde y}_{\alpha,\alpha'}+V^i_{\tilde
z}\sigma^{\tilde
z}_{\alpha,\alpha'} \leqno(3.3.15)$$
where we remind the reader that $i=x,y,z$ are the directions in the TLS
pseudo-spin space for quantization axis $z$, and $\tilde x,\tilde
y,\tilde z$
are the directions in the conduction orbital pseudo-spin space for
quantization
axis $\tilde z$. Now the scaling equations will determine the nine
coupling
coefficients $V^i_{\beta}$.   With this {\it Ansatz}, the leading order
scaling equation
has the form
$${\partial V^m_{\gamma} \over \partial x} = -2\rho_0
V^i_{\alpha}V^j_{\beta}\epsilon^{ijm}\epsilon_{\alpha\beta\gamma}
~~.\leqno(3.3.16)$$

Let us now view the couplings as vectors in the space of conduction
pseudo-spin. Thus, $\vec V^i=(V^i_{\tilde x},V^i_{\tilde y},V^i_{\tilde
j})$
and, for example,
$$\vec V^p \cdot \vec V^s = \sum_{\alpha} V^p_{\alpha}V^s_{\alpha}
~~.\leqno(3.3.17)$$
Using the properties of the Levi-Civitta symbols, it is straightforward
to show
that
$${\partial (\vec V^p \cdot \vec V^s) \over \partial x} = \delta_{ps}
[-8\rho_0 (\vec V^x\times\vec V^y)\cdot \vec V^z]~~. \leqno(3.3.18)$$
We should make a couple of notes about the derivation of Eq.
(3.3.18):\\
(i) The factor of eight follows from one factor of two when
differentiating for
$p=s$, one factor of two already present in the scaling equation, and
one
factor of two for the two possible orderings of $i,j$ in
$\epsilon^{ijp}$ to
give a non-vanishing contribution.\\
(ii) We used the invariance of $(\vec V^i\times \vec V^j)\cdot \vec
V^k$
upon
permuting $i,j,k$ cyclicly to write the RHS of (3.3.18) in a form
manifestly
independent of the choice of $p$.

From Eq. (3.3.18), we may infer that the magnitude of each coupling
constant
vector is independent of its index in the TLS space.  So long as the
product
 $(\vec V^i\times \vec V^j)\cdot \vec V^k>0$, the vectors $\vec V^p$
 increase in
length as $D$ is reduced, and the ratio of the couplings $|\vec
V^p|/|\vec
V^s|$ tends to unity.  Further, since the RHS of Eq. (3.3.18) vanishes
if $p\ne
s$ and since the lengths of the vectors grow towards $\infty$, we see
that the
vectors must become perpendicular upon scaling.  If we start with
$(\vec V^i\times \vec V^j)\cdot \vec V^k<0$, the couplings tend
towards zero, like the ferromagnetic Kondo model.

\begin{figure}
\parindent=2.in
\indent{
\epsfxsize=2.5in
\epsffile{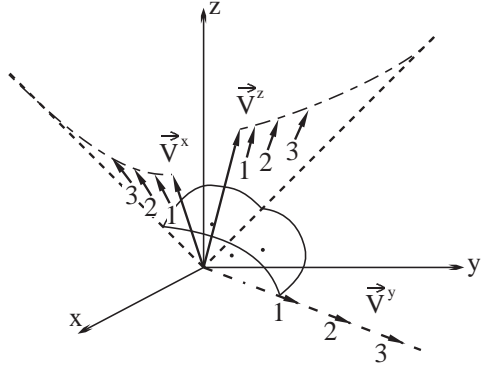}}
\parindent=0.5in
\caption{The scaling trajectories are depicted for Eq. (3.3.16)
in the real space vaiable related
to the lower indices.  The starting vectors $\vec V^x$ and $\vec V^z$
are shown by solid lines,
while $\vec V^y$ starst from $\vec V^y=0$.   $\vec V^x$ and $\vec V^z$
are scaled in the plane
determined by them and $\vec V^y$ is moving in the normal direction to
that plane.  The scaling
trajectories are depicted b dash-dotted lines, the asymptotes of
vectors $\vec V^x$ and
$\vec V^z$ are represented by dotted lines.  The arrows indicatedthe
vectors at different
stages of the scaling, labeled by numbers.  From \zow [1980].}
\label{fig3p6}
\end{figure}

Hence, once we restrict the
conduction orbital degrees of freedom to the
two component space we see that the scaling drives the system towards
isotropic
coupling, allowing a simple rotation in the conduction orbital
pseudo-spin
space to give the Heisenberg form to the exchange coupling between the
TLS and
the electrons.  Given the appropriate sign to the couplings, the
effective
Hamiltonian  form at full
strong coupling is ($V_{iso}(D) = |\vec V^p(D)|$)
$$H_{int}(D) = V_{iso}(D)\vec \sigma_{TLS} \cdot (\vec
\sigma_{c\uparrow}(0) +\vec
\sigma_{c\downarrow}(0)) \leqno(3.3.19)$$
where the quantization axis of the conduction electrons has been
rotated to
line up with the TLS axis, and the real spin labels have been put on
the
conduction spin densities at the TLS site.  Since the coupling is
written
without the conventional minus sign of the Kondo model, our positive
choice of
coupling coencides with the antiferromagnetic Kondo model.
A schematic of the generic scaling trajectories in the coupling space
for this
antiferromagnetic case  are illustrated in Fig.
\ref{fig3p6}.

In the original TLS model, we have $|V^z| >> |V^x|,V^y=0$.  In this
case, upon
initially scaling we will generate $V^y \sim V^z V^x$.  Thus the triple
product
 $(\vec V^i\times \vec V^j)\cdot V^k\sim |V^xV^z|^2 >0$ independent of
 the
signs of $V^x,V^z$ initially.  That means we will {\it always} scale to
strong
coupling (towards the isotropic antiferromagnetic exchange model)
in this extreme anisotropic limit.  

{\it (iii)  Reduction to the two-dimensional subspace}.  The scaling
equations
themselves in the extreme anisotropic limit justify the restriction to
the
two-dimensional subspace of the conduction electron orbital
pseudo-spin.
This will hold provided the couplings $\vec V^{x,y}_{\alpha,\beta}$ in
Eq.
(3.3.13) have magnitudes
much smaller than $V^z_{\alpha,\beta}$, as shown by \vld
and \zow [1983(a)].

In order to prove this, we first perform a unitary transformation
$V^z\to
\tilde V^z$ in the
conduction space such that
$$\rho_0 \tilde V^z_{\alpha,\beta}(0) = \tilde v^z_{\alpha,\beta}(0) =
\tilde
v^z_{\alpha}(0) \delta_{\alpha,\beta} ~~.\leqno(3.3.20)$$
We now linearize the RHS of Eq. (3.3.13) in the small quantities
$v^x,v^z$ to
obtain
$${\partial \tilde v^z \over \partial x} = 0 + {\cal O}(\tilde v^3)
\leqno(3.3.21.a)$$
$${\partial \tilde v^x \over \partial x} = -2i[\tilde
v^x(x),\tilde v^z(x)]_- \leqno(3.3.21.b)$$
and
$${\partial \tilde v^y \over \partial x} = 2i[\tilde v^x(x),\tilde
v^z(x)]_- \leqno(3.3.21.c)$$
where the $v's$ are understood to be matrices and the square brackets
indicate
commutators. We can separate the variables by differentiation with
respect to
$x$ again, and for $i=x,y$ we find
$${\partial^2 \tilde v^i(x) \over \partial x^2} = 4[[\tilde v^i,\tilde
v^z]_-,\tilde v^z]_-~~. \leqno(3.3.22)$$
Using the diagonal form and the fact that $\tilde v^z$ is unchanged in
this linearized form, we see that (putting matrix indices back in)
$${\partial^2 \tilde v^i_{\alpha,\beta} \over \partial x^2} =
4\tilde v^i_{\alpha,\beta}(x) [\tilde v^z_{\alpha}(0)-\tilde
v^z_{\beta}(0)]^2
~~.\leqno(3.3.23)$$
Given the boundary condition $\tilde v^y(0)=0$, the solution to the
linearized
equations may be given as
$$\tilde v^z_{\alpha,\beta}(x) = \tilde
v^z_{\alpha}(0)\delta_{\alpha,\beta} ~~,\leqno(3.3.24.a)$$
$$\tilde v^x_{\alpha,\beta}(x) = \tilde v^x_{\alpha,\beta}(0)
cosh[2(\tilde
v^z_{\alpha}(0) - \tilde v^z_{\beta}(0))x] ~~, \leqno(3.3.24.b)$$
and
$$\tilde v^y_{\alpha,\beta}(x) = i\tilde v^x_{\alpha,\beta}(0)
sinh[2(\tilde
v^z_{\beta}(0) - \tilde v^z_{\alpha}(0))x] ~~. \leqno(3.3.24.c)$$

What we learn from Eqs. (3.3.24.a-c) is that barring unforeseen
degeneracies in
the matrix $v^z$, whichever two elements of $\tilde v^z(0)$ which
produce the
largest difference $|\tilde v^z_{\alpha}(0)-\tilde v^z_{\beta}(0)|$
will
produce the most rapid growth of the $\tilde v^{x,y}$ elements of the
coupling.
Because the functions grow {\it exponentially} with rescaling, any
other
splittings with even slightly smaller differences will grow negligibly
fast
upon scaling compared to the dominant subspace.  In our TLS model, the
axial
character will always in practice restrict this dominant subspace to a
linear
combination of conduction orbitals with $m=0$ and $l=0,1,2$.

We stress that this argument depended upon the assumption of extreme
anisotropy
with zero initial coupling $V^y$, and that it has only been carried out
so far
in the context of second order scaling.  We shall demonstrate that it
is also
valid if we go to all orders in $V^z$ while remaining at lowest order
in
$V^{x,y}$.  We conjecture that it is valid even in instances where the
couplings are closer to isotropic and the perturbative approach breaks
down,
but this remains to be proven.  As we shall explain, this result may
have
important implications for the Kondo models of \ufp~ and \ctp~ ions.
In
our
discussion of these ions, we shall show that in the quadrupolar Kondo
case the
two-channel coupling to an additional symmetry allowed $\gei$ quartet
is
irrelevant through the two leading orders of scaling theory.

In fact, there is a symmetry based reason for this flow to a dominant
two-dimensional subspace which holds at strong coupling.  Namely, at
strong
coupling we have ample evidence that the exchange anisotropy vanishes.
In this
case, one can write down an $SU(2)$ invariant coupling deriving from
the
2-dimensional manifold of the local pseudo-spin.   We may then decouple
this
exchange via a Hubbard-Stratonivich transformation, as discussed in
Sec. 2.3,
and find that there are only two ``exchange fluctuation'' fields,
indexed by the
channel label, which are necessary to decouple the entire interaction.
However,
there is only a 1:1 match of the pseudo-spin index--a bonding
combination of
conduction operators will be selected which has a 1:1 match in
pseudo-spin space
with the labels of the TLS.  The antibonding combination will
decouple.  We
shall demonstrate this explicitly in the quadrupolar Kondo case.

{\it (iv) General Form of the Fixed Point Hamiltonian}.  We will now
offer some
discussion of the form of the fixed point Hamiltonian for our
two-channel Kondo
models, although the discussion should be regarded as extrapolation at
this
point of the review, because the actual fixed point lies outside the
perturbative regime.  The point is that the perturbative scaling
equations at
lowest order provide considerable insight into the structure of the
fixed point
physics.  This kind of analysis is supported in detail by the
considerations of
the numerical renormalization group and conformal field theory
calculations.

Given the flow towards isotropy
as the couplings grow, following \vld~
 and \zow~ [1983(a)]
we may write the couplings in the separated form
$$v^i_{\alpha,\beta}(x) = v^i(x) \tilde V^i_{\alpha,\beta}
\leqno(3.3.25)$$
where the first factor contains the scale dependence and the second
factor
the matrix structure which does not change upon further scaling.  Then
the scaling
equations split into an equation describing the
scaling, and one describing the operator algebra near the fixed point.
The
scaling equation is
$${\partial v^m(x)\over \partial x} = 2v^i(x)v^j(x) ~~, i,j,m~cyclic
~~,
\leqno(3.3.26)$$
and the matrices are specified by the algebra
$$\tilde V^m_{\alpha,\beta} = \sum_{i,j,\gamma} \tilde
V^i_{\alpha,\gamma}
\tilde V^j_{\gamma,\beta} \epsilon^{ijm} ~~,\leqno(3.3.27)$$
from which it is clear we simply scale to some irrep of
$SU(2)$.  Of course, it is possible that in the case of matrices of
higher order,
in a subspace the couplings follow $SU(2)$ symmetry while in
the remaining subspace
the coupling disappears.  It has been shown by \zar [1995]
that the
representation of $SU(2)$ is always two dimensional for
$M>>1$ except in the presence of extra symmetry.  We discuss this
further in
Secs. 3.3.3, 3.4.3 for rare earth and actinide impurity models.
The analysis in the preceding note (iii) above makes
clear we
expect this to be
generally a two dimensional representation to be dominant.
The solution of Eq. (3.3.26) may be written as
$$v^i(x)^2 = \psi^2(x) + v^i(x_0)^2 ~~,\leqno(3.3.28)$$
where
$${\partial \psi^2(x)\over \partial x} = 4v^x v^y v^z \leqno(3.3.29)$$
with boundary condition
$\psi^2(x_0) = 0$.

What is intriguing about this scenario is that we begin with an
apparently
artificial spin variable describing the TLS which is not conserved,
even if we
shut off the spontaneous tunneling terms.  In examining the scaling of
the
exchange terms alone, we develop this picture that at the fixed point
one has a
full isotropy in the combined pseudo-spin space of the TLS plus
conduction
electrons, although this space has no obvious reason to become so
symmetric
from purely high energy considerations.

Thus, the Kondo effect serves to {\it
restore} symmetries which are not present in the bare Hamiltonian,
which is a
fairly remarkable result.  This contrasts with the usual scenario in
high
energy physics, in which symmetry is high at high temperatures and
reduced as
we lower the temperature.

 This increase of symmetry on approach to the fixed point
also holds for the quadrupolar Kondo effect,
where generically only two of the local spin tensors should transform
as an
irrep of the point group that is quadrupolar in character,
yet an extra tensor of octupolar character is generated and at the
two-channel
fixed point there is full isotropy in this unusual combined space.

We close this note  on the general properties of the fixed point with
an open
question about the nature of the scaling.  In the previous note we
argued that
when the bare couplings are extremely anisotropic, the scaling will
preferentially select a two-dimensional subspace in the conduction
orbital
pseudo-spin space as the dominant one.  What is not clear is whether
the couplings in the excluded space will grow or shrink as well when
one moves
out of the perturbative regime and into the fixed point region.  This
growth is
ruled out by the arguments of \zar [1995] for $M>>1$.  While
it is
physically unlikely that such growth occurs for the physically relevant
$M$=2
case,  no rigorous proof yet exists to support this intuition.  \\

{\it (v) Leading order estimate of the Kondo scale.  }\\

If we first consider the fully isotropic limit, we will have to
integrate the
leading order differential equation
$${\partial v(x) \over \partial x} = 4 v(x)^2 \leqno(3.3.30)$$
and identify where the coupling strength grows unity.  This yields the
estimate
$$T_K^{(I)} \approx D_0 \exp(-{1\over 4v(0)}) \leqno(3.3.31)$$
where the superscript $(I)$ denotes leading order.  If we were to use
instead
spin 1/2 matrices, we would replace the exponent by $-1/v(0)$.

In the TLS case where the starting parameters are extremely
anisotropic,
the perturbation theory will break down and the couplings grow to order
unity
when the energy scale from the leading order equations is such that
$v^x(x)\approx v^z(0)$, which produces the leading order estimate for
the Kondo
scale, $T_K^{(I)}$ as
$$T_K^{(I)} \approx D_0 ({v^x(0)\over 4v^z(0)})^{1\over 4v^z(0)}
\leqno(3.3.32)$$
when we use Pauli matrices (see \vld~
and \zow~ [1983(a)]).  If we use spin 1/2 matrices, this expression
changes only in that the singular exponent is modified
according to
$${1 \over 4v^z(0)} \to {1\over v^z(0)} ~~.$$

\subsubsection{Leading Order Scaling for model \ufp~ and \ctp~ ions} 

Much of the work has already been done in the development of the
equations for
the TLS.  We shall present some useful generalizations of these
equations
appropriate to the physics of the models for \ufp and \ctp ions.\\

{\it(a) Leading order Kondo scale}\\

This is the simplest modification to the TLS result. We utilize the
isotropic
result of Eq. (3.3.31) and replace $4v(0)$ by $|g(0)|=N(0)J$, where $J$
is the
maximum (in magnitude) of $J_7,J_8$ in the \ctp~ case.  Hence
$$T_K^{(I)} = D_0 \exp({1\over g(0)})~~.  \leqno(3.3.33)$$\\

{\it(b) Flow to two-dimensional subspace for \ufp ion in cubic
symmetry}\\

We could have generalized the discussion of the quadrupolar Kondo
Hamiltonian in
Sec. 2.2.1 to have allowed for exchange coupling to more than one
$\gei$
quartet.   What we shall demonstrate presently is that at leading order
scaling,
this produces a flow to a dominant two-dimensional subspace as has been
found in
the extremely anisotropic TLS case above.  The interpretation is simply
that one
is selecting the relevant bonding combination of the $\gei$ states.
Our
scaling analysis follows similar reasoning developed for a different
problem by
Cragg, Lloyd, and \noz~ [1980].

The physical source of the second $\gei$ state is the $j=7/2$ angular
momentum
multiplet of the conduction electron partial wave states.  Let us
denote the
$\gei$ level from the $j=5/2$ partial wave manifold with no prime, and
that from
the $j=7/2$ multiplet with a $'$.  The exchange Hamiltonian of Eq.
(2.2.6) is
then generalized to
$$H_{ex} = {-1 \over
N_s}\sum_{l,l'=8,8'}J_{l,l'}\vti\cdot\sum_{k,k',\alpha,\alpha',\mu}\vec
\tau_{\alpha,\alpha'}c^{\dag}_{kl\alpha\mu}c_{k'l'\alpha'\mu}
~~,\leqno(3.3.34)$$
where the couplings $J_{8,8'}=J_{8',8}$.  In fact, since the bare
couplings are
determined by the Schrieffer-Wolff transformation, we may say more
about the
relative size of the bare couplings.  Assume that the strength of
hybridization
of the $8$ quartet with the excited $\gse$ and ground $\gth$ doublets
is $a_8
V_0$, where the hybridization strength $V_0$ is independent of the
labels
$8,8'$, and the corresponding hybidization strength for the $8'$
quartet is
$a_{8'}V_0$.  Then we have that
$$J_{l,l'} = a_la_l' {2V^2_0\over \tilde \ef}  \leqno(3.3.35)$$
so that, e.g, $J_{8,8'} = -\sqrt{J_{8,8} J_{8',8'}} $.  Note that the
appropriate
way to think about the above exchange interaction is in terms of a
conduction
electron spin which is large, i.e., $N_c=4$.  The reason is that the
channel
index is still the two-dimensional magnetic index, and we have simply
enlarged
the space of orbital labels which couple to the impurity.  Thus, this
model is
analogous to the TLS model with $N_c$ actually arbitrary.

\begin{figure}
\parindent=2.in
\indent{
\epsfxsize=2.5in
\epsffile{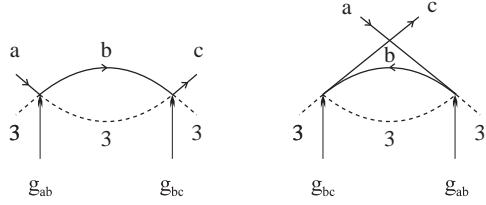}}
\parindent=0.5in
\caption{Leading order multiplicative renormalization group diagrams for the 
quadrupolar Kondo model with two conduction quartets. The lines marked with 3 are pseudo-fermion propagators
for the non-Kramers $\gth$ doublet.  The lines marked with $a,b,c$ are conduction 
electron lines which may come from either $\gei$ quartet (labeled by $a,b,c=$8,8' in the text 
and derived from the $j=5/2,7/2$
partial waves about the impurity).  The vertices are coupling strengths $g_{ab}$ which are symmetric in $a,b$ 
and 
scatter from quartet $a$ to quartet $b$.  In the simplest model, $g_{88}>g_{88'~}>g_{8'~8'~}$, and 
$g_{88'~}^2=g_{88}g_{8'~8'~}$.  }
\label{fig3p7}
\end{figure}

Using the diagrams shown in Fig. \ref{fig3p7}, we obtain the following
leading order
scaling equations for the three dimensionless couplings
$g_{8,8},g_{8,8'},g_{8',8'}$:
$${\partial g_{8,8} \over \partial x} = -g_{8,8}^2 - g_{8,8'}^2 ~~,
\leqno(3.3.36.a)$$
$${\partial g_{8,8'} \over \partial x} = -g_{8,8'}[g_{8,8}+g_{8',8'}]
~~,\leqno(3.3.36.b)$$
and
$${\partial g_{8',8'} \over \partial x} = -g_{8',8'}^2-g_{8,8'}^2
~~.\leqno(3.3.36.c)$$
Notice that we lack a factor of four compared with TLS scaling from our
use of
$S=1/2$ matrices here, and that the different sign convention has
flipped the
sign of the $\beta$-functions relative to the TLS case.

 These scaling equations may be
separated into soluble form in two stages. First,
we identify a constant of the motion.  By subtracting Eq. (3.3.36.c)
from Eq.
(3.3.36.a) and taking the ratio with Eq. (3.3.36.b), we see that
$${d (g_{8,8}-g_{8',8'})\over dg_{8,8'}} = {(g_{8,8}-g_{8',8'})\over
g_{8,8'}}
\leqno(3.3.37)$$
from which we infer that
$$g_{8,8'} = c_0(g_{8,8}-g_{8',8'})~~.  \leqno(3.3.38)$$
We may infer the constant $c_0$ from the initial conditions and Eq.
(3.3.35) as
$$c_0 = {a_8a_{8'}\over (a_8^2-a_{8'}^2)}  ~~.\leqno(3.3.39)$$

Next, we define the coefficients
$$\alpha_8= a_8^2/(a_8^2-a_{8'}^2) \leqno(3.3.40.a)$$
and
$$\alpha_{8'}=-a_{8'}^2/(a_8^2-a_{8'}^2)~~.\leqno(3.3.40.b)$$
Define the linear combinations of exchange couplings
$$\tilde g = \alpha_8 g_{8,8} + \alpha_{8'}g_{8',8'} \leqno(3.3.41.a)$$
and
$$\tilde g' = \alpha_{8'} g_{8,8} + \alpha_{8}g_{8',8'}
~~.\leqno(3.3.41.b)$$
Note that $\alpha_8 + \alpha_{8'}=1$, so that $\tilde
g(0)=g_{8,8}(0)+g_{8',8'}(0)$.
Then the scaling equations, with use of the constant of motion
identified in the
preceding paragraph, decouple giving
$${\partial \tilde g \over \partial x} = - {\tilde g}^2
\leqno(3.3.42.a)$$
and
$${\partial \tilde g' \over \partial x} = - (\tilde g')^2
~~.\leqno(3.3.42.b)$$
In principle, both couplings will grow.  However, with the use of the
initial
conditions, we see that $\tilde g'(0)=0$, so that this set of couplings
just
drops out.   This is not the case for $\tilde g$, and if we solve to
estimate
the leading order Kondo scale, we obtain
$$T_K^{(I)}(\tilde g(0))  \approx D_0 \exp({\tilde \ef \over 2[a_8^2 +
a_{8'}^2] N(0)V_0^2}) ~~. \leqno(3.3.43)$$
Clearly, we can interpret the factor in square braces in this equation
as simply
representing the normalization of the ``bonding combination'' of
$\gei$
orbitals.    We shall show in the next subsection that this result also
holds in
next leading order scaling as well.\\

{\it (c) Effect of excited crystalline electric field levels}\\

We may see using leading order scaling methods that excited crystalline
electric field levels
produce an enhancement of the Kondo scale estimated without the
excited states. Our derivation is restricted to the rescaling of the
$J_7$
exchange for the model \ctp~ ion, and follows the original analysis of
Yamada, Yosida, and Hanzawa,
[1984] and the analogous analysis for the TLS model
of
\zar and \zow~ [1994(a),1994(b)] (see Sec. 3.4.2).

\begin{figure}
\parindent=2.in
\indent{
\epsfxsize=2.5in
\epsffile{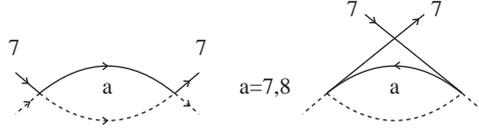}}
\parindent=0.5in
\caption{Leading order multiplicative renormalization group processes in the presence of excited 
crystal field levels for a \ctp impurity.  The dashed line is a pseudo-fermion propagator for the impurity states. 
The label $a$ refers to the intermediate states of both impurity and conduction electrons, which 
can be either $\gse$ (assumed to be the impurity ground state) or $\gei$ (assumed to be the 
impurity excited state). Assuming only $f^0-f^1$ virtual charge fluctuations, all coupling strengths 
are equal.  }
\label{fig3p8}
\end{figure}

The idea is similar to the preceeding subsection. The diagrams are
shown in
Fig. \ref{fig3p8}.
By virtually scattering into
the $\gei$ conduction and $f^1$ manifolds through the coupling
$J_{78}=J_7$
(due to the assumption of mixing only with $f^0,\gon$ excited
configuration) we
obtain the leading order scaling equation
$${\partial g_7 \over \partial x} = -g_7^2[1+2{D \over D+\Delta_8}]
\leqno(3.3.44)$$
where $\Delta_8$ is the splitting between the $f^1\gse$ and $f^1\gei$
levels.
The extra energy appears in the denominator due to the virtual
excitation to
the $\gei$ level, which is much larger than $D$ when we shrink $D$ to
order
$T_K^{(I)}$. The factor of 2 is the ratio of the degeneracies of the
$\gei$
quartet to the $\gse$ doublet, which reflects the greater number of
ways to
scatter into the $\gei$ state.

It is straightforward to integrate these equations, and we find that
$$T_K^{(I)} = D_0 ({D_0\over T_K^{(I)}+\Delta_8})^2 \exp({1\over
g_7(0)}) ~~.\leqno(3.3.45)$$
This represents a self-consistent equation to solve for the Kondo
scale.  In
the limit where $T_K^{(I)}<<\Delta_8$, we obtain an enhancement over
the Kondo
scale without crystal field excitations by the rather large amount of
$(D_0/\Delta_8)^2$.  Given $D_0\simeq 10^4K$, $\Delta_8\simeq 10^2K$)
this
enhancement can be of order 10$^4$!    This is obviously a significant
effect.
In the limit when $T_K^{(I)}>>\Delta_8$, we find that $T_K^{(I)}$ tends
to the
value of the six-fold degenerate multiplet obtained by forcing the
$\gei$ level
to be degenerate with the $\gse$ level.   We will obtain the physics of
the
ground doublet provided that the Kondo scale of the full multiplet is
smaller
than $\Delta_8$, because then the crystal field scale cuts off the
logarithmic
growth associated with the full multiplet.

Hence, the excited
crystal field state produces a
crossover between a high temperature Kondo effect associated with the
entire
multiplet and a low temperature Kondo effect associated with the ground
multiplet, but significantly enhanced over the value with
$\Delta_8=\infty$.
Within the NCA analysis of Sec. 5,
it is easy to generalize this picture from the
simple
example treated here, and below all crossovers one generically finds a
Kondo
scale enhanced by factors of $\sim (D/\Delta_{cef})^{N_{cef}/N_{grd}}$
for all
levels at energy $\Delta_{cef}$ with degeneracy $N_{cef}$ (the ground
level has
degeneracy $N_{grd}$).   Similar results obtain for each excited
spin-orbit
multiplet and $LS$ term in multielectron configurations.\\

{\it (d) Leading order scaling for a \ctp ion with spin/channel-spin
coupling
included}\\

Following Kim [1995], Kim and Cox [1995,1996,1997], and Kim, Oliveira,
and Cox [1996],
we examine the effects of the channel-spin/spin coupling developed in
Eqs. (2.2.34-2.2.38).   We restrict ourselves to the space of $\gei$
coupling
alone, i.e., neglect the scattering between $\gei$ and $\gse$
conduction
partial waves.  In this case, the coupled equations for the
dimensionless
interaction strengths $g_8 =N(0)J_8$ and $\tilde g_8 = N(0) \tilde J_8$
are
$${\partial g_8\over \partial x} = g_8^2- {1\over 2} \tilde g^2_8
\leqno(3.3.47)$$
and
$${\partial \tilde g_8 \over \partial x} = - \tilde g_8 g_8
~~.\leqno(3.3.48)$$
These equations follow from the unusual commutation relations of the
$\tilde S_{c8}$
operators specified in Eqs. (2.2.37) and (2.2.38).  What can be
verified from the
above scaling equations are the following:  \\
1) For $2 g_8> |\tilde g_8|$, $g_8$ grows and $|\tilde g_8|$ shrinks.
\\
2) For $2g_8 = |\tilde g_8|$, both coupling strengths shrink to zero.
\\
3) For $2g_8 < |\tilde g_8|$, $\tilde g_8$ grows without any sign
change,
and $g_8$ is driven towards the lines defined by $2g_8=-|\tilde
g_8|$.\\
4) For $\tilde g_8=0$ and $g_8<0$, there is no growth in $\tilde g_8$
and $g_8$
shrinks to zero.  This is of course the usual ferromagnetic case.  \\

Clearly the lines $2g_8=\pm \tilde g_8$ occupy a special place in this
scaling analysis, and the reason as indicated in Sec. 2.2 is that the
tensors
$I_{c8}^{(i)}(0)-S^{(i)}_{c8}(0)\pm 2\tilde S^{(i)}_{c8}(0)$ obey the
ordinary angular momentum
commutation relations and define a pseudo-spin 3/2 irrep.
Along these lines the effective exchange coupling is $-J_8 \vec
S_{I7}\cdot \vec I_{c8}(0)$.
It is clear that for positive $g_8=|\tilde g_8|/2$
one has ferromagnetic coupling of the pseudo-spin 3/2 conduction
operators
to the impurity spin, while when $g_8=-|\tilde g_8|$ we obtain an
antiferromagnetic
coupling of the impurity pseudo-spin to the conduction pseudo-spin.
Collapse to the
line $g_8=-\tilde g_8/2$ indicates dominant virtual fluctuations to the
$\gfo$ excited state,
while collapse to the line $g_8=\tilde g_8$ indicates dominant virtual
fluctuations to the
$\gfi$ excited state.   We note that the symmetry constraints along
these lines imply that
their special role in the coupling constant phase diagram will be
maintained non-perturbatively.

\subsection{Next Leading Order Scaling} 

The power of the multiplicative renormalization group equations is not
evident
at leading order scaling, because the most important corrections to the
wave
function renormalization factors $z$ occur at next leading order.  In
this
subsection, we shall explore the next leading order scaling equations
for the
Kondo models of interest using the multiplicative renormalization group
equations.  In parallel to the previous subsection on leading order
scaling, we
shall first present the results for the TLS model, which represents the
most
general formulation of the fully anisotropic Kondo model, and then
present
briefly results for the models of \ufp~ and \ctp~ ions.  In the last
part
of this
subsection, we will perform some stability analyses of the next leading
order
scaling equations about the non-trivial fixed point, to illustrate the
destabilizing
effects of channel fields, and the stability  against exchange
anisotropy in the
simplest cases.

\subsubsection{Next Leading Order Scaling for the TLS model} 

{\it(a) Overview}

What we seek are equations relating the original couplings
$V^{i},\Delta^{i}$ and Green's
functions
 at bandwidth $D$ to the rescaled couplings
$V^{i'},\Delta^{i'}$ and Green's functions at bandwidth $D'$ using the
multiplicative renormalization group formalism developed in Sec.
3.2.2.
We will write all
electron-TLS couplings in dimensionless form ($v^i=V^i\rho_0$).  We
shall
perform the scaling analysis at zero temperature, with the external
conduction
electron frequency $\omega$ serving as the infrared cutoff in our
equations.

The multiplicative renormalization group equations for the TLS model
read (\vld and \zow [1983b])
$$G_k(\omega/D',v^{i'},\Delta^{i'}) = G_k(\omega/D,v^i,\Delta^i)~~,
\leqno(3.4.1)$$
$${\cal G}(\omega/D',v^{i'},\Delta^{i'}) = Z_2(D'/D,v^i){\cal
G}(\omega/D,v^i,\Delta^i) ~~,\leqno(3.4.2)$$
$$\tilde \Gamma^i_{\alpha,\beta}(\omega/D',v^{i'}) =
(Z^i_{\alpha,\beta}(D'/D,v^i))^{-1}\tilde
\Gamma^i_{\alpha,\beta}(\omega/D,v^i)
~~,\leqno(3.4.3) $$
and
$$v^{i'}_{\alpha,\beta} =
(Z_2(D'/D,v^i))^{-1}Z^{(i)}_{\alpha,\beta}(D'/D,v^i)v^i_{\alpha,\beta}
\leqno(3.4.4)$$
where the functions $G_k,{\cal G},\tilde \Gamma^i$ only depend upon
dimensionful quantities.  Note that: (i) the renormalization factor
$Z_1$ has
been dropped because corrections to the conduction electron Green's
function in
the dilute limit are of order $1/N_s$.  This point will clearly need
reexamination in lattice models. (ii) The meaning of $\tilde \Gamma^i$
is
that the full interaction vertex is given by
$$\Gamma^i_{\alpha,\beta} = V^i_{\alpha,\beta} \tilde
\Gamma^i_{\alpha,\beta} \leqno(3.4.5)$$
so that $\tilde \Gamma^i_{\alpha,\beta}(D_0) = 1$.  We use $\Gamma^i$
when some
of the $V^i$ are zero, because then $\tilde \Gamma^i$ is not well
defined.
(iii) The pseudo-fermion Green's function ${\cal G}$ is generalized
from the
bare form of Eq. (3.3.4) to include a self-energy term.  Written in
matrix
form for real frequencies at $T=0$,  we have
$${\cal G}^{(\lambda_{ps})}(\omega/D,v^i,\Delta^i) = {1 \over
(\omega+\lambda_{ps}){\bf I} -\Delta^i/2\sigma^i-\Sigma_{ps}(v,\omega)
} \leqno(3.4.6)$$
where ${\bf I}$ is the identity matrix in the $2\times 2$ space, and
$\Sigma_{ps}$ is the pseudo-fermion self-energy.  \\

{\it (b) $\Delta^i=0$ case}\\

To proceed with these equations, let us start with the special case
$\Delta^i=0$.  As mentioned, we needn't compute $G_k$ beyond zeroth
order in
perturbation theory for the purposes of the scaling.   We compute
${\cal
G},\tilde \Gamma^i$ in the leading two orders.  We find that $\tilde
\Gamma^i$
is already renormalized at first order in scaling; in fact, this is
clear from
the discussion surrounding Eq. (3.2.7), in which the renormalization
factor $z$
corresponds to $Z^i$ in our present discussion.  The renormalization of
the
pseudo-fermion Green's function occurs at next leading order, i.e., the
leading
order term in the perturbation expansion is order $v^2$ for
$\Sigma_{ps}$.

The scaling transformation proceeds as follows:\\

\begin{figure}
\parindent=2.in
\indent{
\epsfxsize=2.5in
\epsffile{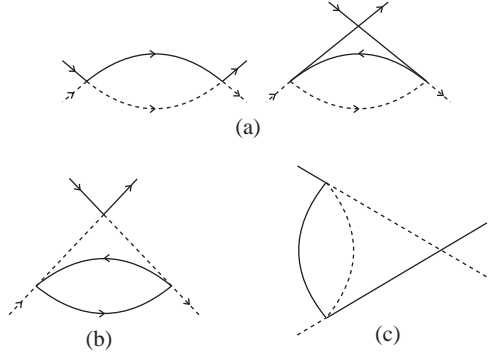}}
\parindent=0.5in
\caption{The vertex diagrams in the leading and next-leading
logarithmic orders.
The solid lines stand for the conduction electron and the dashed line
for the heavy
particle.  Note that (c) 
is not an actual vertex correction.}
\label{fig3p9}
\end{figure}

(i) First the leading order term of the vertex are calculated and then
$Z^i$ is obtained in
the leading logarithmic approximation; this corresponds to the diagram
of
Fig. \ref{fig3p9}(a).  Self consistent determination of $Z^i$ follows by
inserting
the unperturbed ($D=D'$) value of $v$ for
 $\Gamma^i$. What we obtain is just (suppressing matrix indices)
$$(\delta v^i)^{(1)} = -2i\sum_{ij}v^jv^k\epsilon^{jki} \log({D\over
D'})~~ \leqno(3.4.7) $$
where the superscript $(1)$ denotes leading order scaling.  This is
clearly
just a repeat of the
discussion in the previous subsection and if stopped at this point
leads to
precisely the same scaling equations.\\

\begin{figure}
\parindent=2.in
\indent{
\epsfxsize=2.5in
\epsffile{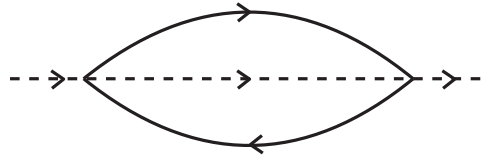}}
\parindent=0.5in
\caption{The first non-vanishing logarithmic self-energy
diagram for the heavy
particle.   The solid line represents the electron and the dashed line
the heavy particle.}
\label{fig3p10}
\end{figure}

(ii) At next leading order we compute $Z_2,Z^i$ being careful not to
include
terms which will be obtained by the solution of the leading order order
scaling
equation.  Since, as remarked earlier, solving that equation is
equivalent to
summing an infinite order of diagrams, we must take care order by order
in the
renormalized couplings that we generate unique {\it non-parquet}
diagrams.  Such a diagram is
illustrated in Fig. \ref{fig3p9}(b).  The test is simple: at a particular higher
than
leading order of the
renormalized coupling, if the diagram can be cut in two by snipping one
pseudo-fermion line and one conduction line, it is a {\it parquet}
diagram.
Once we compute the non-parquet third order diagram of Fig.
\ref{fig3p9}(c) we
must snip
two pairs of conduction/pseudo-particle lines to break the diagram up,
a test
we can apply to higher order diagrams to determine if they are due to
iteration
of this diagram.  Clearly the complexity of the procedure makes it
unfeasible
to apply beyond a few orders of perturbation theory. The lowest order
self
energy term for the pseudo-fermion is shown in Fig. \ref{fig3p10}.  The
corresponding
renormalization factor $Z_2$ is given by
$$Z_2
\simeq 1 + ({\partial
[\Sigma_{ps}(\omega/D',v^{i'})-\Sigma_{ps}(\omega/D,v^i)]
\over \partial \omega})_0 \leqno(3.4.8)$$
and is shown in Fig. \ref{fig3p9}(b).  This figure makes it clear why it appears
only at
next leading scaling: we must multiply the second order correction
coming from
the frequency derivative of the self-energy times one factor of $v^i$
to
produce a third order renormalization of the $v^i$'s. It is
straightforward to
show that the diagram of Fig. \ref{fig3p9}(c) is proportional to
$\log(D/\omega)$, while
the (shifted) pseudo-fermion self-energy goes as $\omega
\log(D/\omega)$, so
that the derivative in Eq. (3.4.8) produces a logarithm.

Note that the dividing out of $Z_2$ in the renormalization of $V$ makes
physical sense when viewed from the perspective of Landau's Fermi
liquid
theory.  We are looking for the interaction between the  conduction
electrons
and the fully dressed pseudo-fermion, or the pseudo-fermion
``quasi-particle.''
Hence, since the quasi-particle has spectral weight $Z_2$, we should
divide the
relevant coupling by that quantity to correctly normalize in the
quasi-particle
space.

Since the next leading order term is more difficult to compute
correctly than
the leading order term, we will spell out a few details here.
The changing of $D\to
D'$  now leads to a total next leading order
correction $(\delta v^i)^{(2)}$ to $v^i$
which is given by (suppressing matrix indices)
$$(\delta v^i)^{(2)} = v^i[({\cal O}(v^i)^2~ from~ Z^i)-({\cal
O}(v^i)^2~
from~Z_2)] \leqno(3.4.9)$$
$$~~~~~=Mv^i[(Tr(v^i)^2-\sum_{j\ne i}Tr(v^j)^2) -(\sum_{j}
Tr(v^j)^2)]\log({D\over D'}) $$
$$~~~~~= 2Mv^i[Tr(v^j)^2+Tr(v^k)^2]\log({D\over D'}) ~~,$$
where the terms in parentheses in the second line have a direct
correspondence
to the terms in parentheses immediately above.   Notice the explicit
appearance
of the number of channels here, which derives from the closed
conduction loop
in Fig. \ref{fig3p9}(c) in which we may freely sum on channel index.

Putting (i),(ii) together, we see that through the leading two orders
we have,
putting all matrix indices back in, that
$${\partial v^m_{\alpha,\beta} \over \partial x} =- 2i\sum_{i,j,\gamma}
v^j_{\alpha,\gamma}(x)v^k_{\gamma,\beta}(x)\epsilon_{ijm}
- 2Mv^m_{\alpha,\beta}(x)
\sum_{l\ne m}Tr[(v^l(x))^2] +.... \leqno(3.4.10)$$
which may be solved with the anticipation of flow to a two-dimensional
subspace
by substituting
$$v^i_{\alpha,\beta}(x) = v^i(x)\sigma^i_{\alpha,\beta}
\leqno(3.4.11)$$
which gives us (for $m\ne i\ne j$)
$${\partial v^m(x)\over \partial x} = 4v^i(x)v^j(x) -
4Mv^m(x)[(v^i(x))^2+(v^j(x))^2] + {\cal O}((v^i)^4) ~~.\leqno(3.4.12)$$
Note that if we had chosen to use spin 1/2 matrices instead of Pauli
matrices,
we would change the prefactor of the ${\cal O}(v^2)$ term to 1 from 4,
and the
prefactor of the ${\cal O}(v^3)$ term to 1/4 from 4.  The flipping of
sign
between the TLS convention and the typical Kondo convention ($v^i>0 \to
J^i<0$)
would flip the sign of the
first term and leave the sign of the second term which expresses the
fact that the couplings grow {\it more negative} as you reduce the
bandwidth
(increase $x$).\\

{\it (c) $\Delta^i\ne 0$ case}\\

Now we are interested in deriving equations for the renormalization of
the
splitting $\Delta^i$.  In view of Eq. (3.4.2), we see that since
$\Delta^i$
appears in ${\cal G}^{-1}$ that it must renormalize according to
$$\Delta^{i'}(D',v^i) = Z_2(D'/D,v^i)\Delta^i(D,v^i) $$
$$~~~~~+
[\Sigma_{ps}(\omega=0,D',v^{i'},\Delta^{i'})-\Sigma_{ps}(\omega=0,D,v^i,\Delta^i)]
\leqno(3.4.13)$$
where we must include the latter term as a price for taking $Z_2$
independent
of the splittings $\Delta^i$.  This term reflects the shifts of the
splittings
through self-energy effects.  It is straightforward to estimate this by
observing that the self-energy diagram of Fig. \ref{fig3p10} vanishes when the
combination $\omega+\lambda_{ps}-\Delta^i\sigma^i/2$ vanishes, so we
may
linearize the Green's function, and employ $\omega$ as the infrared
cutoff of
the resulting logarithmic divergence so that we find
$$\Sigma_{ps}(\omega,D,v^i,\Delta^i) \approx -M\sum_l Tr[(v^l)^2]
\sigma^l[(\omega-\lambda_{ps}){\bf I} - \sum_i\Delta^i\sigma^i]\sigma^l
\leqno(3.4.14)$$
and thus with the use of Eq. (3.4.13) we obtain
$${\partial \Delta^i(x) \over \partial x} = -2M\sum_{j\ne i} \Delta^i
Tr[(v^j)^2] ~~.\leqno(3.4.15)$$

Note that: (i) The sign in front of the RHS for this equation:  the
fact that it is
negative means that as we move in from the weak coupling fixed point,
the
splittings and spontaneous tunneling of the TLS actually renormalize
downwards
with reduced bandwidth.  However, we shall
see that this is not the correct interpretation overall:  the
dimensionless
splitting $\Delta^i/D$ {\it grows} and is thus to be viewed as a
relevant
perturbation near the fixed point
of Eq. (3.4.12). We shall elaborate on this point below.  (ii) In
this case where the impurity pseudo-spin is equivalent to spin 1/2, the
rescaling cannot generate any splittings unless there is a bare
splitting
present.  This corresponds to the fact that a real spin-1/2 doublet
must remain
degenerate in the absence of an applied magnetic field, i.e., it is not
susceptible to quadrupolar splittings.  If we were to allow for higher
pseudo-spin, as in the generalized multi-channel models of Sec. 2.3, we
would
find quadrupolar splittings generated by anisotropic exchange couplings
[H.B.
Pang, 1992].  We
shall return to this point in the numerical renormalization
group (Sec. 4.2) and
conformal
field theory (Sec. 6.2)
discussions and in the fixed point stability analysis
below.  \\

{\it (d) Analysis of the Fixed Point}\\

We note that the RHS (the $\beta$-function) of Eq. (3.4.12) vanishes
when
$$v^x=v^y=v^z=v^* = {1\over 2M} ~~.\leqno(3.4.16)$$
This is then a {\it non-trivial fixed point} of the fourth type,
discussed in
Sec. 3.1.1. Clearly the couplings are isotropic at this fixed point.
 However, caution is required in accepting the validity of this
finding.  Consider these cases:\\

\begin{quote}${\bf M=1}$.  In this case, the fixed point occurs for
$v^*=1/2$ which is certainly outside the perturbative regime. We cannot
trust
the low order expansion of the $\beta$ function.   Strong coupling
calculations first performed by Wilson [1973,1975] show that in this
case,
which corresponds to the single channel anisotropic Kondo model, the
fixed
point is of the strong coupling form, type (3) of Sec. 3.1.1.
\end{quote}

\begin{quote}${\bf M=2}$.  In this case, the fixed point occurs for
$v^*=1/4$, for which the perturbation expansion is still of
questionable
valididity.  Nonetheless, the qualitative
correctness of the fixed point is confirmed by non-perturbative
methods such as the Bethe-Ansatz and NRG.
In this case, the physical arguments presented in the
beginning
regarding the inability of the two conduction channels to exactly
compensate
the impurity pseudo-spin taken together with  non-perturbative
calculations confirm that
the non-trivial fixed point is indeed obtained in the zero temperature
limit,
provided the splittings and spontaneous tunneling are zero.  However,
this is
precisely the marginal case:  rather than power law critical
divergences
 in physical quantities,
logarithmic divergences  are obtained.  \end{quote}

\begin{quote}${\bf M >>2}$.  In this case, the fixed point is small and
well within the bounds of the perturbative expansion of the
$\beta$-function.
In principle, all properties may be obtained as an expansion in powers
of
$1/2M$, the fixed point coupling strength, and therefore one has a
clear
hint of universality.  For a more detailed discussion of this issue, we
direct the
reader to Sec. (3.4.3), which reviews the work of Muramatsu and Guinea
[1986] and
especially Gan, Andrei and Coleman [1993] on the calculation of
physical
properties at the nontrivial fixed point.  Note that a different kind
of
large $M$ expansion is provided by the NCA
analysis of the $SU(N)\otimes SU(M)$ multichannel model in which
$N\to \infty$ while $M/N=\gamma$ is held fixed; see Sec. 5.1 for
a review of this work.  \end{quote}

{\it (e) Linearized Stability analysis of the Fixed point}\\

We may obtain further insight into the physics by linearizing the
multiplicative renormalization group equations around the fixed point
coupling.
Such an approach will be valid for couplings $v^i(x)$ near $v^*$.
This corresponds to an analysis of the stability of the fixed point
against the
introduction of
various fields and couplings.  We shall discuss, in order, exchange
anisotropy,
the relevance of the $\Delta^i$ about the fixed point, and channel
field
splitting.

{\it Exchange Anistropy}.  We write $\delta v^i(x) = v^i(x) - v^*$ and
expand
the $\beta$-function to linear order in the deviations. We begin
scaling at a
bandwidth parameter
value of $x_0$.   The result is
$${\partial \delta v^i \over \partial x} = -{2 \over M} \delta v^i
\leqno(3.4.17)$$
which is readily solved to yield
$$\delta v^i(x) = \delta v^i(x_0) ({x\over x_0})^{-{2\over M}} ~~.
\leqno(3.4.18)$$
Clearly, the introduction of anisotropic couplings is completely {\it
irrelevant}
around the fixed point:  as we increase $x$ from our initial value of
$x_0$,
the couplings actually diminish.
We note that Muramatsu and Guinea [1986]
obtained the same scaling equations using a different diagrammatic
technique.

It is worth noting in this context, following Affleck and Ludwig
[1991a)], that
while the fixed point coupling is {\it non-universal} (clearly we can
insert an
overall arbitrary scale factor in the energy) the slope of the
$\beta$-function
at the fixed point which determines the scaling exponent in Eq.
(3.4.19) must
be universal.

This argument is actually independent of the magnitude of the localized
pseudo-spin [H.B. Pang, 1992].  In the special case of a spin 1/2
localized
pseudo-spin,
then when
the $\Delta^i$ are set to zero, then no further splitting can be
induced by
the exchange anisotropy in this case.  However, if we were to allow the
impurity spin to be greater than 1/2, and keep $M$ large, we would
induce
through the self-energy quadrupolar splittings of the pseudo-spin, and
in this
way the exchange anisotropy would generate a {\it relevant}
perturbation.

{\it Relevance of the splittings $\Delta^i$ about the fixed point}.  If
we
linearize the scaling equation for the renormalized splittings
$\Delta^i(x)$
about the fixed point (Eq. (3.4.15)) we find that since the splittings
are zero
at the fixed point (this is the only way to make the RHS of Eq.
(3.4.15)
vanish) then
$${\partial \Delta^i \over \partial x} \approx -{2\over M} \Delta^i
\leqno(3.4.19)$$
which is integrated to give
$$\Delta^i(x) = \Delta^i(x_0) ({x\over x_0})^{-2\over M}
~~.\leqno(3.4.20)$$
If we were to stop the analysis here, we would incorrectly conclude
that the
splittings $\Delta^i$ are irrelevant about the fixed point.  What is
incorrect
about this statement is that while the physical splittings do actually
shrink,
they do so (for $M>>1$ where the perturbative analysis is valid) at a
rate
much smaller than the bandwidth itself shrinks.  Hence, relative to the
decrease of the bandwidth, the splitting actually grows.  The
dimensionless
splitting $\delta^i = \Delta^i/D$ obeys the scaling equation
$${\partial \delta^i (x) \over \partial x} = (1 - 2\sum_{j\ne
i}Tr[(v^i(x))^2])\delta^i(x) \leqno(3.4.21)$$
which, when linearized about the fixed point,  integrates to give
$$\delta^i(x) = \delta^i(x_0)
({Max\{x,T/D,E/D\} \over x_0})^{1-{2\over M}}
~~.\leqno(3.4.22)$$
We can obtain the same result from Eq. (3.4.20) by simply dividing by
$D$.  As we have discussed in Sec. 3.2.1, all the logarithmic
integrals have an infrared low energy cutoff which is either
the
temperature or the renormalized value of the splitting $E$.

We shall demonstrate using the NCA and conformal theory that Eq.
(3.4.22) gives
the correct scaling to leading order in $1/M$, as we would expect, for
the
local field $\Delta^i$.  This result may also be obtained within the
Bethe-Ansatz approach.

{\it Relevance of fields breaking channel symmetry}.  Channel symmetry
breaking
is practically effected by applying a field that couples to the
electron channel
index so that the exchange integrals are split, i.e., the couplings for
different channels are no longer precisely identical.  The magnetic field does not 
directly appear in the theory, but the magnetic field splits the Fermi energy for up and 
down spin electrons.  Thus the densities of states at the Fermi energy for the 
different spins can differ by the scale of $\mu_B H/E_F\simeq 10^{-5}$ for a 1 T field, 
which results in an equal fractional difference in the dimensionless exchange couplings.  
Such a small change is practically negligible.  Nevertheless, we shall investigate the
effects here.  

For simplicity,
we shall
present
the stability analysis for this perturbation to next leading order in
the $M$=2 isotropic case, although we realize that the analysis is
somewhat
questionable in this limit.  We will obtain the correct qualitative
results as
will be demonstrated when we discuss the non-perturbative approaches,
but our
exponents characterizing the growth of the perturbation will be
incorrect.

The equations in the isotropic limit for the couplings in the two
channels are
$${\partial v_{\sigma} \over \partial x} = 4(v_{\sigma})^2 -
8v_{\sigma}[(v_{\uparrow})^2+(v_{\downarrow})^2] \leqno(3.4.23)$$
where the coupling subscript is the channel label, which is the real
spin of the
conduction states.  This equation follows from observing the decoupling
of
channel labels in the leading order  diagrams, and the coupling of
channel
labels through the conduction electron intermediate state bubble in the
next
leading order diagram.  Linearizing these scaling equations about the
fixed
point $v_{\downarrow}=v_{\uparrow}=v^*=1/4$, we find the scaling
equations
$${\partial (v_{\sigma} - v^*)\over \partial x} = -(v_{-\sigma} - v^*)
= -\delta
v_{-\sigma}
\leqno(3.4.24)$$
which are easily solved to give
$$\delta v_{\pm} = (\delta v_{\uparrow}(x)\pm\delta v_{\downarrow}(x))
\approx (\delta v_{\uparrow}(x_0)\pm
\delta v_{\downarrow}(x_0))({x\over x_0})^{\pm 1} \leqno(3.4.25)$$
where we have assumed that $\delta v_{\uparrow}(x_0) >0,\delta
v_{\downarrow}(x_0)<0$.   What this solution implies is that the sum of
the
linearized couplings tends to zero as we rescale, which implies that as
$v_{\uparrow}$ grows (towards strong coupling, where the perturbative
analysis
surely breaks down) then $v_{\downarrow}$ shrinks towards zero.  This
picture is
completely confirmed by the numerical renormalization group, conformal
field
theory, and NCA analyses.  The physical view is that the stronger
coupling tends
towards an ordinary Kondo effect, while the weaker coupling tends
towards the
zero coupling fixed point.  What we shall see is that at zero
temperature, this
rescaling behavior is discontinuous.\\

{\it (f) Numerical integration of scaling equations}\\

\begin{figure}
\parindent=2.in
\indent{
\epsfxsize=2.5in
\epsffile{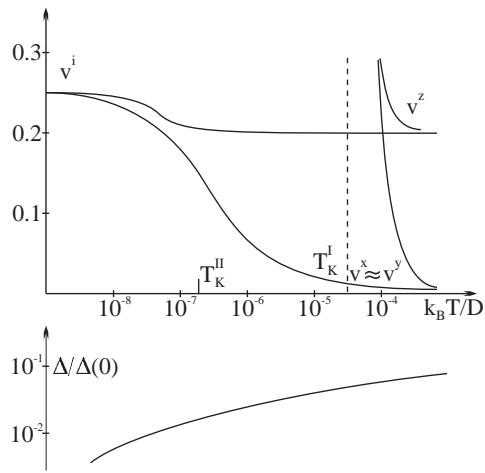}}
\parindent=.5in
\caption{Scaled couplings $v^i$ ($i=x,y,z$) and energy
splitting $\Delta$ as a function
of $k_BT/D$ for $M=2$.  The initial $v^x$ coupling is 0.2 and the
initial ratio $v^x/v^z$ is
10$^{-3}$. The narrow (heavy) lines represent the first-(second-)order
scaling.  $T^I_K$
and $T^{II}_K$ are the Kondo crossover temperatures in the first- and
second-order scaling,
respectively.  The dotted line is the asymptote of the coupling in the
first-order scaling.
The heavy lines for $v^i$ are obtained analytically in second-order
scaling.  The region
where $v^x\approx v^z$ does not hold is not represented.  The ratio of
the scaled and
initial energy splitting $\Delta/\Delta(0)$ is calculated for the
symmetridc TLS by using Eq.
(3.4.13).  The index of $x$ for $\Delta^x$ is dropped.  From \vld and \zow [1992].}
\label{fig3p11}
\end{figure}

Fig. \ref{fig3p11} illustrates the results for the solution of the next leading
order
scaling equations (3.4.12,15) with $v^z(0)>>v^x(0),v^y(0)=0$ and
$M=2$.
The right most curves in Fig. 3.11.a) which blow up at the vertical
line marked
$T_K^{(i)}$ correspond to the leading order equations.  The other two
curves
which show a rise to strong coupling on the next leading order Kondo
scale
$$T_K^{(II)} \approx D_0 ({v^x(0)\over 4v^z(0)})^{{1\over 4v^z(0)}}
[v^x(0)
v^z(0)]^{M\over 4} \leqno(3.4.26)$$
go all the way into the fixed point at $v^*=0.25$ smoothly.  In the
vicinity of
$T_K^{(I)}$, the couplings $v^x(x),v^y(x)$ are nearly equal in
magnitude.
We should contrast Eq. (3.4.26) which corresponds to the fully
anisotropic Kondo
limit solved by Shiba [1970] with the isotropic limit result
$$(T_K^{(II)})_{iso} = D_0 (v(0))^{{M\over 2}} \exp[-{1\over 4v(0)}]
~~.\leqno(3.4.27)$$

The spontaneous tunneling matrix element in the extreme anisotropic
limit is
strongly renormalized downwards, as seen in Fig. \ref{fig3p11}(b).   While the
$\Delta^z$
value is also renormalized downwards, the reduction is less severe
since the
downward renormalization is driven by  the weaker, planar couplings.
In the
limit of full isotropy, the downward renormalization would be
independent of
direction and $\Delta^z$ would thus be strongly reduced in absolute
value.

{\it (g) Dimensionality of the fixed point Hamiltonian}\\

In Sec. 3.3.3, it has been shown using the leading logarithmic
approxmiation that the structure of the matrices
$V^i_{\alpha\beta}~(i=x,y,z)$ at the strong coupling fixed point
is such that they are proportional to a representation of
$SU(2)$.  (Note that the leading logarithmic approximation
generates a flow to the strong [infinite]
coupling fixed point regardless
of channel number since the channel number enters only at next
leading logarithm level.)  The following questions remain,
however, without answers from this discussion:\\
1) Are these fixed point matrices reducible representations or
always irreducible, and \\
2) Are there any restrictions on the dimensions of the
irreducible matrices?  It has been argued following \vld~ and
\zow~ [1983a] that at the beginning of scaling a two-dimensional
subspace emerges due to the scaling.  Nonetheless, it could
still happen that going beyond the leading logarithmic order
of scaling could lead to a larger relevant subspace.  In the limit of 
large number of channels, this possibility has been ruled out by 
\zar [1995].  

These questions cannot be generally answered from the
scaling analysis and require non-perturbative resolution from
such methods as conformal theory and Bethe Ansatz.  To get at
some more rigorous understanding, let us first assume that there
is no additional symmetry which could influence the dimension of
the subspace.  Furthermore, in order to make a rigorous
statement, let us (artificially) increase the channel number
from $M=2$ to $M>>1$ so that the $1/M$ expansion is valid (this
is discussed more extensively in Sec. 3.4.4).  In this case the
scaling equations given by (3.4.12) are correct and the higher
order corrections on the RHS are controllably neglected as they
contribute in higher powers of $1/M$.  By proceeding in the
scaling, the 2-D subspace bcomes dominant and contributed to the
second term of Eq. (3.4.12), which slows down scaling.
In this spirit, \zar [1995] was able to show that this
next leading term suppresses the conduction orbital states with
small amplitude coupling finally only a 2-D subspace remains.

This result can be summarized as follows:  in the two-channel
TLS Kondo problem we conjecture based upon the analysis of \zar
[1995] that the original number of conduction orbitals in the
matrices  $V^i_{\alpha\beta}$ is reduced to 2 at the fixed
point.  Thus the orbital coupling matrices must be simple Pauli
matrices at the fixed point.  On the other hand, the impurity
spin matrix usually is also reduced to a two-dimensional
matrix.  Thus we complete the mapping to the spin 1/2 two
channel Kondo model for this TLS model given that the two-channel
degeneracy is required by time reversal.

\subsubsection{TLS Model Including Excited Atomic States} 

The TLS is defined by truncating the excitation spectrum of the atom in
the double well keeping only the two lowest levels (see Fig.~\ref{fig2p1}).
This projection certainly gives the nature of the low temperature
behavior
correctly, but the higher level may
contribute to the renormalization of the prameters of the truncated
system.
Similar renormalization has already been discussed for the original and
orbital Kondo effect where the influence of the higher states split by
the crystalline electric field is discussed in Sec. 3.3.c.  Such renormalization can be crucial concerning
the
value of the Kondo temperature $T_K^{(II)}$ given by Eq. (3.4.26) which
is
very sensitive to the initial parameter values.  We postpone a more
detailed
discussion related to the ex
periments to Sec. 7.1.  The following theoretical discussion is based
on
the work of \zar and Zawadowski [1994a,b].

\begin{figure}
\parindent=2.in
\indent{
\epsfxsize=2.5in
\epsffile{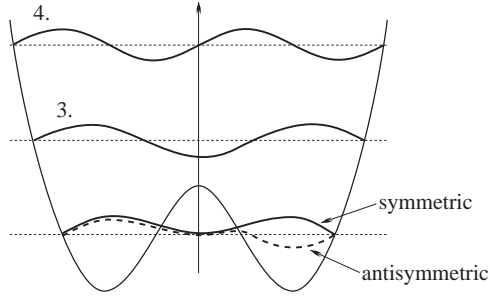} }
\parindent=.5in
\caption{The double potential well is shown by the solid line.
The linear combinations of
the lowest left and right states are presented by heavy solid and
dashed lines.   Two further
excited states labeled by 3 and 4 are also represented by heavy
lines.   The light lines
represent the level of zero amplitudes of the wave function.}
\label{fig3p13}
\end{figure}

For simplicity we treat the situation where all the higher excitations
are extending over both minima of the double well.  Instead of the left
and right states, the exact states are introduced here which are
symmetric
and antisymmetric in the case of a symmetric double well.  
(See Fig.~\ref{fig3p13}).
All of these states are orthogonal, and thus there are no spontaneous
transitions
between them.  The electron density interacting with them is not,
however,
homogeneous in real space and thus the interaction with the electrons
can
induce transitions between the states of the double well.

The wave functions and energies are denoted by $\phi_i$ and
$\epsilon_i$
($i=1,2,....$) with $\epsilon_1<\epsilon_2<\epsilon_3 ....$.  The
generalized
version of Hamiltonian (3.3.3) with creation operators $b^{\dagger}_i$
and
pseudo-particle chemical pote
ntial $\lambda_{ps}$ is
$$H_0 = \sum_i (\epsilon_i -\lambda_{ps})b^{\dagger}_ib_i
~~.\leqno(3.4.28)$$
The electron-tunneling center interaction is
$$H_{int} = \sum_{i,j,k,k',\sigma} V^{i,j}_{k,k'} b^{\dagger}_ib_j
a^{\dagger}_{k\sigma}a_{k'\sigma} \leqno(3.4.29)$$
where the matrix element in the scheme proposed by \zar [1993] is
$$V^{i,j}_{k,k'} = U(\vec k - \vec k') \int dr exp[-i(k-k')r]
\phi_i^*(r)\phi_j(r) \leqno(3.4.30)$$
which is a generalization of Eq. (2.2.14.a,b).  Later notation
$\rho_0 V^{i,j}_{k,k'}=v^{i,j}_{\hat k,\hat k'}$ is introduced
indicating the dependence only on the direction of the momenta
in the vicinity of the Fermi surface.

\begin{figure}
\parindent=2.in
\indent{
\epsfxsize=4in
\epsffile{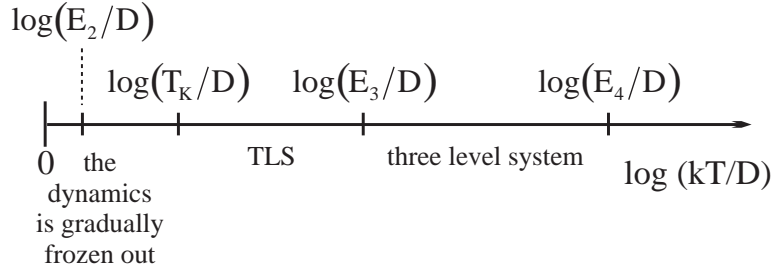}}
\parindent=0.5in
\caption{The different regions of the scaling.  The excited
states with energies $E_3,E_4$
are gradually frozen out.}
\label{fig3p14}
\end{figure}

The first relevant vertex correction is just a straightforward
generalization of
Eq. (3.3.11) which has the form
$$T^{k,l}_{\hat k_1,\hat k_2} = v^{k,l}_{\hat k_1,\hat k_2} + \int
{dS_F(\hat k)\over S_F} \sum_i \ln({max(|\epsilon_i|,T,|\omega|)\over
D})
[ v^{k,i}_{\hat k_1,\hat k} v^{i,l}_{\hat k,\hat k_2}-v^{i,l}_{\hat
k,\hat k_1}
-v^{k,i}_{\hat k,\hat k_2}] ~~.\leqno
(3.4.31)$$
Note that the factor of 2$i$ is missing compared with Eq. (3.3.11)
since we have
not assumed a Pauli spin algebra here.  The scaling equation
is just a
generalization of Eq. (3.3.13) given by
$${\partial v^{kl}_{\hat k_1,\hat k_2} \over \partial \ln(x)} =
\theta(D'-|E_i|)
\int {dS_F(\hat k)\over S_F}
[ v^{k,i}_{\hat k_1,\hat k} v^{i,l}_{\hat k,\hat k_2}-v^{i,l}_{\hat
k,\hat k_1}v^{k,i}_{\hat k,\hat k_2}]  \leqno(3.4.32)$$
where the $\theta$-function assures that only those states contribute
to the
sum which are below the scaled bandwidth $D'$ and $E_i=\epsilon_i -
\epsilon_1$.
As a result of these restrictions at a given energy or temperature only
those states
play a role
 in dynamics for which $E_i < k_BT$ and the others just contribute to
 the
 renormalization of the remaining effective coupling constants (see
 Fig. \ref{fig3p14}).

The solution of these scaling equations were studied numerically and
analytically
in great detail (\zar and \zow [1994(a,b)]).  The Kondo
temperature can be
enhanced or reduced depending on the values of the prameters, but for
most of the
physically
interesting set of couplings, the Kondo temperature is enhanced by a
factor of
2-3 orders of magnitude.

Just to emphasize the role of the excited states, the scaling
calculations were
performed under the fictitious assumption that the direct assisted
tunneling
between the left and right positions are zero, so that  for the
unrenormalized
couplings $V^{12}(0)
=V^{21}(0)=V^{12}=V^{21}=0$ holds.  Even in this case, the large
enhancement of
the Kondo temperature is reduced only by a factor of 0.6-0.7, which
implies the
excited levels are playing a crucial role in the dynamics.

\begin{figure}
\parindent=2.in
\indent{
\epsfxsize=2.5in
\epsffile{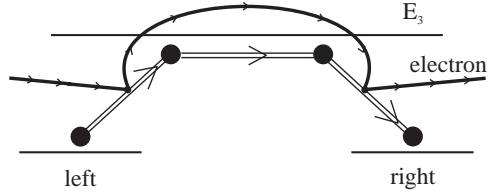}}
\parindent=0.5in
\caption{The left and right states of the TLS with the first
excited state with energy $E_3$.
The double line represents the heavy particle and the electron with
solid line induces
the transitions.}
\label{fig3p15}
\end{figure}

This result is of special importance.  To understand it, consider one
of the
  dominant processes included in the scaling analysis in which the
  electron
  hits the atom, say, on the left and virtually excites the atom to,
  for example,
  the third
level (see Fig.~\ref{fig3p15}).  That level covers both the left and right
minima,
so that from there the atom can fall into the right minimum with the
assistance
of the same electron.  This result normally follows from the behavior
of the
matrix elements given
 by Eq. (3.4.29), which in the limit $k_Fr<<1$ are the dipole matrix
 elements
 between the wave functions $\psi_i$ of the atom.  The interesting new
 feature
 is the following:
Neglecting the overlap between the left and right wave functions, the
induced
effective electron assisted matrix elements are less sensitive on the
distance
despite the fact that  the dipole matrix element even increases with
the
distance; for too large a
 distance the original expression must be used in favor of the dipole
 approximation.  This has the great advantage that the direct hopping
 which
 is neglibible in this case does not split the ground state and first
 excited
 states; nonetheless, the excited
state assisted tunneling is large enough and actually grows with
distance as
shown in Table~\ref{tab3p1}.  Previous to this analysis, any increase in the
assisted
matrix element was associated with the increase of the direct hopping
and
thus the splitting, which 
of course mitigates the possibility of observing the non-trivial fixed
point
physics.  To obtain a sufficiently large Kondo scale
together with a sufficiently small level splitting required in the most
optimistic scenario a delicate
balance of these effects.  The inclusion of the excited levels now
makes the
observability of a TLS with sufficiently large Kondo scale and
negligibly small
splitting appear quite plausible.

\begin{table}
\begin{center}
\begin{tabular}{|l|l|l|l|l|}\hline
$\alpha$ & 1.0 & 1.5 & 2.0 & 2.5 \\\hline\hline
$E_1$ & 331 & 277 & 255 & 245  \\\hline
$E_2$ & 540 & 406 & 337 & 330 \\\hline
$\Delta_0$ & 4.47 & 0.53 & 0.063 & 0.0076 \\\hline
$v^z$ & -0.061 & -0.086 & -0.112 & -0.142 \\\hline
$v^x$ & 0.0005 & 0.0009 & 0.0013 & 0.0018 \\\hline
$\tilde T_K$ & 7.16 $\times 10^{-7}$ & 0.00187 & 0.152 & 2.46 \\\hline
$\tilde T_K^{(1)}$ & $9.14 \times 10^{-9}$ & 0.00342 & 1.29 & 67.2
\\\hline
$\tilde T_K^{(2)}$ & $1.97 \times 10^{-7}$ & 0.005 & 1.29 & 65.1
\\\hline
$\tilde T_K^{(1)}|_{v^x=0}$ & 2.3$\times 10^{-7}$ & 2.02$\times 10^{-5}$
& 0.41 & 44.9 \\\hline
$T_K$ & 4.13$\times 10^{-9}$ & 1.64$\times 10^{-7}$ & 0.00186 & 0.0343
\\\hline
$T_K^{(1)}$ & 3.05$\times 10^{-11}$ & 3.31$\times 10^{-5}$ & 0.026 &
2.76 \\\hline
\end{tabular}
\end{center}
\caption{ Kondo scales for a TLS model including excited states.
Calculations are performed for a symmetric
double square well potential (c.f. Appendix I)
with a variable width following \zar and \zow [1994a,b].
The width of the double well barrier is controlled by the parameter
$\alpha = r_0/2d$, where $r_0$ is the width of an individual well and
$d$ the barrier thickness (See Fig. I.1).  Plane wave conduction states
are assumed.  All energies are measured in
K.  The notation is as follows:  $E_1(E_2)$ is the energy of the first
(second) excited state.   $\Delta_0$ is the splitting of the lowest TLS
states due to spontaneous tunneling through the square barrier.
$v^z,v^x$ are the bare dimensionless TLS coupling
constants which measure the interaction strength with the ground doublet
levels.  $\tilde T_K$ is the Kondo scale in the leading logarithmic
order
neglecting excited states.  $\tilde T_K^{(1)}$ is the Kondo scale in the
leading logarithmic approximation including the first excited state, and
$\tilde T_K^{(2)}$ includes both first and second excited states in
leading
logarithmic order.  $\tilde T_K^{(1)}|_{v^x=0}$ is the leading
logarithmic estimate with the first excited state kept but the bare
$v^x$ artificially set to zero; this illustrates the strong effect of
assisted tunneling
induced by virtual excitations to the first excited state.  $T_K$ is the
next leading logarithmic approximation estimate without excited states,
and $T_K^{(1)}$ is the next leading logarithmic estimate with the first
excited state retained.  We learn from these calculations that for
sufficiently narrow barrier (relative to the square well width)
the excited states enhance the estimated
Kondo scales significantly (by as much as two orders of magnitude).
and that the most crucial state to include
is the first excited state--the second excited state has small
additional effect.  With the additional states Kondo scales within the
correct order of magnitude of experiment to explain the
resistance data on quenched metallic point contact devices
(Ralph and Buhrman [1992,1995], Ralph {\it et al.} [1994,1995],
Upadhyay, Louie, and Buhrman [1996]).}
\label{tab3p1}
\end{table}

The importance of the broad barrier is demonstrated by the data in
Table~\ref{tab3p1},
where the following parameters are chosen for the square well potential
discussed
in App. I and shown in Fig.~\ref{figaip1}:
$1/2mR_0^2=50K$, $V_B=490.5K$, and the width of the barrier is given by
the
parameter $\alpha$ as $d=2r_0\alpha$.  Table~\ref{tab3p1} clearly shows that the
Kondo
temperature in the leading order is strongly reduced in the next to
leading
order due to the factor
$(v^xv^z)^{1/2}$ of Eq. (3.4.26).  The enhancement due to the excited
states is
essential.  The second excited state has a minor effect
due to the reduced coupling strength arising from the presence of more
wave
function nodes.  The splitting $\Delta$ is negligible in the broad
barrier of
interest here.

\subsubsection{Formation of the TLS double well:  Einstein phonon model}

Concerning the anomalous behavior of the A15 compounds, P.W. Anderson and
C.C. Yu proposed a model to clarify the role of the strong electron-phonon
interaction in the formation of a double well potential for a TLS.  The
basic idea is similar to the physics of the Jahn-Teller distortion.  A
single well potential for a single atom can be deformed by the strong
electron-phonon interaction to form a double well potential.  Thus the
coupling to the electronic heat bath is responsible for the TLS formation
similar to the insulating case where the coupling to the phononic bath
plays this role (Sethna [1981]). 

The idea has recently been reinvestigated by H. Kusunose and K. Miyake
[1996], where it is assumed the atomic motion is in a simple parabolic
potential which can be regarded as a localized Einstein phonon described
by the Hamiltonian 
$${\cal H}_{ph} = \Omega(b^{\dagger}b + 1/2)\leqno(3.4.33)$$
on each site, where $\Omega$ is the phonon frequency and $b^{\dagger}$ is
the phonon creation operator.  The phonon operators are described
alternatively in terms of a pseudo-fermion operator $b_n$ which
annihilates the state with $n$ excitations, so we may write equivalently 
$${\cal H}_{ph} = \Omega \sum_n nb^{\dagger}_nb_n ~~. \leqno(3.4.34)$$

Kusunose and Miyake truncated this model to the states $n=0,1$.  
For matrix elements involving the atomic coordinate $Q$ in the direction
along the possible deformation ($Z$-direction) then the harmonic
oscillator wave functions for the $n=0,1$ levels are used.  The atom
electron interaction is similar to the one given by Eq. (3.4.30).  
The $n=0,1$ states are even and odd under $z$ reflection ($z\to -z$).
Then the Hamiltonian (3.4.3) is expanded in $r$ (in the notation of 
Kusunose and Miyake, $Q=q(b+b^{\dagger})$) keeping only the first and
second order terms.  In order to mimic the left and right states for this
starting parabolic potential, the states with pseudo-spin
$\uparrow,\downarrow$ are
introduced as 
$$b_{\uparrow} = {1\over \sqrt{2}}(b_0 - b_1)~~,b_{\downarrow} = {1\over
\sqrt{2}}(b_0 + b_1) ~~,\leqno(3.4.35)$$
and then all the operations for the atomic states can be described by Paul
operators acting in the pseudo-spin space.  Using Eq. (3.4.30), the
interaction $V^{ij}_{kk'}$ can be calculated and decomposed into 
Pauli matrices in the pseudo-spin states in the $ij$ space of conduction
partial waves.  The Hamiltonian then takes the form described in Sec. 
(2.1.2)b, with a spcial initial value of the couplings $v^z,v^x$,
with $v^y=0$.  Kusunose and Miyake then apply the renormalization group
in the form described in Sec. 3.4.2.  When the relatively large value of
the initial splitting $\Delta^x\simeq \Omega$ scales to smaller values,
this is interpreted as the formation of a TLS by effecting smaller direct
coupling and overlap between the states with pseudo-spin
$\uparrow,\downarrow$.  It can be seen that the formation of the
two-channel Kondo ground state requires very special initial parameters
as has been discussed in Sec. 3.4.1.f.  Kusunose and Miyake do not solve
this problem by starting with larger overlaps.  

This scheme represents a good starting point for further work, but for a
more realistic and elaborate theory it will likely be necessary to include
the higher excited states $n\ge 2$ and perhaps the continuum states must
also be taken into account (see Sec. 3.4.2).  More states may lead to
better localization of the atom in the left and right wells and smaller
overlap for the lowest left and right states. 

\subsubsection{Next Leading order Scaling for Model \ufp~ and \ctp~
Ions} 

{\it (a) Kondo scale in the \ufp~ case }\\

As derived above in the isotropic limit, the Kondo scale in this case
where the
bare couplings are isotropic is, for the quadrupolar Kondo model
$$T_K^{(II)} = D_0 |g| \exp({1\over g})  \leqno(3.4.36)$$
with $g=2N(0)V^2/\tilde \ef$.

We shall consider the \ctp~ case separately.  \\

{\it (b) Stability analysis for \ufp~ ions }\\

As in the discussion above for the TLS, we anticipate a non-trivial
fixed point
for the quadrupolar Kondo effect of a \ufp ion.  This fixed point is
stable
against anisotropic exchange, and destabilized by applied spin field
(precisely
analogous to the relevance of the TLS splitting and spontaneous
tunneling
discussed above) and applied channel field (splitting of the exchange
integrals).  In these cases, all of the analysis carries over
completely from
the TLS discussion, modulo the change of sign of coupling constants and
the use
of spin 1/2 matrices.

\begin{figure}
\parindent=2.in
\indent{
\epsfxsize=2.5in
\epsffile{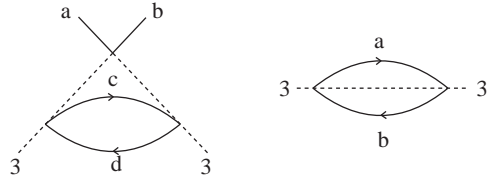}}
\parindent=.5in
\caption{Next leading order scaling corrections to the quadrupolar Kondo model with two 
$\gei$ quartets.  The indices $a,b,c,d$ refer to the possible $8,8'$ labels derived from
the two conduction electron $\gei$ quartets which come from the $j=5/2,7/2$ partial waves (see~\ref{fig3p7}
for the leading order diagrams).  The diagram on the left is a vertex correction, while the
diagram on the right gives the self energy renormalization which must be subtracted as
discussed in Sec. 3.4.1.}
\label{fig3p16}
\end{figure}

We also would like to show that the scaling to a two-fold degenerate
conduction
pseudo-spin space discussed in the leading order scaling analysis
(Sec.
3.3.3.b)  also carries through in next leading order. We employ the
same
notation as Sec. 3.3.3.b.   The generalization of
Eqs. (3.36.a-c) in next leading order follows from the renormalized
parquet diagrams of
Fig.~\ref{fig3p16} 
and gives
$${\partial g_{8,8} \over   \partial x} = -[g_{8,8}^2+g_{8,8'}^2] -
g_{8,8}[g_{8,8}^2 + g_{8',8'}^2+2g_{8,8'}^2] ....\leqno(3.4.37.a)$$
$${\partial g_{8',8'} \over   \partial x} = -[g_{8',8'}^2+g_{8,8'}^2] -
g_{8',8'}[g_{8,8}^2 + g_{8',8'}^2+2g_{8,8'}^2] ....\leqno(3.4.37.b)$$
$${\partial g_{8,8'} \over   \partial x} = -g_{8,8'}[g_{8,8}+g_{8',8'}]
-
g_{8,8'}[g_{8,8}^2 + g_{8',8'}^2+2g_{8,8'}^2] ....\leqno(3.4.37.c)$$
By taking the difference of the first two equations and comparing to
the third,
we again conclude that
$$g_{8,8'}(x) = c_0[g_{8,8}(x)-g_{8',8'}(x)] ~~, \leqno(3.4.38)$$
meaning that the integration constant $c_0 =a_8a_{8'}/(a_8^2-a_{8'}^2)$
once again.
Now, with a little algebra it is possible to show that
$$[g_{8,8}^2+g_{8',8'}^2+2g_{8,8'}^2] = [(\tilde g)^2 + (\tilde g')^2]
\leqno(3.4.39)$$
so that the scaling equations for $\tilde g,\tilde g'$ are
$${\partial \tilde g\over \partial x} = -(\tilde g)^2 - \tilde
g[(\tilde
g)^2+(\tilde g')^2] \leqno(3.4.40.a)$$
and
$${\partial \tilde g'\over \partial x} = -(\tilde g')^2 - \tilde
g'[(\tilde
g)^2+(\tilde g')^2] + ....... \leqno(3.4.40.b)$$

These equations admit the non-trivial fixed points $\tilde g=-1,\tilde
g'=0$, $\tilde g=0,\tilde g'=-1$ (both two-channel fixed points) and
$\tilde
g=\tilde g'=-1/2$ which is a new non-trivial fixed point, which is 
the four-channel Kondo model fixed point (the scaling equations and
fixed
point value are the same at this order).  However, an analogous
analysis to that of the channel field splitting applies here as well
(see Sec. 3.3.3.b) and we see that
that the new non-trivial fixed point is unstable so that whichever
coupling
$\tilde g,\tilde g'$ is larger, we will flow to the two-channel fixed
point
associated with that coupling.  Then a linearized stability analysis
around,
e.g.,
$\tilde g=1,\tilde g'=0$ shows that an added perturbation in $\tilde
g'$ is
irrelevant.  Hence, even if we begin with $\tilde g'(0) \ne 0$, if it
is
smaller than $\tilde g(0)$ we will quickly flow to the subspace where
only
$\tilde g$ is present.  In the present model, since we begin with
$\tilde
g'(0)$ identically zero, we will always simply select out the bonding
combination of $\gei$ orbitals and flow to the two-channel fixed point
associated with that combination.  \\

{\it (c) Next leading order analysis for \ctp ions}\\

\begin{figure}
\parindent=2.in
\indent{
\epsfxsize=2.5in
\epsffile{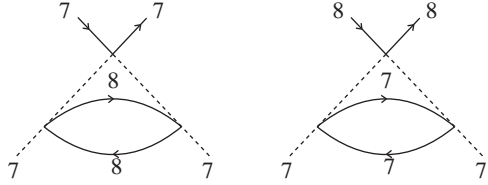}}
\parindent=.5in
\caption{Next leading order vertex corrections for a \ctp ion with both $f^0-f^1$ and 
$f^1-f^2$ valence fluctuations.  Corresponding wave function renormalization factors must
be subtracted as per the procedure of Sec. 3.4.1, and in the case of Fig.~\ref{fig3p16}.}
\label{fig3p17}
\end{figure}

For \ctp ions, the different exchange strengths for the $\gse$ partial
waves
and the $\gei$ partial waves couple at third order, as shown by the
diagrams
in Fig.~\ref{fig3p17},
and the resulting scaling
equations are (Kim [1995], Kim and Cox [1995,1996,1997])
$${\partial g_7 \over \partial x} = -g_7^2 -{1\over 2}
g_7(g_7^2+2g_8^2)
\leqno(3.4.41.a)$$
and
$${\partial g_8 \over \partial x} = -g_8^2 - {1\over 2}
g_8(g_7^2+2g_8^2)
~~,\leqno(3.4.41.b)$$
with $g_{7,8}=N(0)J_{7,8}$.

\begin{figure}
\parindent=2.in
\indent{
\epsfxsize=2.5in
\epsffile{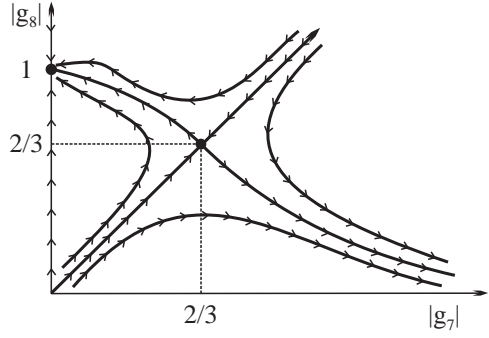}}
\parindent=.5in
\caption{Multiplicative renormalization group flow diagram for the \ctp ion in 
next to leading order.  The horizontal axis is the dimensionless coupling of the 
$\gse$ \ctp spin to the $\gse$ conduction states, and the vertical axis is the 
dimensionless coupling strength of the $\gse$ \ctp spin to the $\gei$ conduction
quartet.  Along the line $g_{7}=g_{8}$, the three channel fixed point is stable. 
A slight deviation from this behavior gives rise to the two-channel fixed point 
(assuming $|g_7|<|g_8|$), or the single channel strong coupling fixed point 
($|g_7|>|g_8|$).  For more details, see Kim [1995], Kim and Cox [1995,1996,1997].}
\label{fig3p18}
\end{figure}

These equations have three possible low temperature behaviors:(i) When
$g_7\to\infty,g_8=0$ then we get the normal Kondo behavior for the \ctp
ion.
(ii) When $g_7=g_8=-2/3$, we obtain the three channel Kondo fixed
point.
(iii) When
$g_7=0,g_8=-1$, we obtain the two-channel Kondo model fixed point.
In case (i), we see from the lowest order results (quadratic term in
Eqs.
(3.4.38.a,b)) that if we start with $|g_7(0)|>|g_8(0)|$ , $g_7$ will
grow
more rapidly than $g_8$ with initial scaling, so we conjecture that we
will
flow to the normal Kondo fixed point.  This is substantiated by the
non-perturbative NCA results we shall discuss in a later section.  If
we make
$g_7=g_8$ initially, we will remain on the $g_7=g_8$ line throughout
the
scaling and thus flow to the three channel Kondo fixed point at low
temperatures.  This fixed point is unstable against any deviation from
the
equality of the couplings:  the linearized stability analysis shows
that if one
applies $\delta g_7 <0,\delta g_8>0$ one will begin to flow towards the
normal
Kondo fixed point, while reversing the inequality will drive one to the
two-channel fixed point associated with the $\gei$ coupling.  Finally,
if we
begin with initial couplings such that $|g_8(0)|>|g_7(0)|$, we will
flow to the
two-channel fixed point with $g_7=0$, and any added $g_7$ coupling is
irrelevant about this fixed point within the linearized analysis.
These results are summarized in the schematic flow diagram depicted in
Fig.~\ref{fig3p18}.

The dividing criterion on the dimensionless couplings corresponds to
the
comparison of the quantum fluctuation weights of $f^0,f^2$ in the
ground state.
Using second order perturbation theory, one can easily show that the
lowest
order estimate for the occupancy deviation of $f^1$ towards $f^0$ is
$\simeq g_7(0)$,
and similarly for $f^1$ towards  $f^2$ the occupancy deviation is
$\simeq
2g_8(0)$.\\

{\it (d) Next leading order scaling for \ctp ion with channel spin/spin
coupling included. }\\

Extending the leading order analysis of the last section, we follow Kim
(Kim [1995]; Kim and Cox, [1996]; Kim, 
Oliveira and Cox, [1996]) and develop next
leading order scaling equations for the \ctp model in which we set
couplings
to the $\gse$ partial waves to zero and retain only the $J_8,\tilde
J_8$ couplings
of Sec. 2.2.  The development of the next leading equations follows
precisely
that of the preceding subsections, and we obtain
$${\partial g_8\over \partial x} = g^2_8 - {1\over 2}\tilde g_8^2 -
g_8[g_8^2 + \tilde g_8^2]
\leqno(3.4.42)$$
and
$${\partial\tilde  g_8\over \partial x} = - g_8\tilde g_8 - \tilde
g_8[g_8^2 + \tilde g_8^2]
~~.\leqno(3.4.43)$$

\begin{figure}
\parindent=2.in
\indent{
\epsfxsize=2.5in
\epsffile{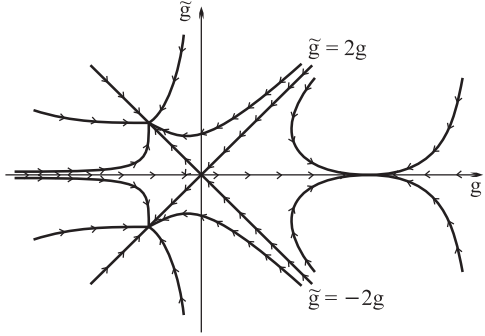}}
\parindent=.5in
\caption{Multiplicative renormalization group flow diagram for \ctp ion when 
channel spin/spin coupling is included (and coupling to the $\gse$ conduction
space is set to zero).  In this case the $\gei$ conduction quartets can either couple as 
two-spin 1/2 objects to the $\gse$ \ctp conduction quartets (dimensionless coupling strength 
$g_8$), or can couple through a mixed spin/channel spin tensor
(dimensionless coupling strength $\tilde g_8$). This leads to competing fixed points.  
The fixed point for positive $g_8$ and zero $\tilde g_8$ is the usual two-channel one. 
The fixed points for negative $g_8$ and finite $\tilde g_8$ are such that the $\gei$ 
manifold acts as a {\it single channel spin 3/2} band coupling to the spin 1/2 impurity.
For further details, see Kim [1995], Kim and Cox [1996], and Kim, Oliveira, and Cox [1997]. 
} 
\label{fig3p19}
\end{figure}

These equations have four fixed points:\\
(i) $g_8=\tilde g_8=0$ (the weak coupling fixed point)\\
(ii) $g_8=1,\tilde g_8=0$ (the two-channel $S_I=1/2$ fixed point)\\
(iii) $g_8=-1/5,\tilde g_8=2/5$ ($S_I=1/2,S_c=3/2$ fixed point)\\
(iv) $g_8=-1/5,\tilde g_8=-2/5$ ($S_I=1/2,S_c=3/2$ fixed point)\\
A linearized analysis about the fixed points (ii-iv) confirms
their local stability in the $\tilde g_8-g_8$ plane, and the
non-trivial
character of the fixed points $(iii,iv)$.  As indicated in the
discussion of
leading order scaling, the special properties of the lines $g_8=\pm
\tilde g_8/2$
are maintained due to the special symmetry properties of the system
along these lines.  A complete schematic scaling diagram 
is displayed in Fig.~\ref{fig3p19} .

\subsubsection{Physical Properties in the $1/M$ Expansion} 

According to the suggestion of \noz~
and Blandin [1980], for large channel number $M>>1$, the fixed point
$v^x=1/2M$
(see Eq. (3.4.16))
is in the weak coupling limit.  As a result, the analytic
renormalization
group calculations based on
a perturbative expansion will be perfectly adequate to treat the
properties at
all temperatures.  The physics may be computed in an expansion of
powers of
$1/M$.  This contrasts to the conformal field theory which is only
applicable
in the low temperature
 regime.
Furthermore, within the perturbative renormalization group treatment,
dynamical
quantities may be computed as well which is beyond the scope of the
Bethe-Ansatz
method.  (We remark that the
large $N,M$ NCA approach discussed in Sec. 5.1 also can treat all
properties.)
Thus, the $1/M$
expansion gives two different types of information:\\
(i) The result can be compared to the conformal field theory approach
to
check the assumed conformal invariance and related hypotheses;\\
(ii) The method can be used to evaluate physical properties in the
entire
temperature regime.

The actual performance of perturbation theory to higher orders in the
coupling
constant and the feedback into the renormalization group equations are
very
technically involved matters, and so we shall quote only some of the
main
results from the literature.
The interested reader may turn to the original articles for further
detail.

Muramatsu and Guinea [1986] were the first to make use of the $1/M$
expansion,
and Gan, Andrei and Coleman [1993] carried out a more extensive set of
calculations up to the next to leading logarithmic order  The former
work
is based upon application of the
bosonization technique developed by D.J. Amit, Y.Y. Goldschmidt, and
G.
Grinstein [1980], while the latter authors used a path integral
method.
The latter authors assumed an isotropic exchange and spin 1/2 impurity
and
conduction spins, as we shall throughout this subsection.
We shall refer to the work of Gan, Coleman,
and
Andrei henceforth.

In the vicinity of the weak coupling fixed point, the effective
exchange
coupling $g$ behaves as (see Eq. (3.4.18))
$$g(x) = {1\over 2M} + [g(0) - {1\over 2M}]({x\over x_0})^{\Delta}
\leqno(3.4.44)$$
where
$$\Delta = {2\over M}(1-{2\over M}) \leqno(3.4.45)$$
as computed to the two leading orders by Gan, Andrei and Coleman
[1993].
Assuming $x$ measures the frequency, this equation may be rewritten
$$g(\omega) = g^* - \zeta ({\omega\over T_K})^{\Delta}\leqno(3.4.46)$$
where to leading order $\zeta = 4/eM^2$, $e$ the natural log basis, and
$T_K= Dg^{M/2} exp(-1/g)$.
Making use of the linked cluster theorem, they obtain the free energy
through
order $g^4$ and from this compute the impurity contributions to the
specific
heat $C_{imp}(T)$, magnetic susceptibility $\chi_{imp}(T)$, and zero
temperature residual entropy $ S_{imp}(0)$.  The results are
$$C_{imp}(T) = {3\pi^2 \over 2} \zeta^2\Delta [{T\over T_k}]^{2\Delta}
\leqno(3.4.47)$$
$$\chi_{imp}(T) = [{M\zeta \over 2}]^2 {1\over T} [{T\over
T_K}]^{2\Delta}
\leqno(3.4.48)$$
$$S_{imp}(0) = \ln 2 - {\pi^2 \over 2 M^2} ~~.\leqno(3.4.49)$$
The resulting Wilson ratio (the dimensionless ratio of susceptibility
to
specific heat coefficient) depends on $M$ as
$$R = lim_{T\to 0} {\chi_{imp} \over C_{imp}}{C_{bulk}\over
\chi_{bulk}} =
{M^3 \over 36} ~~,\leqno(3.4.50)$$
where $C_{bulk} = 2M\pi^2 TN(0)/3$ and $\chi_{bulk} = 2MN(0)$.
The Bethe-Ansatz and conformal theory exact results give
$\chi_{imp}(T),
C_{imp}(T)/T \sim T^{2/(M+2) - 1}$, which clearly agrees to within
leading
 order in $1/M$ in the exponent.
For comparison of the amplitudes, Bethe-Ansatz and conformal theory
give for the
entropy
$$S_{imp}(0) = \ln[2\cos({\pi\over M+2})] \approx \ln 2 + ln[1-
{\pi^2\over 2M^2}]
\approx \ln 2 - {\pi^2\over 2M^2}, ~M\to\infty \leqno(3.4.51)$$
and for the Wilson ratio
$$R = {(4+M)(2+M)^2 \over 36} \approx M^3/36 ~,
M\to\infty~~.\leqno(3.4.52)$$

Gan, Andrei and Coleman also computed several dynamical quantities.
The
electrical resistivity
is found to be
$$\rho(T) = [ {3\pi^2\over 4 M^2}] c\rho_U [1- M\zeta [{T\over
T_K}]^{\Delta}]
\leqno(3.4.53)$$
where $c$ is the impurity concentration and $\rho_U$ corresponds to
unitary
scattering off of the impurity.  The exponent  of the next leading
correction
obtained by conformal theory is $2/(M+2)$ which agrees to leading order
in
$1/M$ with the above expression.
Notice that the prefactor in square braces may be interpreted as spin
disorder scattering--computing the spin disorder scattering from an
impurity
with dimensionless coupling strength $1/2M$ gives precisely this
estimate.
Clearly, this saturation
 of $\rho(T)$ contrasts with the $T^2$ saturation found in the
 conventional
 Kondo problem.

The lack of Fermi liquid behavior is made more concrete by studying the
dynamic spin susceptibility $\chi''$. It is found that the spin
fluctuation
power spectrum
$${\chi''(\omega,T)\over \omega} \approx {3\over 4T}
[{Max(T,\omega)\over
T_K}]^{2\Delta} {\Lambda(T) \over \omega^2+\Lambda(T)^2}
\leqno(3.4.54)$$
where $\Lambda(T) = 4\pi T/M$.  In the single channel Kondo problem, a
Lorentzian form works rather well to describe $\chi''/\omega$, but
$\Lambda \simeq T_K$.  The vanishing width at low
$T$ found here contrasts with the local Fermi liquid theory of the
single
channel model, and agrees
with results found for the two- and three-channel model using the NCA
method
(Kim [1995]; Kim and Cox [1995,1997]) (see Secs. 5.1,5.2).

We will further compare these large $M$ results with the NCA (Sec.
5.1),
Conformal Field Theory (Sec. 6.1), and the Bethe-Ansatz (Sec. 7) later
in the paper.

\subsubsection{Next Leading Logarithm Results in the $SU(N)\otimes
SU(M)$
Coqblin-Schrieffer  Model} 

In this subsection we wish to just briefly point out that with a
Hamiltonian of the form
in Eq. (2.3.2) is used, the origin of the next to leading order term in
the $\beta$ function
has a different origin than in the spin exchange form.

\begin{figure}
\parindent=2.in
\indent{
\epsfxsize=2.5in
\epsffile{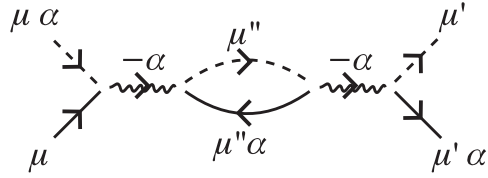}}
\parindent=.5in
\caption{
Leading order multiplicative renormalization group diagram for the 
$SU(N)\times SU(M)$ Coqblin-Schrieffer Model ($N$=spin degeneracy, $M$=channel degeneracy).  
The wavy line is an $SU(N)$ exchange vertex, the dashed line a pseudo-fermion line for
the local $SU(N)$ impurity spin, and the solid line a conduction electron line (these 
carry both channel and spin).  }
\label{fig3p20}
\end{figure}

The first point is that at second order in the coupling strength, the
only diagram possible is
that of Fig.~\ref{fig3p20}.  Note that we have drawn the exchange vertex as an
extended
wavy line. This diagram corresponds to a dynamical dressing of  the
exchange interaction
through pseudo-fermion particle-conduction hole pairs. The contribution
to the $\beta$ function
from this diagram is simply $-Ng^2$ where $g=N(0)J$ is the
dimensionless coupling strength.

The next point is that at third order in the coupling strength, no
vertex correction can be
written down for the Coqblin-Schrieffer Hamiltonian.  Thus, all
corrections to third order
arise solely from the pseudo-fermion self-energy [Coleman, 1983].  When
this is evaluated
for the Coqblin-Schrieffer model we find
$$\Sigma(\omega,T) \approx -MNg^2\omega \ln({D\over max\{\omega,T\}})
\leqno(3.4.55)$$
and the resulting scaling equation for the coupling is
$${\partial g\over \partial x} = -N[g^2 + Mg^3 + ....]
~~.\leqno(3.4.56)$$
If we solve for the Kondo scale at this
order, we find
$$T_K = D|g(0)|^{\gamma}\exp({1\over Ng(0)})~~. \leqno(3.4.54)$$
Eq. (3.4.57)
gives the fixed point coupling strength $g^* = -1/M$; the slope of
the $\beta$ function at
the fixed point is $\beta'(g^*) = -N/M$, so that $g(x) \approx g^* +
\delta g (x/x_0)^{-N/M}$.
Defining $\gamma=M/N$, we see that an expansion in $1/\gamma$ is
possible about the fixed point.
We see that the leading deviations from the fixed point are irrelevant,
since as $x$ grows the
second term shrinks.
Clearly this
results  agrees with Eq. (3.4.45) when $N=2$, so the physics
is the same as the
more conventional Kondo exchange model in that case.

From this perspective, it is clear that the NCA to be discussed in
Sec.
5,
which produces self consistent analytic solutions
for the self-energy of the pseudo-fermion and pseudo-boson of the
$SU(N)\otimes SU(M)$
Anderson model is in some sense just the proper analytic continuation
of the third order scaling
theory.  The pseudo-boson self-energy should just be viewed in the
Coqblin-Schrieffer
context as dynamical dressing of the exchange by pseudo-fermion
particle, conduction-hole pairs
(or particle-particle, depending on which configuration the
pseudo-boson resides in).

\subsection{Path Integral Approach to the TLS Problem} 

A separate approach which has been frequently used is the path integral
method first pioneered
by Anderson, Yuval and Hamann (Hamann [1970], Yuval and Anderson
[1970], Anderson, Yuval,
and Hamann [1970]).  This approach yields renormalization group
differential equations which
have many of the features of the leading and next
leading scaling theory but are different in a number of respects we
shall make clear.
The basic idea is to map the problem to a classical one dimensional
Coulomb gas problem.  Renormalization group equations are then derived
for this Coulomb gas.  The screening interaction $V^z$ gives rise to
phase shifts which will be expressed in terms of
the ``charges'' of the Coulomb gas,
while the tunneling processes give rise to ``fugacities,'' and
may make contributions to the charges as well.  The contributins
to the charges are determined by the electronic transitions
combined with the tunneling processes.  In making
this mapping,
the local conduction green's function is approximated with a long time
form
which allows one to pick up leading logarithmic behavior to all orders
in $V^z$.
Hence, the path integral approach allows the phase shift to be
introduced in the place of
$V^z$ for terms that are of the leading logarithmic order.  However,
the
tunneling terms which give the fugacity must still be treated
perturbatively,
and the long time approximation does not allow next leading logarithms
associated with $V^z$ to be picked up.  Hence, the RG equations are
something of a mix of leading and next leading log theory treated both
perturbatively and non-perturbatively.  While it is a great advantage
to have $V^z$ treated non-perturbatively at least in leading logs,
the phase shift is ill defined at the non-trivial fixed point.
Hence, the present method is useful for giving some overall view of
the scaling flow tendencies, and is not to be relied upon in the
vicinity of the non-trivial fixed point.

The path integral approach to the TLS model is very elaborate and
cannot be followed without
studying the original papers (\zow~ and \zim~ [1985], \vld, \zow, and
\zim [1988a,b]).  The present section provides only a brief outline by
pointing out some of the crucial steps and the additional complications
in comparison to the treatment of the Kondo problem by Yuval and
Anderson
[1970].

In the previous treatment of Secs. 3.3 and 3.4, the renormalization
group
was constructed by determining the $\beta$-function up to the third
order
in perturbation theory.  This is an internally consistent approach in
the weak coupling limit.  In physical applications, however, the
screening coupling $v^z$ can be large while the other two remain
small $v^x,v^y<<1$.  This situation is thus a hybrid of weak and strong
coupling.
A systematic treatment has been worked out in which the phase shift
associated
with an arbitrarily large $v^z$ is combined with small $v^x,v^y$.  The
renormalization group is constructed by using the path integral method
only in the leading order of $v^x,v^y$ and thus the next leading
results given by Eq. (3.4.12) cannot be reproduced exactly (the
terms are similar, but there are differences in the
coefficients).

\begin{figure}
\parindent=2.in
\indent{
\epsfxsize=2.5in
\epsffile{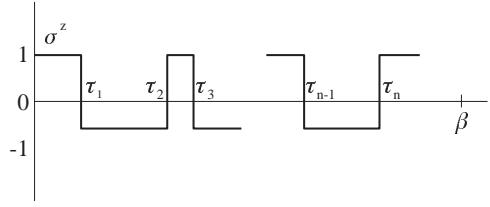}}
\parindent=.5in
\caption{The classical path of the pseudo-spin $\sigma^z$ is
shown as a function
of the imaginary time $\tau$.   $\tau_1,\tau_2....$ are the times of
the jumps.}
\label{fig3p21}
\end{figure}

We now describe the main procedure following Anderson, Yuval, and
Hamann (AYH) (Hamann [1970], Yuval and Anderson [1970], Anderson, Yuval
and Hamann [1970]).  The TLS flips between two states. Following each
flip the screening cloud starts to build up.  If the flips are far
enough apart in time (see Fig.~\ref{fig3p21} ), the the screening can be
described by the long time
shift.  Assuming that the flips are spontaneous, the problem was solved
by Yu and Anderson [1984] using the AYH method, where the free energy
is constructed with the imaginary time path integral method.  The
electron
variables are integrated out so that the free energy only depends on
the
classical path of the TLS on the time interval $(0,\beta)$ (see Fig.
\ref{fig3p21}).
The interaction along the time axis between the flips is logarithmic as
in a one dimensional
Coulomb gas.  We note that Yu and Anderson also [1984]
considered how the TLS
forms beginning with a single well potential and assuming strong
electron-phonon
coupling.

\begin{figure}
\parindent=2.in
\indent{
\epsfxsize=2.5in
\epsffile{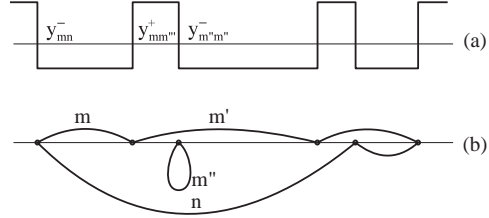}}
\parindent=.5in
\caption{(a) The jumps of the quasi-spin are shown and the
fugacities are
also indicated. 
(b) A particular choice of the connecting of the different fugacities
by electron lines
is shown.  }
\label{fig3p22}
\end{figure}

The current problem is more complex than that considered by Yu and
Anderson, as discussed by \vld, \zow, and \zim~ [1988a,b], since even
if the electrons are integrated out the partition function still depends
explicitly on the particular choice of the interaction at the different time
points and the summation over the different realizations of those interactions
cannot be given in a closed form.  The problem without assisted hopping was
studied by Black and Gy\"{o}rffy [1978] using AYH techniques.
In what follows, we will summarize some of the main ideas to give an
overview to the work in the original papers, leaving the technical details to
those texts.

For the sake of simplicity, assume that a linear combination of
angular momentum
channels is used which diagonalizes the screening interaction so that
$V^z_{mm'} =
\delta_{mm'}$ (see Eq. 2.1.29 for the definition of $V^z_{mm'}$.  Now,
consider
a the classical path $\sigma^z(\tau)$ of the tunneling atom shown in
Fig.~\ref{fig3p21} where the states of the atom are characterized by the Ising
variable $\sigma_z=\pm 1$.  (We note that the representation in terms
of these
Ising variables may be effected directly by the discrete
Hubbard-Stratonovich
transformation introduced by Hirsch [1983] and Hirsch and Fye [1986] [
Zhang, 1994].  If one considers only the $V^z$ interactions, then
different
time dependent potentials $V^z_m(\tau)=V^z_m\sigma^z(\tau)$ are acting
on
the electrons of angular momentum $m$.  This potential is shown in
Fig.~\ref{fig3p22}(a).
Each flip of $\sigma^z$ is associated with either a spontaneous flip or
an assisted hopping process that turns the pseudo-spin
either up or down, and
which
must also be characterized by the angular momentum channels of the
incoming
and outgoing electrons.  In statistical physics, the probabilities of
these
hoppings are introduced as fugacities, which we define here as
$$ y = \Delta_0\tau_0,~~ y^+_{mm'} =
V^+_{mm'}\rho_0,~~y^-_{mm'}=V^-_{mm'}\rho_0~~ \leqno(3.5.1)$$
where $\tau_0$ is a short time cutoff replacing the bandwidth,
i.e., $\tau_0 \simeq \hbar/D_0$.
Notice that even if the interaction $V^z$ is diagonal, the interactions
$V^{\pm}$ need not be diagonal in the angular momentum indices.
In Fig.~\ref{fig3p22}(b), we illustrate the electron hops
by projection of the Ising variable flips down to interaction points
labelled
by the angular momentum index.
Note that the $V^z$ interactions are not represented
by points as they must be treated as time dependent potentials.  These
are constant between points along the projection line.

The continuous electron lines which connect the interaction points in
Fig.~\ref{fig3p22}(b) 
can have either up or down real spin.  The continuous lines must
form
loops with the number of loops depending on the way the connections
between
points are made.

Let us first consider the effect of the time dependent field on an
electron in
angular momentum state $m$.  In the presence of the potential
$V^z_m(\tau)$,
the Dyson equation of the one electron Green's function is solved.
Applying the
technique of \noz~ and de Dominicis [1969], the Green's function
depends
on the time
positions $\tau_i$ of flip $i$. The on-site Green's function in angular
momentum
state $m$ has the form for long times $|\tau-\tau'|>>\tau_0$ (where
$\tau_0$
is a short time cutoff)
$$G_m(\tau,\tau') = {\rho_0\over \tau-\tau'} \cos^2\delta_{\mu}
\exp[\sum_i c_i^m
\ln|{\tau_i-\tau'\over \tau_i-\tau}|] ~~.\leqno(3.5.2)$$
The denominator $(\tau-\tau')^{-1}$ is due to the long time
approximation of an
unperturbed electron leaving and arriving at the TLS.  The phase shift
$\delta_m$
is used instead of $V^z_m$ with
$$\delta_m = -tan^{-1}(\pi\rho V^z_m) ~~.\leqno(3.5.3)$$
The coefficients of $c^{\mu}_i$ are introduced to indicate whether the
spin flips
up or down at flip time $i$, so that
$$c^\mu_i = {\delta_\mu(\tau_i+O^+)\over \pi}-
{\delta_\mu(\tau_i-0^+)\over \pi} ~~.\leqno(3.5.4)$$

Calculation of the part of the electron line contribution to the
diagram of Fig.~\ref{fig3p22}(b)
from the exponential factor on the right hand side of Eq. (3.6.3) gives
a result independent of the way in which the different interaction
points
are connected.  This does not hold for the products of the first factor
$1/(\tau-\tau')$.   Anderson and Yuval [1970] noticed that for a given
channel $M$ the sum of the various contributions of the products of
those
factors forms a Cauchy determinant of order $N$ which is
$$det{(\tau_i - \bar\tau_j)^{-1}} = {\prod_{i<i'}(\tau_i-\tau_i')
\prod_{j>j'}(\bar\tau_j-\bar\tau_j')\over
\prod_{ij}(\tau_i-\bar\tau_j)}
\leqno(3.5.5)$$
where $\tau_1,\tau_2,....\tau_N$ and
$\bar\tau_1,\bar\tau_2,...\bar\tau_N$
are the time orderings of the times where the electron is created and
annihilated, respectively.  This is valid only if the electrons are
created and
destroyed at different times which we assume for the time being.

With this assumption, the partition function can be written in the form
of
a sum over configurations which we explain presently.  The interaction
points $\tau_i$ are on the imaginary
time axis in the interval $(0,\beta)$ as shown in Fig.~\ref{fig3p22}(a) .  Each
point is associated with
one of the fugacities $y,y^{\pm}_{mn}$.
(There are other fugacities $y^z,y^z_{\pm}$ generated for
technical reasons we shall not discuss here.)
That association is called a configuration $\{\alpha\}$.  If the
fugacity
descibes the assisted process, the the angular momentum indices of the
outgoing
and incoming electron lines $(m,n)$ must also be added, which forms
another
configuration $\{m,n\}$.  Finally we need a combinatorial factor $R=\pm
1$
which is defined in the original paper of \vld, \zow, and \zim~
[1988a].
The resulting partition function is summed over loops of length $N$
and the configurations, giving after much work the result
$$Z = \sum_{N=0}^{\infty}
(-1)^N\sum_{\{\alpha\}}\sum_{\{m,n\}}\prod_{j=1}^N
y_j Tr_{\sigma}\prod_{i=1}^N[\sigma(\alpha_i)] \leqno(3.5.6)$$
$$\times
\tau_0^{-N}\int_0^{\beta}d\tau_N...\int_0^{\tau_{i+1}-\tau_0}d\tau_i
.....\int_0^{\tau_2-\tau_0} d\tau_1 R \exp[\sum_{i<j}\vec C_i\cdot\vec
C_j
\ln|{\tau_i-\tau_j\over \tau_0}|]  $$
where the real spin degeneracy has been set to 1 for the moment for
simplicity.
The time $\tau_0\sim D^{-1}$ serves as a short time cutoff for the
problem.
The matrices $\sigma(\alpha_i)$ are $\sigma^{+},\sigma^-,\sigma^z$ at a
given
time point for the fugacity configuration $\{\alpha(\tau_i)\}$ depend
on
whether the TLS pseudo-spin is flipped up,flipped down, or does not
change.
The vector ``charge'' $\vec C_i$ has components
indexed by the angular momentum
label $m$
and is given by
$$C_{im} = {[\delta_m(\tau_i+0^+)-\delta_m(\tau_i-0^+)]\over \pi}
\leqno(3.5.7)$$
for spontaneous tunneling, and
$$C_{im} = {[\delta_m(\tau_i+0^+)-\delta_m(\tau_i-0^+)]\over\pi}
+
\delta_{m,m} - \delta_{m,n} \leqno(3.5.8)$$
for assisted tunneling, with the index pair $m,n$ associated to the
charge.  (For technical reasons in the initial papers of \vld, \zow,
and \zim~ [1988a,b] charge matrices were used instead of vectors
in order to avoid the diagonalization of $V^z_{mn}$.)
The phase shifts are due to the exponents in Eq. (3.5.2), while the
Kronecker delta terms take into account the determinants.  Note the
analogy to the 1D Coulomb gas with a logarithmic interaction; the
vectors $\vec C_i$ take the place of the charges.

Clearly at this point a great complication has occured relative to the
ordinary
spin Kondo problem.  In the latter case the configurations $\{\alpha\}$
and $\{m,n\}$ don't occur because the spin flips occur in a
simple alternating way (+-+-+-....).
Because of this, the more complex charge factor of Eq. (3.5.6)
is simply replaced
by $(-1)^{i-j}\delta/\pi$ in the ordinary Kondo problem.

A second complication is even more serious. It can happen that at a
time
point $\tau_i$ of an assisted tunneling process, the electron is
annihilated and
created with the same angular momentum index $m=n$.  In that case,
$\tau_i=\bar\tau_i$,
so a divergence appears in the expression of the Cauchy determinant in
Eq. (3.5.5).  The divergence can be eliminated by the short time cutoff
$\tau_0$, by splitting the time of annihilation and creation by
$\tau_0$ (i.e., as $\tau_i=\bar\tau_i+\tau_0$).
This complication results in a separate treatment of the Hartree-Fock
(HF)
terms in which a single electron line is attached to the aforementioned
time point.  This leads to the introduction of HF fugacities.

The main consequence of this complication is that a charge at time
$\tau_j$
can interact with two charges at the points $\bar\tau_i$ and
$\tau_i=\bar\tau_i+\tau_0$.
Since the interaction of the 1D Coulomb gas is logarithmic, the
following
expression will appear in the sum over interaction terms:
$$\ln|{\tau_j-\tau_i+\tau_0\over\tau_0}| - \ln|{\tau_j-\tau_i\over
\tau_0}|
\approx {\tau_0\over \tau_j-\tau_i} \leqno(3.5.9)$$
where $|\tau_i-\tau_j|>>\tau_0$.  Thus the charge at point $\tau_j$
interacts with the opposite charges forming a dipole at $\tau_i$.
Therefore, in the $m=n$ case, dipoles must be introduced which lead to
charge-dipole interactions.  For a complete discussion of all these
complexities,
we refer the reader to the original papers \vld, \zow, and \zim~
[1988a,b] where the complete
expression for the partition function is given.

The derivation of the scaling equations follows a delicate and
cumbersome
procedure.  The idea is to eliminate the short time behavior in small
steps
or in other words to replace $\tau_0$ by $\tau_0+d\tau_0$ where
$d\tau_0$ is the
increment in the short time cutoff.  In that case if a pair of flips
are eliminated
they must be replaced by a single flip or an interaction without
a flip.  This leads to the essential
renormalization
of the fugacities and the phase shifts.

Restricting to the case when only two angular momentum states are
important but
the channel number (real spin here) is allowed to be arbitrary ($M$),
the scaling equations are expressed in terms of the phase shift
$\delta$,
the fugacity for spontaneous tunneling $y$ and the fugacity for
assisted
tunneling $y_a$ where the $m,n$ dependence of Eq. (3.5.1) is now
removed by
decomposing the fugacities in Pauli matrices times the amplitude.
The scaling equations are given by
$${d(\delta/\pi)\over d\ln\tau_0} = 4y_a^2(1-2M\delta/\pi) -
2y^2\delta/\pi \leqno(3.5.10.a)$$
$${dy_a\over d\ln\tau_0} = 4y_a(\delta/\pi)(1-M\delta/\pi)
\leqno(3.5.10.b)$$
$${dy\over d\ln\tau_0} = y(1-4M(\delta/\pi)^2) ~~.\leqno(3.5.10.c)$$
These are similar to the scaling equations derived previously assuming
noting
that $v^x=v^y$ and the correspondence $\Delta_0\tau_0\approx
\Delta_0/D=y$, and that
to linear order $\delta = -\pi v^z + O([v^z]^2)$.  To within leading
logarithmic order and the expansion of the phase shift, they agree
with the results from the multiplicative renormalization group
treatment
(see Sec. 3.3.1).  Considering the next leading order, some terms are
missing in the scaling equations presented above.  This means that this
derivation is not completely systematic, which can be traced back to
the use of the long time approximation for the Green's function.  The
other main difference from the scaling equations derived previously is
that the spontaneous tunneling occurs in (3.5.10.a) which describes
the renormalization of the screening interaction. In the multiplicative
renormalization group treatment, the electronic phase space is reduced
and the spontaneous tunneling only occurs as an infrared cutoff in the
logarithmic integrals. That is the origin of the difference in
the scaling equations which has already occurred in the
commutative model (see Black and Gy\"{o}rffy) [1978], and Black,
\vld, and \zow~ [1987]).

We now want to discuss the most interesting features of the scaling
equations (3.5.10.a-c).  First, ignoring the spontaneous tunneling
in Eq. (3.5.10.a), we see that the phase shift renormalizes to the
value
$$\delta^* = {\pi \over 2M} \leqno(3.5.11)$$
which seems to have some connection to the fixed point given by
Eq. (3.4.16) if we expand the phase shift to linear order in $v^z$.
However, we must remind the reader that at the non-trivial fixed point
the phase shift loses physical meaning since the projection to outgoing
one particle processes is zero (see Fig.~\ref{fig1p4} in Sec. 1, and 
Sec. 6.1.3). 
Clearly, however, the
flow to a non-trivial phase shift ($\delta\ne 0,\pi/2$) is indicated
by this procedure.  The same result emerges from a Friedel sum rule
estimate of the number of electrons tied on average to the tunneling
center, $Z=1=2M\delta/\pi$.  Indeed, for an infinitesimal applied
spin field in the two channel problem, the phase shift experienced by
up and down spin electrons is $\pm \pi/4$ (see Secs. 4.2,6.1.2).
Next, if we consider the scaling of the fugacities, the primary results
are that: (i)$y_a$ increases if $\delta(0)<\pi/M$, which is always
satisfied for $M=2$ since $\delta(0)<\pi/2$ is assumed.  (ii) $y$
increases if $\delta(0)<\pi/(2\sqrt{M})$, but decreases otherwise
hence giving localization to one well ($y=0$) if the condition is
reversed (see Yu and Anderson [1984]).  For more channels
($M>2$), $y_a\to 0$ may be the case so that localization in one
of the potential wells occurs.

\begin{figure}
\parindent=2.in
\indent{
\epsfxsize=5in
\epsffile{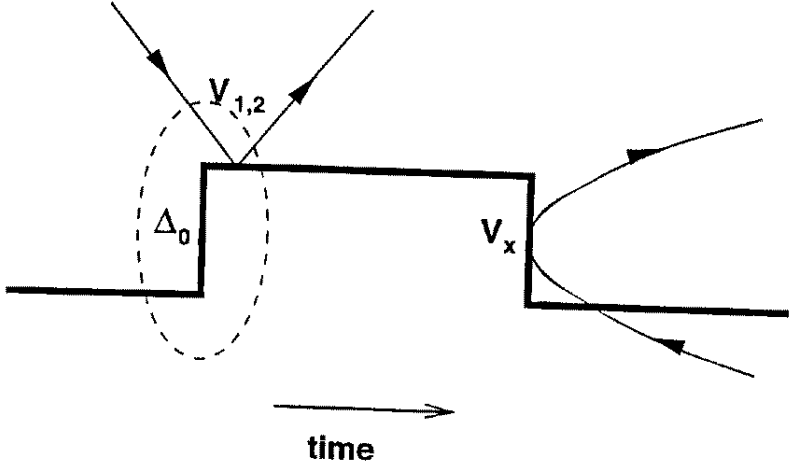}}
\parindent=0.5in
\caption{Diagram resulting in the generation of artificial interaction terms in the elimination
scheme (i).  The heavy line represents the motion of the heavy particle.  The 
artificial interaction is genrated by the elimination of a spontaneous tunneling
event $\Delta_0$ and a potential scattering close to it ($V_{1,2}$).  The lines with 
arrows represent the conduction electrons.}
\label{fig3p23}
\end{figure}

In a recent paper, Moustakas and D. Fisher [1995,1996] used the path
integral approach to study the
commutative model of TLS ($V^z\ne 0,~V^x=V^y=0$) with an
additional potential scattering term at the TLS site (see
$V^0_{\vec k,\vec k'}$ in Eq. (2.17)).  This problem has already
been studied by Kagan and Prokof'ev [1983].  Suppressing the
spin indices, the interaction
part of the Hamiltonian is written as
$$H_1 = V_1 (c^{\dagger}_ec_e + c^{\dagger}_oc_o) + V_1
(c^{\dagger}_ec_e -  c^{\dagger}_oc_o) + V_3
\sigma^z(c^{\dagger}_ec_o+ c^{\dagger}_oc_e) \leqno(3.5.12)$$
where $e,o$ refer to even and odd parity combinations of
conduction states about the TLS center (see Appendix II for a 
discussion of how to generate these states).  Moustakas and D. Fisher
[1995,1996] pointed out that the presence of both $V_1$ and $V_2$ can be
relevant in determining physical quantities through a
modification of the local conduction density of states and in
the modification of states near the band edges in the
renormalization process.   Indeed, one can sum up the diagrams
involving the one particle Green's functions diagonal in the
$e,o$ indices (\zow {\it et al.} [1997]).  The associated two
spectral functions are smooth, but the density of states is
depressed by the increase in bandwidth. The remaining coupling
$V_3$ can be treated in a rotated representation of the TLS
where $-\sigma^x$ replaces $\sigma^z$.  Application of the
multiplicative renormalization group shows that the resulting
model belongs to the commutative class, so that the coupling
$V_3$  remains marginal after making it dimensionless by
multiplying by the density of states that has been renormalized
by $V_1,V_2$.  In this formulation, no new coupling is
generated.  The formulation of Moustakas and D. Fisher [1995,1996]
using the path integral approach derives a generalization of
the scaling equations (3.5.a-c) by introducing an infinite
number of new couplings.  Among them the assisted tunneling coupling 
$V^x$ is also generated by eliminating a close pairing of a 
$V_2$ potential scattering with a spontaneous hopping $\Delta_0$ 
as shown in Fig.~\ref{fig3p23} .  This generates an assited hopping 
term of the order of $V_2\Delta_0$ in lowest order.  
The serious consequence of this result
is that the induction of this interaction $V^x$ which fails to
commute with $V^z$ changes the universality class to the fully
noncommutative model, away from the marginal line of fixed points
associated with $V^x=V^y=0$.  

Considering that essential differences arise between the results 
obtained by the leading logarithms diagrammatic approach and the
program of Moustakas and Fisher [1995,1996], the following questions
are raised:\\
(i) The introduction of the coupling $V_2$ which breaks electron hole
symmetry renders the $V^z=const.$, $V^x=V^y=0$ line of fixed points
unstable as mentioned above; assuming we start with bare $V^x=0$, 
can we get close to the TLS two-channel Kondo fixed point?\\
(ii) There are two essentially different path integral approaches
which may be adopted.  In the space-time scaling, one integrates out 
short regions of time between interaction lines.   In the first approach, 
only those times eliminated in the space-time scaling 
in which both ends are connected to electron lines (\vld,\zow, and \zim
[1988a,b,c].  In the second approach of 
Moustakas and Fisher [1995,1996], as discussed 
above, a spontaneous tunneling process combined with potential scattering
at short times generates an effective $V^x$, even though there is an 
absence of interaction between the dynamical degrees of freedom of the 
TLS and the conduction electrons.  The question is which approach generates
the physical scaling equations, where the generated couplings can be directly used 
to calculate the quantities of physical interest (e.g., the resistivity), 
since at most one approach can be correct unless 
a mathematical equivalence is established.  

We may answer the first question with assistance from symmetry 
considerations (\zow {\it et al.} [1997]).  Consider a 
commutative model with $V^x=V^y=0$, $V^z \ne 0$, $\Delta^z=\Delta_0\ne 0$, 
and $\Delta^z=\Delta=0$.  The Hamiltonian is invariant under the 
following combined symmetry operation: \\
\noindent{$\to$ Interchange the $L,R$ indices of the tunneling TLS.\\
$\to$ Apply electron-hole symmetry in both the $e,o$ channels, 
{\it viz.}, $c_{e,o} \to c^{\dagger}_{e,o}$ (where we have 
suppressed spin indices).  Note that without
explicit spin dependence, we needn't introduce the customary 
$(-1)^{1/2-\sigma}$ factors because they cancel in pairs. (The electron-hole transformation
is defined in this way where the electrons and holes are at the same sector of the Fermi surface, and the 
density of states and couplings are replaced by their values taken at the Fermi surface.  That transformation 
is equivalent to one of the transformations used by Affleck, Ludwig, and Jones [1995] (see Sec. 9.3.1). 
An additional symmetry
is that of time reversal, which gives $c^{T}_{e(o)}=c_{e(o)}$ and $d^T_i=d_i$, $i=1,2$, assuming that all the
atomic wave functions are real.  Furthermore, all $c$-numbers are replaced by their complex conjugates under
time reversal.  The Hamiltonian of Eq. (3.5.12) is also invariant under that transformation.  Finally, 
going over to the left and right states defined by Eqs. (A.2.7a-b) of App. II, the symmetries discussed above
lead to the conservation of total number of electonrs separately for left and right states, as is 
pointed out by Mostakas and Fisher [1995].)} \\
\noindent{It is clear that the spontaneous  hopping interaction 
and the $V^z$ coupling are unchanged under this transformation, as
is the conduction band provided it begins particle hole symmetric.
However, the assisted hopping terms proportional to 
$$\sigma^x(c^{\dagger}_ec_e + c^{\dagger}_oc_o) $$
and
$$i\sigma^y(c^{\dagger}_oc_e - c^{\dagger}_ec_o) $$
change sign under the transformation, as does any potential 
scattering (specifically the $V^1,V^2$ terms).  A dependence
of fixed point stability on particle-hole symmetry is also found in the
two-impurity one- and two-channel Kondo models; see Sec. 9.3.1 and Affleck, Ludwig, and Jones [1995] for 
some details.}

Now, clearly the marginal line of fixed points associated with this
high symmetry Hamiltonian is unstable if  small $V^x$ and/or $V^y$ couplings
are introduced that generate a flow towards the TLS two-channel Kondo
fixed point.  We now demonstrate that any potential scattering leads
to the same instability by breaking the same symmetry.  This also hinges
on the non-zero spontaneous tunneling term.  A simple diagrammatic 
summation of multiple scattering processes results in the renormalization
of the local electronic density of states in both the even and odd 
channels giving 
$$\rho_{e,o} = \rho_0[1+\alpha_{e,o} {\epsilon \over E_0}] \leqno(3.5.13)$$
where $\alpha_{e,o}\sim V_1$ and $V_2$ to leading order measures the asymmetry
and $E_0$ is of the order of the bandwidth.  We note that the $e,o$ ``bands''
may have initial asymmetry as discussed in Appendix II (see also Sec. 9.3.2
on the two-impurity model). 

\begin{figure}
\parindent=2.in
\indent{
\epsfxsize=5.in
\epsffile{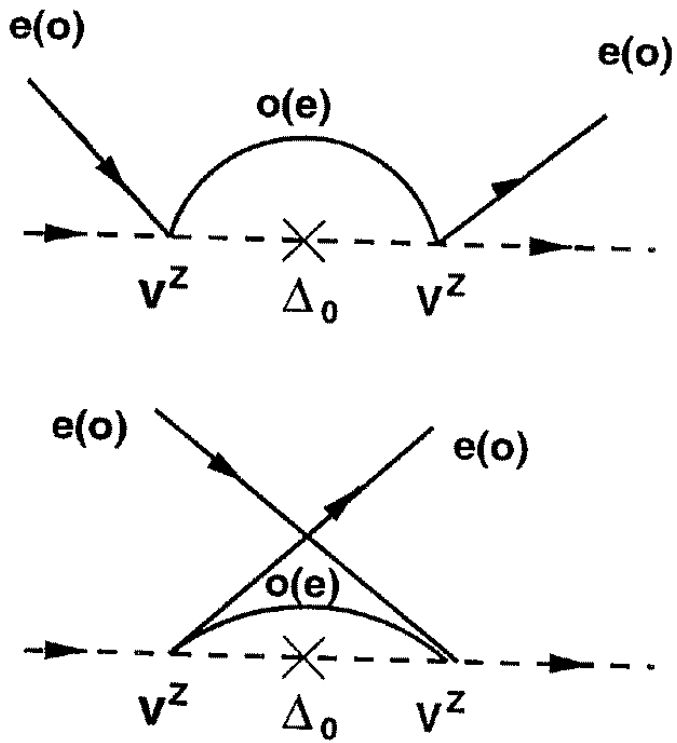}}
\parindent=0.5in
\caption{Time ordered diagrams generating the assisted tunneling process in the presence 
of appropriate electron-hole symmetry breaking.  The cross indicates a
spontaneous tunneling between 
the two positions of the heavy particle.  The labels on the electron lines refer to the even and odd parity electron channels with repsect to
reflection through the center of the system. }
\label{fig3p24}
\end{figure}

\begin{figure}
\parindent=2.in
\indent{
\epsfxsize=5.in
\epsffile{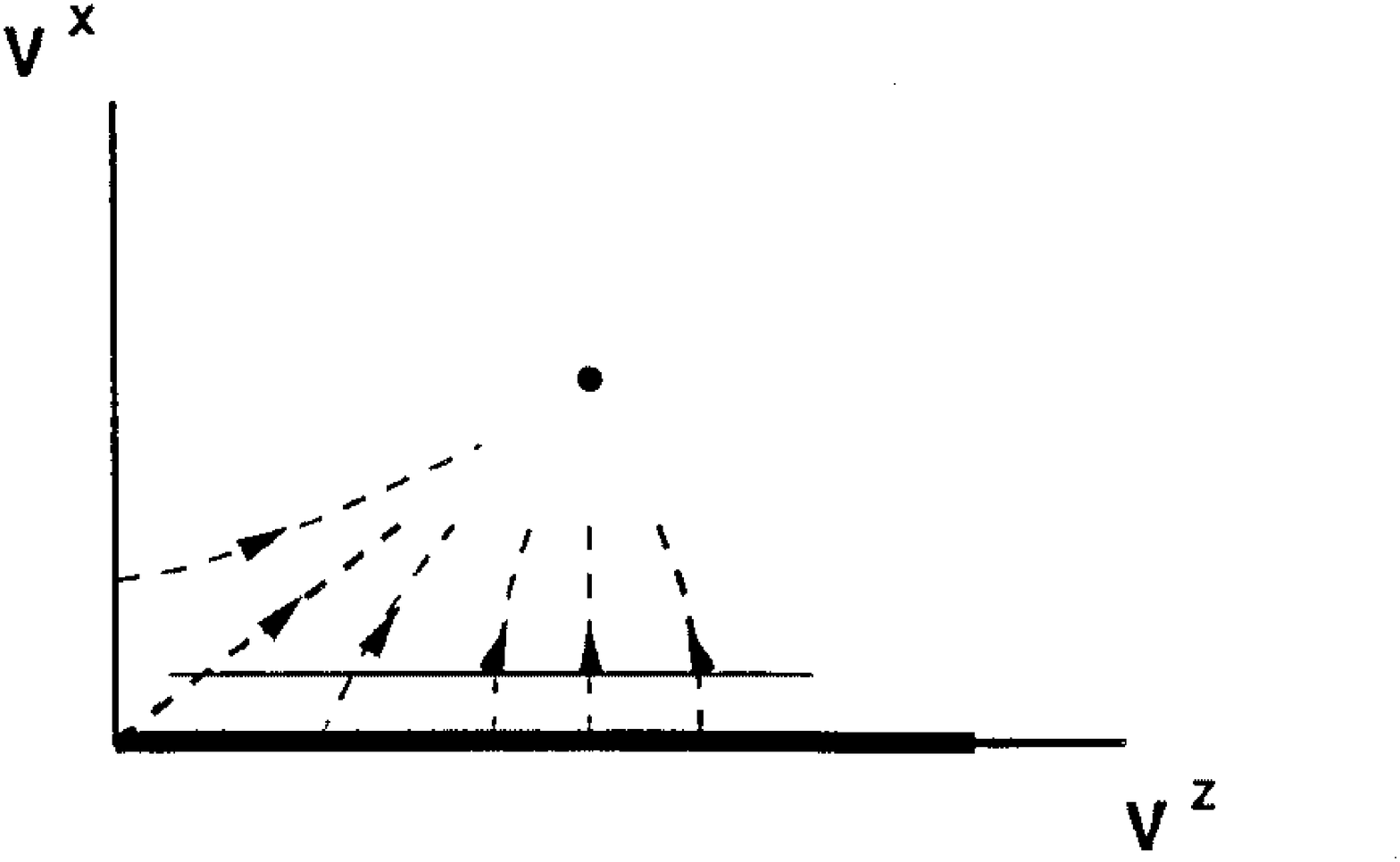}}
\parindent=.5in
\caption{Sketch of the scaling trajectories of the TLS.  The appropriate 
electron-hole symmetries drives the TLS away from the marginally stable
fixed line $V_x=0$($\Delta_1=0$) towards the two-channel Kondo fixed point.  The scaling
is stopped by the renormalized splitting, and the final ground state is a Fermi 
liquid.  The freezing out of the TLS is indicated by a light continuous line.}
\label{fig3p25}
\end{figure}

In a diagrammatic renormalization group analysis 
the generation of the assisted hopping takes place as suggested by 
Moustakas and Fisher [1995,1996], but at a lower order than in their approach, namely, in 
the leading logarithmic order. 
Refer to the diagrammatic scaling analysis depicted in 
Fig.~\ref{fig3p24}. 
which displays 
second order vertex corrections 
which include one spontaneous hopping renormalization of the internal 
TLS pseudofermion line.   The contributions of the diagrams shown to 
the dimensionless coupling $v^x = \rho_0 V^x$ are
$$ 
(v^z)^2 \Delta^x\{\int_T^D {d\epsilon\over \epsilon^2} \rho_0(1+\alpha
{\epsilon\over E_0}) - \int_{-D}^{-T} {d\epsilon\over \epsilon^2} 
\rho_0(1+\alpha {\epsilon\over E_0}) \} $$
which leads to
$$\delta v^x \simeq (\rho_e\alpha_e-\rho_o\alpha_o)(v^z)^2 \alpha ({\Delta^x\over E_0}) \ln ({D\over T}) 
\leqno(3.5.14)$$
where $\rho_{e(o)},\alpha_{e(o)}$ may be calculated straightforwardly as an
extension of App. II. 
The initial strength of this generated interaction is very weak, as
typically $\Delta^x/E_0 \simeq 10^{-5}$, $v^z(0) \simeq 0.2$, and 
$\alpha << 1$. Hence, at $T\sim 1K$, the generated dimensionless 
coupling strength would be of order $v^x \simeq 5\times 10^{-6}$, 
which is very small.  Below this temperature the generation of the
non-commutative couplings will be stopped due to the splitting $\Delta^x
\simeq 1K$ which serves as an infrared cutoff (see Fig.~\ref{fig3p25}). 

Thus the breaking of the artificial combined particle-hole symmetry 
by introducing $V_2$ does indeed render the marginal line of fixed points
of the commutative model unstable as suggested by Moustakas and Fisher [1995,1996].
However, we do not expect this symmetry breaking instability to lead 
in parameter space to the vicinity of the TLS two-channel Kondo fixed 
point.  Thus even with the symmetry breaking present, it appears to 
be difficult to observe experimentally relevant consequences.  

We now briefly 
return to question (ii) from above 
which concerns which path integral scaling approach is appropriate 
(for more details see \zow {\it et al.} [1997]).  The
free energy which is the logarithm of the path integral can be 
computed by either approach, and demanding invariance of the partition
function (free energy) then leads to different scaling equations 
between the two methods.  In principle this is not a problem as it
may correspond to a simple reorganization of variables. However,
when computing any physical quantity with the obtained renormalized
couplings we must of course get equivalent physical results if each
method is correct.  A check is whether the scaling equations can be
made to correspond to the diagrammatic method associated with the
multiplicative renormalization group (leading logarithms 
approximation) at the appropriate level of approximation.  In this way
it has been shown that the approach in which only those time intervals 
are eliminated with endpoints connected to electron lines gives the same 
scaling as the multiplicative renormalization group; physically inequivalent
scaling equations are generated by the second approach of Moustakas 
and Fisher [1995,1996].  The first space-time scaling method also
has the feature in common with the multiplicative RG that in a noninteracting
TLS-electron system no new couplings are generated.  

A more general discussion of the scaling trajectories in the model of 
Moustakas and Fisher [1995,1996] was presented by Ye [1996e], applying 
bosonization methods to the problem.  This will be briefly discussed in 
Sec. 6.3.  His key finding is that all scaling goes finally to a Fermi 
liquid fixed point.  

Finally, we close this section noting that 
as will be discussed in Sec. 6.2, the two-channel Kondo problem
can, for a particular value of the coupling $V^z$ (phase shift
here) be mapped to a resonant level model as pointed out by
Emery and Kivelson [1992].  That model has a strong similarity
with the model suggested by Toulouse [1970].  In these works
the mapping is established by bosonization. Recently, Fabrizio,
Gogolin, and \noz~ [1995] have established the mapping to the
resonant level model by starting with the path integral
formulation.  They studied the partition function by studying its
expansion in terms of the perpendicular coupling of the
two-channel Kondo problem and compared with a similar expansion
of the resonant level model of Emery and Kivelson [1992].  They
have shown that the the two expansions agree term-by-term for
the special value of the longitudinal coupling $V^z$ (which
corresponds to $J^z$ in Sec. 6.2).  The virtue of the path
integral approach in this context is the ability to study the
cross-over regime from high to low temperatures as opposed to
the asymptotic fixed point regime accessible with the
bosonization approach.

\section{Numerical Renormalization Group Approach} 

Our goal in this section will be to provide an overview of the NRG
method
pioneered by Wilson [1973,1975] for the ordinary Kondo model.  We will
stress
the results obtained from this approach which provide insight on the
two-channel Kondo model and the influence of symmetry breaking fields
on that
model.  Specifically, we
will consider the effects of exchange anisotropy, ($J^{(1)}\ne
J^{(2)}=J^{(3)}$), the influence
of local and bulk fields which couple to the spin, and the influence of
a
channel field which splits the exchange  integrals for the electrons in
the two
different channels.  We will also discuss a simple ``shell model''
approach
which works remarkably well for characterizing the lowest few states of
the
non-trivial fixed point and appears to generalize to more complex
models not
treatable by the NRG.

The original NRG work on the two-channel model was performed by Cragg,
Lloyd and
Nozi\`{e}res [1980], who provided the first non-perturbative
confirmation of the
Nozi\`{e}res and Blandin arguments for a non-trivial fixed point in
multichannel models with
$M>2S_I$. These calculations were performed for strong bare coupling
values and for one case where the exchange in one channel differed from
that in
the other.   Subsequent calculations have been performed by Pang and
Cox
[1991], and Affleck {\it et al.} [1992] to explore the weak coupling
approach
to the transition and the influence of exchange
anisotropy, applied local and bulk spin fields, and channel fields
(through the
exchange splitting between channels) in greater detail.

\subsection{Logarithmic Discretization, Hamiltonian, and 
Renormalization Group
Transformation } 

{\it (a) Logarithmic Discretization of the Hamiltonian}\\

\begin{figure}
\parindent=2.in
\indent{
\epsfxsize=6.in
\epsffile{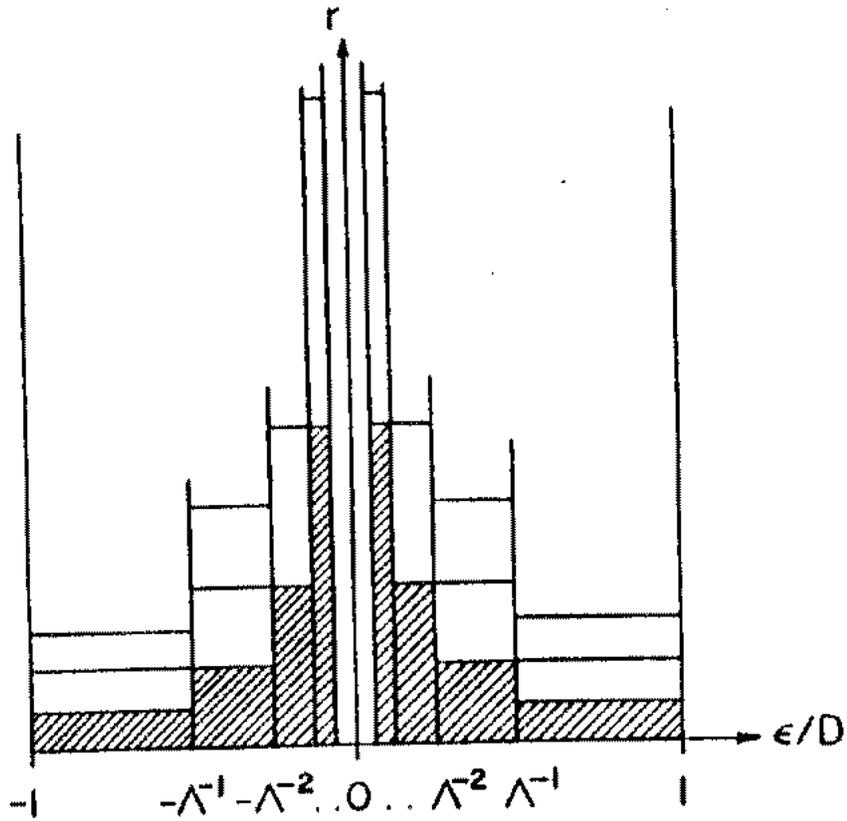}}
\parindent=.5in
\caption{Logarithmic discretization of the conduction band for Numerical
Renormalization Group (NRG) calculations.  The conduction band is divided
up into logarithmic bins $[\Lambda^{-(n+1)},\Lambda^{-n}]$ for electrons 
and $-[\Lambda^{-n},\Lambda^{-(n+1)}]$ for holes, with $0<n<\infty$.  The
states are Fourier analyzed on each bin, and only the average Fourier components
are retained (as only these couple to the impurity operators). After 
Krishna-murthy, Wilkins, and Wilson [1980].}
\label{fig4p1}
\end{figure}

We begin with the two-channel Kondo model given by Eq. (1.1). As in the
introduction, we label spin with $\mu =\uparrow,\downarrow$ and
channel
index with $\alpha=\pm$.   Following the
pioneering work of Wilson [1973,1975],
we transform the Hamiltonian in the following ways:\\
(1) We logarithmically discretize the conduction band as
shown in Fig.~\ref{fig4p1}.  This means we split the band up into a sequence of
intervals between $\Lambda^{-n}D >| \ek| > \Lambda^{-(n+1)}D$, with
$\Lambda
>1$ the logarithmic discretization parameter.\\
(2) Within each logarithmic
interval, we Fourier analyze the conduction states, but retain only the
average
components.  There are two justifications for this seemingly gross
neglect: (i)
only the average states couple to the impurity spin given the assumed
contact
form of the Kondo exchange interaction, and (ii) in the full continuum
limit
($\Lambda \to 1$) only these average states survive.  \\
(3) For numerical convenience, we employ the Lanczos algorithm to
convert the Hamiltonian to a tridiagonal basis which has a
meaning in position space.  The Lanczos states correspond to electron
creation
operators $f^{\dag}_{n,\alpha,\mu}$ which have approximate radial
extent of
$\Lambda^{n/2}/k_F$ measured from the impurity.  In order to get
these wave functions the hoping amplitudes $\epsilon_n$ must
also depend upon $N$.  As the electronic energy grows, so does
the energy spacing.  Concommitantly, the
amplitudes at the impurity site are increased due to localization
of the electron into smaller volumes.  The corresponding wave
functions have higher numbers of nodes for higher $n$ so that
orthogonality
holds.  The state created by $f^{\dag}_{0,\alpha,\mu}$ is just the
on-site or
Wannier orbital at the impurity position.

Following these steps, the two-channel Kondo Hamiltonian takes the form
$${H_{NRG} \over D} =
\sum_{n=0,\alpha,\mu}^{\infty}\epsilon_n[f^{\dag}_{n,\alpha,\mu}f_{n+1,\alpha,\mu}
+ h.c.] -{J\over D}\vspi \cdot \sum_{\mu,\alpha,\alpha'} {\vec
S}_{\alpha,\alpha'}
f^{\dag}_{0,\alpha,\mu}f_{0,\alpha,\nu}  ~~,\leqno(4.1.1)$$
where
$$\epsilon_n = {\Lambda^{-n/2}(1+\Lambda^{-1})(1-\Lambda^{-(n+1)})\over
2[(1-\Lambda^{-(2n+1)})(1-\Lambda^{-(2n+3)})]^{1/2}} ~~.\leqno(4.1.2)$$
The $\epsilon_n$'s correspond to the radial hopping matrix elements
which
transfer electrons from a state in ``shell'' $n$ to ``shell'' $n+1$.
Thus in a
way rather different from the conformal field theory approach and
Bethe-Ansatz
approaches, we arrive at an effective one dimensional problem, with
only the
radial dimension being important.  \\

{\it (b) Renormalization Group Transformation}\\

We can define a sequence of finite size
dimensionless effective Hamiltonians which will reproduce $H_{NRG}$ in
the
thermodynamic limit, and
which define a renormalization group transformation.  The discretized
Hamiltonians are
$${\cal H}_N = \Lambda^{N/2}\{\sum_{n=0,\alpha,\mu}^{N}\epsilon_n
[f^{\dag}_{n,\alpha,\mu}f_{n+1,\alpha,\mu}
+ h.c.] -{J\over D}\vspi \cdot \sum_{\mu,\alpha,\alpha'} {\vec
S}_{\alpha,\alpha'}
f^{\dag}_{0,\alpha,\mu}f_{0,\alpha,\nu}\} \leqno(4.1.3)$$
and the limit which reproduces $H_{NRG}$ is
$$H_{NRG} = D\lim_{N\to \infty}[\Lambda^{-N/2}{\cal H}_N] ~~.
\leqno(4.1.4)$$
The renormalization group transformation is
$${\cal H}_{N+2} = \Lambda{\cal H}_N +
\Lambda^{N/2+1}\sum_{n=N+1,\alpha,\mu}^{N+2}\epsilon_n[f^{\dag}_{n,\alpha,\mu}
f_{n+1,\alpha,\mu}
+ h.c.] ~~.\leqno(4.1.5)$$

Notice:\\
(i) The multiplication by $\Lambda^{N/2}$ in Eq. (4.1.3) has the effect
of making the smallest hopping matrix element of order 1, which means
the
smallest resulting dimensionless excitation energy will be of order 1
in the
corresponding
spectrum.  \\
(ii) The diagonalization of ${\cal H}_N$ corresponds to finding the
effective
Hamiltonian to describe the physics at length scale
$$L_N\approx\Lambda^{N/2} k_F^{-1} \leqno(4.1.6)$$
and temperature scale
$$T_N\approx \Lambda^{-N/2} D ~~.\leqno(4.1.7)$$.  \\
(iii) The need for a step size of 2 in the transformation (4.1.5) is
because there
are, generically, different fixed points for even and odd number of
shells.
Physically, we rescale the system size by an amount $\Lambda k_F^{-1}$
and
inquire about the properties on that scale compared to the previous
length
scale.

The practical implementation of the RG transformation (4.1.5) is
carried out by
numerically diagonalizing the truncated Hamiltonians ${\cal H}_N$ for
each $N$
and using the states at level $N$ to construct the basis for states at
level
$N+1$.  The diagonalization process is repeated iteratively. A well
defined
procedure exists for the construction of these basis states using
Clebsch-Gordan technology, and we refer the interested reader to
Krishna-murthy, Wilkins, and Wilson [1980a)] and Jones [1988] for
further
details.    A fixed point of
the transformation (4.1.5) is obtained if the lowest lying spectrum of
eigenvalues for the successive ${\cal H}_N$ is unchanged.

The approximation of logarithmic discretization
thins out the number of states to keep for a particular system size
quite
substantially with respect to the customary exact diagonalization
approaches.
Taken together with block diagonalization using
the symmetries we will discuss in the next subsection, the
problem becomes considerably more manageable.
Even still, the large degeneracy of states
makes full diagonalization of ${\cal H}_N$ usually impossible for
$N\ge2$.
Explicitly, the degeneracy  is $\simeq 16^{N}$ at the $N$-th
shell, the factor of 16 deriving from the product of four conduction
states for
channel one times four conduction states for channel two
including all occupancies.
A
practical way of implementing the transformation while keeping
manageable
numerics is to retain only the lowest few hundred states at each
iteration.
While there is no direct proof that this approximation is reliable,
convergence
tests in the original works [Wilson, 1973,1975; Krishna-murthy,
Wilkins, and
Wilson, 1980a),1980b)] and
{\it a
posteriori} comparison with other exact methods has confirmed the
general
validity of the method.  In calculations for the two-channel model
performed on
a Sun workstation, typically the lowest 250 states were retained at
each $N$
value, while larger runs performed on a CRAY-YMP48 showed no
significant
difference from the runs with fewer states.

While the original efforts on the single channel Kondo model and
$s$-wave spin
1/2 Anderson Hamiltonian had sufficiently many states within this space
to
reliably compute thermodynamic quantities, the higher degeneracy in the
present
problem means that practically we may only compute eigenvalues.
Nevertheless,
a considerable amount of information may be obtained from examining the
spectrum of eigenvalues, as we shall discuss further below.  \\

{\it (c) Use of symmetry to reduce the basis size}

For the isotropic model, one may obviously exploit the $SU(2)$ symmetry
under
spin rotations and use the total spin $S_{tot}^2$ and
$z$-component
$$S^{(3)}_{tot} = \sum_{n=1,\alpha,\mu}^{\infty} \mu
f^{\dag}_{n,\mu,\alpha}f_{n,\mu,\alpha} +S^3_I\leqno(4.1.8)$$
as conserved quantities.  $S^{(3)}_{tot}$ remains useful even in the
presence of an
applied spin field along the $z$-direction or an axial breaking of
exchange
isotropy $J^3\ne J^1=J^2$.  In these cases the $SU(2)$ symmetry is
broken down
to a $U(1)$ symmetry for which $S^3_{tot}$ is the conserved charge.

There are two ways to handle the charge and channel degrees of freedom
of the
conduction electrons.  First, we have an obvious $SU(2)$ symmetry from
rotations in channel space, and we could  employ the total channel spin
operators $S^2_{ch}$ and $S^{(3)}_{ch}$
$$S_{ch}^{(3)} = \sum_{n,\alpha,\mu} \alpha
f^{\dag}_{n,\mu,\alpha}f_{n,\mu,\alpha}
\leqno(4.1.9)$$
as good quantum numbers.  In addition, we can use the conduction charge
operator
$$\hat Q= \sum_{\mu} Q_{\mu} =
\sum_{n,\alpha,\mu}[f^{\dag}_{n,\alpha,\mu}
f_{n,\alpha,\mu} -{1\over 2}] \leqno(4.1.10)$$
which is constructed to be zero in the ground state.

Alternatively, we may employ the ``axial charge symmetry'' or
``isospin''
symmetry first found by Jones [1988; Jones and Varma, 1988; Jones,
Varma, and
Wilkins, 1988].  Jones observed that the single channel
particle-hole symmetric Kondo model
enjoys an additional global $SU(2)$ symmetry specified by the ``axial
charge''
generators
$$j^{+}_{\mu} = \sum_{n} (-1)^n
f^{\dag}_{n\uparrow\mu}f^{\dag}_{n,\downarrow,\mu} \leqno(4.1.11.a)$$
$$j^{-}_{\mu} = \sum_{n} (-1)^n
f_{n\downarrow\mu}f_{n\uparrow\mu} \leqno(4.1.11.b)$$
and
$$j^{(3)}_{\mu} = {Q_{\mu} \over 2} ~~,\leqno(4.1.11.c)$$
where $Q_{\mu}$ is the charge in each channel.
The total axial charge $j^2_{\mu}$ and $j^{(3)}_{\mu}$ in each channel
$\mu$ may
be used as good
quantum numbers.  This has a clear numerical advantage over the use of
the
$SU(2)\times U(1)$ symmetry of channel spin times charge discussed in
the previous
paragraph because it leads to a greater reduction of basis size.
Moreover, in
the presence of channel symmetry breaking fields, this axial charge
symmetry
may still be used.

The axial charge operators, with the $(-1)^n$ generalized to
$(-1)^{n_x+n_y}$
in two-dimensions are precisely the $SU(2)$ generators found later
by Yang for the two-dimensional Hubbard Model [Yang, 1988] and employed
by Zhang [1989],
Yang and Zhang [1989], and Singh, Scalettar, and Zhang [1991] to
speculate on
the excitation spectra and ``eta'' pairing ($<j^{+}>\ne 0$) in the
Hubbard
model away from half filling.

It is clear that the different symmetry choices of $SU(2)_{spin}\times
SU(2)_{channel}\times U(1)_{charge}$ should we
employ channel spin and charge or $SU(2)_{spin}\times
SU(2)_{axial,1}\times
SU(2)_{axial,2}$ if we use axial charge in
each channel can only be compatible if these are subgroups of some
larger
group.  Affleck {\it et al.} [1992] have noted that even in the
logarithmically
discretized form, the $M$-Kondo spin-1/2 model will have a full
symmetry
group of $SU(2)_{spin}\times Sp(M)$, where $Sp(n)$ is the so called
symplectic group.  This $Sp(M)$ symmetry
is a hidden symmetry of the problem.
(In the paper of Affleck {\it et al.} [1992], this is denoted
$Sp(2n)$.) This is the group which results from the symmetry breaking
of the
$SU(2M)$ group of the free conduction electrons by the spin coupling to
the impurity.   It was noted that in the single channel Kondo model,
$Sp(1)$ is
isomorphic to $SU(2)$, which corresponds to the axial charge of the
lone
conduction channel.  In the two-channel case, the only invariant
subgroups of
$Sp(2)$ (which is isomorphic to $SO(5)$)
are $SU(2)\times U(1)$ which corresponds to the choice of
channel spin
and charge symmetries, and $SU(2)\times SU(2)$ which corresponds to the
separate axial charge symmetries.   In general, for $M$ arbitrary, one
always has $SU(M)\times U(1)$ and $[SU(2)]^{M}$ as invariant
subgroups of $Sp(M)$ so that the channel spin/charge and axial charge
symmetries may
always be used.
It was noted by Affleck {\it et al.} [1992]
that the $Sp(M)$ symmetry holds for the generalization of
the $M$ model to a bipartite lattice form, which may be important for
future work.

\subsection{Overview of results from NRG studies} 

{\it (a) Concepts and Reference Points}\\

\begin{figure}
\parindent=2.in
\indent{
\epsfxsize=6.in
\epsffile{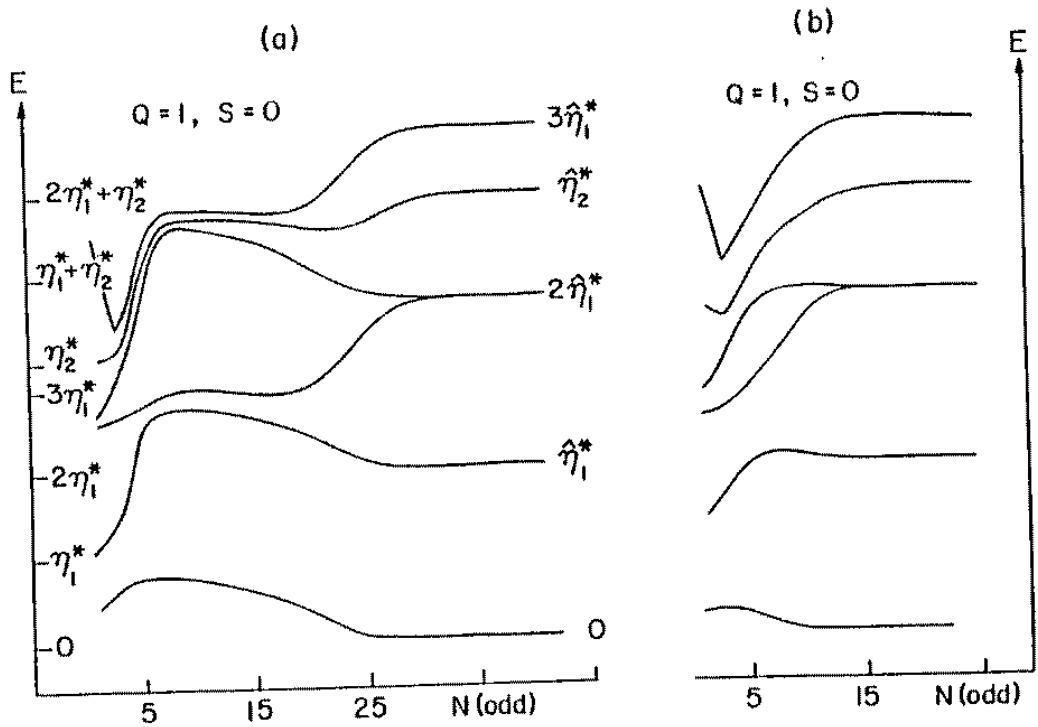}}
\parindent=0.5in
\caption{Odd spectra from the NRG calculations of the single channel
spin 1/2 symmetric Anderson 
model for the charge 1 spin zero sector 
as a function of iteration number $N$. The levels begin near the 
free orbital fixed point, reach a plateau near the local moment 
fixed point, and finally approach the strong coupling Kondo fixed 
point for a ratio $U/\pi\Gamma=5.63$ where $U$ is the local 
Coulomb repulsion in the Anderson model and $\Gamma$ is the
hybridization width. The dimensionless exchange coupling is 
$N(0)J=8\Gamma/\pi U$.  Taken from Krishna-murthy, 
Wilkins, and Wilson [1980a].}
\label{fig4p2}
\end{figure}

To orient the reader, in Fig.~\ref{fig4p2}, we show the odd eigenvalues
calculated by Krishna-murthy, Wilkins, and Wilson [1980a] for the 
single channel symmetric Anderson model in the Kondo limit
beginning with
an initial weak coupling $|J|/D<<1$.  The lines connect the eigenvalues
of the
sequence of dimensionless Hamiltonians ${\cal H}_N$ for 
odd $N$.
The labels reflect the spin and charge values of the many body states.

The reader should note the following features of these curves:\\

\begin{figure}
\parindent=2.in
\indent{
\epsfxsize=6.in
\epsffile{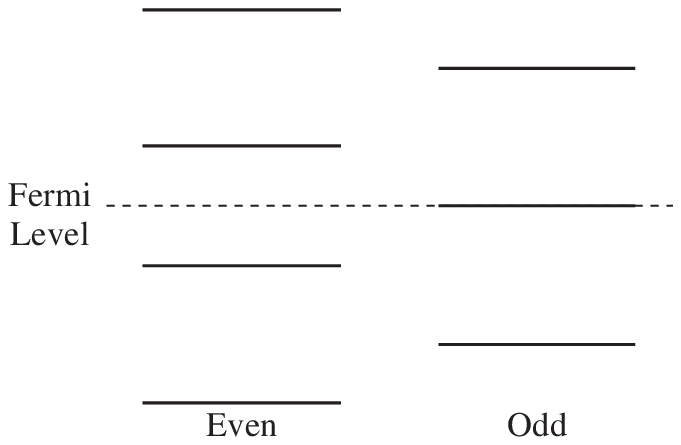}}
\parindent=0.5in
\caption{Even and odd energy levels for the noninteracting ($J=0$) fixed
point of the single channel Kondo model in the NRG logarithmically 
discretized scheme.  }
\label{fig4p3}
\end{figure}

{\it (i) Even-odd alternation}.  The spectra for an even number of
shells
differ from those for an odd number of shells.  The reason is shown in
Fig.~\ref{fig4p3}
in the limit $J=0$.  When we have
an even number of shells, we have an odd number of electrons at
half-filling in
the non-interacting limit
and thus the Fermi level in the non-interacting limit cuts right
through a
level which is two-fold degenerate because of spin.  For an odd number
of
shells, we have an even number of electrons, and the Fermi level passes
through
a gap above a non-degenerate Fermi sea.

{\it(ii) Cross-over }.  There is a crossover evident in the spectra of
Fig.~\ref{fig4p2} 
in that the spectra for even $N$  crossover at large $N$ to resemble
the spectra for small odd $N$, while the spectra for odd $N$ cross over
at
large $N$ to resemble the spectra for small even $N$.  The scale of
this
crossover is a measure of the Kondo temperature $T_K$.  The reason is
that in
view of Eq. (4.1.6), we see that we may convert iteration number $N$ to
temperature through
$$N(T) = 2 {\log(D/T) \over \log(\Lambda)} \leqno(4.2.1)$$
so that the scale of the crossover $N_K$ must correspond to a
temperature
$T_K$.

{\it (iii) Equal level spacing:  local Fermi liquid behavior}.
Provided we
accept the reversal of spectra for even and odd $N$ after the
crossover, we see
that there is a uniformity to the level spacing and a precise $1:1$
connection
of large $N$ even spectra to small $N$ odd spectra and vice versa.
This
implies our low temperature spectrum is that of a Fermi liquid, since
the
electrons are essentially decoupled from the impurity at high $T$, and
thus the
high $T$ spectrum is that of a free Fermi gas.

{\it (iii) Meaning of the Large $N$/Low $T$ spectra}.  The shifting of
the low
$T$ spectra has a simple interpretation: due to the strong coupling to
the
impurity, each spin channel experiences a magnitude $\pi/2$ phase
shift.

To
elaborate, in the presence of a phase shift
$\delta$ measuring
 scattering from a spherically symmetric target, radial quantization of
 the
scattered waves on a sphere of size $L$ implies a shift of wave numbers
$$k \to k - {\delta \over \pi L} ~~.\leqno(4.2.2)$$
For the linearized spectrum of conduction electrons near the Fermi
energy, this
implies a corresponding shift of energy levels by an amount
$$ \delta E = -{v_F \delta \over \pi L}  \leqno(4.2.3)$$
with the convention that an attractive potential gives a positive phase
shift
and thus shifts the levels down in energy.

Turning back to the Kondo problem, the sequence of Hamiltonians ${\cal
H}_N$ is
dimensionless, so the factor of $v_F/L$ in $\delta E$ is removed.  The
shift by
1/2 unit of the fundamental spacing corresponds to exactly a $\pi/2$
phase
shift of electrons with each spin value.  This means the effective
coupling is
infinitely strong.

Note that this agrees with the Friedel sum rule also.  The Friedel sum
rule,
valid for a Fermi liquid, relates the screening moments of the
conduction
electrons about an impurity to the scattering phase shifts.
  Since the impurity has
only spin but no charge difference from the background, the sum rule
insures
that the screening charge around the impurity $\Delta
Q=\delta_{\uparrow}/\pi+\delta_{\downarrow}/\pi=0$.  This implies
$\delta_{\uparrow}=-\delta_{\downarrow}$.   Imagine applying now an
infinitesimal field in the positive $z$ direction.  For the ground
state
singlet case ($N$ even) the total induced
magnetization must be zero, but that results from a screening of the
$S^z_I=1/2$ contribution by the conduction electrons.  Hence
$$S^z_{tot}=0 = {1\over 2} +
{\delta_{\uparrow}-\delta_{\downarrow}\over 2\pi}~~.$$
Solving for $\delta_{\sigma}$, we find
$$\delta_{\sigma} = -\sigma\pi ~~.$$
where $\sigma =\pm 1/2$.  \\

{\it (b) Isotropic Non-trivial Fixed Point}\\

\begin{figure}
\parindent=2.in
\indent{
\epsfxsize=5.in
\epsffile{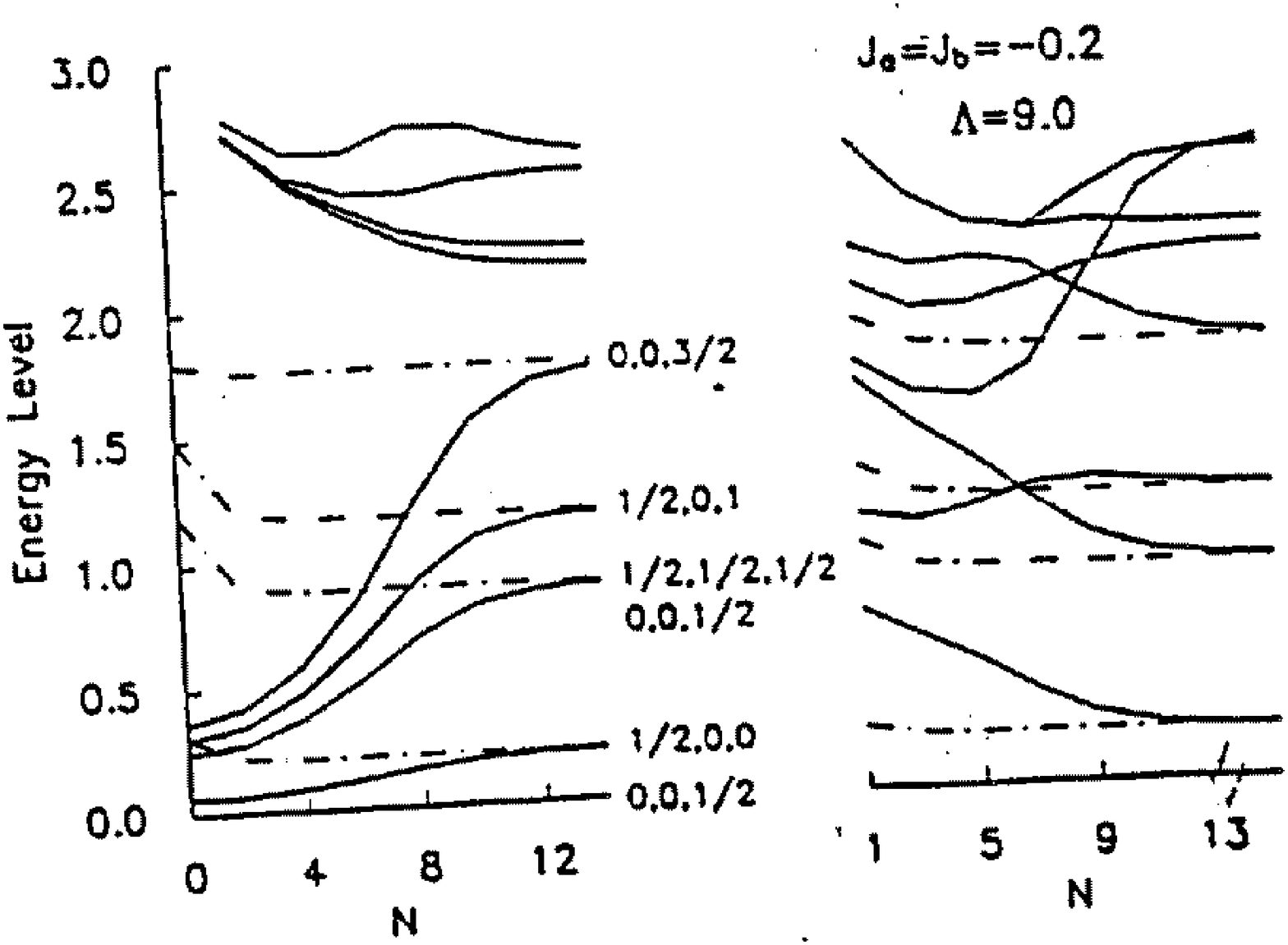}}
\parindent=0.5in
\caption{Lowest NRG energy levels for the weak-coupling isotropic two-channel 
spin 1/2 Kondo model, taken from Pang and Cox [1991]. The coupling strengths
for the two-channels were taken equal to $-0.2$ and the logarithmic discretization 
parameter $\Lambda$ was taken to be 9.0 for rapid convergence.  There are
discernible even-odd alternations for the first few NRG iterations, due to the
proximity to the non-interacting fixed point, but the spectrum eventually goes to 
a single vixed point with increasing iteration number $N$.  For comparison, 
the dashed line  represents the energy levels for the case where the exchange
coupling is set to $-1.0$ in each channel (also for $\Lambda=9.0$).
States are labeles by axial charge or isospin (one for each channel) and
spin. }
\label{fig4p4}
\end{figure}

With the background of Wilson's calculations to guide us, we now turn
to the
more complicated spectra of the two-channel Kondo model.
Fig.~\ref{fig4p4}   shows the results of Pang and Cox [1991] for the approach to
the
non-trivial fixed point.  What is plotted are the eigenvalues of ${\cal
H}_N$
for sequential $N$ connected by a line.  In the case of the two-channel
model,
there is no even-odd alternation of energy levels in the region of the
fixed
point, and the curves in Fig.~\ref{fig4p4}(a) displays the results for even $N$,
the
curves for odd $N$ being in Fig.~\ref{fig4p4}(b).  The eigenvalues,
are labeled by the quantum numbers $(j_1,j_2,S)$ (see Eqs.
4.1.8,4.1.11).   (Note that a very
large
$\Lambda$ value was chosen to allow the weak-coupling crossover to
occur in a
manageable amount of computer time.  The level spacings are $\Lambda$
dependent
for these spectra, though less dependent than in the Fermi liquid
case.)
The dashed curves are for an initial large value of $J/D$, while the
solid
curves are for an initial small value of $J/D$.  In each case, the
eigenvalues
for large $N$ tend to a fixed structure independent of the original
coupling
value.  This confirms in a non-perturbative way that the two-channel
spin 1/2
model has a non-trivial fixed point which is stable in the absence of
fields
which break the full $SU(2)$ invariance of the exchange coupling.
 The rise of the low lying levels for the
weak coupling starting case takes place on a crossover scale which
corresponds
to the Kondo temperature $T_K$.  By studying a variety of couplings, we
estimate, for $\Lambda=3$, that $J_c \simeq -0.7D$.  It is interesting
that the
eigenvalues for
the
initial strong coupling $J/D$ value settle almost immediately to the
fixed
point value.

The other crucial point to notice in comparison to Wilson's spectra for
the
single channel model is that the spectra of the two-channel model
have a non-uniform spacing.
While it
does not obviously follow from this that one has a non-Fermi liquid
excitation
spectrum, this is in fact the case.  One may verify this by noting that
the quantum numbers of the free states cannot be that of a Fermi
liquid.
Further discussion of this point is given in Sec. 6.1.2.
(One could imagine a complicated
combination of spectra from several independent Fermi gases producing
the
spectra for Fig.~\ref{fig4p4}; this does not in fact work.)  In fact, the
concept of
the phase shift used in the previous subsection is completely
irrelevant
here.  The utility of the phase shift rests on the assumption that one
has outgoing single particle states when incoming electrons scatter off
the impurity which certainly holds in the ordinary Kondo model.
However,
as Ludwig and Affleck [1991]  have emphasized, the projection of
the
full $S$-matrix on outgoing single particle states is identically zero
for
the two-channel Kondo model.  Once a symmetry breaking field is applied
to drive the system to a Fermi liquid fixed point, the phase shift
analysis
again becomes relevant.

The results for the non-trivial fixed point spectra are in quantitative
agreement with conformal field theory finite size spectra.  We postpone
a
discussion of that until the conformal field theory section (Sec.
6.1.2).  \\

\begin{figure}
\parindent=2.in
\indent{
\epsfxsize=5.in
\epsffile{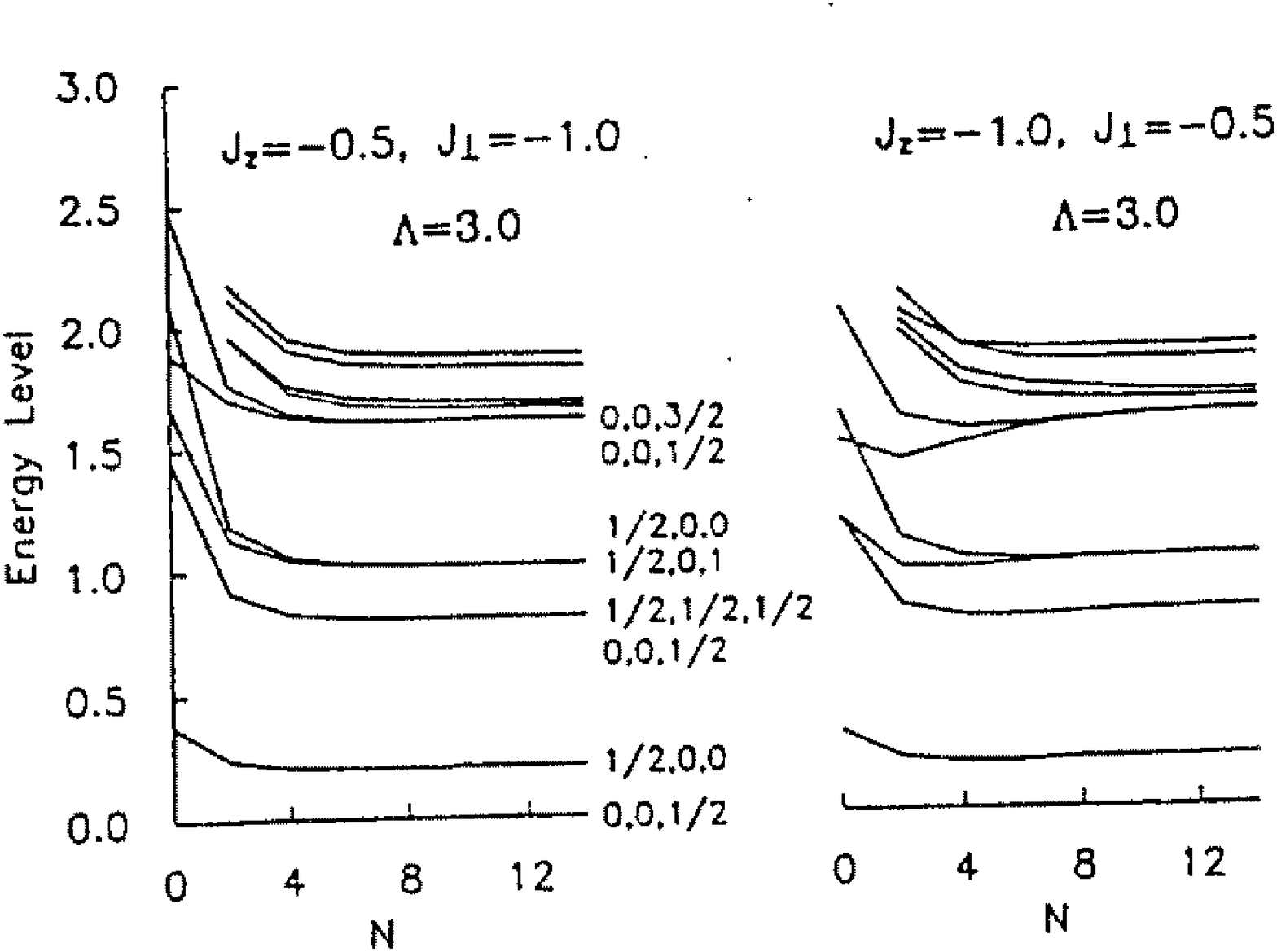}}
\parindent=.5in
\caption{Irrelevance of exchange anisotropy for the two-channel spin
1/2 Kondo model.  For both $J_z>J_{\perp}$ and $J_z<J_{\perp}$, 
the energy levels flow to the isotropic fixed point. The states here 
are labeled by axial charge or isospin for each channel (first two numbers) and 
total $z$-projected spin $S_z$ (last number).  Taken from Pang and 
Cox [1991]}
\label{fig4p5}
\end{figure}

{\it (c) Stability of non-trivial fixed point
against exchange anisotropy}.\\
Fig.~\ref{fig4p5}  shows the NRG
spectrum for the weak coupling side of the fixed point with an initial
easy
axis anisotropy in the exchange, such that $J^3=2J^1=2J^2$.  The last
label is
now simply $S^{(3)}_{tot}$.  The arrows mark the positions of the
non-trivial
fixed point eigenvalues.  We see that the initial anisotropy does
indeed relax
away as anticipated from our scaling arguments in the previous
section.

This irrelevance of exchange anisotropy has also been shown to work in
strong
coupling and for easy plane anisotropy ($|J^3|<|J^2=J^1|$).

The NRG approach provides a way to understand this irrelevance of
exchange
anisotropy for the $M=2$ case which generalizes to $M>2$ as well, and
we shall explore that picture in the last part of this subsection.  \\

{\it (d) Instability of non-trivial fixed point against
application of a Channel Field}.\\
We apply a channel field to the
model through the perturbation
$${\cal H}_{ch} = \Lambda^{N/2} -{\Delta J \over D} \vspi \cdot
\sum_{\alpha,\alpha',\mu} \vec S_{\alpha,\alpha'}\mu
f^{\dag}_{0,\alpha,\mu}f_{0,\alpha',\mu} ~~, \leqno(4.2.4)$$
where the channel index $\mu=\pm 1$.
As discussed in previous subsections, the origin of this symmetry
breaking may
be a magnetic field along a principal axis or a pure rhombohedral
stress for
the quadrupolar Kondo effect, and a uniaxial stress along a principal
axis for
the two-channel magnetic Kondo effect in cubic symmetry.

We have anticipated in our discussion of the next leading order
multiplicative
renormalization group equations that the application of channel
symmetry
breaking will produce a flow away from the non-trivial fixed point.  We
could
not completely trust the scaling analysis in the two-channel case,
however,
since the fixed point coupling is of order unity.  Hence the NRG
provides a
reliable non-perturbative approach for checking the scaling theory.
The
expectation of the scaling analysis was that the more strongly coupled
channel
(e.g., $\mu=+$ if $\delta J,0$) will provide the ordinary Kondo effect,
while
the weakly coupled channel will produce a free Fermi gas (no phase
shift).

\begin{figure}
\parindent=2.in
\indent{
\epsfxsize=5.in
\epsffile{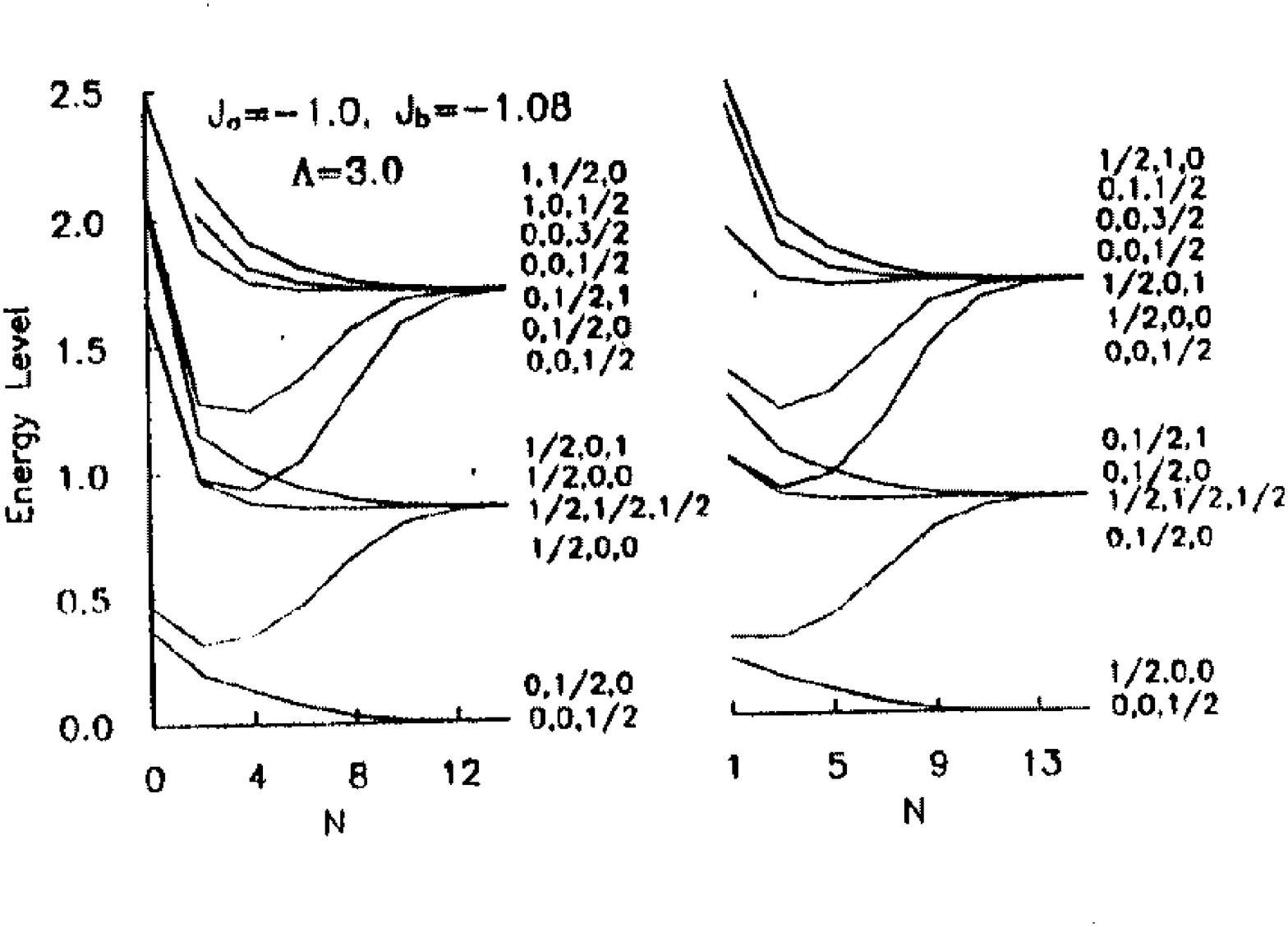}}
\parindent=.5in
\caption{Lowest NRG energy levels for the case of channel anisotropy. Taken
from Pang and Cox [1991].  States are labeled by axial charge for each
channel (also known as isospin) and by total spin. }
\label{fig4p6}
\end{figure}

Examination of the spectra in Fig.~\ref{fig4p6} shows that this expectation is
precisely
met.  We see uniform level spacing characteristic of the Fermi liquid.
There
is no even-odd alternation, but that is because the fixed point spectra
are
formed from the vector space for both channels.  For definiteness,
assume
$\Delta J>0$.  Then we expect a $\pi/2$ phase shift at the fixed point
for
$\mu=+$,
 so that
even $N$ correspond to zero coupling odd $N$, and vice versa.  On the
other
hand, the $\mu$=- spectra are zero coupling spectra.  Hence the
combined
excitations for both even or odd $N$ are the sum of even $N$ and odd
$N$
spectra derived for a single fermi gas.

The NRG allows a determination of the crossover scale.  We can define
$T_{ch}$
as that temperature where the first excited state shrinks to 1/10 of
the
splitting at the non-trivial fixed point value under application of the
channel
field splitting (Eq. (4.2.4)). (Recall that temperature and iteration
number
are related through Eq. (4.2.1).) By plotting
$\Delta J/D$ vs.
$T^x_{ch}$
on a log-log plot and measuring the slope we get the  crossover
exponent.  Pang
and Cox
[1991] found that
$\Delta J \sim (T^x_{ch})^{1/2}$.  The third order scaling analysis of the
previous
section predicted $\delta J \sim (T^x_{ch})^1$, but the exponent is
expected to be
correct only to leading order in $1/M$ which is 1/2 here.  The exponent
of
1/2 agrees exactly with conformal field theory and NCA arguments to be
presented in subsequent sections.

Recently, some controversy has arisen over whether the fixed point
in the presence of channel symmetry breaking is in fact a Fermi liquid
based upon Bethe-Ansatz calculations (Jerez and Andrei, [1995]) and
Majorana fermion techniques (Coleman and Schofield, [1995]); a bosonization 
method yields different results altogether, providing a Fermi liquid fixed point 
(Fabrizio, Gogolin, and \noz [1995a,b]).  We defer
discussion of these issues to Secs. 6 and 7, and 9.2.  \\

{\it (e) Application of a local spin field}.\\
If we apply a local spin
field, we will
also flow away from the non-trivial fixed point.  The perturbation to
be added
to ${\cal H}_N$ is
$${\cal H}_{sp} = -\Lambda^{N/2} {H_{sp}\over D} S^{(3)}_I
~~.\leqno(4.2.5)$$
This perturbation corresponds to the level splitting $\Delta^z$ in the
TLS
model, to an applied {\it local} magnetic field in the two-channel
magnetic
Kondo model, and to an applied {\it local} uniaxial stress in the cubic
quadrupolar Kondo model, or an applied magnetic field along the
$c$-axis for the
tetragonal and hexagonal quadrupolar Kondo models.

\begin{figure}
\parindent=2.in
\indent{
\epsfxsize=5.in
\epsffile{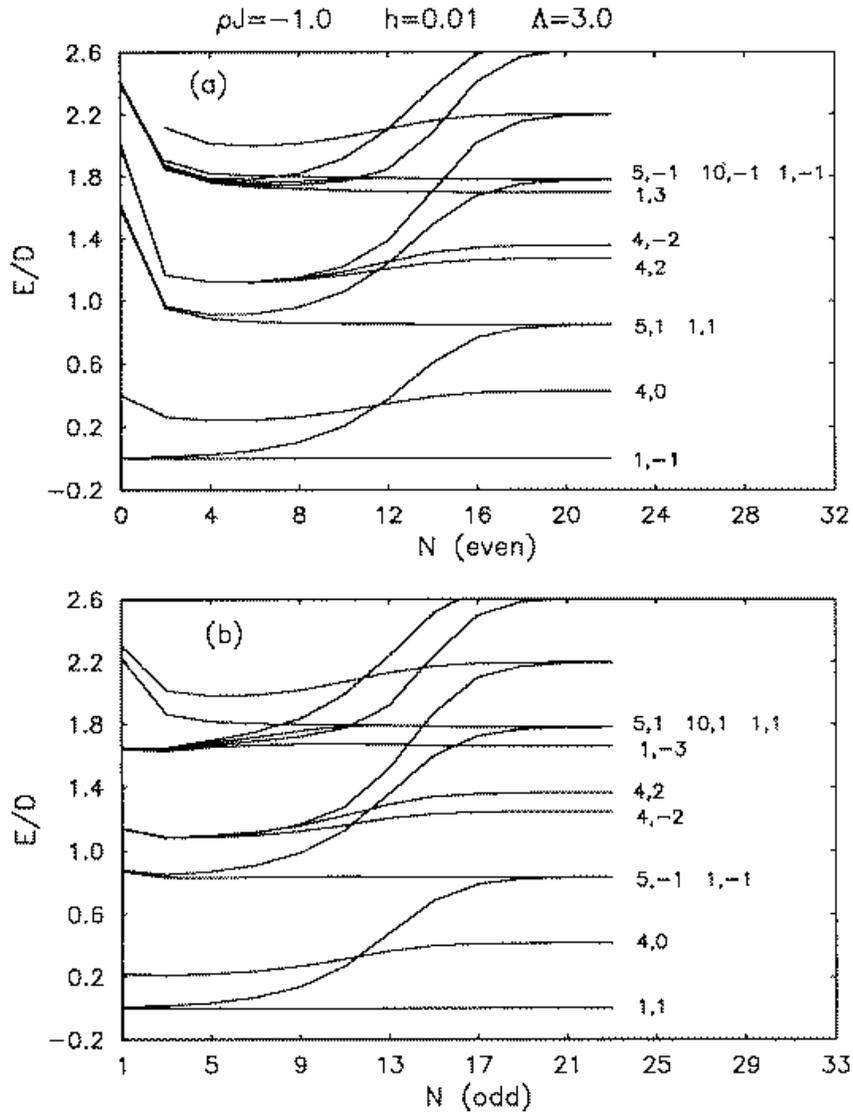}}
\parindent=0.5in
\caption{Lowest NRG energy levels in the presence of an applied spin field. 
From Pang and Cox [1991]. States are labeled by axial charge or isospin
for each channel and by total $z$ projected spin. }
\label{fig4p7}
\end{figure}

For a particular choice of $H_{sp}$ in the strong coupling limit, Fig.~\ref{fig4p7}
shows the NRG flows obtained in the presence of the perturbation in
(4.2.5) as
obtained by Pang and Cox [1991].
The resulting spectra are rather mysterious and clearly spaced
unevenly.
However, a consistent analysis of the spectra may be performed by
assuming that
electrons have only an effective Ising coupling to the
impurity so that the phase shifts for scattering are equal and opposite
for the
different spin values (for a complete test of this
hypothesis, see Table VI of Affleck {\it et al.},  [1992]).  The
spectra are
then the
sum of excitations from up and down spin free Fermi gases scattering
off a
polarized impurity.

The phase shift, which determines the fixed point, may be
computed in this manner as a function of magnetic field.  This produced
the
curves in Fig.~\ref{fig3p3}(a), taken from Affleck {\it et al.} [1992]. 
It should be noted that:\\
(i) For small field, the phase shift behaves as
$$|\delta(H_{sp})| = {\pi \over 4} -sgn(J-J^*)A {H_{sp}\over
T_K}\log({T_K \over
H_{sp}}) \leqno(4.2.6)$$
where $J^*$ is the fixed point coupling strength and A is a pure
number. The
leading order term vanishes for  $J=J^*$.  \\
(ii) The value for $H_{sp}=0^+$ of $|\delta(0^+)|=\pi/4$ may be
understood from
Friedel sum rule arguments [Pang, 1992].   The
generalization from
the single channel Kondo
case is straightforward; taking $\delta Q=0$ again demands
$\delta_{\uparrow,\mu}=-\delta_{\downarrow,\mu}$, while in the absence
of
channel symmetry breaking the phase shifts for fixed spin and opposite
channel
must be equal.  Hence the induced spin polarization in an infinitesimal
positive field is given by
$$S^{(3)}_{tot}=0={1\over 2} + {2\over
2\pi}(\delta_{\uparrow,+}-\delta_{\downarrow,+}) \leqno(4.2.7)$$
where the $1/2$ again reflects the induced $S^{(3)}_I$ value and the
phase
shift dependent term is due to the conduction electron polarization.
This gives
the $\pi/4$ phase shift on solution.  The special value of $\pi/4$ was
also
singled out in the path integral approach to the scaling equations
[\vld,
\zow, and \zim,  1988a,b].  \\
(iii)  The phase shift dependence implies a $H_{sp}lnH_{sp}$ behavior
to the
low field magnetization, in agreement with exact Bethe-Ansatz and
conformal
field theory results, and a leading order $H_{sp}lnH_{sp}$ saturation
to the
magnetoresistance proportional to $sgn(J^*-J)$,
which is quite different from the ordinary Kondo model which always
produces an
$H_{sp}^2$ saturation that is negative, i.e., saturates from below.
Note that the phase shift is meaningful when used in the
one-channel model and in the two-channel model with symmetry
breaking fields, but not at the isotropic two-channel fixed
point.

In analogy to the crossover study of the $\Delta J$ perturbation, we
may study
the crossover exponents for the applied local spin field.  Namely, we
apply the
local field and ask for the temperature $T_{sp}$ (related to iteration
number
through Eq.
(4.2.1)) when a certain set splitting has reached
10\% of
the splitting at the isotropic fixed point.  Affleck {\it et al.}
obtained the
crossover behavior $H_{sp} \sim T_{sp}^{1/2}$, which agrees with
conformal
field theory and NCA results.

 We may also apply a bulk magnetic field to the RG Hamiltonians, and
 provided
the field is small compared with the bandwidth similar results are
obtained
[Pang 1992].  There are some technical issues
involved
for the NRG method in dealing with the bulk field, for which we direct
the
reader to the references.

The conformal field theory results for the analysis of these spectra
are in
excellent agreement with the NRG results, as we shall discuss in Sec.
(6.1.2.c). \\

{\it (f) Shell Model}.

\begin{figure}
\parindent=1.0in
\vspace{.4in}
\indent{
\psfig{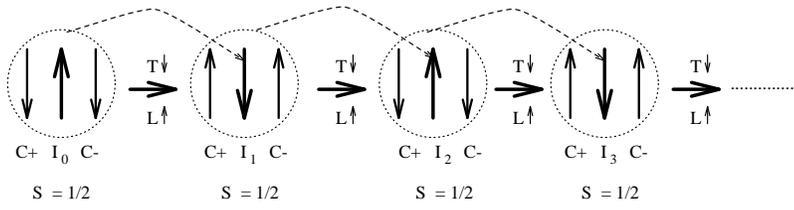}}
\parindent=0.5in
\caption{Cartoon visualization of the NRG process for the two-channel Kondo 
model.  At high temperatures and short length
scales $L$, the local moment ($I$)  is
weakly aligned  antiparallel to the two conduction electron channels
$(C \pm)$ of spin.
However, the binding process of Fig. 3 leads to another doublet
which is antiferromagnetically coupled to the two channels of
conduction spin outside that length scale.  Eventually, this
process continues till a fixed point finite coupling strength is
attained. Taken from Cox and Jarrell [1996]}
\label{fig4p8}
\end{figure}

One can understand some of the physics shown in the NRG level spectra
through a
simple shell model. This picture is motivated by Fig.~\ref{fig4p8}.   Namely,
suppose
that we take seriously the idea
that the physics at length scale is that of an effective spin-1/2 core
object
coupled to conduction electron states in the surrounding adjacent
shell.  We can
readily find the energy by writing
$$S_{tot}^2 = S_I^2 + (\vec S_{c1} + \vec S_{c2})^2 + 2\vec
S_I\cdot(\vec
S_{c1}+\vec S_{c2}) $$
where $S_I=1/2$ is the effective spin of the impurity, and $S_{c1,2}$
is the
spin of each channel.  By solving for the dot product in terms of the
total spin
and the total spin in the conduction sector we can find the energy
through
$$E = -{J_{eff}\over 2D}[S^2_{tot} - S_I^2 - S_{cond}^2]
\leqno(4.2.8)$$
where $S_{cond}$ is the total conduction spin.
 Clearly, we minimize the shell model energy of Eq. (4.2.8) by making
 the largest
 conduction spin $S_{cond}$ for the
smallest total spin $S_{tot}$.

\begin{table}
\begin{center}
\begin{tabular}{|c|c|c|c|c|c|c|}\hline
$Q$  &$S_{tot}$ & $S_{c}$ & $S_{I}$ & $E/|g^*|$ & $\Delta E/|g^*|$ &
CFT \\\hline\hline
0 & 1/2&  1 & 1/2&  -1&  0&  0 \\\hline
$\pm 1$ & 0  &1/2 & 1/2&  -3/4 & 1/4 & 1/8 \\\hline
0 & 1/2&  0&  1/2 & 0 & 1 & 1/2 \\
$\pm$2 & 1/2&  0 & 1/2&  0&  1 & 1/2 \\\hline
$\pm$1 & 1 & 1/2 & 1/2&  5/4 & 5/8 & 5/8\\\hline
\end{tabular}
\end{center}
\caption{ Comparison of Shell Model Energies with Conformal
Theory for
$S_I=1/2,~M=2$.  The shell model exchange formula (Eq. (4.2.8))
reproduces the lowest
three splittings of the conformal field theory finite size spectrum
provided
one takes the fixed point coupling $|g^*|=1/2$, which agrees with the
fixed
point coupling $\lambda_K=2/(M+2)=1/2$ in the notation of Affleck and
Ludwig
[1991b)].  Energies above these in the table are not reproduced by the
shell model.}
\label{tab4p1}
\end{table}

The Shell Model  exchange Hamiltonian gives the
following results for the lowest few states listed in
Table~\ref{tab4p1}.
Beyond this, we find detailed discrepancies with the NRG levels and
with the
conformal field theory finite size spectra, which is not surprising
since
the excited levels correspond to particle hole excitations which must
take
us outside the core shell.  If we take $|J|/D\approx
0.8$, we
get excellent agreement of these results with our first three
splittings in Fig.~\ref{fig4p4} 
As we shall show, if we take $|J|/D$=0.5, we get perfect agreement
with
the lowest four states of the conformal field theory finite size
spectra.  The
discrepancy here in the choice of normalization
factors is likely a $\Lambda$ dependent renormalization.

It also turns out that this gives good agreement with the lowest levels
of the
finite size spectra for conformal theory for three channels and spin
1/2 which is shown in Table~\ref{tab4p2}.

\begin{table}
\begin{center}
\begin{tabular}{|c|c|c|c|c|c|}\hline
$Q$ & $S_{tot}$ & $S_{cond}$ & $E/|g^*|$ & $\Delta E/|g^*|$ &
CFT \\\hline\hline
0 & 1/2 & 3/2  & -5/2 & 0 & 0 \\\hline
$\pm 1$ & 0  & 1 & -1 & 1/2 & 1/5 \\\hline
0 & 1/2 & 1/2 & -1 & 3/2 & 3/5 \\
$\pm$1& 1 & 1 & -1 & 3/2 & 3/5 \\
$\pm$2 & 1/2 & 0 & -1 &3/2 & 3/5 \\\hline
\end{tabular}
\end{center}
\caption{Comparison of Shell Model Energies with Conformal field
theory
finite size spectrum for $S_I=1$, $M=3$.  The shell model exchange
formula
for the energy (Eq. (4.2.8)) agrees with the the lowest few splittings
in the conformal field
theory spectrum provided we take the fixed point coupling $|g^*|=2/5$
which
agrees with the non trivial fixed point coupling of the conformal
theory
$\lambda_K=2/(M+2)=2/5$.  For higher energy states the shell model
disagrees with the exact spectrum.}
\label{tab4p2}
\end{table}

The shell model is useful for understanding the irrelevance of exchange
anisotropy which turns out to generalize to the $SU(2)\times SU(M)$
model with
$S_I=1/2,(M-1)/2$, and the relevance of exchange anisotropy for
$S_I>1/2$.
To see this, start out with $S_I=1/2$, $M$ arbitrary.   Then after one
RG iteration
to
the extent the shell model is correct, we will have a ground state with
$S_{tot}= M/2-1/2$.   At the next iteration, application of the shell
model
to the effective impurity spin equal to $(M-1)/2$ coupled to $M$
electrons gives back an effective spin of 1/2.  Thus we alternate
between these
two spin values.  Now, everytime we are on an iteration where the
effective
impurity spin is
1/2, the lowest few states are
either total spin zero or total spin 1/2.  Exchange anisotropy cannot
lift the
degeneracy of these levels.  Hence, since these states are used to
construct the
spectrum for the next iteration, we expect the anisotropy to decay away
with
increasing iteration number.

On the other hand, consider for example $S_I=1$, and four channels.  In
this
case, the effective impurity spin will always be $S=1=4/2-1$.  Exchange
anisotropy can generate a ``crystal field'' like splitting of the $S=1$
state
that will be propagated through the RG iterations.  This is seen
because the
self-energy diagram corresponding to Figs.~\ref{fig3p10},\ref{fig3p16} will now have an induced
splitting of the $S=1$ levels through the quadrupolar field of the
conduction
electrons.

These simple ideas developed from the shell model turn out to be
completely
supported by conformal field theory, as we shall discuss in Sec.
(6.1.2.c).

The shell model also provides the basis for a strong coupling expansion
in the
inverse exchange coupling,
first discussed by \noz~ [1974] and \noz~ and Blandin [1980].  The idea
is
that the shell model gives the exact eigenvalues in the limit of
infinite exchange
coupling.  One can then derive a perturbative expansion in powers of
$t/J$ where
$t$ is the hopping to the next Wilson onion-skin shell.  In this way,
one can
determine that the $J=\infty$ limit is stable for the ordinary Kondo
model, and
unstable for the multichannel model, which together with the
perturbative analysis
from weak coupling assures the existence of a non-trivial, intermediate
coupling
fixed point.

\section{Non-Crossing Approximation (NCA) } 

In this section, we survey some of the most important applications of
the
Non-Crossing
Approximation, or NCA, to the multi-channel Kondo model.  We shall
first
discuss application
of the NCA to the $SU(M)\times SU(N_I=N_c=N)$ Anderson model for which
the results
are rigorous as $N\to \infty$ with $\gamma=M/N$ fixed.  Despite the
large
$N$ character to the theory, the physical properties
computed for finite $N,M$ are found to be in good agreement with exact
results
for most properties.  Next, we shall discuss an application to a simple
three
configuration model for a Ce$^{3+}$ ion which illustrates 1,2, and 3
channel
Kondo physics in appropriate regimes.
An important physical result which emerges from this analysis is that
the
sign of the low temperature thermopower is a diagnostic for the
emergence of
the two-channel ground state
as required by the dynamic selection rule 5 of Sec. 2.2.4.  Finally, we
shall
apply
the NCA to a model for the U$^{4+}$ ion which includes both ground state
and
excited state
crystal field levels.   These results show how the two-channel physics
can
emerge at low
temperatures even with significant overlap between crystal field states
due
to conduction
electron damping.

Since exact results, such as exist for the $SU(N)\otimes SU(M)$ model,
are
preferrable to
approximate ones, the need for the NCA must be clarified.  It has two
principle
virtues: First, it is quite simple to develop and use, and as a result
the
physical motivation is quite clear.  Second, while the Bethe-Ansatz,
conformal
field theory, NRG, bosonization, and for the most part,
Quantum Monte Carlo methods are currently limited to
pristine models that lack much of the realistic physics such as crystal
field
excitations and interconfiguration fluctuations. The NCA is not limited
in this
way and still appears to give quite reasonable results and in
some cases some useful new physics.  Moreover, the NCA is able to 
calculate dynamics (such as the inelastic neutron scattering cross section) 
and transport properties over a much wider range 
of paramters and temperature regimes than any of the exact methds, 
and offers a ready extension to {\it non-equilibrium} properties which 
is still rarely possible for the exact methods.  
Hence, the first sub-section largely serves to calibrate the value of
the
method by comparing repeatedly to conformal field theory and
Bethe-Ansatz
results.  This section also lays out the formalism and essential
concepts of the NCA.  The next two
subsections are devoted to realistic applications to
model \ctp~ and \ufp~ impurities.  As we shall see, considerable new
physics
emerges from these applications.

\subsection{$SU(M)\otimes SU(N)$ Model:  The NCA as a Large $N$ Limit} 

In this subsection, we shall follow the work of Cox and Ruckenstein
[1993].
Following the discussion of Secs. 2.3 and 3.3, we write down an
Anderson
Model in the
pseudo-particle representation as
$$H = \sum_{k\mu\alpha} c^{\dagger}_{k\mu\alpha}
c_{k\mu\alpha}
+ \epsilon_f \sum_{\mu} f^{\dagger}_{\mu}f_{\mu}
\leqno(5.1.1)$$
$$ ~~~~~-{\tilde V\over \sqrt{NN_s}} \sum_{k\alpha\mu} sgn(\alpha)
[f^{\dagger}_{\mu}b_{-\alpha}c_{k\mu\alpha} + h.c.]  -
\lambda_{ps}
[\sum_{\mu} f^{\dagger}_{\mu}f_{\mu}+
\mu_{\alpha} b^{\dagger}_{\alpha}b_{\alpha} - 1]$$
where we have anticipated the large $N$ limit and normalized the
hybridization
such that
$\tilde V = \sqrt{N} V $ is well defined in the $N\to \infty$ limit,
and
$\lambda_{ps}$ is to be taken
to $-\infty$ to project to the physical Hilbert space (c.f. Sec.
3.3.1, Eq. 3.3.3). The indices
$\sigma$
run from $-(N-1)/2,-(N-3)/2,...,(N-1)/2$, and the indices $\alpha$ from
$-(M-1)/2,-(M-3)/2,...,(M-1)/2$.  We have inserted  the phase factor
$-sgn(\alpha)$ relative to the hybridization term defined in Cox and
Ruckenstein [1993], which will not alter the
physics of the model.   A Schrieffer-Wolff transformation on this model
produces the $SU(M)\otimes SU(N)$ Coqblin-Schrieffer model discussed in
Sec.
2.3, with exchange
coupling $J = \tilde V^2/N\epsilon_f$ assuming $\epsilon_f<0$ and
large.
We shall pass to $N\to\infty$ by holding the ratio $\gamma
=M/N$ fixed.

We remark that a curious feature of the NCA is that it only appears to
be useful for
models which have the full $SU(N)$ symmetry so that a single boson is
required
to decouple each channel.

\begin{figure}
\parindent=2.in
\indent{
\epsfxsize=3.in
\epsffile{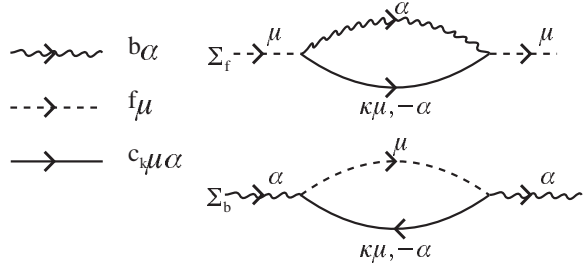}}
\parindent=.5in
\caption{Order $(1/N)^0$ self-energy diagrams of the $SU(N)\times SU(M)$
multichannel Anderson model.  The lowest ionic configuration is represented by 
a pseudo-fermion (dashed) line which carries spin index $\mu$ which runs 
over $N$ values, and the excited ionic 
configuration by a pseudo-boson (wavy) line carrying channel index $\alpha$ which runs over 
$M$ values.  Conduction electrons carry both spin and channel (solid lines). 
Vertices represent the hybridization, which is scaled by a factor of 
$1/N^{.5}$ in 
this approach.  Note that the $N\to \infty$ limit is taken 
by holding $N/M$ fixed. As a result, the closed spin and channel loops in these diagrams
give compensating factors of $N,M$ to the $1/N$ from the vertices, rendering them 
order $1/N^0$. These self-energies (and the corresponding propagators) are not in themselves 
physically observable quantities. After Cox and Ruckenstein [1993]}
\label{fig5p1}
\end{figure}

The leading order (order 1)
diagrams for the self-energy of the pseudo-fermion
and
pseudo-boson propagators
may be obtained now from the diagrams of Fig.~\ref{fig5p1}.
Each diagram is
quadratic
in the hybridization.
The pseudo-boson self-energy $\Sigma_b$
contains a sum over internal spin labels, and hence acquires a factor
of $N$ which cancels the $1/N$ in the denominator.  The pseudo-fermion
self
energy $\Sigma_f$ contains a sum over internal channel labels and hence
acquires a factor of $M$ which yields a net factor of $\gamma$
out front when divided by $N$.  Because each self-energy is $O(1)$, a
self-consistent solution of
the resultant coupled equations is required, since we can insert an
infinite
series of self-energy corrected propators into each of the diagrams.
The leading order vertex corrections are shown in Fig. 5.1.b.
These are down by order $1/N^2$ relative to the $O(1)$ diagrams,
which is seen by counting vertices (6) giving a net count of
$N\times (1/N)^3$ where the $N$ comes from the sum over internal
degrees of freedom.

Assuming
we have particle-hole symmetry of the conduction band, the resulting
$O(1)$ integral
equations corresponding to the self-energy diagrams of Fig.~\ref{fig5p1}
are given by
$$ \Sigma_f(\omega) = {\gamma \tilde \Gamma\over \pi} \int d\epsilon
f(\epsilon)
 {1\over \omega + \epsilon - \Pi_b(\omega+\epsilon)} = {\gamma \tilde
 \Gamma\over \pi} \int d\epsilon f(\epsilon) {\cal
 G}_b(\omega+\epsilon)
 \leqno(5.1.2.a)$$
$$\Sigma_b(\omega) = {\tilde\Gamma \over \pi} \int d\epsilon
f(\epsilon)
{1\over \omega + \epsilon - \epsilon_f - \Sigma_f(\omega+\epsilon)}=
{\tilde\Gamma \over \pi} \int d\epsilon f(\epsilon)
{\cal G}_f(\omega+\epsilon)\leqno(5.1.2.b)$$
where $\tilde\Gamma = \pi N(0) \tilde V^2$ and $f(\epsilon)$ is the
Fermi-Dirac function.  These equations also define the
pseudo-boson and pseudo-fermion propagators ${\cal G}_b,{\cal
G}_f$.   We note that the conduction electron self-energy is of
order $1/N_s$ for this single impurity problem.

These equations are called the ``non-crossing approximation'' (NCA)
because
the corresponding diagrams
in a particular representation (diagrams on a cylinder) have no crossed
conduction lines.  See
Bickers [1987] for an extended discussion of this diagrammatic
approach.
We observe that if the
pseudo-boson propagator is viewed as a dynamically dressed exchange
coupling,
as is appropriate for the Coqblin-Schrieffer model, then these
equations are
none other than a fully self-consistent
form of the third order scaling theory as discussed in Sec. (3.4.4).
The scaling theory retains only leading logarithms (and thus
neglects the imaginary parts of propagators),
but the NCA retains the full analyticity of the
various propagators.  Note that: (i) the projection to
$\lambda_{ps}=-\infty$
has taken place and so the superscripts of Eqs. (3.3.4) and (3.3.10)
have been
dropped; (ii) we are interested for the moment in the physical
situation of
zero spin and channel
fields and so have dropped the spin and channel subscripts
on ${\cal G}_b,{\cal G}_f$.

Alternatively, instead of a diagrammatic approach, one may derive
Eqns.
(5.1.2.a,b) from a path
integral formulation as a saddle point condition.  We sketch this
derivation
here.   First, from the
action corresponding to the Hamiltonian of Eq. (5.1.1), the conduction
electron
fields are integrated out.  This produces an effective interaction
between the
pseudo-boson and pseudo-fermion given
by the term
$$S_{int} = -{ \tilde V^2 \over N} \sum_{\sigma\alpha} \int d\tau \int
d\tau'
f^{\dagger}_{\mu}(\tau)
f_{\mu}(\tau') G^0(\tau -\tau')
b^{\dagger}_{-\alpha}(\tau')b_{-\alpha}(\tau)
\leqno(5.1.3)$$
where
$$G^0(\tau)  = -{1\over N_s} \sum_k {1\over \partial/\partial\tau +
\epsilon_k}
\leqno(5.1.4)$$
is the conduction green's function at the impurity site.  This
interaction term
may be decoupled with
collective Hubbard-Stratonovich fields$\Phi_b(\tau,\tau'),
\Phi_f(\tau,\tau')$
which
obey $\Phi_{f,b}(\tau',\tau) = \Phi_{f,b}^*(\tau,\tau')$.   The
resulting
effective action term which
is quadratic in the $\Phi$ fields is
$$\tilde S_{\phi} = - \tilde V^2 \int d\tau \int d\tau'
G^0(\tau-\tau')[
N\Phi_f(\tau',\tau)\Phi_b(\tau,\tau')
- \sum_{\mu}
f^{\dagger})_{\mu}(\tau)\Phi_b(\tau,\tau')f_{\mu}(\tau')
-\sum_{\alpha}
b^{\dagger}_{\alpha}(\tau')\Phi_f(\tau',\tau)b_{\alpha}(\tau)
] ~~.\leqno(5.1.5)$$
This may now be followed by an integration over the $f,b$ fields to
produce an
effective action solely
in terms of the $\Phi$ fields.  Variation of $\Phi_{f,b}$ to determine
the
saddle point yields Eqs. (5.1.2.a,b) as the
extremum conditions, provided we note that: (i) the $\Phi$ fields are
time
translation invariant at the
saddle point, and (ii) $\Phi_{f,b}(\omega)$ are the projected
$\lambda_{ps}$
pseudo-fermion and pseudo-boson
propagators in this limit.

The physical properties of the system can be expressed in terms of the
spectral functions of the pseudo-particles.   Keeping our notation here
consistent with that of Cox and
Ruckenstein [1993], we define
$${\cal A}_{f,b}(\omega) = {Im\Phi_{f,b}(\omega - i0^+)\over \pi}
={Im{\cal G}_{f,b}(\omega-i0^+)\over \pi} .
\leqno(5.1.6)$$
At $T=0$, these spectral functions vanish below the ground state energy
$E_0$.  We thus define
occupied state spectral functions
$${\cal A}_{f,b}^{(-)}(\omega) = e^{\beta(E_0-\omega)}{\cal
A}_{f,b}(\omega)
~~.\leqno(5.1.7)$$
Upon multiplication of the exponential factor through Eqns.
(5.1.2.a,b), it
is seen that these
occupied state spectral functions satisfy the self-consistency
equations
$${{\cal A}_f^{(-)}(\omega)\over |{\cal G}(\omega)|^2} = {\gamma\tilde
\Gamma
\over \pi} \int d\epsilon f(-\epsilon) {\cal
A}_b^{(-)}(\omega+\epsilon)
\leqno(5.1.8.a)$$
$${{\cal A}_b^{(-)}(\omega)\over |{\cal D}(\omega)|^2} = {\tilde
\Gamma\over
\pi} \int d\epsilon f(-\epsilon) {\cal A}_f^{(-)}(\omega+\epsilon)~~.
\leqno(5.1.8.b)$$

It is important also to keep track of the partition function which
enters
a calculation of all physical
quantities.  The total partition function factorizes into a product of
the
bare conduction band partition
function times the pseudo-particle partition function ${\cal Z}_f$
given by
$${\cal Z}_f = \int d\omega [N{\cal A}^{(-)}_f(\omega) + M{\cal
A}^{(-)}_b
(\omega)] \leqno(5.1.9)$$
where we have assumed spin and channel isotropy for the moment.

\begin{figure}
\parindent=2.in
\indent{
\epsfxsize=3.in
\epsffile{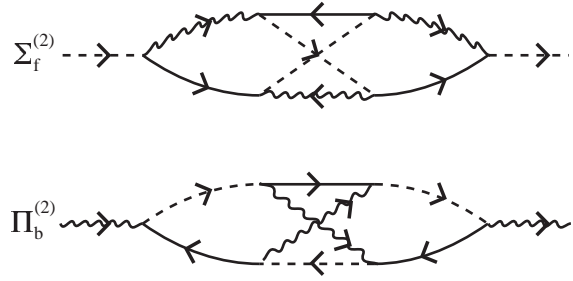}}
\parindent=.5in
\caption{Order $1/N^2$ vertex corrections to self energy diagrams of the
$SU(N)\times SU(M)$ Anderson model. Because there is only one closed loop
for either spin or channel in each diagram, the presence of six vertices each
scaled by $1/N^{1/2}$ leads to an overall suppression of these diagrams by 
order $1/N^2$ relative to the diagrams of Fig.~\ref{fig5p1}. }
\label{fig5p2}
\end{figure}

\begin{figure}
\parindent=2.in
\indent{
\epsfxsize=3.in
\epsffile{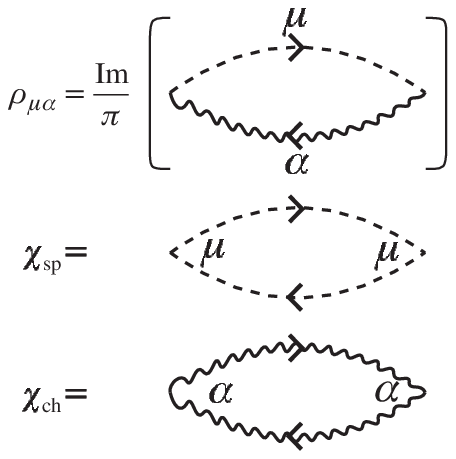}}
\parindent=.5in
\caption{Leading order diagrams for physically measurable quantities
in the $N\to \infty$ limit.  Physically observable quantities are represented as 
convolutions of the pseudo-particle propagators.  The local spin susceptibility to 
leading order is a self-convolution of the pseudo-fermion propagator.  Likewise, the 
channel susceptibility is a self-convolution of the pseudo-boson propagator to leading
order.  The physical $f$-electron addition/removal propagator is a convolution of the
pseudo-fermion and pseudo-boson propagators.  Each physical quantity must be properly
projected onto the constrained subspace as detailed in the text.  }
\label{fig5p3}
\end{figure}

The reason the partition function enters is due to the projection
procedure:
one assumes a Grand
ensemble in the charge $Q_f$ of Eq. (3.3.6), and then projects to the
physical
$Q_f = 1$ subspace.
Practically, this means we must divide any observable quantity by the
$Q_f=1$
canonical partition function.  We shall be interested in three
quantities in
this subsection, the physical one-particle
spectral function $\rho_{\sigma\alpha}(\omega,T)$ which determines the
conduction
electron $t$-matrix, the
spin susceptibility $\chi_{sp}(\omega,T)$, and the channel
susceptibility,
$\chi_{ch}(\omega,T)$.
The leading order diagrams for these quantities within the NCA are
shown in
Fig.~\ref{fig5p2} (vertex corrections are down by $O(1/N^2)$.
The diagram for $\rho_{\sigma\alpha}$ is the physical
propagator
${\cal G}_{\sigma\alpha} $ with  $\rho_{\sigma\alpha}(\omega,T)=
Im{\cal G}_{\sigma\alpha}(\omega-i0^+,T)/\pi$
which measures the density of states for adding and removing electrons
of
spin $\sigma$ and channel index $\alpha$ at the impurity site.  When
these diagrams are evaluated ($\rho(\omega,T)$ is the imaginary part of
Fig.~\ref{fig5p3}. divided by $\pi$) we obtain
the expressions
$$\rho_{\sigma\alpha}(\omega,T) = {1\over {\cal Z}_f} \int d\omega'
[{\cal A}^{(-)}_{f}(\omega')
{\cal A}_b(\omega+\omega') + {\cal A}^{(-)}_b(\omega')
{\cal A}_f(\omega'-\omega)] \leqno(5.1.10)$$
$$\tilde \chi''_{sp}(\omega) =Im\tilde\chi_{sp}(\omega-i0^+) =  {\pi
N\over
{\cal Z}_f} \int d\omega' [{\cal A}^{(-)}_{f}(\omega')
{\cal A}_f(\omega+\omega') -{\cal A}^{(-)}_f(\omega')
{\cal A}_f(\omega'-\omega)] \leqno(5.1.11)$$
$$\tilde \chi''_{ch}(\omega) = Im\chi_{ch}(\omega-i0^+) =
{\pi M \over {\cal Z}_f} \int d\omega' [{\cal A}^{(-)}_{b}(\omega')
{\cal A}_b(\omega+\omega') -{\cal A}^{(-)}_b(\omega')
{\cal A}_b(\omega'-\omega)]~~. \leqno(5.1.13)$$
The tilde over the spin and channel susceptibilities signifies a
definition
wherein we assume a linear coupling
of the form $-\mu_{sp}\sigma H_{sp}$ to a spin field and
$-\mu_{ch}\alpha H_{ch}$ for the channel field with
$\mu_{sp}^2 (N^2-1)/12=\mu_{ch}^2(M^2-1)/12$
set to unity.  We shall employ these expressions in the
following analysis.  Note that the  denominators of
$N$ and $M$ in the corresponding
definitions of Cox and Ruckenstein [1993] should be
removed to obtain approximate
equality in the RHS of Eq. (6) therein.

\subsubsection{Differential Form of the NCA Equations at $T=0$} 

As discussed by Kuramoto and Kojima [1984] and
M\"{u}ller-Hartmann [1984], the NCA equations may be
converted to differential
equations at zero temperature.  The procedure is thoroughly discussed
there and in the review article of Bickers
[1987], and we shall simply outline it here.  We note that these
analyses were applied to the single channel Kondo model for which the
NCA
provides a pathological low temperature behavior that is non-Fermi
liquid
like in contrast to the known Fermi liquid
excitation spectrum of the single channel Kondo model.  It is amusing
that
the ``pathology'' obtained
in applying the NCA to the single channel case is exactly what provides
excellent results for the multi-channel model!

We assume the conduction electrons occupy a broad flat density of
states from
$-D$ to $D$ and
that this is half-filled.
At zero temperature, Eqs. (5.1.2.a,b) may thus be rewritten as
$$\Sigma_f(\omega) = {\gamma\tilde \Gamma\over \pi}
\int_{-D+\omega}^{\omega}
d\omega'{\cal G}_b(\omega') \leqno(5.1.14.a)$$
$$\Sigma_b(\omega) = {\tilde\Gamma\over \pi} \int_{-D+\omega}^{\omega}
d\omega'
{\cal G}_f(\omega')~~. \leqno(5.1.14.b)$$
Provided the bandwidth $D>>\tilde\Gamma$ which sets the scale of
$\Sigma_f,
\Sigma_b$ over most
of the frequency range, we may neglect the frequency dependence
of the lower integration limit.   The integral equations may then be
differentiated with respect to
$\omega$ to produce differential equations.  It is most convenient to
express
these in terms of the
inverse green's function variables
$$g_f(\omega)  = -{\cal G}_f(\omega)^{-1}~~, g_b(\omega) = -
{\cal G}_b(\omega)^{-1} \leqno(5.1.15)$$
in terms of which the differential equations are
$${dg_f \over d\omega} = -1 - {\gamma\tilde \Gamma\over\pi  g_b}
\leqno(5.1.16.a)$$
$${dg_b\over d\omega} = -1 - {\tilde\Gamma\over \pi g_f}
\leqno(5.1.16.b)$$
subject to the boundary conditions
$$g_f(-D) = \epsilon_f + D,~~g_b(-D)=D \leqno(5.1.17)$$
which are compatible with the neglect of $\omega$ in the lower limit of
Eqs.
(5.1.14.a,b).
We must also employ differential equations for the negative frequency
spectral
functions;
by a similar analysis we obtain
$${d\over d\omega}[{\cal A}^{(-)}_f(\omega) g_f^2(\omega)] =
-{\gamma\tilde\Gamma\over \pi }{\cal A}^{(-)}_b(\omega)
\leqno(5.1.18.a)$$
$${d\over d\omega}[{\cal A}^{(-)}_b(\omega) g_b^2(\omega)] =
-{\tilde\Gamma\over \pi }{\cal A}^{(-)}_f(\omega)
\leqno(5.1.18.b)$$
subject to the boundary conditions
$$[{\cal A}^{(-)}_f(E_0) g_f^2(E_0)] =[{\cal A}^{(-)}_f(E_0)
g_b^2(E_0)] =0~~.
\leqno(5.1.19)$$

To solve Eqs. (5.1.16.a,b), it is first important to note that they
possess a
constant of integration ${\cal C}$ that allows a connection to the
Kondo scale.
Namely, it is easy to verify by dividing the two equations and
integrating that
$$g_f + {\tilde\Gamma\over \pi}\ln(({g_f\over D}) - g_b -
{\gamma\tilde\Gamma\over\pi} \ln({g_b\over D}) = {\cal C}  = \epsilon_f
\leqno(5.1.20)$$
provided we assume $\ef<<D$.  The equality on the far RHS of the
above equation follows from the boundary conditions of Eq. (5.1.15).
This equation may be rewritten in a form more convenient near $E_0$ as
$${ g_f /T_0 \over (\pi g_b/\tilde\Gamma)^{\gamma}} =
\exp[\pi(g_b-g_f)/\tilde\Gamma] \leqno(5.1.21)$$
with the Kondo scale $T_0$ defined by
$$T_0 = D({\gamma \tilde\Gamma \over \pi D})^{\gamma}
\exp(\pi\epsilon_f/\tilde\Gamma) ~~.\leqno(5.1.22)$$
This agrees with the corresponding scale identified from third order
scaling theory to within a factor of order unity (c.f., Eqs.
3.4.27, 3.4.48).

The NCA differential equations may now be solved in a power series
expansion in the variable
$$\Theta = \{[{1+\gamma\over\gamma}]{(E_0-\omega)\over T_0}\}^{1\over
1+\gamma} ~~.\leqno(5.1.23)$$
The results are, for $\omega<E_0$,
$$g_f (\omega) \approx T_0 \Theta ^{\gamma} + ... \leqno(5.1.24.a)$$
$$d_b(\omega)\approx {\tilde \Gamma\over \pi} \Theta + ...~~.
\leqno(5.1.24.b)$$
These expressions
may be analytically continued above $E_0$ to give the positive
frequency spectral functions,
and may also be used to obtain the negative frequency spectral
functions below $E_0$.  The
results for the positive frequency spectra are
$${\cal A}_f(\omega) = \theta(\omega-E_0) {1\over \pi T_0}
\sin({\pi\gamma\over 1+\gamma}) |\Theta|^{-\gamma}
+ ...\leqno(5.1.25.a)$$
$${\cal A}_b(\omega) = \theta(\omega-E_0) {1\over \tilde\Gamma }
\sin({\pi\over
1+\gamma})| \Theta |+ ...\leqno(5.1.25.b)$$

We need a couple of tricks to obtain the negative frequency spectra.
First, an
{\it Ansatz} is made that near threshold ${\cal A}_f^{(-)}\sim \alpha
/g_f$,
${\cal A}^{(-)}_b \sim \alpha/g_b$.  Because
of the identity
$${d\over d\omega}[M{\cal A}_b^{(-)}(\omega)g_b(\omega)+
N{\cal A}_f^{(-)}(\omega)g_f(\omega)]
=M{\cal A}_b^{(-)}(\omega)+N{\cal A}_f^{(-)}(\omega) \leqno(5.1.26)$$
which follows from Eqns. (5.1.16.a,b), (5.1.18.a,b) with a
little algebra,
we have
$${\cal Z}_f = [M{\cal A}_b^{(-)}(E_0)d_b(E_0)+N{\cal
A}_f^{(-)}(E_0)g_f(E_0)]
~~.\leqno(5.1.27)$$
From this it follows that $\alpha = {\cal Z}_f/(N+M)$; noting that
${\cal Z}_f$
at $T=0$ is none other
than the expectation value of $Q_f=1$, we see that $\alpha = 1/(N+M)$.
In consequence,
$${\cal A}^{(-)}_f(\omega) = \theta(E_0-\omega) {1\over (N+M)T_0}
\Theta^{-\gamma} + ...\leqno(5.1.28.a)$$
$${\cal A}_b^{(-)}(\omega) = \theta(E_0-\omega) {\pi\over(N+M)
\tilde\Gamma }
\Theta+ ...~~.\leqno(5.1.28.b)$$
(Note that ${\cal Z}_f$ may be normalized to unity by introducing a 
simultaneous chemical potential for the pseudo-particles prior to projection
onto the physical subspace.  This fixes the threshold energy of the 
auxiliary particle spectral functions to $E_0$=0 and greatly facilitates
numerical evaluations [for a detailed discussion, see Appendix D of 
T.A. Costi {\it et
al.}, 1996].  Note also that the exponents for the spectral functions
may be evaluated numerically by application of the NRG, as shown 
T.A. Costi {\it et al.} [1994,1996].)

\subsubsection{Scaling Dimensions} 

The scaling dimension of an operator $O$, denoted $\Delta_0$, for a
zero
temperature impurity critical point as we have in the multi-channel
Kondo
model, indicates how the correlation function of the operator
decays in the long time limit.  Namely, ${\cal G}_O(\tau) = -<T_\tau
O(\tau)
O^{\dagger}(0)> \sim \tau^{-2\Delta_O}, ~~\tau\to \infty$. This implies
that
the corresponding Green's function in the frequency domain will behave
as
${\cal G}_O(\omega) \sim |\omega-E_0|^{2\Delta_O-1}$ as may be
readily verified through Fourier transformation.  Alternatively, the
scaling
dimension tells us what scale factor to multiply the operator by under
an
arbitrary rescaling of time.   The scaling dimension concept is
particularly
useful for connection to the Conformal Field theory work (Sec.
6.1), where the singular properties of various
quantities are expressed in terms of the scaling dimensions of
operators furnished by the theory.

We may read off the NCA expressions for the scaling dimensions of
various
operators straightforwardly from  a knowledge of the above solutions to
the
zero temperature differential equations.
If we express the frequency dependence of the pseudo-particle spectral
functions as
${\cal A}_{f,b} (\omega) \sim |E_0-\omega|^{2\Delta_{f,b} -1} $ we read
off
the scaling
dimensions of the operators $f_{\sigma},b_{\alpha}$ as
$$\Delta_b = \gamma \Delta_f = {\gamma \over
2(1+\gamma)}~~.\leqno(5.1.29)$$
The spin field, channel spin field, and physical fermion field are all
quadratic in the $f,b$ operators
and so the corresponding scaling dimensions may be readily obtained as
$$(Spin): ~~\Delta_{sp} = 2\Delta_f = {1\over 1+\gamma}
\leqno(5.1.30.a)$$
$$(Channel): ~~\Delta_{ch} = 2\Delta_b = {\gamma\over 1+\gamma}
\leqno(5.1.30.b)$$
$$(Fermion): ~~\Delta_F = \Delta_f+\Delta_b = {1\over 2} ~~.
\leqno(5.1.30.c)$$
The latter result is consistent with unitarity, i.e., the conduction
electron scattering rate which
is proportional to the spectral function of the physical fermion
operator ($\rho_{\sigma\alpha}$)
must be bounded by unitarity limit scattering at the Fermi energy.
This implies that the low
frequency leading behavior must be a constant, independent of $N,M$.
Clearly, $\Delta_F=1/2$
yields this result.

The crucial point we may take from the above paragraph is that the
scaling dimensions $\Delta_{sp,ch,F}$ are correctly determined by the
NCA for all $N\ge 1$ and $M\ge 2$  as may be determined from
comparison with the conformal field theory, an exact approach that we
discuss in detail in the next section.  This is a remarkable result.
It is diminished somewhat by the observation that
if we take unitarity of the scattering matrix
as a given constraint on the scaling dimensions, then
the NCA determines really only one independent scaling dimension.  As
a result, certain operator
scaling dimensions will not be correctly obtained by the NCA, a
particular case of interest being the
local pair field operators. These have an explicit $N$ dependence
unlike $\Delta_{sp,ch,F}$.  Nevertheless, we shall see that the
agreement of the NCA with exact results is remarkably good
even in the unlikely region where $N=2$!

\subsubsection{Physical Properties at $T=0$} 

We now summarize the results for calculation of physical properties.

{\it One Electron Spectral Function.}  Substitution into the
convolution formula of Eq. (5.1.9) gives
$$\rho_{\sigma\alpha}(\omega) \approx {\pi\over (1+\gamma)^2
N\tilde\Gamma}
[1+\theta(\omega)f_+(\tilde\omega) + \theta(-\omega)f_-(\tilde\omega) +
....]
\leqno(5.1.31)$$
with $\tilde\omega = (1+\gamma)\omega/\gamma T_0$, and
$$f_{\pm}(\tilde\omega) = a_{\pm}|\tilde\omega|^{\Delta_{sp}}+ b_{\pm}
|\tilde\omega|^{\Delta_{ch}}\leqno(5.1.28.a)$$
$$a_- = -{4\gamma\over (2+\gamma)\pi} B(2\Delta_{sp},\Delta_{ch})
\leqno(5.1.28.b)$$
$$a_+ = -cos(\pi\Delta_{ch}) a_-\leqno(5.1.28.c)$$
$$b_- = -{4W_{ch}\over (1+2\gamma)\pi} sin(\pi\Delta_{ch})
B(2\Delta_{ch},
\Delta_{sp}) \leqno(5.1.28.d)$$
$$b_+ = cos(\pi\Delta_{ch}) b_+ ~~.\leqno(5.1.28.e)$$
Here $W_{ch} = \pi T_0/\tilde\Gamma$ measures the fluctuation weight of
the channel configuration
in the ground state, and $B(x,y)$ is the Beta function.  In the case
$M\ge N$, both the leading
and next leading frequency dependence of Eq. (5.1.25.a,b) agree with
the
results obtained from
conformal field theory.  This spectral function explicitly breaks
particle hole symmetry due to the
non-particle hole-symmetric Hamiltonian.

Since the one-particle $T$-matrix describing scattering of conduction
electrons off the impurity in an Anderson Hamiltonian is given by
$t(\omega,T) = V^2{\cal G}_{\sigma\alpha}$(Langreth, 1966),
using the optical theorem we see that the conduction electron
scattering
rate is given by
$${1\over \tau(\omega,T) } = {2\tilde\Gamma
\rho_{\sigma\alpha}(\omega,T)
\over N(0)N} ~~.\leqno(5.1.33)$$

\begin{figure}
\parindent=2.in
\indent{
\epsfxsize=6.in
\epsffile{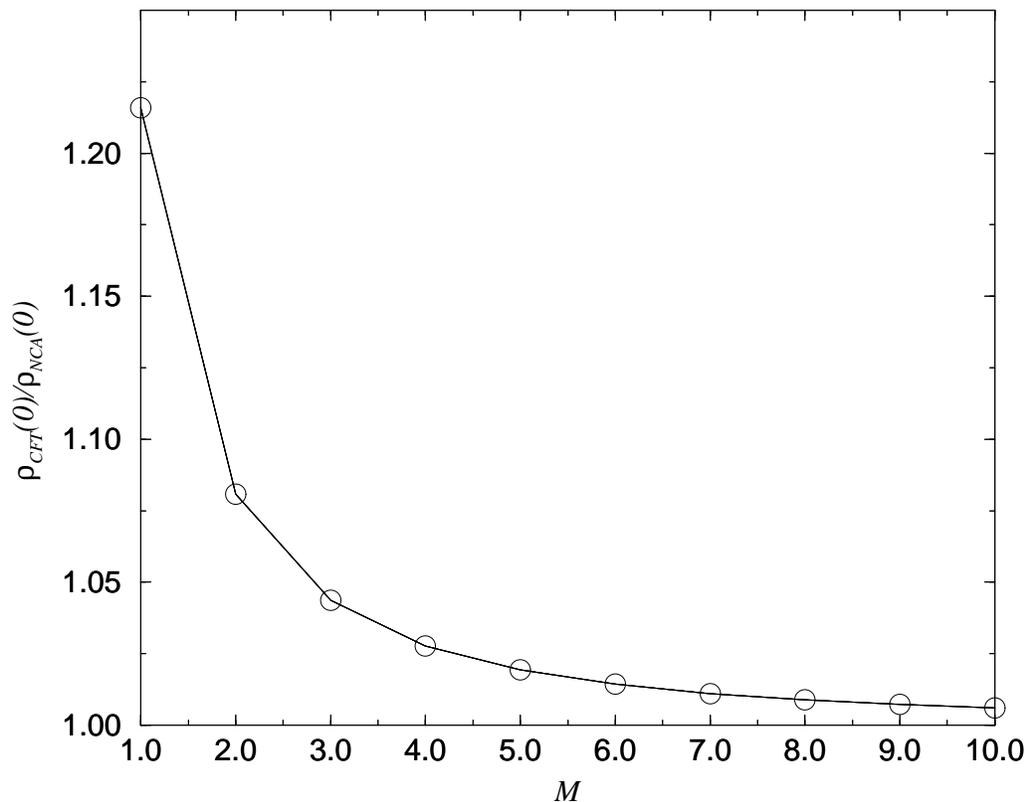}}
\vspace{.3in}
\parindent=0.5in
\caption{Ratio of NCA to Conformal Field Theory Resistivity vs. $M$ for $N$=2
multi-channel Model in the Kondo limit.  Potential scattering from the NCA has
been subtracted.  Although the NCA is strictly a large $N$ limit, it clearly 
does well for $N=2$ and arbitrary $M$, with agreement improving as $M\to\infty$. }
\label{fig5p4}
\end{figure}

The electrical resistivity $\rho(T)$ may be obtained from the
scattering
rate assuming dominant scattering in the angular momentum channels
corresponding to the pseudo-particles and using the standard transport
theory formula (see, e.g., Bickers, Cox, Wilkins, 1987)
$$\rho(T) \sim [\int d\epsilon (-{\partial f\over \partial \epsilon})
\tau(\epsilon,T)]^{-1}~~. \leqno(5.1.34)$$
On dimensional grounds, we can see that the NCA will give
$${\rho(T)\over \rho(0)}  \sim 1 - c({T\over T_0})^{min(\Delta_{sp},
\Delta_{ch})} + ....\leqno(5.1.35)$$
where $c$ is a pure number determined from the full temperature
dependence of $\tau$ which is
beyond the scope of the zero temperature NCA.  This agrees with the
conformal theory, in particular yielding a $\sqrt{T}$ correction for
the special case $N=M$ in which $\Delta_{sp}=\Delta_{ch}=1/2$.
The two-channel spin 1/2 model is a special case of this limit
($N=M=2$).  Moreover, for $N=2$,
the zero temperature scattering rate (when corrected for potential
scattering present in this model)
gives
$${\pi N(0)\over 2\tau(0,0)} = {3\pi^2\over 4(2+M)^2}
~~.\leqno(5.1.36)$$
As may be seen from Fig.~\ref{fig5p4}, this formula agrees with the exact result
from conformal theory
to within 8\% for all $M\ge 2$. Clearly, it also agrees with Eq.
(3.4.44)
from the $1/M$
expansion when expanded to the leading order in $1/M$.

We note that because of the built in particle-hole asymmetry of our
model
Hamiltonian, the thermopower will also display a $\sqrt{T}$ behavior at
low
temperatures as has been found in the conformal theory.

{\it Local Spin and Channel Susceptibilities} The term local here means
that the diagrams we keep from Fig.~\ref{fig5p3}  correspond to applied fields
coupling
linearly only to the impurity spin and channel operators and not the
conduction
electron spin and channel
 operators.  Using the results of the differential equation solutions
 together
 with Eqs. (5.1.10,11), we obtain
$$\tilde\chi_{sp,ch}''(\omega) \approx {C_{sp,ch} \over T_0}
sgn(\omega)
|\tilde\omega|^{(\Delta_{sp,ch} - \Delta_{ch,sp})} \leqno(5.1.37)$$
with
$$C_{sp} = \gamma\Delta_{sp}^2 \sin(\pi\Delta_{sp}) B(\Delta_{sp},
\Delta_{sp}) \leqno(5.1.38.a)$$
$$C_{ch} = W_{ch}^2 \Delta_{ch}^2 \sin(\pi\Delta_{ch}) B(\Delta_{ch},
\Delta_{ch}) ~~.\leqno(5.1.38.b)$$
Note that $\Delta_{sp,ch}-\Delta_{ch,sp} = 2\Delta_{sp,ch}-1$ in view
of Eq. (5.1.26.c).
The leading behavior of $\tilde\chi''_{sp,ch}$ is in full agreement
with conformal theory for all $N,M$.
Next leading corrections to Eq. (5.1.36) go as
$|\tilde\omega|^{(2\Delta_{sp,ch}-\Delta_{ch})}$.
In the special case $N=M$ which includes the two-channel model, both
susceptibilities reduce to
the form
$\tilde\chi''(\omega) \sim sgn(\omega) [1- B\sqrt{{|\omega|\over T_0}}
+ ...] $
which corresponds to a static susceptibility which is logarithmically
divergent in temperature.  This
follows simply with the application of the Hilbert transform to
$\chi''$.
This leading behavior in $\chi''$, first noted by Cox
[1988(a),1990,1994]
may provide a link to the marginal Fermi liquid phenomenology [Varma
{\it et al.}, 1989, Kotliar {it et al.}, 1990] developed
to understand the unusual normal states of the copper oxide
superconductors.

The precise temperature dependence of $\chi_{sp,ch}$ must be determined
numerically, but we
can roughly estimate this along with relevant zero temperature values
through Kramers-Kronig
analysis, which implies ($\lambda=sp$ or $ch$ below)
$$\tilde\chi_{\lambda}(T) = {1\over \pi} \int_{-\infty}^{\infty}
{\tilde\chi_{\lambda}''(\omega,T)\over \omega}~~. \leqno(5.1.39)$$
We would like to put in the power law form of $\tilde\chi''$ from
Eq. (5.1.32); this holds to an upper
cutoff of order $T_0$ since the power law behavior only sets in
below the Kondo temperature.  The
lower cutoff must be of order $T$ since finite temperature will
round the Fermi functions in the integral equations and hence round
all power laws determined from these.  With these limits of
integration we see that
$$\tilde\chi_{\lambda}(T) \approx {2 C_{\lambda} \over \pi
(2\Delta_{\lambda}-1) T_0} [1-({T\over T_0})^{2\Delta_{\lambda}-1}]
~~for~\Delta_{\lambda} \ne {1\over 2} \leqno(5.1.40.a)$$
$$\tilde\chi_{\lambda}(T) \approx {2 C_{\lambda} \over \pi T_0}
\ln({T_0\over T}) ~~for~\Delta_{\lambda}={1\over 2}
~~.\leqno(5.1.40.b)$$

\begin{figure}
\parindent=2.in
\indent{
\epsfxsize=3.in
\epsffile{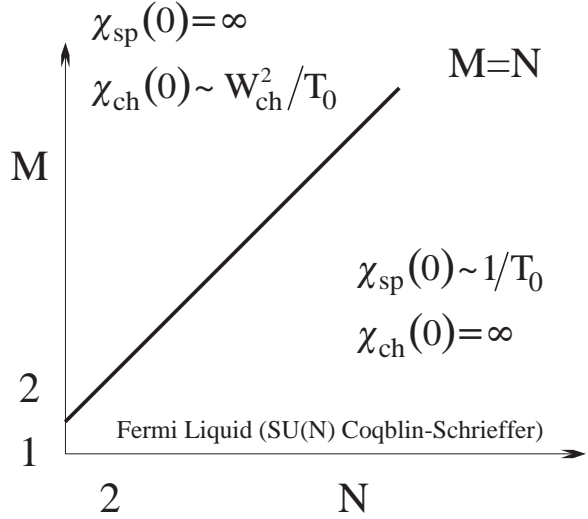}}
\parindent=.5in
\caption{Phase diagram for $SU(N)\times SU(M)$ multichannel Kondo/Anderson
Model.  For $M>N$, spin fluctuations dominate, and the low temperature spin
susceptibility is divergent.  The only low energy scale is the Kondo scale. 
For $1<M<N$, channel fluctuations dominate, and the low temperature physics
has two scales, the Kondo scale and a final energy scale where channel fluctuations dominate. 
The low temperature channel susceptibility diverges in this case. 
For $M=N$, both channel and spin susceptibilities diverge logarithmically, and there
is only one low energy scale (the Kondo scale).  For $M=1$, a Fermi liquid line results
and there is only one low energy scale ( the Kondo scale). }
\label{fig5p5}
\end{figure}

The above results for the spin and channel susceptibilities allow us to
produce
a phase diagram in the $N,M$ plane, shown in Fig.~\ref{fig5p5}.  For $M>N$,
which is the
customary over-compensated situation shown by \noz~ and Blandin [1980]
to lead to a non-trivial fixed point, the spin susceptibility is
divergent at $T=0$, diverging with a power law that follows from Eq.
(5.1.40.a).
On the other hand, the leading singular behavior of $\chi_{ch}(T)$
vanishes as
$T^{(\gamma-1)/(\gamma+1)}$ which implies a finite value to
$\chi_{ch}(0)\sim
W_{ch}^2/T_0$ as
the leading term.   In this region, all non-Fermi liquid response is
driven
dominantly by the spin fluctuations.   For $N=M$, of which the
two-channel
spin 1/2 model $N=M=2$ is a special case, we see that
$\Delta_{sp}=\Delta_{ch}=1/2$ so that each susceptibility diverges
logarithmically. Both spin and channel fluctuations contribute
to the non-Fermi liquid behavior at low temperatures, although the
dominant
effects are due to spin fluctuations due to the much smaller energy
scale
$T_0$ to
be compared with $\tilde \Gamma$.  This is reflected in the suppression
of
$\chi_{ch}$ by the factor $W_{ch}^2\approx T_0^2/\Gamma^2$.  For $N>M$,
which is still
overscreened, we are in a new
physical
situation, not discussed by \noz~ and Blandin.  In this case, Eq.
(5.1.40.a)
implies that $\tilde\chi_{ch}(T)$ diverges as
$T^{(\gamma-1)/(\gamma+1)}$,
while the spin susceptibility is finite at $T=0$,
$\tilde\chi_{sp}(0)\sim
1/T_0$.  In this regime, the low temperature
deviations from Fermi liquid behavior are ultimately driven by the
channel
spin fluctuations. However,
this can only occur to the extent that their strength exceeds that of
the
spin fluctuations.  This implies that a new temperature scale is
present
in the problem, which we shall call $T_{ch}$.
This may be estimated by seeing where the leading channel driven term
in
$\rho_{\sigma\alpha}$
equals the leading spin driven term, yielding the energy scale
$$T_{ch} \approx T_0 W_{ch}^{1\over \Delta_{sp}-\Delta_{ch}} \approx
T_0
({\pi T_0\over \tilde\Gamma}) ^{ 1+\gamma \over 1-\gamma } ~~.
\leqno(5.1.41)$$
We note that $T_{ch}$ evolves into the ``pathology'' temperature $T_p$
discussed by M\"uller-Hartmann [1984] and Bickers [1987; c.f. Eq.
(5.53)]
when $M=1$.  This corresponds to the $SU(N)$
Anderson and Coqblin-Schrieffer model, designated by the horizontal
$M=1$
axis in Fig.~\ref{fig5p5}.  The
low temperature physics here is of course Fermi-liquid like and beyond
the
scope of the NCA.
However, as discussed extensively in Bickers, Cox, and Wilkins [1987]
and
Bickers [1987], the NCA does an excellent job on finite temperature
properties
provided $T_p << T << T_0$.  Clearly the difficulty with applying the
NCA in
this context is in attempting to extrapolate to the line of zero slope;
it does very well
in describing the physics on all lines of non-zero slope in the $N,M$
plane.
Note that the scale
$T_{ch}$ is a real one provided $M>1$.  However, we know of no physical
model
for which $N>M>2$ can be realized in practice.

\subsubsection{Crossover Effects in Applied Spin and Channel Fields} 

The application of a spin or channel field will induce a crossover
to new
physics at low temperatures as discussed in previous sections.
Specifically,
we anticipate Fermi liquid behavior
of two different sorts.  For the applied spin field, at zero
temperature,  the
physics becomes that
of a Fermi gas interacting with a polarized scattering center, so that
Fermi
liquid physics must set
in below the crossover scale denoted as $T^x_{sp}$.  For the applied
channel
field, the couplings
to all but the lowest energy pseudo-boson are expected to become
irrelevant,
and this will lead to
the $SU(N)$ Coqblin-Scrieffer model for that coupling with all other
channels
having zero scattering.
The $SU(N)$ model will give a Fermi liquid excitation spectrum with a
scale
set by the crossover
temperature denoted $T^x_{ch}$.  The NCA cannot describe the low
temperature
physics well below
the crossover scale well, since it cannot describe Fermi liquid
physics.
However, the crossover region will be well described and the NCA
produces a
correct estimate for the crossover scales
$T^x_{sp,ch}$.

To estimate these crossover scales, consider first the case of an
applied
channel field.  With the
normalization $\mu_{ch}\sim 1/M$ to obtain a sensible large $N,M$
limit,
we see that the overall
splitting of channel energies is of order $H_{ch}$, assumed to be small
compared to the channel fluctuation scale $\tilde\Gamma$.  Each of the
pseudo-boson propagators now acquires a channel label $\alpha$.  The
NCA
differential equations (5.1.16.a,b) are
modified to
$${d g_f \over d\omega} = -1 - {\gamma \tilde\Gamma\over
M\pi}\sum_{\alpha}
 {1\over d_{\alpha}} \leqno(5.1.42.a)$$
$${d g_{b\alpha} \over d\omega} = -1 - {\tilde\Gamma\over \pi g_f}
\leqno(5.1.42.b)$$
with the new boundary conditions for the $d_{\alpha}$
$$g_{b\alpha}(-D) = D - \alpha_L\mu_{ch}H_{ch} ~~.\leqno(5.1.43)$$
Denote the $\alpha$ index corresponding to the lowest energy $b$ state
as
$\alpha_L$.  It is easy
to see that $dg_{b\alpha}/dd_{\alpha_L} = 1$,  so that $g_{b\alpha} =
g_{b\alpha_L} + |\alpha - \alpha_L|\mu_{ch}H_{ch}$.  Hence, the spread
of
$g_{b\alpha}$ values is no more than order $H_{ch}$, given the above
remarks.
Now, at sufficiently high energies above the crossover scale, the
lifting of the channel degeneracy should be irrelevant, meaning that
all
$g_{b\alpha}$ still go as
$\sim \tilde\Gamma |(\omega-E_0)/T_0|^{1/(1+\gamma)}/\pi$ in this
energy
region.  The crossover scale is determined by the equation
$$g_{b\alpha_L}^>(|\omega-E_0| = T^x_{ch}) = H_{ch} \leqno(5.1.44)$$
where the superscript $>$ means we utilize the high energy form for the
inverse propagator $g_{b\alpha}$.
This gives
$$T^x_{ch} = {\gamma T_0 \over (1+\gamma)}
({\pi H_{ch}\over \tilde \Gamma})^{1+\gamma} \leqno(5.1.45)$$
with the crossover exponent $1+\gamma$ in agreement with conformal
field
theory analysis, as we shall see in the next section.
In particular, for the special case $N=M=2$, we get $T^x_{ch}\sim
H_{ch}^2$,
which was also found in the NRG and Bethe-Ansatz treatments.
We observe that we could also determine the crossover exponent from
below, using the NCA forms for the one-channel $SU(N)$ Anderson model
and an
effective single channel Kondo scale given by the expression
$$T_0(H_{ch}) = D({\tilde \Gamma\over \pi D})^{1\over N}
\prod_{\alpha\ne\alpha_L}
({D\over |\alpha-\alpha_L|\mu_{ch}H_{ch}})^{1\over N}
\exp({\pi\epsilon_f \over
\tilde\Gamma})~~. \leqno(5.1.46)$$
In this case, $g_{b\alpha_L} \simeq \tilde\Gamma
|(\omega-E_0)/T_0(H_{ch})|^{N/(N+1)}$, and we
would equate this expression to $H_{ch}$ to obtain $T^x_{ch}$.  Now,
however,
$H_{ch}$ appears
on each side of the equation through the functional dependence in
$T_0(H_{ch})$.
It may readily
be verified that as a result we obtain the same estimate for
$T^x_{ch}$.

We may follow a similar procedure for estimating $T^x_{sp}$.  Here we
assume
$H_{sp}$ to be smaller than the spin fluctuation scale $T_0$. We equate
the
high energy form for $g_f$ to the spin field $H_{sp}$ which
gives
$$ T^x_{sp} = {\gamma T_0 \over (1+\gamma)} ({H_{sp}\over
T_0})^{1+1/\gamma}
~~.\leqno(5.1.46)$$
This estimate agrees with NRG, Bethe-Ansatz, and conformal theory for
$N=M=2$,
and in any
case gives the right dependence on $H_{sp}$ for all $N,M$.

From a knowledge of the crossover behavior, one can compute the spin
magnetization $M_{ch}$
and channel spin magnetization $M_{ch}$.  For $\gamma > 1$, $M_{sp}
\sim (H_{sp}/T_0)^{1/\gamma}$
as expected from conformal field theory and the Bethe-Ansatz. For
$\gamma<1$, $M_{ch}
\sim(\pi H_{ch}/\tilde\Gamma)^{\gamma}$ as can be
inferred from conformal field theory.  For $\gamma = 1$,
$M_{sp} \sim (H_{sp}/T_0)\ln(T_0/H_{sp})$ and $M_{ch}\sim (\pi H_{ch}
/\tilde\Gamma)
\ln(\tilde\Gamma/\pi H_{ch})$, both in agreement with conformal
theory (sec 6.1.3.c).  The method for computing
the magnetization
is based on a form for the ground state energy in terms of the inverse
green's functions.  We refer
the reader to Appendix III for details.

\subsubsection{Vertex Corrections} 

The success of the NCA in producing the critical exponents for
$\rho_{\sigma\alpha},\tilde\chi_{sp,ch}$ and the crossover exponents
correct
for all $N,M$, suggests  that the vertex corrections are somehow
unimportant
in modifying the physics of the
problem from these simple diagrams.  The lowest order vertex
corrections are
illustrated in Fig.~\ref{fig5p2} for the pseudo-particle self-energies, and
these are
of order $1/N^2$ as argued previously.
A little thought indicates the following scenario
for the
vertex corrections:  since the exponents are obtained correctly for all
orders
in $1/N$ for the quantities considered, the most singular contribution
from the
vertex corrections cannot modify these exponents.  The vertex
corrections can, however,
correct
the {\it amplitudes} of the leading singular contributions by terms of
order
$1/N^2$ and higher.

Let us apply this reasoning to the self-energy equations, which, as we
have
seen, determine all
the physics of the above quantitities.
This scenario may be checked self-consistently by assuming the
pseudo-particle
propagators retain the same critical behavior at threshold $E_0$ as in
the order
1 solutions.  Explicit evaluation of
the pseudo-fermion self-energy of Fig.~\ref{fig5p1} then shows that the
dominant low
energy contribution
vanishes again as $|\omega-E_0|^{\gamma/(1+\gamma)}$.  A detailed
analysis is deferred to Appendix III.

With power counting arguments, we can see that this
result is
more general.  Consider a generic contribution to $\Sigma_f$ which has
$L$ loops and independent energy integrations, and hence contains $L$
pseudo-boson propagators and $L-1$ pseudo-fermion propagators.
By converting each energy integration to
dimensionless
form, we can read off the power law dependence on $\omega-E_0$.  Each
boson propagator diverges as $|\omega-E_0|^{2\Delta_b -1}$,
each fermion propagator as $|\omega-E_0|^{2\Delta_f - 1}$.  From
de-dimensionalizing
the integrations we obtain $L$ factors of $\omega-E_0$ from the
differentials, $L$
powers of $\omega-E_0$ from the boson propagators, and $L-1$ powers of
$\omega-E_0$ from the fermion propagators.  The net power is then
$L + L(2\Delta_b-1)+(L-1)(2\Delta_f-1)
= 2L(\Delta_b+\Delta_f) -2\Delta_f - L+1$ which, since
$\Delta_f+\Delta_b=1/2$, is just $1-2\Delta_f$.  But this is precisely
the leading order power of the self-energy, so we have
self-consistently
demonstrated the scenario.

This is analogous to result for the quasiparticle lifetime in a Fermi
liquid--
the correct $\omega^2+\pi^2T^2$ behavior can be found in a diagram
second order
in the interaction strength.
Higher order diagrams may readjust the strength of the effective
interaction--
the amplitude of the leading order power--but will not modify the form
of the
lifetime, that is, the leading order power itself.  As in Fermi liquid
theory,
where the renormalized interaction is of the order of the bandwidth and
so no strict perturbation
theory is applicable, we find ourselves in a situation here where the
regime of
strict validity of the theory (large $N$ only) appears far smaller than
its
regime of applicability.

The observations of the last two paragraphs led Cox and Ruckenstein
[1993] to
speculate that in systems with non-Fermi liquid ground states which
typically
show some form of spin-charge separation, extended to spin-channel spin
separation here, some form
of self-consistent perturbation theory can capture the essential
physics
even when no obvious small parameters are available, barring some kind
of phase
transition which binds spin and charge (or spin and channel spin
here).

In fact, there is evidence that just such a phase transition occurs in
the Kondo problem between
the overcompensated regime for which the NCA is obviously well suited
to
the single channel
Fermi liquid line of Fig.~\ref{fig5p5}.  The work of Kroha {\it et al.} [1992,1996],
together with that of Anders and Grewe [1994] and Anders [1995a,1995b] moves
 beyond the NCA to work on the single channel model.  The work of Anders 
 and Grewe [1994] and Anders [1995a,1995b] demonstrates that a tendency to restore
 Fermi liquid properties is indeed produced by a self-consistent inclusion
 of vertex corrections through order $1/N^2$.  Kroha {\it et al.} [1992,1996]
 demonstrate that the vertex corrections display a tendency towards 
 bound state formation between the conduction electron and the pseudo-fermion
 state, closely related to the spin-screened Fermi liquid state.  In 
 particular, in a saddle point evaluation of the conduction electron-pseudo-fermion 
 two-particle $T$-matrix, a pole is obtained on resumming an infinite 
 number of repeated particle-particle interactions between the electrons
 and pseudo-fermions.  However, due to fluctuations beyond this saddle 
 point approximation, a true bound state is not in fact realized.  Instead, the 
 pole contribution serves to renormalize the threshold exponents of  
 the pseudo-particle propagators from 
 the NCA values of $\alpha_f=1/(N+1)=1-\alpha_b$ to exact values which
 have been deduced both from a combined Bethe-Ansatz/Conformal Field Theory
 approach [Fujimoto {\it et al.}, 1996] and an analytic argument 
 [Menge and M\"{u}ller-Hartmann, 1988].  These exponents are dependent 
 upon the occupancy of the pseudo-fermion level, in contrast to the 
 NCA values, and are indeed characteristic of a Fermi-liquid ground state. 
 Thus, the Fermi liquid ground state appears to be reinstated by an 
 appropriate recombination of spin and charge degrees of freedom 
 induced by complicated interactions between fermionic and bosonic degrees
 of freedom.  
 
\subsubsection{Properties at Finite Temperature} 

The NCA is not limited, of course, to zero temperature where the
differential
equation approach is applicable.   The integral equations may be
self-consistently
solved at finite temperature and the
properties calculated.  In this subsection, we wish to point out that
the
comparison is astonishingly good of magnetic susceptibility, specfic
heat, and entropy curves
with
exact calculations from the Bethe-Ansatz. In particular, the
discrepancy
for the susceptibility is barely visible, while a discrepancy with the
$T=0$ entropy at the several percent level is seen, but expected given
the large $N$ character of the analysis.
The calculations are based on the \ctp model to be discussed in the
next
subsection, which has
parameter regimes described by the two and three-channel spin 1/2
model.

\begin{figure}
\parindent=2.in
\indent{
\epsfxsize=3.in
\epsffile{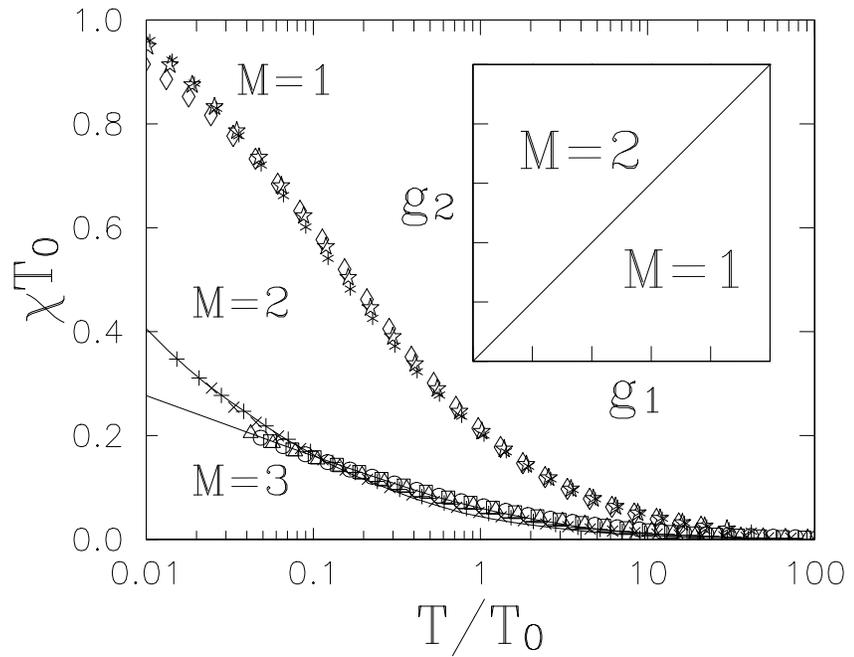}}
\parindent=.5in
\caption{Spin susceptibility for $SU(2)\times SU(M)$ multichannel Anderson 
models in the Kondo regime obtained from the 
NCA. Of particular note here is the excellent agreement between the exact 
Bethe-Ansatz results for $M=2,3$ and the NCA results.  From Kim and Cox [1997].}
\label{fig5p6}
\end{figure}

In Fig.~\ref{fig5p6}, we show results taken from Kim and Cox [1995,1997] for
various
parameter sets in the two-channel Kondo regime of the simplified model. 
What is apparent is
that all
the computed NCA $\chi(T)$ curves agree almost perfectly with the exact
Bethe-Ansatz results of Sacramento and Schlottmann [1991].  Also shown in Fig.~\ref{fig5p6} 
is shown the corresponding calculation for the 3-channel regime.
The agreement is clearly excellent.  It needs to be mentioned that this
is
actually a two-parameter
comparison; $T_0$ is adjusted to $T_K$ from Sacramento and Schlottmann
by
sliding along the
logarithmic temperature axis.  A mild scale factor adjustment is also
required to bring the vertical
axis into alignment.  Kim and Cox [1995,1997] argue that this scale
factor is
related to the crossover effects in the \ctp model they study.

\begin{figure}
\unitlength .3mm
\begin{picture}(200,600)(0,0)
\put (100,100){
\epsfxsize=5.in
\epsffile{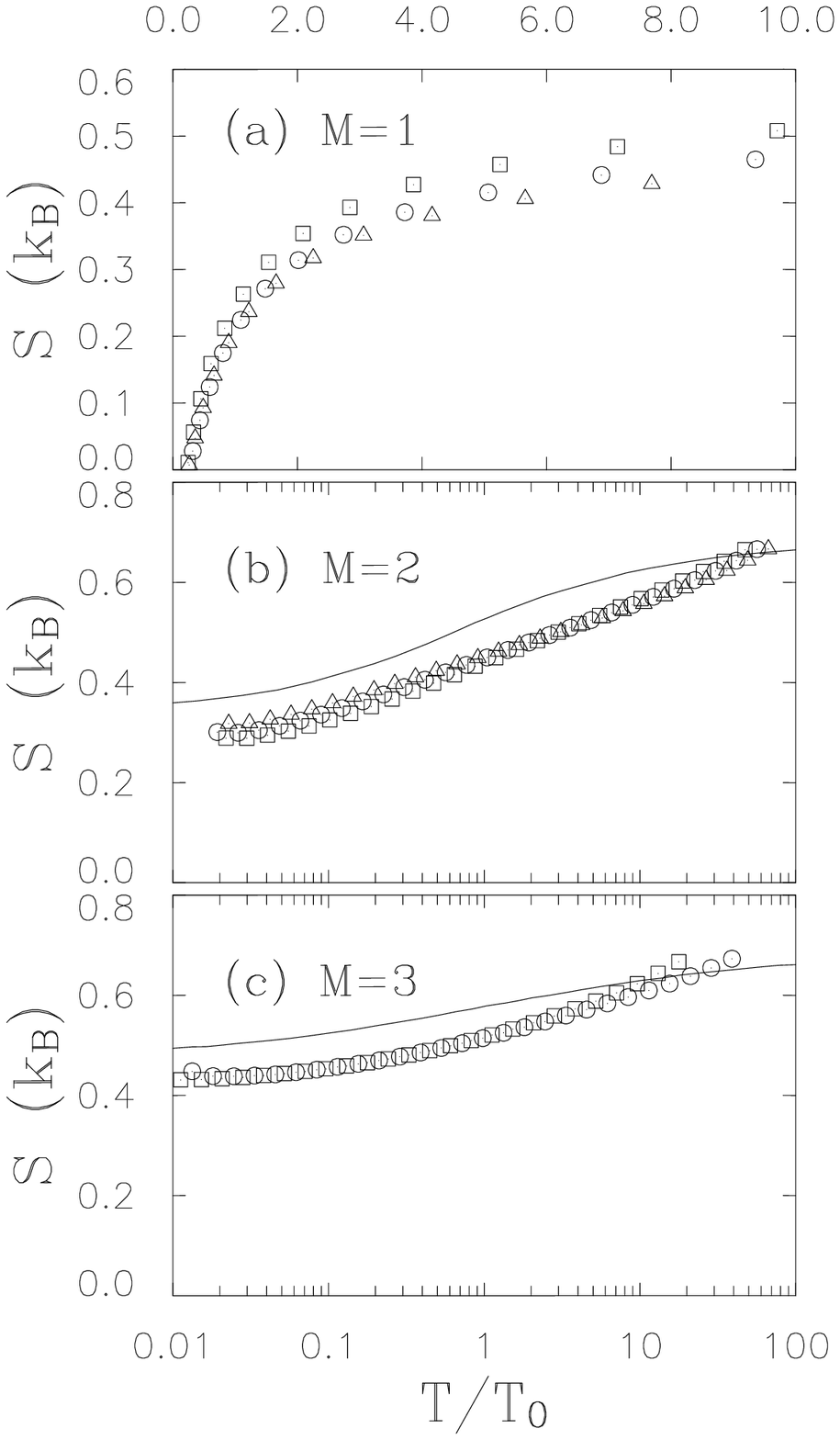}}
\end{picture}
\vspace{.5in}
\caption{Numerically calculated entropy curves from the NCA for 
the $SU(2)\times SU(M)$ Anderson models in the Kondo regime.  The $M=1$
curve shows the clear (and expected) tendency towards $S(0)=0$ as expected
for the single channel model.  The $M=2,3$ curves agree well with exact
Bethe-Ansatz results (solid lines) and clearly display a tendency towards
a finite residual entropy.  Agreement for $S(0)$ is evidently at the 10-15\% level. 
From Kim and Cox [1997].}
\label{fig5p7}
\end{figure}

Attempts to compute the specific heat have also proven successful.  The
only method developed to compute the specific heat from the NCA
consists
of numerical differentiation of ${\cal Z}_f$ (Bickers, Cox, and Wilkins
[1987]).  This works well for sufficiently large specific heat signal,
but in the multichannel models, the
residual entropy robs the finite temperature specific heat of
considerable
integrated intensity.  As a result, it is essential to have high
numerical precision to obtain the specific heat and entropy results.
A recent numerical advance in the NCA codes
by Kim [1995] allowed reliable computation of entropy and specific
heat.
The entropy curves for $N=2$ and $M=1,2,3$ are displayed in Fig.~\ref{fig5p7} 
[Kim and Cox, 1995,1997].
The $M=1$ curve shows the expected extrappolation to $S=0$ as $T\to 0$,
although the  NCA curve is not to be trusted in the low temperature
region for this Fermi liquid case.  The $M=2,3$ curves show temperature
dependences in excellent agreement with Bethe-Ansatz. The residual
entropies are within 10-15\% of the expected values for $M=2,3$.  We
don't
expect exact agreement of the residual entropy given the explicit $N$
dependence found in Bethe-Ansatz [Tsevlik, 1985; Sacramento and
Schlottman, 1991] and conformal theory treatments [Affleck and Ludwig,
1991c].  However, the overall agreement is clearly exceptional for this
simple theory.

\begin{figure}
\parindent=2.in
\indent{
\epsfxsize=5.in
\epsffile{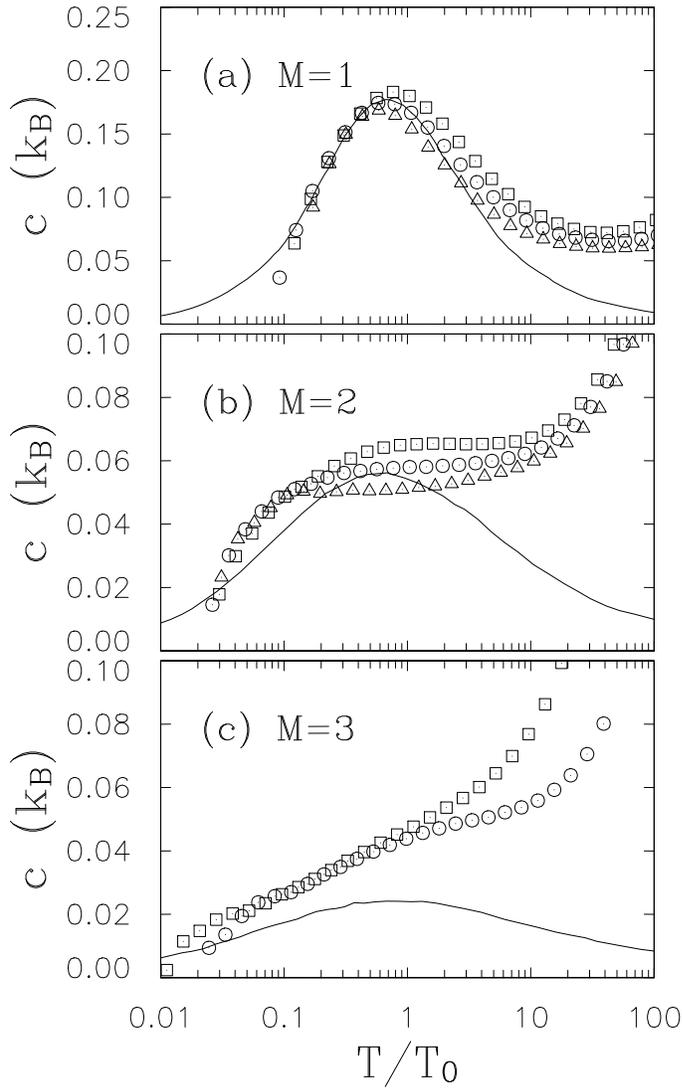}}
\parindent=0.5in
\vspace{.7in}
\caption{Specific heat for $SU(2)\times SU(M)$ Anderson models in the Kondo limit
with $M=1,2,3$ as calculated from the NCA.  Points are from the NCA, lines are from
exact Bethe-Ansatz results.  The high temperature enhancements are due to the presence
of the excited configurations in the model, which are not present in the Bethe-Ansatz
model.  From Kim and Cox [1997].}
\label{fig5p8}
\end{figure}

The specific heat curves for $N=2$ and $M=1,2,3$ are shown in Fig.~\ref{fig5p8}
[Kim and Cox, 1995,1997].
Clearly, the $M=1$ curve is in excellent agreement with the exact
results.  The slight underestimate of the magnitude for $M=2,3$ is
understandable from the slight overestimate of the  residual entropy
evident in Fig.~\ref{fig5p7}, since the net high temperature entropy must be
$R\ln2$.

\subsubsection{Alternative Large $N$ formulation}

Recently Parcollet and Georges [1997] have developed an alternative
large $N$ approach to the multi-channel model which represents the spins
in terms of constrained Schwinger boson fields. This theory is able to 
treat overcompensated, undercompensated, and compensated cases exactly,
and is expressed in terms of saddle point equations which are strongly
reminiscent of the non-crossing approximation.  The physical properties 
computed within this approach agree well with other methods, and in
contrast to the NCA, an analytic formula for the zero temperature
entropy is obtained for the over- and undercompensated models.  The
method would appear to hold considerable promise for a lattice
generalization.

\subsection{Application of the NCA to a model \ctp~ impurity} 

This subsection reviews the work of Cox [1993], Kim [1995], and Kim
and
Cox [1995,1997] in studying a model \ctp~ impurity which encompasses
the
possibilities of one, two, and three-channel spin 1/2 Kondo effects.
The model
is
intended to be a simplified version of
a more complete model for Ce in LaCu$_{2.2}$Si$_2$.

\subsubsection{Pseudo-Particle Hamiltonian and NCA equations} 

\begin{figure}
\begin{picture}(200,200)(0,0)
\put (100,100){
\epsfxsize=5.in
\epsffile{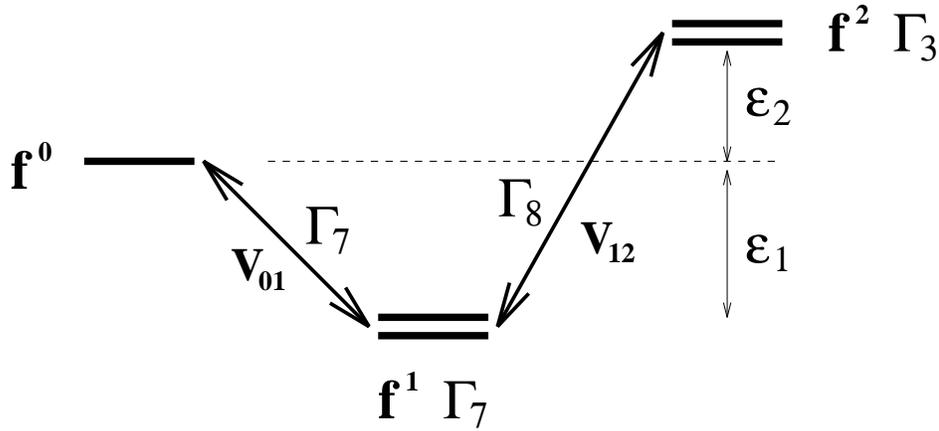}}
\end{picture}
\vspace{.7in}
\caption{Energy levels of the $f^0,f^1,f^2$ states for the simplest \ctp ion Anderson 
model. Only a magnetic doublet is kept in the $f^1$ configuration, and only a non-Kramers' doublet 
is kept in the $f^2$ configuration. Taken from Kim and Cox [1997].}
\label{fig5p9}
\end{figure}

The assumed level spectrum of the \ctp~ ion is shown in Fig.~\ref{fig5p9}.  The
Hamiltonian
is then given by Eqs. (2.2.25,26).  We wish to rewrite this Hamiltonian
in
pseudo-particle form.  We represent the
$f^1$ doublet by a pair of  pseudo-fermion operators $f_{7\mu}$, and
the
$f^0,f^2$
states by pseudo-bosons $b_1,b_{3\alpha}$ respectively.  To enlarge the
Hilbert space
from the two-configuration model, the $f$-charge $Q_f$ of Eq. (3.3.6)
is
modified to
$$Q_f = \sum_{\alpha} b^{\dagger}_{3\alpha}b_{3\alpha} + \sum_{\mu}
f^{\dagger}_{7\mu}
f_{7\mu} + b^{\dagger}_1b_1 ~~\leqno(5.2.1)$$
The resulting pseudo-particle Hamiltonian is
$$H=\sum_{k\Gamma_c\alpha_c} \epsilon_k c^{\dagger}_{k\Gamma_c\alpha_c}
c_{k\Gamma_c\alpha_c} + \epsilon_1 \sum_{\mu}
f^{\dagger}_{7\mu}f_{7\mu}
\leqno(5.2.2)$$
$$~~~~~ + \epsilon_2 \sum_{\alpha}
b^{\dagger}_{3\alpha}b_{3\alpha} +
{V_{17}\over \sqrt{N_s}} \sum_{\mu} [f^{\dagger}_{7\mu} b_1 c_{7\mu} +
h.c.] $$
$$~~~~~ - {V_{37}\over \sqrt{N_s}} \sum_{\alpha\mu} sgn(\mu)
[b^{\dagger}_{3\alpha} f_{7,-\mu} \caei + h.c.] - \lambda_{ps}(Q_f - 1)
 ~~.$$
We denote the hybridization width corresponding to $V_{17}(V_{37})$ by
$\Gamma_{17}=\pi N(0)V^2_{17}(\Gamma_{37}=\pi N(0)V^2_{37})$.  Here
$\epsilon_1=\epsilon_f$ and
$\epsilon_2 = \tilde\ef + \ef =
2\epsilon_f + U_{ff}$ where $U_{ff}$ is the Coulomb
repulsion.

As discussed in Sec. 2.2.2, for $\Gamma_{37}/\pi|\tef| <
\Gamma_{17}/\pi|\ef|$,
we anticipate a one
channel Kondo effect. The NCA will correctly describe the approach to a
Fermi
liquid fixed point here, though not the actual fixed point.  However,
the
scaling dimensions in this instance will be uniquely specified by the
NCA
and serve as a measure of the
 universality class within the method.
For $\Gamma_{37}/\pi|\tef| > \Gamma_{17}/\pi|\ef|$, we anticipate a
two-channel
Kondo effect to describe the low temperature physics.  This will be
obtained
essentially correctly through the NCA.
For $\Gamma_{37}/\pi|\tef| = \Gamma_{17}/\pi|\ef|$, we anticipate the
three-channel Kondo effect to regulate the low temperature physics,
which
again is obtained essentially correctly through the NCA.  The above
conditions on the dimensionless Schrieffer-
Wolff coupling constants will be slightly modified in the full NCA
analysis,
but the essential physics is unchanged.

\begin{figure}
\parindent=2.in
\indent{
\epsfxsize=3.in
\epsffile{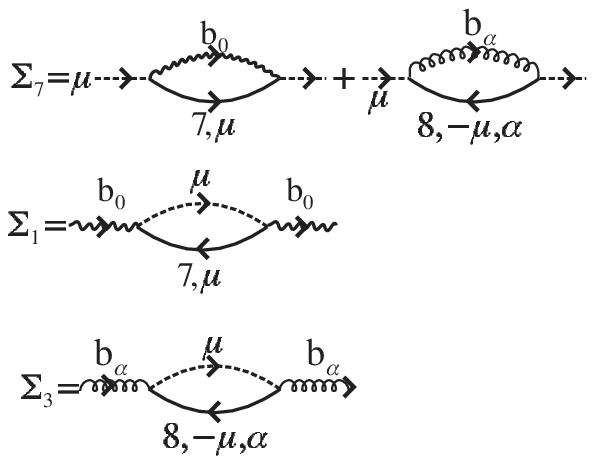}}
\parindent=0.5in
\caption{Pseudo-particle self energy diagrams for the simple \ctp ion 
Anderson model.  Dashed lines are for the $f^1$ pseudo-fermion propagator
which carries spin (magnetic) index $\mu$. Wavy lines are for the $f^0$ pseudo-boson
propagator.  Curly lines are for the $f^2$ pseudo-boson propagator which 
carries channel (orbital) index $\alpha$. Solid lines are for conduction 
electrons which may carry only spin ($\gse$, denoted by $7$), or both 
spin and channel ($\gei$, denoted by 8).  See Kim and Cox [1995,1997].}
\label{fig5p10}
\end{figure}

Fig.~\ref{fig5p10} illustrates the self-energy diagrams retained in the NCA.
The resulting self energies are denoted $\Sigma_7$ for the $f^1$
doublet,
$\Pi_1$ for the $f^0$ singlet, and $\Pi_3$ for the $f^2$ doublet. The
corresponding propagators are
$${\cal G}_7(\omega) = {1\over \omega -\ef -\Sigma_7(\omega)}
\leqno(5.2.3)$$
$${\cal G}_1(\omega) = {1\over \omega - \Sigma_1(\omega)}
\leqno(5.2.4)$$
and
$${\cal G}_3(\omega) = {1\over \omega - \epsilon_2 - \Sigma_3(\omega)}
~~.\leqno(5.2.5)$$
 Assuming, for convenience, particle-hole symmetry to the conduction
 band,
 the integral equations of the NCA are
$$ \Sigma_7(\omega) = {\Gamma_{17}\over\pi}\int d\epsilon f(\epsilon)
{\cal G}_1(\omega+\epsilon)
{2\Gamma_{37}\over \pi}\int d\epsilon f(\epsilon)  {\cal
G}_3(\omega+\epsilon)
\leqno(5.2.6.a)$$
$$ \Sigma_1(\omega) = {2\Gamma_{17}\over\pi}\int d\epsilon f(\epsilon)
{\cal G}_7(\omega+\epsilon) \leqno(5.2.6.b)$$
$$ \Sigma_3(\omega) = {2\Gamma_{37}\over\pi}\int d\epsilon f(\epsilon)
{\cal G}_7(\omega+\epsilon) ~~.\leqno(5.2.6.c)$$
Notice that $\Sigma_3 = (\Gamma_{37}/\Gamma_{17})\Sigma_1$.

This model has a special feature relative to other three configuration
models
studied previously with the NCA, such as the s-wave spin 1/2 Anderson
Hamiltonian [Pruschke 1990,1992].  In the latter case, there is a
vertex
correction which in effect mixes the $s^0,s^2$ configurations.  The
origin of the simplification for this model
is not mysterious when viewed from
the
standpoint of the Schrieffer-Wolff transformation:  the $s^0,s^2$
states in
the s-wave model each transform as fully symmetric representations and
so
contribute terms of the same form in the effective exchange
Hamiltonian.
Hence, there can be only one Kondo scale associated with the
two excited configurations.  The vertex corrections used by Pruschke
have a
direct corresondence
to the scaling theory diagrams which couple the $s^0$-driven exchange
to the
$s^2$-driven exchange.   In this model for a \ctp ion, however, the
Schrieffer-Wolff exchange interactions induced by virtual $f^0$ and
$f^2$ fluctuations have entirely different form and couple to
different symmetry conduction partial wave states.  Hence,
there can be no such cross-couplings of effective exchange
interactions in low order scaling theory, and no vertex corrections
within the
NCA analysis.

\subsubsection{NCA Differential Equations} 

Following the same procedure of Sec. 5.1.1, and defining the inverse
propagators
$g_7 = -{\cal G}_7^{-1}$, $g_1=-{\cal G}_1^{-1}$, and $g_3 = -{\cal
G}_3^{-1}$
we obtain the following differential equations for the $T=0$ NCA:
$${dg_7\over d\omega} = -1 - {\Gamma_{17}\over \pi  g_1} -
{2\Gamma_{37}\over
\pi g_3} \leqno(5.2.7.a)$$
$${dg_1\over d\omega} = -1 - {2\Gamma_{17}\over \pi g_7}
\leqno(5.2.7.b)$$
$${dg_3\over d\omega} = -1 - {2\Gamma_{37}\over \pi g_7}
\leqno(5.2.7.c)$$
subject to the boundary conditions
$$g_7(-D) = D+\ef,~~g_1(-D) = D,~~g_3(-D) =
D+\epsilon_2~~.\leqno(5.2.8)$$

These equations are clearly more complex than for the $SU(N)\otimes
SU(M)$
model of the previous subsection.  Indeed, if
$\Gamma_{17}\ne\Gamma_{37}$,
a complete low frequency analytic solution cannot be obtained, as we
shall
discuss below.  The reason is
that there is no integration constant corresponding to $\tilde C$ of
Eq.
(5.1.7) which connects low and high energy regimes.
This does not prevent the derivation of some analytic results, nor does
it prohibit a full numerical solution of the NCA equations.

\subsubsection{Solution in the Special Case $\Gamma_{17}=\Gamma_{37}$} 

The special case $\Gamma_{17}=\Gamma_{37}=\Gamma$ is fully soluble in
the spirit of the previous subsection, and provides considerable
insight to
the physics of the model (Kim [1994]).  Accordingly, we shall devote
most
of our attention to this case.  We can see immediately that the
equality of
$\Sigma_3$ and $\Sigma_1$ implies that $dg_3/dg_1 = 1$, so that $g_3 =
g_1 +
const.$.  From Eq. (5.2.8),
the integration constant is seen to be $\epsilon_2$.  We consider three
separate
cases, for $\epsilon_2=0,\epsilon_2>0,$
and $\epsilon_2<0$.

{\it (i) $\epsilon_2$=0}.  In this case we anticipate the physics to be
that
of
the three channel model.  Clearly, the NCA differential equations imply
$g_1=g_3$.  Substituting into Eqs. (5.2.7.a,b) we find the integration
constant
$$\tilde C = \exp({\pi\over 2\Gamma}[g_7 - g_1])({g_7\over D})^2
({D\over
g_1 })^3 ~~. \leqno(5.2.9)$$
Following the previous analysis for the $SU(N)\otimes SU(M)$ model, we
evaluate
this expression at $-D$ to obtain $\tilde C = \exp(\pi\ef/2\Gamma)$.
This
implies that at low energies approaching the threshold $E_0$,
$$1 \approx {[g_7/ T^{(3)}_0]^2 \over [\pi g_1/ \Gamma]^3}
\leqno(5.2.10)$$
with $T_0^{(3)}$ being the NCA estimate for the three-channel Kondo
temperature
$$T^{(3)}_0 = D({\Gamma\over \pi D})^{3/2} \exp({\pi\ef\over 2\Gamma})
~~.\leqno(5.2.11)$$
Comparison with the previous section or direct substitution into Eqs.
(5.2.7.a,b) confirms that
the low temperature behavior is precisely that of the $N=2,M=3$ model.
This
means that
$g_7 \sim T_0|\omega-E_0|^{3/5}$ and $g_1,g_3 \sim \Gamma
|\omega-E_0|^{2/5}$
for $\omega\to E_0$.

{  \it (ii) $\epsilon_2>0$}.
We expect the low temperature physics here to
be the
same as in the three channel case.  Now we eliminate $g_3$ in favor of
$g_1$.
Substituting into Eqs. (5.2.7.a,b) we find the integration constant
$\tilde C$ given by
$$\tilde C = \exp({\pi\over 2\Gamma}[g_7 - g_1])({g_7\over D})^2 {D^3
\over g_1 (g_1 + \epsilon_2)^2} ~~. \leqno(5.2.12)$$
Following the previous analysis for the $SU(N)\otimes SU(M)$ model, we
evaluate this at $-D$ to
obtain $\tilde C=\exp(\pi\ef/2\Gamma)$.  This implies that at
sufficiently
low energies
when $g_1 <<\epsilon_2$ as $\omega-E_0 \to 0$, we have the relation
$$1 \approx {[g_7/T_0^{(1)}]^2 \over \pi g_1/\Gamma} \leqno(5.2.13)$$
where the Kondo scale $T_0^{(1)}$ is given by
$$T_0^{(1)} = D({\epsilon_2\over D}) ({\Gamma\over \pi D})^{1/2}
\exp({\pi\ef
\over 2\Gamma}) ~~.\leqno(5.2.14)$$
The low frequency relation of Eq. (5.2.10) implies asymptotically that
$g_7 \sim |\omega-E_0|^{1/3}$
and $d_1\sim |\omega-E_0|^{2/3}$.  These are precisely the asymptotic
forms
expected for the
single channel spin 1/2 model (see M\"uller-Hartmann [1984] and Bickers
[1987]). The superscript
(1) in the Kondo scale refers to the single channel character.  T
his implies further that ${\cal G}_7$ diverges as
$|\omega-E_0|^{-1/3}$, and
${\cal G}_1$
diverges as $|\omega - E_0|^{-2/3}$.  In contrast, ${\cal G}_3$ is
finite at
threshold,
which implies that ${\cal A}_3(\omega) \sim |\omega-E_0|^{2/3}$ close
to
threshold.  Since ${\cal G}_3$ corresponds to the dynamically screened
exchange
between the
$\gse$ \ctp~ doublet and the $\gei$ conduction quartet, this vanishing
corresponds to the
irrelevance of that coupling when the bare $\gse-\gei$ exchange is
smaller than
the bare $\gse-\gse$ exchange.

We next identify the crossover energy scales at which the low
temperature
single channel behavior begins to dominate.
This proceeds in much the same spirit as in the identification of
crossover
scales for applied spin and channel fields
in the $SU(N)\otimes SU(M)$ model of the previous subsection.  The
relevant
comparison here
is between the magnitude of $g_1$ and $\tef$.
The maximal scale of $g_1$ is set by $\Gamma$ in the low energy
region.  Hence,
if $\Gamma << \epsilon_2$, we will always pass
to the
single channel Kondo physics on lowering from high energy/temperature
scales
without
seeing the three channel Kondo physics corresponding to essentially
degenerate
$f^0,f^2$ states.
However, if $\epsilon_2 < \Gamma << |\ef|$, we have a more interesting
situation.
 For $E_0>\omega$, we have, following M\"uller-Hartmann [1984],
$$g_1(\omega) \simeq {\Gamma\over \pi} [{3(E_0-\omega)\over T_0}]^{2/3}
\leqno(5.2.15)$$
which should be equated to $\tef$ to determine the crossover scale
$T^x_{(1)}$.
We thus
find
$$T^x_{(1)} \simeq {1\over 3} T^{(1)}_0 ({\pi \epsilon_2\over
\Gamma})^{3/2}
~~.
\leqno(5.2.16)$$
This clearly tends to zero as $\epsilon_2 \to 0$.  For frequencies
above this
scale,
$g_1$ exceeds $\epsilon_2$ and the physics becomes that of the three
channel
model.
An alternative approach is to compare the three channel form for $g_1$
above
$T^x_{(1)}$ with
$\epsilon_2$.  This yields
$$T^x_{(1)} \simeq {3\over 5} T^{(3)}_0 ({\pi \epsilon_2\over
\Gamma})^{5/2}
~~.
\leqno(5.2.17)$$
Using  the expression for $T^{(3)}_0$, we see that this result differs
from
that of Eq. (5.2.16) only by order unity.

The most interesting feature of the discussion in the preceding
paragraph is
that even for vanishingly small $\epsilon_2$,
the physics will ultimately be
that
of the single channel model at temperatures below
$T^x_{(1)}$, with a Kondo scale fixed by $T^{(1)}_0$.  Indeed, to
cleanly
see the single channel physics numerically, we infer
that $\epsilon_2$ must be
at least of order $\Gamma$ since $T^x_{(1)} \sim
(\epsilon_2/ \Gamma)^{5/2}$.  Clearly more energy scales are
present for
a
model \ctp~ ion than are evident from the bulk of theoretical
approaches
taking $U_{ff}\to\infty$!

{\it (iii) $ \epsilon_2<0$} In this case, we expect the low temperature
physics
to be governed by the two-channel spin 1/2 fixed point.  Now we
eliminate
$g_1$ in favor of $g_3$ and obtain the integration
constant
$$\tilde C = \exp({\pi \over 2\Gamma}[g_7-g_3])({g_7\over D})^2
{D^3\over
g_3^2(g_3+|\epsilon_2|)} \leqno(5.2.18)$$
which now gives $\tilde C=\exp(-\pi[\ef+U_{ff}]2\Gamma)$ and implies
near
threshold
$$1 \approx {[g_7/T_0^{(2)}]^2\over [\pi g_3/\Gamma]^2}
\leqno(5.2.19)$$
with
$$T^{(2)}_0 = D({|\epsilon_2| \over D})^{1/2} ({\Gamma\over \pi
D})\exp(-{\pi\tef
\over 2\Gamma})
~~.\leqno(5.2.20)$$
Notice the new exponential factor in this Kondo scale.  Use of Eq.
(5.2.19) in
solving the NCA differential equations confirms that this is the
$N=M=2$ limit
of the $SU(N)\otimes SU(M)$ model.
Hence, each of $g_7,g_3$ vanish as $|\omega-E_0|^{1/2}$ near threshold,
while
${\cal A}_1(\omega)$ vanishes as $|\omega-E_0|^{1/2}$ near threshold
corresponding to the irrelevance of
the single channel coupling in this instance.

We may discuss the crossover physics in exactly the same manner as for
the
single channel case.
For $|\epsilon_2|>>\Gamma$, the three-channel fixed point
is never approached from
high energies and we simply flow directly to the physics of the
two-channel
fixed point.  For $|\epsilon_2|<\Gamma$, we can determine the
energy scale at which we crossover from three channel to
two-channel physics by equating
$$g_3(\omega) \simeq {\Gamma \over \pi}[{2|\omega-E_0|\over T_0}]^{1/2}
\leqno(5.2.21)$$
to $|\epsilon_2|$ which yields
$$T^x_{(2)} \simeq {1\over 2 } T_0^{(2)}({\pi |\epsilon_2|\over
\Gamma})^2
\leqno(5.2.22)$$
or by equating the three channel form for $d_3$ with $|\tef|$ that
gives
$$T^x_{(2)} \simeq {3\over 5}  T_0 ({\pi |\epsilon_2|\over
\Gamma})^{5/2}
\leqno(5.2.23)$$
which agrees with Eq. (5.2.21) to within factors of order unity.  Once
again,
even for vanishingly small $\epsilon_2$, the low temperature physics
below
$T^x_{(2)}$ will be governed by the two-channel fixed point.

\subsubsection{Remarks on the General Case $\Gamma_{17}\ne
\Gamma_{37}$} 

As we mentioned in Sec. 5.2.2, when $\Gamma_{17}\ne \Gamma_{37}$, no
integration constant can be found corresponding to $\tilde C$ and a
full
solution must be numerical.  However, we can
make some statements about universality classes with confidence since
there are approximate integration constants there obtained by
neglecting the
-1 terms in Eqs. (5.2.6a,b).  This is valid near threshold assuming
$g_7$
diverges which in turn implies that
at least one of $d_1,d_3$ diverge near threshold.  Specifically, we see
that $dg_3/dg_1 \approx {\Gamma_{37}/\Gamma_{17}}$.  Thus
$$ g_3 \approx ({\Gamma_{37}\over \Gamma_{17}}) g_1 + C
\leqno(5.2.24)$$
and $C$ can be identified from the values at threshold to be
$$C= \epsilon_2^* = \epsilon_2 + ({\Gamma_{37}\over \Gamma_{17}}-1)E_0
~~.\leqno(5.2.25)$$
The meaning of $\epsilon_2^*$ is that this integration constant
essentially plays the role of a hybridization
renormalized value of $\epsilon_2$ near the threshold.  Clearly, the
conditions of the previous section
regarding the sign of $\epsilon_2$ can now be replaced by
the corresponding conditions on the sign of
$\epsilon_2^*$.  Unfortunately, $E_0$ is not known
{\it a priori}, so that this
condition is not as useful in
determining the low temperature physics.

The approximate integration constant relation of Eq. (5.2.24) will hold
up to
some energy cutoff $D^* $.  Consider the case $\epsilon_2^*$=0.
Then in the same
energy region we can identify a new approximate integration constant
$\tilde C^*$ given by
$$ \tilde C^*  \approx ({g_7\over D^*})^2 ({D^*\over g_1})^3
\leqno(5.2.26)$$
which implies
$$1 \approx {[g_7/T^{(3)*}_0]^2\over [\pi g_1/\Gamma_{17}]^3}
\leqno(5.2.27)$$
with
$$T^{(3)*}_0 \approx D^*\tilde C^* ({\Gamma_{17}\over\pi D^*})^{3/2}
~~.
\leqno(5.2.28)$$
Note that Eq. (5.2.23) ensures that $\Gamma_{37}$ cancels from
$dg_7/dg_1$.
Thus the form of the low temperature physics must be the same as in the
simpler case
with equal hybridizations, but the Kondo scale must be determined
numerically for a given
cutoff $D^*$ through evaluation of $\tilde C^*$ at the cutoff.  Similar
analyses apply for the one
and two channel limits.  The crossover analysis can be done in terms of
$\epsilon_2^*$ and $T^{(3)*}$ in the same way as discussed
in Sec. 5.2.2.

Finally,
we note that we can easily rewrite the condition $\epsilon_2^*>,<,=0$
in terms of renormalized coupling constants
$\tilde g_7,\tilde g_8$ given by
$$\tilde g_7 = {\Gamma_{17}\over \pi |E_0|},~~\tilde g_8 =
{\Gamma_{37}\over
\pi|E_0-\epsilon_2|} ~~. \leqno(5.2.29)$$
Let $\delta = {\tilde g_7}^{-1}-{\tilde g_8}^{-1}$.  Then for $\delta
>0$, we
will obtain two-channel physics, for $\delta=0$ three channel physics,
and
for $\delta <0$ we will obtain one channel physics at low
temperatures.  The
estimates of the crossover temperatures are precisely those of Eqs.
(2.2.40,41) discussed in Sec. 2.2.4.
Clearly, if $\Gamma_{17}=\Gamma_{37}$, we revert to
the sign of $\epsilon_2$
determining the physics as in Sec. 5.2.2, and to the crossover scale
formulae Eqs. (5.2.16,20).

\subsubsection{Physical Properties at Finite Temperature} 

Kim and Cox [1995,1997] have evaluated the physical properties of
this model at
finite temperatures, and we shall survey their results here.

\begin{figure}
\epsfxsize=4.5in
\epsffile{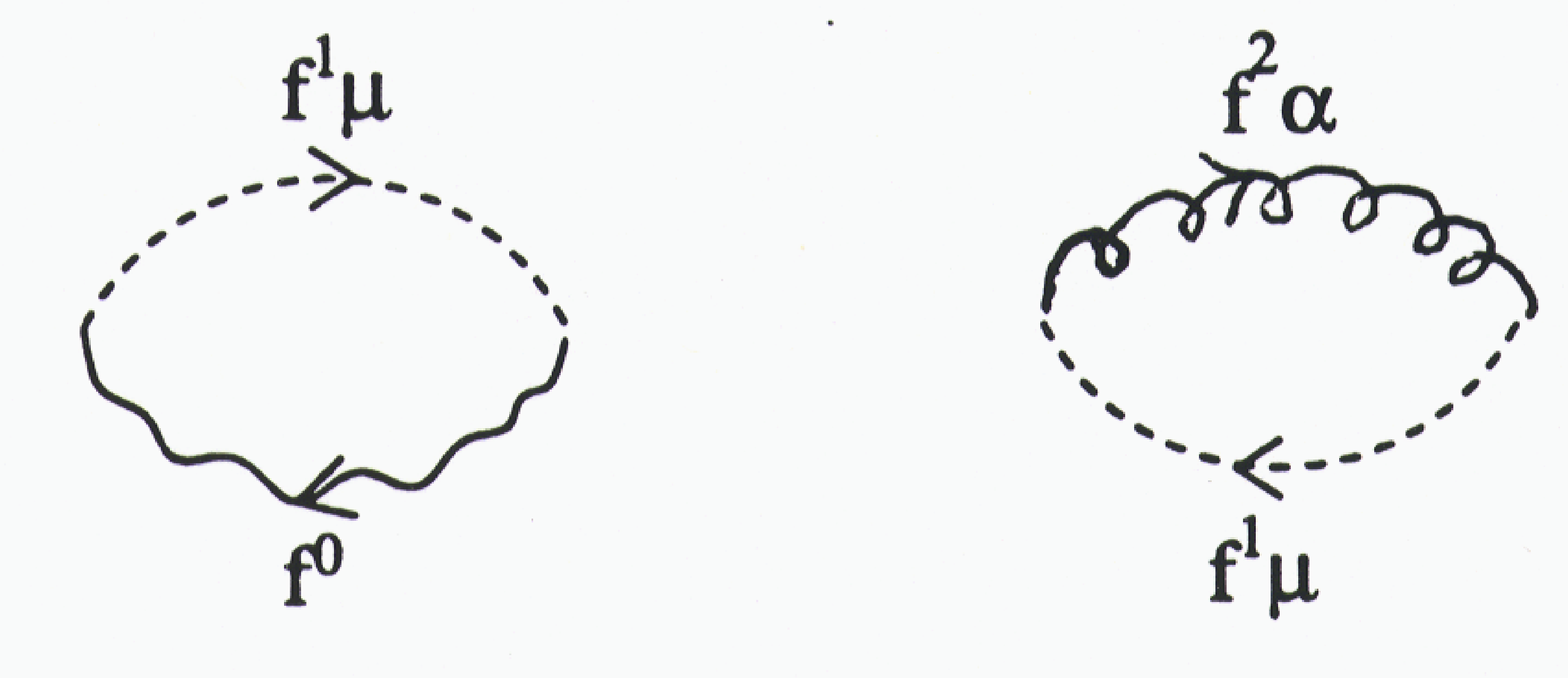}
\caption{Diagrams for full 4$f$ propagators for the simplest \ctp
Anderson model. The diagram at left is for the $\gse$ symmetry full $f$
propagator (which hybridizes with the corresponding conduction
propagator), and the diagram at right for the $\gei$ symmetry full $f$
propagator (which hybridizes with the conduction quartet propagator.}
\label{fig5p11}
\end{figure}

We shall focus on the magnetic susceptibility and the 4f spectral
function
here.  The susceptibility diagram is precisely that of Fig.~\ref{fig5p3}, with
the
$\gse$ propagators put in.  The spectral function diagram includes two
contributions show in Fig.~\ref{fig5p11}.
 One term arises from $\gse$ symmetry interconfiguration transitions
 arising from $f^0-f^1$ processes, and the other from $\gei$ symmetry
 transitions arising from $f^1-f^2$ processes.  Notice the reversed
 order of the $\gse$ propagator in the two diagrams.

\begin{figure}
\parindent=4.in
\indent{
\epsfxsize=5.in
\epsffile{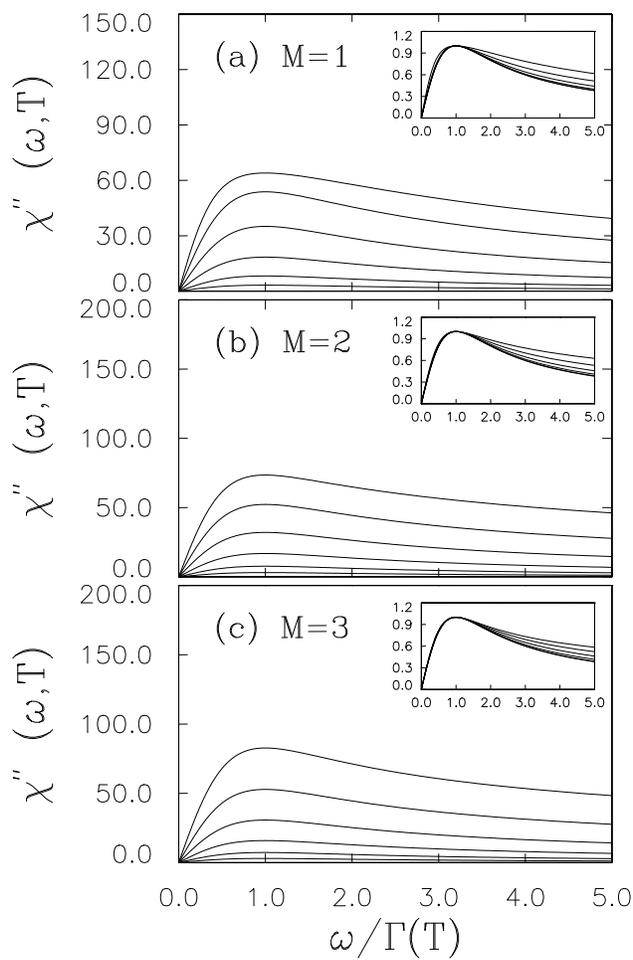}}
\parindent=.5in
\vspace{.5in}
\caption{Dynamic susceptibility of the simplest \ctp Anderson model in the 
$M=1,2,3$ channel regimes.  The insets show scaling behavior in which the curves
are divided by their maximum value and centered at their maximum position.  The
maximum position defines the linewidth $\Gamma$ of Fig.~\ref{fig5p13}.  From Kim and Cox [1997].}
\label{fig5p12}
\end{figure}

\begin{figure}
\parindent=4.in
\indent{
\epsfxsize=5.in
\epsffile{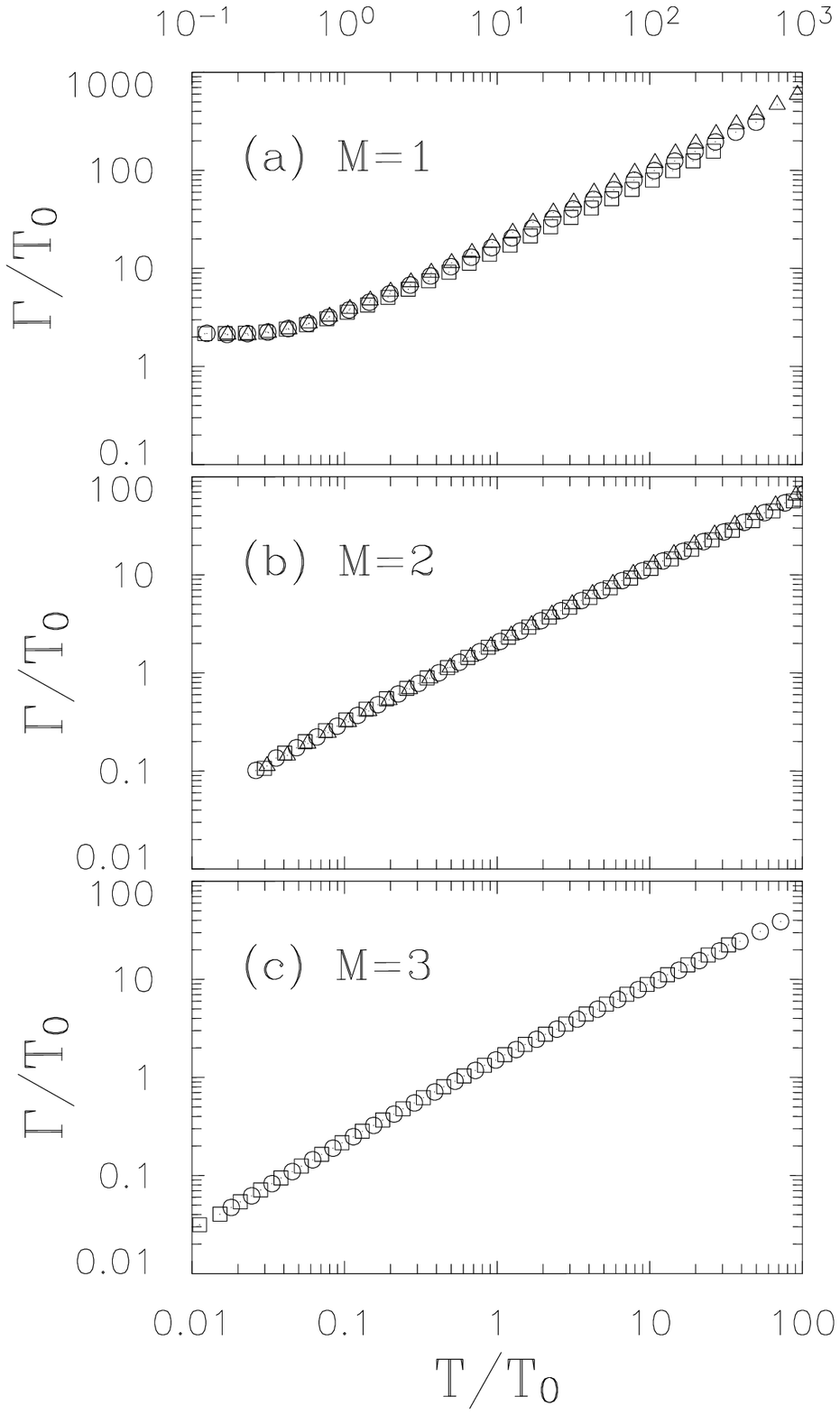}}
\parindent=.5in
\vspace{.8in}
\caption{Magnetic relaxation rate of the simplest \ctp Anderson model in the $M=1,2,3$
channel regimes.  These rates are determined from the maximum position of the dynamic susceptibility curves 
in Fig.~\ref{fig5p12}.  Reflecting the residual ground state degeneracy (``degenerate spin screening cloud''),
the linewidth vanishes linearly in temperature for $M=2,3$, whereas it is finite for $M=1$, compatible with
Fermi liquid behavior in that limit.  From Kim and Cox [1997].}
\label{fig5p13}
\end{figure}

The susceptibilities in the two- and three-channel regimes are shown in
Fig.~\ref{fig5p6} .  As discussed previously, these agree very well with the
corresponding Bethe-Ansatz curves of Sacramento and Schlottmann
[1991].
There is however a necessary vertical scale adjustment which depends
upon parameters.  The origin of this
scale factor, which is of order unity, is likely a residue of the
crossover physics.  Namely, the full one-parameter universality of the
two-channel model, for example, may not show
up until $\tilde g_8 >>\tilde g_7$.
The dynamical spin susceptility in the three different regimes is shown
in Fig.~\ref{fig5p12}.  In each case the inset shows the peak position
as a function of temperature.  This peak position
is a rough measure of the
spin relaxation rate, which is then illustrated in Fig.~\ref{fig5p13}.  For the one
-channel case, this saturates to a constant value of order $T_0$ at low
$T$,
compatible with the Fermi liquid ground state (although $\chi'' $ is of
course
singular at lower frequency).  For the two- and three-channel regimes,
the
effective relaxation rate vanishes as $T\to 0$, doing so linearly in
the two-channel case.  This is
of course compatible with the marginal Fermi liquid hypothesis which
postulates
that $T$ sets the low temperature energy scale (Varma {\it et al.},
[1989]).

The specific heat and entropy curves have already been discussed
in Figs.~\ref{fig5p7},\ref{fig5p8} and section 5.1.6.  Of importance in
understanding the curves is the
presence of the interconfiguration  peak which gives a large
background to the specific heat and entropy.

\begin{figure}
\parindent=4.in
\indent{
\epsfxsize=6.in
\epsffile{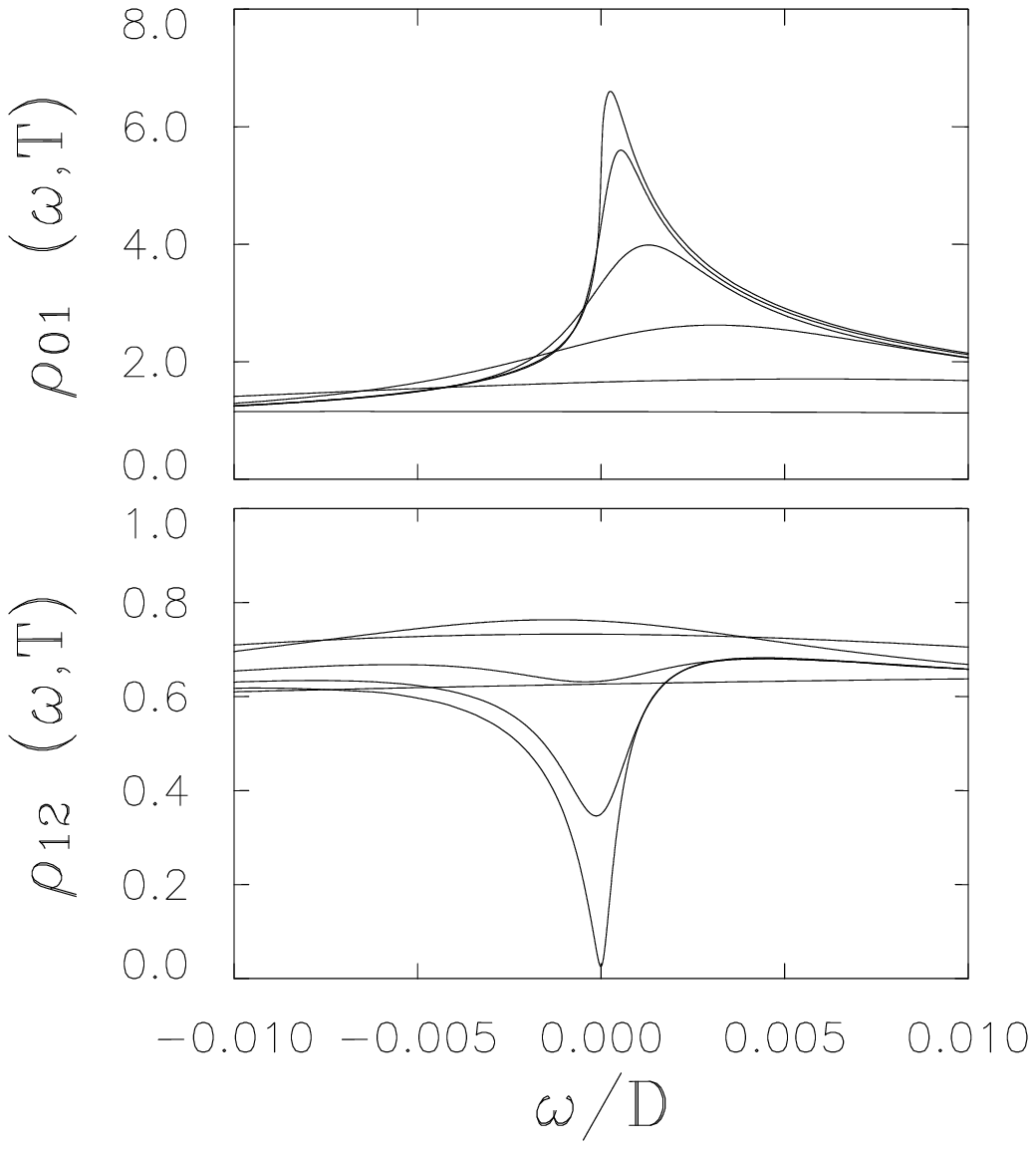}}
\parindent=.5in
\caption{Atomic spectral functions in the one-channel regime of the simplest
\ctp Anderson model.
$\rho_{01}$ is the interconfiguration spectral function which is
obtained from the convolution between $f^0$ and $f^1 \Gamma_7$ states.
$\rho_{12}$ is the interconfiguration spectral function which is
obtained from the convolution between $f^1 \Gamma_7$ and
$f^2 \Gamma_3$ states. The one-channel Kondo effect leads to the
Kondo resonance development in $\rho_{01}$ just above the
Fermi level and the spectral depletion in $\rho_{12}$ right at
$\omega =0$. Spectral functions are displayed for model set 8.
The temperature variations are $T/D=3.678\times 10^{-2}$,
$1.077\times 10^{-2}$,
$3.155\times 10^{-3}$, $9.239\times 10^{-4}$,
$2.706\times 10^{-4}$, $7.924\times 10^{-5}$,
$2.321\times 10^{-5}$.
From Kim and Cox [1997].}
\label{fig5p14}
\end{figure}

\begin{figure}
\parindent=4.in
\indent{
\epsfxsize=6.in
\epsffile{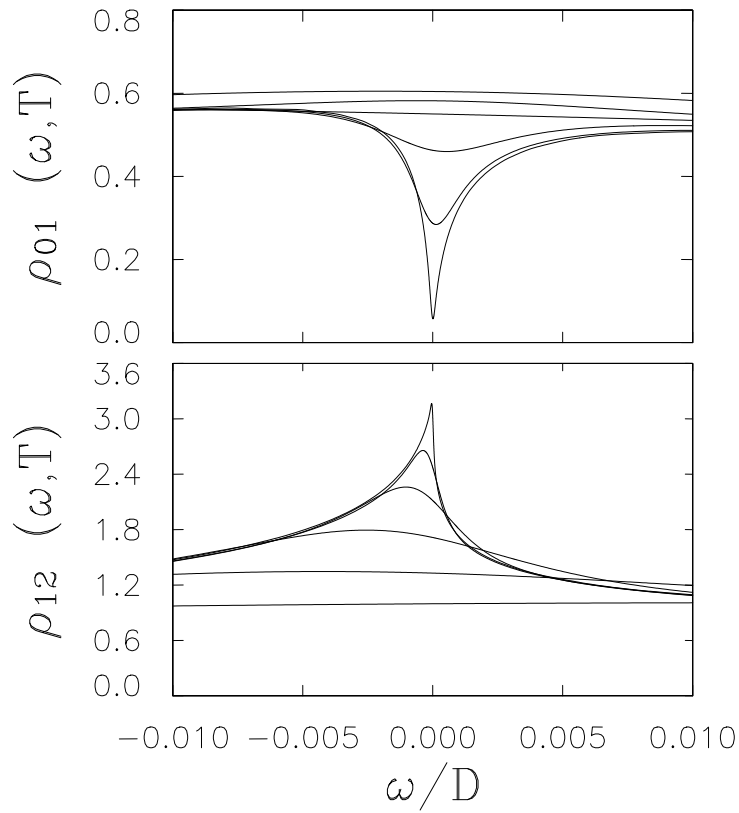}}
\parindent=.5in
\caption{Atomic spectral functions in the two-channel regime for the
simplest \ctp ion Anderson model. The two-channel Kondo effect leads to the
Kondo resonance development in $\rho_{12}$ at the
Fermi level ($T=0$) and the spectral depletion in $\rho_{01}$ right at
$\omega =0$. Spectral functions are displayed for model set 1.
The temperature variations are the same as in Fig.~\ref{fig5p14}.
From Kim and Cox [1997].}
\label{fig5p15}
\end{figure}

\begin{figure}
\parindent=4.in
\indent{
\epsfxsize=6.in
\epsffile{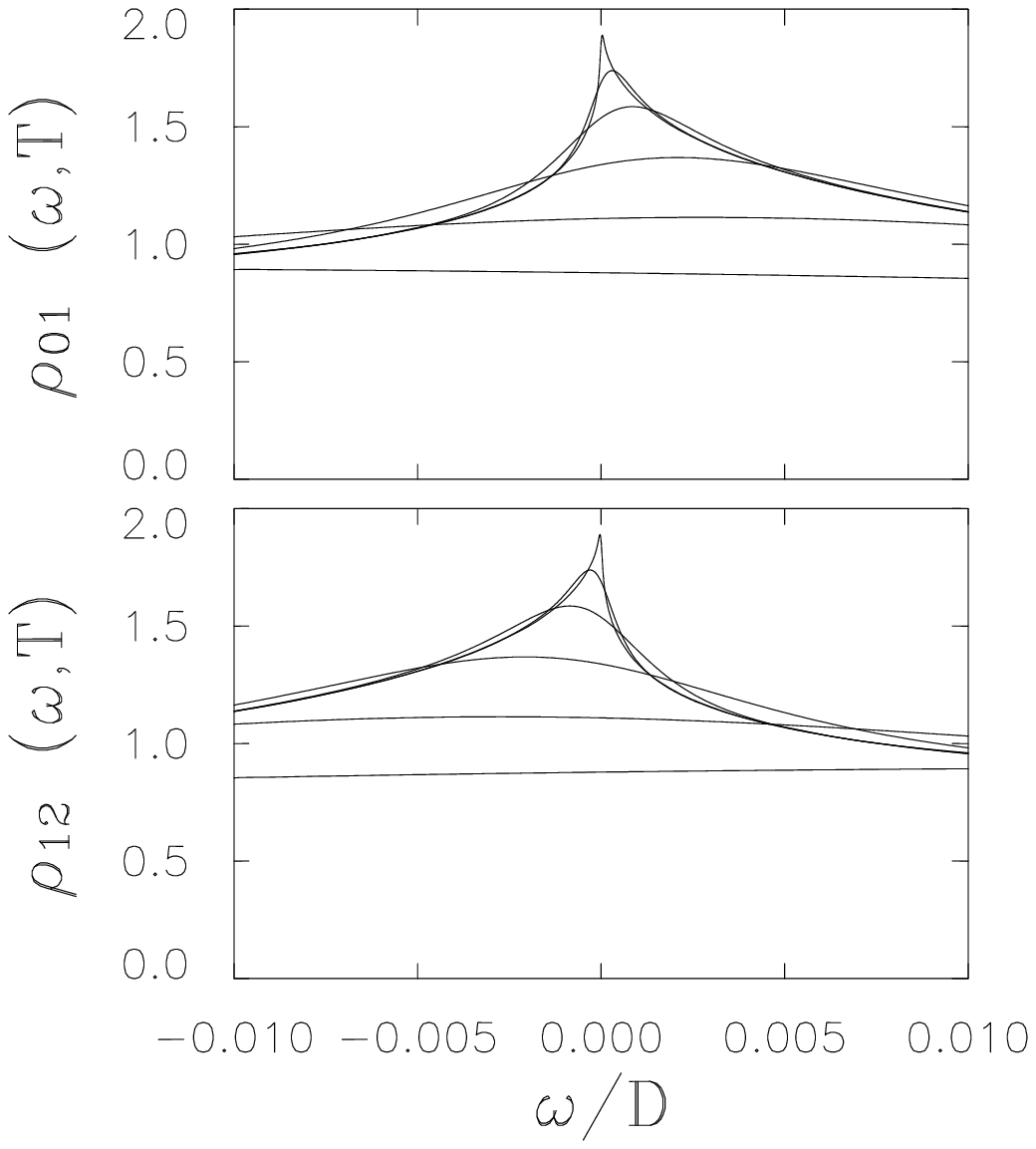}}
\parindent=.5in
\caption{Atomic spectral functions in the three-channel regime of the simplest 
\ctp ion Anderson model. For this parameter regime (model set 4), the
two spectral functions $\rho_{01}$ and $\rho_{12}$ are equivalent
in the asymptotic limit after a particle-hole transformation.
The temperature variations are the same as in Fig.~\ref{fig5p14}.
From Kim and Cox [1997].}
\label{fig5p16}
\end{figure}

The most interesting new physics to emerge from this treatment concerns
the
transport coefficients, specifically the thermoelectric power. The
temperature
dependent spectral functions in the 1,2, and 3-channel regimes are
shown in
Figs.~\ref{fig5p14},\ref{fig5p15},\ref{fig5p16}.  For the 1 and 2-channel cases, the inset
shows the separate contributions
of the $f^0-f^1$ and $f^1-f^2$ diagrams at low temperatures.  We notice
that
the contribution from either $f^1-f^2$ transitions in the single
channel case
or $f^0-f^1$ transitions in the two-channel case vanishes at zero
frequency.
This corresponds to the
irrelevance of the smaller coupling constant in the renormalization
group sense;
the width of the region over which each spectral function vanishes is a
measure
 of the corresponding crossover scale defined in the previous
 subsection.

What is quite clear from these figures is that in the one-channel
regime is that
the Kondo resonance weight is shifted predominantly to positive
frequencies.
For the two-channel regime, the Kondo resonance weight is shifted
predominantly to
negative frequencies.  The physical origin is clear--in the one channel
case,
virtual charge fluctuations to the $f^0$ configuration dominate and so
the $f$-occupancy is less than 1, meaning
we should shift spectral weight above the Fermi energy relative to the
$n_f=1$
case.  For the two-channel regime, virtual charge fluctuations to $f^2$
dominate, and hence we expect the $f$-occupancy to exceed one.
Corresponding
to this, we should shift spectral weight below the Fermi energy.

\begin{figure}
\parindent=2.in
\indent{
\psfig{file=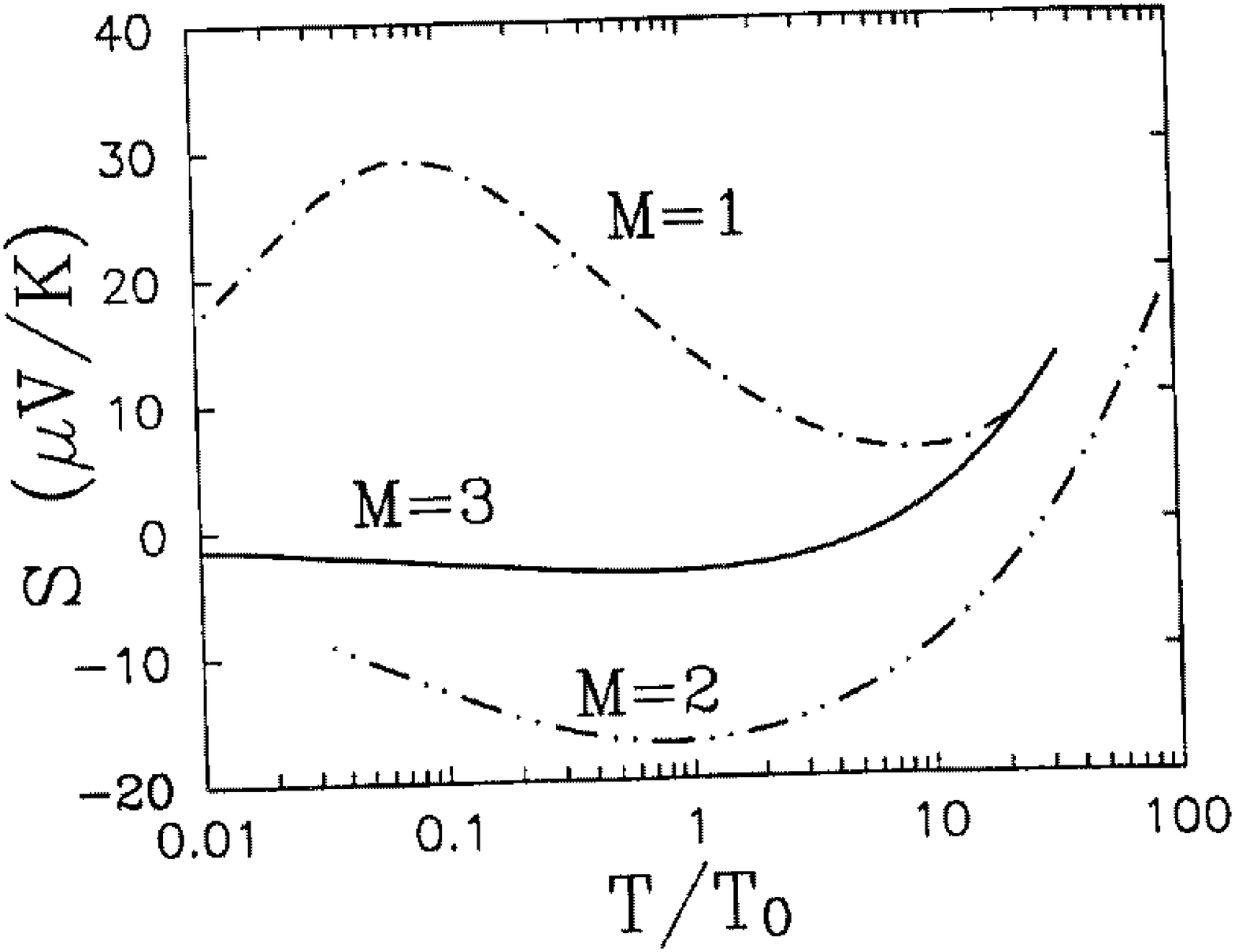,height=4.in,angle=270}}
\parindent=.5in
\caption{Thermopower $S(T)$ for the simplest \ctp ion Anderson model. 
The thermopower is
positive at low $T$ for the $M=1$ parameter regime,
strongly
negative for the $M=2$ parameter regime, and weakly negative for the $M=3$ parameter
regime.
Dominant $f^0-f^1$($f^1-f^2$)  virtual
charge fluctuations give positive(negative) $S(T)$.
From Kim and Cox [1995].}
\label{fig5p17}
\end{figure}

Since the scattering rate $1/\tau$ is proportional to to the full
spectral
function (modulo corrections from anisotropic hybridization matrix
elements
--see Kim and Cox [1995,1997] for a discussion), the discussion of the
preceding
paragraph has a direct bearing on the thermoelectric power which is
proportional to the transport integral
$I_1 = \int d\epsilon (-\partial f/\partial \epsilon) \tau(\epsilon)$.
The one-channel regime scattering rate will lead to a stronger
scattering of
unoccupied (particle) states
meaning that occupied(hole) states will dominate the heat transport in
$I_1$
and hence we expect a
positive sign to the thermopower in this regime.  For the two-channel
case,
holes are scattered more strongly overall than particles, so we
anticipate
more effective heat transport by particles and a negative sign to the
thermopower.  In the three channel
 case, the scattering rate is approximately symmetric but slightly
 dominated
 by hole scattering since the excited $f^2$ state is a doublet.  As a
 result
 we expect the thermopower to be slightly negative at low
 temperatures.
The full numerical calculations bear out the physical discussion of the
previous paragraph, as shown in Fig.~\ref{fig5p17}.  This implies that the
thermopower
is a sensitive probe of the possible {\it universality class} for \ctp
impurities!  As we have noted previously, CeCu$_{2.2}$Si$_2$ has a
negative thermopower below 70K, well above
any possible lattice coherence effects.  This is strong support for a
model
in which the $f^2$ fluctuation weight in the ground state exceeds the
$f^0$
weight.  However, as discussed in Sec. 8.2, for dilute Ce in LaCu$_2$Si$_2$, 
the thermopower regains a positive sign, which is problematic for an interpretation 
in terms of the two-channel Kondo model. 

\subsubsection{Critique and Directions} 

The simple model analyzed in this subsection may be criticized in its
treatment
of the $f^2$ configuration.  Specifically, it is known for \ctp ions
that
$\epsilon_2 \approx +1-2$eV. On the other hand, $\Gamma_{17}$ is known
to be
$\simeq 0.1-0.3 eV$, and due to fractional parentage effects, we
anticipate that for the lowest multiplet in
$f^2$ that $\Gamma_{37} < \Gamma_{17}$.   Thus, naively, unless
it would seem that the two- and three-channel possibilities
can never be realized in practice.  Also, excited state $f^2
\Gamma_4,\Gamma_5$
levels give rise to the additional $\tilde J$ coupling discussed in
Sec. 2.2.2.

The NCA analysis can in principle be extended to include the lowest
crystal
field excitations of the
excited configuration, but this will require inclusion of vertex
corrections
of the sort discussed by Pruschke [Pruschke, 1989]
to reliably describe the low
temperature
physics.  The effect here is not yet known, but given the scaling
analysis
of Sec. 3.4, we anticipate that the low energy scale physics will still
be
determined by the $\Gamma_3$ virtual charge fluctuations provided the
coupling
is sufficiently weak
to the $\Gamma_4,\Gamma_5$ levels.

The first critique is more serious, but a possible answer is as
follows: all
the excited state $\Gamma_3$ levels will produce a contribution to the
two-channel coupling.  While the hybridization
to any given $\Gamma_3$ is too weak to give an effective coupling
exceeding
the one-channel coupling driven by the $f^0$ charge fluctuations, we
anticipate
that the sum over the manifold of
$\gth$ states may well produce a significantly larger coupling.  Since
this
appears already at second order in the hybridization, we expect the NCA
extended to the entire configuration to produce an answer to the
question.
The additional couplings will likely be brought in through
admixture of all the $d_3$ propagators, which occurs at second order in
the
hybridization.

This argument can be made precise by considering a model with two
excited
state $\gth$ doublets and examining the first iteration of the NCA,
which
is sufficient to determine the Kondo temperature.
Denote the corresponding hybridization matrix elements by
$V^{(i)}_{37}$
and energies as $\tef^{(i)}$.
Define the quantity $\tilde\pi(\omega) = \ln|(\omega-\ef)/D|/\pi$.
Because
the excited $\gth$ doublets can mix as shown in Fig. 5.12, the $\gth$
green's
function is now a $2 \times 2$
matrix.  The secular equation determining $T_0$
is found
by requiring that the
denominator of the inverse Green's function vanishes, which gives
$$[\omega - \tef^{(1)} - \Gamma^{(1)}_{37}\tilde \pi(\omega)][\omega -
\tef^{(2)} - \Gamma^{(2)}_{37}\tilde \pi(\omega)] - \Gamma^{(1)}_{37}
\Gamma^{(2)}_{37}\tilde\pi^2(\omega) ~~.\leqno(5.2.29)$$
This function $\tilde\pi$ becomes large and negative near $\omega=\ef$,
and to leading exponential accuracy we may thus rewrite this equation
as
(putting $\omega-\ef = -T_0$)
$$1 = [{\Gamma^{(1)}_{37}\over \pi(\ef-\tef^{(1)})} +
{\Gamma^{(2)}_{37}\over
\pi(\ef-\tef^{(2)})}]\ln({T_0\over D}) \leqno(5.2.30)$$
so that the effective couplings driven by the different $\gth$ levels
are
clearly summed.  This analysis may be readily extended to an arbitrary
number
of excited $\gth$ levels.  Hence, the model with just
one
$\gth$ level captures the essential physics; the lone $\gth$
should just be taken as an effective
representative of all the excited $\gth$
levels.

\subsection{Application of the NCA to a Model \ufp Impurity} 

In this section we discuss results obtained from applying the NCA to a
model
\ufp impurity
that have been reported in Cox[1987b], Cox [1988(a)], Cox and Makivik
[1994],
and Kim {\it et al.} [1997].  The most important physical conclusions
are:
(i) that even for arbitrarily small crystal field splitting in the
ground
$J=4$ multiplet, when the
$\gth$ level lies lowest the low temperature physics will be that of
the
two-channel Kondo model below a suitable crossover temperature.  Hence,
even when the crystal field levels overlap substantially, as appears to
be the case for UBe$_{13}$ to the extent that this model
applies for this material, one may be confident of
two-channel Kondo physics arising at low temperatures.
(ii) the singular character of the ground non-magnetic doublet may
reflect in the magnetic susceptibility producing a $\sqrt{T}$ law.
This result appears to be highly parameter dependent and most likely
occurs in the regime where the crystal field levels
 overlap substantially.

\subsubsection{Pseudo-particle model and NCA equations} 

We shall consider only a simple model in which we retain the lowest
$J=4\gth$ doublet of
the ground configuration together with the first excited crystal field
level (assumed to be
a $\gfo$ triplet) and the lowest $J=5/2\gse$ doublet of the excited
$f^1$ configuration.
We will assume only $J=5/2\gei$ conduction partial wave states.
Additional crystal field states
and conduction partial waves may easily be included with altering the
essential conclusions.
 At the
end of the section we shall briefly remark about the effects of
including an excited

\begin{figure}
\parindent=2.in
\indent{
\epsfxsize=3.in
\epsffile{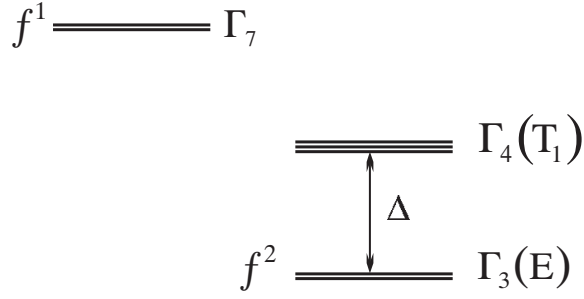}}
\parindent=.5in
\caption{Level scheme for the simplest \ufp ion Anderson model studied within the NCA.  }
\label{fig5p18}
\end{figure}

In this case, we simply introduce a pseudo-fermions for the $\gth,\gfo$
states of
the $f^2$ configuration, and a pseudo-boson doublet for the $\gse$ of
the excited $f^1$
configuration.  (Alternatively, we could assume a $\gsi$ doublet in an
excited $f^3$
configuration since $\gsi\otimes\gth=\gse\otimes\gth=\gei$.)
The level scheme is shown in Fig.~\ref{fig5p18}.  These are
then put into
the Hamiltonian in equations (2.2.30) and (2.2.32) in precisely the
same manner
we did for the Ce case.  The reader should not be disturbed by the use
of a pseudo-fermion operator for an even numbered configuration and a
pseudo
-boson operator for an odd numbered configuration since the statistics
are
lost once the full projection is made.   The $f$ charge is written as
$$Q_f = \sum_{\Gamma(f^2)\eta(f^2)} f^{\dagger}_{\Gamma(f^2)\eta(f^2)}
f_{\Gamma(f^2)\eta(f^2)}
+ \sum_{\mu} b^{\dagger}_{\gse\mu}b_{\gse\mu} \leqno(5.3.1)$$
and the full pseudparticle Hamiltonian is
$$H = \sum_{k\eta_8} \epsilon_k c^{\dagger}_{k8\eta_8} c_{k8\eta_8} +
\sum_{\Gamma(f^2)\eta(f^2)}
(\tef +
\Delta_{\Gamma(f^2)})f^{\dagger}_{\Gamma(f^2)\eta(f^2)}f_{\gamma(f^2)\eta(f^2)}
\leqno(5.3.2)$$
$$~~~~~+ H_{hyb} -\lambda_{ps}(Q_f-1) $$
with
$$H_{hyb} = {V\over N_s} \sum_{k,\eta_8,\mu}\sum_{\Gamma(f^2)\eta(f^2)}
\Lambda(\Gamma_7\mu;\Gamma(f^2)\eta(f^2);\gei\eta_8)
[f^{\dagger}_{\Gamma(f^2)\eta(f^2)}
b_{\gse\mu} c_{k8\eta_8} + h.c.] ~~.\leqno(5.3.3)$$
Note that $\eta_8$ is shorthand for the spin/channel notation of Eqs.
(2.2.18.a,b).
We have taken $E(f^2)=\tef$, and shall denote henceforth
$\Delta_{\gth}=0,\Delta_{\gfo}=\Delta$.

\begin{figure}
\parindent=2.in
\indent{
\epsfxsize=3.in
\epsffile{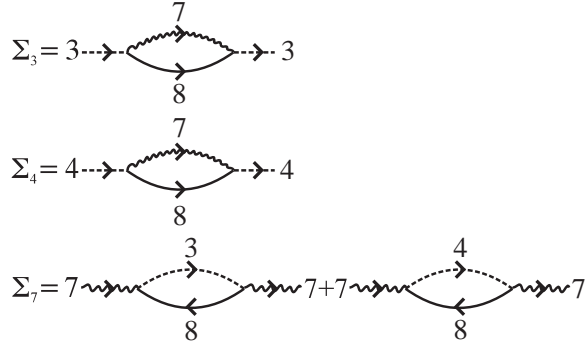}}
\parindent=.5in
\caption{NCA pseudo-particle self energy diagrams for the simplest \ufp ion Anderson model.
}
\label{fig5p19}
\end{figure}

The NCA self-energy diagrams for this model are shown in Fig.~\ref{fig5p19}.
Prior to writing down the
integral equations, it is useful to define the group theoretic coupling
strengths
$$c(\Gamma(f^2),\Gamma(f^1)) ={1\over \nu(\Gamma(f^1))}
\sum_{\Gamma_c\eta_c\eta(f^1))}
\Lambda^2(\Gamma(f^1)\eta(f^1);\Gamma(f^2)\eta(f^1);\Gamma_c\eta_c)
\leqno(5.3.4.a)$$
and
$$c(\Gamma(f^1),\Gamma(f^2))=c(\Gamma(f^2),\Gamma(f^1))= {1\over
\nu(\Gamma(f^2))} \sum_{\Gamma_c\eta_c\eta(f^2))}
\Lambda^2(\Gamma(f^1)\eta(f^1);\Gamma(f^2)\eta(f^1);\Gamma_c\eta_c)
\leqno(5.3.4.b)$$
where $\nu(\Gamma)$ is the degeneracy of irrep $\Gamma$.  For a given
coupling model ($LS$,
$jj$, or intermediate) and crystal field Hamiltonian,
the $c$ coefficients may be determined once and for all.  Depending on
the crystal field,
the coefficients may also have to be recomputed in an applied magnetic
field.  This definition
allows for a ready generalization of the NCA to arbitrary crystal field
schemes.  The
definition is unmodified by the inclusion of all conduction partial
waves.   The hybridization
factors appearing in the integral equations for the state indexed
by irrep $\Gamma_e$ and arising from a diagram with internal impurity
irrep $\Gamma_i$
will always just be of the form
$\nu(\Gamma_i)c(\Gamma_e,\Gamma_i)\Gamma$.
The coupling coefficients have an implicit dependence on the $f^2$
multiplet. They obey a simple
sum sule
$$\sum_{\Gamma(f^1)} c(\Gamma(f^2),\Gamma(f^1)) = 2 $$
independent of $\Gamma(f^2)$,
where the sum includes the crystal field states of the $j=7/2$
multiplet as well. This measures
the total number of ways to make a transition from $f^2\to f^1$, which
is just two, since we may
remove one or the other of the electrons.  If we are instead
considering a model with the $f^3$
excited configuration, the RHS of the above sum rule is changed to 12
provided we sum over {\it all}
states of the excited $f^3$ configuration.

The integral equations of the NCA for this model are then
$$\sigma_3(\omega) = {2c(\gth,\gse) \Gamma \over \pi} \int d\epsilon
f(\epsilon)
{\cal D}_7(\omega+\epsilon) \leqno(5.3.5.a)$$
$$\sigma_4(\omega) = {2c(\gfo,\gse) \Gamma \over\pi} \int d\epsilon
f(\epsilon)
{\cal D}_7(\omega+\epsilon) \leqno(5.3.5.b)$$
and
$$\pi_7(\omega) = {2c(\gth,\gse) \Gamma \over \pi} \int d\epsilon
f(\epsilon)
{\cal G}_3(\omega+\epsilon) + {3\Gamma c(\gfo,\gse)\over \pi} \int
d\epsilon
f(\epsilon) {\cal G}_4(\omega+\epsilon) ~~.\leqno(5.3.5.c)$$
We have again assumed particle-hole symmetry to the conduction density
of states.

Following the procedures of the previous two sections, we may convert
these integral
equations to differential form at $T=0$
with the usual definition of inverse propagators, which gives
$${dg_3 \over d\omega} = -1 - {2c(\gth,\gse)\Gamma \over \pi d_7}
\leqno(5.3.6.a)$$
$${dg_4 \over d\omega} =  -1 - {2c(\gfo,\gse)\Gamma \over \pi d_7}
\leqno(5.3.6.b)$$
and
$${dd_7\over d\omega} = -1 - {2c(\gth,\gse)\Gamma \over \pi g_3}
- {3c(\gfo,\gse)\Gamma \over \pi g_4} \leqno(5.3.6.c)$$
subject to the boundary conditions
$$g_3(-D) = D+\tef,~~g_4(-D)=D+\tef+\Delta,~~d_7(-D)=D
~~.\leqno(5.3.7)$$

\subsubsection{Solution of the NCA equations for the Model \ufp
Impurity} 

Clearly Eqs. (5.3.6a-c) have a very similar structure to Eqs.
(5.2.7a-c).  As in the case of
those equations for the model \ctp impurity, we cannot solve Eqs.
(5.3.6a-c) analytically without
a further simplifying assumption.  For convenience (and not reality) we
shall take
$c(\gth,\gse)=c(\gfo,\gse)$.  We shall then denote the product of the
coupling
coefficient and $\Gamma$ by $\tilde \Gamma = \Gamma c(\gth,\gse)$.
No qualitative differences arise in this modified model
relative to the original one.
Typically the coupling coefficients are of order unity; for the $LS$
coupling scheme and all
conduction partial waves included, $c(\gth,\gse)=0.64$.  This reflects
fractional parentage
coefficients.  Notice that the effective reduction of the hybridization
is important for
understanding why we can get reasonably small Kondo scales in U
materials when compared
with Ce based materials despite
the expectation of smaller interconfiguration energy differences and
larger hybridization
due to the greater spatial extent of the 5f wave-function.

With this simplifying assumption, we see that $g_4=g_3 + C$, with the
constant $C=\Delta$.
This allows us to find the second integration constant $\tilde C$ given
by
$$\tilde C = \exp[{\pi\over \tilde\Gamma}(g_3 - d_7)]({g_3\over
D})^2({g_3+\Delta\over D})^{3/2}
({D\over d_7})^2 ~~.\leqno(5.3.8)$$
By evaluation at $-D$, as usual, we find $\tilde
C=\exp(\pi\tef/2\tilde\Gamma)$.
In the low frequency limit for $g_3<<\Delta$, we can rewrite this as
$$1 \approx {[g_3 /T_0]^2\over [\pi d_7/\tilde\Gamma]^2}
\leqno(5.3.9)$$
with
$$T_0 = D({\tilde\Gamma\over \pi D})({D\over \Delta})^{3/2}
\exp({\pi\tef\over 2\tilde\Gamma})
~~.\leqno(5.3.10)$$
Notice the crystal field enhancement factor appears just as discussed
in Sec. 3.3.3.c.
Also, the hybridization multiplied by the coupling coefficient,
$\tilde\Gamma$, appears, allowing
in principle for a smaller Kondo scale despite increased $\Gamma$
values as discussed
previously.

In this low temperature region, it is clear that the physics is that of
the $N=M=2$ Kondo model.
As a result, we expect the {\it quadrupolar} susceptibility and
specific heat coefficient to diverge
logarithmically, and the electrical resistivity to saturate with a
$\sqrt{T}$ law as the temperature
is lowered.

As the temperature or frequency
is raised, a crossover can occur to the Kondo model corresponding to
degenerate $\gth,\gfo$ multiplets.  The relevant comparison here is of
$g_3\sim
T_0 (2|\omega-E_0|/T_0)^{1/2}$ to $\Delta$.
If $\Delta >> T_0$, the crossover to the ground two-channel Kondo
effect
simply occurs at $T_0$.  If on the other hand $\Delta \le T_0$, then
the crossover
occurs when $g_3\approx \Delta$.  This occurs at an energy scale
$T^x_{CEF}$
given by
$$T^x_{CEF} \approx {1\over 2} T_0 ({\Delta \over T_0})^2
~~.\leqno(5.3.11)$$
What is important to notice here is that even for arbitrarily small
crystal field splitting,
the low temperature physics will still be that of the two-channel
quadrupolar Kondo effect.
This is important, since in UBe$_{13}$, for example, for which the
quadrupolar Kondo effect
was first proposed as a possible explanation for the heavy fermion
behavior, the excited crystal
field levels appear to be very broad and overlap strongly with the
ground state.  Nonetheless,
a quadrupolar Kondo effect can still occur below a crossover
temperature as defined above.

We briefly comment on two other issues before discussing physical
properties:\\
{\it (1) Inclusion of Excited Crystal Field States in the $f^1$
configuration}.  Here another kind
of crossover can occur.  Let us assume that the $\Gamma_4$ level is
taken to $\infty$, for
simplicity, and that a $\gei$ quartet is included at energy $\Delta'$
in the excited configuration.
Maintaining the equal coupling coefficient limit, the relation between
$d_7$ and $g_3$ at low
energies is still given by Eq. (5.3.9), but with $T_0$ now modified to
$$T_0 = D({\tilde\Gamma\over \pi D} )({\Delta'/D})^2\exp({\pi\tef\over
2\tilde\Gamma})
 ~~.\leqno(5.3.12)$$
Notice that excited state crystal field splittings {\it reduce} $T_0$.
Arriving at the above
equation required assuming $d_7\approx
(\tilde\Gamma/\pi)(2|\omega-E_0|/T_0)^{1/2} << \Delta'$.
 If $\Delta'>>\tilde\Gamma$, this is always satisfied and the crossover
 to the low temperature
two-channel fixed point occurs at $T_0$.  If $\Delta'\le\tilde\Gamma$,
then the crossover scale
$T^x_{CEF'}$ below which the two-channel physics sets in is given by
$$T^x_{CEF'} \approx {T_0\over 2} ({\pi \Delta'\over \tilde\Gamma})^2
~~.\leqno(5.3.13)$$
Clearly, the crossover physics of crystal field states in ground and
excited configurations
(together with the similar physics arising from
all the higher lying angular momentum multiplets in  ground and excited
configurations) must
be included in a complete treatment of the \ufp impurity and will
produce a single crossover scale
determined by the lowest energy scale determined from all the various
crossover criteria.
The Kondo scale will also require modification to respect all the
excited levels in both the
ground and excited configurations.  Such a complete treatment has not
yet  been performed. Once
any more excited crystal field levels are included in either
configuration we must
resort to full numerical solutions of the NCA equations.   \\
{\it (2) Effect of Non-equal Coupling Coefficients}.  As in the
corresponding example for the \ctp
impurity, we can no longer solve the model analytically if we relax the
assumption of equal
coupling coefficients.  We can once more find approximate constants of
integration at low
frequencies since $dg_4/dg_3 \approx c(\gfo,\gse)/c(\gth,\gse)$ so that
$$g_4 \approx {c(\gfo,\gse)\over c(\gth,\gse)} g_3 +
\tilde\Delta,~~\tilde\Delta = \Delta
+ (\tef-E_0)({c(\gfo,\gse)\over c(\gth,\gse)}-1) \leqno(5.3.14)$$
which follows from requiring $g_3(E_0)=0$.  Depending on the ratio of
coupling coefficients,
the crystal field level may be renormalized upwards or downwards.  When
all states are included,
we generically find a downward renormalization as expected on the basis
of the orthogonality
catastrophe (c.f. the discussion in Sec. 3.4.1.d).  Then a the Kondo
scale may be written in
terms of a cutoff $D^*$ over which the approximate integration constant
relation holds (to be
determined numerically) in a manner similar to that discussed in
deriving Eq. (5.2.28).  The
crystal field crossover scale is changed to
$$T^{x*}_{CEF} = {T_0 \over 2} ({c(\gth,\gse)\tilde\Delta\over
c(\gfo,\gse)T_0})^2 ~~.\leqno(5.3.15)$$
No qualitative modifications of the physics will emerge in the
subsequent analysis.

\subsubsection{Physical Properties} 

In this subsection we will discuss those physical properties which are
novel to the
quadrupolar Kondo effect and for which the NCA gives useful
information.  An extensive
review of calculations on model \ufp impurities appears in Kim {\it et
al.} [1995].

\begin{figure}
\parindent=2.in
\indent{
\epsfxsize=6.in
\epsffile{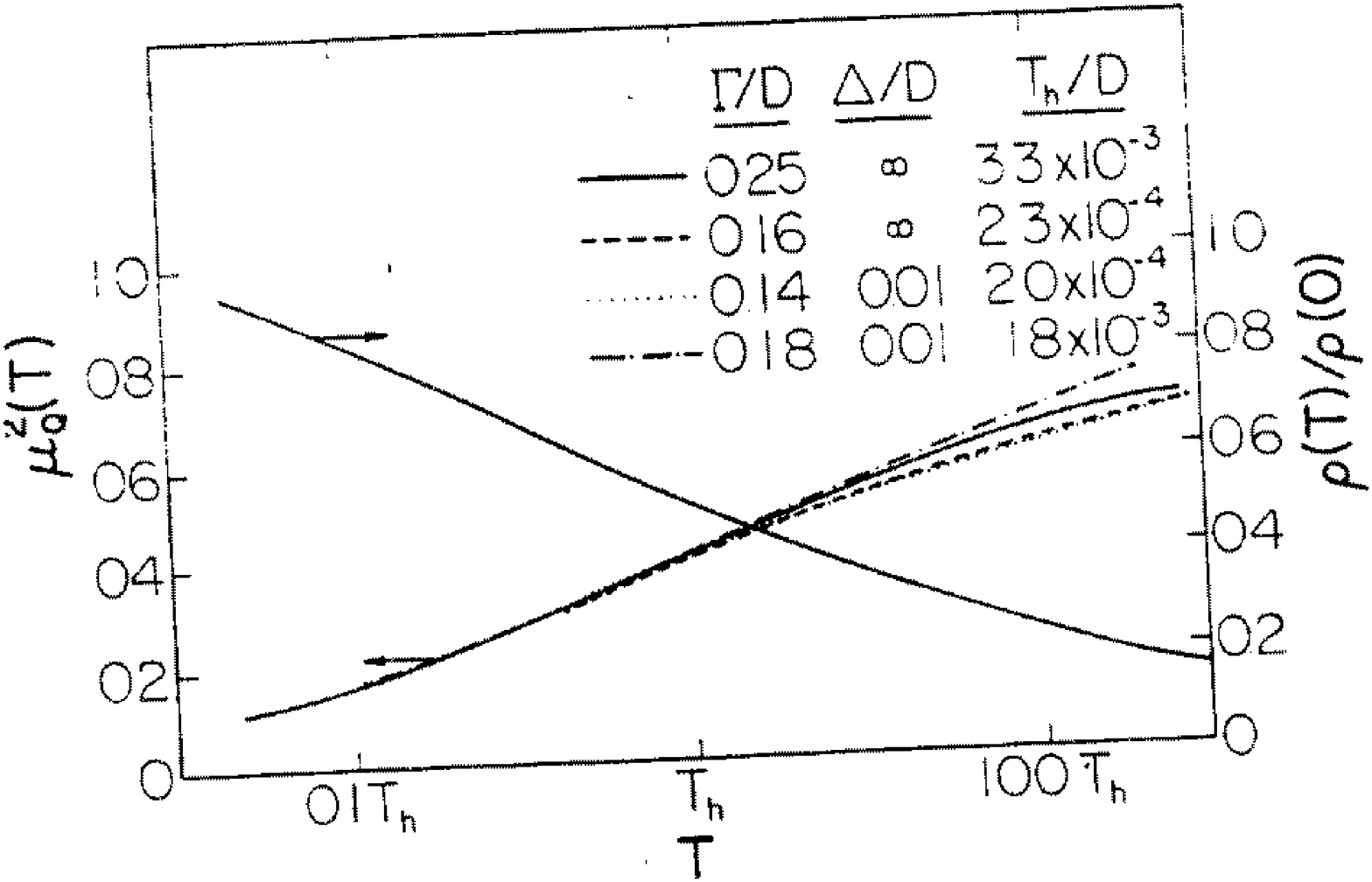}}
\parindent=.5in
\caption{Static quadrupolar susceptibility vs. temperature for a \ufp
ion model calculated by the NCA.  Also shown (right hand axis) is the
resistivity scaled by the appropriate zero temperature value obtained
from an extrapolation.  The temperature $T_h$ is measured from the 
resistivity midpoint and is of order the Kondo temperature. 
From Cox [1987(b)].}
\label{fig5p20}
\end{figure}

\begin{figure}
\parindent=2.in
\indent{
\epsfxsize=6.in
\epsffile{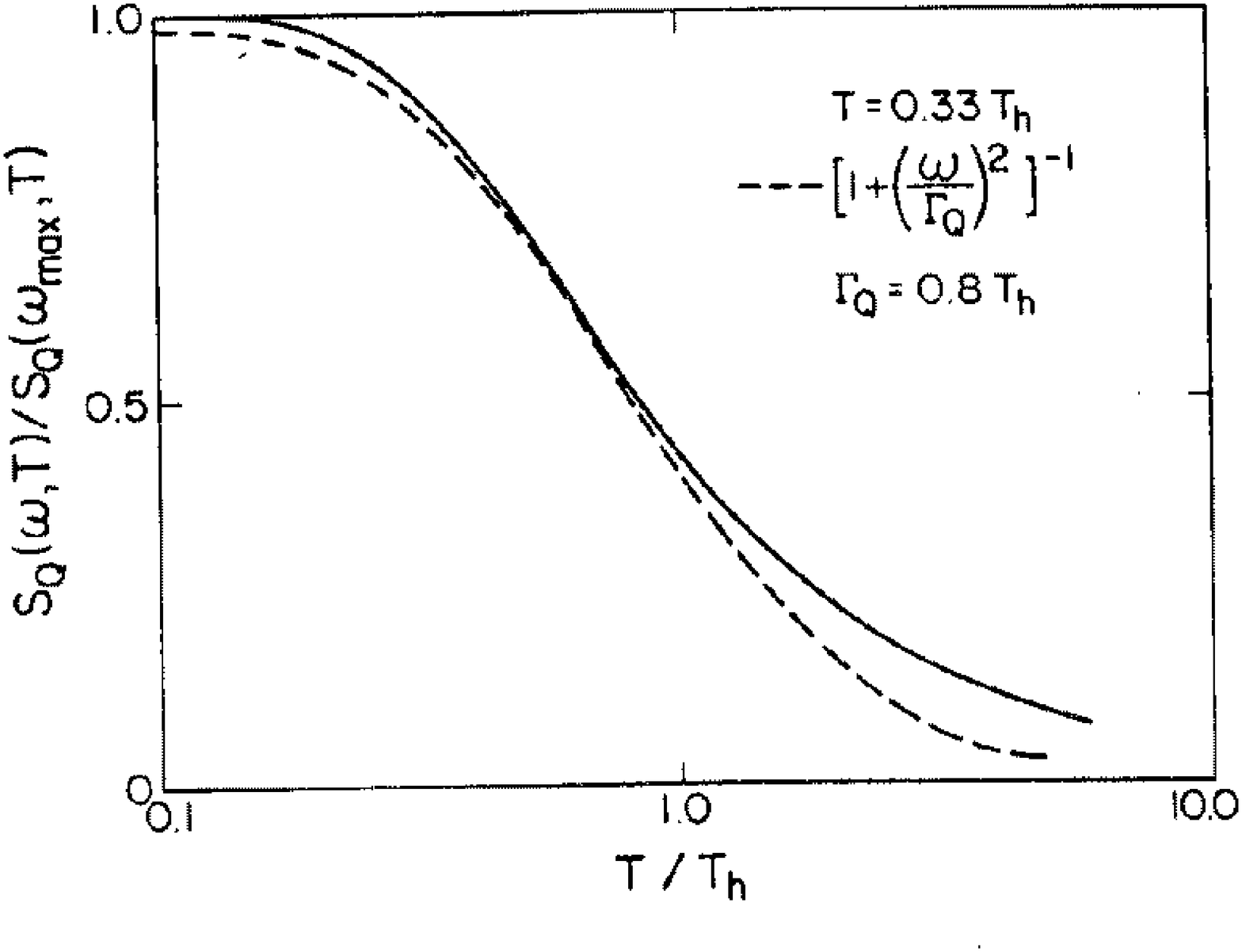}}
\parindent=.5in
\caption{Quadrupolar dynamic structure factor 
of a model \ufp ion calculated within the NCA.  For low $T$, this is 
the same as the positive frequency evaluation of the dynamic quadrupolar 
susceptibility.  A Lorentzian fit to the structure factor works
reasonably well, anticipating the Marginal Fermi Liquid theory of 
Varma {\it et al.} [1989], and the outcome of conformal field theory 
(Tsvelik [1990]; Ludwig and Affleck [1991,1994]) and
bosonization results (Emery and Kivelson [1992]).  
From Cox [1988(a)].}
\label{fig5p21}
\end{figure}

{\it Quadrupolar Susceptibility}.  Since the ground doublet is
quadrupolar here, the
strongly divergent susceptibility relevant for the Wilson ratio is the
quadrupolar susceptibility
$\chi_Q$.  This is obtained from the dashed bubble diagram of Fig.~\ref{fig5p3}
with the $\gth$
propagators put in as pseudo-fermions.
The calculated $\chi_Q$ curves from the NCA show a universality over a
wide range of parameter
sets as shown in Fig.~\ref{fig5p20} taken from Cox[1987b].  It is important
that the $\gth$ occupancy
must be divided out to produce the universality when excited crystal
field levels are included.
The data also agree well with the universal Bethe-Ansatz results of
Sacramento and Schlottmann
[1992].
The dynamic quadrupolar susceptibility shows the anticipated Marginal
Fermi liquid form with
a step function at the origin, and may
be roughly fit to the form $sgn(\omega) {T_0/(\omega^2+T_0^2)}$ as
shown by Cox [1988(a)] and
in Fig.~\ref{fig5p21}.

{\it Magnetic Susceptibility: van Vleck contribution}.  Given a
quadrupolar doublet ground state,
the dominant source of magnetic response must be of the type considered
by van Vleck, namely,
due to virtual transitions to excited crystal field levels. Considering
only the excited $\gfo$
state for the moment, in the
ionic (zero hybridization)
limit, this goes as $\chi_{vV}(0) \simeq g_J^2
\mu_B^2|<\gth|J_z|\gfo>|^2/2\Delta$.
As we shall discuss in Sec. 8.2, it was
noted by Cox[1987b] that because of large magnetic moments the van
Vleck susceptibility
is sufficiently large to explain the measured $\chi(T)$ values in
UBe$_{13}$.  An open theoretical
question is whether the form of $\chi_{vV}(T)$ is modified in an
interesting (singular) way
due to the singular character of
the ground $\gth$ doublet.  The temperature dependence in the ionic
limit is exponential in that
the saturation with diminishing $T$ goes as $\exp(-\Delta/T)$.  This is
too weak a temperature
dependence to fit any experiments on the relevant materials.

\begin{figure}
\parindent=2.in
\indent{
\epsfxsize=3.in
\epsffile{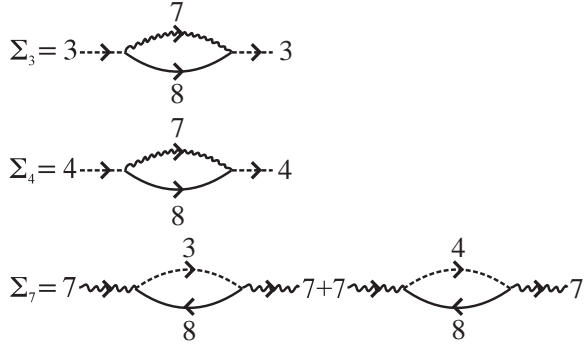}}
\parindent=.5in
\caption{Van Vleck susceptibility diagrams within the NCA for the simplest 
\ufp Anderson model. 
}
\label{fig5p22}
\end{figure}

\begin{figure}
\parindent=2.in
\indent{
\epsfxsize=3.in
\epsffile{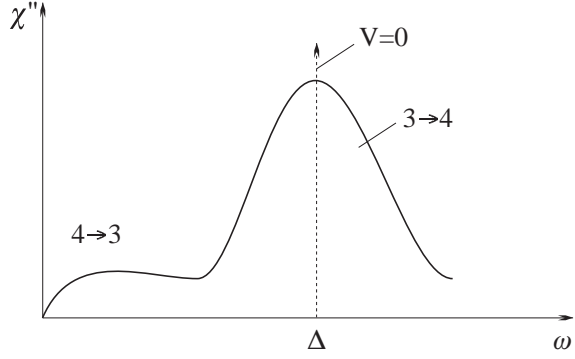}}
\parindent=.5in
\caption{Schematic dynamic van Vleck susceptibility in the ionic and finite hybridization
limits of the simplest \ufp Anderson model.
}
\label{fig5p23}
\end{figure}

A qualified ``yes'' may be given as an answer to the question of
whether $\chi_{vV}$ is modified
in a singular way by the behavior of the $\gth$ ground doublet.  To see
why, we must delve in
detail into the NCA diagrams for $\chi_{vV}$
which appear in Fig.~\ref{fig5p22}.  Using a $\tilde\chi_{vV}$ for the
susceptibility
with the magnetic matrix elements divided out,
we have that
$$\tilde\chi_{vV}''(\omega,T) = \tilde\chi_{vV,3\to 4}''(\omega,T)
+   \tilde\chi_{vV,4\to 3}''(\omega,T) \leqno(5.3.16)$$
with
$$\tilde\chi_{vV,3\to 4}''(\omega,T) = {\pi\over \pi{\cal Z}_f} \int
d\epsilon {\cal A}^{(-)}_3(\epsilon,T)[
{\cal A}_4(\omega+\epsilon,T)-{\cal A}_4(\epsilon-\omega)]
\leqno(5.3.17.a)$$
and
$$\tilde\chi_{vV,4\to 3}''(\omega,T) = {\pi\over \pi{\cal Z}_f} \int
d\epsilon {\cal A}^{(-)}_4(\epsilon,T)[
{\cal A}_3(\omega+\epsilon,T)-{\cal A}_3(\epsilon-\omega)]
~~.\leqno(5.3.17.b)$$
The first term corresponds to transitions from occupied $\gth$ states
to $\gfo$ states,
while the second
corresponds to transitions from occupied $\gfo$ states to $\gth$
states.
In the ionic limit, the former would give delta function absorption
lines at $\omega=\pm\Delta$, while
the latter would give no intensity at all since the $\gth$ state sits
precisely at zero frequency.
Moreover, the weight of $\gfo$ in the ground state is zero since it may
only be thermally occupied.
This ionic limit result is sketched in Fig.~\ref{fig5p23}..

The quantum broadening due to the Kondo effect allows for more
interesting physics.  As shown
in Fig.~\ref{fig5p23}, the delta-function lines of the ionic limit will be
broadened out.  These will still
contain the bulk of the spectral response.  However, since now the
$\gfo$ states acquire finite
width, some quasielastic response due to the occupied $\gfo\to \gth$
transitions may now arise.
The net weight in this quasielastic response will be proportional to
the weight of $\gfo$ in the
ground state, as may be seen by integrating
$(N_B(\omega)+1)\tilde\chi_{vV,4\to 3}(\omega,T)$
where $N_B(\omega)=(\exp(\omega/T)-1)$ is the Bose-Einstein factor.
This gives the net
spectral intensity in the dynamic structure factor for the
$\gfo\to\gth$ transitions.

The question of the singularity in $\chi_{vV}$
now hinges on the behavior of the various spectral functions
near to threshold.  For simplicity, let us assume the limit of equal
coupling coefficients.
 Applying the results of the $N=M=2$ case from Sec. 5.1
together with $g_4=g_3+\Delta$, we see
that the positive frequency spectral functions are
$${\cal A}_3(\omega,0) \approx \theta(\omega-E_0)
{1\over T_0 }({T_0 \over 2|\omega-E_0|})^{1/2} \leqno(5.3.18.a)$$
and
$${\cal A}_4(\omega,0) \approx \theta(\omega-E_0)
{T_0 \over \Delta}^2 ({2|\omega-E_0|\over T_0})^{1/2}
~~.\leqno(5.3.18.b)$$
Below $E_0$, the green's functions are purely real and given by
$${\cal G}_3(\omega) \approx {1\over T_0 }({T_0 \over
2|\omega-E_0|})^{1/2} \leqno(5.3.19.a)$$
and
$${\cal G}_4(\omega) \approx {1\over \Delta} ~~.\leqno(5.3.19.b)$$
What of the occupied state spectral functions?  If we make
M\"{u}ller-Hartmann's {\it Ansatz}
(M\"{u}ller-Hartmann [1984]), then we have ${\cal A}^{(-)}_i =\alpha
{\cal G}_i$ with
$\alpha = {\cal Z}_f(0)/(2+3+2)=1/7$ here, the denominator being the
sum
of all the level degeneracies.
(see the discussion above Eq. (5.1.22)).  With this {\it Ansatz},
$${\cal A}^{(-)}_3(\omega) \approx {1\over 7} \theta(E_0-\omega)
{1\over T_0 }({T_0 \over 2|\omega-E_0|})^{1/2} \leqno(5.3.20.a)$$
and
$${\cal A}^{(-)}_4(\omega) \approx {1\over 7\Delta} \theta(E_0-\omega)
~~.\leqno(5.3.20.b)$$

The applicability of the {\it Ansatz} is a crucial question.  Clearly,
it is nonsense for
the {\it Ansatz} to apply to all excited levels because this would
increase the denominator
of $\alpha$ beyond any reasonable bound.  Hence, it must fail for
states which are too high
in energy.  Clearly it must work for the two propagators which are most
divergent, since
the model in that case is entirely equivalent to a multichannel model.
It seems reasonable
at least that it may apply to excited states which have sufficient
overlap with the ground state
in the sense of their quantum fluctuation induced occupancy.

If we accept the validity of the {\it Ansatz}, we obtain the following
results for
$\tilde\chi_{vV}''$:
$$\tilde\chi_{vV,3\to4}''(\omega,0) \approx {\pi^2 \omega \over
14\Delta ^2} \leqno(5.3.21)$$
$$\tilde\chi_{vV,4\to3}''(\omega,0) \approx sgn(\omega)
{\sqrt{2}\pi\over 7 \Delta}
\sqrt{{|\omega|\over T_0}} ~~.\leqno(5.3.22)$$
The $3\to 4$ response
 is regular in $\omega$, and if we calculate the corresponding
 $\chi(0)$ value,
we obtain approximately $\tilde\chi_{vV,3\to 4}(0) \approx \sqrt 2\pi
/7\Delta$ which is very
close to the ionic limit.  The deviations at finite $T$ are expected to
go as $T^2$ since the
dynamic response is analytic for $\omega\to 0$.  If we roughly estimate
the contribution from
the $4\to 3$ response, we obtain $\tilde\chi_{vV,4\to 3}(T) \approx
(\sqrt 2\pi/7\Delta)
[1-\sqrt{(T/T_0)}]$.  Hence, a $\sqrt{T}$ singularity seems to be
observed which is a novel
signature of the quadrupolar Kondo effect in the magnetic
susceptibility.

\begin{figure}
\parindent=2.in
\indent{
\epsfxsize=6.in
\epsffile{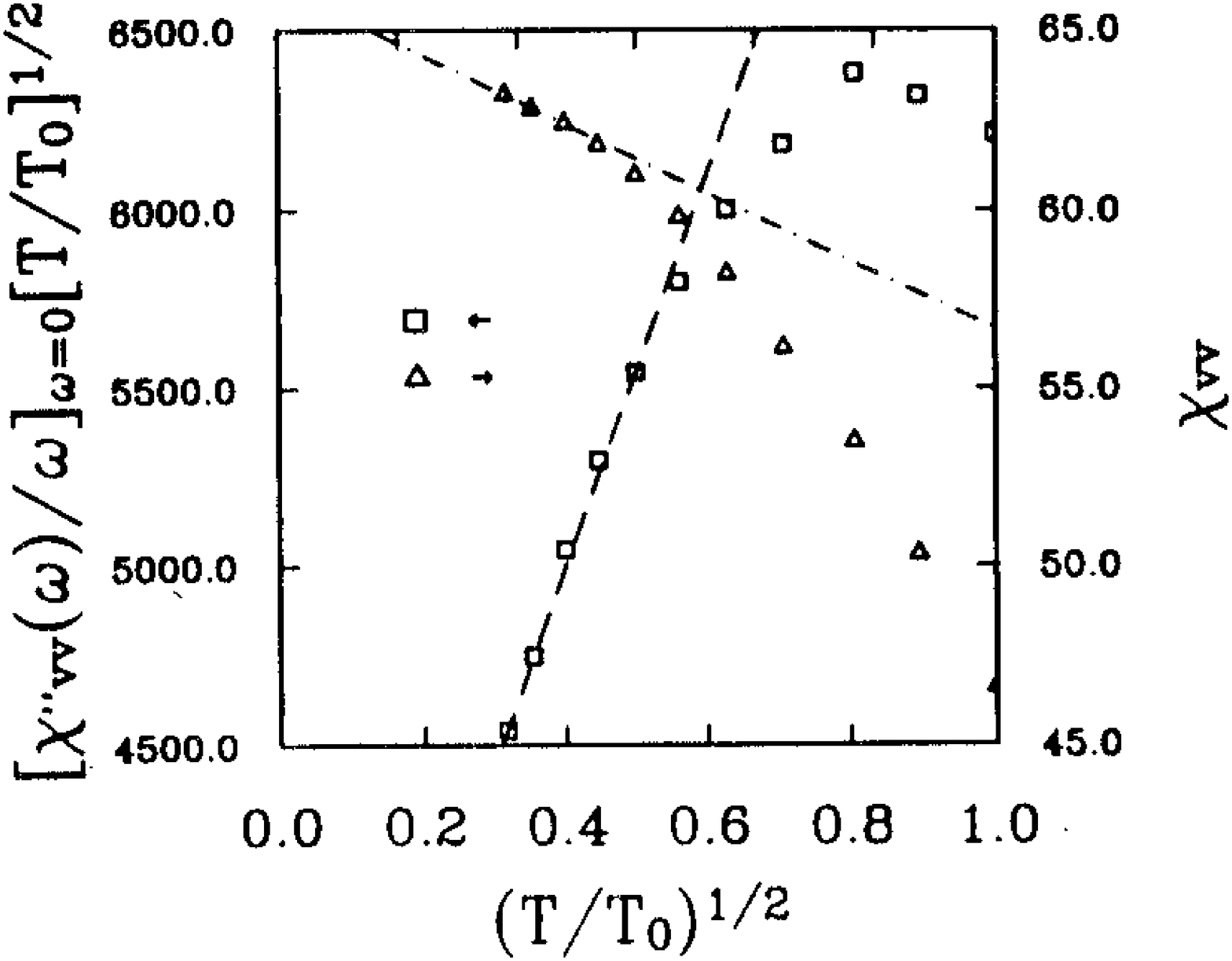}}
\parindent=.5in
\caption{ NCA calculations of $\chi_{vv}$ for the simplest \ufp ion model.  
Squares are the zero frequency limit of the
imaginary part of the van Vleck susceptibility divided by frequency and multiplied by $(T/T_0)^{1/2}$,
triangles are the static van Vleck susceptibility.  Here the bare crystal field
splitting is
$\Delta(\Gamma_4)=0.01D$ and $T_0\approx 0.002D$.
The matrix element coupling the $\Gamma_3,\Gamma_4$
states has been divided out.  The dashed line is given by 
$(0.056/\Delta(\Gamma_4)k_BT_0)[1+1.95(T/T_0)^{1/2}]$
and the dash-dot line by $\chi_{vv}(0)[1-0.14(T/T_0)^{1/2}]
=(0.66/\Delta)[1-0.14(T/T_0)]$.
The ground state occupancies of the doublet and triplet are $n_3=0.56,n_4=0.23$.  From 
Cox and Makivic [1994].
}
\label{fig5p24}
\end{figure}

The numerical evidence for the $\sqrt{T}$ behavior is mixed.  For a set
of calculations based on
the present model with a single excited triplet level, the square root
appears to be present in
$\chi_{vV}(T)$ and $\lim_{\omega\to 0}(\chi_{vV}''(\omega,T)/\omega
\sim 1/\sqrt{T}$.  The
latter quantity is important for an understanding of nuclear relaxation
rates as we shall
discuss further in Sec. 8.2.  This is
shown in Fig.~\ref{fig5p24} taken from Cox and Makivik [1994]. The model
parameters for this
run were chosen to give significant overlap between ground and crystal
field levels in an effort
to mock UBe$_{13}$, which we shall discuss further in Sec. 8.2.
 However, in runs with well separated crystal field levels, the static
 susceptibility appears to
saturate as $T^2$ as shown in Fig. 5.23 from Kim {\it et al.}
[1996].  The systematics
of the $\sqrt{T}$ behavior remain unclear at the present time.
Experimentally, the $\sqrt{T}$
form appears to describe the low temperature response in a number of
materials
as we shall discuss further in Sec. 8.2.  It is quite clear that the
quadrupolar Kondo effect does not substantially modify the zero
temperature van Vleck
susceptibility from the ionic limit.

{\it Magnetic susceptibility:  Contribution from the excited
configuration}.  Because the excited
configuration carries magnetic character, it will contribute a singular
log-divergent term to the
susceptibility.  Of course, as we know from Sec. 6.1.4, this has a
small pre-factor so that
$$\tilde\chi_7(T) \simeq {W_{ch}^2\over T_0}\ln(T_0/T)
~~.\leqno(5.3.23)$$
This must eventually overtake the leading constant behavior of the van
Vleck susceptibility
at sufficiently low temperatures.  We can determine the temperature
$T^x_{M}$ at which this
occurs by demanding $\tilde\chi_7(T) \approx \tilde\chi_{vV}(0) \approx
1/\Delta$.
The result is
$$T^x_M =T_0 \exp({T_0\over W_{ch}^2\Delta})
\approx  T_0 \exp(-{\Gamma^2\over\pi^2T_0 \Delta}) ~~.\leqno(5.3.24)$$
It would appear that unless $\Delta$ is quite large, this scale is too
small to observe
the expected logarithmic divergence.  The only possible way out is if
somehow $W_{ch}$ is
replaced by $\Gamma/\pi|\tef|$, the perturbative estimate which checks
with the discussion
of channel field splitting in the NRG analysis (c.f. Sec. 4.2.c.).
Then the logarithmic
divergence might be observable.

\section{Conformal Field Theory and Abelian Bosonization Methods} 

In this section we shall discuss two powerful analytic methods for
describing the low
temperature physics of the multi-channel Kondo model. Both rely
upon the ability to convert between bosons and fermions in
1(space)+1(time) dimensions.
The spatial dimension here is the radial direction away from the
impurity.
The conformal field theory approach
pioneered by Affleck and Ludwig in this context (Affleck [1990a];
Affleck and Ludwig
[1991a,b,c;1992;1993]; Ludwig and Affleck [1991,1994]; Affleck {\it et
al.} [1992];
 Ludwig [1994a,b]; see also
Tsvelik [1990])
is applicable to all versions of the Kondo model, Fermi liquid fixed
point or not.  This
rests upon a description of low temperature states in terms of spin,
channel, and charge
degrees of freedom, and the ability to write conduction electron spin
and channel
operators in terms of
``bosonic currents'' which obey non-Abelian commutation relations (the
Kac-Moody algebra).
The charge operators are written in terms of conventional bosonic
operators.  The method
may access both thermodynamic and dynamical quantities, but is
restricted to low temperatures
(asymptotically close to the critical point).  We shall discuss this
method in Sec. 6.1.  A good review of the application of conformal
field theory to
a number of condensed matter problems is given in Affleck [1990b], and 
a recent review of the application to the Kondo problem may be found
in Affleck [1995].  A recent reformulation of the problem by Maldacena
and Ludwig [1996] in terms of Majorana fermions recovers the results 
of the previous work and makes explicit connection to the Abelian bosonization
approach of Emery and Kivelson [1992] discussed in Sec. 6.2.   We 
shall briefly discuss the Maldacena and Ludwig approach in Sec. 6.3.  

For a particular highly anisotropic limit of the two-channel model,
Emery and Kivelson [1992]
(see also Sengupta and Georges [1994]) have shown that a purely Abelian
bosonization scheme
is possible to describe the physics.  This approach leads to many
results which overlap with the
conformal theory, in particular providing a nice interpretation of the
residual entropy and a simple
expression for the dynamic susceptility of the impurity which agrees
well with previous numerical
NCA results [Cox, 1988(a)].  We shall discuss the abelian bosonization
approach in Sec. 6.2.   The application to transport coefficients
is not transparent, but with the Majorana fermion formulation of the 
conformal theory produced by Maldacena and Ludwig [1996] the difficulties
are resolved.  
We shall discuss this approach
briefly in Sec. 6.3.

\subsection{Conformal Field Theory Approach to the Kondo Model} 

Conformal field theory (CFT) has arisen in a number of contexts
for describing two-dimensional critical phenomena, superstrings and
other
$1+1$ dimensional relativistic quantum field theories.  Essentially,
the theory exploits
the absence of a length scale which occurs at critical points or in
relativistic
field theories together with the two dimensionality which assures
invariance at the
critical point under arbitrary conformal transformations which have
spatially
dependent scale factors.  Hence, the conformal symmetry is much larger
than the
simpler dilatation invariance employed in the renormalization group.
As a result, it can
be used to fully specify critical exponents, correlation functions, and
finite size spectra
for a number of interesting models in field theory and two-dimensional
critical phenomena.
As shown by Cardy [1984,1986a,b], the CFT approach is not only useful
for bulk problems, but also
for problems with a boundary present.  This is precisely the case for
the Kondo model,
which may be mapped to a $1+1$ dimensional relativistic
field theory on the space+time  half plane
($r>0$, $r$ the radial direction about the impurity site) where the
`speed of light' is the
Fermi velocity $v_F$ set by the linear dispersion of conduction
electrons.
 The conformal invariance of the
boundary condition (it is invariant under arbitrary conformal
transformations with position
and time dependent scale factors) assures the utility of the CFT
technology.  In a number
of papers, Affleck and Ludwig (Affleck [1990a]; Affleck and Ludwig,
[1991a,b,c];
Ludwig and Affleck [1991]; Affleck
{\it et al.} [1992]; Affleck and Ludwig [1993]; Ludwig [1994a,b])
 employed CFT in the presence of a boundary to work out
the low temperature properties of the multi-channel Kondo model.

We shall divide our discussion of the conformal theory into three
parts.
First, in Sec. 6.1.1, we lay out the ideas of bosonization and
non-Abelian bosonization in
particular and show how this leads to a complete if complicated
description
of free fermions.  Next, in Sec. 6.1.2, we show how the complicated
formulation
for free fermions is quite natural for the Kondo problem, revealing in
a remarkably
simple way the ``absorption'' of the impurity spin into the many body
spin density.
Also, we show how  finite size spectra and non-trivial operators at the
impurity are readily generated within this approach through suitably
chosen
boundary conditions and ``fusion'' rules governing the absorption of
the impurity
spin into the many body spin density.  Third, in Sec. 6.1.3,
we outline
how various thermodynamic quantities are obtained within the conformal
theory  approach.
Finally, in Sec. 6.1.4, we focus on dynamical quantities such as the
one-particle
electron Green's function.  Our intention in this section is to
motivate the
key ideas of the theory without reproducing in detail all the
calculations
presented in the papers of Affleck and Ludwig.

\subsubsection{Non-Abelian Bosonization and Free Fermion Spectra} 

At the core of the conformal theory approach is the rewriting of the
original
Kondo Hamiltonian in terms of charge, spin, and channel-spin densities
or
``currents.''  This exploits the effective one-dimensional character of
the problem.
To review how this works, we follow Affleck [1990a] and Affleck and
Ludwig [1991a]
in the following order: \\
(a) First we review the mapping of three dimensions to one-dimension
described
in terms of left and right moving fermions on the half-axis (radial
direction);\\
(b) We remind the reader of free fermion spectra in one-dimension for
spinless and spin 1/2
one-channel fermions expressed in terms of the charge and spin of the
excitations;\\
(c) Following Haldane [1981] we review how the spectrum of spinless
one-dimensional
fermions may
be expressed in an Abelian bosonization approach;\\
(d) We then show how the spectrum of one dimensional
spin 1/2 fermions may be recovered in a
rather complicated way as a sum over commuting spin and charge
Hamiltonians.
The Kac-Moody algebra emerges naturally in this approach.  In this free
fermion
case, spin and charge excitations are bound together in a way to
reproduce free
fermion excitations.  \\
(e) We then outline how the multi-channel free fermion Hamiltonian may
be written
in terms of mutually commuting spin, charge, and channel Hamiltonians,
where the
spin and channel densities or currents both obey appropriate Kac-Moody
commutation
relations.  \\
(f) Finally, we briefly comment on the generalization to arbitrary spin
and channel degeneracy. \\

{\it (a)  Mapping of three-dimensions to one-dimension}\\

This discussion follows closely Ludwig [1994a].   A similar discussion
occurs in Krishna-Murthy, Wilkins, and Wilson [1980a], but here strict
attention is paid to the boundary conditions and the appropriately
defined one-dimensional fermion states.
In the Kondo problem, only a particular set of
conduction electron partial waves couples to the impurity spin
or pseudo-spin.  Although for real rare earth and actinide impurities
these
are likely to be dominantly of $f$ symmetry about the ion, and for TLS
sites they are considerably more complicated as indicated in Secs. 2
and 3,
we shall make the simplifying assumption here that the impurity has an
$s$-wave symmetry.

The $s$-wave projected part of a free electron annihilation operator
 with momentum $\vec k$ and spin and channel
indices $\mu$ and $\alpha$ is given by
$$c_{k\mu\alpha} = {k\over \sqrt{4\pi}} \int d\hat k c_{\vec
k\mu\alpha}~~.
\leqno(6.1.1)$$
The operators so defined obey continuum commutation relations.
Corresponding to the momentum space operator is a real space operator
$$c_{\mu\alpha}(\vec r) = {1\over i2\sqrt{2\pi}r} c_{\mu\alpha}(r) ~~.
\leqno(6.1.2)$$

The energy spectrum of the free electrons may be linearized about the
Fermi momentum as is usual within some cutoff $\pm \Lambda$ about
$k_F$ so that $\epsilon_k = v_F(|k|-k_F)$.  The physically relevant
size of the
cutoff scale is set by the Kondo temperature, {\it viz.},
$\Lambda \approx k_BT_K/\hbar v_F$.   This reminds us that a continuum
theory
approach such as the CFT is useful only asymptotically below the low
temperature
energy scale.

We want to define left moving and right moving fields
$\Psi_{p\mu\alpha}(r)$ (with
``chirality index''
$p=\pm$ for right or left) in the radial spatial dimension
which have the rapid oscillations of order $\exp(\pm ik_Fr)$ removed.
Physically,
a left-mover corresponds to an incoming spherical wave front, while a
right mover
corresponds to an outgoing spherical wave front.  These operators
are defined in terms of  the $c_{k\mu\alpha}$ by
$$\Psi_{p\mu\alpha}(r) = \int_{-\Lambda}^{\Lambda} dk e^{-ipkr}
c_{k_F+k,\mu\alpha} ~~.
\leqno(6.1.3)$$
In terms of these fields,
$$c_{\mu\alpha}(r) = e^{ik_Fr}\Psi_{-,\mu\alpha}(r) -
e^{-ik_Fr}\Psi_{+,\mu\alpha}(r) ~~.
\leqno(6.1.4)$$
With this definition the free electron Hamiltonian near the Fermi
energy can be written
as
$$H_0 = {iv_F\over 2\pi} \int dr(\Psi_{-,\mu\alpha}^{\dagger}(r){d\over
dr}\Psi_{-,\mu\alpha}(r)
- \Psi_{+,\mu\alpha}^{\dagger}(r){d\over dr}\Psi_{+,\mu\alpha}(r))
\leqno(6.1.5)$$
in position space and
$$H_0 = v_F\sum_{p,\mu\alpha} \int_{-\Lambda}^{\Lambda}
 dk pk \Psi_{kp\mu\alpha}^{\dagger}(r)\Psi_{kp\mu\alpha}
 \leqno(6.1.6)$$
in momentum space.

Given the linear spectrum corresponding to massless
fermions, the effective `speed of light' of the problem is $v_F$.
Moreover, the
absence of a mass scale implies the absence of a length scale, which
assures the
equivalence of space and time axes and conformal invariance, the
invariance of the
system under arbitrary conformal transformations in the $1+1$
dimensional plane.
We shall think of the time axis in terms of imaginary time, so that it
becomes infinite only
at $T=0$.
At first site it might appear that conformal invariance is violated by
the boundary,
(only positive $r$ is allowed).   However, the half-plane is indeed
conformally invariant
and may be mapped back to the full plane.  Also, any finite strip may
be mapped reversibly
to the half plane, a fact which is useful for determining finite size
spectra, allowed operators
in the problem, and finite temperature correlation functions as we
shall discuss in Secs.
(6.1.2,3,4).

To specify the theory further it is necessary to introduce boundary
conditions.
First, the left and right moving fermions are not independent.  At the
origin,
$\Psi_{+,\mu\alpha}(0)=\Psi_{-,\mu\alpha}(0)$.  Thus one may eliminate
the
right moving fermions completely and express the physics in terms of
the left
movers by artificially extending to negative $r$ and
setting $\Psi_{+,\mu\alpha}(r) = \Psi_{-,\mu\alpha}(-r)$.  Assuming the
system to
have a size $l$ in the radial direction, Affleck and Ludwig [1991a,b]
then typically assume
$$\Psi_{-,\mu\alpha}(l) = -\Psi_{-,\mu\alpha}(-l) ~~.\leqno(6.1.7)$$
This boundary condition is typically denoted $F^-$ by Affleck and
Ludwig,
 and with the opposing sign on the right hand side
the boundary condition is referred to as $F$ (see Ludwig [1994a]).
This specifies completely the effective one-dimensional theory.
We may take as a convention that 
only the left moving branch couples to the impurity.  We will typically
write the
momentum integral in discrete form to follow closely the conventions of
Affleck
and Ludwig.  We depart from their convention on notating left and right
movers
slightly (they use simply $L$ to denote left and $R$ to denote right).
This choice
is of some convenience in the discussion of bosonization algebras that
follows
below.  When we discuss the Kondo problem in more detail in Sec. 6.1.2,
we
will return to their convention.

Note that the linearization approximation is common in one-dimension.
We shall
usually not explicitly write the cutoff in which makes this treatment
equivalent to
the Luttinger model for the effective 1D system (Haldane [1981]).   \\

{\it (b) Spectrum of non-interacting spinless and spin 1/2 fermions in
one dimension}\\

The energy levels and states of one-dimensional free fermions may be
specified completely
in terms of the charge and spin relative to the ground state together
with the number of
excited electron hole pairs relative to the ground state.  In the
following discussion,
we follow Affleck [1990a] and Affleck and Ludwig [1991a].

\begin{figure}
\parindent=2.in
\indent{
\epsfxsize=4.in
\epsffile{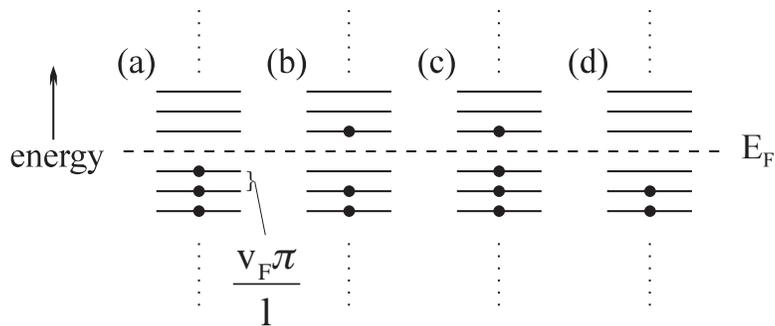}}
\parindent=.5in
\caption{Low lying excitations of a single branch for a one-dimensional 
spinless Fermion gas. The shown levels are only
for left (or right) moving electrons.  The boundary conditions of Eq. (6.1.7)
are chosen so that the ground state (filled Fermi sea) shown in (a) is non-degenerate. The 
excitations may be charge neutral, such as the particle-hole pair excitation of (b).  
The two lowest charge 1 excitations are shown in (c), where an electron is added, and
(d) where a hole is added (electron removed).  }
\label{fig6p1}
\end{figure}

To remind the reader, we first focus on the case of spinless fermions.
With the boundary
condition of Eq.(6.1.7), the allowed k vectors for left moving fermions
are given by
$k_n = -\pi v_F/l (n+1/2)$ with $n=0,\pm 1, \pm 2,....$. The
corresponding single particle
energies are $\epsilon_n = -v_Fk_n$.  The ground state is obtained
by filling all one particle levels below zero energy at which the Fermi
level resides.  This
situation is shown in Fig.~\ref{fig6p1}(a).  We may excite relative to the
ground state in two
ways.  First, we can create a particle hole excitation.  This is shown
in Fig.~\ref{fig6p1}(b), where
an electron below the Fermi energy is promoted above the Fermi energy.
These
excitations always raise the energy by integral multiples of $\pi
v_F/l$.  The second
kind of excitation is effected by adding or removing charge from the
Fermi sea.  The simplest
processes are shown in Figs.~\ref{fig6p1}(c,d).  The lowest energy to add or
remove charge $Q$ is
obtained when states are sequentially filled, leading to an elementary
sum for evaluation
of the energy which gives $E(Q,0)=\pi v_F Q^2/2l$.
Putting these two kinds of excitations together, we see that the energy
spectrum for free left moving spinless fermions subject to the boundary
condition Eq. (6.1.7) on
a system size $2l$  are given by
$$E(Q,n_Q) = {\pi v_F\over l}[{Q^2\over 2} + n_Q] \leqno(6.1.8)$$
where $Q$ counts the added charge (which may be positive or negative)
and $n_Q$ the
number of particle-hole pairs.    The states corresponding to these
excitations are of
course simple Slater determinants.

This kind of argumentation may now be straightforwardly extended to
spin 1/2 fermions.
Clearly, we can view the spectrum for spin 1/2 as a direct product of
up- and down-spin
electron spectra so that the energies add for the two independent
branches.  Hence we
can define a charge $Q_{\mu}$ for each branch together with
particle-hole excitation
quantum numbers.   Now, define the total charge of an excitation as
$Q=Q_{\uparrow}+Q_{\downarrow}$ and the $z$ component of the spin as
$S_z = (Q_{\uparrow}-Q_{\downarrow})/2$.  Since $\sum_{\mu}Q^2_{\mu}/2=
Q^2/4 + S_z^2$, and $S_z^2=S(S+1)/3$ for an excitation of spin $S$, we
see
that the total energy can be written as
$$E(Q,S,n_Q,n_s) = {\pi v_F\over l}[{Q^2\over 4} + {S(S+1)\over 3} +
n_Q + n_S] \leqno(6.1.9)$$
where $n_Q+n_s$ is a combination of particle hole excitations in the
different spin
branches.  Note that not all values of $Q,S$ are allowed.  Strictly
speaking, the $S$
values must be restricted to 0,1/2.  For integer spin, all other
$S_z^2$ values can be reached from $S=0$ by adding integers to zero
spin, and all other $S_z^2$ values for half integer
spin can be reached from $S=1/2$ by adding integers.  Hence, these are
equivalent to
particle hole excitations from the fundamental values.  Also, there is
a ``gluing''
condition on the excitations:  if $Q$ is even, then $2S_z$ must be an
even integer, while if
$Q$ is odd, then $2S_z$ must be an odd integer.

\begin{figure}
\parindent=2.in
\indent{
\epsfxsize=2.in
\epsffile{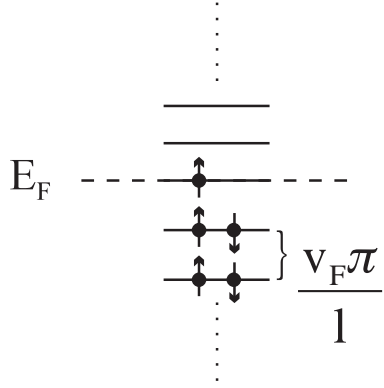}}
\parindent=.5in
\caption{Ground state for spin 1/2 fermions in one-dimension
for the boundary condition $\Psi_{\mu}(l)=\Psi_{\mu}(-l)$. In this case, 
the highest most occupied level is a zero mode, and we choose the charge reference
to correspond to single occupancy of this level, whence charge and spin quantization 
conditions are reversed relative to the boundary conditions of Fig.~\ref{fig6p1}.}
\label{fig6p2}
\end{figure}

For a different choice of boundary condition,
$\Psi_{\mu}(l)=\Psi_{\mu}(-l)$, the $k$ values
are shifted to $-\pi n/l$ and now the upper most occupied energy is at
zero.  This corresponds
to a $\pi/2$ phase shift of the levels considered in the previous
paragraph.  The corresponding
level spectrum for the left moving branch is shown in Fig.~\ref{fig6p2}.  Note
that there is a four fold
degeneracy ($S=1/2,Q=0$ or $S=0,Q=\pm 1$)
to the ground state because of the presence of a zero energy state.
  The charge may
be shifted by one unit to accomodate this phase shift, and the charge
spin gluing conditions are
reversed as a result.  Even charge requires odd values of $2S_z$, and
odd charge requires
even values of $2S_z$.   We note that the NRG free fermion Hamiltonians
for odd $N$ correspond
to the boundary condition of Eq. (6.1.7) and for even $N$ they
correspond to the reversed
boundary condition.  \\

{\it (c) Abelian Bosonization approach to the spinless fermion
spectrum}\\

The goal in this subsection is to show how the spectrum of Eq. (6.1.8)
can be reproduced
with an appropriate choice of bosonic operators which simply coincide
with the momentum
space charge density operators.

Here we closely follow Haldane [1981], who uses a system length $L$, to
be
compared with our $2l$.   We now include both left and right moving
fermions
indexed by $p=\mp$, respectively.   With $k$ measured relative to the
Fermi energy, the
free Hamiltonian is given by
$$H_0 = \sum_{k,p} v_Fpk \Psi^{\dagger}_{kp}\Psi_{kp}~~.
\leqno(6.1.10)$$

We now introduce the density operators $J_{qp}$ of momentum $q$ and
chirality index
$p$ which are given by
$$J_{qp} = \sum_k [\Psi^{\dagger}_{k+q,p}\Psi_{k,p} -
<\Psi^{\dagger}_{k+q,p}\Psi_{k,p}>_0] \leqno(6.1.11)$$
where the second term ``normal orders'' the density operator, and
refers to the ground
state expectation value of the operator with respect to $H_0$.  This
term regularizes the
density operator in such a way as to remove any divergences due to the
occupied ``positron
sea'' in our Luttinger model spectrum (the right and left movers are
allowed to have positive
and negative momenta).  This particular regularization choice can be
maintained at
finite temperature and for an interacting model.
 Our notational choice of $J$ to designate the density
 follows
the convention of Affleck and Ludwig.   The density operators in real
space at $l,-l$ must have
the same value and thus $q$ runs over integer multiples of $\pi/l$.
For $q$=0, $J_{qp}$ is
simply $Q_p$, the charge in each branch.

The density operators obey a simple commutation relation which is exact
in this
Luttinger formulation (though it will have band edge corrections in
general--see Mahan
[1990] pp.  324-343) which is
$$[J_{qp},J_{q'p'}] = \delta_{p,p'} \delta_{q+q',0} {pq l\over \pi}
~~.\leqno(6.1.12)$$
This relation is easily verified by considering simply the operator
parts of (6.1.11)
which give a sum that must be regularized due to the occupied sea, {\it
viz.}
$$[J_{qp},J_{q'p'}] = \delta_{p,p'} (J_{q+q',p} - J_{q+q',p}) +
\delta_{p,p'} \delta_{q+q',0}
\sum_k [<n_{k+q',p}>_0 - <n_{k,p}>_0] ~~\leqno(6.1.13)$$
The first term follows from the operator output of the commutator, and
the second term from the regularization procedure.  The expectation
values simply give
zero temperature Fermi functions, and the difference is non-vanishing
only for a width
$q$ in momentum space, meaning that the sum gives $pql/\pi$, since the
spacing between
$q$ points is $\pi/l$.  Although the first term in Eq. (6.1.13) is
clearly zero, we write it
suggestively to indicate what will happen when the density operators
involve internal
degrees of freedom.  In that case the second term will still arise but
the first term will
no longer cancel, and this is the source of the Kac-Moody algebra for
the spin and
channel degrees of freedom.  The vanishing of this first term leads to
the `Abelian'
nomenclature, while the non-vanishing of this first term when internal
degrees of freedom
are present is what leads to a `Non-Abelian' label, corresponding to a
Non-Abelian
group symmetry for the Hamiltonian.    The nomenclature is in precise
accordance with that used in gauge field theories.

We note that Eq. (6.1.12) is reminiscent of the commutation relations
for free bosons.
To continue this observation it is straightforward to verify by use of
(6.1.10), (6.1.11) that
$$[H_0,J_{qp}] = v_Fpq J_{qp} ~~.\leqno(6.1.14)$$
This is reminiscent of the commutation relation of a boson raising
operator.  In particular, if
the product $qp$ is positive, this is exactly the case.  However, some
normalization is
required to define a suitable boson creation operator.  Following
Haldane, this is given
for $|q|>0$ as
$$a^{\dagger}_q = \sqrt{{\pi\over |q|l}}[\theta(-q)J_{q,-} + \theta(q)
J_{q,+}] ~~.\leqno(6.1.15)$$
This operator satisfies $[H_0,a^{\dagger}_q]= v_F|q|a^{\dagger}_q$, and
together with
its Hermitian conjugate obeys the canonical boson commutation relation.
Physically, these
operators create particle-hole pairs of momentum $q$.  This does not
specify the energy
due to addition or removal of charge
with respect to the ground state.  This can be written in terms of
$J_{0,\pm}$.
As a result, the now ``bosonized'' Hamiltonian can be written as
$$H_0 = \sum_{|q|>0} v_F|q| a^{\dagger}_qa_q + \sum_p {v_F\pi \over 2l
} J^2_{0,p} \leqno(6.1.16)$$
which can be rewritten as an unrestricted $q$ sum
$$H_0 = {\pi v_F\over 2l} \sum_q [\theta(-q) J_{-q,-}J_{q,-} +
\theta(q)J_{q,+}J_{-q,+}]
$$
$$~~~~~~= {\pi v_F\over 2l} \sum_{q,p} :J_{qp}J_{-qp}: \leqno(6.1.17)$$
which defines the ``normal ordering'' of the density operators.
Clearly, the structure of this Hamiltonian reproduces faithfully the
finite size energy
spectrum of Eq. (6.1.8) for spinless fermions.

Strictly speaking, to prove that this is a faithful representation of
the Hamiltonian, we need
to prove that the states specified by the $J_{qp}$ quantum numbers
(occupancies of $q\ne 0$
states and charges $Q_p$) are complete.  To do this, Haldane [1981]
further introduces a
ladder operator $U^{\dagger}_p$ which adds a unit of charge to $Q_p$
each time it is
applied.  The states specified by Eqs. (6.1.12), (6.1.14), and (6.1.15)
are then
$$|\{Q_p\},\{n_{q,p}\}> \sim \prod_{qp>0} [J_{qp}]^{n_{q,p}}\prod_p
[U^{\dagger}_p]^{Q_p}|0> ~~.\leqno(6.1.18)$$
Haldane gives a simple and elegant proof of completeness by comparing
the partition
function for the free fermion case with that found from Eq. (6.1.17).  They
are in exact agreement
indicating that no states are omitted from Eq. (6.1.18).  \\

{\it (d) Non-Abelian Bosonization of the Spin 1/2 Free Fermion
spectrum}\\

Now we follow the various papers of Affleck and Ludwig (particularly
Affleck [1990a],
Affleck and Ludwig [1991a,b], Ludwig [1994a]).  Some key  technical
details of the
decomposition are arrived at in the work of Altsch\"{u}ler, Bauer, and
Itzykson [1990].  As  in the previous section, we introduce
left and right moving densities, now for both charge and spin:
$$J_{qp} = \sum_{k\mu} \Psi^{\dagger}_{k+q,p\mu}\Psi_{kp\mu}
\leqno(6.1.20)$$
$$\vec J_{qp} = \sum_{k\mu\nu} {\vec \sigma_{\mu\nu}\over 2}
\Psi^{\dagger}_{k+q,p\mu}\Psi_{kp\nu} ~~.\leqno(6.1.21)$$
Note that $[J_{qp},\vec J_{q'p'}]=0$, so the charge and spin degrees of
freedom are decoupled in these densities.   As in the spinless case,
the operator $J_{0,p}$ simply counts the total charge in branch $p$,
and
the operator $J^c_{0,p}$ measures the total $z$ component of spin in
branch $p$.

Because  of the spin degrees of freedom, the commutation relation
(6.1.12) is modified to
$$[J_{qp},J_{q'p'}] = \delta_{q+q',0}\delta_{pp'} {2pql\over \pi}
\leqno(6.1.22)$$
where the factor of 2 arises from the spin sum in the ground state
expectation
value term of Eq. (6.1.13).  On the other hand, the commutator with the
Hamiltonian
is unchanged, as may be verified by direct calculation with the free
fermion form
of Eq. (6.1.10) (augmented by a sum over spin degrees of freedom).
Hence there
is a charge term in the Hamiltonian still, but with a different
normalization than
Eq. (6.1.17) because of the factor of 2 in the commutation relation
(6.1.22). We may
write this as
$$H_{Q0} = {\pi v_F\over 4 l } \sum_{qp} :J_{qp}J_{-qp}:
~~.\leqno(6.1.23)$$
We note that the $q$=0 terms in this sum reproduce the $Q_p^2$ term in
Eq. (6.1.9) for
the energy of spin 1/2 free fermion excitations, and the $q\ne 0$ terms
generate the
particle-hole excitations corresponding to the $n_Q$ term in (6.1.9).

Turning to the spin density, we may evaluate its commutation relation
in a way
precisely analogous to that of Eq. (6.1.13).  However, in this case the
operator
dependent term does not cancel out because of the presence of the
non-commuting
Pauli matrix factors in Eq. (6.1.21).  The result is (with Einstein
summation convention
on the spin components)
$$[J^a_{qp},J^b_{q'p'}] = i\epsilon_{abc} \delta_{pp'} J^c_{q+q',p} +
\delta_{ab}\delta_{pp'}
\delta_{q+q',0} {pql\over 2\pi} ~~.\leqno(6.1.24)$$
This is the so called `level 1' $SU(2)$ Kac-Moody algebra.  The `level'
is read off from the
second term (which arises from the ground state subtraction) and is
simply the overall numerical
coefficient of the factor $pql\over 2\pi$.   When we add channel
degrees of freedom this will
be modified to $M$ due to the free sum over channel degrees of freedom
in the ground
state subtraction. The $SU(2)$ of course refers to the spin algebra of
the density operators
reflected in the first term of (6.1.24).   ( Note that the factor of
$2$ in Eq. (6.1.22) from the correponding
ground state subtraction term is cancelled here by the factor of 1/4
from the product of
spin matrices.)

Just as for the ordinary angular momentum algebra one can specify
irreducible representations by the construction of raising and lowering
operators, one
can do the same with the Kac-Moody algebra.
   The irreducible representations of the $SU(2)$ level
$M$ Kac-Moody algebra specified by
Eq. (6.1.24) consist of `primary' states for each branch $p$
with allowed spin $s_p$ restricted by  $0\le s_p \le M/2$ along with
`descendants'  which contain particle-hole pair excitations generated
by acting on the
primary states with operators $J^a_{qp}$ where $qp$ is positive.  In
the present
case with $M=1$, the allowed $s_p$ values are $0,1/2$, consistent with
our
discussion in Sec. (6.1.1.b).  The set of
a given primary state together with its descendants is known as a
`conformal tower.'
 High spin states can only be contained in the descendant sectors.

As with the charge density, one can compute the commutator of the
current density with
the free Hamiltonian to find if a bosonized structure is possible.  By
direct computation
with the free fermion form of (6.1.10) (augmented by spin) one finds
$$[H_0,J^a_{qp}] = v_FqpJ^a_{qp} \leqno(6.1.25)$$
which suggests that the spin part of the free Hamiltonian can be
written as a quadratic
form $H_{sp,0}=A\sum_{qp}:\vec J_{qp}\cdot\vec J_{-qp}:$.  As in the
case of the
charge, the $q=0$ terms generate the added spin energy $\sim s(s+1)/3$
in
Eq. (6.1.9), and the $q\ne 0$ terms generate the particle-hole pairs
that carry net spin.
The normalization $A$ can be
fixed by commuting this form with the spin density and matching to Eq.
(6.1.25).  This gives
$$[H_{sp,0},J^a_{qp}] = A\{{pql\over \pi} J^a_{qp} +
i\epsilon_{bac} \sum_{q'} J^b_{-q'p}J^c_{q+q',p} [\theta(pq') -
\theta(p[q'+q])\} $$
$$~~~~~~~~~~ = A{(1+2) pql\over \pi} J^a_{qp} = {3A pql\over \pi}
J^a_{qp} \leqno(6.1.26)$$
where the `1' in the numerator follows from the ground state
subtraction part of the Kac-Moody
commutator in (6.1.24)  and the `2' follows from the operator part.
Note that
$2\delta_{cd}=\epsilon_{abc}\epsilon_{abd}$ generates the `2'.  This is
important for
Sec. (6.1.1.f) where we remark on the generalization to arbitrary spin
and channel.

By comparing Eq. (6.1.26) with (6.1.25) we specify $A$ and hence can
write the
free Hamiltonian as the sum of spin and charge parts with
$$H_0 = {\pi v_F\over l}[ {1\over 4} \sum_{qp} :J_{qp}J_{-qp}: +
{1\over 3} \sum_{qp}:\vec J_{qp}\cdot
\vec J_{-qp}:]~~. \leqno(6.1.27)$$
This representation of $H_0$ in terms of quadratic forms in the density
operators is known as the `Sugawara form' of the Hamiltonian.
Clearly, these quadratic forms reproduce the energies of Eq. (6.1.9)
obtained by elementary
considerations, where the $q=0$ terms generate the $Q^2$ and $S(S+1)$
terms in
that equation, and the $q\ne 0$ terms generate the particle-hole pair
excitations.
Obviously the non-Abelian bosonization is not the easiest way to obtain
these
energies in the free case; its merit lies in the interacting case we
discuss in Sec. (6.1.2).
The corresponding states can be constructed in complete analogy with
Eq. (6.1.18) for the
spinless case, where we now have a direct product of boson Fock spaces
spanned by the
charge and spin excitations.  The complete specification of allowed
states must satisfy the
fermion gluing rules discussed in Sec. (6.1.1.b), so that odd(even) $Q$
is matched with odd(even) $ 2S_z$.  This corresponds to the boundary
condition of Eq. (6.1.7),
and as remarked in Sec. (6.1.1.b) these are reversed if we change the
sign in that boundary
condition corresponding to a $\pi/2$ phase shift.\\

{\it (e) Generalization to Include Channel Degrees of Freedom}\\

If we now include channel degrees of freedom, we need to introduce
additional density operators
for these channel excitations.  We now use a subscript $s$ to denote
spin density operators,
and a subscript $c$ for channel density operators.  The definitions of
the charge, spin, and
channel densities become
$$J_{qp} = \sum_{q\mu\alpha}
\Psi^{\dagger}_{k+q,p\mu\alpha}\Psi_{kp\mu\alpha} ~~,
\leqno(6.1.28)$$
$$\vec J_{sqp}=\sum_{q\mu\nu\alpha}
{\vec\sigma_{\mu\nu}\over 2}
\Psi^{\dagger}_{k+q,p\mu\alpha}\Psi_{kp\nu\alpha} ~~,
\leqno(6.1.29)$$
and
$$\vec J_{cqp}=\sum_{q\mu\alpha\beta}
{\vec\sigma_{\alpha\beta}\over 2}
\Psi^{\dagger}_{k+q,p\mu\alpha}\Psi_{kp\mu\beta} ~~.
\leqno(6.1.30)$$
The commutation relations for the operators amongst themselves are
$$[J_{qp},J_{q'p'}] = \delta_{q+q',0}\delta_{pp'} {4pql\over \pi}~~,
\leqno(6.1.31)$$
$$[J^a_{sqp},J^b_{sq'p'}] = i\epsilon_{abc} \delta_{pp'} J^c_{s,q+q',p}
+ \delta_{ab}\delta_{pp'}
\delta_{q+q',0} {pql\over \pi} ~~,\leqno(6.1.32)$$
and
$$[J^a_{cqp},J^b_{cq'p'}] = i\epsilon_{abc} \delta_{pp'} J^c_{c,q+q',p}
+ \delta_{ab}\delta_{pp'}
\delta_{q+q',0} {pql\over \pi} ~~.\leqno(6.1.33)$$
Hence both the spin and channel densities satisfy level 2 $SU(2)$
Kac-Moody algebras.
As such, the primary states of spin are restricted to values
$S_p=0,1/2,1=M/2$ and those of
channel spin (also an $SU(2)$ field) to $S_{cp}=0,1/2,1=N/2$ where
$N=2$ is the conduction
spin degeneracy.
The commutation relations of $J_{qp},J^a_{sqp}$ with the free particle
Hamiltonian (6.1.7)
(augmented to include spin and channel degrees of freedom) are
unchanged from
Eqs. (6.1.14) and (6.1.25).  To this we add the commutation relation of
$J^a_{cqp}$ with
the free Hamiltonian which has the same form
$$[H_0,J^a_{cqp}] = v_Fqp J^a_{cqp} ~~.\leqno(6.1.34)$$

Hence, proceeding along the lines of Secs. (6.1.1.c,d), we write the
free particle
Hamiltonian in terms of mutually commuting charge, spin, and channel
quadratic
forms (the `Sugawara' representation) as
$$H_0 = {\pi v_F\over l} \sum_{qp} [{1\over 8}:J_{qp}J_{-qp}: +
{1\over 4} :\vec J_{sqp}\cdot \vec J_{s,-q,p}: + {1\over 4} :\vec
J_{cqp}\cdot \vec J_{c,-q,p}:]
~~.\leqno(6.1.35)$$
This creates free-fermion excitations which are tensor products of
states in the charge,
spin, and channel  spin boson Fock spaces in each of the left and right
moving
branches, created from the primary states and
descendants generated by application of density operators with positive
$qp$ values.
The states are classified in terms of their primary charge ($Q_p$),
spin
$S_p$, and channel spin $S_{cp}$ for each branch, along with integers
characterizing
the number of bosonic excitations, with energies give by
$$E = {\pi v_F\over l} \sum_p [{Q^2_p\over 8} + {S_p(S_p+1)\over 4} +
{S_{cp}(S_c+1)\over 4}
+ n_{Qp} + n_{Sp} + n_{Cp}] \leqno(6.1.36)$$
which is an obvious generalization of Eq. (6.1.9).   The spectrum is
again subject to
a fermion gluing condition and boundary conditions which can shift the
spin/charge/channel
gluing conditions.

Note that this representation of the free Hamiltonian is not unique.
It is quite convenient
for treating the Kondo problem, however.  We could, for example, write
down a Sugawara
form for the larger $U(1)\times SU(4)$ symmetry deriving from overall
rotations in spin and
channel-spin space (Affleck and Ludwig, [1991b]).
 However, the coupling to the impurity will break this symmetry down
into the  $U(1)\times SU(2)\times SU(2)$  considered in Eq. (6.1.35).
Obviously we could also
write down an Abelian bosonization scheme for each branch, spin, and
channel-spin index
in terms of the charge densities restricted to each independent fermion
branch.
A different representation for the energies
 could then be generated by
rewriting the charge  for each value of spin and channel-spin indices
in each branch $p$
in terms of $Q,S^z,S^z_c$ and the `double-tensor'  spin-channel
operator $S^zS^z_c$.
This scheme does not appear manifestly rotationally invariant in either
the spin or channel
spin sectors, however.  The interesting thing is that this
representation is
closely related to the Abelian bosonization scheme developed by Emery
and Kivelson.  That
approach has extra complications associated with `re-fermionization' as
we discuss in
Sec. 6.2.  \\

{\it (f) Generalization to Arbitrary Spin and Channel Degeneracy}\\

If we now assume the fermion spins are $N$-fold degenerate and the
channel spins
are $M$ fold degenerate, we must generalize our definitions of the spin
and channel
densities to $SU(N),SU(M)$ form.  These are best specified in terms of
the $N^2-1$
generators of $SU(N)$ ${\cal T}^a_N,~a=1,2,...N^2-1$ with normalization
condition
$Tr({\cal T}^a_N{\cal T}^b_N = \delta_{ab}/2$ and commutation relations
$[{\cal T}^a_N,{\cal T}^b_N] = if^{(N)}_{abc} {\cal T}^c_N$ where
$f^{(N)}_{abc}$ is the structure
factor of the $SU(N)$ Lie algebra (see, for example, Hammermesh
[1961]).  Similar
generators ${\cal T}^a_M$ should be introduced for the $SU(M)$ channel
spin
symmetry.  Note that the structure constants satisfy
$f^{(N)}_{abc}f^{(N)}_{abd}= N\delta_{cd}$.
In the $SU(2)$ case, $f^{(2)}_{abc}=\epsilon_{abc}$.

The new density operators are thus defined as
$$J_{qp} = \sum_{q\mu\alpha}
\Psi^{\dagger}_{k+q,p\mu\alpha}\Psi_{kp\mu\alpha} ~~,
\leqno(6.1.37)$$
$$J^a_{sqp}=\sum_{q\mu\nu\alpha}
{\cal T}^a_N \Psi^{\dagger}_{k+q,p\mu\alpha}\Psi_{kp\nu\alpha} ~~,
\leqno(6.1.38)$$
and
$$J^a_{cqp}=\sum_{q\mu\alpha\beta}
{\cal T}^a_M \Psi^{\dagger}_{k+q,p\mu\alpha}\Psi_{kp\mu\beta} ~~.
\leqno(6.1.39)$$
The commutation relations for the operators amongst themselves are
$$[J_{qp},J_{q'p'}] = \delta_{q+q',0}\delta_{pp'} {2MNpql\over \pi}~~,
\leqno(6.1.40)$$
$$[J^a_{sqp},J^b_{sq'p'}] = if^{(N)}_{abc} \delta_{pp'} J^c_{s,q+q',p}
+ \delta_{ab}\delta_{pp'}
\delta_{q+q',0} {Mpql\over 2\pi} ~~,\leqno(6.1.41)$$
and
$$[J^a_{cqp},J^b_{cq'p'}] = if^{(M)}_{abc} \delta_{pp'} J^c_{c,q+q',p}
+ \delta_{ab}\delta_{pp'}
\delta_{q+q',0} {Npql\over 2\pi} ~~.\leqno(6.1.42)$$
The spin(channel-spin) densities thus obey $SU(N)(SU(M))$ Kac-Moody
algebras of
level $M(N)$.  There are restrictions on allowed primary states which
are not however
as readily written down as in the $SU(2)$ case described in Sec.
(6.1.1.e).

The commutation relations of the charge, spin, and channel-spin
densities with the suitably
generalized free particle Hamiltonian are unchanged from the discussion
in the preceding
sections.  The appropriately defined Sugawara Hamiltonian may be
determined precisely as
before, and the result is
$$H_0 = {\pi v_F\over l} \sum_{qp} [{1\over 2MN} :J_{qp}J_{-qp}: +
{1\over M+N} :\hat J_{sqp}\cdot \hat J_{s,-q,p}: +
{1\over M+N} :\hat J_{cqp}\cdot \hat J_{c,-qp}:] \leqno(6.1.43)$$
where $\hat J_{sqp}$($\hat J_{cqp}$) are vectors in the $N(M)$
dimensional tensor
space transforming under $SU(N)(SU(M))$ rotations. The states are again
tensor products of vectors from the charge, spin, and channel-spin
boson Fock spaces
for each branch $p$.  These are subject to suitably generalized fermion
gluing conditions.
The single particle energies corresponding to Eq. (6.1.36) are exactly
analogous except that
the $S_p(S_p+1)$ and $S_{cp}(S_{cp}+1)$ terms are generalized to the
quadratic
Casimirs for each of the primary states.  These may be worked out with
suitable Lie
group technology as has been done, for example, for the $M=3,N=2$ model
(Affleck and
Ludwig [1991b]).  \\

\subsubsection{Non-Abelian Bosonization Formulation of the Kondo
Hamiltonian} 

In this subsection we shall review how Affleck and Ludwig developed a
Non-Abelian
bosonization formulation of the Kondo Hamiltonian.  The outline of the
subsection is
as follows: in (6.1.2.a) we show how at special values of the exchange
coupling,
the impurity spin is `absorbed' by the conduction
electrons yielding a Hamiltonian which is still of the Sugawara form
discussed in the
previous subsection with new current operators which still obey the
Kac-Moody algebra.
The working hypothesis is that the special coupling values correspond
to the low
temperature fixed points of the Kondo model.  The states however are
subjected to new
{\it non-fermionic}
`fusion' rules developed by Affleck and Ludwig [1991b; Ludwig, 1994]
which precisely state how the spin, charge,
and channel-spin degrees of freedom are restricted.  While these
quantum numbers
are restricted to fermionic gluing conditions for exactly compensated
$S_I=M/2$ models,
they are `freed up' in the overcompensated case, yielding a finite size
spectrum which
is no longer that of a Fermi liquid.  Hence, as in the case of
interacting one-dimensional
metals which display non-Fermi liquid behavior, spin-charge-channel
separation occurs
in the multichannel overcompensated Kondo model.  In Sec. (6.1.2.b) we
discuss
how the operator spectrum of the model may be determined by a finite
size calculation
with modified boundary conditions.  This allows one to generate the
allowed non-trivial
operators which can effect the critical behavior on approach to the
fixed point as well
as cross-over phenomena in the presence of applied fields that break
the spin and
channel-spin symmetry.  In Sec. (6.1.2.c), we discuss the effects of
various perturbations
on the finite size spectra following the work of Affleck {\it et al.}
[1992].  \\

{\it (a) Sugawara form of the Kondo Hamiltonian:  Absorption of the
Impurity Spin}\\

As discussed in (6.1.1.a), the lack of independence of left and right
moving fermions in the
one-dimensional radial half-space of the Kondo model means that we can
eliminate the
right movers from the problem with the suitable mirroring into the
$r<0$ half-plane.  This means
the Kondo exchange interaction can be written solely in terms of the
impurity spin together
with the left-moving spin current.  Restricting to spin 1/2 conduction
electrons, the result is
$$H_K = {\pi v_F\over l} \lambda \sum_q \vec J_{sqL}\cdot \vec S_I
\leqno(6.1.44)$$
where $\lambda=N(0)|J|$ is the dimensionless coupling constant.   Note
that we have now
favored the $L,R$ notation of Affleck and Ludwig over the $p$ notation
used in the previous
subsection.  Right movers will henceforth be dropped from the
discussion.   The free
Hamiltonian for left movers is written in the Sugawara form discussed
in the previous subsection.

Perhaps the most central component  of Affleck and Ludwig's [Affleck,
1990a; Affleck and
Ludwig, 1991a,b,c; Ludwig, 1994] work is the observation
that for certain special
values of $\lambda$, it is possible to `complete the square' of the
spin term in the free
Hamiltonian to absorb the impurity term. leaving behind a trivial
constant in the energy.
Specifically, when $\lambda = 2/(2+M)$, we can write
$$H_{sp} = H_{0sp} + H_K = {\pi v_F\over l} {1\over 2+M} \sum_q :\vec
{\cal J}_{sqL}
\cdot \vec {\cal J}_{s,-q,L}: \leqno(6.1.45)$$
where the shifted spin density operator is simply
$$\vec {\cal J}_{sqL} = \vec J_{sqL} + \vec S_I ~~.\leqno(6.1.46)$$
These new spin densities {\it still} obey the Kac-Moody algebra
specified by Eq. (6.1.41)
(with $N$=2).   Because only the spin degrees of freedom are modified,
the
charge and channel parts of the free Hamiltonian in Eq. (6.1.43) are
left unchanged.

Because the Hamiltonian with the absorbed impurity spin remains a
quadratic
form in the charge, spin, and channel-spin densities, the form of the
energies is
unchanged from Eq. (6.1.36) in the two-channel case ($M=2$), and indeed
in general from Eq. (6.1.43).  However, what
differs are the constraints on allowed states.   Affleck and Ludwig
[1991(b); Ludwig, 1994] made
a ``fusion rule'' hypothesis to determine which states would be allowed
in the new Hamiltonian with the absorbed spin.  First, they apply a
generalization of
the $SU(2)$ triangle rule appropriate for adding spins in the Kac-Moody
algebra. Restricting
attention to $N=2$, define
$S$ as the total spin of a primary state in a conformal tower in the
shifted Hamiltonian specified by
Eqs. (6.1.45,46), and $S'$ as the conduction part of that spin.  The
generalized triangle
rule is
$$|S'-S_I| \le S \le min\{S'+S_I,M-S'-S_I\} ~~.\leqno(6.1.47)$$
Note that since the maximal primary spin in a conformal tower is $M/2$
that the
right hand side of (6.1.47) assures that the primary spin in the new
conformal towers of
$\vec {\cal J}_{sqL}$ will not exceed $M/2$.  Next, define the free
fermion fusion factor
$n^{QS'S_c}_0$ to be 1 if  a free fermion primary state of charge $Q$,
spin $S'$, and
channel-spin $S_c$ is allowed in the free spectrum of Eq. (6.1.43), and
zero otherwise.
Define the corresponding quantity for the system with absorbed spin as
$n^{QSS_c}_*$.
The fusion rule of Affleck and Ludwig [1991b; Ludwig, 1994] states that
$$n^{QSS_c}_* = \sum_{S'} N^S_{S_I,S'} n_0^{QS'S_c} \leqno(6.1.48)$$
where $N^S_{S_I,S'}$ is one if (6.1.47) is satisfied for $S,S_I,S'$, and
zero otherwise.
Taking this fusion rule together with the general rules for
constructing states within
the Kac-Moody algebra allows a complete generation of the spectrum at
the fixed point.
Corrections to scaling arise as one moves away from the fixed point (by
going to higher
energies for example).

\begin{table}
\begin{center}
\begin{tabular}{|c|c|c|c|c|}\hline
$Q$ & $S$ & $S_c$ & $SO(5)$ & $(El/\pi v_F)$ \\\hline
0 & 0 & 0 & 1 & 0\\\hline
$\pm$1 & 1/2 & 1/2 & 4 & 1/2 \\\hline
0 & 1 & 0 & 1 & 1 \\\hline
$\pm$2 & 1 & 0 & 5 & 1 \\
0 & 1 & 1 & & 1 \\\hline
$\pm$2 & 0 & 1 & &  \\
0 & 0 & 0  & 10' & 1\\
0 & 0 & 1 & & \\\hline
$\pm$1 & 1/2 & 1/2' & 4' & 3/2 \\\hline
$\pm$1 & 1/2' & 1/2 & 4  & 3/2 \\\hline
$\pm$1 & 3/2' & 1/2 & 4  & 3/2   \\\hline
$\pm$3 & 1/2 & 1/2 & & 3/2 \\
$\pm$1 & 1/2 & 3/2' & 16' & 3/2\\
$\pm$1' & 1/2 & 1/2 & & 3/2 \\\hline
\end{tabular}
\end{center}
\caption{Free fermion energy levels for the $M=2$ channel model
(c.f.
sec. 6.1.d).
The states are labeled by charge $Q$, spin $S$, channel spin $S_c$, and
the combined $Q,S_c$ indices
are also designated by their $Sp(2)\sim SO(5)$ labels (which indicates
the combined charge/channel degeneracy); different $SO(5)$ blocks are
delineated by horizontal lines.  Energies are measured
in dimensionless units of $l/\pi v_F E$ and the $F^-$ boundary
condition $\Psi_L(l) =
-\Psi_L(-l)$ is assumed so that the fermion wave functions are forced
to zero at
the boundary. (This is a combination of Tables I and III of Affleck
{\it et al.} [1992].)
Primes indicate descendant states generated by exciting particle hole
pairs. }
\label{tab6p1}
\end{table}

\begin{table}
\begin{center}
\begin{tabular}{|c|c|c|c|c|c|}\hline
$Q$ & $S$ & $S_c$ & $SO(5)$ & $l E/\pi v_F$ & $E_{NRG}$ \\\hline
0 & 1/2 & 0 & 1 & 0 & 0 \\\hline
$\pm$1 & 0 & 1/2 & 4 & 1/8 & 0.125 \\\hline
$\pm$2 & 1/2 & 0 & 5 & 1/2 & 0.505 \\
0 & 1/2 & 1 & &1/2  &0.505 \\\hline
$\pm$1 & 1 & 1/2 & 4 & 5/8 & 0.637 \\\hline
0 & 3/2' & 0 & 1 & 1 & 1.013 \\\hline
$\pm$2 & 1/2 & 1 & & 1 & 1.035 \\
0 & 1/2 & 1 & 10' & 1 & 1.035 \\
0 & 1/2 & 0 & & 1 & 1.035 \\\hline
$\pm$1 & 0 & 3/2' & & 9/8 & 1.147 \\
$\pm$1 & 0 & 1/2 & 16' & 9/8 & 1.147 \\
$\pm$3 & 0 & 1/2 & & 9/8 & 1.147 \\\hline
$\pm$1 & 1' & 1/2 & 4 & 9/8 & 1.179 \\\hline
$\pm$1 & 0 & 1/2 & 4' & 9/8 & 1.232\\\hline
\end{tabular}
\end{center}
\caption{Spectrum of the $M=2,S_I=1/2$  Kondo model.  The
notation
is the same as for Table~\ref{tab6p1}.  The fifth column shows the energies
generated by conformal field theory (CFT); the last column is the
numerical renormalization group
(NRG)
energy (see Sec. 4.2) with appropriate normalization.  This table is
based on
Table V of Affleck {\it et al.} [1992] and Table 1b of Ludwig [1994a].
As discussed
in Secs. 4.2, 6.1.2.a, the NRG respects the $SO(5)$ symmetry but not
the full conformal
invariance.  Hence, $SO(5)$ blocks all have the same energy as shown
below.
Spectra can be generated from Table~\ref{tab6p1} by using the single fusion rule
(Eq. (6.1.48)) in
conjunction with the Kac-Moody triangle rule (Eq. (6.1.47)).}
\label{tab6p2}
\end{table}

To understand how this fusion process
works, the reader is directed to Tables~\ref{tab6p1} (free fermions for $M=N=2,
S_I=1/2$) and~\ref{tab6p2} (spectrum of $N=M=2,S_I=1/2$ Kondo model).  These are
constructed from
corresponding tables in
Affleck {\it et al.}[1992] and
Ludwig [1994a].   Taking $S'=0$ from the ground state $Q=S'=S_c=0$
 of the free fermion problem, it is easy
to see from Eq. (6.1.47) that $S$ is uniquely constrained to be 1/2 and
hence $n^{0,1/2,0}_*=1$ by Eq. (6.1.48).  Now at first one might think
this
corresponds to the spin charge reversal of the $M=1$ model at the
strong coupling
fixed point as discussed in Sec. 4.  However, the lone fermion with
$S=1/2$ would also
have to have $S_c=1/2$, which is clearly not the case.  Next consider
the first excited
state of the free fermion spectrum with $Q=\pm 1(mod
4),S'=1/2,S_c=1/2$.
 The triangle rule (6.1.47) constrains $S$ uniquely to zero, and yields
 an energy of
$\pi v_F/8l$.  This differs from a Fermi liquid theory in two key
aspects.  First, the level
spacing relative to the ground state is a fraction of the minimum
free fermion value of $\pi v_F/l$.  Second, the quantum numbers cannot
be that of free
fermions with simple reversed spin and charge quantum numbers because
the creation of
a free $S_c=1/2$ excitation would necessitate a spin $S=1/2$ to
accompany it.

Table~\ref{tab6p2} also contains a comparison with the lowest 76 NRG energy
levels
(Affleck {\it et al.} [1992]).  It can be seen that
the agreement is quite good, with discrepancies arising as one moves to
higher energies.  This disagreement can arise from three sources: (1)
The logarithmic
discretization of the NRG treatment breaks the conformal symmetry; (2)
The numerical
truncation procedure in the NRG can introduce systematic numerical
error; (3) As one
moves away from the low energy spectrum, corrections to scaling arise
from the effects
of the irrelevant operators.  In practice, it is believed that (1,2)
contribute more to the
differences observed based upon an extrappolation of the NRG numerics
to the
continuum limit $\Lambda=1$, where $\Lambda$ is the logarithmic
discretization parameter
of the NRG.  Notice that the NRG levels respect the $Sp(2)\sim SO(5)$
symmetry discussed
in Sec. 4.  The higher conformal symmetry renders a number of the
$Sp(2)$ energy
levels degenerate.

We end this subsection by noting that the fusion rule can be applied to
exactly compensated
and undercompensated models as well, as was discussed particularly
clearly in Affleck [1990a].
As a couple of examples, consider first the $M=1,S_I=1/2$ model.  The
fusion rule then simply
generates the reversed spin/charge relations corresponding to a $\pi/2$
phase shift as discussed
in Secs. (4.1) and (6.1.1).  For example, for the free fermion ground
state with $S'=0,Q=0$, the
fusion rule generates the state $S=1/2,Q=0$ (with energy zero, after
effecting a shift of
$\pi v_F/4l$ to the spectrum), and for the first excited free fermion
state with
energy $\pi v_F/2l$ and $S'=1/2,Q=\pm 1$, the fusion rule generates the
state $S=0,Q=\pm 1$
with energy zero.  The second excited state with $Q=\pm 2$, $S'=0$, and
energy $\pi v_F/l$ is mapped to the state $Q=\pm 2, S=1/2$ with energy
$\pi v_F/l$. These
lowest two states correspond precisely to the spectrum of free fermions
with the boundary condition
$\Psi_{L\mu}(L)=\Psi_{L\mu}(-L)$ and reversed
spin/charge relations, or equivalently to a $\pi/2$ phase shift.
Affleck [1990a] also demonstrates
that this simple absorption of the impurity spin
holds for the $M/2=S_I$ compensated Kondo model and the
$SU(N),M=1$ Coqblin-Schrieffer model which also has a compensated
(singlet) ground state.
For the undercompensated model with $S_I>M/2$,  Affleck demonstrates
that the
absorption holds for $M/2$ of the impurity spin, leaving behind a local
moment with
net value $S_I-M/2$ and residual ferromagnetic coupling to the impurity
spin.  Hence, the fixed
point spectrum is a direct product of a decoupled local moment with a
Fermi liquid, as was
suggested by \noz and Blandin [1980].  Corrections to scaling from the
marginally irrelevant
ferromagnetic coupling can generate ``non-Fermi liquid'' thermodynamics
which are
related to those in the charge-only model discussed by Giamarchi {\it
et al.} [1993].  \\

{\it (b) Relation of Boundary Operator Spectrum to Finite Size
Spectra}\\

One of the key features of boundary CFT as developed by Cardy
[1984,1986a,1986b]
is the existence of boundary operators with non-trivial scaling
dimensions. The existence
of these boundary terms gives rise to singular
contributions to the free energy of the quantum impurity problem
which are independent of $l$.  These boundary operators live only at
$r=0$ and thus
are dependent only upon the time variable.

Each boundary operator $\phi$ will have a distinct behavior under
rescaling of
space and time coordinates, which is described in terms of the
scaling index $\Delta_{\phi}$ specified by its long time green's
function, {\it viz.}
$$<\phi(\tau)\phi(\tau')> \sim |\tau-\tau'|^{-2\Delta_{\phi}}
\leqno(6.1.49)$$
for $|\tau-\tau'|\to \infty$.  The scaling index also gives the
renormalization group
eigenvalue of the field $h_{\phi}$ conjugate to the boundary operator
$\phi$.
This follows from adding to the Lagrangian ${\cal L}$ associated with
the Sugawara
Hamiltonian specified by Eqs. (6.1.43,45) a term of the form
$$\delta {\cal L}(h_{\phi}) = h_{\phi}\int d\tau \phi(\tau)
~~.\leqno(6.1.50)$$
Consider a uniform rescaling of space and time by a factor $b$.  Then
from the integral
we see that the field $h_\phi$ scales like $b^{1-\Delta_\phi}$ to leave
the term
scale invariant.  Hence operators with scaling dimensions less than one
have
relevant conjugate fields.  As we shall see below, this includes the
primary spin
and channel spin fields at the fixed point of the two-channel model.

Any operator with a scaling index $\le 1/2$ may have
a diverging susceptibility defined by the equation
$$\chi_{\phi}(T) = \int^\beta_0 d\tau <\phi(\tau)\phi(0)> \sim
T^{2\Delta_{\phi}-1}
~~.\leqno(6.1.51)$$
An exponent of $1/2$ is a special case, clearly, and in this case one
will either have
logarithmic or constant behavior at long times and low energies.  If
the operator is
bosonic in character, it will have a log divergence at low temperature
because the spectral function
of the $\phi$ green's function must change sign at zero frequency.  If
the operator is
fermionic in character, it will be a constant at low temperature
consistent with
unitary bounds on the scattering amplitude for fermions.   For a
related discussion, see
Sec. 5.1.2.

\begin{figure}
\parindent=2.in
\indent{
\epsfxsize=5.in
\epsffile{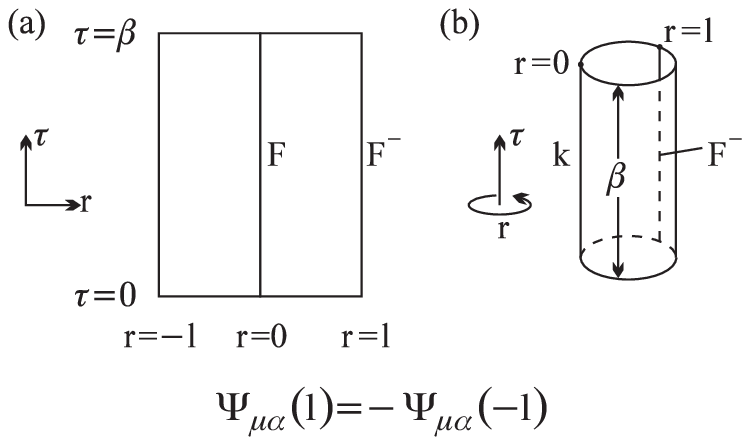}}
\parindent=.5in
\caption{Boundary conditions for free fermions (a) and system with a
Kondo impurity at the origin (b) for use in the conformal theory.  
After Affleck and Ludwig [1991a,b,c].}
\label{fig6p3}
\end{figure}

\begin{figure}
\parindent=2.in
\indent{
\epsfxsize=3.in
\epsffile{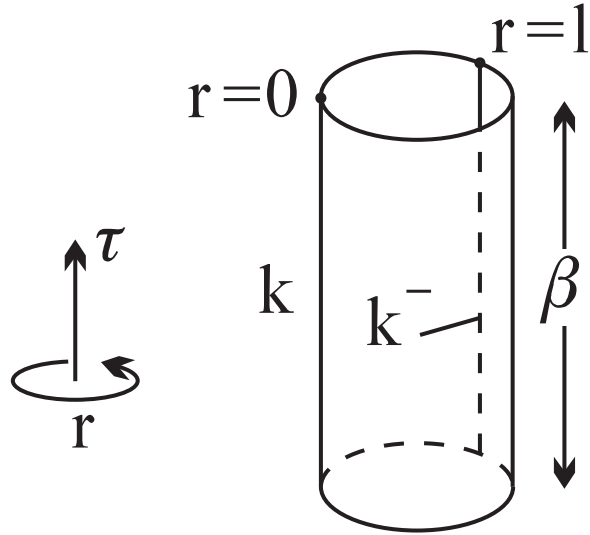}}
\parindent=.5in
\caption{Boundary conditions for operator spectrum of Kondo impurity
model.  In this case a Kondo impurity is fused in at each boundary
($KK^-$ boundary conditions).  After Affleck and Ludwig [1991a,b,c].}
\label{fig6p4}
\end{figure}

A key point made by Cardy [1986b]
and utilized by
Affleck and Ludwig [1991b], Affleck {\it et al.} [1992], and Ludwig
[1994a] is that for
a particular choice of boundary conditions on a finite strip, there is
a one-to-one
correspondence of the boundary operators to the low lying states and
the corresponding
energies (in units of $\pi v_F/l$) are precisely the scaling dimensions
of the boundary
operators.   Let us briefly review the boundary conditions,
referring to Figs.~\ref{fig6p3},\ref{fig6p4}.    In this section, we have described
free left moving fermions which obey the $F$ boundary condition at
$r=0$ and the $F^-$
boundary condition at $r=l$ ({\it c.f.} Eq. (6.1.7). We refer to this
as $FF^-$
boundary conditions, following Ludwig [1994a], and represent this as
the finite
width strip in
Fig.~\ref{fig6p3}(a). Alternatively, given the mirroring of $\Psi_L$ in the
$r<0$ plane, we may view
this as a cylinder of circumference $2l$ with a `seam' along which
$\Psi_L$ is forced to
zero.  In the presence of the impurity spin,
the boundary conditions are shifted to $KF^-$, where the $K$ denotes
absorption of the
impurity spin at the $r=0$ boundary.  This is represented in Fig.~\ref{fig6p3}(b) .

According to the work  of Cardy [1986],  to obtain the scaling
dimensions of the boundary operators
we should look at the spectrum corresponding to the boundary conditions
$KK^-$, namely,
the non-trivial boundary is placed at each edge with a wall at the
$r=l$ edge.  It is important
to note that this finite strip can be related back to the original
half-plane with the conformal
transformation shown in Fig.~\ref{fig6p4}.

To implement this $KK^-$ boundary condition, Affleck and Ludwig [1991b]
(Ludwig [1994a])
hypothesized that the appropriate description was the `double fusion
rule'.  Namely, first
the impurity spin is absorbed at the $K$ boundary, corresponding to the
$KF^-$ boundary
conditions discussed in Sec. 6.1.2.a.  Then the impurity spin must be
absorbed at the
far boundary $r=l$ {\it again}!   In practice, this means that the
conduction spin density
for left movers is shifted (in real space) by $\vec J_{sL}(x) \to
\vec {\cal J}^{op}_{sL}(x) + \delta(x) \vec S_{I1} + \delta(x-l) \vec
S_{I2}$ and in $k$ space to
$\vec {\cal J}^{op}_{sqL} + \vec S_{I1} + (-1)^{lq/\pi}\vec S_{I2}$.
Here we have included the
superscript $op$ to remind the reader this is for calculating the
spectra of allowed boundary
operators.  It may be readily verified that this doubly shifted density
operator still obeys
the Kac-Moody algebra.  Hence, the double fusion may be viewed first as
conduction spin
$S''$ absorbing impurity spin $S_{I1}$ at $r=0$ yielding composite spin
$S'$, with $S'$ then
absorbing $S_{I2}$ at the boundary $r=l$ yielding spin $S$.  Allowed
states in the finite size
spectra are characterized by integers $n^{QSS_c}_{KK^-}=0$(unallowed)
or 1(allowed)
which are calculated from the rule
$$n^{QSS_c}_{KK^-} = \sum_{S'S''} N^S_{S_I,S'}N^{S'}_{S_I,S''}
n^{QS''S_c}_0 \leqno(6.1.52)$$
where $n^{QSS_c}_0$ and $N^{S}_{S_I,S'}$ were defined in Sec.
6.1.2.a.   Each generated
state which is then a primary state of the Kac-Moody algebra correponds
to a so-called
primary field boundary operator.  One may also generate descendants by
operating with
the charge, spin, and current densities for positive $qp$ (i.e.,
negative $q$ for left movers).

\begin{table}
\begin{center}
\begin{tabular}{|c|c|c|c|c|}\hline
$Q$ & $S$ & $S_c$ & $SO(5)$ & $\Delta = El/v_F\pi$ \\\hline
0& 0 & 0 & 1 & 0 \\\hline
0 & 1 & 0 & 1 & 1/2 \\\hline
0 & 0 & 1 & 5 & 1/2 \\
$\pm$2 & 0 & 0 && 1/2 \\\hline
$\pm$1 & 1/2 & 1/2 &4& 1/2 \\\hline
$\pm$1 & 1/2 & 1/2 &4& 1/2 \\\hline
0 & 1 & 1 & 5 & 1 \\
$\pm$2 & 1 & 0 & & 1 \\\hline
0 & 0 & 1 & & 1 \\
$\pm$2 & 0 & 1 & 10 & 1 \\
0 & 0 & 0 & & 1 \\\hline
0 & 0' & 0 & 1 & 3/2 \\\hline
0 & 2' & 0 & 1 & 3/2 \\\hline
\end{tabular}
\end{center}
\caption{Operator spectrum of the $M=1/2,S_I=1/2$ Kondo model
from the  conformal theory double
fusion rule (c.f. Sec. 6.1.2.b).  Notation is the same as for Table~\ref{tab6p1}.
Here $\Delta$ is the scaling index of the operators.
Table is adopted from Table 1.c of Ludwig [1994a].  Spectra can be
generated
from the free-fermion spectrum of Table 6.1 with the double fusion rule
(Eq. (6.1.50))
or from Table~\ref{tab6p2} with the single fusion rule (Eq. (6.1.48)) in
conjunction with
the Kac-Moody triangle rule (Eq. (6.1.47)).  The table is not complete
for all operators
with scaling index 3/2.  The operators shown with scaling
index 3/2 are the leading
irrelevant operator about the fixed point (spin 0)
and the quadrupolar spin operator (spin 2--see Sec. 6.1.2.c for
discussion).}
\label{tab6p3}
\end{table}

The lowest several states for the $KK^-$ boundary conditions generated
by the double
fusion rule are displayed in Table~\ref{tab6p3}.  Again, the energies divided by
$\pi v_F/l$ give
the scaling dimensions, which may then be read off from Eq. (6.1.36)
for non-zero
values of $n^{QSS_c}_{KK^-}$.  Note that the constraints of the
Kac-Moody algebra
imply that no primary fields can exist with $S$ or $S_c$ greater than
one.
From Table~\ref{tab6p3} we see that apart from the trivial constant operator
there are four kinds
of operators with scaling dimension $1/2$.  The physical meaning of
each is as follows:\\
{\it $Q=\pm 1,S=1/2,S_c=1/2 $ Operators}:  These are the quantum
numbers associated with
a free fermion excitation, so the scaling dimension must be
$\Delta_F=1/2$ as discussed above and in
Sec. 5.1.2.  It is of interest that there are two such fields; the
physical interpretation of
this is unclear.  \\
{\it $Q=0,S=1,S_c=0$ Operator}:  This is $\vec \phi_s$,
a primary operator transforming as a vector triad of
local spin tensors.
This object describes the effective core spin at the fixed point.  The
scaling dimension of
$\Delta_s$=1/2 implies that this operator has a logarithmically
divergent susceptibility which will be
explicitly calculated in Sec. 6.1.3.  The scaling dimension value of
1/2
also implies that the spin field
(magnetic field for magnetic impurities, strain field for quadrupolar
Kondo
 ions or TLS sites) is a relevant field with
renormalization group eigenvalue of 1/2.  \\
{\it $Q=0,S=0,S_c=1$ Operator}:  This is $\vec
\phi_{sc}$, a primary operator transforming as a vector triad of
local channel spin tensors.  The scaling dimension of $\Delta_c$=1/2
implies that the corresponding
channel spin susceptibility is logarithmically divergent at low
temperatures, and the application
of an external channel field (strain field for magnetic impurities,
magnetic field for quadrupolar
Kondo or TLS impurities) is a relevant perturbation. \\
{\it $Q=\pm 2,S=0,S_c=0$ Operator:}  This is a primary operator
describing an electron pair
around the impurity site (Ludwig and Affleck [1991], Ludwig [1993]).
The corresponding
local pair field susceptibility is logarithmically
divergent, and in principle a local source of such pairs is
a relevant perturbation.   This pair field is unusual, in that the
antisymmetry of spin and
channel spin labels implied by $S=S_c=0$ means that something else must
be done to
the pair operator to render it antisymmetric under the Pauli
principle.  As discussed by
Ludwig and Affleck [1991c], antisymmetrizing in the $L,R$ indices
between the electron
fields will assure satisfaction of the Pauli principle.  This is
equivalent to odd-parity in the one-dimensional spatial index in view
of the mirroring
condition $\Psi_L(r)=\Psi_R(-r)$.   After
Emery and Kivelson [1992]
noted the log divergence associated with an odd-in-time pair field in
their Abelian
bosonization approach, it was realized that antisymmetrizing in
 temporal arguments for $x=0$ will do  just as well
(Ludwig [1993]) as a spatial gradient, given the equivalence of space
and time axes in
the conformal space.  Alternatively, one may construct a dot product of
the spin field
$\vec \Phi_s$ with a local triplet pair field possessing quantum
numbers $Q=\pm 2,
S=1,S_c=0$.  Although the triplet field by itself has scaling exponent
$1$ and thus
has a non-diverging susceptibility, in combination with the local spin
tensor a spin
singlet is formed which has the same quantum numbers as those with
inserted
gradients.   The equivalence is readily seen by explicitly carrying out
the time and
space derivatives.  We postpone a more elaborate discussion of the pair
fields
 until Sec.  9.4.

Some other interesting operators are: (1) the double tensor operator,
with quantum
numbers $Q=0,S=1,S_c=1$, which is relevant to the generalized model
considered
by Pang [1992,1994] and \zar [1995], and which has scaling dimension 1;
(2) the spin
quadrupole operator (not shown in Table~\ref{tab6p3}) which has $Q=0,S=2,S_c=0$
and
scaling dimension 3/2 (Affleck{\it et al.} [1992]).
This operator is relevant to our discussion of perturbations
about the fixed point, specifically exchange anisotropy, which we shall
discuss in the
next subsection 6.1.2.c.

Notice that no operators with $Q=0,S=1/2,S_c=0$ or $Q=0,S=0,S_c=1/2$
appear
in the $KK^-$ spectrum even though these are in principle allowed
primary operators
coming from the Kac-Moody algebras which permit, e.g., $S=0,1/2,...M/2$
primary
field spins for a level $M$ spin algebra.  The double fusion rule
legislates against
these possibilities.

Appliction of the Kac-Moody densities to the primary field operators
generates the
so called descendant operators with higher scaling dimensions.  As an
explicit example,
put $q=n\pi/l>0$ and consider the operator $\vec {\cal J}_{s,-q,L}\cdot
\vec\phi_s$.  This
operator has $S=0$ (since it is a scalar product of two spin operators)
and scaling
dimension $n+\Delta_{\phi_s}= n+1/2$ for the two-channel case.

By way of slight generalization, we note that the primary spin and
channel-spin operators
transform as the fundamental representations of $SU(N)$ and $SU(M)$ for
the
generalization to the $SU(N)\times SU(M)$ model. The resulting scaling
dimensions are
$\Delta_s=N/(N+M)$ for the spin field, and $\Delta_c=M/(M+N)$
for the channel field, in precise agreement
with  Eqs. (5.1.30.a,b) (recalling that $\gamma=M/N$ is held fixed in
the large $N$
limit of Sec. 5.1). \\

{\it (c) Perturbations about the Fixed Point}\\

In this section we give an overview of the effects of various
perturbations about the two-channel
fixed point following the discussion in Affleck and Ludwig [1991c] and
Affleck {\it et al.}
[1992].  The point is that the conformal theory gives a natural basis
for identifying
relevant, irrelevant, and marginal operators about the fixed point.
In order, we shall
discuss perturbations due to: (1) The leading irrelevant operator; (2)
an external spin
field; (3) an external channel field; (4) exchange anisotropy.

{\it (1) Leading Irrelevant Operator}\\
As discussed by Affleck and Ludwig [1991b] and Affleck {\it et al.}, a
central role is played
by the leading irrelevant operator about the fixed point Hamiltonian.
The restrictions on this
operator are that it must be a singlet in spin and channel indices, so
as to preserve the
$SU(2)\times SU(2)$ symmetry, and must be chargeless to preserve the
$U(1)$ symmetry.
Finally, on physical grounds we expect this operator to involve only
the primary spin field
$\vec \phi_s$ since the non-trivial coupling induced by the Kondo
impurity occurs only in the spin
sector.  By the remarks of the previous section, a reasonable candidate
operator is
$\vec {\cal J}_{s,-q,L}\cdot \vec \phi_s$ for $q=\pi/l$ which has
scaling dimension
$1+\Delta_s=
1+ N/(N+M)$=3/2 when $N=M=2$.  In the case of $N=M$, the operator $\vec
{\cal J}_{c,-q,L}\cdot
\vec \phi_c$ also has the same scaling dimension, but since there is no
impurity channel field
is not a physically plausible candidate for the leading irrelevant
operator.

 The way this operator is added to the Hamiltonian is in the form
$\delta\lambda = \lambda - \lambda^*$ where
$\lambda^*=2/(2+M)$ for $N=2$ is the fixed point coupling.  In the next
section we shall see
how this term enters into the calculation of the specific heat. (Note
that in Affleck and Ludwig
[1991b], the coupling
$\lambda$ used in the discussion of thermodynamics corresponds to
$\delta\lambda$
here.)

{\it (2) Application of a Spin Field}\\
If we introduce a field $\vec h_s$
which couples linearly to the primary boundary spin operator $\vec
\phi_s(\tau)$.
Physically, this corresponds to a magnetic field for the magnetic
impurity, and
a stress field for the quadrupolar impurity or TLS impurity.  Also,
electric field
gradients and magnetostriction induced stress fields (order $H^2$, $H$
the magnetic
field) will split the quadrupolar Kondo impurity, and the TLS impurity
is subject to
spontaneous tunneling matrix elements and well asymmetry which have the
same
effect.  The presence of a spin field produces a
term in Lagrangian of the form in Eq. (6.1.49) with
$\vec h_s\cdot \vec \phi_s(\tau)$ inside the integral.   Hence the spin
field $\vec h_s$ is
relevant with a renormalization group eigenvalue of $1/2$.  This
implies that in the
presence of a spin-field, there will be a crossover to a new fixed
point with crossover
exponent $1/2$, i.e., low temperature properties will be universal
functions of $h^2_s/T$.
This implies the existence of a new energy scale in the problem given
by $h_s^2/T_K$,
as is verified in NRG and  Bethe-Ansatz treatments.  The conformal
field theory cannot
specify the nature of the new fixed point, but it is reasonable to
guess that it is a free
fermion fixed point in the presence of a polarized scatterer (the
Zeeman split primary
field).   This is born out by the NRG calculations discussed in Sec.
4.2.e
(see Fig.~\ref{fig3p3}(a)). We note that the crossover physics is  in precise
agreement with the
discussion of  Sec. 5.1 where the large $N$ NCA approach was used.

{\it (3) Application of a Channel Field}\\
A channel field $\vec h_c$ couples linearly to the primary channel spin
operator
$\vec \phi_c(\tau)$.  This field corresponds to a stress field or
electric field gradient for
the magnetic impurity, and a magnetic field for the quadrupolar or TLS
impurity.
In practice, the channel field probably arises from a splitting of the
exchange integrals
in the presence of applied stress (magnetic impurity) or applied
magnetic field (quadrupolar
Kondo impurity).  It remains something of a mystery how to effectively
obtain a channel field splitting for the TLS since
there is no obvious magnetic coupling to the impurity.  By considering
a perturbation to the
Lagrangian of the form (6.1.49), with $\vec h_c\cdot \vec \phi_c(\tau)$
in the integrand, we
see that
the renormalization group eigenvalue of the field $\vec h_c$ is 1/2.
This implies that the perturbation is relevant
with a crossover exponent of 1/2, i.e., low temperature properties are
universal
functions of $h_c^2/TT_K$, and $h_c^2/T_K$ is a new energy scale which
sets the
crossover temperature on passing first through the Kondo scale and then
at lower
temperatures to this new energy scale.  The crossover behavior is in
good agreement
with that discussed in Sec. 5.1.  The conformal theory again does not
provide
an answer as to what the crossover goes to, but the obvious guess in
view of the analysis
of Secs. 3.1.2, 4.2, and 5.1 is that the crossover is to the ordinary
Kondo fixed point for
whichever channel couples more strongly to the impurity and the
weak coupling fixed point for the other channel.  This is indeed seen
in the NRG
spectra, which suggest a direct product of phase shifted fermions with
unshifted fermions
at the fixed point.  Jerez and Andrei [1995] and Coleman and 
Schofield 
[1995] have
questioned the Fermi liquid character of the fixed point recently.
Using a combination of Abelian bosonization and path integral methods, 
Fabrizio, Gogolin, and \noz [1995a,b] have reasserted that the new fixed
point is indeed a Fermi liquid.  

{\it (4) Exchange Anisotropy}.  \\
Exchange anisotropy breaks the $SU(2)$ spin rotational invariance to a
$U(1)$ subgroup
which is however {\it even} under spin reversal (time reversal for the
magnetic impurity)
unlike the magnetic field.  If it is a relevant perturbation, there
should be an appropriate
primary field operator in the $KK^-$ spectrum.  The logical candidate
is a spin quadrupole
operator with quantum numbers $Q=0,S=2,S_c=0$.
This makes physical sense as argued by Pang [1992] and discussed in
Sec. 3.3.1,3.4.1,
since the exchange anisotropy will induce a local quadrupolar splitting
of the impurity
spin through the two-loop order contribution to the pseudo-fermion self
energy.
Thus, although the anisotropy term ${\cal J}^3_{s,-1,L}\phi^3_s$ has an
irrelevant
scaling dimension, it generates a relevant term in next order.

As noted by Affleck {\it et al.}
[1992], the only primary operator with $Q=0,S_c=0$ for free fermions
has $S=0$.
So it is logical to apply the double fusion rule to this state, which
implies that the
only primary field operator with $S=2$ must satisfy the constraint
implied by the
Kac-Moody triangle rule, so that
$$2 \le 2S_I, ~~~~2\le M- 2S_I ~~.\leqno(6.1.53)$$
This implies that $1/2<S_I<M/2-1/2$.  The scaling dimension is
$\Delta_{quad}=6/(M+2)$,
and so it is relevant for $M\ge 5$, and marginal for $M=4$, provided
$1/2<S_I<M/2-1/2$.
For $S_I=1/2$ or $S_I=k/2-1/2$, the exchange anisotropy enters only
through the
perturbation $J^z_{s,-1,L}\phi_s^z$ which is part of the leading
irrelevant operator
about the fixed point and so cannot be relevant.  In physical terms, as
discussed in
Secs. 3.3.1,3.4.1, and 4.2, one can understand this because for $S_I=1/2$
or $k/2-1/2$
and $M>2$, the ground state spin alternates between $k/2-1/2$ and $1/2$
for even or
odd number of sites in the NRG, or alternatively one can always arrange
to have a
$S=1/2$ ground state with appropriate boundary conditions in the CFT
analysis.
$S=1/2$ states can never have a quadrupole moment;  generically, the
next few states
in such a case have often $S=1/2$ or $S=0$ which also of course
experiences no
anisotropy to leading order.  \\

\subsubsection{Calculation of Thermodynamic Properties in the CFT
Approach} 

In this subsection we show how Affleck and Ludwig [1991b,c] were able
to compute
various thermodynamic properties within the conformal theory approach.
There are
three parts.  First, we identify the conformal transformation and
mathematics which
allows finite temperature calculations to be performed
straightforwardly given a knowledge
of the physics at the fixed point (Sec. 6.1.3.a).  Next, we outline
explicitly how the
calculation of the specific heat and magnetic susceptibility with this
finite temperature technology is carried out in Sec. 6.1.3.b,
yielding an estimate of the Wilson ratio together with explicit
confirmation of the
singular behavior of the low temperature quantities.  Finally, we
switch to a discussion
of how the ground state residual entropy is calculated which involves
sophisticated use
of the conformal invariance to relate spatial boundaries to temporal
boundaries (Sec. 6.1.3.c). \\

{\it (a) Conformal Mapping for Finite Temperature Physics}\\

As noted by Affleck and Ludwig[1991b], the low
temperature free energy of the continuum Kondo model can be written
in two different ways, first as a trace over the exponential of the
inverse
temperature times the Hamiltonian living on the half plane
$-\infty<\tau<\infty$, $0<r$, and second in a Lagrangian formulation
where
the temperature explicitly enters the imaginary time integrals.
Including a spin field $h_s$ along the $z$ axis, the Hamiltonian
description gives
$$F(T,\delta \lambda,h_s) = {-1\over \beta} \ln[Tr\exp\{-\beta(\hat
H(\delta\lambda) - h_s
\hat {\cal J}^3_{s0L})\}] \leqno(6.1.54)$$
and in the Lagrangian formulation
$$F(T,\lambda,h_s) = F(T,0,0) - {1\over \beta} ln<\exp\{\int_0^{\beta}
d\tau[\delta \lambda
(\tilde {\cal J}^a_{s,-1,L} \tilde\phi^a_s)(\tau) + h_s
\tilde{\cal J}^3_{s0L}(\tau)]\}> \leqno(6.1.55)$$
where the hat refers to the half plane geometry ${\cal G}_0$ with the
Kondo impurity living at
$r=0$, and the tilde refers to the semi-infinite cylinder geometry
${\cal G}_T$ specified by
$0<\tau<\beta$,
$0<r<\infty$ where bosonic operators are periodic in $\beta$.  Note
that the spin field has been
coupled to the total (conserved) spin operator ${\cal J}^3_{s0L}$.  As
in the discussion of
Secs. 6.2.b,c, we put $q=n\pi/l$ and index the spin density operators
by $n$ rather than $q$.
The above expectation value is with reference to the critical
Hamiltonian ($\delta \lambda=h_s=0$).
These expressions have employed the identity ${\cal J}^3_{s0L} =
(1/2\pi)\int {\cal J}^3_L(x)$
which is just the Fourier transform rule.  Hence this treatment differs
slightly from Affleck and
Ludwig [1991b] where the spin density in real space is written down.

Affleck and Ludwig [1991b] point out that the free energy given by Eqs.
(6.1.54,55) must obey
standard finite size scaling relationships.   Specifically, if we
restrict the spatial dimension
such that $0<r<l$ and $l/v_F>>\beta$, then $F$ separates into a bulk
piece that scales with
$l$, and an impurity piece independent of $l$, {\it viz.}
$$F(T,h_s,\delta\lambda ) = {l \over a} f_{bulk}(T,h_s) +
f_{imp}(T,h_s,\delta\lambda) \leqno(6.1.56)$$
where $a$ is a minimum length scale (of order $\hbar v_F/k_BT_K$
here).
The finite size scaling hypothesis (Barber, [1983]) says that the bulk
and impurity parts
may be written for $\beta\to\infty$ as
$$f_{bulk}(\beta,h_s) \approx E^{(0)}_{bulk} + {1\over \beta^2}
Q_{bulk}(\beta h_s) \leqno(6.1.57)$$
and
$$f_{imp}(T,h_s,\delta\lambda) \approx E^{(0)}_{imp} + {1\over \beta}
Q_{imp}(\beta h_s;A\beta^{-
\Delta}\delta\lambda) \leqno(6.1.58)$$
where $\Delta=1+\Delta_s$
is the scaling dimension of the leading irrelevant operator.  In
principle, all
bulk and impurity irrelevant couplings may be included in Eqs.
(6.1.57,58), but these produce
subleading corrections compared to the leading irrelevant operator.
Our interest is in
computing $Q_{imp}$ which may be evaluated about the fixed point by
expanding Eq.
(6.1.55) in powers of $h_s$ and $\delta\lambda$.  This involves
calculation of correlation
functions between various combinations of ${\cal J}^3_{s0L}$ and ${\cal
J}^a_{s,-1,L}\phi_s^a$
on the geometry ${\cal G}_T$.

\begin{figure}
\parindent=2.in
\indent{
\epsfxsize=4.in
\epsffile{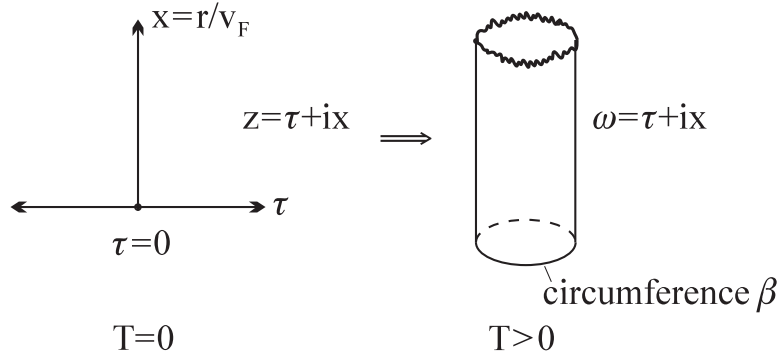}}
\parindent=.5in
\caption{Conformal mapping of zero temperature half-plane to finite
temperature cylinder.  By applying this mapping of $z\to w$, one may 
straightforwardly evaluate finite temperature correlation functions in terms
of zero temperature ones. }
\label{fig6p5}
\end{figure}

To evaluate the finite temperature correlation functions one may
exploit the conformal
transformation shown in Fig.~\ref{fig6p5} which maps ${\cal G}_0$ into ${\cal
G}_T$.
Defining
$z=\tau + ix$ in the half-plane and $w=\tau + ix$ in the
semi-infinite cylinder ($x=r/v_F$), the transformation is
$w=(\beta/\pi)\tan^{-1}(z/\tau_0)$
where $\tau_0 =a/v_F \approx \hbar /k_BT_K$ sets the short time or high
frequency cutoff
of the model.  (Note that Affleck and Ludwig [1991b] include the
$\tau_0$ factor in this
transformation only implicitly.)
As a result, at long times any boundary operator $\hat O$ in geometry
${\cal G}_0$ is related to the
corresponding operator $\tilde O$ in
geometry ${\cal G}_T$ by the identity
$$\tilde O(w) = ({dw\over dz})^{-\Delta_O} \hat O(z) = ({\beta\over
\pi\tau_0}
{1\over 1+(z/\tau_0)^2})^{-\Delta_O}
\hat O(z) \leqno(6.1.59)$$
where $\Delta_O$ is the scaling dimension of $\hat O$.  This means that
correlation functions
in the long time regime which are simple power laws in geometry ${\cal
G}_0$ may be
related straightforwardly to correlation functions at finite
temperature.  For example,
$$<\tilde O(w_1) \tilde O(w_2)> = ({\beta \over
\pi\tau_0})^{-2\Delta_O} (1+(z_1/\tau_0)^2)^{\Delta_O}
 (1+(z_2/\tau_0)^2)^{\Delta_O} <\hat O(z_1)\hat O(z_2)> $$
$$~~~~~~~~ \approx ({\beta \over \pi\tau_0})^{-2\Delta_O}
(1+(z_1/\tau_0)^2)^{\Delta_O}
 (1+(z_2/\tau_0)^2)^{\Delta_O} {A_O\tau_0^{2\Delta_O}
\over (z_1-z_2)^{2\Delta_O}} \leqno(6.1.60)$$
where $A_O$ is a normalization factor for the $O-O$ green's function.
Through the conformal mapping the $z$ variables on the
right hand side can be related to the $w$ variables on the left hand
side in a straightforward
manner.   Specifically, repeated use is made of the relation
$${(1+(z_1/\tau_0)^2)(1+(z_2/\tau_0)^2) \tau_0^2 \over (z_1-z_2)^2} =
{1\over \sin[{\pi\over\beta}(w_1-
w_2)]} \leqno(6.1.61)$$
which follows from elementary trigonometric identities.\\

{\it (b) Evaluation of the Specific Heat and Magnetic Susceptibility}
\\

The specific heat is calculated by evaluating the first non-vanishing
term in
the impurity free energy for $h_s=0$ and $\delta\lambda$ finite.  This
term is of
order $\delta\lambda^2$ because the expectation value of Eq. (6.1.55)
is with
respect to the fixed point Hamiltonian at which the average of the
leading irrelevant
operator is zero.  Hence, the leading contribution to the free energy
at zero field
is
$$\delta f^{(2)}_{imp} (T) = -{\delta \lambda^2 \over 2\beta}
\int_0^{\beta}d\tau_1
\int_0^{\beta} d\tau_2 <(\tilde{\cal J}^a_{s,-1,L}\tilde
\phi_s^a)(\tau_1) (\tilde {\cal J}^b_{s,-1,L}
\tilde \phi^b_s)(\tau_2)> $$
$$~~~~~~~ = -3({M\over 2}+2)
\delta\lambda^2 ({\pi\tau_0\over \beta})^{2(1+\Delta_s)}
\int_{\tau_0}^{\beta/2} {d\tau \over
[\sin{\pi\tau\over \beta}]^{2(1+\Delta_s)}} $$
$$~~~~~~~ =  -3({M\over 2}+2)
\delta\lambda^2 ({\pi\tau_0\over \beta})^{2(1+\Delta_s)}
I(2(1+\Delta_s),\beta)\leqno(6.1.62)$$
where Eqs. (6.1.60,61) were used, $\tau_0 \approx \hbar/k_BT_K$ is a
short time
cutoff, and the factor $3[(M/2)+2]$ corresponds to $A_O$ in Eq.
(6.1.60).  This factor
corresponds to the usual green's function numerator for equal times and
may be
evaluated straightforwardly from the Kac-Moody algebra as noted by
Affleck and
Ludwig [1991a].  For the special case of $M=2$ where , the integral in
(6.1.62)
is straightforwardly evaluated with the substitution
$z=arsh[cot(\pi\tau/\beta)]$ to
give
$$I(3,\beta) = {\beta \over \pi}  \int_0^{arsh[\beta/\pi\tau_0]} dz
cosh^2z $$
$$~~~~~~~ = {\beta\over 2\pi}
[{\beta\over\pi\tau_0/}\sqrt{1+({\beta\over\pi\tau_0})^2} +
\ln({\beta\over\pi\tau_0})+\sqrt{1+({\beta\over\pi\tau_0})^2})]~~.
\leqno(6.1.63)$$
As a result, for $\beta\to\infty$, we see that the impurity
contribution to the specific
heat goes as
$$C_{imp} = -T{\partial^2f_{imp}\over\partial T^2} $$
$$~~~~~ \approx k_B [9 (\pi \delta\lambda \tau_0)^2 {T\over T_K}
\ln({2T_K\over \pi
\sqrt{e} T})] ~(M=2)~~. \leqno(6.1.64)$$
This displays the expected $T\ln T$ singular behavior.  For $M>2$, the
specific heat
behaves as $T^{4/(2+M)}$ as was earlier calculated by the Bethe-Ansatz
( Andrei and
Destri [1985], Wiegman and Tsvelik [1985]).

Turning now to the susceptibility, we wish to extract the leading
impurity contribution to
the free energy which is also order $h_s^2$.    This must be quadratic
in both $\delta\lambda$
and $h_s$ because terms linear in $h_s$ and $\delta\lambda$ have
vanishing
expectation values from the free Hamiltonian by construction.
Hence the relevant term in the impurity free energy is found by the
expansion of
Eq. (6.1.55) to quadratic order in $h_s$ and $\delta\lambda$ yielding
$$\delta f^{(2,2)}_{imp} = {-1\over 4\beta} \delta\lambda^2 h_s^2
[\prod_{i=1}^4 \int_0^{\beta}]
<\tilde{\cal J}^3_{s0L}(\tau_1)\tilde{\cal J}^3_{s0L}(\tau_2) (\tilde
{\cal J}^a_{s,-1,L}
\tilde\phi^a_s)(\tau_3)
(\tilde {\cal J}^b_{s,-1,L}\tilde\phi^b_s)(\tau_4)>_{conn}
\leqno(6.1.65)$$
which gives $\chi_{imp} = -(\partial^2 f_{imp}/\partial h_s^2)(h_s=0)$
as
$$\chi_{imp} = {1\over 2\beta} \delta\lambda^2 [\prod_{i=1}^4
\int_0^{\beta}]
<\tilde{\cal J}^3_{s0L}(\tau_1)\tilde{\cal J}^3_{s0L}(\tau_2) (\tilde
{\cal J}^a_{s,-1,L}\phi^a_s)(\tau_3)
(\tilde {\cal J}^b_{s,-1,L}\phi^b_s)(\tau_4)>_{conn}
~~.\leqno(6.1.66)$$

The green's function in the above equation may be evaluated with the
use of the operator
product expansion (OPE) which pulls out
the singular behavior in a product of operators.  Heuristically,
this amounts to a kind of generalized Wick's theorem in the following
sense:
the bare Hamiltonian in Eq. (6.1.55)
is quadratic in the densities which obey (generalized) canonical
commutation
relations (the Kac-Moody algebra);  hence, apart from possible
non-trivial normalization factors
for the Green's functions, we might expect Wick's theorem rules to
hold.  Thus we
expect for the corresponding ${\cal G}_0$ Green's function to that of
Eq. (6.1.66) that
$$<\hat{\cal J}^3_{s0L}(\tau_1)\hat{\cal J}^3_{s0L}(\tau_2)
(\hat {\cal J}^a_{s,-1,L}\hat\phi^a_s)(\tau_3)
(\hat {\cal J}^b_{s,-1,L}\hat\phi^b_s)(\tau_4)>_{conn} =$$
$$A[{1\over (\tau_1-\tau_3)^2(\tau_2-\tau_4)^2 } + {1\over
(\tau_1-\tau_1)^2(\tau_2-\tau_3)^2 }]
{1\over (\tau_3-\tau_4)^{2\Delta_s}} \leqno(6.1.67)$$
which follows by contracting the density operators together (they
always give scaling dimension
1) and the two $\hat\phi^a_s$ operators together.  This identity does
indeed hold, and the
normalization constant $A=(2+M/2)^2$ may be fixed by a rigorous  OPE
calculation.

Employing Eq. (6.1.59) to convert the ${\cal G}_0$ correlation
functions to ${\cal G}_T$
correlation functions, and shifting limits of integration, Eq. (6.1.65)
may be manipulated
into the form
$$\chi_{imp} = 8(2+{M\over 2})^2\delta\lambda^2
({\pi\tau_0\over\beta})^{2(2+\Delta_s)}
[I(2,\beta)]^2 I(2\Delta_s,\beta) \leqno(6.1.68)$$
where
$$I(x,\beta) = \int_{\tau_0}^{\beta/2} {d\tau \over [\sin(\pi
\tau/\beta)]^x} ~~.\leqno(6.1.69)$$
Let us specialize to $M=2$.
The relevant integrals here are, for large $\beta$,
$$I(2,\beta) \approx \tau_0 ({\beta \over \pi \tau_0})^2
\leqno(6.1.70)$$
and
$$I(1,\beta) \approx {\beta\over \pi} \ln({2\beta\over \pi\tau_0})
~~.\leqno(6.1.71)$$
As a result, the estimate for the impurity susceptibility is, for $M=2$
and large $\beta$,
$$\chi_{imp} \approx 72 \delta\lambda^2 \tau_0^3 \ln({2\beta\over
\pi\tau_0})~(M=2) ~~.\leqno(6.1.72)$$
For $M>2$,
it is apparent that the integral $I(2\Delta_s,\beta)$ goes as $\beta$
for large $\beta$,
so that overall $\chi_{imp}\sim \beta^{1-2\Delta_s}\sim T^{4/(2+M)-1}$
in agreement with
the Bethe-Ansatz.

Next we turn to a calculation of the Landau-Wilson ratio.  Although the
overcompensated
model does not have a Fermi liquid fixed point, the fact that the
specific heat coefficient and
susceptibility have the same singular low temperature properties
suggests that there should
be a well defined Landau-Wilson ratio.  Calculation of this ratio
requires a knowledge of the
normalization of the bulk Hamiltonian specific heat and susceptibility
 with no impurity present.  Since there are $M$ channels,
the bulk specific heat follows from the usual Sommerfeld calculation
for $\beta\to
\infty$ as
$$C_{bulk} \approx {2\pi^2 M k_B\over 3} {k_BT\over D} \leqno(6.1.73)$$
where $D\simeq v_Fk_F$ is the bandwidth of the conduction electrons.
Note that
the normalization of Eq. (6.1.73) follows the usual solid state physics
conventions and
not the conformal theory conventions (compare to
eq. (3.6) in Affleck and Ludwig [1991b]; note also in their equation
$D$ is set to 1).
The corresponding calculation for the Pauli susceptibility with the
effective moment $\mu$
absorbed in $h_s$ gives
$$\chi_{bulk} = {2 M\over D} ~~;\leqno(6.1.74)$$
again, this normalization differs slightly from Affleck and Ludwig
[1991b] (compare with Eq.
(3.7) of this reference).  We can now specialize to the case $M=2$ and
find the Landau-Wilson Ratio as
$$R=\lim_{\beta\to\infty} {(\chi_{imp}/\chi_{bulk})\over
(C_{imp}/C_{bulk})} = {8\over 3} ~~(M=2).
\leqno(6.1.75)$$
This agrees with numerical results from the Bethe-Ansatz.  Affleck and
Ludwig [1991b]
go further to note that for arbitrary $M$ that
$$R_W = {(2+M/2)^2(2+M)\over 18} \leqno(6.1.76)$$
which follows with explicit evaluation of the integrals in Eqs.
(6.1.62) and (6.1.68).

\begin{figure}
\parindent=2.in
\indent{
\epsfxsize=4.in
\epsffile{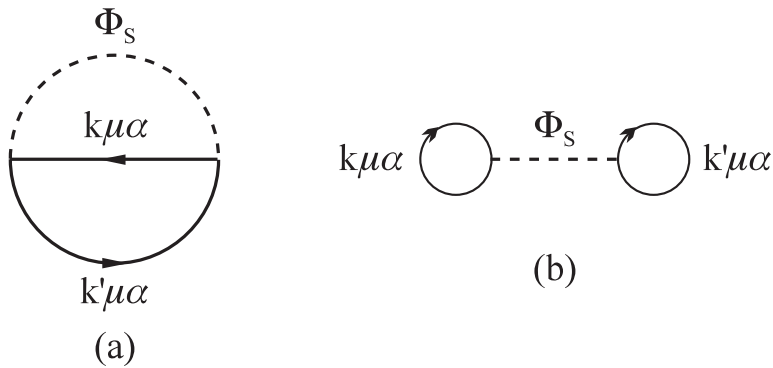}}
\parindent=.5in
\caption{Diagrammatic visualization of free energy contributions yielding 
specific heat $C_{imp}$ and susceptibility $\chi_{imp}$.  Solid lines are 
conduction electron propagators, dashed lines are local spin tensor 
$\Phi_s$ propagators.  Dots are leading irrelevant operator vertices $\delta\lambda$, 
crosses are applied spin-field vertices $h_s$. Fig. (a) is in zero applied field
and differentiated twice yields the specific heat.  Fig. (b) is in finite applied 
field and differentiated twice yields the susceptibility.  These diagrams have a close
connection to the corresponding ones in the Abelian bosonization approach discussed in 
Sec. (6.2). }
\label{fig6p6}
\end{figure}

To close this discussion of $C_{imp}$ and $\chi_{imp}$, we note that a
diagrammatic view
of Eqs. (6.1.62) and (6.1.65) for the free energy is possible. The
conceptual point behind this
is that it will illustrate a very close link to the discussion of these
quantities within the
Abelian bosonization approach discussed in Sec. 6.2.2.c.  The
appropriate diagrams are
shown in Fig.~\ref{fig6p6}, where solid lines represent density-density green's
functions, and the
dashed lines represent the $\phi_s$ green's function; dots represent
the $\delta\lambda$
vertex, and crosses represent an $h_s$ vertex.  Fig~\ref{fig6p6}(a) for the
specific heat shows that
the relevant free energy diagram is a bosonic `bubble' with one
line 
representing the $\phi_s$
propagator (which is the dynamic susceptibility for a localized
magnetic field) and one
line representing the conduction spin density propagator, which in turn
must be a continuum
electron bubble for appropriately renormalized continuum fields.  Fig.~\ref{fig6p6}(b) 
for the susceptibility
shows that the $\phi_s$ propagator gets two spin density bubbles
attached at the end.
Formally, these bubbles would have to correspond to double bubbles of
suitably redefined
continuum fermion fields. \\

{\it (c) Residual Entropy}\\

The calculation of the residual entropy requires different concepts
from the conformal theory,
drawing in particular on some mathematical properties of the so-called
``modular S-matrix''
worked out in the literature.  We shall briefly outline the calculation
here, as it appears in
Affleck and Ludwig [1991c] and Ludwig [1994a].

The entropy of the multi-channel model in any finite size calculation
will be the logarithm of
an integer, where the integer represents the ground state degeneracy.
Here finite size
means effectively that we have lowered the temperature to be comparable
to the level
spacing so that $\beta v_F/l \ge 1$.  For example, in the NRG spectra
for the two-channel
model displayed in Fig.~\ref{fig4p4}, the degeneracy is always two, which
would give an entropy
or $R\ln 2$.  In the continuum or ``high temperature limit'' where $\beta
v_F/l << 1$ relevant
to a  macroscopic system, this is not the case any longer.  In this
case, by high temperature
we only mean with respect to the quantum spacing of the levels, not to
the
Kondo scale itself (indeed we are interested in temperatures $T$
satisfying
$(\pi v_F/l)<<k_BT<<k_BT_K$).

\begin{figure}
\parindent=2.in
\indent{
\epsfxsize=4.in
\epsffile{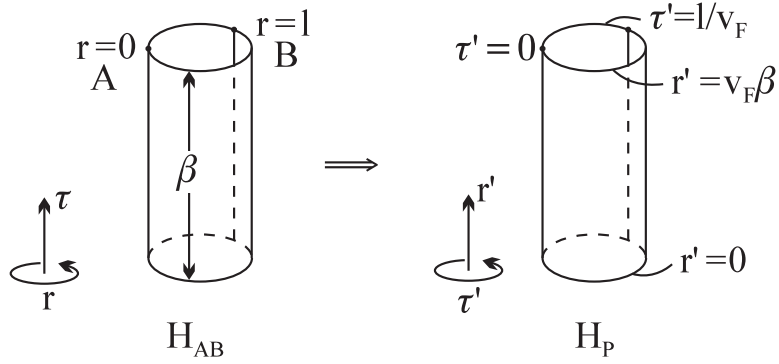}}
\parindent=.5in
\caption{Conformal transformation exploiting the modular invariance of the 
partition function.  To cross from the low temperature limit in which the 
temperature is less than the level spacing $v_F\pi/\ell$ to the ``high 
temperature limit'' in which $v_F\pi/\ell << k_BT << k_BT_K$, it is convenient
to interchange space and Euclidean time coordinates.  Because of the equivalence of 
space and Euclidean time coordinates, the transformation must preserve the 
partition function.  This trick enables a calculation of the residual entropy 
for the thermodynamic limit. }
\label{fig6p7}
\end{figure}

The conformal theory provides
an elegant approach for switching from the low temperature limit to the
``high temperature
limit'' which was introduced first by Cardy [1989].  The idea is
illustrated in Fig.~\ref{fig6p7}.  It goes
as follows. We normally calculate the partition function by tracing
over $exp(-\beta H_{AB})$
where $A,B$ are the boundary conditions on the system at $r=0,l$ in
this one dimensional
example.  As is usual, this may be viewed as a path integral summing
over imaginary time
in a periodic way, namely we `propagate' with $exp(-\beta H_{AB})$
around the
cylinder of Fig.~\ref{fig6p7} in the time direction, always winding up back at
the state at which we started.
However, the conformal invariance implies that we can interchange space
and time directions and
that we must get equivalent physical results.  Namely, we can
interchange
space and time axes, to get a new system `length' equal to
$v_F\beta$, and a new system `time'  of $2l/v_F$.  The states are
specified by a
Hamiltonian $H_P$ which has periodic boundary conditions in the new
`space' direction.

The partition function in the original case can be written in terms of
`characters' of the
conformal towers in the following form
$$Z = Tr[e^{-\beta H_{AB}}] = \sum_{QSS_c} n^{QSS_c}_{AB}
\chi_{QSS_c}(q) \leqno(6.1.76)$$
where the sum is only over primary values of the spin, channel-spin,
and charge,
where $q=\exp(-\beta \pi v_F/l)$, and the character $\chi_{QSS_c}$ is
given by
$$\chi_{QSS_c} (q) = q^{c/24} \sum_{m=0}^{\infty} d_m(QSS_c)
q^{\Delta(QSS_c)+m} ~~.
\leqno(6.1.77)$$
In Eq. (6.1.77), $c$ is the `central charge' of the theory which for
this model simply
counts the degeneracy of the conduction electrons, i.e., $c=2M$, and
$d_m(QSS_c)$ is the
degeneracy of the $m$th descendant of the primary state labeled by
$QSS_c$.  The character
$\chi$ factorizes into a product of characters for the $U(1)$,$SU(2)$,
and $SU(M)$ algebras
of the charge, spin, and channel-spin sectors.    The
integers $n^{QSS_c}_{AB}$ have the same meaning as in Secs. 6.1.2.a,b.

In the basis where space and time are interchanged, the partition
function propagates along
the new `time' axis from a state characterized by boundary condition
$A$ denoted $|A>$ to
a state characterized by boundary condition $B$ denoted $|B>$.  This
replaces the periodicity
familiar from the standard trace formula of Eq. (6.1.75).  The discrete
spacing between
states of $H_P$ is now set by the scale $2\pi/\beta$ and as a result
the partition function
is given by
$$Z=<A|\exp(-2l H_P/v_F)|B>=\sum_{\tilde Q\tilde S\tilde S_c} <A|\tilde
Q\tilde S\tilde S_c>
\chi_{\tilde Q\tilde S\tilde S_c}(\tilde q^2) <\tilde Q\tilde S\tilde
S_c|B>
\leqno(6.1.78)$$
where $\tilde q=\exp(-2\pi l/\beta v_F)$.  Clearly, as $l/\beta
\to\infty$, then only the lowest
term of the above sum will remain, and if we set the ground state
energy of $H_P$ to zero,
then we see that
$$S(0) = \ln Z(T=0) = \ln<A|0> + \ln<0|B> \leqno(6.1.79)$$
and the calculation of the ground state entropy reduces to the
calculation of the matrix
elements in Eq. (6.1.79).    Note that the mathematical transformation
between Eq. (6.1.76)
and (6.1.78) is reminiscent of a kind of generalized Poisson summation
formula.

The characters for the different realizations $H_{AB}$ and $H_P$ are
related by a
linear transformation specified by the `modular S-matrix'; here we
restrict attention to
just the spin sector which is all that is modified by the absorption of
the impurity spin,
and define the modular S-matrix by
$$\chi_{\tilde S}(\tilde q^2) = \sum_S S^{\tilde S}_s \chi_s(q)
~~.\leqno(6.1.80)$$
In general, the modular S-matrix factorizes into products of S-matrices
for the
spin, charge, and channel-spin sectors. For an $SU(2)$ level $M$
Kac-Moody algebra,
the modular S-matrix is given by (Kac and Peterson [1984])
$$S^j_{j'} = \sqrt{2\over 2+M} \sin[{\pi (2j+1)(2j'+1)\over 2+M}]
~~.\leqno(6.1.81)$$
A useful  identity involving the modular S-matrix is
$$\sum_a S^{\tilde a}_a n^a_{AB} = <A|\tilde a><\tilde a|B>
\leqno(6.1.82)$$
which follows straightforwardly from substitution of Eq. (6.1.74) on
the left hand side
of (6.1.78) and Eq. (6.1.80) for $\chi_a(\tilde q)$ on the right hand
side.
Another important result is the Verlinde formula (Verlinde [1988])
$$\sum_j S^{\tilde j}_j N^j_{s,j'} = {S^{\tilde j}_s S^{\tilde
j}_{j'}\over S^{\tilde j}_0} ~~.
\leqno(6.1.83)$$
Armed with these mathematical
results, it is now possible to obtain the ground state entropy.

The strategy is to pick $A=K$, the Kondo boundary condition, $B=F$, the
free fermion
boundary condition, and $\tilde Q=Q=0$, $\tilde S_c=S_c=0$, and $\tilde
S=0$.  Then
using the fusion rule of Eq. (6.1.48) together with Eqs. (6.1.82) and
(6.1.83) allows one
to show that
$$<K|000><000|F> = <F|000><000|F>({S_{S_I}^0\over S_0^0})
~~.\leqno(6.1.84)$$
The free fermion matrix element $<F|000>$ is one, so in view of Eqs.
(6.1.79), (6.1.81),
and (6.1.84)
we see that the residual entropy is given by
$$S(0) = \ln<K|000> = \ln[{\sin[\pi(2S_I+1)/(M+2)]\over
\sin[\pi/(M+2)]}] \leqno(6.1.85)$$
which is exactly the same result obtained from the Bethe-Ansatz (see
Sec. 7).

\subsubsection{Dynamical Properties} 

{\it (a) One Electron Green's Function and $T$-matrix}\\

We follow Ludwig and Affleck [1991c], Affleck and Ludwig [1993], and
Ludwig [1994a]
here.  The $T$-matrix is useful in calculating the electrical
resistivity since the scattering
rate is related by $\tau^{-1}(\hat k,\omega,T) = -2c_iIm T(\hat k\hat
k,\omega+i0^+,T)$, where
$c_i$ is the concentration of impurities and $\hat k$ is the
propagation
direction of an electron
with momentum $\vec k$. This expression holds to leading
order in the impurity concentration.
For $s$-wave scattering as we have assumed here, the
$\hat k$ dependence drops out.  The
$T$-matrix for an impurity at position $\vec R_i$ is defined by
$$G(\vec r_1,\vec r_2,\omega) = G_0(\vec r_1-\vec r_2,\omega) +
G_0(\vec r_1-\vec R_i,)T(\omega)G_0(\vec r_2-\vec R_i) \leqno(6.1.86)$$
where $G_0$ is the unperturbed one-electron green's function and
$G$ is the perturbed one.

In terms of the effective one-dimensional problem, the $T$ matrix
arises from
scattering of left moving electrons (incoming spherical waves) to right
moving
electrons (outgoing spherical waves) and thus shows up in the mixed LR
electron green's function.  The LL, RR electron green's functions are
unperturbed by the impurity.  Non-trivial physics shows up in the LR
green's
function due to the presence of the boundary at which the conduction
operators
mix with the local fermion operators discussed in Sec. 6.1.2.b.

At zero temperature and energy, where the effects of the leading
irrelevant operator
discussed in Sec. 6.1.2.c are unimportant,
the LR green's function is also related to the $S$-matrix projected
onto outgoing
single particle states, denoted $S_{(1)}$.  At the Fermi energy,
$T(0,0) =
(1-S_{(1)})/2i\pi N(0)$ according to standard scattering theory, so
that a
calculation of $S_{(1)}$ gives the Fermi surface value of $T(0,0)$.
In the
case of trivial potential scattering, $S_{(1)}= \exp(2i\delta)$, where
$\delta$
is the phase shift.  This also goes through for the $M/2=S_I$
compensated
Kondo model, and the unit modulus of this says that any incoming single
particle state has probability one of being scattered into an outgoing
single
particle state.  In contrast, for the multi-channel model, $|S_{(1)}|$
is generally
less than one, reflecting the unbinding of spin, charge, and channel
degrees
of freedom so that an incoming electron can generate outgoing many body
states.

A formal definition of the one-particle projected $S$-matrix follows
from the
LR green's function.  Put $z=v_F\tau + ir$,$\bar z=v_F\tau-ir$.  Then
$$G^{LR}_{\mu\alpha}(z_1,\bar z_2) =
<\Psi_{L\mu\alpha}(z_1)\Psi^{\dagger}_{R\mu\alpha}(\bar z_2)> $$
$$~~~~~~~= {S_{(1)} \over z_1 - \bar z_2} \leqno(6.1.87)$$
where we have used the fact that left movers can only depend on $z$
and right movers on $\bar z$. This follows from the fact that the
electron
fields are primary fields with scaling dimension $\Delta_F=1/2$
together
with the conformal invariance.

Eq. (6.1.85) is a special case of a more general boundary relation for
primary fields. Consider a primary field operator $O^a$ with quantum
numbers specified by $a$.  The green's function for a combination of
left moving operators is
$$<O^a_L(z_1)O^a_L(z_2)> = {1\over (z_1-z_2)^{2\Delta_a}}
\leqno(6.1.88)$$
where $\Delta_a$ is the scaling dimension.  On the other hand, the
boundary $A$ characterized by boundary state $|A>$ (c.f, Sec.
6.1.3.c)  can mix left and right
movers giving the green's function
$$<O^a_L(z_1)O^a_R(\bar z_2)>= {<a|A>\over <0|A>} {1\over (z_1-\bar
z_2)^{2\Delta_a}}
\leqno(6.1.89)$$
where $<0|A>$ is a normalization factor measuring the effect of the
identity operator on the boundary state.  Since a fermion operator has
the quantum numbers $Q=1,S=1/2$ and transforms as the fundamental
representation of $SU(M)$ which we denote simply $M$, then we see that
for $A=K$, the Kondo boundary condition
$$S_{(1)} = {<1,1/2.M|K>\over <0,0,0|K>} ~~.\leqno(6.1.90)$$

This ratio of matrix elements can be evaluated with the same kinds of
methods
employed to calculate the residual entropy of Sec. 6.1.3.c, and again
Ludwig and Affleck [1991], Affleck and Ludwig [1993], and Ludwig
[1994a] have
employed the methods of Cardy [1986a,b;1989] for boundary critical
phenomena.  Using the factorization of the modular $S$-matrix (not to
be confused here with $S_{(1)}$), together with Eq.(6.1.80), we see
that
for $A=F$, $B=K$
$$\sum_{QSS_c} S^Q_{Q'}S^S_{S'}S^{S_c}_{S_c'} n^{Q'S'S_c'}_{FK} =
<F|QSS_c><QSS_c|K>
~~.\leqno(6.1.91)$$
If we now substitute in the fusion rule of Eq. (6.1.48), and re-employ
Eq.(6.1.82)
together with the Verlinde formula Eq. (6.1.83) (Verlinde [1988]), we
can obtain the relation
$${<QSS_c|K>\over <QSS_c|F>} = {S^S_{S_I}\over S^S_0} \leqno(6.1.92)$$
which generalizes Eq.(6.1.84) used for the entropy.
We now note that for a free fermion boundary $S_{(1)}=1$, so that the
ratio
$<1,1/2,M|F>/<0,0,0|F>=1$. In view of Eq. (6.1.91) this implies
$${<1,1/2,M|K> \over <1,1/2,M|F>}{<0,0,0|F>\over <0,0,0|K>} =
{<1,1/2,M|K> \over
<0,0,0|K>} = {S^{1/2}_{S_I}S^0_0\over S^{1/2}_0
S^0_{S_I}}~~.\leqno(6.1.93)$$
Employing the formula for the modular $S$-matrix of an $SU(2)$ level
$M$
Kac-Moody algebra given in Eq. (6.1.81) together with some minor
trigonometric
manipulations, we obtain
$$S_{(1)} = {\cos[\pi(2S_I+1)/(2+M)]\over \cos[\pi/(2+M)]}
~~.\leqno(6.1.94)$$
As a result, the $T$-matrix at the Fermi level is purely imaginary and
the
zero temperature scattering rate is given by
$${1\over \tau(0,0)} = -2ImT(\omega=0^+,T=0) = {c_i (1-S_{(1)})\over
\pi N(0)} \leqno(6.1.95)$$

We now consider this formula in some special cases: \\
{\it (1) Compensated Kondo Problem ($S_I=M/2$)}.  In this case it is
easy to see that $S_{(1)}=-1$, corresponding to a phase shift of
$\pi/2$
as expected from Sec. 4, Sec. 6.1.1,Sec. 6.1.2.  Hence, $1/\tau(0,0)$
obtains the unitarity limit of $1/\pi N(0)$.  \\
{\it (2) $M>>1$}.  In this case, the perturbative treatment of Gan {\it
et al} [1993]
discusssed in Sec. 3.4.5 should hold.  We expect that the $T$-matrix
should just
represent spin-disorder scattering off an effective impurity with spin
$S_I$ with
dimensionless exchange coupling strength $N(0)J=2/M$.  If we expand Eq.
(6.1.95)
for large $M$, we obtain
$${1\over \tau(0,0)} \approx {\pi \over 4N(0)} ({2\over M})^2
S_I(S_I+1) + O({1\over M^3})
\leqno(6.1.96)$$
which fulfills our expectations. As noted in Sec. 5.1, the large $N$
NCA restricted in
an uncontrolled approximation
to $N=2$ and $M$ arbitrary gives excellent agreement
with Eq (6.1.95)
for all values of $M$.   \\
{\it (3) M=2}.  In this case, $S_{(1)}$ vanishes, so $1/\tau(0,0)$
reaches half the
unitarity limit.  What is remarkable, as stressed by Ludwig and Affleck
[1991],
is that this implies the strongest possible violation of Fermi liquid
theory:  an
incoming single particle state is scattered completely into many body
states!

In fact, the result is much stronger:  scattering of an incoming 
particle(hole) into any state with 
$2n+1$ uncorrelated particles(holes) and $2n$ uncorrelated 
holes(particles) has zero amplitude 
(Maldacena and Ludwig [1996]).  
Thus, on the surface, unitarity of the scattering amplitude is completely
violated.  This ``unitarity paradox'' is resolved when a different basis
for the conformal theory is chosen in terms of majorana fermions 
(Maldacena and Ludwig [1996]).  The 
essence of the idea is that the Hilbert space containing just physical
particle/hole occupation is incomplete, and when extended, a new fermion 
arises into which a physical fermion my scatter with unit projection. 
This new fermion has fractional occupancy in terms of the physical 
states, and is thus not detectable with any external probes.  We shall
give a brief discussion of this approach in Sec. 6.3.  

Affleck and Ludwig (Ludwig and Affleck [1991], Affleck and Ludwig
[1993]) have
also computed the leading corrections to the scattering rate in
temperature and
frequency.  This calculation involves perturbing the one-electron
green's function
linearly in the leading irrelevant operators $\vec {\cal
J}_{s,-1,L}\cdot \vec \phi_s$
(and, for $M=2,S_I=1/2$, $\vec {\cal J}_{c,-1,L}\cdot \vec \phi_s$ ).
The full calculation is
quite complex and exploits the operator product expansion (OPE) method
extensively; since it is too long and involved to outline here, we
refer the reader
to Affleck and Ludwig [1991] for details.
The main result is that the resistivity will experience a correction
proportional to $\delta\lambda T^{\Delta_s}$ (which power law is in
agreement with the NCA
treatment of Sec. 5.1), so that the sign of the deviation is determined
by
whether one approaches the fixed point coupling strength from above or
below.
In the special case of $M=2,S_I=1/2$ they find that the resistivity is
given by
$${\rho(T)\over \rho(0)} \approx [1 + 4\delta\lambda \tau_0 \sqrt{\pi
T\tau_0}
 + O(T)] ~~.\leqno{6.1.97}$$
There is thus a universal amplitude relation between the square of
$T^{1/2}$
term and the coefficient of the $\ln T$ divergence in $C_{imp}/T$.

Affleck and Ludwig [1993] also consider the effects of potential
scattering, and
show that while for the $M=2,S_I=1/2$ case it does not affect the zero
temperature resistivity, it does affect the magnitude of the $\sqrt{T}$
term and
can also induce a thermopower which goes as $\sqrt{T}$.   \\

{\it (b) Local Field Dynamical Susceptibility}\\

Here we employ the conformal theory technique to calculate the local
spin and
channel spin susceptibilities for the $S_I=1/2,M=2$ case to confirm the
suspicions
that they have marginal fermi liquid form (Cox [1988]; Varma {\it et
al.}
[1989]; Tsvelik [1990]; Emery and Kivelson [1992]:
Cox and Ruckenstein [1993]; Gan, Coleman, and Andrei [1993]).
We simply show the calculations for the $\phi_s$ case; they are
completely
analogous for the channel field.  This calculation has been carried out
in Ludwig and Affleck [1994].  

The idea is to consider a dynamical local field $h_s(\omega)$ which is
coupled only
to $\phi^3_s$. At finite temperature, the response to this field will
simply be the $\phi_s$ green's function given by
$$<\phi^3_s(\tau)\phi^3_s(0)> = { \pi\tau_0 \over\beta} {1\over
sin[\pi\tau/\beta]} ~~.\leqno(6.1.98)$$
It is trivial to analytically continue this to real times, and if we
Fourier transform to
obtain the absorptive response function $\chi''_s(\omega,T)$
we obtain
$$\chi''_s(\omega,T) = { 2\pi\tau_0\over \beta} \int_0^{\infty} dt
{sin(\omega t) \over
sinh[\pi t /\beta]} ~~.\leqno(6.1.99)$$
This integral is easy to evaluate, and the result is
$$\chi''_s(\omega,T) = 2\tau_0 Im\psi({1\over 2} - {i\beta\omega\over
2\pi}) $$
$$~~~~~~ = \pi \tau_0 tanh({\beta\omega\over 2}) \leqno(6.1.100)$$
where $\psi(x)$ is the digamma function.  This is precisely the
marginal fermi
liquid form of the local dynamic susceptibility (Varma {\it et al.}
[1989]).  When
analytically continued to the real axis and set to zero frequency it
reproduces the
log divergence of the static susceptibility.  \\

{\it (c) Other Dynamic Response Functions}\\

In addition to the single electron green's function, Affleck and Ludwig
(Ludwig and Affleck [1991], Affleck and Ludwig [1994a,b]) have computed
a number of two-particle electron response functions.  The
functions have complicated and non-intuitive forms reflecting the
non-trivial
boundary condition and non-Fermi liquid fixed point.   We simply survey
the results
here obtained for the $M=2,S_I=1/2$ case:\\
{\it (1) Spatial-temporal dependent spin polarization response}  This
function contains
non-Fermi liquid response that could in principle be sampled by Knight
shift measurements
for a magnetic impurity and nuclear
electric-field gradient measurements for a quadrupolar
kondo impurity.  \\
{\it (2) Spatial-temporal dependent pair field response function}.
This function displays
the expected singularities for the $Q=2,S=0,S_c=0$ field obtained by
antisymmetrizing
in the spatial index $r$, which corresponds to antisymmetrizing in
$L,R$ indices near
the impurity site in view of the mirroring condition
$\Psi_L(r)=\Psi_R(-r)$.  Equivalently
we may go right to the impurity site and antisymmetrize in imaginary
time in view
of the conformal invariance.
 Specifically, in terms of the
three dimensional field operators $\Psi_{\mu\alpha}(\vec r)$, the full
three
dimensional pairing field is an orbital $p$-wave which is a
singlet in spin and channel indices with
$$P^a = \sigma^{(2)}_{\mu\nu}\sigma^{(2)}_{\alpha\beta}
\Psi_{\mu\alpha}(\vec r)
{\partial \over \partial r_a} \Psi_{\nu\beta}(\vec r) = {ik_F r^a \over
8\pi^2r^3}
[\psi_{L\mu\alpha}(r)\psi_{R\nu\beta}(r) -
\psi_{R\mu\alpha}(r)\psi_{L\nu\beta}(r)]
~~ \leqno(6.1.101)$$
We shall provide a more complete discussion of pairing correlations
in Sec. 9.4.

\subsection{Abelian Bosonization Approach to the Two-Channel Kondo
Model} 

In this subsection, we describe the Abelian bosonization approach to
the
two-channel Kondo model developed by  Emery and Kivelson.   The central
idea which makes
this work is that when using a spinless Fermion representation for the
impurity spin, the $x$ and
$y$ components corresponding to Majorana or ``real fermions''.  The
Kondo Hamiltonian becomes,
for a special value of the longitudinal exchange coupling, a one
particle resonant level model
in the space of conduction electrons plus one of the Majorana
variables.  This model is
exactly soluble.  The other
Majorana variable decouples.  All singularities and anomalies in
this approach are seen to be due to the presence of the uncoupled
field.   In a counting sense which
can be made mathematically precise, each Majorana fermion is half a
full fermion.  This leads to a
nice interpretation of the $R/2\ln2$ entropy (this corresponds
to the decoupled variable)
and (possibly) of the scattering rate value at the Fermi energy  to one
half
the unitarity limit (deriving from unitarity scattering from half a
fermion!).  However, as we shall
discuss, a full interpretation of transport properties remains
problematic in this approach.  (Recently Fabrizio and Gogolin [1994]
have
applied similar ideas to the four channel spin 1/2 model to illustrate
its equivalence to a model where two spin 1 conduction channels couple
to the impurity and to calculate the low temperature properties.)

\subsubsection{Model, Mapping to a Resonant Level Hamiltonian} 

Emery and Kivelson begin with the two-channel Kondo model in the
anisotropic (xxz) limit,
taking only left moving fermions living on $-\infty<x<+ \infty$
(the incoming s partial wave states reflected about the origin) coupled
to
the impurity, following Affleck and Ludwig [1991(b)].  Thus, $x<0$ is
for incoming waves
and $x>0$ is for outgoing waves. Following their
notation, the Hamiltonian is
$$H = iv_F\sum_{\mu\alpha}\int_{-\infty}^{\infty} dx
\psi_{\mu\alpha}^{\dagger}(x) {\partial \psi_{\mu\alpha}(x)
\over \partial x} + {1\over 2} \sum_{\mu\nu\alpha\lambda}J_\lambda
\tau^{\lambda}
\sigma^{\lambda}_{\mu\nu}
\psi_{\mu\alpha}^{\dagger}(0)\psi_{\nu\alpha}(0) \leqno(6.2.1)$$
where $\mu,\nu$ are conduction spin indices, $\alpha$ is the channel
index,
$\sigma^{\lambda}$
are Pauli matrices, and the $\tau^\lambda$
are the spin 1/2 impurity operators.  We assume $J_x=J_y$, $J_z>0$ to
ensure the Kondo
effect (c.f. Sec. 3.3).  In the presence of a bulk magnetic field $h$,
we add to this
the term (Eq. 2.10 of Emery and Kivelson [1992] and Eq. (13) of
Sengupta and Georges
[1994])
$$H_{Zeeman} = -h\{\tau^z + {1\over 2}
\sum_{\mu\alpha} \int_{-\infty}^{\infty} dx \sigma^{z}_{\mu\mu}
\psi_{\mu\alpha}^{\dagger}(x)\psi_{\mu\alpha}(x)\}  ~~.\leqno(6.2.2)$$
We will also consider the possibility of an impurity field in which the
second term of Eq. (6.2.2)
is absent.

Now a sequence of transformations are applied to Eq. (6.2.1) to map
this into a resonant
level Hamiltonian, which go as follows:\\
{\it (1) Bosonization}.  Following the conventions of Bander [1976],
the ``massless''
(linear dispersion relation) fermion fields are replaced
by bosons, through the relation
$$\psi_{\mu\alpha}(x) = {1\over \sqrt{2\pi
a}}\exp(-i\Phi_{\mu\alpha}(x)) \leqno(6.2.3)$$
where
$$\Phi_{\mu\alpha}(x) = \sqrt{\pi}
\int_{-\infty}^{x}dx'\{\Pi_{\mu\alpha}(x) - \phi_{\mu\alpha}(x)\}
\leqno(6.2.4)$$
and the Bose fields $\phi,\Pi$ obey canonical commutation relations
$$[\phi_{\mu\alpha}(x),\Pi_{\nu\beta}(x')] =
i\delta_{\mu\nu}\delta_{\alpha\beta} \delta(x-x') ~~.
\leqno(6.2.5)$$
The length $a$ appearing in Eq. (6.2.3) is essentially the lattice
constant, which we take to
zero in the full continuum limit.  We shall look for physical
quantities independent of $a$.
The prescription of (6.2.3) should be familiar to any reader used to
the Jordan-Wigner
transformation.  Using it, for example, it is easy to see that by
applying the Baker-Hausdorff
lemma (c.f., sec. 4.3 of Fradkin [1991]).
$$\psi_{\mu\alpha}(x)\psi_{\mu\alpha}(x') =
\psi_{\mu\alpha}(x')\psi_{\mu\alpha}(x)
e^{(-\pi[\Phi_{\mu\alpha}(x),\Phi_{\mu\alpha}(x')])}=
\psi_{\mu\alpha}(x')\psi_{\mu\alpha}(x)
e^{(i\pi[\theta(x'-x)-\theta(x-x')])} \leqno(6.2.6)$$
so that the Fermion anticommutation relation is satisfied so long as
$x\ne x'$; clearly
from idempotence of the $\psi$ fields, the ambiguity in the exponent
for $x=x'$ is
irrelevant in this case.  However, for proper derivation of all the
commutation relations we
must take care at equal spatial separations as singular contributions
will
arise (Bander [1976]).  Using the boson operators, the free Fermion
hamiltonian becomes
a free boson Hamiltonian (sum of harmonic oscillators),
$$H_{free} = {v_F\over 2} \sum_{\mu\alpha}\int dx
[\Pi^2_{\mu\alpha}(x)+
({\partial \phi_{\mu\alpha}(x)\over \partial x})^2] \leqno(6.2.7)$$
and the exchange term becomes
$$H_{Kondo} = {J^z\over 2\pi} \tau^z\sum_{\mu\alpha}\sigma^z_{\mu\mu}
 [{\partial \Phi_{\mu\alpha}\over \partial x}]_{x=0} + {J_x \over 4\pi
 a}
\sum_{\lambda=x,y}\sum_{\mu\nu\alpha} \tau^{\lambda}
\sigma^{\lambda}_{\mu\nu}
\exp[i(\Phi_{\mu\alpha}-
\Phi_{\nu\beta})] ~~.\leqno(6.2.8)$$
The reason that the $\Phi$ gradient comes in the $z$-axis exchange
coupling
is because of the
singularities in products of the exponential operators at zero
separation (Bander [1976]).
Specifically,
$$\lim_{x\to x'}\psi^{\dagger}_{\mu\alpha}(x)\psi_{\mu\alpha}(x') \sim
\{\exp(i\sqrt{\pi}[
\Phi_{\mu\alpha}(x)-\Phi_{\mu\alpha}(x')]) - 1\} \leqno(6.2.9)$$
from which the origin of the derivative term is clear.  All fermion
number and current operators,
in fact, can be expressed as linear forms of bosonic operators.  \\
{\it (2) Canonical Transformation to Collective Coordinates}.  Under a
canonical transformation of
boson coordinates that preserves the commutation relation structure of
Eq. (6.2.5), the free boson
Hamiltonian will be unchanged.  However, with a suitable transformation
we may greatly
simplify the interaction term $H_{Kondo}$.  The choice made by Emery
and Kivelson [1992] is to write
the original bose fields in terms of ``collective'' coordinates
describing charge ($\Phi_c$),
spin ($\Phi_s$), channel or
flavor ($\Phi_{sf}$), and mixed spin/flavor degrees of freedom
($\Phi_{sf}$).   The definitions are
$$\Phi_c = {1\over 2} \sum_{\mu\alpha} \Phi_{\mu\alpha}
\leqno(6.2.10.a)$$
$$\Phi_s = {1\over 2} \sum_{\mu\alpha} \sigma^z_{\mu\mu}
\Phi_{\mu\alpha} \leqno(6.2.10.b)$$
$$\Phi_f = {1\over 2} \sum_{\mu\alpha} \sigma^z_{\alpha\alpha}
\Phi_{\mu\alpha} \leqno(6.2.10.c)$$
$$\Phi_{sf} = {1\over 2} \sum_{\mu\alpha} \sigma^z_{\mu\mu}
\sigma_{\alpha\alpha} \Phi_{\mu\alpha} \leqno(6.2.10.d)$$
which may readily be inverted to give
$$\Phi_{\uparrow+} = {1\over 2} [\Phi_c + \Phi_s + \Phi_f + \Phi_{sf}]
\leqno(6.2.11.a)$$
$$\Phi_{\uparrow-} = {1\over 2} [\Phi_c + \Phi_s - \Phi_f- \Phi_{sf}]
\leqno(6.2.11.b)$$
$$\Phi_{\downarrow+} = {1\over 2} [\Phi_c - \Phi_s + \Phi_f -
\Phi_{sf}] \leqno(6.2.11.c)$$
$$\Phi_{\downarrow-} = {1\over 2} [\Phi_c - \Phi_s - \Phi_f +
\Phi_{sf}] ~~.\leqno(6.2.11.d)$$
In terms of these operators, the conduction spin densities
$s^{\lambda}(0)$ are given by
$$s^z(0) = {1\over 2}\sum_{\mu\alpha} \sigma^z_{\mu\mu}
\psi^{\dagger}_{\mu\alpha}(0)
\psi_{\mu\alpha}(0) = [{\partial \Phi_s\over\partial x}]_{x=0}
\leqno(6.2.12.a)$$
$$s^x(0) = 2\cos(\Phi_s(0))\cos(\Phi_{sf}(0)) \leqno(6.2.12.b)$$
$$s^y(0) = 2\sin(\Phi_s(0))\cos(\Phi_{sf}(0)) \leqno(6.12.c)$$
so that
$$H_{Kondo} = {J_z \over \pi} \tau^z  [{\partial \Phi_s\over\partial
x}]_{x=0} +
{J_x \over \pi a} [\tau^x \cos(\Phi_s(0)) + \tau^y
\sin(\Phi_s(0))]\cos(\Phi_{sf}(0))~~. \leqno(6.2.13)$$
Notice that the sines and cosines in the above equation are already
suggestive
of ``real'' fermions, in that we know pure complex exponentials obey
fermionic
commuation relations, so the idea is to write, schematically,  $\psi
\sim \psi_R + i \psi_I$,
which would correspond to the sine and cosine factors.  Notice also
that the transverse coupling
in the above equation is in a position for simplification by a
rotation about the $z$-axis, which
in fact is the next step.\\
{\it (3) Unitary Transformation}.  $\Phi_s$ can be eliminated from the
transverse Kondo coupling by
rotating the impurity pseudospin about the z-axis through the angle
$-\Phi_s(0)$.  This corresponds
to applying the unitary transformation $UHU^{-1}$ to the Hamiltonian
with
$U=\exp(i\tau^z\Phi_s(0)/2)$.  When applied to the free particle term
we obtain
$$UH_{free}U^{-1} = H_{free} - v_F\tau^z [{\partial \Phi_s\over\partial
x}]_{x=0} \leqno(6.2.14)$$
$$UH_{Kondo}U^{-1} = -{J_z\over \pi} \tau^z[{\partial \phi_s\over
\partial x}]_{x=0}
+ {J_x\over \pi a} \tau^x \cos\phi_{sf}(0) \leqno(6.2.15) $$
and when applied to the Zeeman energy we obtain (Sengupta and Georges
[1994])
$$UH_{Zeeman}U^{-1} = -{h\over 2}
\sum_{\mu\alpha} \int dx_{-\infty}^{\infty} \sigma^{z}_{\mu\mu}
\psi_{\mu\alpha}^{\dagger}(x)\psi_{\mu\alpha}(x) = -{h\over 2\pi}
\int_{-\infty}^{\infty} dx
 {\partial \phi_s(x)\over \partial x} \leqno(6.2.16)$$
so that the impurity coupling to the field ($-\tau_z$) drops out!  This
result shall be used below in
discussing thermodynamics.\\
{\it (4) Fermionization}.  Now  the Hamiltonian can be
``re-fermionized''
in terms of the
fermion operators corresponding to the collective coordinates, given by
e.g.,
$\psi_c = \exp(  -i\Phi_c)/\sqrt{2\pi a}$.
The Hamiltonian may be written as a sum of  terms from  each of the
$c,s,f,sf$ sectors. The $c$ and $f$ sectors decouple from the impurity
and are simply
free fermion Hamiltonians.  In addition to the re-fermionization of the
conduction fields,
the impurity spin field operators may be expressed in terms of a
spinless
fermion representation with
creation and annihilation operators $d^{\dagger},d$ such that the
occupied fermion state
corresponds to up spin and the empty state to down spin. In terms of
these operators,
$$\tau^z = d^{\dagger}d - 1/2 \leqno(6.2.17.a)$$
$$\tau^+ = d^{\dagger} \leqno(6.2.17.b)$$
$$\tau^- = d ~~.\leqno(6.2.17.c)$$
It is clear that the commuation relations
$[\tau^i,\tau^j]=i\epsilon_{ijk}\tau^k$ are faithfully
reproduced by this choice.
The operators $\tau^x,\tau^y$ are then proportional to the real or
Majorana fermion variables $\hat a
,\hat b$ given by
$$\hat a = {1\over\sqrt{2}} [d+ d^{\dagger}] = \sqrt{2}\tau^x
\leqno(6.2.18.a)$$
and
$$\hat b = {1\over i\sqrt{2}} [d^{\dagger}-d] = \sqrt{2}\tau^y
~~.\leqno(6.2.18.b)$$
The normalization conditions on $\hat a,\hat b$ are $\hat a^2 = \hat
b^2 = 1/2$.  In terms of
these operators $\tau^z = -i\hat a\hat b$.

 In terms of these new fermion operators, the $sf,s$
 sector
Hamiltonians may be written as
$$H_{sf} = \int dx \psi_{sf}^{\dagger}(x){\partial \psi_{sf}(x)\over
\partial x} + {i J_x \over \sqrt{\pi a}}
[\psi_{sf}(0)+\psi_{sf}^{\dagger}(0)] \hat b \leqno(6.2.19)$$
and
$$H_s = \int dx \psi_s^{\dagger}(x) {\partial \psi_{s}(x)\over \partial
x} - 2i(J_z - \pi v_F)
\hat a\hat b \psi^{\dagger}_s(0)\psi_s(0) ~~.\leqno(6.2.20)$$
We make three notes about Eqs. (6.2.19,20):\\
(i) If we work at the special point $J_z=\pi v_F$, the coupling of the
impurity to
$[\partial \phi_s/\partial x]_0]$ drops out (see eqns.
(6.2.15,6.2.16)).
As a result, only $H_{sf}$
includes any coupling to
the impurity, and at that only to the $\hat b$ Majorana fermion.  Since
this hamiltonian is a
quadratic form (it is a particular realization of the resonant level
model),
the properties may be exactly solved for at this special point in
coupling space.  Nearby points in coupling space can be reached by
performing perturbation theory in $\lambda = J_z- \pi v_F$. \\
(ii) A technical point about (6.2.19) which is glossed over by Emery
and Kivelson [1992]
 is that to arrive at this
properly Hermitian form, another unitary transformation
$U=\exp(-i\pi\tau^z/4)$ must be performed,
since $\hat b \sim \tau^y$, not $\tau^x$.  This unitary transformation
doesn't affect any of the
conduction fields. The need for this transformation is related to the
desire for $\psi_{sf},\hat b$ to
anticommute.  If instead we keep the $\cos\Phi_{sf}(0)$ term in Eq.
(6.2.15),
we would not need the
additional unitary transformation.  Eq. (19) of Sengupta and Georges
[1994] must be amended
in this regard.\\
(iii) The coupling to Majorana fermions is quite novel; one Majorana
unit of $\Psi_{sf}$ fermion
is hybridized with the $\hat b$ fermion.  In contrast, the single
channel spin 1/2
model through similar tricks may be mapped to a spinless resonant level
model with an ordinary
hybridization term $\sim \Psi^{\dagger}(0)d + d^{\dagger}\Psi(0)$
(Toulouse [1970],
Schlottmann [1979], Wiegman and Finkelshtein [1979]).
All impurity degrees of freedom
couple into the conduction electrons in this latter case, while the
$\hat a$ field is left over in
the two-channel case.

\subsubsection{Thermodynamics } 

We now review the derivation of the thermodynamic properties within
this approach.

{\bf (a) Green's Functions}\\

The thermodynamics of the model to the extent that we perturb in powers
of $\lambda$ may
be specified completely in terms of the Green's functions of $H_{sf}$.
These may be obtained
by equations of motion quite straightforwardly, which we present in
App. IV.   An important feature is that anomalous Green's
functions appear for the $\psi_{sf}$ fields because $\hat b$ couples to
both $\psi_{sf},\psi^{\dagger}_{sf}$.  Also, the $b$ fermion acquires a
width $\Gamma = J_x^2/(\pi v_F a)$ \\

{\bf (b)  Entropy}\\

A crucial point in dealing with
the Majorana variables is that whenever we count their spectral weight
we must include a factor
of 1/2 relative to ordinary fermionic degrees of freedom.  To
understand this, we shall evaluate the
the impurity free energy
in the $J_x=J_z=0$ limit by the complicated procedure of using the
spinless fermion representation
Green's functions.
The imaginary time $d$ fermion propagator is given
by
$$G_d(\tau) =
-<T_{\tau} d(\tau)d^{\dagger}(0)> = {1\over \beta}
\sum_{\omega_n}{e^{i\tau\omega_n}\over i\omega_n} \leqno(6.2.21)$$
where $\omega_n=(2n+1)\pi/\beta$ is a Fermionic Matsubara frequency,
and
from which we infer the spectral density $A_d(\epsilon) = \pi
\delta(\epsilon)$ and the
free energy
$$F_d = -k_BT \int {d\epsilon \over \pi} A_d(\epsilon) \ln(1 +
e^{-\beta\epsilon}) \leqno(6.2.22)$$
$$~~~~~ = -k_BT \ln 2~~.$$
Now, on the other hand,
$G_d$ may be expressed in terms of the Majorana Green's
functions $G_a(\tau) = -<T_\tau \hat a(\tau)\hat a>,$
$G_b(\tau)
= -<T_\tau \hat b(\tau)\hat b>$ we have $G_d = (G_a+G_b)/2$, so that
each Majorana fermion
contributes
$(k_B/2) \ln 2$ of entropy to the free energy of Eq. (6.2.22).

Turning now to the case $\lambda=0,J_x\ne 0$, we denote the spectral
functions of the $\hat a,\hat
b$ fields as $A_a(\omega) = \pi \delta(\omega)$, and $A_b(\omega) =
\Gamma/(\omega^2+\Gamma^2)$.  Using fermion statistics and the weights
identified above, we can write the impurity entropy as
$$S(T) = S_a(T) + S_b(T) = -k_BT \sum_{i=a,b} \int {d\epsilon \over
2\pi}
A_i(\epsilon) [f(\epsilon)\ln f(\epsilon) + (1-f(\epsilon))
\ln(1-f(\epsilon)) ] \leqno(6.2.23)$$
$$~~~~ = {k_B\over 2}\ln 2 - k_B\int {d\epsilon\over 2\pi} {\Gamma\over
\epsilon^2+\Gamma^2}
[f(\epsilon)\ln f(\epsilon) + (1-f(\epsilon))
\ln(1-f(\epsilon)) ] ~~.$$
The first term above is due to the $\hat a$ field, the second term from
the $\hat b$ field.  It is easy
to see that for $T>>\Gamma$, the second term tends to $(k_B/2)\ln 2$,
so the full $k_B\ln 2$
entropy of the impurity is recovered.  On the other hand, for
$T<<\Gamma$, we rewrite the
second term in a form amenable  to the Sommerfeld expansion
$$S_b(T) = k_B \int {d\epsilon\over 2\pi} ({-\partial f\over \partial
\epsilon}) (\beta \epsilon) tan^{-1}
({\epsilon\over \Gamma}) \leqno(6.2.24)$$
$$~~~~ \approx {\pi^3 k_B\over 6} ({k_BT\over \Gamma}), ~~T\to 0 ~~.$$
The above expression for the entropy may be obtained by differentiation
of Eq. (3.5) in
Emery and Kivelson [1992] with the field $H$ set to zero.

Hence, for $\lambda=J_z-\pi v_F = 0$ we see that the unusual residual
entropy of the
two-channel model is
understood as arising from the decoupled Majorana degree of freedom
($S_a$).  However, the
specific heat is analytic in the temperature, which will be remedied in
Sec. 6.2.2.c by
performing perturbation theory in $\lambda$.

An interesting question is the extent to which this idea can be pushed
with other multichannel
models to understand the residual entropy.  It is clear that a Majorana
representation alone
will not be sufficient.  For example, as mentioned previously, for the
three channel spin 1/2 model the residual entropy is
$-(R/2)\ln([\sqrt{5}+3]/8])$ which is not
simply related to the entropy of a single Majorana field
$R\ln\sqrt{2}$.    However, it is conceivable
that a different kind of composite field might be developed to describe
the three channel case,
which may give a description with a similar flavor to the above
decomposition.  \\

{\bf (c) Thermodynamics at $\lambda\ne 0$}\\

The thermodynamics at $\lambda\ne 0$ has been considered by Emery and
Kivelson [1993, see
footnote [6]; 1994], by D. Clarke {\it et al.} [1993], and by Sengupta and
Georges [1994].  We
follow the more extensive discussion of the latter paper here.  Now we
include the bulk field term
$H_{Zeeman}$ and recall that after the unitary transformation only the
coupling to the conduction
degrees of freedom remains.

\begin{figure}
\parindent=2.in
\indent{
\epsfxsize=4.in
\epsffile{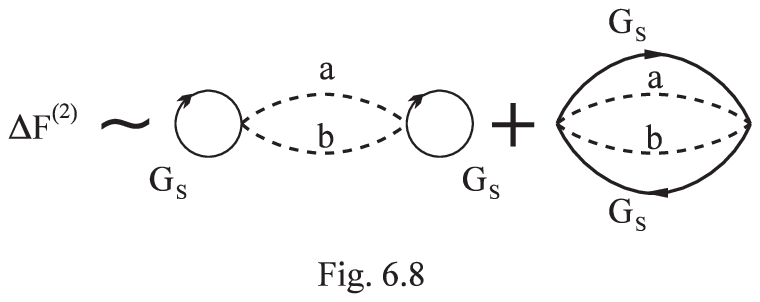}}
\parindent=.5in
\caption{Leading order in $\lambda$ contributions to the impurity free
energy in the Abelian bosonization scheme of Emery and Kivelson [1992,1993]. 
Solid lines are conduction ``spin fermion'' propagators, evaluated in the
free ($\lambda=0$) limit.  Dashed lines are local Majorana fermion propagators. 
The first term is second order in the applied field and gives rise to the 
susceptibility (which is zero at the Emery and Kivelson point), while the 
second term gives the leading logarithmic contribution to the specific heat
away from the Emery and Kivelson point. Since the convolution of the $a,b$
Majorana propagators gives the local $z$-axis spin susceptibility, it is clear that 
these free energy 
diagrams have a close correspondence to those of the conformal theory (Fig.~\ref{fig6p6})
modulo the assumed anisotropy here.}
\label{fig6p8}
\end{figure}

Using standard perturbation theory methods, two terms arise in the free
energy to second order
in $\lambda$. These are shown in Fig.~\ref{fig6p8}, and are given by
$$\Delta F_{imp} = -{\lambda^2 \over 2} [ G_s^2(0,h)
\int_0^{\beta}d\tau G_a(\tau)G_b(\tau)
+ \int_0^{\beta} d\tau G_s^2(\tau,h) G_a(\tau)G_b(\tau) ]
~~.\leqno(6.2.25)$$
It is easy to see that
$$G_s(0,h) = {1\over \beta} \sum_{\omega} \int {dk\over 2\pi a}
{e^{i\omega 0^+}\over i\omega
-v_F k - h} \approx {h \over 2\pi v_F} \int d\epsilon ({-\partial
f(\epsilon)\over \partial \epsilon}) =
{h\over 2\pi v_F} ~~.\leqno(6.2.26)$$
Note that $1/2\pi v_F$ is simply the Fermi level density of states
$N(0)$.  The first term in Eq. (6.2.25) will
contribute the leading $h$ dependence, while the second term will
contribute the leading $T$
dependence at $h=0$; henceforth we set $h$=0 in the second term.

The first term in Eq. (6.2.25) is easy to evaluate in frequency space.
Decomposing
in Fourier modes, the integral can be
written
$$\int_0^{\beta}d\tau G_a(\tau)G_b(\tau) = -{1\over \beta}
\sum_{\omega} {1\over i\omega(
i\omega + i\Gamma sgn\omega )} \leqno(6.2.27)$$
$$~~~~ = {1\over 2\pi\Gamma} [\Psi({1\over 2} + {\beta \Gamma\over
2\pi}) - \Psi({1\over 2})] \approx {1\over 2\pi \Gamma} \ln(1.13\beta
\Gamma) ~~.$$
The second integral is easier to evaluate in the time domain (Sengupta
and Georges [1994]),
using
$$G_a(\tau) = -{1\over 2} sgn(\tau),~~G_b(\tau) \approx {1\over
\beta\Gamma \sin(\pi\tau/\beta)},~~
G_s(\tau) \approx {1\over 2\beta v_F \sin(\pi\tau/\beta)}
\leqno(6.2.28)$$
where the expressions for $G_b,G_s$ are valid for $T<<\Gamma$ and
$\pi/\Gamma
<<\tau<<\beta$. Clearly the expressions for $G_b,G_s$ are symmetric
about $\tau = \beta/2$, and
so we cutoff at $\pi/\Gamma$ below and $\beta-\pi/\Gamma$ above in the
integral.  This
arbitrariness will not affect the amplitude of the most singular piece
in the second term, but will
affect the argument of the logarithm.   The
result is
$$I_2 = \int_0^{\beta} G_a(\tau)G_b(\tau)G_s^2(\tau,0) \approx
{(k_BT)^2 \over 4\pi v_F^2 \Gamma}
\int_{\pi/\beta\Gamma}^{\pi/2} dx {1\over \sin^3(x)}
~~.\leqno(6.2.29)$$
The latter integral may be evaluated using
$$\int_y^{\pi/2} {dx\over \sin^3(x)} = {1\over 2} arsh[cot(y)] +
{1\over 2} cot(y)csc(y) \leqno(6.2.30)$$
to give
$$I_2 \approx {1\over 4\pi^3 v_F^2\Gamma}[{\pi^2(k_BT)^2\over
2}\ln({\Gamma\over k_BT})
+ {\Gamma^2\over 2}] ~~.\leqno(6.2.31)$$

Putting these results together, we see that the most singular terms in
the
impurity contribution to the free energy from the $s$
sector to order $\lambda^2$ are  given by
$$\Delta F_{imp} = - {[N(0)\lambda]^2\over 2\pi\Gamma} [h^2
\ln(1.12\beta\Gamma) +
{\pi^2 (k_BT)^2 \over 2} \ln(\pi\beta\Gamma)]  \leqno(6.2.32)$$
from which we may immediately read off the impurity susceptibility
$\chi_{imp}$
and specific heat $C_{imp}$ as
$$\chi_{imp}(T) = -{\partial ^2\Delta F_{imp}\over \partial h^2}
\approx {[N(0)\lambda]^2\over
\pi\Gamma} \ln(1.13\beta\Gamma) \leqno(6.2.33)$$
and
$$C_{imp}(T) = -{\partial ^2\Delta F_{imp}\over \partial T^2}  \approx
k_B {[N(0)\lambda]^2 \over
2\pi \Gamma} k_BT \ln(\pi\beta\Gamma) ~~.\leqno(6.2.34)$$

This allows us to compute the Wilson ratio; we see that
$$\lim_{T\to 0} {T\chi_{imp}\over C_{imp}} = {2\over \pi^2 k_B^2}
~~.\leqno(6.2.35)$$
The corresponding bulk value which sets the free fermion scale is
$3/4\pi^2k_B^2$ (Affleck
and Ludwig [1991(b)]).
Hence, the Wilson ratio of the diverging susceptibility and specific
heat coefficients is
$$R = \lim_{T\to 0} {\chi_{imp}\over \chi_{bulk}}{C_{bulk}\over
C_{imp}} = {8\over 3} \leqno(6.2.36)$$
in perfect agreement with the general results obtained by the
Bethe-Ansatz and conformal field theory. Note that the present
derivation is
valid only for $\lambda/v_F<<1$ (see also Emery and Kivelson [1993]).
\\

{\bf (d) Impurity Susceptibility}\\

It is useful also to consider the case of a magnetic field which
couples only to the impurity site,
so the coupling is of the form $-H\tau^z = -iH\hat a\hat b$.  We denote
the corresponding
susceptibility $\chi_I$.  Within linear response, we see that
$$\chi_I(T) = \int_0^{\beta} d\tau G_a(\tau)G_b(\tau) = {1\over
\pi\Gamma} \ln(1.13\beta\Gamma)
\leqno(6.2.37)$$
where we used the results of the previous subsection.  Hence, the
response to a purely local
field is divergent as well for $T\to 0$.  \\

\subsubsection{Dynamical Properties} 

{\bf (a) Impurity Dynamic Susceptibility}\\

A quantity which shall play a recurring role in discussion which
follows is the dynamic
susceptibility for an applied longitudinal magnetic field which couples
only to the impurity.
Following the discussion of Sec. 6.2.2.d, we may write down the dynamic
susceptibility in
Matsubara space as  ($\nu = 2\pi n/\beta$ a bose Matsubara frequency,
$\omega$ is a fermi
Matsubara frequency)
$$\chi_I(\nu) = {1\over \beta} \sum_{\omega}
G_a(i\omega)G_b(\omega+\nu)  \leqno(6.2.38)$$
$$~~~~  = \int {d\zeta\over \pi} {\Gamma\over \zeta^2+\Gamma^2} {1\over
\beta}\sum_{\omega}
{1\over i(\omega-\nu)(i\omega - \zeta)} $$
$$~~~~ = \int  {d\zeta\over \pi} {\Gamma\over \zeta^2+\Gamma^2}
{\tanh(\beta\zeta/2) \over
i\nu - \zeta} ~~. $$
Analytic continuation of this result implies that the absorptive part
of the dynamical
susceptibility has the
extremely simple form
$$\chi''_I(\omega) = {\Gamma\over \omega^2+\Gamma^2}
\tanh(\beta\omega/2)~~. \leqno(6.2.39)$$
Given an identification of $\Gamma$ with the cutoff $\omega_C$
this is precisely the form anticipated from Marginal Fermi liquid
theory (Varma {\it et al.} [1989]),
and identifying $\Gamma$ with $T_K$, this
also agrees with the form found numerically from NCA calculations by
Cox [1988(a)], and
with the form postulated from conformal field theory arguments by
Tsvelik [1990].  Note that
in Cox [1988(a)] a Lorentzian fit was made for positive frequencies to
the dynamic spin
structure factor $S_I(\omega) = (N_B(\omega)+1)\chi''_I(\omega)$ at low
temperatures
with the understanding that
this suffices to give $\chi_I'' \sim sgn(\omega)
\Gamma/(\Gamma^2+\omega^2)$ at zero
temperature.

What is clear in $\chi_I''$ is that the singular structure arises from
the convolution of a regular
Majorana field propagator $G_b$ with the singular propagator of the
decoupled field $G_a$.
Hence again, as stressed in the previous subsection, it is the physics
of coupling one of the
two Majorana variables from the spinless fermion representation of the
spin variable which gives
rise to the interesting critical physics in this picture.  \\

{\bf (b) Self-Energies}\\

As remarked previously, after averaging over a random array of
impurities, the self-energy of
itinerant electrons to leading order in the impurity concentration $c$
is given by the concentration times the one particle $t$ matrix for
scattering
off a single impurity.

When the fermionic degrees of freedom are written in collective
coordinates, only $\psi_{sf}$ and
$\psi_s$ couple to the impurity, and so we can develop self energies
only for these fields.  We
can obtain the retarded $t$-matrix for $\psi_{sf}$ from Eqns.
(A.4.4.a-c) as
a Nambu matrix $\hat t$ with
$$\hat t(\omega) = {1\over 4\pi N(0)} {\Gamma\over \omega + i\Gamma }
[1 + \sigma^{(1)}]
\leqno(6.2.40)$$
where $\sigma^{(1)}=\delta_{i,-j}$ is a Pauli matrix in Nambu indices.
From this we can see that
the Fermi level scattering rate for the $\psi_{sf}$ electrons is given
by
$${1\over 2\tau(0)}  = c[-2 Im(\hat t(+i\eta))_{11}] = {c\over \pi
N(0)} \leqno(6.2.41)$$
which is precisely {\it half} the unitarity limit.  The reduction from
unitarity may be traced to the
spectral weight factor of 1/2 associated with the Majorana character of
the $b$-field.
It is very tempting to compare this with the result
from the conformal field theory, which says that the total Fermi level
scattering rate for an incoming
electron scattering off a two-channel site is half the unitarity
limit.
However, the comparison is
problematic, for reasons we shall discuss further below.

\begin{figure}
\parindent=2.in
\indent{
\epsfxsize=4.in
\epsffile{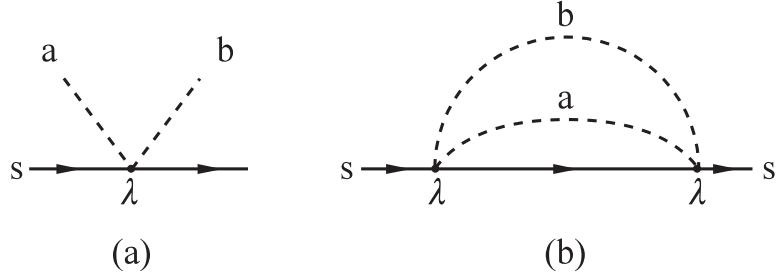}}
\parindent=.5in
\caption{Leading order in $\lambda$ self-energy diagrams for the conduction spin fermion 
of Emery and Kivelson [1992].  There is no self-energy correction which is linear
in $\lambda$, since as shown in (a) this would yield unclosed Majorana fermion legs. 
There is a contribution at order $\lambda^2$ shown in (b), which has the marginal 
Fermi liquid leg.  Suitable re-fermionization from the transformed fields may resolve this 
apparent discrepancy between the Abelian and non-Abelian bosonization approaches.}
\label{fig6p9}
\end{figure}

The coupling to the $\psi_s$ field is perturbative in $\lambda$.
Following the conformal theory,
where the leading order imaginary part of the self-energy is,
surprisingly linear in the deviation from
the fixed point coupling, we can look for a term linear in $\lambda$.
Because this has dangling
$\hat a,\hat b$ legs, no such term exists (see Fig.~\ref{fig6p9}(a)).  The
first non-vanishing diagram
is shown in Fig.~\ref{fig6p10}(b).  This corresponds to the exchange of a
local spin fluctuation boson
with spectral weight $\chi_I''$.  This is precisely the kind of
self-energy diagram considered in
the Marginal Fermi liquid phenomenology (Varma {\it et al.} [1989]) and
which was used by
Cox to produce a heuristic estimate for linear in $T$ scattering in
interpreting the resistivity
data for
Y$_{1-x}$U$_x$Pd$_3$ (Seaman {\it et al.} [1991]).
It is straightforward to evaluate this diagram
giving the imaginary
part of the retarded $s$
self-energy as
$$Im\Sigma_s(\omega,T) =-c\pi N(0)\lambda^2 \int {d\zeta\over \pi}
\chi_I''(\zeta,T) [N_B(\zeta)
+ 1 - f(\omega-\zeta)] \leqno(6.2.42)$$
$$~~~~ \approx -c{\pi\over N(0)} [N(0)\lambda]^2 {1\over
\pi\Gamma}[|\omega| +
2k_BT],~~(|\omega|,k_BT<<\Gamma)~~.$$
Corresponding to this self-energy, the real part behaves as
$Re\Sigma_s(\omega,T) \sim
\omega \ln(max\{|\omega|,k_BT\}/\Gamma)$ which gives a logarithmically
diverging effective
mass for the $s$-fermions.

The difficulty in connecting these self energies to the self energies
computed by, e.g., the
NCA and conformal field theory, is that in the latter approaches the
self energies are computed
directly for the propagators of the original fermion fields.  Loosely
speaking, one of the original
fermion fields is a square root of four collective fermion fields, as
we can see by writing out
$\psi^{\dagger}_{\uparrow +} \sim
\exp(-i\Phi_{\uparrow +}) = \exp(-i[\Phi_c+\Phi_s+\Phi_f+\Phi_{sf}]/2)
$, for example. Each of the
exponentiated collective boson factors gives a ``square root'' of the
corresponding collective
fermion field operator because of the factor of 1/2 in the exponent.
Thus, it is difficult to
interpret the self energies of collective coordinate fermion fields in
terms of physical self energies
of electrons which scatter off of the impurity.  This situation is
reminiscent of the Bethe-Ansatz approach in which the wavefunctions of
the
real incoming and outgoing electrons cannot be easily constructed in
terms
of the exactly determined eigenstates.  \\

{\bf (c) Conductivity}\\

The Abelian bosonization scheme also produces a problematic
interpretation of transport
coefficients.   First, as discussed in the previous subsection, there
is no obvious way to go
from a calculation of $\Sigma_{sf},\Sigma_s$ to a calculation of
$\Sigma_{\mu\alpha}$ where
$\mu\alpha$ are the original spin and channel labels.  Thus we cannot
immediately transcribe
the self-energy results of the previous subsection to resistivity
results through the transport
integral formulae discussed in the NCA and conformal field theory
reviews.

A second difficulty arises in the connection of the physical three
dimensional current operator
to the current operators of the effective one dimensional model.  As
discussed by
Emery and Kivelson [1992], following Bander [1976], the collective
fermion fields have currents
which may be expressed by gradients of the corresponding collective
bose fields. Thus, for
example, the one dimensional charge current of left moving states is
proportional to
$\partial \Phi_c/\partial x$, the spin current to $\partial
\Phi_s/\partial x$, the flavor current to
$\partial \Phi_f/\partial x$, and the spin-flavor current to $\partial
\Phi_{sf}/\partial x$.  Of these,
only $\partial \Phi_{s,sf}/\partial x$ couple to the impurity.  This
would suggest that the impurity
cannot affect the electrical conductivity, because of the following
argument:  while the three
dimensional current operator is not identical to $\partial
\Phi_c/\partial x$, this is the only
effective one dimensional current operator which is a singlet in spin
and channel indices like
the three dimensional current, and hence the only candidate effective
one-dimensional current
operator to match to the three dimensional one.

An exception to this identification is in the case when the local
$z$-component of the
impurity
``spin'' has the same transformation
properties as the current operator under parity and time reversal,
which is the case for example
in the electric dipole Kondo effect discussed by Emery and Kivelson
[1993(a),1993(b)].   In that
case, the contribution to the conductivity from the impurity is
proportional to  the spin conductivity,
given by $Re\sigma_s \approx \chi_I''(\omega,T)/\omega$.  This quantity
behaves precisely
as the marginal Fermi liquid phenomenology conductivity (Varma {\it et
al.} [1989]).  Emery and
Kivelson [1993(a)] have shown that for a one dimensional array of such
electric dipole Kondo
scatterers, that there is no Drude contribution (following the
discussion of the preceding paragraph)
and that the above term gives a contribution to the electrical
conductivity due to the bound
dipole charge of the array.

As discussed in the next subsection, the difficulties can be reconciled
conceptually, at least, by considering the Majorana fermion reformulation
of conformal field theory introduced by Maldacena and Ludwig [1996].  
This theory contains three eight dimensional fields which may be 
explicitly expressed in terms of the Abelian boson fields of Emery and
Kivelson.  The conductivity may be computed in terms of these fields, 
and there are no discrepancies between the results obtained in this 
method and those of the non-Abelian bosonization scheme discussed in 
Sec. 6.1.  

\subsubsection{Finite size spectra and scattering states}

Recently von Delft and \zar [1997] have shown that the spectrum of 
states for the Emery-Kivelson
of the two-channel model can be solved analytically for arbitary
spin-flip 
coupling and magnetic field strength.  Using this, they produced a
complete spectrum of eigenvalues, states, and scattering states, and a
new interpretation of the unitarity paradox (discussed more extensively 
in section 6.3.1 below).  In particular, for zero magnetic field and
infinite spin flip coupling, the fixed point spectrum of Affleck and 
Ludwig [1991c] is obtained, and this is taken to be an analytic
confirmation of the fusion hypothesis, to be contrasted with the less
direct confirmation obtained by comparison with NRG or Bethe-Ansatz 
spectra.  It should be noted that the scattering states for the
Emery-Kivelson line and its generalization to channel anisotropy were
first constructed in a slightly different context by Schiller and 
Hershfield [1995,1997], who used the Emery-Kivelson approach to solve a
nonequilibrium Kondo problem.


\subsection{Additional Developments}

\subsubsection{Reformulation of the Conformal Theory 
with Majorana Fermions} 

In this subsection we will briefly review the results of Maldacena and
Ludwig [1996] who reformulated the conformal theory in terms of Majorana
fermions.  As the work depends upon some rather technical aspects of 
Lie group theory, we will focus on summarising the philosophy and key 
results of the work.  

Motivated in part by the success of the Abelian bosonization scheme of 
Emery and Kivelson discussed in the previous subsection and by the unitarity
puzzle described in the introduction to Sec. 6 and in Sec. 6.1.4, 
Maldacena and Ludwig [1996] undertook to write the free Fermion 
hamiltonian in terms of eight Majorana fermion fields.  The form is 
$${\cal H}_{free} = -{i\over 2\pi} \int dx \chi_{a}(x){d\over dx}\chi_a(x) 
\leqno(6.3.1)$$
where $a$ runs over the spin, channel, and complex (imaginary or real) 
indices of the Majorana fields.  This Hamiltonian has a manifest $SO(8)$
symmetry under rotations in the space indexed by $a$.  Rotations in 
the space are generated by currents 
$$j^A(x) = \chi_a(x)(T^A)_{ab}\chi_b(x) \leqno(6.3.2)$$
where $T^A$ are appropriately defined 8x8 matrices associated with the
generators of $SO(8)$.   The currents obey a Kac-Moody algebra with 
four irreducible 
representations, a singlet, an eight dimensional vector representation,
and two eight dimensional 
spinor representations.  The Hilbert spaces of the representations
are distinguished by the boundary conditions which must apply to the
representations, namely, the singlet and vector representations are 
antisymmetric under $x\to x+l$, while the spinors are symmetric.  
Similarly, under translation in imaginary time $\tau$ by $\beta$, the
singlet and vector irreps are antisymmetric, while the spinors are
symmetric.  The free Hamiltonian is transparent to the two spinor 
irreps, but in principle they are present awaiting only some coupling
to the free fermion space to make their presence felt.  

To foreshadow the key results, we note that the $SO(8)$ set of 
eight dimensional representations obey a unique symmetry called `triality'
in which we can really flip around which of the three we choose as a
fermionic vector and which two we choose as spinors.  The bosonization
scheme of Emery and Kivelson [1992] corresponds to one particular choice of
fermionic vector.  When we make this choice, the free fermion field becomes
a spinor representation, and may in principle mix with the other spinor.
The effect of the impurity at the boundary is to convert the free physical
fermions with this other field.  The scattering takes one free fermion
incoming wave into precisely one outgoing spinor wave, which resolves
the unitarity puzzle (the original Hilbert space was not the full Hilbert
space of the problem).  When the two point green's function is computed
with this approach, the electron self energy acquires a $\sqrt{T}$ term.  
The problem with considering the self energies of the fermions in the 
Abelian bosonization scheme is that they have no simple relation 
to the original free fermions, as discussed in the previous subsection.  
A beautiful aspect of this approach is that no $SU(2)$ spin symmetry
breaking is required, unlike the Emery and Kivelson [1992]
expansion around the Toulouse limit.  

The free fermion fields can be bosonized by writing the normal ordered
products 
$$\psi^{\dagger}_{\mu\alpha}(x)\psi_{\mu\alpha}: = i \partial_x 
\phi_{\mu\alpha}(x) \leqno(6.3.3)$$
so that 
$$\psi^{\dagger}_{\mu\alpha}(x) = \exp[-i\phi_{\mu\alpha}(x)] \leqno(6.3.4)$$
which is unchanged by a $2\pi$ rotation in phase, i.e., addition of 
$2\pi$ to the boson field.  In contrast, the spinor fields written 
as exponentiated boson operators contain
linear combinations of the $\phi$ fields multiplied by $\mp i/2$ ($-$ for
a ``creation'' operator, $+$ for an ``annihilation'' operator) and hence
change sign under the corresponding rotation.  

Now, if we follow the Emery and Kivelson prescription to linearly
transform to the $\phi_c,\phi_s,\phi_f,\phi_{sf}$ bose fields, what 
happens is that the original fermion field becomes a spinor while 
one of the original spinor fields (type II) 
becomes a fermion or vector irrep
of $SO(8)$ in that it becomes expressed as simple exponentials of the
transformed $\phi$ fields while the original fermion field and other
spinor (type I) exponentials of linear combinations multiplied by $\pm i/2$.  
The impurity interaction basically becomes a boundary condition which
expresses the interconversion of free fermion and type I spinor field 
interconvert.  The type I spinor field has the same spin, channel, and
charge quantum numbers as the original fermion fields, but fractional
occupancy in terms of the original fields.  In contrast, the type II
spinor field has fractionalized spin, charge, and flavor quantum numbers
and actually can be viewed as carrying the non-trivial primary fields 
discussed in Sec. 6.1.2 (spin, channel spin, and pair fields, a total 
of eight) when appropriately multiplied by grassman numbers.  

Maldacena and Ludwig [1996] also 
show with this formalism that the two impurity one channel Kondo model
and the Callan-Rubakov problem of four fermion species scattering off
an $SU(5)$ magnetic monopole have equivalent non-trivial fixed points,
and that the fixed point correlation functions of the three different
models (two channel, two impurity one-channel, and Callan-Rubakov)
may be mapped into one another.  This latter trick employs the triality
symmetry of the $SO(8)$ group explicitly.  

Using a somewhat different approach with a Majorana fermion scheme
similar to that of Coleman, Ioffe, and Tsvelik [1995], Zhang, Hewson,
and Bulla [1997] have reached similar conclusions and presented an
explicit realization of the finite size spectra and leading irrelevant
operator in the Majorana fermion basis.  

\subsubsection{Conformal Theory of the Large Conduction Spin Single 
Channel Model}

Fabrizio and \zar [1996], 
Sengupta and Kim [1996] and Kim, Cox, and Oliveira [1996] have studied
the large conduction spin single channel model with conformal field theory
techniques (Fabrizio and \zar [1996] have also employed 
$1/M$ and $1/S_c$ expansions, where $S_c$ is the conduction spin,
through the framework of the multiplicative renormalization group).  
In this model, the conduction electrons are allowed to 
have arbitrary spin $j$.   The effective channel number which is read off
as the rank of the spin Kac-Moody algebra is calculated to be 
$M(j)= 2j(j+1)(2j+1)/3$. (Fabrizio and \zar [1996] have established a 
more general correspondence allowing the channel number of the large
spin electrons to vary; see the discussion in Sec. 7.1 for further
details.)  For example, when $j=3/2$, $M(3/2)=10$.  
This suggests that any spin satisfying $S_I< M(j)/2$ will be overcompensated
in this problem.  This model can arise as an unstable fixed point of
the TLS Kondo effect and for Ce$^{3+}$ ions in cubic symmetry as 
discussed in Secs. (3.3.3) and (3.4.3).  

\begin{figure}
\parindent=2.in
\indent{
\epsfxsize=2.in
\epsffile{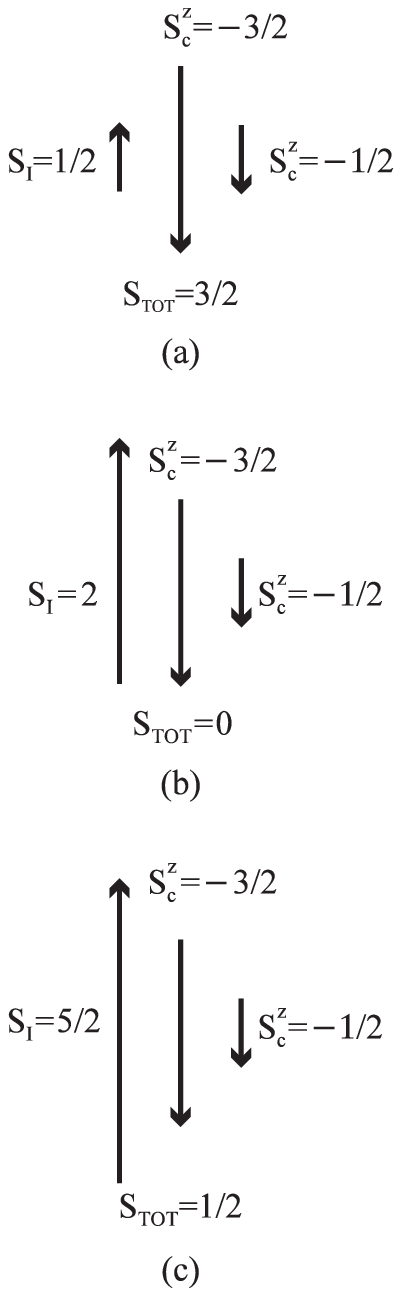}}
\parindent=.5in
\caption{Strong coupling pictures for the single band $S_c=3/2$ Kondo model.  
(a) displays an example of the overcompensated model when the impurity spin takes the
value $S_I=1/2$.  In this case, the $S_z=-3/2,-1/2$ conduction channels overscreen
the impurity, leading to an non-Fermi liquid fixed point.  
For $S_I=2$, there is just enough conduction spin to compensate the
impurity, as shown in (b), and this is expected to lead to a Fermi liquid fixed 
point.  For $S_I=5/2$, as shown in (c), there is too much impurity spin to be compensated 
by the conduction electrons, and we anticipate that an undercompensated fixed point 
will ensue.}
\label{fig6p10}
\end{figure}

The simple physical reason for this overcompensation is as follows. (See 
also  Secs. (3.3.3,3.4.3).)  For definiteness, consider
the $j=3/2$ case.  Proceed to the strong coupling limit, i.e., shut off
the hopping.   As shown in Fig.~\ref{fig6p10}(a), it is energetically favorable
about a single site to draw in {\it two} electrons with $j_z=+3/2,+1/2$, 
assuming the local moment to have down spin.  From this simple picture, 
we would anticipate that for impurity spin up to $S_I=3/2$ that an 
overcompensated ground state would result. For $S_I=2$, exact compensation
should occur, and for $S_I>2$ undercompensation would occur.  As in the
arguments for the multichannel models, this strong coupling limit is 
unstable for $S_I\le 3/2$, because a residual antiferromagnetic coupling 
will remain with electrons in the next spatial RG shell (Fig.~\ref{fig6p10}(b)).  
For $S_I>2$, 
the residual coupling is ferromagnetic, as illustrated in Fig.~\ref{fig6p10}(c). 

For general $j$, adding up only positive $j_z$ values,  we would
expect the total screening spin to be $j_{max}=(j+1/2)^2/2$ for 
half integer $j$, 
And $j_{max}=j(j+1)/2$ for integer $j$. The condition in general for 
over-compensation is then $S_I < j_{max}$.  This disagrees with the
results of the conformal theory (Fabrizio and \zar [1996], 
Sengupta and Kim [1996], Kim, Oliveira, and Cox [1997]) as discussed 
below.  

For the large $j$ single channel model, 
the Hamiltonian written in terms of left moving fields is, in the 
position space domain,  
$${\cal H} = \sum_{\alpha}\int dx \psi^{\dagger}_{\alpha}(x){\partial \over 
\partial x} \psi_{\alpha}(x) + {\cal J}\vec S_I\cdot \vec J(x=0) 
\leqno(6.3.5)$$
where $\psi^{\dagger}_{\alpha}(0)$ creates an electron with z-component 
spin $\alpha=-j,-j+1,....,j$, and 
$$\vec J(0) = \sum_{\alpha\beta} \psi^{\dagger}_{\alpha}(0)
\vec S_{\alpha\beta}\psi_{\beta}(0) \leqno(6.3.6)$$
with $\vec S$ spin matrices of the spin $j$ representation of $SU(2)$.  

Restricting attention now to just the $j=3/2$ case, 
the Hamiltonian of Eq. (6.3.5) was cast by Sengupta and Kim [1996]
and Kim, Oliveira, and Cox [1997] in Sugawara
form for the generation of finite size spectra from conformal theory. 
As in the multichannel model analysis of Affleck and Ludwig [1991b], 
the single fusion hypothesis was 
assumed to generate finite size spectra relevant for comparison to the
NRG, and the double fusion hypothesis was assumed for generating
the operator spectrum and scaling indices.  

The first observation is that the Kac-Moody algebra corresponding to 
the conduction spin current $\vec J(x)$ has a large rank.  When written
in momentum space, the commutation relation is 
$$[J^a_{q},J^b_{q'}] = i\epsilon_{abc} J^c_{q+q'} -
\delta_{ab} \delta_{q+q',0} {k(j)ql\over 2\pi} ~~.\leqno(6.3.7)$$
which follows from the fact that 
$$\sum_{\alpha\beta}S^a_{\alpha\beta}S^b_{\beta\alpha}= 
\delta_{ab} \sum_{\alpha=-j}^j \alpha^2
 = \delta_{ab}{k(j)\over 2} \leqno(6.3.8)$$
as is familiar from $SU(2)$ spin algebra.  Hence the rank of the spin 
current Kac-Moody algebra is given by $k(j)$ in contrast to 
the $j=1/2$ case 
where it is the channel number $M$. For $j=3/2$, $k(j=3/2)=10$. 
One has a choice of which quantum numbers to use here; we follow 
Kim, Oliveira, and Cox [1996], who used a generalized axial charge 
(Jones [1988]; Jones and Varma [1988]; Jones, Varma, and Wilkins
[1988]), or
isospin defined by (in their reference, $Q$ is used in favor of $I$)
$$I_{q}^{z} =
\sum_{k,\alpha}\left[c^{\dagger}_{k,\alpha}c_{k+q,\alpha}-\case1/2\delta_{n,0}\right] \leqno(6.3.9)$$
and
$$I_{q}^{+} =\left(I_{q}^{-}\right)^{\dagger}=
\sum_k\left(c^{\dagger}_{k,3/2}c^{\dagger}_{-k-q,-3/2}-
c^{\dagger}_{k,1/2}c^{\dagger}_{-k-q,-1/2}\right)~~.\leqno(6.3.10)$$
$\vec I$ satisfies the Kac-Moody algebra ($I^x=(I^++I^-)/2$, 
$I^y=(I^+-I^-)/2i$)
$$\left[I_q^{a}, I_{q'}^{b}\right] = i\epsilon^{abc}I^{c}_{q+q'} +
\delta_{ab}\delta_{q+q',0}{ql\over \pi} \leqno(6.3.11)$$
from which we infer a rank $k_I=2$.  By following the same techniques
to construct a Sugawara Hamiltonian as in Sec. 6.1, we obtain for the
free $S_c=3/2$ conduction electrons 
$${\cal H}_0 = {v_F\pi\over \ell}
\left[\sum_{q=-\infty}^{\infty}{1\over 12}:\vec J_{-q}\cdot\vec J_{q}:
+\sum_{n=-\infty}^{\infty}{1\over 4}:\vec I_{-q}\cdot\vec I_{q}:\right] 
\leqno(6.3.12)$$
and the impurity may be absorbed into new conduction spin currents 
$$\vec{\cal J}_q = \vec J_q + \vec S_I \leqno(6.3.12)$$
which obey the same Kac-Moody algebra as Eq. (6.3.8) provided
the coupling strength $\lambda =N(0){\cal J}=1/6$ which we identify
as the fixed point coupling within the conformal theory.  
In this case, the Sugawara Hamiltonian becomes 
$${\cal H} = {v_F\pi\over \ell} 
\left[\sum_{q=-\infty}^{\infty}{1\over 12}:\vec {\cal J}_{-q}\cdot
\vec {\cal J}_{q}: 
+\sum_{n=-\infty}^{\infty}{1\over 4}:\vec I_{-q}\cdot\vec I_{q}:\right] 
\leqno(6.3.13)$$
(Sengupta
and Kim [1996] choose a different basis with quantum numbers of spin, 
charge, and an Ising variable.)  

In this basis, one can  readily construct the spectrum of the system. 
For the free electrons, the primary (highest weight) states must have
$i\le k_I/2=1$, and $j\le k(j)/10 = 5$ here.  The spectrum of eigenvalues
is then 
$$E_0(i,j,m) = {v_f\pi\over \ell}[{i(i+1)\over 4} + {j(j+1)\over 12} + m]
\leqno(6.3.14)$$
where $m$ is an integer measuring the number of excitations induced by
operating with $I^a_{-q},J^b_{-q}$, $q>0$ on the primary states. These
states are constrained by the conditions that they correspond to 
excitations from the free fermi sea of $S_c=3/2$ electrons.   Note that
one must construct different spectra for a system with a degenerate 
ground state (periodic boundary conditions in the continuum or even 
interation number in the numerical renormalization group) or a
non-degenerate ground state (antiperiodic boundary conditions in the 
continuum or an odd NRG iteration number).  

For the interacting system, we must apply the fusion rule of Affleck and
Ludwig [1991b], which reads here, for those states satisfying fermion 
``gluing conditions'' in the free spectrum, that  
$$i \to i~~, j \to j'~with ~ |j-1/2|\le j'\le min\{j+1/2,k(j)-j-1/2\}~~.
\leqno(6.3.15)$$
One may similarly generalize the double fusion rule for operator spectra. 
Kim, Oliveira, and Cox [1996] find excellent agreement between the 
conformal field theory spectrum generated in this way and the NRG energy
levels, provided the CFT antiperiodic boundary conditions are identified
with odd numbered NRG iterations, and periodic boundary conditions with 
even numbered NRG iterations.  The inferred operator spectrum possesses
a primary spin operator $\vec \Phi$
with a scaling dimension of 1/6, which will induce a $T^{-2/3}$ divergence in 
the spin susceptibility and the specific heat coefficient. (The leading 
irrelevant operator has a scaling dimension of 7/6.)  This scaling
dimension to $\vec \Phi$ produces
a scaling exponent for the magnetic field of $\Delta_h=5/6$, implying 
a crossover to Fermi liquid behavior in a polarized scattering potential
in the variable $h^{6/5}/T$.  Exchange anisotropy is relevant in this
model.  Sengupta and Kim [1996] reached the same conclusions about the 
lowest few energy levels and the leading irrelevant operator about the
fixed point.  

\subsubsection{Conformal Field Theory Treatment of the Anisotropic Two-Channel
Model, Spin-Flavor Two-Channel Model, and TLS Kondo model.}

Recently, Ye has extended the applications of conformal field theory and
abelian bosonization in a series of papers (Ye [1996a,b,c,d,e]).  Among 
his accomplishments are: generalizing from the isotropic fixed point of
Affleck and Ludwig to study the anisotropic ``Emery Kivelson'' line 
first identified in the work of Emery and Kivelson using Abelian
bosonization (Ye [1996a]); a solution of the spin-channel Kondo model 
within conformal theory in which the impurity possesses both spin and
channel degrees of freedom which couple to the electrons--in this work, 
the spectrum of Fermi liquid and non-Fermi liquid fixed points was 
worked out fully (Ye [1996b]); 
a general paper on the Abelian bosonization approach for quantum
impurity problems (Ye [1996c]); 
a comparison of the two-channel Kondo and compactified one channel Kondo
model (mentioned above and in Sec. 9.2 (Ye [1996d]) which leads to the 
conclusion that for channel anisotropy the resultant fixed point is
a Fermi liquid one (in contrast to the claims of Andrei and Jerez [1995] 
and Coleman and Schofield [1995] but in agreement with Fabrizio,
Gogolin, and \noz [1995a,b]); and a paper (Ye
[1996e]) which
applies a combination of scaling theory, abelian bosonization, and
conformal theory to study the two-level system Kondo model beginning
from the framework of Moustakas and Fisher [1995,1996]. The latter paper
concludes that the two-channel Kondo non-Fermi liquid fixed point is 
connected by a continuous line of unstable non-Fermi liquid fixed points to the 
new one of Moustakas and Fisher [1995,1996], and the latter is
equivalent to the two-impurity single channel Kondo fixed point
(discussed in Sec. 9.3.1). Any amount of spontaneous tunneling or 
TLS asymmetry will send the system to a Fermi liquid fixed point 
through the spin-field crossover described extensively in secs. 3,4,5,7
and above in this section.  We refer the 
reader to these works for further details.


\section{Bethe-Ansatz Method} 

The importance of the Bethe-Ansatz is not just for providing exact wave
functions,
spectra and thermodynamics for models to which it is applicable.  As the 
energy spectrum is obtained for the entire energy range, the 
thermodynamic results are therefore valid for the entire range of temperature
and magnetic field.   The
application
of conformal field theory relies upon the hypotheses of conformal
invariance (that the
conformally invariant model corresponding to the special value of
$\lambda$
corresponds with the non-trivial fixed point of the multichannel Kondo
model) and
fusion rules, as discussed in Sec. 6.1, which are valid only at low temperatures
and energies in contrast to the Bethe-Ansatz.  Strictly speaking these cannot
be proved
{\it a priori}, but by making these assumptions and deriving results
one can then
compare to the exact Bethe-Ansatz results to verify the hypotheses.
The conformal
theory can then be extended to calculate dynamical properties for which
the Bethe-Ansatz
fails.  Similar remarks apply to the NCA:  as discussed in Secs. 5.1
and 5.3, the
thermodynamic properties of the NCA are in good agreement with exact
Bethe-Ansatz
results for the overcompensated Kondo models, giving confidence in the
dynamics
results which are inaccessible to the Bethe-Ansatz.  (The problem with
dynamics
is that while the Bethe-Ansatz provides exact many body wave functions,
it is
an unsolved problem to properly express the operators which couple to
external probes, such
as the electrical current, in terms of this diagonal basis.  In
particular, it is not known
how to construct scattering state solutions for the Hamiltonian.)

In this section we present a brief overview of the Bethe-Ansatz
approach to the
multichannel Kondo model.  Because there are extensive reviews
available of the
Bethe-Ansatz (Andrei, Furuya, and Lowenstein [1983], Wiegman and
Tsvelik [1983b],
Schlottman and Sacramento [1993], Hewson [1993]) we shall focus
primarily on
summarizing some of the key aspects of the method as applied to this
problem
together with some of the main results.

\subsection{Methodology for the $M$-channel Kondo problem. } 

The Bethe-Ansatz method was originally used by Bethe to study the
one-dimensional
nearest neighbor antiferromagnetic spin 1/2 chain.  It has also proven
to be a powerful
method for treating the electron gas in one-dimension and quantum
impurities embedded
in three dimensional hosts (where the mapping back to one dimension
discussed in
Secs. 4, Sec. 6.1.1.a is used).  The essence of the Ansatz is to make a
`guess' for the
many body wave function which is kind of determinant of generalized
plane wave states.
Clearly this can work only for interactions which are very short ranged
so that the particles
propagate rather freely between interactions.   The Ansatz will not
work for all models; an
assessment of whether it will work rests upon testing it for 1,2, and 3
particles which is
sufficient to prove that it will work for all particles.

The first applications were to spinless fermions and bosons (through
the Jordan-Wigner
transformation, clearly the spin 1/2 chain can be mapped to a spinless
boson or fermion
problem).  Later came breakthroughs by Yang [1967] and Gaudin
[1967] which allowed
spin-ful particles to be treated: the `plane wave' product terms
acquire spin-matrix pre-factors
which have certain consistency requirements.  Yang and Gaudin showed
that the
consistency requirements map onto a separate Bethe-Ansatz solution for
the spin 1/2
case. In general, spin
$S$ particles can be solved by $N=2S+1$ nested Bethe-Ansatz

The original $S_I=1/2,M=1$ Kondo model was solved
with the Bethe-Ansatz independently by Andrei [1980] and Wiegman
[1980].  Later,
the multi-channel model was solved by Andrei and Destri [1984] and
Wiegman and Tsvelik
([1983a], Tsvelik and
Wiegmann [1984,1985], Tsvelik [1985]).
In order for the Ansatz to work for the impurity
models, one typically needs contact interactions together with a linear
dispersion for
conduction electrons.

Generalization of the original method for the $M=1$ model to the
arbitrary $M$ case
was not straightforward.  Namely, as we discussed in the introduction
(Sec. 1), the
ground-state of the $M$ channel spin $S_I$ Kondo model consists of $M$
electrons
with parallel spin glued together to form a net spin $M/2$ which
couples anti-parallel
to the impurity spin $S_I$.  Straightforward extension of the contact
interaction
picture with linear dispersion in this case leads to a bare $S$-matrix
which is diagonal
in the channel indices.  Clearly this cannot be true if the physical
picture is to hold.  Consider the $M=2,S_I=1/2$ example:  in this case
the `big' spin
one of the conduction electrons requires a channel spin
antisymmetrization.
Moreover, for large $M$ this clearly contradicts the rigorous results
computed perturbatively
in the two-loop expansion of the beta function (see Sec. 3.4.5).   This
defect in the
simple extension is due to the linear dispersion assumption, which
implies the lack
of a cutoff or short distance scale in the model.

In order to get physical results from the Ansatz for the multi-channel
model, the
different groups applied modifications to the original Kondo model.
Andrei and Destri [1984]
introduced a non-linear dispersion, $\epsilon_k = v_F[(k-k_F) +
\Lambda^{-1}(k-k_F)^2]$
where $\Lambda$ is a cutoff parameter.  At the end of the calculation
$\Lambda$ is taken
to $\infty$, and results are independent of $\Lambda$.   To keep the
model integrable,
local counter-term potentials must be introduced into the Hamiltonian
which follow the
work of Rudin [1983].  The counterterm introduces coupling between the
channels, but
the strength of the counterterm drops out in the $\Lambda\to\infty$
limit and is therefore irrelevant. The presence of the cutoff allows the dynamic
formation of the spin complex with spin $s_c=M/2$ (``dynamic fusion''). 
This dynamic fusion is observed both in the structure of the wave 
function and in the corresponding fused Bethe-Ansatz equations.  In 
the former the momenta develop imaginary parts leading to the binding
of $M$ electrons to form a spin $s_c=M/2$ and flavor singlet, while 
the Bethe-Ansatz equations describbe the coupling of this composite
spin to the impurity.  

Tsvelik and Wiegmann [1984],[1985] use a different approach to resolve
this difficulty.  They
first introduce a generalized $M$-channel Anderson Model with
parameters for which a
local moment of spin $S_I$ is formed and for which the ground state is
exactly compensated
($M=2S_I$).  The integrable $M$-channel Anderson model has the form
$$H=\sum_{k\sigma m} v_F (k-k_F) c^{\dagger}_{km\sigma}c_{km\sigma} +
\sum_{m\sigma}
\epsilon_d d^{\dagger}_{m\sigma} d_{m\sigma} $$
$$ + {U\over 2} \sum_{mm'\sigma\sigma'} n_{dm\sigma}n_{dm'\sigma'} +
V\sum_{km\sigma} (c^{\dagger}_{km\sigma}d_{m\sigma} + h.c.)
\leqno(7.1.1)$$
where $m$ is the orbital index ranging from one to $M$,
$\sigma=\uparrow,\downarrow$
is the spin index, and $d^{\dagger}_{m\sigma}$ creates a local electron
of spin $\sigma$
in orbital $m$.  This model does not represent a realistic extension of
the Anderson
Hamiltonian since the Coulomb interaction is diagonal in the density
operators
(see \noz~ and Blandin [1980] and Mihaly and \zow~ [1978] for further
discussion).
This model, studied with diagrammatic methods by Yoshimori [1976] does
give the
correct universal behavior of the spin-compensated $M$-channel Kondo
model at
low temperatures and for large $U$, as can be expected from a
Schrieffer-Wolff [1966]
transformation to eliminate the hybridization at order $V$ generating
an effective
exchange coupling.

Tsvelik and Wiegman [1984],[1985] then separately consider a
generalized
exchange model with arbitrary local spin $S_I$ with electrons that have
spin
$M/2$.  The
Hamiltonian has the form
$${\cal H} = \sum_{ka} v_F(k-k_F) c^{\dagger}_{ka}c_{ka} +$$
$$~~~~~ J\sum_{kk',aa'} c^{\dagger}_{ka} P(\vec S_I\cdot S_c,J)_{aa'}
c_{k'a'} \leqno(7.1.2)$$
where $a$ indexes the conduction spin states and $P(x,y)$ is a
polynomial of order
$min(2S_I,M)$.  This model is integrable (i.e., the Bethe-Ansatz works)
if the polynomial
has the explicit form
$$P(x,J) = \sum_{l=|M/2-S_I|}^{M/2+S_I}\prod_{k=0}^l {1-ikJ\over 1+ikJ}
\prod_{p=0,p\ne l}^{min(M,2S_I) } {x-x_p\over x_l-x_p} \leqno(7.1.3)$$
where
$$x_p={p(p+1)\over 2} -{S_I(S_I+1)\over 2} - {M(M/2+1)\over 4}
~~.\leqno(7.1.4)$$
In the special case $S_I=1/2$, the polynomial
reduces to the form
$$P(\vec S_I\cdot \vec S_c,J) = a + b(\vec S_I\cdot\vec S_c)
\leqno(7.1.5)$$
where $a,b$ are constants which may be determined from (7.1.4).

The claim of Tsvelik and Wiegman [1984,1985] is that the Hamiltonian of
Eq. (7.1.2)
is equivalent to that of Eq. (7.1.1) when charge fluctuations are
projected out and
$M=2S_I$ as they have identical energy spectra.  
They further conjecture that the solution of the
Bethe-Ansatz of
Eq. (7.1.2) holds for the multichannel Kondo model for arbitrary $S_I,M$.   The appealing
feature of  this picture is that it says one must simply form a large
conduction spin
as suggested by the \noz and Blandin [1980] picture and the NRG
results (Sec. 4)
and solve the Bethe-Ansatz for one channel of electrons with that large
spin.  In contrast, Andrei and Destri [1984] do not assume the formation of the 
electron complex with total spin $M/2$, but it follows from their treatment.  
Tsvelik and Wiegman establish the equivalence of the different models
by explicit
Bethe-Ansatz solution (and exact comparison of the excitation spectra)
which
confirm that the results are identical for
$M=2S_I$.  Also, for the overcompensated case $M>2S_I$, they obtain
indentical
Bethe-Ansatz spectra as Andrei and Destri [1984].  
As Tsvelik and Wiegman start with the electron complex with $s_c=M/2$, 
they do not therefore need the sophisticated cutoff scheme with counter 
term employed by Andrei and Destri [1984] which leads to the formation of the
electron complex.  
For a
full discussion of the different cutoff schemes, see Andrei, Furuya,
and Lowenstein
[1983].

For $S_I$=1/2, the conjectured mapping of Tsvelik and Wiegman [1984,1985] of the $M$ channel model 
to the single channel, conduction spin $M/2$ model has been called 
into question by Fabrizio and Gogolin [1994], 
Fabrizio and \zar [1996], Kim and Cox [1996], and Kim, Oliveira, and 
Cox [1997].   Concretely, the conjectured mapping may be summarized in 
the form $MS_c = S^*_c$, where $S_c$ is the conduction spin of the original 
multichannel model, and $S^*_c$ is the conduction spin of the single channel
large spin model.  (This may be generalized to $MS_c=M^*S^*_c$ for effective 
models with arbitrary numbers of channels, as per Fabrizio and \zar [1996].)  
Explicitly, Fabrizio and Gogolin [1994], have shown that the $M=4$, $S_c=1/2$ model 
maps to the $M^*=1$, $S^*_c=1$ model, and Kim, Oliveira, and Cox have found that 
the critical properties of the single channel $S_c$=3/2 model are in complete 
disagreement with the $M=3$,$S_c=1/2$ model and rather agreed with the $M=10,S_c=1/$ model.  
Both of these results are in contradiction with the Tsvelik and Wiegman [1984,1985]
conjecture.   This latter result for $S^*_c=3/2$ was also obtained within conformal theory by Sengupta and Kim [1996]. 

The most thorough treatment of this issue was given by Fabrizio and \zar [1996].  
Utilizing $1/M$ and $1/S_c$ expansions as well as conformal field theory arguments, they have 
suggested that the correct equivalence mapping is summarized by 
$$MS_c(S_c+1)(2S_c+1) = M^*S^*_c(S^*_c+1)(2S^*_c+1)~~. \leqno(7.1.6)$$
For the case $M^*=1,S^*_c=1$, the RHS of the above equation gives 6, while 
for $M=4,S_c=1/2$, the LHS gives 6 as well, in agreement with the findings of 
Fabrizio and Gogolin [1994]. For $M^*=1,S^*_c=3/2$, the RHS of Eq. (7.1.6) gives 
15, while the LHS gives 15 also for the $M=10,S_c=1/2$ model, in agreement with 
Kim, Cox, and Oliveira [1997], and Sengupta and Kim [1996].  The origin of the discrepancy 
between the Bethe-Ansatz results and these other approaches will require further investigation.  

Once the solubility of a model by the Bethe-Ansatz is established,
physical properties
can be computed.  This is practically restricted to thermodynamic
quantities (dynamical
properties such as the electrical conductivity
require a writing of, e.g, current operators in the diagonalized many
particle basis which
is a non-trivial and largely unsolved problem).  The thermodynamic
quantities are expressed
in terms of excitation densities for spin(spinon), channel(channelon),
and charge(holon)
 degrees of freedom which are
solutions to coupled non-linear integral equations.  Note that the
`unbinding' of charge, spin,
and channel degrees of freedom which arises naturally in the NCA (Sec.
5), Conformal
Theory (Sec. 6.1) and Abelian Bosonization (Sec. 6.2) approaches arises
quite naturally
in the Bethe-Ansatz approach.

\subsection{Results from the Bethe-Ansatz Method} 

\subsubsection{Coupled Integral Equations and Numerical Procedure} 

In addition to the analytical expressions for various quantities at low
and zero temperature,
it is important for the computation of thermodynamics to understand the
numerical
results.  Here we refer only to the final equations to be solved, and
direct the reader to
the original references for details.  Our discussion here follows
Sacramento and Schlottmann
[1991].  One solves the Ansatz for arbitrary spin by flipping over
electron spins from a fully polarized state.  The ground state is given by
a ``sea of 2-strings'' in the center of mass rapidity $\Lambda$.  
 The resulting excitations are  spin 1/2 `spinons' formed by creating holes 
in the ground state distribution and 
characterized by a one-particle
rapidity $\lambda$ which describes their momentum and energy.  These
spinons may form bound states with center-of-mass rapidity $\Lambda$
(not to
be confused with the above cutoff parameter in Eq. (7.1.1)) with
population factors
$$\eta_{l}(\Lambda) = \exp(\epsilon_l(\Lambda)/T)
~~l=1,2,.....\leqno(7.2.1)$$
where $\epsilon_l$ is the energy of a bound-state of $l$ spin waves
with rapidity
$\Lambda$.  These thermal occupancy factors obey an infinite recursion
sequence
$$\ln(\eta_k(\Lambda)) = \int_{-\infty}^{\infty} d\Lambda'
G(\Lambda-\Lambda')
\ln[(1+\eta_{k-1}(\Lambda'))(1+\eta_{k+1}(\Lambda'))]
-\delta_{kn}\exp(\pi\Lambda/2)
~~k=1,2,.......~~\leqno(7.2.2)$$
with
$$G(\Lambda) = {1\over 4\cosh[\pi\Lambda/2]} ~~.\leqno(7.2.3)$$
These recursion relations are completed by the boundary condition
$\eta_0=1$ and
the asymptotic requirement
$$\lim_{k\to\infty}[({1\over k})\ln(\eta_k(\Lambda))] = {H\over T} =
2x_0 \leqno(7.2.4)$$
where $H$ is the magnetic field (or the asymmetric level splitting in
the TLS Kondo case).
Physical properties can now be expressed by these quantities $\eta_l$.
For example,
the impurity contribution to the free energy is given by
$$F_{imp} = - T\int_{-\infty}^{\infty} d\Lambda G[\Lambda - ({2\over
\pi})\ln(T_K/T)]
\ln[1+\eta_{2S_I}(\Lambda)] \leqno(7.2.5)$$
where $T_K$ is the Kondo temperature.

In the limit $\Lambda\to\pm\infty$, the asymptotic solutions as given
by Andrei and Destri [1984], Desgranges [1985]
and Sacramento and Schlottmann [1991] are
$$\ln[1+\eta_k(\infty)] = 2\ln\{sin[\pi(k+1)/(M+2)]/sin[\pi/(M+2)]\},
~~k<M$$
$$~~~~~~~ = 2\ln\{sinh[(k+1-M)x_0]/sinh[x_0]\},~~k\ge M
\leqno(7.2.6.a)$$
and
$$\ln[1+\eta_k(-\infty)] = 2\ln\{sinh[(k+1)x_0]/sinh[x_0]\}
\leqno(7.2.6.b)$$
which holds for all values of $k$.   The $\eta_k(\Lambda)$ functions
are
monotonic decreasing, interpolating smoothly between the limits at $\pm
\infty$.
For the special case $k=M$, $\eta_M(\infty)$=0, which implies
$\epsilon_M(\infty)=-\infty$,
signalling the formation of a bound state of $M$ spinons. For all other
$k$ values $\eta_k$
is finite everywhere.

For finite $\Lambda$, the following numerical procedure can be used to
solve Eq. (7.2.2).
First, truncate the $k$ values to some large discrete value $k_c$;
typically, $k_c$ of order
50 suffices.  Above $k_c$, only the asymptotic solution given by Eq.
(7.2.4) is employed.
Further, truncate $\Lambda$ (typically $|\Lambda|<56$ is sufficient)
and use the asymptotic
forms specified by Eqs. (7.2.6.a,b) to characterize the large $\Lambda$
behavior.  The
integral equations are then solved numerically on a discretized mesh,
and thermodynamic
quantities are obtained by taking numerical derivatives of the free
energy.  These numerical
solutions are well controlled unless $H/T>>1$, for which special care
must be taken since
the limits $T\to 0,H\to 0$ do not commute.

\subsubsection{Thermodynamic Properties} 

We now summarize the various results.  The low temperature
thermodynamic properties
may be extracted from the asymptotic behavior of the coupled integral
equations. \\

\begin{figure}
\parindent=2.in
\indent{
\epsfxsize=5.in
\epsffile{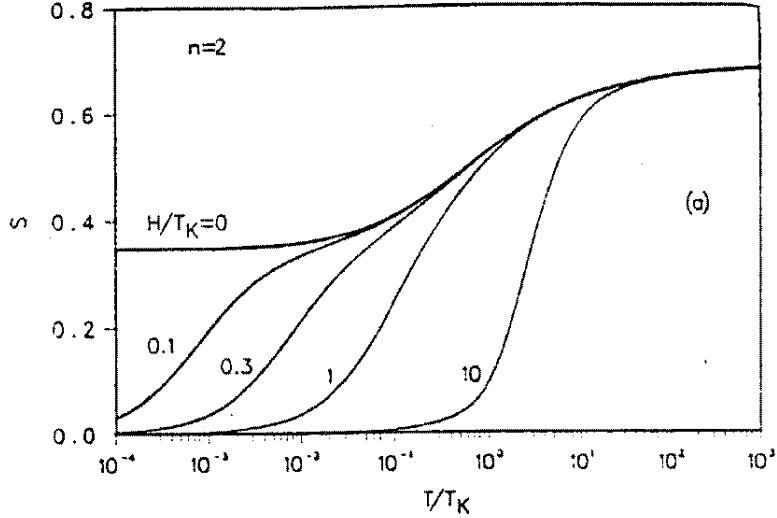}}
\parindent=.5in
\caption{Entropy for two-channel Kondo models computed with the Bethe-Ansatz
method, in zero and applied magnetic field. 
Note that $n$ here corresponds to our $M$ for channel number, and that $H$ corresponds 
to the spin field $H_{sp}$. The magnetic field reduces the residual
entropy to zero.   
From Sacramento and Schlottmann [1991].}
\label{fig7p1}
\end{figure}

{\it (a) Residual Entropy }\\
 The
residual entropy (with the limits $H\to 0$ then $T\to 0$ taken) is, for
$S_I=1/2$, (Andrei and Destri [1984], Tsvelik [1985])
$$S(0) = \ln\{2\cos[\pi/(M+2)]\} ~~. \leqno(7.2.7)$$
This result can be obtained from the free energy of Eq. (7.2.5) where 
$\eta_{2S_I} =\eta_1$ when $S_I=1/2$.  In the $T\to 0$ limit, the large
$\Lambda$ values dominate the integral (see Eqs. (7.2.3) and (7.2.5)), so
that $ln(1+\eta_1(\Lambda)) \approx 2\ln\{sin[2\pi/(M+2)]/sin[\pi/(M+2)]\}$
can be taken out of the integral.  The remaining integral yields $1/2$,
so Eq. (7.2.7) results. 
This result agrees with the conformal theory result quoted in Eq. (6.1.83)
and is non-zero when $M>1$.
With the limits taken in the order $T\to 0,H\to 0$, the entropy is
zero, indicating that the
external field generates a non-degenerate singlet ground state and
hints at the return to
Fermi liquid behavior discussed in Secs. 4,5, and 6.  In general, the
new scale
$T_s = T_K(H/T_K)^{1+2/M}$ is introduced in the presence of field for
$H<<T_K$
which sets the scale of the crossover from non-Fermi liquid behavior to
Fermi liquid
behavior as the temperature is lowered through $T_s$. For the
special case of $M=2$, this scale is $T_s=H^2/T_K$.   Clearly this
scale definition agrees
with the discussion of Sec. 4 (for $M=2$), Sec. 5.1, and Sec. 6.1.2.c.
In contrast, for the
fully compensated model, the only scale is $H$ itself.
The crossover from non-Fermi liquid to Fermi liquid behavior is well
illustrated in Fig.~\ref{fig7p1}
which shows the entropy calculated by Sacramento and Schlottmann [1991]
for a
number of different $M$ values.  It can be seen that as $M$ is raised
so that $T_s\to H$
the range over which the non-Fermi liquid behavior is apparent shrinks
relative to $M=2$.

For $M=2$, when one considers the TLS Kondo model, the level splitting
$\Delta$ plays the role
of $H$ so that the crossover occurs at $\Delta^2/T_K$.  For splitting
much smaller than
$T_K$, the region over which non-Fermi liquid occurs is much smaller
than the splitting
itself, which may have considerable relevance to the experimental
situation in small
point contacts (Ralph and Buhrman [1988,1991], Ralph {\it et al.}
[1994]).  

An alternative approach to recovering the $T=0$ entropy (and
in principle the entire quasi-particle spectrum and 
thermodynamics) has been presented by Fendley [1993].  He solves
directly for the $S$-matrix for scattering off of the impurity and
interprets the entropy fractionalization in terms of ``kinks'' or 
solitons interpolating between $k+1$ 
degenerate minima of the ground state.  
A complete discussion is beyond the scope of this paper and we refer 
the reader to the original reference for more detail.  \\

{\it (b) Zero Temperature Magnetization}\\

The zero temperature forms of the magnetization and susceptibility for
the $S_I=1/2$
case are given by  (Wiegman and Tsvelik [1983a], Andrei and Destri
[1984],
Tsvelik and Wiegman [1984])
$$M_{imp}(T=0,H) \sim ({H\over T_K})^{2/M},~~\chi_{imp}(T=0,H)
\sim({H\over T_K})^{2/M-1} \leqno(7.2.9)$$
for $M>2$, and for $M=2$ (Sacramento and Schlottmann [1989])
$$M_{imp}(T=0,H) \sim ({H\over T_K})\ln({H\over
T_K}),~~\chi_{imp}(T=0,H)\sim \ln({H\over T_K}) ~~.\leqno(7.2.10)$$
These results agree with the NRG results of Sec. 4.2 (for $M=2$), the
NCA results of Sec. 5.1.4, the CFT results of Sec. 6.1.3, and for large
$M$ with the
$1/M$ expansion of Sec. 3.4.5.\\

{\it (c) Low Temperature Specific Heat and Susceptibility}\\

For $S_I=1/2$, the heat capacity and susceptibility for $H=0$ and low
$T$ are found to be
(Andrei and Destri [1984], Tsvelik and Wiegman [1984])
$${C_{imp}\over T} \sim \chi_{imp}(T) \sim T^{4/(2+M)-1}
\leqno(7.2.11)$$
and  for $M=2$ (Sacramento and Schlottmann [1989])
$${C_{imp}\over T} \sim \chi_{imp}(T) \sim \ln({T_K\over T})
~~.\leqno(7.2.12)$$
These results are in agreement with the discussion of Secs. 5.1.3
(NCA),
6.1.3 (CFT), and 6.2.3 (Abelian bosonization, $M=2,4$ cases).
For large $M$, the results also check with the large $M$ expansion of
Sec. 3.4.5.  Note that the different power laws in $\chi$ at $T=0,H\ne
0$
and $T\ne 0,H=0$ are associated with the non-trivial scaling of $H$
with
$T$ in the overcompensated model (for $M>2$, 
$\chi(T,H=0)\sim T^{4/(2+M)-1}$,
while ($\chi(0,H)\sim H^{2/M-1}$; for $M=2$, the behavior is
logarithmic in both $T,H$).  \\

{\it (d) Full Numerical Solutions for $C_{imp},\chi_{imp}$}\\

\begin{figure}
\parindent=2.in
\indent{
\epsfxsize=5.in
\epsffile{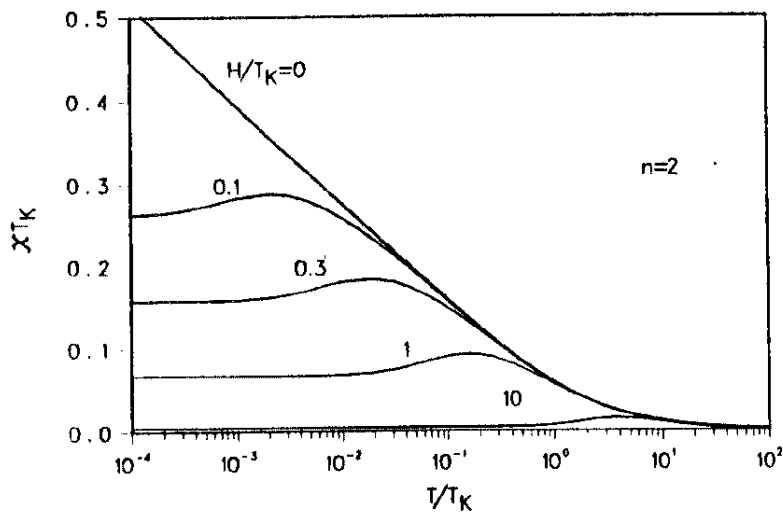}}
\parindent=.5in
\caption{Magnetic susceptibility for the two-channel 
multichannel Kondo models in 
applied spin field computed with the Bethe-Ansatz
method. 
Note that $n$ here corresponds to our $M$ for channel number, and that $H$ corresponds 
to the spin field $H_{sp}$.  The magnetic field removes the logarithmic 
singularity in the susceptibility.  
From Sacramento and Schlottmann [1991]}
\label{fig7p2}
\end{figure}

Desgranges [1985] and Sacramento and Schlottmann [1991] have solved
for the thermodynamic properties over the full temperature range (above
and
below $T_K$).  We display the $\chi_{imp}$ results from Sacramento and
Schlottmann [1991] in Fig~\ref{fig7p2}.  The essential instability of the
multi-channel
fixed point is readily understood from the diverging $\chi_{imp}$
curves in
this figure (for $H=0$).  Namely, an arbitrarily small asymmetry in the
TLS,
electric field gradient or strain field in the quadrupolar  Kondo
model, or magnetic
field in the magnetic impurity model will roll the divergence over
below $T^x_{sp}(\approx H^2/T_K$ for $M$=2). 
to a Fermi liquid behavior.  Heuristically, the presence of other
impurities which
produce a self-consistent
molecular field due to the Ruderman-Kittel coupling mediated by
the conduction electrons will also produce this kind of effect. This
reasoning can be useful in interpreting features in the susceptibility
data of
Th$_{1-x}$U$_x$Ru$_2$Si$_2$ (Amitsuka
{\it et al.} [1993a,b]).    See the discussion
of Cox[1987b,1988a], Sacramento and Schlottmann [1989], and Gogolin
[1995]
for related remarks on the Jahn-Teller effect in the quadrupolar Kondo
model.
Note that the compensated Kondo model requires a `critical' field
strength to
depolarize the impurity, of the order of $T_K$ (Doniach [1976],
Jayaprakash,
Krishna-Murthy, and Wilkins [1981], Cox [1987a]).

\begin{figure}
\parindent=2.in
\indent{
\epsfxsize=5.in
\epsffile{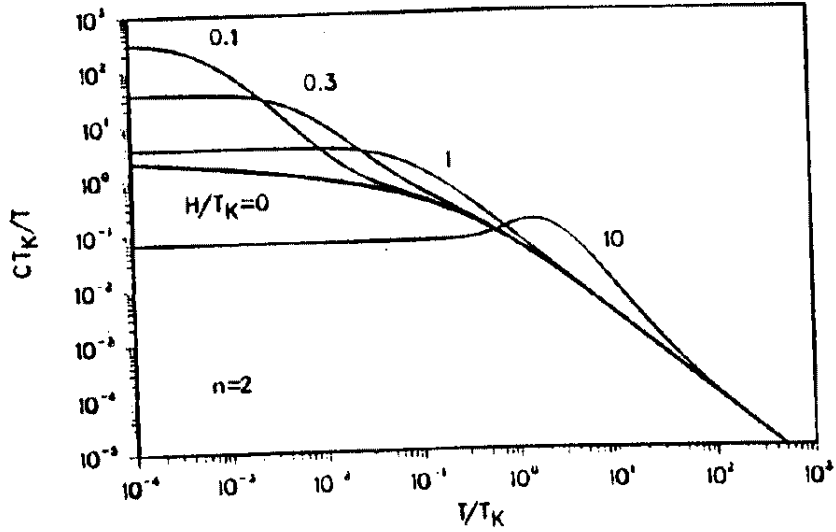}}
\parindent=.5in
\caption{Specific heat for the two-channel multichannel Kondo models in applied 
spin field computed with the Bethe-Ansatz
method. 
Note that $n$ here corresponds to our $M$ for channel number, and that $H$ corresponds 
to the spin field $H_{sp}$.  The applied field pushes out a Schottky 
anomaly seen very clearly as the peak in $C/T$ here.  
From Sacramento and Schlottmann [1991]}
\label{fig7p3}
\end{figure}

The specific heat curves of Sacramento and Schlottmann [1991] are shown
in
Fig.~\ref{fig7p3} .  Note the presence of the second peak below $T^x_{sp}$ for the
$H\ne 0$
case, which again indicates the crossover to the Fermi liquid.    A
heuristic interpretation
of this peak is that it is a `Schottky' like anomaly corresponding to
the removal
of the ground state residual entropy in the applied field.  This
appears to be useful
in interpreting features in the specific heat data of
Y$_{1-x}$U$_x$Pd$_3$
(Seaman {\it et al.} [1991,1992]; Cox, Kim, and Ludwig [1995]).

\begin{figure}
\parindent=2.in
\indent{
\epsfxsize=5.in
\epsffile{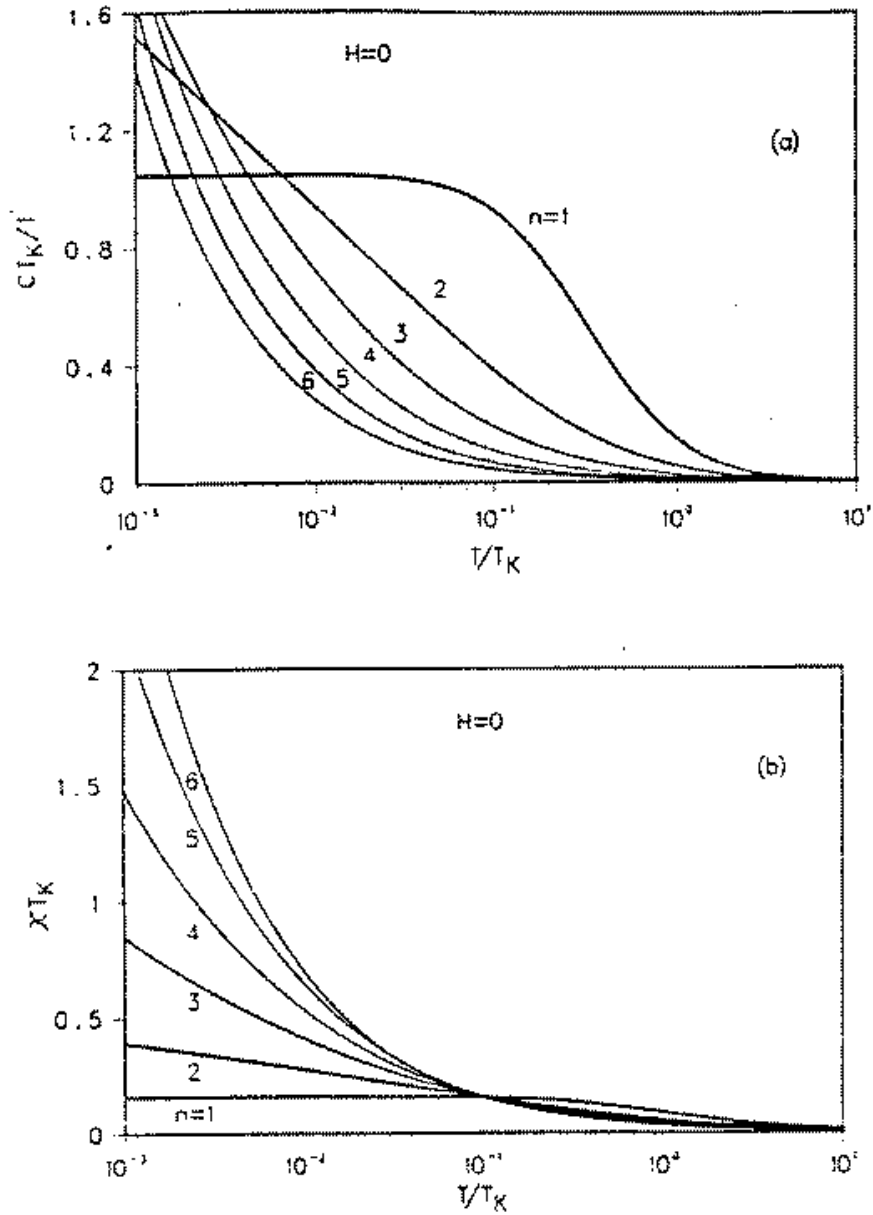}}
\parindent=.5in
\caption{Dependence of zero field magnetic susceptibility and specific heat on channel number 
for overcompensated 
multichannel Kondo models computed with the Bethe-Ansatz
method. 
Note that $n$ here corresponds to our $M$ for channel number. 
From Sacramento and Schlottmann [1991]}
\label{fig7p4}
\end{figure}

The overall dependence of the specific heat and the susceptibility on
the
channel number in zero magnetic field is shown in Fig.~\ref{fig7p4}.\\

{\it (e) Electrical Resistivity:  Magnetoresistance}\\

The electrical resistivity cannot be calculated within the Bethe-Ansatz
approach for arbitrary temperature.  This requires a knowledge of cross
sections for states with arbitrary numbers of excited particles and holes, and
construction of the exact wave functions for outgoing many body states.  
This problem has not yet been solved.  

At zero temperature, however, the Bethe-Ansatz can be used at least in 
the presence of finite magnetic field.  For example, in the one-channel 
problem at zero temperature, the magnetic field excites 
a single spinon and holon 
which corresponds physically to a one fermion excitation.  At zero temperature 
and arbitrary spin field, the scattering may be parameterized for 
both the compensated and over-compensated models in terms of phase shifts
for each spin channel (Andrei [1982]).   For example, for one channel 
$\delta_\uparrow(H)=-\delta_\downarrow(H) = \pi/2-\pi M(H/T_K)$ where 
$M = (n_\uparrow-n_\downarrow)/2$ is the magnetization. The 
magnetoresistance is proportional to $\sum_{\sigma}\sin^2\delta_{\sigma}(H)$.  
 The situation
for the two-channel model is different in that at least two spinons are
generated (Andrei [1995]); nevertheless, a phase shift parameterization 
still holds allowing calculation of the magnetoresistance and magnetization 
(Jerez [1995]).

\section{Experimental Results} 

In this section, we review the status of experimental understanding of
possible
TLS Kondo systems and two-channel spin and quadrupolar Kondo candidates
among
heavy fermion
alloys and compounds.

A general comment is that the theory described in this
paper is rigorous only
for a single TLS site or \ufp~ or \ctp~ ion.  Due to the divergent
length scale about
the multi-channel Kondo site, it is clear that there must always be
some crossover
away from the non-trivial fixed point for even two impurities (see sec.
9.3.1 for a
discussion of the two impurity model).  Hence the attitude here is that
one may flow
close to the non-trivial fixed point over some regime of temperature
and other parameters
at the finite concentrations which are always present in experiment,
and then eventually
will flow away.   If the crossover regime occurs over an extended
temperature range
below the Kondo scale (which identifies the weak coupling to strong
coupling crossover
as the temperature is lowered) we may be confident that the physics is
well described
by the non-trivial fixed point.

\subsection{Experiments on Possible TLS Kondo Systems} 

For  this class of candidate systems, an excellent review of earlier
experimental
and theoretical work appears in J.L. Black [1981].  A more recent review
of experimental work appears in von Delft {\it et al.} [1997a] for the
specific results on copper nanoconstrictions.  

The contributions of the TLS to the low temperature and low energy
dynamics of amorphous
materials and materials
 with defects are due to transitions between the ground state and the
 first excited
state.  Thus, according to the discussion of Sec. 2.1.1, the TLS must
be slow or fast to have the
energy splitting $\Delta$ of these TLS below or comparable to the
characteristic energy of these
experiments.  The ultrafast TLS with large splittings are frozen out
and those behave as rigid
defect centers.  The typical concentration of the TLS in amorphous
materials is about 10$^{18}-10^{19}$ per cm$^3$, thus about 10-100 ppm.

Kondo type behavior only occurs in those cases where the Kondo
temperature exceeds the
TLS splitting ($T_K>\Delta$); the Kondo state does not form in the
opposite case because
the splitting ``saturates'' the pseudo-spin of the TLS.  In the limit
$\Delta=0$ non-Fermi liquid
behavior dominates the low temperature physics.  For finite splitting,
however, there is a crossover
to Fermi liquid behavior as the characteristic energy of the experiment
is reduced below $\sim
\Delta^2/T_K$ (c.f., secs. 4.2,5.1.4,6.1.3.c,7.2), where $\Delta$ is the 
unrenormalized asymmetry splitting and $\Delta <T_K$ is assumed.
 In most of the samples, the parameters of the TLS,
i.e., the splitting and
the Kondo scale, are characterized by broad distributions.  Thus the
Fermi liquid and non-Fermi
liquid behaviors are mixed in macroscopic samples where there are many
TLS impurities.
In principle, that can be avoided in two cases:\\
(i) materials with very well defined, uniform TLS's in a crystalline
environment, and \\
(ii) small mesoscopic samples where there are only a few centers and by
sample selection
the two behaviors (Fermi liquid and non-Fermi liquid) can be separately
studied.  \\

In most instances, the TLS are formed in amorphous material or
amorphous regions of the samples.
The latter may also be represented by the vicinity of dislocations or
by surfaces and interfaces
between two different kinds of materials.    The surfaces and
interfaces are especially
important in mesoscopic devices where they represent a large fraction
of the samples.

There are however very few cases where the TLS are very well defined.
The very well defined
crystalline material Pb$_{1-x}$Ge$_x$Te will be discussed below.  This
material shows Kondo
like resistivity anomalies associated with the Ge atoms, which may hop
between several
quasi-equilibrium positions.  A closely related and
well defined  problem is the diffusion of a light atom (e.g., hydrogen
or a muon) through a metal.  In this case, the interstitial sites
between which
 the light atom moves are regularly
distributed in the crystals. The hopping between two of those sites can
be approximated by
a TLS.  A very extensive review of these phenomena has been given by
Kagan and
Prokof\'{e}v [1992].  The most important process is the reconstruction
of the
electronic screening
after the hopping which shows infrared divergent character.  However,
the non-commutative
terms characterizing the Kondo effect do not play an important role,
and so this fascinating
area of research will not be discussed further in the present review.

The rest of the discussion in this section will focus on the electrical
resistivity of bulk materials
and the I-V characteristics of small mesoscopic devices, where the
scattering rates of the electrons
due to TLS's can be measured.  The thermodynamic properties will be
discussed only briefly,
because in thsoe cases both the slow and fast TLS from the
distribution of TLS sites play a
role and as a consequence it is difficult to extract any reliable
information about those sites
which display a Kondo effect.

\begin{figure}
\parindent=2.in
\indent{
\epsfxsize=3.in
\epsffile{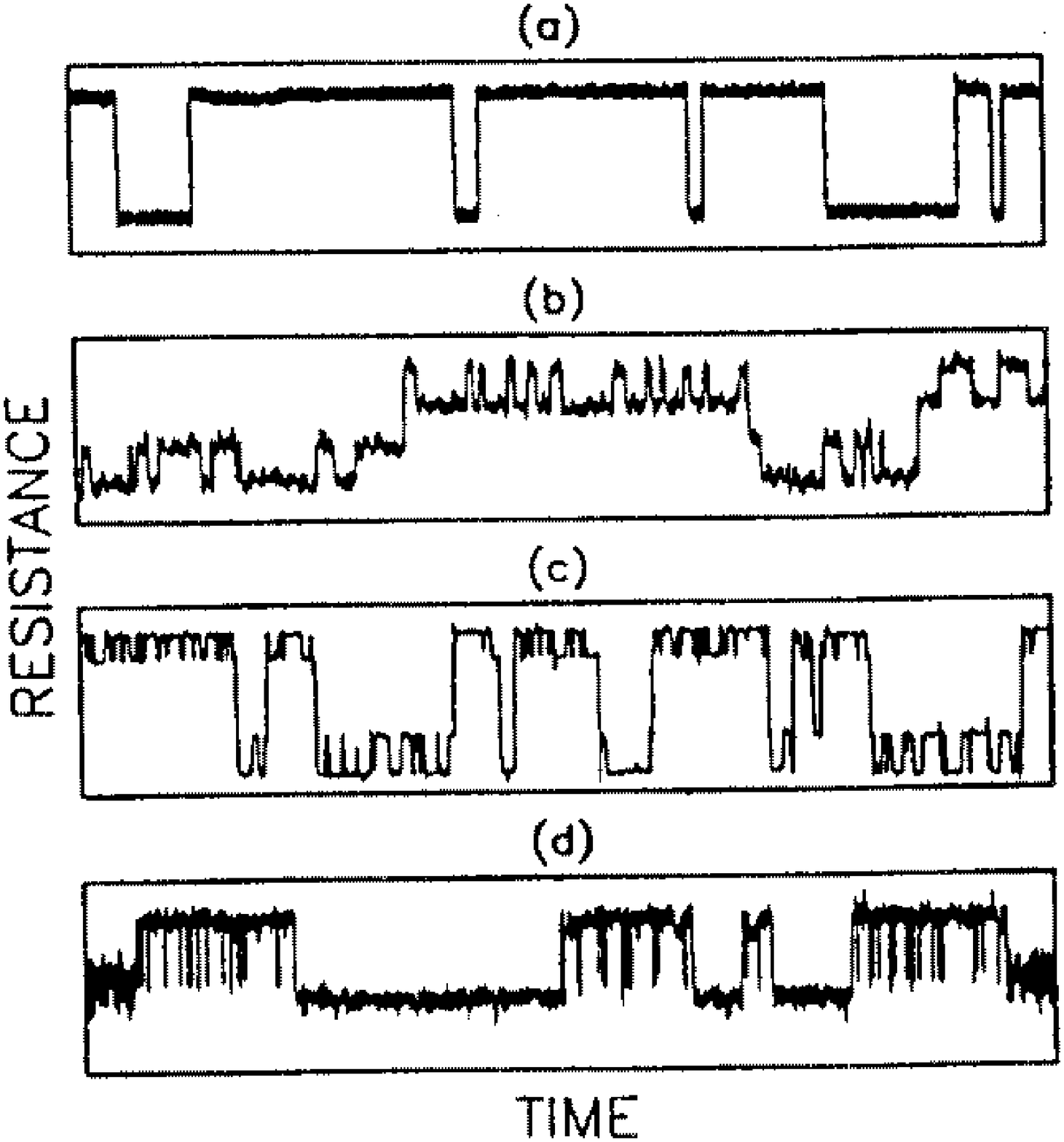}}
\parindent=0.5in
\caption{Telegraph noise displayed in time dependent resistivities of 
copper nanobridges for temperatures smaller than 150K (but still thermally activated)
from Ralls and Buhrman [1988]. (a) shows the modulation of the resistance from a single TLS; (b) shows two independent TLS; (c) 
shows modulation of the amplitude by one TLS acting on another; (d) shows the modulation of the frequency by one TLS acting on 
another.  }
\label{fig8p1}
\end{figure}

\begin{figure}
\parindent=2.in
\indent{
\epsfxsize=3.in
\epsffile{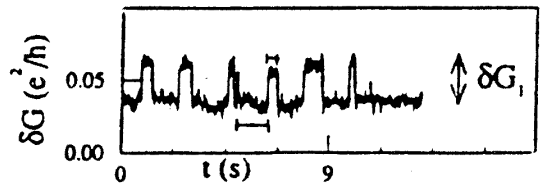}}
\parindent=0.5in
\caption{Time dependent electrical resistivity of a submicron bismuth sample near 1K where only tunneling 
is possible from Zimmerman, Golding, and Haemmerle [1991].  }
\label{fig8p2}
\end{figure}

{\it Slow Centers}.  For slow centers $V_x\approx V_y\approx 0$ and
$V_z\ne 0$ characterizing
the screening by conduction electrons.  The slow centers show up in the
fluctuations of the
electrical resistivity (see Fig.~\ref{fig8p1},\ref{fig8p2}). If there are several centers
then they can modify each other's parameters, so that the average frequency of the 
transition and the amplitude depend on the states of the other centers, as 
measured on nanobridges at high temperatures by Ralls and Buhrman [1988,1991], where the
transitions are thermally activated, and in the measurements of Zimmerman, Golding, and 
Haemmerle [1991], which were carried out at very low temperatures where only tunneling is
possible.  Such modulation of the conductance/resistance is called ``telegraph noise''.  (For a review, see N. Giordano
[1991]. )
The average transition rate in metals is reduced by the screening of the
electrons according
to
$$\Delta_0(T)  = \Delta_0({max\{T,\Delta_0(T)\}\over D})^{-K}
\leqno(8.1.1)$$
for a symmetric TLS, with splitting $\Delta_0$ in the absence of TLS-electron 
interactions.  This formula applies for thermally activated and tunneling induced 
transitions. The exponent $K$ is, for weak coupling, quadratic
in $V_z$ and may
be found through application of the renormalization group equation
(3.4.15) to the commutative
model which gives
$$K = 4 v^z(0)^2 ~~.\leqno(8.1.2)$$
More generally, $K$ may be related to the phase shifts of the different
angular momentum
scattering channels off of the atom (Anderson [1967], Kagan and
Prokof\`{e}v [1992]), {\it viz.}
$$K = \sum_l (2l+1)({\delta_l\over \pi})^2 ~~.\leqno(8.1.3)$$
In most of the papers from the metallic theory side, the notation $K$
is used for the exponent,
while from the macroscopic quantum tunneling theory initiated by
Caldeira and Leggett [1981,1983]
$K$ is denoted $\alpha$.  At low temperatures, the tunneling rate
always exceeds $T$, and
thus provided $K<1$ the solution to Eq. (8.1.1) is
$$\Delta^*_0 = \Delta_0 ({\Delta_0 \over D})^{K\over 1-K}
~~.\leqno(8.1.4)$$
This corresponds to a temperature independent hopping rate at low
temperatures.
In the case $K>1$, the $\Delta_0(T)\to 0$ as $T\to 0$ and the particle
localizes.  Such a large
value of $K$ is not likely in a realistic system as the couplings are
replaced by the phase shifts
and $K$ has an upper bound dependent on how many angular momentum
channels participate
in the screening (see, e.g., Kagan and Prokof\`{e}v [1992],
\vld~[1993]).

Measuring the distribution of the time between two tunneling events $K$
can be determined.
A typical value is about $K=0.2-0.3$, showing that the measured TLS are
far from being localized
(B. Golding {\it et al.} [1992], S.N. Coppersmith [1992], N.M.
Zimmerman {\it et al.} [1991], K.
Chun and N.O. Birge [1993]).  These experiments can give information
also about the asymmetric
energy splitting of a particular TLS because the ratio of the time
spent
in the different states is determined by thermal equilibrium
conditions.

{\it Fast Centers}.  Eq. (3.4.26) gives the Kondo temperature for the
two channel Kondo
problem as
$$T_K^{(II)} = D_0 (v^x(0)v^z(0))^{1/2} ({v^x(0)\over 4v^z(0)})^{1\over
4v^z(0)} ~~.\leqno(8.1.5)$$
In order to estimate $T_K^{(II)}$, we take $D_0 \approx 10 eV \approx
10^5K$.  A reliable
estimate
of $v^z(0)$ requires experimental input--it can be sampled through
ultrasonic absorption or
internal friction measurements (sound velocity shifts can provide
further useful information).
A TLS absorbs ultrasound and may be saturated for sufficiently high
power (Golding
{\it et al.} [1978], Black [1981]).  The saturation power is much
higher in metals than in
insulators because the relaxation time characterizing the transition
from the excited state of
the TLS back to the ground state is much shorter due to the
Korringa-type process relaxing
the pseudo-spin by the creation of particle-hole pairs.  This process
for the TLS pseudo-spin
is completely analogous to the Korringa relaxation of nuclear and
electronic spins in metals.
The Korringa process dominates over the channels existing in the
insulator phase (e.g., phonon relaxation of the TLS).  We discuss
this in detail in the following paragraph.  Most of the relevant
experiments were performed over a decade ago and the results are
summarized in \vld and \zow
[1983(c)].   For more recent data, see Esquinazi {\it et al.} [1986],
[1992] and references
therein.

Above $T_K$,
the Korringa relaxation rate is proportional to the temperature times
the
renormalized coupling strength $v^2 =
(v^z)^2+(v^x)^2
+(v^y)^2$, which has a  temperature dependence $v^2\sim 1/\ln^2(T/T_K)$
in the vicinity of the
Kondo temperature.  If we go to sufficiently high temperatures,
however, $v^z$ dominates the
coupling strength and is only mildly renormalized except in the most
extreme cases.  Hence,
a rough estimate of $v^z(0)$ may be found by taking $v^2 \approx
(v^z(0))^2$.  The estimates
found in this way give the range $v^z(0) \approx 0.02-0.25$, depending
on the material
(see Table I of  \vld~ and \zow [1983(c)]).  That estimate is in
accordance with the value of
$K=\alpha=0.2-0.3$ measured for the slow center, since by Eq. (8.1.2)
we see that $v^z(0)
= (\alpha/4)^{1/2}$, which yields $v^z(0) \approx 0.2-0.3$.

If we consider the original TLS model with only the lowest two states
of the TLS retained,
Eq. (2.1.26) gives
$${v^x(0)\over v^z(0)} \sim 10^{-3}-10^{-4} ~~.\leqno(8.1.6)$$
Taking the largest value of this ratio (10$^{-3}$) and for $v^z(0)$ of
0.3, we come up with the
most optimistic upper bound on $T_K^{(II)}$ of about 0.3K.  Clearly, we
can make this go
down by orders of magnitude with the smaller estimates (and a smaller
value for $D_0$).
Experimental observability would require a larger $T_K$ value.
Recently, the work of
\zar~ and \zow [1994(a,b)] has solved this problem by taking into
account the further excited
states of the TLS, as discussed in Sec. 3.4.2.  The essential point is
that the higher levels
induce a significant enhancement of the Kondo scale quite analogous to
the enhancement of
$T_K$ for magnetic impurities by the presence of higher lying crystal
field states and angular
momentum multiplets.  With realistic parameter choices, it was found
that Kondo scales of the
order of a few Kelvin were well within reach.

\begin{figure}
\parindent=2.in
\indent{
\epsfxsize=5.in
\epsffile{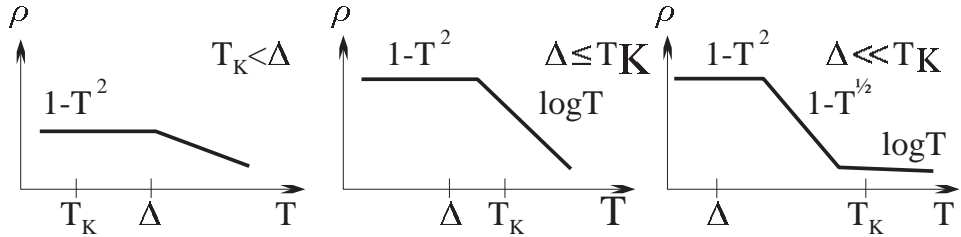}}
\parindent=0.5in
\caption{The temperature dependence of the electrical resistivity
is shown in a schematic way for different
cases $T_K<\Delta$ (left), $\Delta\le T_K$ (center), and $\Delta <<T_K$ (right).  }
\label{fig8p3}
\end{figure}

{\it Electrical Resistivity}. The electrical resistivity measures the
electronic scattering rate off of
the TLS.  That subject has been first discussed by Cochrane {\it et
al.} [1975] who introduced
an ill-defined model with two sets of conduction electrons
heuristically provided.  The first
calculation was performed by Kondo [1976(b)] up to fourth order
introducing the assisted
tunneling. Because $v^y(0)=0$, and $v^y$ is thus generated with a
logarithmic factor at second
order in perturbation theory (see Eq. (3.3.26)), the first
non-vanishing logarithmic correction to
the scattering rate is of order $(v^x v^z)^2 \ln^2(D/T)$. This
contrasts,
for the original Kondo problem where
the leading log correction in the scattering rate shows up at third
order (Kondo [1964]).
The smallness of $v^x(0)$ led Kondo to conclude that no Kondo
resistivity anomaly could be seen
due to the TLS.  Only the large renormalization discussed in Sec. 3
makes the effect observable.
The resistivity behavior expected at different temperatures depends on
the ratio of $\Delta_0/T_K$,
and this is illustrated in Fig.~\ref{fig8p3}.  In the left most curve of 
Fig.~\ref{fig8p3}, $T_K<\Delta_0$ so
that the Kondo correlated
state is not developed and we pass to the Fermi liquid state.  In the center and right hand curves of
Fig.~\ref{fig8p3} we see the Fermi
liquid behavior developing eventually below $T_K$, but with a non-Fermi
liquid region possible
provided $\Delta_0<<T_K$ (see the right hand curve of Fig.~\ref{fig8p3}.

\begin{figure}
\parindent=2.in
\indent{
\epsfxsize=3.in
\epsffile{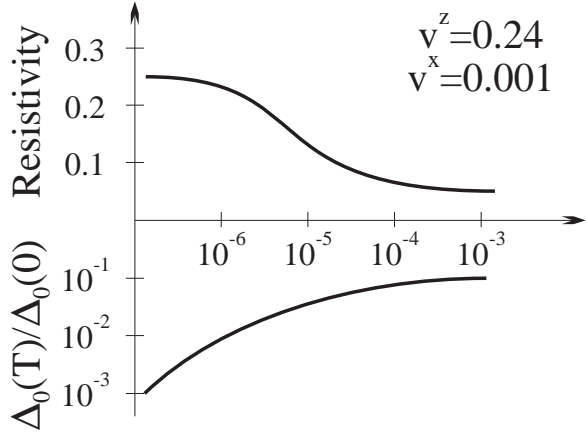}}
\parindent=0.5in
\caption{ The resistivity and the renormalized tunneling rate are
shown as a function
of temperature in the next-leading logarithmic order of the
renormalization group.  The
bare parameters are $v^z=0.24$ and $v^x=0.001$.  }
\label{fig8p4}
\end{figure}

At high temperatures, summation of the logarithmically divergent terms
gives the correct
logarithmic rise in $\rho(T)$.  A crude estimate for the incremental
scattering strength per impurity
$\Delta \tau^{-1}$ is provided by evaluation of the scattering rate
at zero temperature and finite frequency using the coupling strengths
evaluated at frequency
scale $x=\omega/D$ in a golden rule estime.  This gives
$$\Delta\tau^{-1}(\omega,0)= {4\pi \over
\rho_0}[(v^z)^2+(v^x)^2+(v^y)^2]_{x=\omega/D} ~~.
\leqno(8.1.7)$$
We then estimate the additional resistivity due to the TLS as
$$\Delta \rho(T) \approx {m\over ne^2} c_{TLS}
 \Delta\tau^{-1}(\omega)|_{\omega=k_BT} \leqno(8.1.8)$$
where $n$ is the electronic density, $e$ the electron charge, $m$ the
electron band mass, and
$c_{TLS}$ the concentration of TLS. More appropriate formulae can be
found in \vld~ and \zow
[1983(c)] where the transport lifetime is properly evaluated.  The
result of the expression (8.1.8)
is shown in Fig.~\ref{fig8p4} with the calculated TLS splitting $\Delta_0$ for a
symmetrical level using
$\Delta=0$ (c.f. Eq. (3.4.15)).

At low temperatures for zero splitting of the levels, the non-Fermi
liquid excitation spectrum
produces an anomalous saturation of the resistivity.  According to
conformal field theory [Ludwig and Affleck [1991], Affleck and Ludwig
[1993]]
and the NCA [Cox and Ruckenstein [1993]] $\delta \rho(T) \approx
\Delta \rho(0)[1- aT^{1/2}]$ in the weak-coupling limit
($v^x(0),v^z(0)<<1$), where $a$ is a pure
number that may depend on the presence of ordinary potential scattering
at the impurity site.
So far, reliable calculation of the crossover from high to low
temperatures can be done only
with the NCA (a large $N$ technique) or the direct $1/M$ expansion
discussed in Sec. 3.4.5
(Gan {\it et al.} [1993]).

\begin{figure}
\parindent=2.in
\indent{
\epsfxsize=5.in
\epsffile{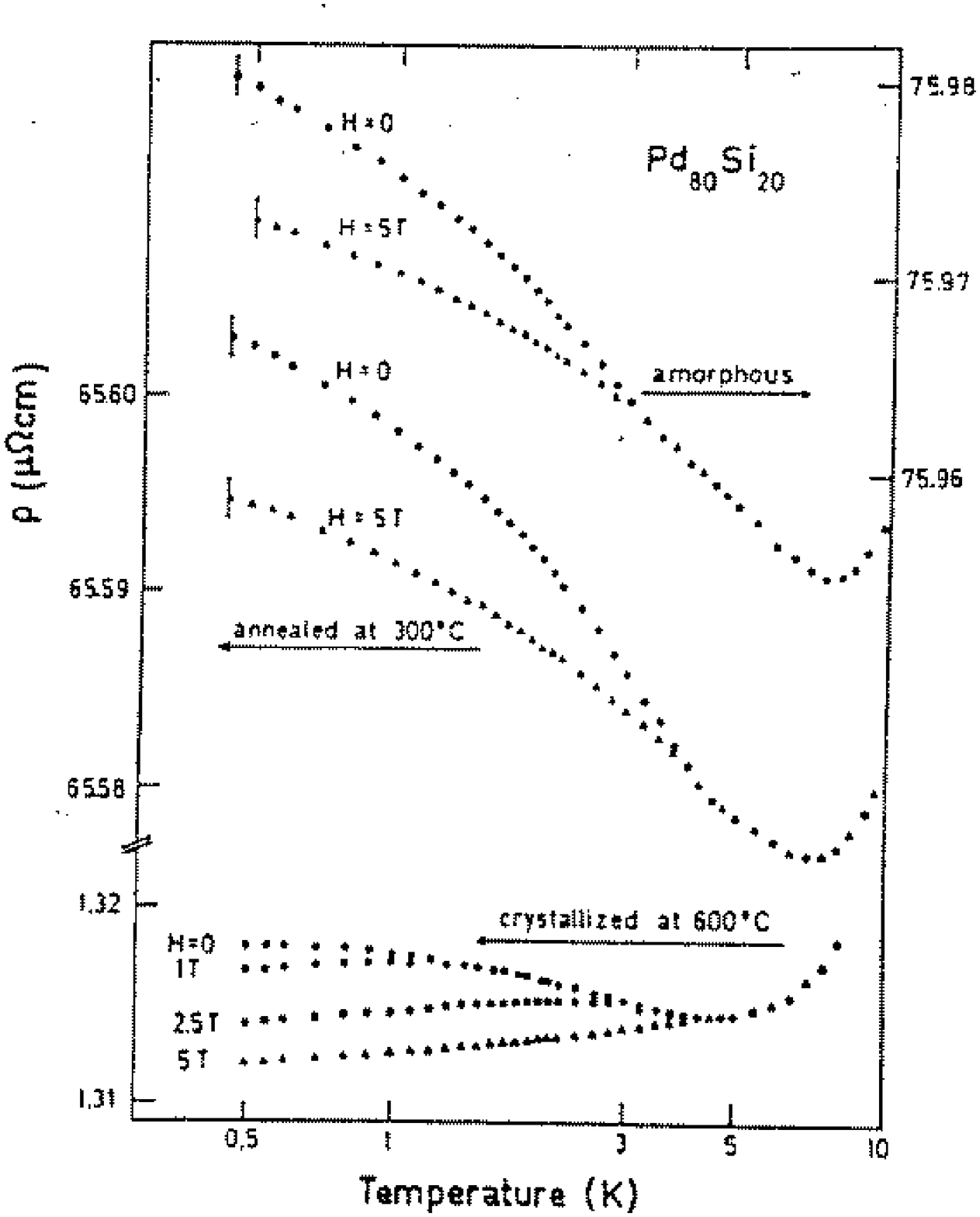}}
\parindent=0.5in
\vspace{.3in}
\caption{Resistivity vs. temperature of Pd$_{80}$Si$_{20}$ in the amorphous, annealed, and crystallized state for
magnetic fields $0\le H\le 5T$.  The positive magneotresistivity has always been subtracted.  
From K\"{a}stner {\it et al.} [1981] }
\label{fig8p5}
\end{figure}

{\it Examples of logarithmic resistivity signatures for TLS
candidates}.  In amorphous materials,
there are many experiments which show a logarithmic increase of the
resistivity with decreasing
temperature at low temperatures.  The most convincing are those in
which the amplitude of the
maximum resistivity attained at low temperature decreases with sample
annealing and disappears
with re-crystallization.  In order to rule out the spin Kondo effect as
an origin for this anomaly,
any dependence on applied magnetic field must be weak.  Even in these
cases, it is hard to rule
out localization as a source of similar logarithmic temperature
dependence.  Typical resistivity
data of this type are shown in Fig.~\ref{fig8p5}, taken from K\"{a}stner {\it et al.} [1981] from a study of
Pd$_{80}$Si$_{20}$ alloys.

A completely different example is the resistivity in highly doped
conducting polymers at low
temperatures studied by Ishiguro {\it et al.} [1992].  One example is
highly doped  polyacetylene.
In this case, the authors argue against the competing localization
mechanism mentioned above
as a source for the logarithmic upturn in the resistivity.  An
alternative
explanation in terms of magnetic impurities associated with certain
carbon groups has been put forward (Cruz {\it et al.} [1995]).

{\it Resistivity of Pb$_{1-x}$Ge$_x$Te}.  PbTe is a type $II-VI$
crystalline semiconducting
compound in which a small amount of Pb can be replaced by Ge.  The
material thus obtained
is either a $p$-type or $n$-type degenerate semiconductor.   The
ambiguity arises because the
concentration of charge carriers is different from sample to sample and
cannot be controlled
in a systematic way.  Because the ionic radius of Ge$^{2+}$  of
0.73$\AA$ is much smaller than
the 1.2$\AA$ radius of Pb$^{2+}$, the Ge atoms do not stay at the lead
positions, but slide
around in the various body diagonal directions ([111]) giving rise to
eight possible equilibrium
positions.  The randomness of the Ge atoms induces local strain fields
that lift the eight-fold
degeneracy, plausibly leaving only two close in energy at each Ge site
and thus giving rise to
a TLS.  The relevant experimental work was performed by Takano {\it et
al.} [1984].

\begin{figure}
\parindent=2.in
\indent{
\epsfxsize=3.in
\epsffile{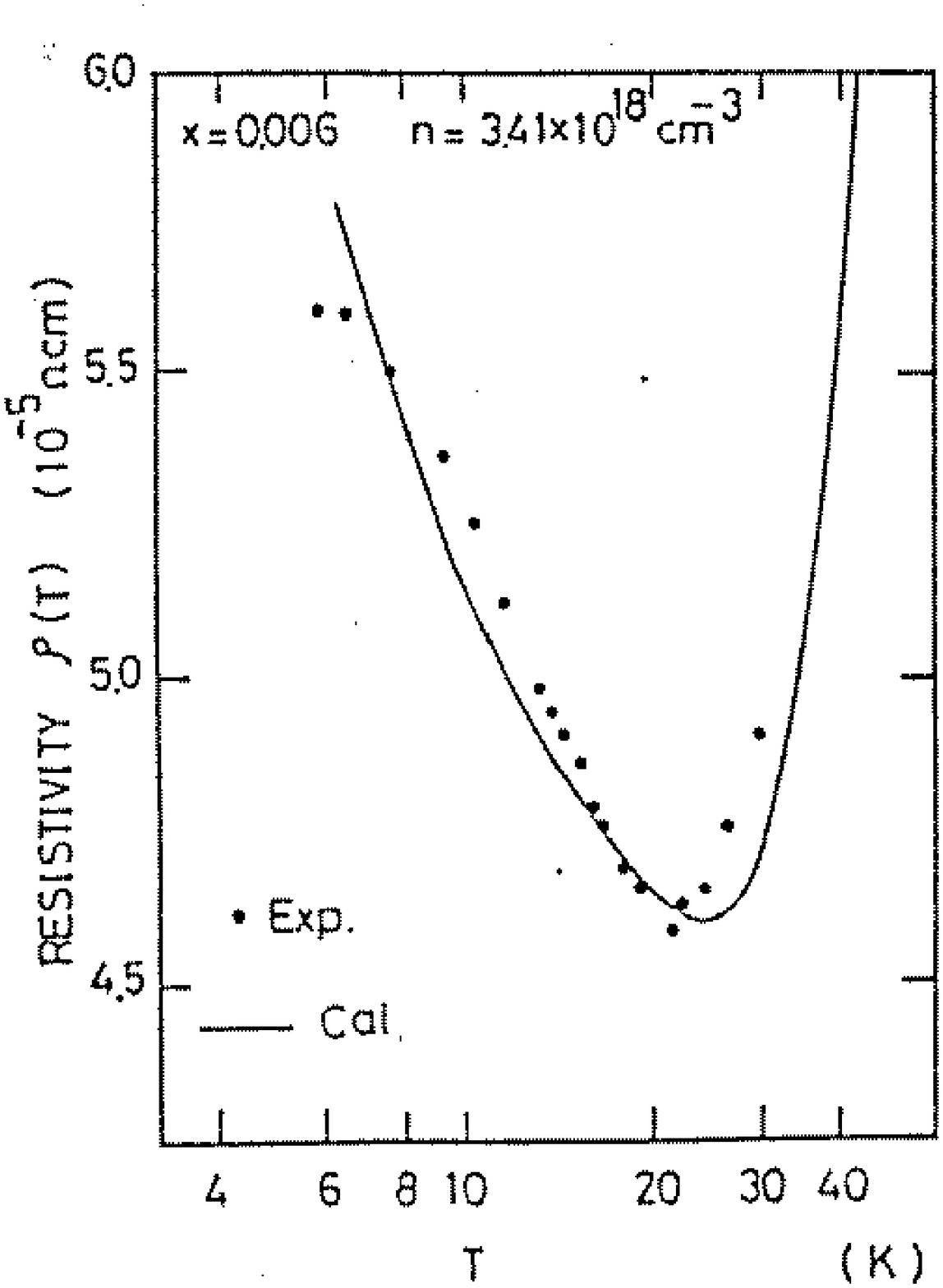}}
\parindent=0.5in
\caption{Total resistivity $\rho(T)$ vs. temperature for
Pb$_{1-x}$Ge$x$Te.  The closed circles
correspond to the experimental results by Takano {\it et al.} [1989].}
\label{fig8p6}
\end{figure}

The experimental data concerning resistivity minima was analyzed by
Katayama {\it et al.}[1987].
The theoretical fit to the data of Takano {\it et al.} [1984]
 is based upon use of Eqs. (8.1.7,8), and is shown in Fig.~\ref{fig8p6}.  The
 curve clearly shows a
logarithmic temperature dependence over about a decade of temperature.
The composition of
the sample for which the fit was performed has $x=0.006$, a carrier
density of 3.41$\times
10^{18} cm^{-3}$, an effective band mass of $m^*=0.053$,  a Fermi
energy of $E_F=718K$, and
an estimated bandwidth of $D=665K$.  The coupling strength
$v^z(0)=0.33$ was estimated from
a knowledge of the Ge positions, and $v^x(0)/v^z(0)$ was taken to be
0.001 in order to give the
correct Kondo scale $T_K=0.89K$.
However, the leading logarithmic formulae were used to estimate $T_K$,
and with this estimate for $v^x(0)$ the more correct estimate provided
by $T_K^{(II)}$ will be
too small.  This can be remedied by making $v^x(0)=0.01$, which seems
perhaps too large.
A more likely explanation is that the ratio of $v^x(0)/v^z(0)$ may be
about right, but that
the higher lying levels must be properly taken into account as per the
discussion of Sec.
3.4.2 and the work of \zar and \zow [1994(a,b)].  In order to get more
accurate estimates,
ultrasonic attenuation studies  are needed in the future.

While this is a good material for study given that the exact source of
TLS is known, there are
complications.  The random strain field induces a distribution of
splittings, as discussed above.
Also, the Ge ions not only introduce carriers and, apparently,
 the TLS Kondo effect, they also give rise at
higher concentrations to ferroelastic transitions which intervene in
the Kondo upturn.  Thus there
is practically no way to sample cleanly the non-Fermi liquid behavior
at lower temperatures.

{\it TLS at Dislocations}  TLS may be formed at dislocations.  Endo
{\it et al.} [1988]
studied different dilute aluminum alloys where dislocations where
introduced by
shock loading or extension at different temperatures.  The increase of
the density of
dislocations was associated with a logarithmic increase in the
resistivity at low
temperatures.

\begin{figure}
\parindent=2.in
\indent{
\epsfxsize=3.in
\epsffile{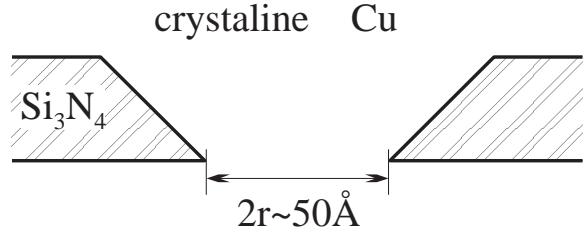}}
\parindent=0.5in
\caption{The point contact prepared by Ralph and Buhrman
[1992,1994]. The gate (shaded)
is made  of Si$_3$N$_4$.  The diameter of the orifice is $2r$.  The
contact is crystalline
copper.   }
\label{fig8p7}
\end{figure}

{\it Point Contact Spectroscopy of Nanoscale  Junctions}.  The most
convincing measurement of
TLS scattering amplitudes to date has been in the study of nanoscale
junctions or gates
through point
contact spectroscopy sampling of $dI/dV$ of the gates (Ralph and
Buhrman [1992],
Ralph {\it et al.} [1994], Ralph and Buhrman [1994]).  The method of
fabrication of such a junction
is shown in Fig.~\ref{fig8p7}.  It starts with electron beam lithography as
developed by Ralls and Buhrman
[1988,1991] and reactive ion etching to make a hole in a silicon
nitride membrane.  The minimum
size of the hole is 3-15 nm.  Then the membrane is rotated and covered
by different metals
(Al,Ag,Pt,Ti,V and primarily Cu have been used) on both sides.
The electron mean free path of the electrodes
is about 180 nm at 4.2K.  In the region of the gate very likely TLS are
formed.  The low frequency
noise can be recorded by measuring the conductance of that gate as a
function of time (
Ralls and Buhrman [1988]).  The recorded conductivity is similar to the
one shown in Figs.~\ref{fig8p1},\ref{fig8p2} and
those slow fluctuations are due to slow TLS.

The effect of the fast TLS cannot be seen through the resistivity
fluctuations and can only be
sampled by $dI/dV$ characteristics.  In that case there is a voltage
difference between
two electrodes attached on either side of the gate.  On one side the
Fermi energy is higher and
thus the gate works something like a tunnel junction.  Current is
driven by electrons propagating
ballistically through the small gate orifice.  Because there is no
barrier but actually a geometrically
limited path for the electrons, the experiment is equivalent to point
contact spectroscopy in which
a sharp metal tip is placed against a material.  In that case, the tip
contact serves as the small
orifice.  Two reviews of point contact spectroscopy are Yanson and
Shklyarevskii [1986] and
Jansen {\it et al.} [1980].

The essential idea is as follows: even in the absence of scattering in
the contact, the small orifice
produces a geometrically limited resistance $R_{B0}$ to the incoming
electrons because not
all the electrons can make it through the gate. In addition, the
electrons experience
scattering processes within the contact giving rise to a resistance
electrons.  To the extent these are frequency dependent, they will be
sampled at a characteristic
energy $eV$ for each temperature because added electrons flying through
the gate
come with energy $eV$ above the Fermi level.
As a result, the dynamic resistance of the junction is given
by (Jansen {\it et al.} [1980])
$${dV\over dI} = R(V) \approx {h \over e^2} {8 \over (k_Fd)^2}[1 + 0.4
({d\over v_F\tau(ev)})]
\leqno(8.1.9)$$
where $h$ is Planck's constant and $d$ is the orifice diameter.  This
equation is actually the result
for a circular hole in a membrane.

\begin{figure}
\parindent=2.in
\indent{
\epsfxsize=5.in
\epsffile{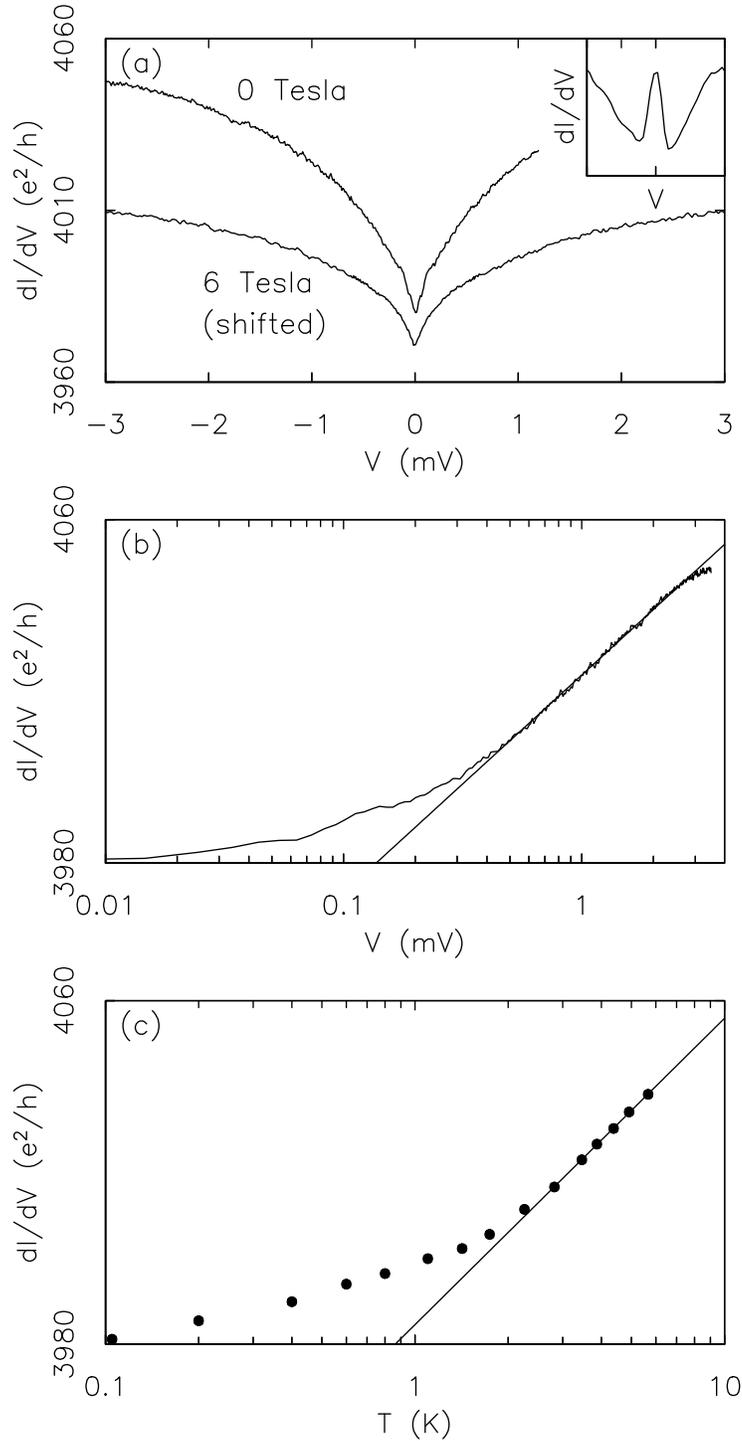}}
\parindent=0.5in
\vspace{.3in}
\caption{(a) Differential conductance of a Cu constriction at $T$=100mK. 
The curve measured in 6 T magnetic field  is shifted down by 20$e^2/h$ for clarity.  
(b) $V$ dependence of the differential conducatance for $B=0$ T and $T$=100 mK. (c) 
$T$ dependence of th conductance for $B=0$ T and $V$=0 mV.  Straight lines illustrate
regions of logarithmic $V,T$ dependencees.  Inset to (a):  Conductance of Cu constriction with 200
ppm Mn at 100mK, 4 T, showing suppression of magnetic-impurity scattering by an 
applied magnetic field.  From Ralph and Buhrman [1992].  }
\label{fig8p8}
\end{figure}

For the hypothesized TLS  in Ralph and Buhrman's data, typical
conductance curves $G(V,T) =
dI(V,T)/dV$ are shown in Fig.~\ref{fig8p8}.  The conductance is minimum at $V=0$
indicating a zero
bias anomaly, and showing that the relaxation rate decreases with
decreasing voltage. The
conductance shows a clear region of logarithmic drop with voltage,
indicative of a Kondo effect.
To rule out a magnetic Kondo effect, applied magnetic field was also
measured--a Zeeman splitting
of the central zero bias anomaly would be expected, but none is
observed in contrast to the devices controllably doped with
magnetic impurities.  The voltage dependence
of the observed relaxation rate is similar to frequency dependent rate
computed by \vld~
and \zow [1983(c)] for a TLS, and are also very similar to resistivity
results of Fig.~\ref{fig8p3}.  In Fig.~\ref{fig8p8}(c)
the temperature dependence of the zero bias conductance is shown.  This
also shows a logarithmic
temperature dependence with a break in that behavior at lower $T$ that
could was discussed in
the original work as either a new $\ln T,\ln V$ dependence or a
possible $\sqrt{T},\sqrt{V}$ dependence at lower temeperature, which
is
characteristic of the low temperature two-channel Kondo model
resistivity.
At even lower temperature and bias, the two-channel Kondo scaling must
be stopped by the presence of the renormalized TLS splitting.  This
occurs at a crossover scale we have denoted $T_x$ in this review, and
which is called $T_{\Delta}$ in the experimental paper of Upadhyay, Louie,
and Buhrman [1997].  Fermi liquid behavior is recovered below this 
temperature scale.  Thus, for $T<T_x=T_{\Delta}$, $T^2$ behavior
should be seen in the zero bias resistance, and the
$\sqrt{V}$ behavior saturates below $eV=k_BT_{\Delta}$.  This
crossover has been observed in Ti and V point contacts as we shall
discuss further below (Upadhyay, Louie, and Buhrman [1997]).

In  order to confirm the behavior arose from defects of the TLS
variety, Ralph and Buhrman
[1994] have
performed a number of experimental tests.   First, the effect is very
reproducible--it shows up in
half of the fabricated junctions and while initial studies focussed on
Cu junctions, similar
conductance curves have been found in silver, vanadium, 
titanium, and platinum
devices.  It is interesting that the effect of $T_x$ is shown
only by Ti and V samples (Upadhyay, Louie, and Buhrman [1997]).  
Furthermore, application of a high current to the junction leads to a
shift of the TLS parameters measured after the high current is switched off and 
then used to fit the data, which strongly
suggests that the current rearranges the atomic positions (electromigration) 
and modifies
the TLS in the point contact (Upadhyay, Louie, and Buhrman [1997]). 
Second, the effect
only arises in unannealed junctions, in which the samples are cooled to
cryogenic temperatures
hours after fabrication. Annealing removes the anomaly.
Third, to rule out static disorder as a source of the anomaly, six
percent Au was co-evaporated
with Cu; no zero bias anomaly was found. (Moreover, it was observed
that
the conductance amplitude is too large to correspond to candidate
energy dependent effects
such as weak localization and disorder enhanced electron-electron
scattering).
Finally, to further rule out magnetic impurities as a source of
the anomaly, small amounts of Mn and Cr ions were controllably
co-deposited with the host junction metal.
Not only do the Kondo anomalies associated with these impurities show a
sizeable magnetic
field dependence, the resulting conductance signals are stable over
periods of months indicating
that the magnetic impurities do not anneal away from the junction.

Ralph and Buhrman noted
that the magnitude of the conductance anomaly indicated the presence of
several scattering
centers within the junction region (based upon the scattering being at
half of
the unitarity limit at $T=0$ (c.f. Sec. 6.1.4) in conjunction with
Eq. (8.1.9)) as the contribution to the conductance of a single scatterer in these
junctions is of the order of $e^2/h$.  They also ascribe the likely TLS
sites to dislocation jogs or
kinks.  They observe that these same kind of zero bias conductance
anomalies have been observed
in many types of metal point contacts for many years and suggest the
earlier measurements
have also been of TLS Kondo scattering.

\begin{figure}
\parindent=2.in
\indent{
\epsfxsize=5.in
\epsffile{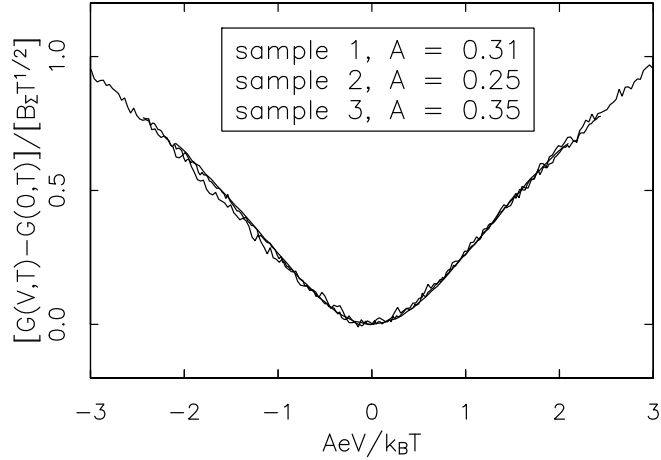}}
\parindent=0.5in
\vspace{.3in}
\caption{Scaling curves for the three Cu nanoconstriction samples of 
Ralph {\it et al.} [1994].  }
\label{fig8p9}
\end{figure}

Perhaps the most compelling evidence that the observed anomalies
correspond to TLS Kondo
centers is a demonstration of non-Fermi liquid scaling behavior at low
temperatures (Ralph {\it
et al.} [1994]; Ralph and Buhrman [1994]).  Ralph {\it et al.} focussed
 on the conductance in regions below the
apparent Kondo scale inferred from the logarithmic regions seen first
as voltage and temperature
are lowered. Motivated by  conformal field theory results for the
electronic self-energy
(Ludwig and Affleck [1991], Affleck and Ludwig [1993]), a scaling form
for the conductance was assumed
in which the conductance contributions from different TLS within the
junction was assumed to
be additive.  This is reasonable provided the different sites only
interact weakly.  The resulting
scaling form is
$$G(V,T) - G(0,0) = T^{\alpha} \sum_i B_i\Gamma({A_i eV\over
[k_BT]^{\alpha/\beta}})
\leqno(8.1.10)$$
where the summation is over the TLS sites, $\Gamma(x)$ is a universal
function, and
 $B_i,A_i$ are non-universal amplitudes which
may vary, for instance, as a function of the position of the defect
from the narrowest point of
the constriction.  A test of this scaling for $\alpha=\beta=1/2$ is shown in 
Fig.~\ref{fig8p9}. The function $\Gamma(x)$ must go as $x^{\beta}$ for
$x\to \infty$ so that
$G(V,T)$ is independent of $T$ for $eV>>k_BT$.  To normalize,
$\Gamma(0) = 1$, and
$d\Gamma(x)/dx^\beta \to 1$ for $x\to \infty$.  We can immediately
infer $\alpha=\beta$ since
the voltage only enters in the Fermi functions of the leads in the
combination $V/T$.  Since
the conformal theory gives $\alpha=1/2$ in bulk, this is the
expectation for the above scaling
form; however, in attempting to describe the data, $\alpha$ was allowed
to vary.  We note that
the above scaling must crossover to Fermi liquid behavior below $T_x=
T_{\Delta}
\approx \Delta^2/k_B^2T_K$ for $\Delta < k_BT_K$
as discussed in Affleck {\it et al.} [1992] and in Secs. 4.2.e, 5.1.4,
6.1.2.c, and
7.2.  No such
deviations from {\it non Fermi liquid} behavior were found for the data
of Ralph and Buhrman for Cu junctions,
indicating very small splittings for the putative TLS in the
junctions.   Recently, a direct theoretical
calculation of the scaling function was made using the NCA treatment of
the two-channel
Kondo model (Hettler {\it et al.} [1994]), while an extensive discussion
of the corresponding effort based upon a conformal field theory approach
appears in von Delft, Ludwig, and Ambegaokar [1997b].  We note that it is quite
puzzling that
order ten scattering centers exist in the sample with identical $T_K$
values
as evidenced by the scaling curves.

\begin{figure}
\parindent=2.in
\indent{
\epsfxsize=5.in
\epsffile{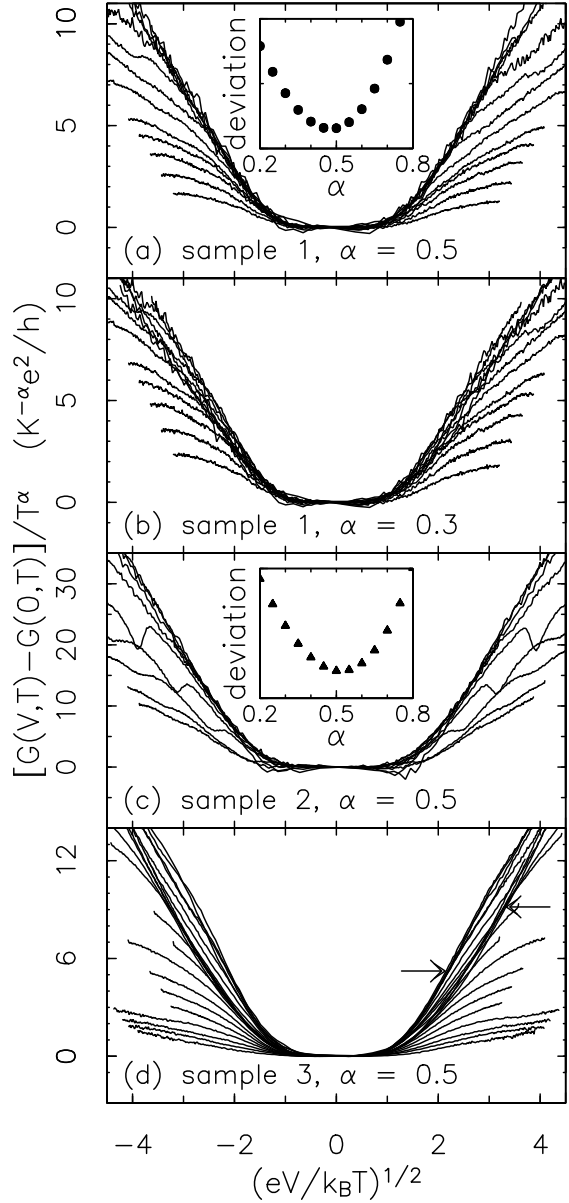}}
\parindent=0.5in
\vspace{.4in}
\caption{Scaling behavior of conductance for three different 
Cu nanoconstrictions at low voltage for $eV,k_BT<k_BT_K$.  
(a),(b) Rescaled conductance of sample 1 for (top to bottom) 
$T=0.4,0.6,0.8,1.1,1.4,1.75,2.25,2.8,3.5,3.9,4.3,4.9, and 5.6$K.  
(c) Rescaled conductance of sample 2 at same $T$ values up to 4.3K. 
(d) Rescaled conductance of sample 3 from 125mK to 7.6K.  Arrows show 2 separate scaling 
curves.  Insets:  Integrated mean square deviation from the average scaling curve for $T\le 1.4$K
and -8$\le eV/k_BT \le 8$.  The scale of the deviation axis is in (a) from 0 to 4 $(e^2/h)^2$
and in (c) from 0 to 25$(e^2/h)^2$.  Residual deviations for $\alpha=0.5$ are consistent
with the amplifier noise. From Ralph {\it et al.} [1994].}
\label{fig8p10}
\end{figure}

To test the scaling hypothesis, Ralph {\it et al.} manipulated Eq.
(8.1.10) to eliminate $G(0,0)$ and
considered
$${G(V,T)-G(0,T)\over T^{\alpha}} = \sum_i B_i[\Gamma(A_ix)-1]
\leqno(8.1.11)$$
where $x=eV/k_BT$.  
The quantity on the left hand side of this equation
may be plotted for
different temperatures as a function of $x$ with $\alpha$ varied.  The
result must be universal
to the extent that $A_i$ can be approximated as a single constant,
which holds experimentally
with $A_i \approx 0.25-0.35$.  Then $\sum_i B_i$ can be found from
scaling the curves together
for large $x$ and is found to be $\approx 10-30 e^2/h$.   The results
normalized together are
shown for several samples in Fig.~\ref{fig8p10}, where different $\alpha$ values
are used in a,b.  The
fit with $\alpha=0.5$ looks perfect, while the universality is lost
with $\alpha=0.3$.  The Fermi
liquid value ($\alpha=2$) is ruled out, as is apparently the
possibility that $\alpha=1$.  Deviations
from universal behavior are expected for $T,V\ge T_K$ (a range of
 $T_K$ values of  0.6K$\le T_K\le$5K is suggested).
Sample 2 data are shown in Fig.~\ref{fig8p10}(c) for which $T_K \approx 0.5K$.
For Sample 3 (Fig.~\ref{fig8p10}(d)). 
at two different temperature ranges universality can be found
indicating very likely that there
is a distribution of parameters with two groups of TLS likely
dominating.  The idea is to
determine $\alpha$ from those curves where universality is well
satisfied so that the
scattering centers have uniform parameters.

The scaling hypothesis is further tested in applied magnetic field
$|H|$.  Applied field serves as a
channel field in this case, lifting the degeneracy of up and down spin
electrons.  In this case, the
zero bias scaling function for a single TLS site must be generalized to
the two parameter form
$T^{1/2}\Gamma(V=0,|H|/T^{1/2})$ which follows from the considerations
of channel field
scaling (see Secs. 4.2.c,5.1.4,6.1.2.c).
Since this must be independent of $T$ as $T\to 0$, we infer that
$\Gamma(x=0,y) \sim y,~~y\to\infty$ so that the subtracted conductance
should scale as $|H|$ for
$H\to 0$.  This is observed for small $H$ and $T=0.1K$.  Nonlinear
conductance spikes appear for
fields above 1 Tesla which are clearly non-universal phenomena.  In
contrast, the ordinary Kondo
resistivity would be expected to show scaling with $H^2$ for small
$H$.  At the theoretical level,
the microscopic origin of the channel splitting in the TLS model is
unclear, unlike the quadrupolar
Kondo model for example.  As discussed in Sec. 3.4.1.e, the Zeeman
splitting of the conduction
states alone cannot produce a discriminating channel asymmetry unless
the applied field is of order
the conduction bandwidth.  Nonetheless, the observed field dependent
scaling offers strong support
for the applicability of the two-channel Kondo model description.

The success of the above scaling analysis places a constraint on the
values of $\Delta$ for the
TLS in the junction.  Given the estimate for the crossover scale
$T_x=T_{\Delta}=\Delta^2/k_B^2T_K$ (for $\Delta < k_BT_K$), for
$T_K$=5K, the absence of fermi liquid behavior down to $T=0.4K$ implies
$T_x=T_{\Delta}<0.4K$ and
$\Delta<1.4K$.

The recent experiments by Upadhyay, Louie, and Buhrman [1996] on Ti 
and V point contacts appear to show the effect of the renormalized
splitting below $T_{\Delta}=T_x$.  For example, one Ti junction studied 
down to 76mK appears to have $T_{\Delta}=1.4K$ and renormalized
splitting $\Delta = 0.4 $meV.  In another junction, two such
crossovers appear, very likely due to two different TLS, with
$\Delta=0.47$ meV and 1.6 meV.  Here $\Delta$ is estimated from 
the formula $T_{\Delta}= T_x = \Delta^2/(k_B^2T_K)$ with the two 
different $T_K$ estimates  of 6.2K and 28K and $T_x$ read off from the
crossover away from $\sqrt{V}$ behavior.  To confirm the origin of
these phenomena as arising from TLS Kondo scattering, it was observed
that after the treatment of the junction with a large current flow for 
10 seconds the inferred splittings were seen to change from 0.5$\to$ 0.7 meV, and
1.6$\to$ 1.54 meV.  This is explainable by rearrangement of atomic
positions and hence changed parameters of the TLS.  We note that 
the alternative explanation of Altshuler, Wingreen, and Meir [1995] is
unable to explain the crossover to $T^2$ behavior at a fixed
temperature scale.  
The observation of the additional scale $T_{\Delta}=T_x$ gives us 
three distinct regions for the point contact resistance:\\

\indent{$\to$  $T>T_K$,  $\rho(T,V)$ is logarithmic in $T,V$}
 
\indent{$\to$  $T_K>T>T_{\Delta}=T_x$  $\rho(T,V)$ displays non-Fermi
liquid $\sqrt{T,V}$ behavior}

\indent{$\to$  $T_x=T_{\Delta}>T$ $\rho(T,V)$ displays $T^2,V^2$
saturation (Fermi liquid)}\\

We note that the inferred $\Delta$ values are relatively large.  Since
the distribution of TLS splittings is {\it apriori} flat in energy, it is more 
probable to observe TLS with higher $\Delta$ values.  It remains still
to explain the origin of the small (or zero) estimated splittings in 
Cu based junctions as compared to Ti and V based contacts.  As mentioned above
the overall size of the zero bias conductance is between 10-100 $e^2/h$.  However, the
current induced change is already about $e^2/h$ indicating that the splitting of 
a single TLS plays the role.

Analogous phenomena (apart from the non-Fermi liquid behavior) should
appear for magnetic
impurities.  Apart from the kind of controlled doping experiments with
Mn and Cr impurities
in the metal junctions described above that were performed by Ralph
[1993], some early
point contact experiments on Kondo alloys were performed (Jansen {\it
et al.} [1980], Yanson [1995]) which
showed characteristic logarithmic Kondo anomalies in the resistivity.
Earlier experiments were performed with tunnel junctions.  In this
case, the magnetic impurities
may reside either in the tunneling barrier or in the electrode near the
barrier (on a scale less than
50$\AA$.  In the first case, the resonant density of states near the
impurity induces near the
Fermi energy actually assists the tunneling process and thus enhances
the conductance,
producing a maximum at zero bias (Wyatt [1964], Appelbaum and Shen
[1972]).  On the other
hand, if the impurities are nearbye the interface, similar to the point
contact configuration the
junction resistance  shows a maximum at zero bias (Bermon and So
[1978], Mezei and
\zow [1971]).  This can be understood because the electrode impurities
backscatter incoming
electrons away from the junction thus reducing the tunnel current.

Another interesting feature of the point contact experiment with
magnetic
impurities (Yanson [1995]) is that the intensity of the zero bias
anomalies
are larger by one or two orders of magnitude for small contact
orifices.
In a recent theory, \zar and Udvardi [1996a,b] have shown that the local
density
of states fluctuation for the electronic density is enhanced for
reduced orifice size.
Since $T_K \sim \exp(-1/J[\rho+\delta\rho])$ this will strongly enhance
$T_K$ for
those
centers for which the density of states is enhanced, and strongly
suppress $T_K$
(below observability) for other sites.  According to that theory this enhancement must occur at the Femri energy in 
an energy range with width not more than about 10\% of the Fermi energy or even less.  
Naturally, the enhanced $T_K$
sites will
be preferred.  Obviously something similar can carry through for the
TLS
Kondo scale.

Another point contact experiment where the zero-bias anomaly is
attributed to TLS
is in the work of Keijsers {\it et al.} [1995].  In the original theory of Kozub and
Kulik [1986]
only the excitations of the TLS to higher states were taken into
account, without
Kondo corrections.  However, it looks as though the TLS Kondo picture
fits the
data better as the temperature dependence looks predominantly like a
thermal
smearing in contrast to the other explanation (\zar and \zow [1995]).
In these experiments
the apparent Kondo temperature (measured from the low temperature peak
position
of the derivative of the differential resistance) increases with
decreasing contact
size (larger contact resistivity) just like in the magnetic case discussed above
(Yanson {\it et al.} [1995]).  To the the large $T_K$ values
inferred from
this interpretation requires the \zar and \zow [1994a,b] enhancement
arising from
the excited states.

Finally, we mention that recently Wingreen, Altshuler, and Meir [1995]
have criticized
the TLS interpretation of the point-contact experiment by Ralph and
Buhrman [1992]
(Ralph {\it et al.} [1994], Ralph and Buhrman [1994]).  They argue that
in tunneling
experiments with disordered material zero bias anomalies have been
observed with square root
behavior (for a review see Altshuler and Aronov [1985]) reflecting the
renormalization of
the local density of states and that similar density of states effects
can occur in
the present experiments.  Ralph and Buhrman [1995] however presented
evidence
against that argument by introducing controlled disorder into their
contacts as mentioned
earlier.

Furthermore, Wingreen, Altshuler, and Meir [1995] challenged the TLS
concept itself
by considering the effect of extrinsic disorder from impurities on the
TLS parameters
through the couplings $V^x,V^z$ (see sec. 2.1).  In particular, they
calculate the TLS
self-energy diagrams of a Hartree type with the TLS on one end of  a
conduction electron
bubble and a non-TLS impurity on the other end.  Without the disorder
interruption this
diagram vanishes on tracing out the electron orbital pseudo-spin.
This physically
corresponds to a Ruderman-Kittel charge interaction between the TLS and
non-TLS
impurity.   They then calculate the root mean square distribution of
splittings $\Delta$ and find
that it goes as $E_F v^z/\sqrt{k_Fl}$ which is about 50-100K for the
estimated mean free
path in Ralph and Buhrman's experiments [1992,1994].   This same
estimate for the
magnitude of the splitting can be obtained by replacing the electron
spectral functions
in the Hartree diagram by the local density of states about the TLS
site.    Separately,
one can estimate an upper bound on the contribution to
the spontaneous tunneling matrix element $\Delta_0$ as $\rho_0\int
V_x(\omega)
d\omega$$\simeq 50K$(\zar and \zow [1995]).  This is easily
renormalized downwards by scaling.
(Calculating the root mean square spread in $\Delta_0$ induced by the extrinsic
impurities from the
diagram considered by Wingreen, Altshuler and Meir [1995] produces a
small estimate,
of the order of a few tenths of a Kelvin.)

With regard to the large estimated RMS spread in $\Delta$ two remarks
are to be
made:\\
1) The modification of $\Delta$ corresponds to a modification of the
{\it shape } of the
potential and thus must influence $\Delta_0$.  The TLS of interest are
only those few
for which the the splitting $\Delta$ is small.  The selection of these
TLS must be made
on the basis of the {\it renormalized} (by scaling)
potential and is unrelated to the unrenormalized one.  (An asymmetric
potential may be
made symmetric after the renormalization and vice versa.)  The
existence of TLS with small
splitting and asymmetry is an experimental fact in amorphous
materials.  This most striking
evidence for this is found in
the existence of linear heat capacities in amorphous superconductors
below the
superconducting transition temperature (Graebner {\it et al.} [1977],
L\"{o}hneysen {\it et al.} [1980], Lasjaunias and Ravex [1993]).\\
2) Recently Kozub [1995] has pointed out that in the argument of
Wingreen, Altshuler, and
Meir [1995], the largest contribution to the RMS splitting comes from
those impurities close
(i.e., nearest neighbor or next-nearest neighbor) to the TLS site (see also Smolyarenko
and Wingreen [1997]).  In
a dilute case, most TLS
do not have such a neighbor so that the actual splitting modification
by the disorder is likely
to be much smaller than reported.  For those which do have a non-TLS
impurity nearby, there
will be a `wipeout':  the large splitting will destroy the TLS
character and these simply will
not be seen.  Hence some more detailed consideration of the non-TLS
impurity configurations
is needed than provided by Wingreen, Altshuler, and Meir [1995].

Keijsers, Shklyarevskii, and van Kampen [1996] carried out another set of experiments using metallic
glasses in break junctinos.  At zero bias resistivity measurements revealed the same kind
of telegraph nois due to a slow tunneling center as shown in
Figs.~\ref{fig8p1}\ref{fig8p2}. When the
slow center switches, then the zero bias anomaly is also changed, and actually two different 
curves are measured for the zero bias anomaly.  This was interpreted as a demonstration of 
the interaction between a slow and a fast tunneling center.  When the slow center 
changes then the parameters of the fast ones are also tuned, which results in a change of the 
zero bias anomaly.  (Note that in Figs.~\ref{fig8p1},\ref{fig8p2} the interaction between two 
slow centers is displayed.)

\begin{figure}
\parindent=2.in
\indent{
\epsfxsize=5.in
\epsffile{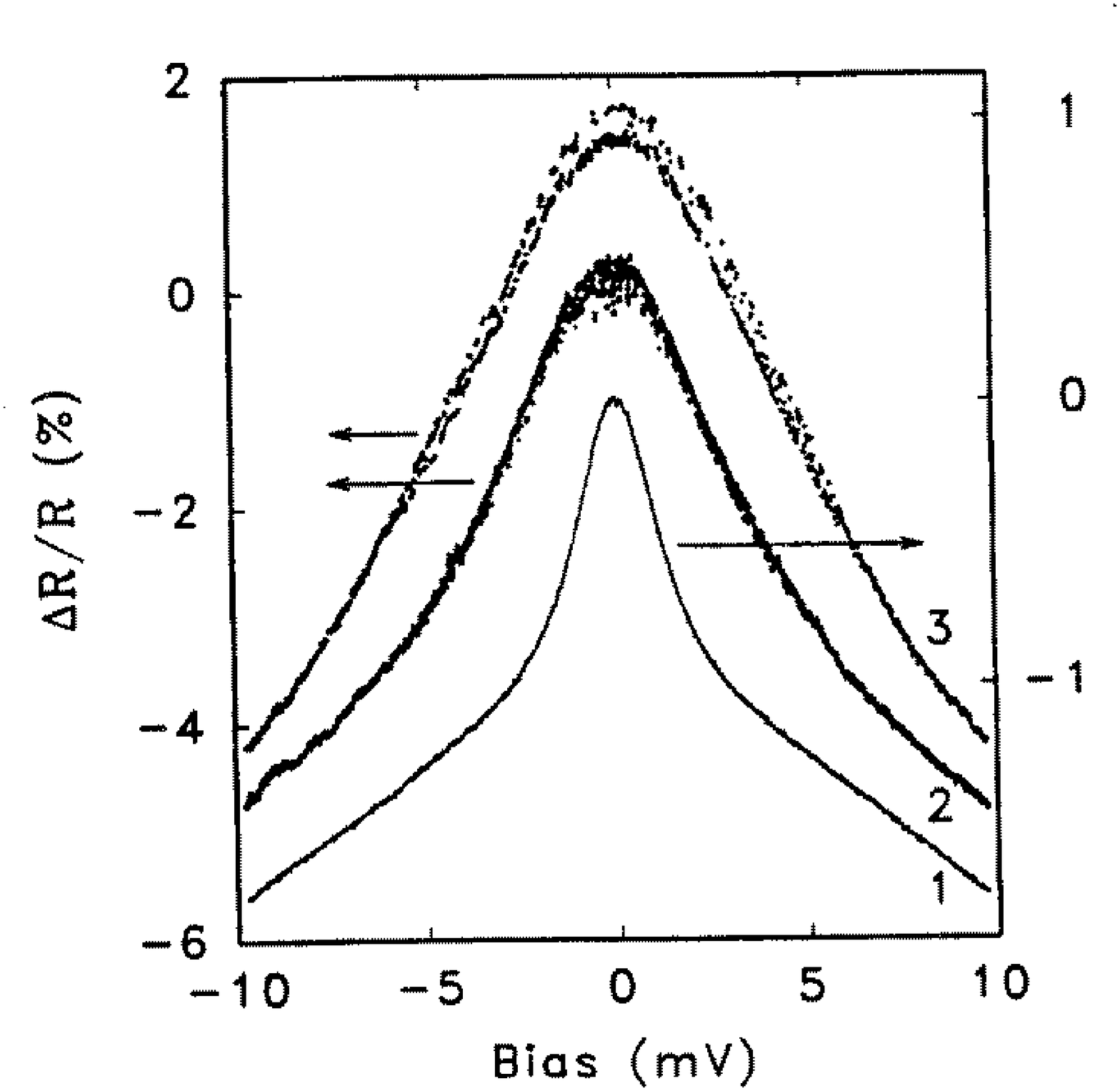}}
\parindent=0.5in
\caption{Experimental evidence for modulation of fast TLS by slow TLS (telegraph noise) in 
mechanically controlled metallic glass break junctions.  The 
differential resistance $R_d$ is measured as a function of bias 
voltage for Fe$_{80}$B$_{20}$ (1 and 2) and Fe$_{32}$Ni$_{36}$Cr$_{14}$P$_{12}$B$_6$ (3)
break junctions at $T$=1.2K.  (1) A 6.6$\omega$ contact, showing almost no noise.  
(2) A 366$\Omega$ contact that shows clear noise around zero bias.  The noise 
amplitude decreases as the bias voltage increases. (3) A 145$\omega$ contact, showing 
a clear two-level switching behavior between two different $R_d$ peaks.  
From Keijsers, Shklyarevskii, and van Kampen [1996].  }
\label{fig8p11}
\end{figure}

\begin{figure}
\parindent=2.in
\indent{
\epsfxsize=5.in
\epsffile{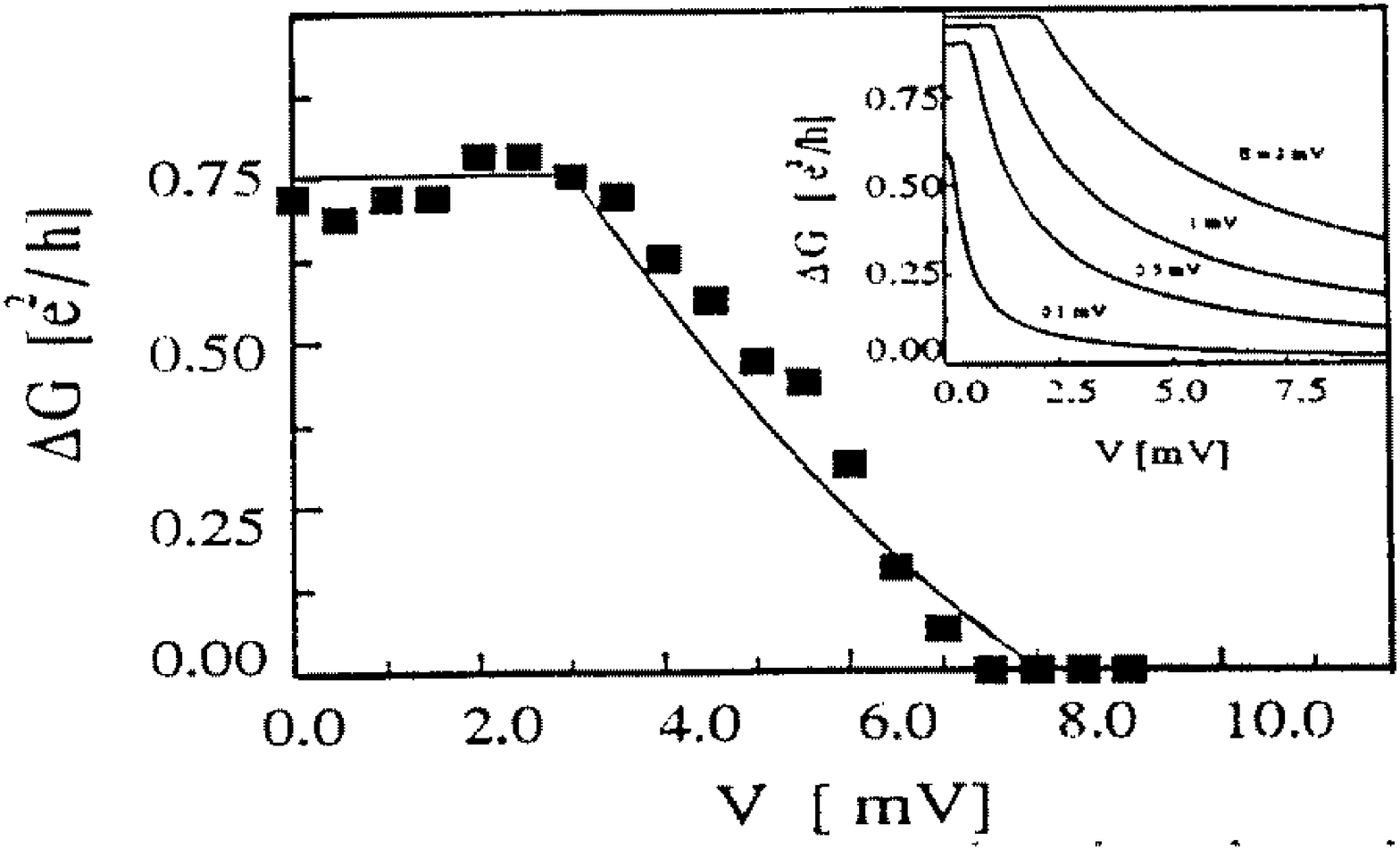}}
\parindent=0.5in
\caption{Conductance difference for mechanically controlled metallic glass break junctions.  The dots are obtained 
from the measured resistance change values shown in curve (3) of Fig.~\ref{fig8p11}, with 
uncertainties of order 0.1$e^2/h$.  The solid line is obtained from the theory of 
\vld and \zow [1983a,b,c] using a Kondo scale of $T_K\approx$35K, and two different splittings
for the fast TLS of $E_1=3.5$meV, $E_2$=8meV, with the assumption that the modulation by the
slow center induces these two different splitting values.  The inset shows conductance difference
curves generated by the theory of Kozub and Kulik [1986].  This figure is from \zar, von Delft, and 
\zow [1997].}
\label{fig8p12}
\end{figure}

The curve indicates that the low energy cutoff determined by the asymmetry $\Delta$ is tuned, and that
the other parameters must have remained almost the same so as not to modify the width of the zero 
bias anomalies, which actually measure the Kondo temperature.  This means that the $v^{\alpha}$ interactions
are only very slightly tuned, which is the case when the assisted tunneling processes are induced via
transitions through the excited states of the potential well.  The first excited state overlaps with both 
minima of the TLS (see Sec. 3.4.2 and \zar and \zow [1994a,b]).  Assuming the orbital Kondo effect
picture, the top of the anomaly is chopped off at different heights after the $\Delta$ parameter 
of the TLS is tuned. The difference in the conductance is shown in Fig.~\ref{fig8p12} for a single
junction (curve (3) of Fig.~\ref{fig8p11}).  The theoretical fit, given by the solid line, assumes that 
the scattering resonance has the shape given by the second order renormalization group equations discussed in 
Sec. 3.4.1, and a sharp cutoff is used for $eV<\Delta$.  The fit is remarkably good.  On the other hand, the 
theory of Kozub and Kulik [1986] woudl provide the difference between two curves shown in the inset.  The 
experimental data clearly does not show the long tail expected from the Kozub and Kulik [1986] theory, 
and thus offers strong support for the origin of the zero bias anomaly in terms of the orbital Kondo 
effect associated with TLS.  An important feature of the experimetnal data is that the changes in the conductance are smaller than 
$e^2/h$ which makes it possible to interpret them as arising from a single modulated TLS.  In similar experiments, 
two such transitions are superposed, indicating that at least two fast TLS are affected by the slow TLS.  

Recently a new experiment was performed by Balkashin {\it et al.} [1997] to distinguish between the TLS 
Kondo theory and the theory of Kozub and Kulik [1986] which includes direct excitation of the TLS without
the Kondo effect.  The two interpretations have two different characteristic energy scales, namely 
the Kondo scale $T_K$ and, for Kozub and Kulik [1986], the Korringa-linewidth $\Gamma_K = (\rho V^z)^2(\Delta_0/E)^2 E$ where
$E=\sqrt{\Delta^2_0+\Delta^2}$ is the full splitting of the two-level system including asymmetry 
and spontaneous tunneling.  This latter result follows provided the temperature is low compared to $E$ so 
that $k_BT$ is replaced by $E$, and that $v^x=v^y=0$ so the $V^z$ interaction induces the transition 
in the presence of spontaneous tunneling and asymmetry (\vld and \zow [1983c]).  To fit the zero bias
anomalies, $T_K$ or $\Delta_0$ must be in the range of 1 meV.  Thus, knowing that $(\rho_0V^z)^2 \simeq 0.1$, 
the two characteristic frequencies are very different, of order 1 meV for the Kondo case, and 10$^{-2}$ meV for
the direct excitation model of Kozub and Kulik [1986].

In these more recent experiments, an a.c. bias was superimposed on a d.c. bias in the form 
$$V=V_0+V_1 cos(\omega t)~~.$$
For $\omega <<\omega_0$, where $\omega_0$ is the characteristic energy of either theory, then at each
time during the oscillation period the zero bias anomaly is determined by the instantaneous bias.  Thus 
the time average of the current is 
$$\bar I = \bar {I(V_0+V_1 cos\omega t)} \approx I(V_0) + {1\over 4} {\partial^2 I\over \partial V^2} V_1^2 $$
provided that $V_1<<V$, and no significant frequency dependence is expected.  

On the other hand, when $\omega>\omega_0$, the oscillating voltage averages out during the characteristic
time scale $t_0= 1/omega_0$, and the change in conductance must vary as $1/\omega$ with an amplitude proportional
to $V_1^2$.  The experiment was performed at two different frequencies, $\omega=0.6GHz,60GHz$, where the latter
frequency corresponds to 0.25 meV.  No significant dependence upon the amplitude or $\omega$ was found.  Hence,
the relevant energy scale must be significantly larger than 0.25 meV.  As argued above, this can only be true 
for the Kondo effect, so the experiments offer further support for interpretation in terms of the TLS
Kondo effect. 

Summarizing, it can be claimed on solid ground that there are several cases where only the orbital Kondo effect
of a TLS can explain the observed zero bias anomalies, and in several cases the effect observed is of order 
$e^2/h$ in the conductance, indicating that only one or at most a few TLS are playing the relevant role.  

The above considerations notwithstanding, it remains an open question
why several almost
symmetric ($\Delta \le $1.4K by estimates above) TLS with {\it
identical } $T_K$ values
exist in the devices of Ralph and Buhrman [1992,1994].

{\it Other Candidate Systems and Thermodynamics}.  Another promising
candidate system
of the mesoscopic variety has been studied by Gregory [1992] who finds
Kondo
anomalies which don't split in magnetic field in oxide coated tungsten
wires in a crossed geometry.

Finally, we comment on the thermodynamics and other measurements on
more
concentrated systems such as the metallic glasses.  In these the TLS
have a broad distribution
and for a large number of them the noncommutative terms can be
neglected.  Certainly the most
important effect in this case is the renormalization of the splitting
(see Eq. (8.1.4)).  That
downward renormalization may change an initially uniform distribution
$P(\Delta)=constant$
 to one which is peaked
at low energy scales.  The effect of the TLS on the electrons
 can also be probed in the superconducting states
of amorphous metals where the conduction electrons and therefore the
infrared divergences
are cut off below by the superconducting gap.  Tuning the gap by
external magnetic field to
smaller values will allow the suppressed infrared renormalization to
return and the energies of
the TLS to be pushed to smaller values.  Such a renormalization effect
has been observed,
where the tunneling particle is atomic hydrogen in a superconducting
metal (Yu and Granato
[1985]).  The splitting renormalization shown in Fig.~\ref{fig8p4} is enhanced
by the noncommutative
terms and that enhancement should be verified by further experiments.

\subsection{Experimental Data on Two-Channel Quadrupolar and Magnetic
Candidate Heavy
Fermion Materials} 

Questions about the possibility of non-Fermi liquid physics in the heavy
fermion materials are not new, and occurred concommitant with similar
observations for the cuprate superconductors (on the theory side, see 
Cox [1987b]; on the experiment side see Ott [1987]).

Since 1991, a 
number of heavy fermion materials have come to light which 
display logarithmic upturns in
the specific heat and more occasionally the susceptibility.  Concommitant
with
these log upturns is usually a non-Fermi liquid resistivity.  
In Table~\ref{tab8p1}, we summarize the results for various materials in 
which evidence exists for non-Fermi liquid behavior.

\begin{table}
\begin{center}
\begin{tabular}{|l|c|c|c|c|c|}\hline
{\bf Alloy/Compound} & $T_K$  & $C/T$  & $\chi(T)$ & $\rho(T)$ & Single Ion? 
 \\\hline\hline
Y$_{1-x}$(Th$_{1-y}$,U$_y$)$_x$Pd$_3^{(*)}$ & $\sim$40K & $\ln T$ &
$1-aT^{1/2}$ & $1-AT$ & Yes  \\\hline
Th$_{1-x}$U$_x$Ru$_2$Si$_2^{(*)}$ & 12K & $\ln T$ & $\ln T$ &
$1+BT^{1/2}(?)$ & Yes  \\ 
&     &         & ($H\parallel c$) &        &      \\\hline
Th$_{1-x}$U$_x$Pd$_2$Si$_2^{(*)}$ & 12K & $\ln T$ & $\ln T$ &
$1+BT^{1/2}(?)$ & Yes  \\
&     &         & ($H\parallel c$) &        &      \\\hline
La$_{1-x}$Ce$_x$Cu$_2$Si$_2^{(*)}$ & $\sim$10K & $\ln T$ & $\ln T$ & $1-AT$ & 
Approx. \\\hline
Th$_{1-x}$U$_x$M$_2$Al$_3$ & $\sim 20$K & $\ln T$ & $\ln T(?)$ & $1-AT$
& ?  \\\hline
Th$_{1-x}$U$_x$Be$_{13}$ & $\sim 10$K & $\ln T$& $1-aT^{1/2}$ &
$1+BT^{1/2}$ or & Yes($\chi(0)$) \\
&&&& $1+AT$ & No($C/T$)  \\\hline\hline
UBe$_{13}$ & 10K & $\ln T$ at $H=12$T & $1-aT^{1/2}$ & $1+AT$ & -
\\\hline
CeCu$_2$Si$_2$ & 10K & $\ln T$ & ? & $1+AT$ & - \\\hline
PrInAg$_2$ & 2K & const. & ? & $AT$? & - \\\hline
\end{tabular}
\end{center}
\caption{ Non-Fermi Liquid Heavy Fermion Alloys and Compounds. This
table lists the relevant properties of all non-Fermi liquid
heavy fermion alloys and compounds for which a two-channel Kondo model
description (in either dilute or concentrated limits) 
may be an appropriate starting place.  
The columns for specific heat,
susceptibility, and resistivity indicate the low temperature
behavior. All (but possibly CeCu$_2$Si$_2$ and PrInAg$_2$) 
have logarithmic in $T$
specific heat coefficients over an extended temperature range.
Those alloys marked with an asterisk show evidence for significant
residual
entropy at low temperatures.
The coefficients $A,B$ listed in
the resistivity column are assumed positive, as is $a$ in the
susceptibility
column.  The column under
`Single Ion?' answers whether single ion scaling has been
observed.  Adapted from Cox and Jarrell [1996].}
\label{tab8p1}
\end{table}

These non-Fermi liquid behaviors can be broadly summarized in three
categories:\\
{\it (1) Dilute or Local}  In this case doping on the rare
earth/actinide sublattice away
from a fully concentrated compound reveals the non-Fermi liquid
behavior.  The
examples are:  Y$_{1-x}$U$_x$Pd$_3$ (Seaman {\it et al.} [1991],
Andraka and Tsvelik
[1991]), Th$_{1-x}$U$_x$Ru$_2$Si$_2$ (Amitsuka {\it et al.} [1993,1994]),
Th$_{1-x}$U$_x$Pd$_2$Si$_2$ (Amitsuka {\it et al.} [1995]), 
La$_{1-x}$Ce$_x$Cu$_{2.2}$Si$_2$ (Andraka [1994]),
Th$_{1-x}$U$_x$Pd$_2$Al$_3$
(Maple {\it et al.} [1994]), Th$_{1-x}$U$_x$Ni$_2$Al$_3$ (Kim, Andraka,
and Stewart [1993]),
Th$_{1-x}$U$_x$Be$_{13}$ (Aliev {\it et al.} [1994]). It is of interest
that all of these
materials except Y$_{1-x}$U$_x$Pd$_3$ are heavy fermion superconductors
when
$x=1$.   These are the systems for which the single impurity
multi-channel
Kondo  model which is the focus of
this paper have the best chance of working.  Thus a compelling proof of
single impurity
behavior is important, and we shall discuss the evidence for this in
each case.  \\
{\it (2) Concentrated and Ordered} Two compounds appear substantially
non-Fermi liquid like
at the fully concentrated limit: the heavy fermion superconductors
UBe$_{13}$ (Aronson
{\it et al.} [1989], McElfresh {\it et al.} [1994], Cox [1995]) and
CeCu$_2$Si$_2$ (Steglich {\it et al.} [1995]). A third candidate has
recently arisen, PrInAg$_2$ (Yatskar {\it et al.} [1996]) which 
has $4f^2$ Pr ions in a $\gth$ ground state (confirmed by neutron 
scattering studies).  This material shows some behavior incompatible
with a Fermi liquid.  
Two criteria specify
the assignment of non-Fermi liquid behavior: (i) a
continual rise of $C/T$ at and below the superconducting transition
(under application of
a magnetic field (UBe$_{13}$: Steglich [1996]; CeCu$_2$Si$_2$: Steglich {\it
et al.} [1995])-- PrInAg$_2$ does not superconduct, but does show a 
region of logarithmic temperature dependence in $C/T$ prior to a low
temperature saturation;
(ii) anomalously large residual resistivities and linear in $T$
resistivity behavior which
is suppressed in magnetic field and pressure (for UBe$_{13}$)
(UBe$_{13}$: Aronson {\it
et al.} [1989]; CeCu$_2$Si$_2$: Steglich {\it et al.} [1995])--PrInAg$_2$
does not have a particularly large residual resistivity but never shows
a region of Fermi liquid like $T^2$ behavior.  \\
{\it (3) Concentrated and Disordered}  Three systems, UCu$_{5-x}$Pd$_x$
(Andraka
and Stewart [1992]), U(Pt$_{1-x}$Pd$_x$)$_3$ (Kim,
Stewart and Andraka [1992]), and CeCu$6-x$Au$_x$ (L\"{o}hneysen {\it et
al.} [1994]) 
fit into this category.   Each is near an antiferromagnetic instability
which is accessed by tuning the
concentration of dopants off the rare earth/actinide 
 sublattice.  The non-Fermi liquid behavior appears
closely associated with this proximity to a $T=0$ magnetic transition,
which has been
extensively discussed recently by Millis [1994] (in an ordered
system--the role of disorder
in this theory requires further study).   In the case of
UCu$_{5-x}$Pd$_x$,
the disorder may play a role in inducing a spread of Kondo temperatures
which extends
all the way down to $T=$0K based upon interpretation of
inhomogeneous line broadening  measurements in copper nuclear magnetic
resonance lines
(Bernal {\it et al.} [1995]; MacLaughlin, Bernal, and Lukefahr [1996]).  
These authors relied on the work of 
Dobrosavljevi\'{c},
Kotliar, and Kirkpatrick [1992] who argued that non-Fermi liquid behavior may 
be mimicked by this distribution of local fermi liquids in the vicinity 
of a metal insulator transition with a log normal distribution of the 
local conduction density of states about the impurity sites. 
An unusual {\it local}
dynamic scaling of the
neutron scattering cross section with frequency was observed by Aronson
{\it et al.} [1995]
over a wide range of $x$ including $x=1$ which is likely an ordered
compound.  This suggested that 
other mechanisms may be at work to induce the non-Fermi liquid behavior.
The work of Miranda, Dobrosavljevi\'{c}, and 
Kotliar  [1996,1997] has produced a different form of the
probability distribution than Dobrosavljevi\'{c}, Kotliar and
Kirkpatrick [1992], derived in infinite spatial dimensions, which 
holds away from the metal insulator transition. This form is in complete
agreement with the form used by Bernal {\it et al.} [1995] in fitting 
experimental data.  With this form, they 
reproduce the simple Gaussian distribution of coupling constants used by 
Bernal {\it et al.} [1995], and they are able with this distribution to 
(i) fit the linear temperature dependence of the resistivity, and (ii)
the dynamic susceptibility of Aronson {\it et al.} [1995].  Hence, taken
together with the experimental work of Bernal {\it et al.} [1995], it
genuinely appears that the UCu$_{5-x}$Pd$_x$ system is described by 
the modified Kondo disorder theory as put forward by Miranda,
Dobrosavljevi\'{c}, and Kotliar [1996,1997].   In related work, 
the optical conductivity of UCu$_{3.5}$Pd$_{1.5}$ has been studied (DeGeorgi
and Ott [1996]).  This can be fit with the same disorderd 
distribution of Kondo scales used by Bernal {\it et al.} [1995] (also 
MacLaughlin, Bernal, and Lukefahr [1996]) and Miranda,
Dobrosavljevi\'{c}, and  
Kotliar  [1996,1997] as was shown by Chattopadhyay and Jarrell [1996]
(see also Jarrell {\it et al.} [1996c]).

Another material which defies the above categorization yet displays 
non-Fermi liquid behavior is Ce$_{1-x}$Th$_x$RhSb (Andraka, [1994b]). 
This material is a Kondo insulator for $x\to 0$, but for the range 
$0.2\le x\le 0.4$ shows possible non-Fermi liquid behavior in the form 
of a specific heat coefficient which appears logarithmically divergent
over approximately a half decade of temperature.  

Our attention shall focus here on a brief review of materials in
categories (1) and (2) with
regard to their possible explanation in terms of the two-channel Kondo
effect.  We discuss
each system in order.

\begin{figure}
\parindent=2.in
\indent{
\epsfxsize=6.in
\epsffile{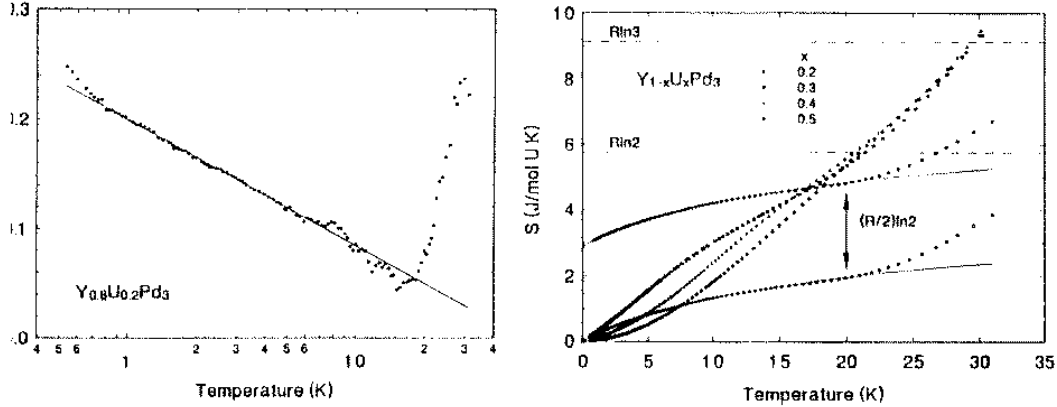}}
\parindent=0.5in
\vspace{.3in}
\caption{Specific heat and entropy of Y$_{1-x}$U$_x$Pd$_3$. (a) shows the specific heat
per mole for $x=0.2$ (with YPd$_3$ subtracted off), and 
displays a $\ln T$ behavior over about a decade and a half of temperature. 
The specific heat integrated from the lowest temperature yields the entropy 
curve shown in (b); notice that for $x>0.2$, of order $R\ln 2$ entropy is 
available below $T=30K$; hence, entropy appears to be missing from 
the $x=0.2$ curve. From Seaman {\it et al.} [1991].  }
\label{fig8p13}
\end{figure}

{\bf Y$_{1-x}$U$_x$Pd$_3$}.  The striking properties of this material
have garnered considerable
attention.  Fig.~\ref{fig8p13}(a) shows the specific heat for the composition
$x=0.2$ with the entropy
shown in Fig.~\ref{fig8p13}(b) as taken from Seaman {\it et al.} [1991].
The specific heat shows logarithmic behavior over about
one-and-one-half
decades of temperature, with an upturn clearly visible below 0.5K which
we shall return to
dicuss later in this subsection.  The specific heat difference when integrated
to yield the entropy difference 
has a clear shoulder near
$R/2\ln 2$ per mole U near $T=20$K.  At higher concentrations
apparent spin glass order sets in (evidenced by a hysteresis in the
magnetization) and in the
corresponding temperature range $R\ln 2$ entropy is recovered.
This suggests that $R/2 \ln 2$ entropy per U site should appear below
the lowest temperature
in Y$_{0.8}$U$_{0.2}$Pd$_3$.
In addition, the resistivity in the same temperature range goes as
$1-AT$ after showing a
linear in $\ln T$ behavior at higher temperatures, and
the magnetic susceptibility appears to go as $1-BT^{1/2}$
(once a subtraction of apparent large moment
paramagnetic impurities is made).  Taken together, these data put
forward a compelling
view of this material as a non-Fermi liquid metal.

A number of facts make the quadrupolar Kondo effect (Cox [1987b,d]) 
a worthwhile model
to begin with
to try to understand the physics of this system:\\
{\it Tetravalence of U ions}.  The quadrupolar Kondo effect requires
tetravalent U ions.
From photoemission experiments clear evidence of ``Fermi
level tuning'' is observed (Kang {\it et al.} [1992]).  This means that
the U 5f occupied peak
shifts with doping.  If the U ions were trivalent, they would not shift
the occupancy because
the Y ions are also trivalent.  However, tetravalent U ions would shift
the f-level provided the
underlying density of states is small, which is the case for YPd$_3$.
This Fermi level
tuning complicates the proof of single ion behavior, though this may be
remedied by combined
Th,U doping as we discuss further below.  The tetravalence is further
supported by the absence
of significant lattice constant change with doping; there is a very
close size match between
U$^{4+}$ and Y$^{3+}$ ions which is not the case for U$^{3+}$ ions.  \\
{\it Cubic Symmetry} The quadrupolar Kondo effect requires a site of
either
cubic, hexagonal, or tetragonal symmetry.   Cubic symmetry is indeed
maintained
in this material.
 YPd$_3$ has a cubic Cu$_3$Au structure, as does antiferromagnetically
 ordered
UPd$_4$ (which has 50/50 disorder of U and Pd ions on the Au sublattice
of the
Cu$_3$Au structure).  It is also informative that UPd$_3$ has
a dhcp structure with one site that is pseudo-cubic for the U ions.  It
is known in
the UPd$_3$ case that the U ions are tetravalent and have well resolved
crystal
field splittings (Buyers {\it et al.} [1980], McEwen {\it et al.}
[1994]; indeed,
UPd$_3$ shows a complex triple-Q vector quadrupolar order--see Walker
{\it et al.}
[1995]). \\
{\it Logarithmic behavior of $C/T$ and residual entropy}.   The data
obviously support
the two-channel Kondo model in this case.  For tetravalent U ions in
cubic symmetry,
only the non-magnetic $\gth$ doublet can provide the two-channel Kondo
effect.
The fact that order $R \ln 2$ entropy is available also points towards
the $\gth$ doublet
ground state.  Employing the thermodynamics calculations of Sacramento
and
Schlottmann [1989,1991] yields an estimated Kondo scale of 40K for
$x=0.2$ and
about 200K for $x=0.1$.  The increase is explainable in terms of the
Fermi level
tuning hypothesis.  \\
{\it Different temperature dependence to $\chi$ and $C/T$}  As
discussed extensively
in the Secs. 5,6, and 7, the susceptibility relevant to the impurity
pseudo-spin must have
the same temperature dependence if the two-channel Kondo effect is to
prevail.  The fact
that the temperature dependence is significantly different argues for
the quadrupolar
Kondo effect. \\
{\it Low temperature upturn in $C/T$}  Below about 0.6K the specific
heat rises above
the logarithmic behavior, which has been studied more extensively by
Ott [1993], who
rules out a nuclear Schottky anomaly as an explanation (only small
abundance Pd isotopes
are candidates in zero magnetic field and there is simply too much
entropy involved in the upturn).
This upturn is quite reminiscent of the upturns visible in Fig.~\ref{fig7p3}  in
the calculations of
Sacramento and Schlottmann [1991] representing the Schottky-like anomaly
arising from the
removal of ground state residual entropy.  In the quadrupolar Kondo
case the `magnetic
field' analogous to that of the pure spin Kondo model can either be
strain fields or electric
field gradients.  The Y,U sizes are close which suggests little strain
contribution; however,
the charge difference can lead to sizeable random electric field
gradients.  An estimate of
the mean electric field gradient splitting from Thomas-Fermi theory
gives
$\Delta\simeq 5K$ at $x=0.2$ (Cox, Kim, and Ludwig [1997]).  In
conjunction with the
spin crossover temperature $T^x_{sp}= \Delta^2/T_K$ discussed in Secs.
5.1,4.2, 6.1.3.c, and 7.2,
this produces a crossover scale estimate of about 0.6K in good
agreement with the location
of the upturn. \\

Two crucial questions arise about the ground state of this system.
First, is the non-Fermi
liquid physics associated with collective effects associated with the
proximity to antiferromagnetic
and/or spin glass ordering, or with single ion physics?  Second, is the
identification of a ground
state non-magnetic doublet on the U ions appropriate?  Alloying
experiments shed considerable
light on these questions, strongly affirming the single ion picture and
the doublet ground state
assignment.  There are two classes of alloying experiments:\\

\begin{figure}
\parindent=2.in
\indent{
\epsfxsize=6.in
\epsffile{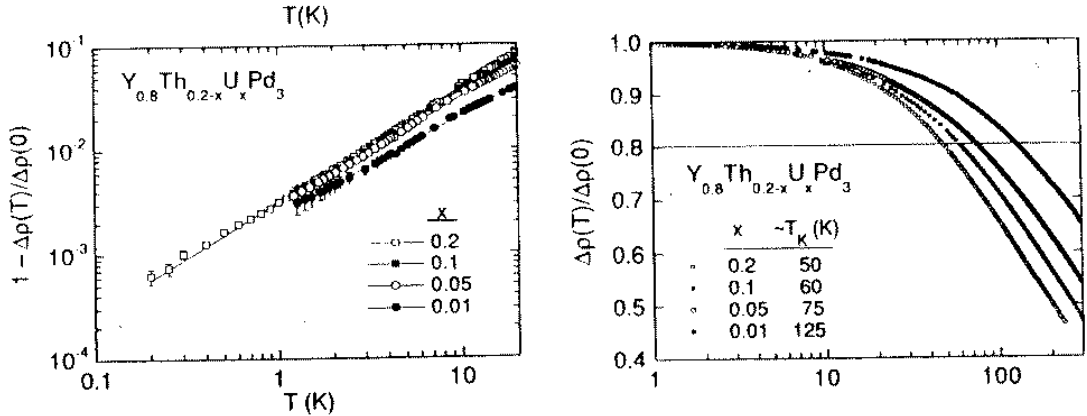}}
\parindent=0.5in
\vspace{.4in}
\caption{Resistivity of Y$_{1-x}$(Th$_{1-y}$,U$_y$)$_x$Pd$_3$, from 
Seaman and Maple [1994].  }
\label{fig8p14}
\end{figure}

{\it 1) Y$_{1-x}$(U$_{y}$Th$_{1-y}$)$_x$Pd$_3$ Alloys}  Seaman {\it et
al.} [1994] have
studied this system to shed light on whether 
the origin of the non-Fermi liquid
behavior is to be found in single ion
physics.  The idea is that the Fermi level tuning introduces an
intrinsic doping dependence to
the Kondo scale through the $f$ energy $\tilde \epsilon_f$
 ($T_K \sim \exp(-\pi|\tilde\epsilon_f|/2\Gamma)$) and so simply
 reducing the number of
U ions is insufficient to test the single ion hypothesis in this case.
However, by introducing
tetravalent Th ions of nearly the same size as the U ions, the
$f$-level can be kept approximately
constant and the U concentration can be diluted.   In this way it is
found that for $x=0.1,0.2$
very nearly the same Kondo scale and low temperature resistivity
behavior are maintained down
to low concentrations of U ions (order 1-5\%).  However, the distance
from the spin-glass and/or
antiferromagnetic ordering is increased.  This provides strong support
for a single ion
interpretation.  Resistivity curves for this system are shown in Fig.~\ref{fig8p14}.  \\

\begin{figure}
\parindent=2.in
\indent{
\epsfxsize=6.in
\epsffile{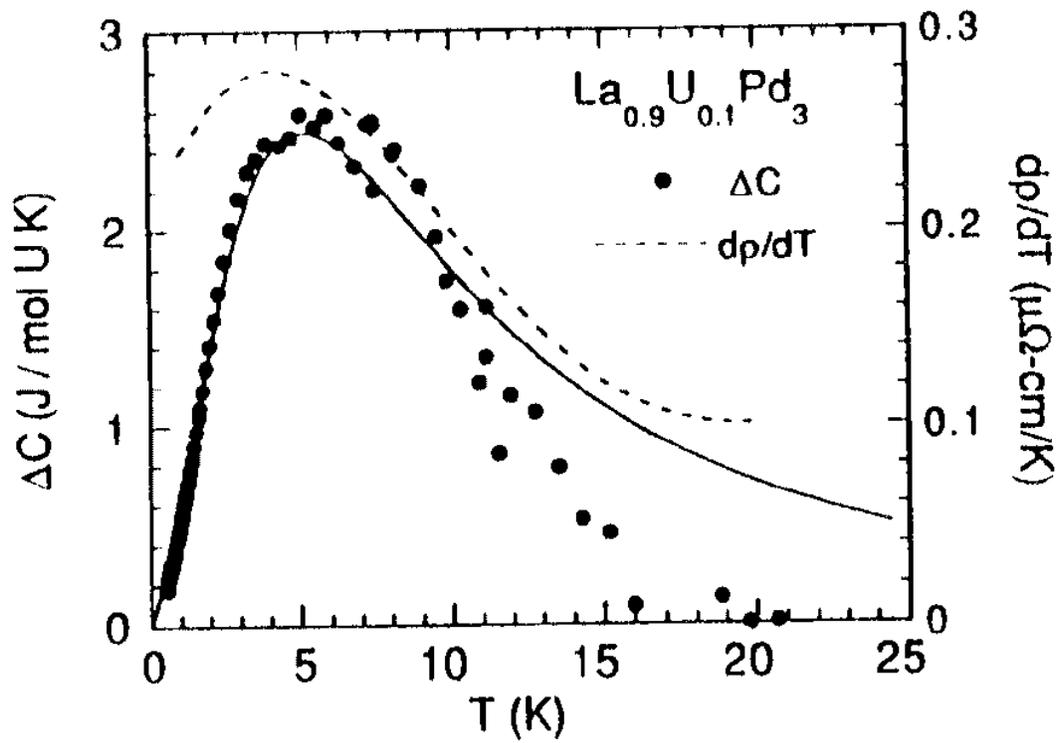}}
\parindent=0.5in
\vspace{.3in}
\caption{Specific heat of La$_{0.9}$U$_{0.1}$Pd$_3$ (from Seaman and Maple [1994]). 
}
\label{fig8p15}
\end{figure}

{\it 2) M$_{1-x}$U$_x$Pd$_3$ alloys, M=Sc,Pr,La}  Here the trivalence
of the $M$ ions
is kept fixed but the ionic volume is shifted systematically (Gajewski
{\it et al.}[1994]).
The volume primarily affects the hybridization:  for the largest ion
(La) the effect is to
expand the lattice providing negative chemical pressure which would
hypothetically
drive down the hybridization width $\Gamma$
and reducing the Kondo scale in turn relative to
Y$_{1-x}$U$_x$Pd$_3$ for a given $x$ value.
For the smallest ion (Sc, radius smaller than Y)
the lattice would be compressed, which increases the hybridization
width
$\Gamma$, driving the Kondo scale up for fixed $x$.   In the case of
large ions, if
$T_K$ is sufficiently reduced it should be possible to see the full
$R\ln 2$ entropy
in the quadrupolar Kondo picture due to the random field gradient
splitting of the
non-magnetic $\gth$ doublet.  These expectations are born out.  For
M=Sc, the
region of non-Fermi liquid behavior is extended all the way out to
$x=0.3$, where
the estimated Kondo scale is comparable to that of M=Y at $x=0.2$,
while for
$x=0.2$ the estimated Kondo scale for M=Sc is quite large (order
200K).  For
M=Pr, a mixed behavior is seen in the specific heat, as a pronounced
peak in $C/T$
is visible in the few K range. There is still a logarithmic upturn in
the resistivity at
higher temperatures but a downturn visible at low temperature in the
same region
where $C/T$ peaks.   For M=La, no Kondo anomaly is seen in the
resistivity.  At the
same time, a clear peak arises in $C$ at about 6K.  This peak may be
fit by a Gaussian
broadened Schottky anomaly, and clearly has $R \ln 2$ entropy per U
ion.  We note
that the peak position gives an estimated average splitting of the
doublet ground state
by $11K$ which is within a factor of two of the random field gradient
estimate
of Cox, Kim, and Ludwig [1997].  The size mismatch of the La and U ions
will certainly induce random strain fields that can enhance the
splitting above the random field
gradient splitting.  The specific heat curve for M=La and $x=0.1$ is
shown in Fig.~\ref{fig8p15}.  \\

Hence, there are a number of reasons to believe that the non-Fermi
liquid physics
of this alloy are driven by the one-impurity quadrupolar Kondo effect.
However, a
number of criticisms can be made of this interpretation based upon
other experimental
data:\\

{\it Critique (1): Resistivity}.  The electrical resistivity shows a linear
behavior even below 1K where,
on the basis of NCA calculations one would expect a rollover to
$T^{1/2}$ behavior
 (See Secs. 5.1 and 5.3 and
Cox and Makivic [1994], Kim and Cox [1997]).  \\

{\it Critique (2): Field Dependent Specific Heat}.  Andraka and Tsvelik [1991]
pointed out that the
magnetic field dependence of the specific heat does not follow what
would be expected
for a quadrupolar Kondo system.  At low fields, the magnetic field acts
as a channel field
which should produce a large increase of the specific heat coefficient
as calculated by
Andrei and Jerez [1995] (see Sec. 7.2 for a discussion).  At higher
values the
$\gth$ level splits quadratically in the field which would also have
the effect of increasing the specific heat coefficient.    The detailed
scaling function
fit to $C(T,H)/T-C(0,T)/T$ is that it appears to go as $f(H/T^{\beta})$
with
$\beta=1.3$ whereas the channel-field
scaling would yield $\beta=0.5$ and the spin-field scaling (through
quadratic in $H$
splitting of the $\gth$ doublet) would yield $\beta=0.25$. Moreover, 
the specific heat coefficient pretty much just drops with magnetic field, 
whereas one would expect from the calculations of Andrei and Jerez 
[1995] an enhancement as the residual entropy is shoved out of 
the ground state.  Similar behavior is seen in U(Cu$_{1-x}$Pd$_x$)$_5$
(Kim, Stewart, and Andraka [1992].) A possible
complication with
this picture is that in the quadrupolar Kondo picture there is a
substantial background
specific heat associated with excited magnetic triplet levels which
would presumably
drop in the applied magnetic field.  Mysteriously, the
magnetoresistance of this material is very
small (Seaman and Maple [1994]). 
Only detailed numerical analysis within the NCA or
other suitable method can be used to address this point on the 
theory side. \\

{\it Critique (3): Neutron Scattering Cross Section}.  Unpolarized
neutron scattering studies have been carried out
by Mook {\it et al.} [1993] and McEwen {\it et al.} [1994] which
subtract Y$_{1-x}$Th$_x$Pd$_3$
as a reference.  A subsequent polarized neutron study was carried out
by Dai {\it et al.} [1995].
For high concentrations, the McEwen {\it et al.} [1994] and Dai {\it et
al.} [1995] studies
suggest that the U ground state is a $\gfi$ magnetic triplet level.
Indeed, the Dai {\it et al.} [1995]
study finds evidence for antiferromagnetic Bragg peaks at $x=0.5$ which
have the same
structure as UPd$_4$.   Moreover, for $x=0.2$ they find evidence for
near critical scattering
in $S(\vec q,\omega)$ near these same Bragg peaks.  For $x=0.2$,
both studies present evidence for
two inelastic transitions which is compatible with the $\gth$ doublet
ground state (this would
have magnetic transitions to each of the excited triplets), at energies
of about 6 meV and
39 meV.  McEwen  {\it et al.} [1994] take the $\gth$ interpretation.
However, Dai {\it et al.}
[1995] argue from the polarized data that significant quasielastic
scattering exists below 1 meV
in energy and have made a preliminary case for a $\gfi$ ground state.

There are two
difficulties with this assignment in reconciling with the experimental
thermodynamics: (a)
the intensity of this peak yields a $T\to 0$ magnetic susceptibility
which exceeds the
experimental value by a factor of at least 20, and a $T=10K$ entropy
which exceeds the
experimental value by a factor of 3-5; (b) the energy scale associated
with the peak is
at most 10K, and the only characteristic energy scale from the data is
40K.  Clearly more
work is needed to understand the origin of this quasielastic
scattering; the presence of
concentration gradients in U ions and large moment paramagnetic
impurities as evidenced in the magnetization need to be
examined closely.  \\

{\it Critique (4): Ultrasonic Sound Velocity Measurements}  Sound velocity
measurements on
polycrystalline samples show no appreciable softening at any
temperature (Amara {\it et al.} [1995]), which not
only calls to question the quadrupolar Kondo effect (for which a
logarithmic transverse anomaly
would be expected at low temperatures) but also the Kondo effect itself
(as most Kondo
systems show a significant temperature dependence in the measured sound
velocities
due to the strong volume dependence of $T_K$-- see L\"{u}thi and
Thalmeier [1988]).
A possible drawback is the polycrystalline character of the samples
measured thus far
(Mandrus [1995]).\\

{\it Critique (5): Concentration Gradients}.  
The Y$_{1-x}$U$_x$Pd$_3$ is subject
to signficant
concentration gradients (Mydosh {\it et al.} [1993], Seaman {\it et
al.} [1993]) which
may affect the ability to sort out single ion from concentrated physics
effects.  Recent studies show that this is a relatively small effect, 
in that the mean square concentration deviation is about 10
near 0.2 the concentration would range from about 0.18 to 0.22--see
Maple {\it et al.} [1996]).  \\

Clearly an unambiguous assignment of the quadrupolar Kondo effect as
the source of
non-Fermi liquid physics cannot be made at this time, though there are
a number of
good arguments on behalf of this picture for this alloy system.  \\

\begin{figure}
\parindent=2.in
\indent{
\epsfxsize=6.5in
\epsffile{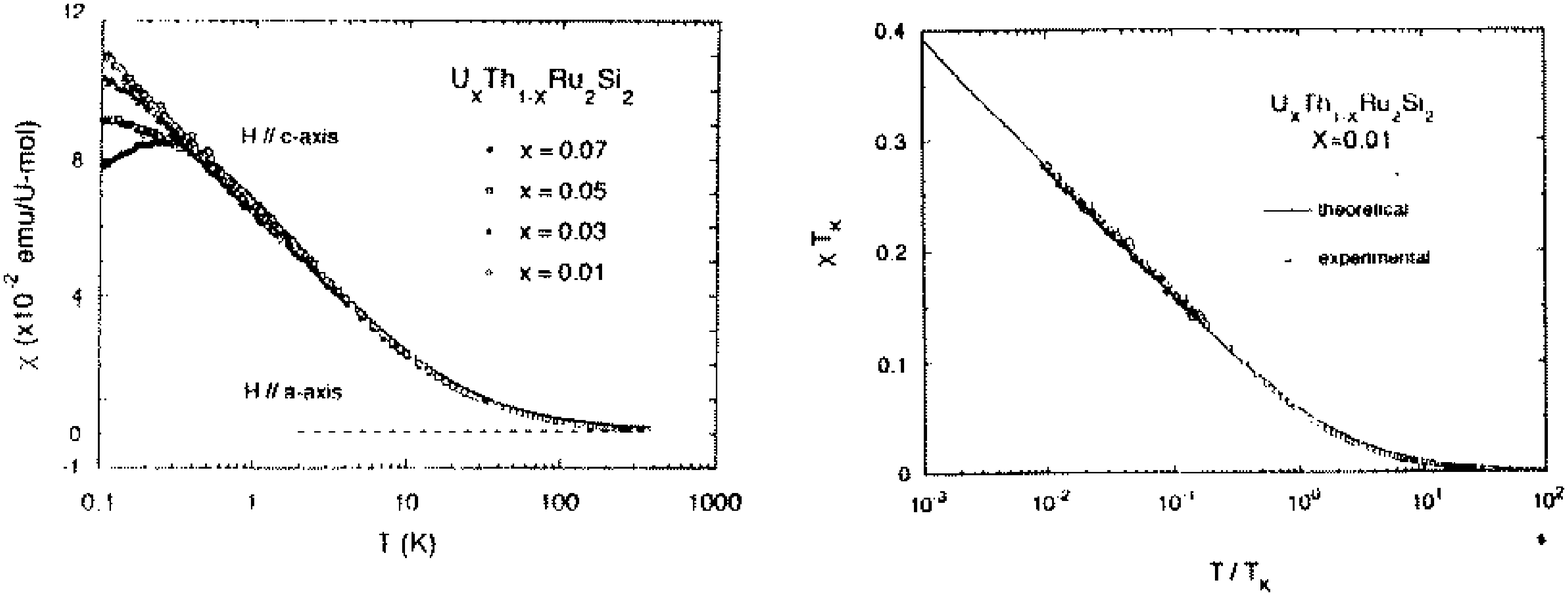}}
\parindent=0.5in
\vspace{.3in}
\caption{Magnetic susceptibility of Th$_{1-x}$U$_x$Ru$_2$Si$_2$ for several values of 
$x$ from Amitsuka {\it et al.} [1993].  The fit to the scaled curves is from the 
Bethe-Ansatz calculations of Sacramento and Schlottman [1991].}
\label{fig8p16}
\end{figure}

{\bf Th$_{1-x}$U$_x$Ru$_2$Si$_2$ and Th$_{1-x}$U$_x$Pd$_2$Si$_2$}  
Single crystals of tetragonal Th$_{1-x}$U$_x$Ru$_2$Si$_2$
have been studied by Amitsuka {\it et al.}
[1993a,b] from $x=0.01-0.07$.  The magnetic susceptibility, electronic
specific heat, and
electrical resistivity were measured.   The magnetic susceptibility for
$x=0.01$ is
shown in Fig.~\ref{fig8p16}(a) , where it is clear that for in plane fields the
susceptibility has weak
temperature dependence, while for $c$-axis fields there is a
logarithmic divergence apparent
over about 2 decades of temperature.  The solid curve is a fit of
Sacramento and
Schlottman's Bethe-Ansatz calculations for $M=2,S_I=1/2$ to the
$c$-axis data, and
the fit is quite good over four decades of temperature.  The estimated
Kondo scale is 11K.
 This is clearly a single site
effect, as the $c$-axis concentration dependent susceptibility is shown
in Fig.~\ref{fig8p16}(b).
Below about 0.5K, rounding is visible in the $\chi_c(T)$ curves, but
above this temperature
the curves are identical for four different concentrations up to
$x=0.07$.  The rounding
is reminiscent of that seen for applied field in Fig.~\ref{fig7p2} from
Sacramento and Schlottmann [1991].
A heuristic interpretation is that interaction effects between the U
ions are producing a
self-consistent magnetic field that induces the crossover.  Actually,
only a self-consistent {\it fluctuation} is likely important since the 
crossover scale goes as $H^2$.  

\begin{figure}
\parindent=2.in
\indent{
\epsfxsize=6.in
\epsffile{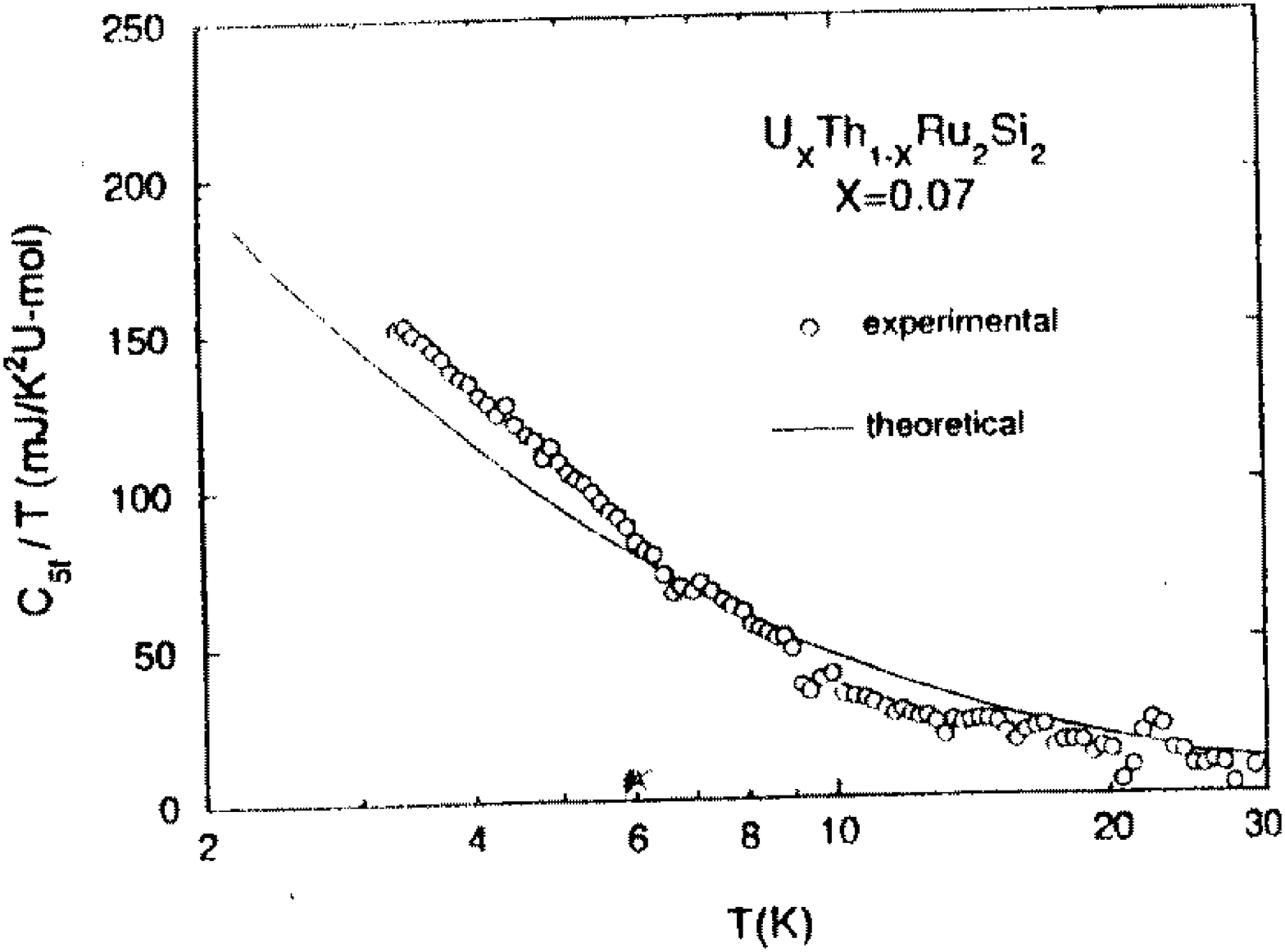}}
\parindent=0.5in
\vspace{.3in}
\caption{Specific heat of Th$_{1-x}$U$_x$Ru$_2$Si$_2$ for $x$=0.07 from 
Amitsuka {\it et al.} [1993].  The fit is to the theoretical calculation 
of Sacramento and Schlottman [1991] with the value $T_K=11K$ determined 
from the magnetic susceptibility.  }
\label{fig8p17}
\end{figure}

Taking the $T_K$ value from the $\chi_c$ fits, Amitsuka {\it et al.}
produce a satisfactory
zero parameter fit to the $x=0.07$ $C/T$ data, which is shown in Fig.~\ref{fig8p17}, where the
solid curve is taken from Sacramento and Schlottmann [1989,1991].
Because the theory
curve integrates to $(R/2)\ln 2$ entropy per mole $U$, that implies
that a residual entropy
of $(R/2)\ln 2$ remains below the lowest temperature measured which
also fits with the
$M=2,S_I=1/2$ theory.

Because the U ions are presumably tetravalent, the crystal symmetry is
tetragonal, and the
$c$-axis susceptibility obeys a Curie law well above $T_K$, it is
reasonable to suggest
that this physics arises from the quadrupolar Kondo effect in
tetragonal symmetry as
discussed in Sec. 2.2.3.  This would arise from a non-Kramers doublet
which can be
Zeeman split by an applied field along the $c$-axis.

\begin{figure}
\parindent=2.in
\indent{
\epsfxsize=4.in
\epsffile{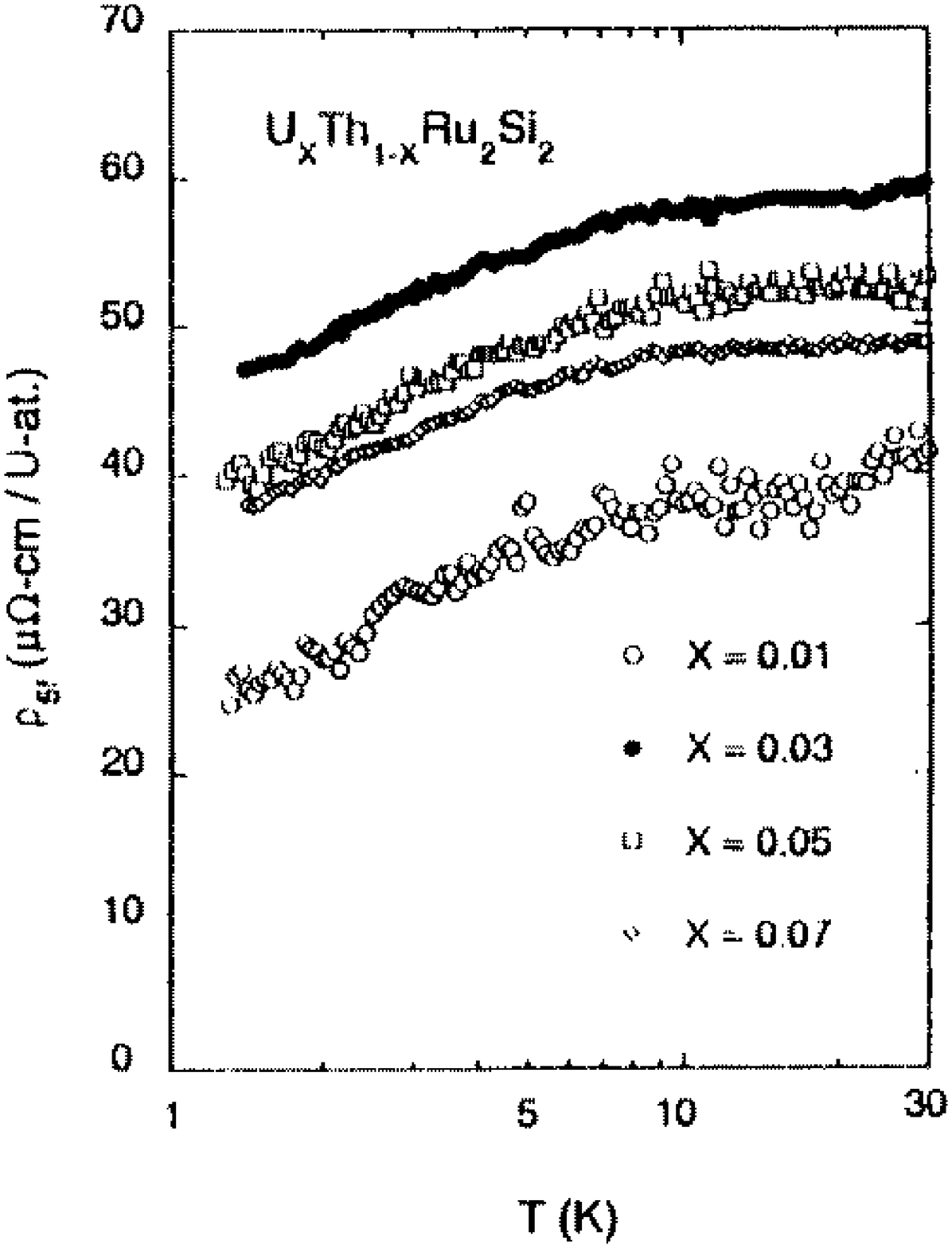}}
\parindent=0.5in
\caption{Resistivity of Th$_{1-x}$U$_x$Ru$_2$Si$_2$ from Amitsuka {\it et al.} [1993].
}
\label{fig8p18}
\end{figure}

An experimental difficulty with this proposal is the resistivity, shown
in Fig.~\ref{fig8p18}.   The
shape of the resistivity curves for several different concentrations is
clearly the same
(the magnitude is subject to systematic correction from the different
crystal geometry).
While the magnitude is sufficiently large (order 40 $\mu-\Omega$-cm) to
warrant consideration
as a Kondo system, there are two differences from the two-channel Kondo
theory:
(1) There is no high temperature linear in $\ln T$ region present in
the data; (2) There is
a low temperature {\it downturn} below $T_K$, which, over limited regions,
may be fit with $T$,$T^{1/2}$, or $\ln T$ behavior.  

It is difficult to conceive of simple
explanations for these resistivity discrepancies, but we do note a number of
important considerations
which could modify the behavior: \\
(i) The resistivity is significantly dependent on the details of the
hybridization matrix
elements between conduction states and the U 5$f$ states.  For example,
in the simplest
$M=1,S_I=1/2$ model for \ctp~ ions, a low temperature downturn in the
resistivity is
possible due to the ``hot-spots'' along principle axis directions
at which the $\gse$ conduction partial wave states must
have vanishing hybridization matrix elements (see Cox [1987c], and
Trees [1993,1995] (Trees and Cox [1994]) for a discussion
of the hybridization hot-spots and Kim and
Cox [1995b] for a discussion of their influence on resistivity). \\
(ii) The $T^{1/2}$ coefficient expected at sufficiently low
temperatures can experience a
sign reversal for sufficiently large potential scattering on the U site
which breaks
particle-hole symmetry (Affleck and Ludwig [1993]--see
eqn. (4.6)).  If this is true there should be a substantial
thermoelectric power present
(this can only be non-zero for the $M=2,S_I=1/2$ model in the presence
of particle-hole
symmetry breaking).  \\
(iii) A possible culprit for the potential scattering is excited
crystal field levels, which will
clearly break particle-hole symmetry, and which complicate the search
for $\ln T$ upturns
at high temperatures. (See Cox and Makivic [1994] for a discussion.) \\
(iv) The resistivity $\sqrt{T}$ coefficient 
can reverse sign in the strong coupling regime since the term is 
proportional to the leading irrelevant operator which is proportional
to the deviation from the fixed point coupling.   Arguing against this 
are the small Kondo scale and the excellent fits to Bethe-Ansatz 
results which are computed in the weak coupling limit.  

In Th$_{1-x}$U$_x$Pd$_2$Si$_2$ (Amitsuka {\it et al.} [1995]) the qualitative
and quantitative features are similar.  Again, in $\chi_c(T)$ (per U ion) 
 a region of 
single ion behavior (overlapping curves for different concentrations) 
with a rounding of $\chi_c$ that increases with increasing $x$, and the 
in plane susceptibility has a smaller value and negligible temperature 
dependence.  For the lowest concentration ($x=0.03$) a fit to $\chi_c(T)$
with the Bethe-Ansatz results is similarly good.  The specific heat per 
$U$ ion shows
a $\ln T$ behavior for low concentrations and drops as $x$ is increased. 
This is actually mysterious since it is not clear where the entropy goes. 
The authors note that the estimated residual entropy for the $x=0.03$ 
sample is actually somewhat larger than $R\ln 2/2$.  
The resistivity does show a Kondo like minimum at temperatures of order 
20K, but below a maximum again drops in a manner that may be fit over limited 
regions by $T$, $T^{1/2}$, or $\ln T$ behaviors.  

It should be mentioned that the dilute system La$_{1-x}$U$_x$Ru$_2$Si$_2$
{\it does not} show any evidence for non-Fermi liquid behavior for $x\le 0.15$ (Marumoto, Takeuchi, and Miyako 
[1996]).  

Given the sheer number of details for these two materials
 which check with the $M=2,S_I=1/2$
model, 
further experimental investigation of these materials is clearly warranted despite the
difficulties in understanding
the resistivity.  In particular, it would be desireable to have 
the specific heat measured under conditions of uniaxial stress in 
the basal plane and c-axis fields.  This should shove out the residual
entropy and confirm the assignment to a tetragonal quadrupolar Kondo
model.

\begin{figure}
\parindent=2.in
\indent{
\epsfxsize=6.5in
\epsffile{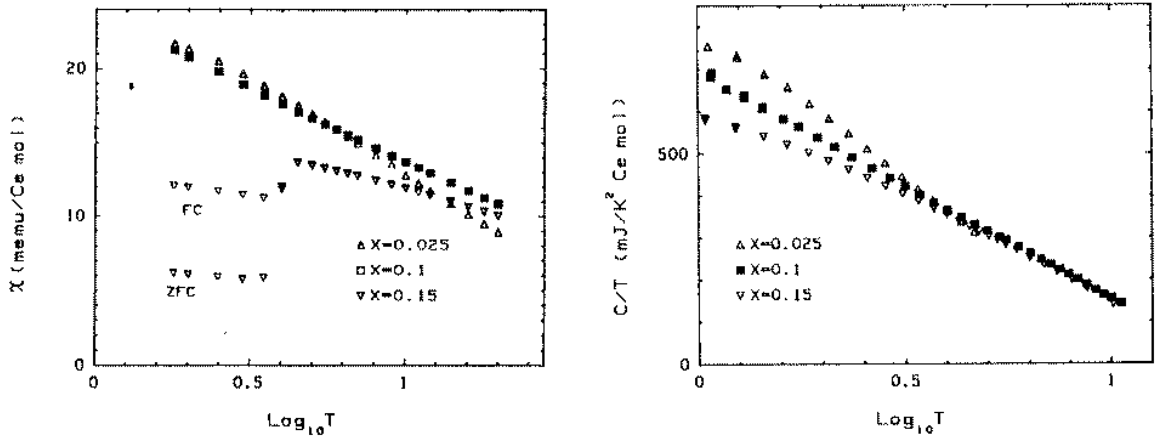}}
\parindent=0.5in
\vspace{.3in}
\caption{Susceptibility (left plot) and specific heat (right plot) per Ce for 
 La$_{1-x}$Ce$_x$Cu$_{2.2}$Si$_2$.  From Andraka [1994].  }
\label{fig8p19}
\end{figure}

{\bf La$_{1-x}$Ce$_x$Cu$_{2.2}$Si$_2$}  This system has been studied by
Andraka [1994]
and shows great promise as a dilute $M=2,S_I=1/2$ {\it magnetic} Kondo
system.  The added 
magnetic susceptibility and specific heat per Ce ion 
for $x=0.05,1,0.2$ are shown in Fig.~\ref{fig8p19}. 
We note that there is rough single ion behavior in that the $\chi$ 
curves for $x=0.05,0.1$ are very close, and the specific heat curves are
all close though deviations occur at lower $T$.  For $x=0.2$  there is 
hysteresis in $\chi$ which is taken as evidence for spin glass behavior. 

We focus on $x=0.1$ results.  Clearly both $\chi$ and $C/T$ 
exhibit a $\ln T$ behavior at low temperatures.  From a fit to
the $\ln T$ slope of
the $C/T$ curve, one estimates $T_K\approx 12K$.   Separately, one may
compute the
Landau-Wilson ratio.  To do this requires a knowledge of the effective
moment of the
presumed doublet ground state of the \ctp~ ion.  Fortunately we know
from neutron
scattering, susceptibility, and specific heat measurements (see Grewe
and Steglich [1991]
for details and references as well as Gorymychin and Osborne [1994])
that despite the tetragonal crystal structure, the crystal field
on the Ce site is pseudo-cubic with a $\gse$ doublet ground state and
excited $\gei$
quartet at about 360K.  To then compute the Landau-Wilson 
ratio, we should compare the logarithmic
slopes of the
$\chi$ and $C/T$ curves since there can be background constant
terms in each quantity.   Using the $\gse$ moment of
$\mu_{eff}=10\mu_B/7$ gives
$R_W=2.7(1)$, to be compared with the expected value of 8/3 from Secs.
6.1.3, 6.2.3, and
7.2.  The agreement is obviously excellent.

\begin{figure}
\parindent=2.in
\indent{
\epsfxsize=6.in
\epsffile{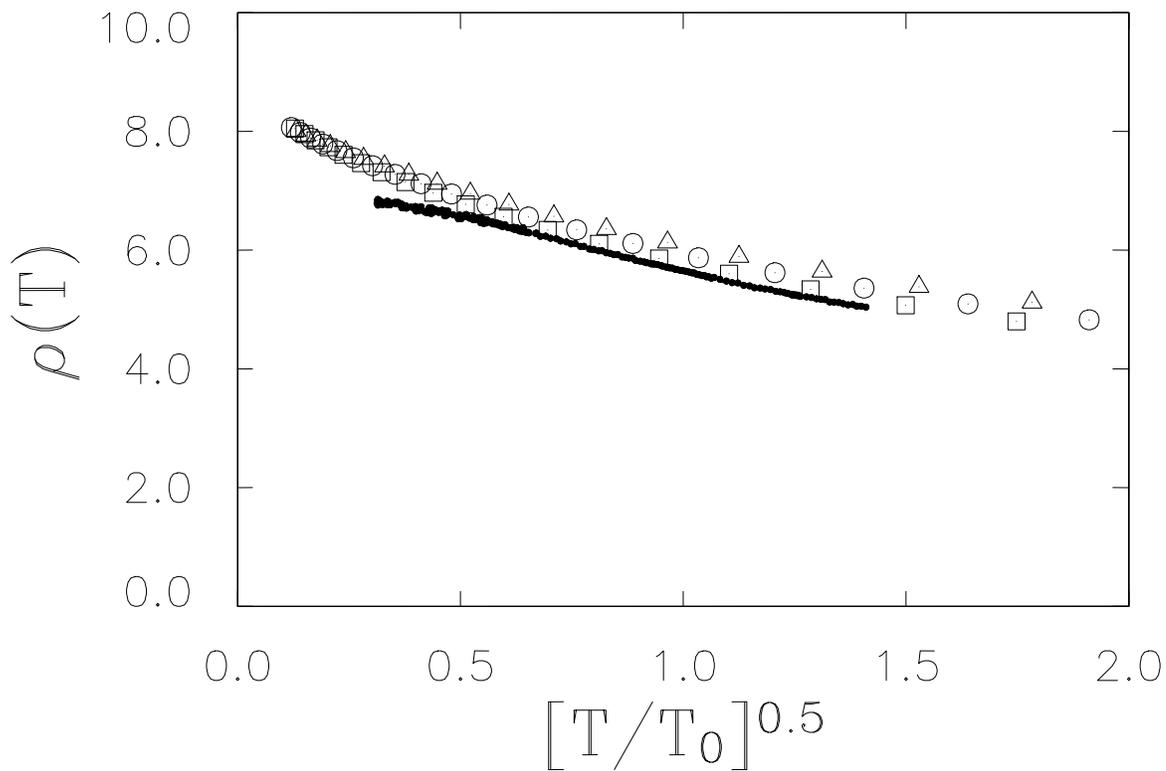}}
\parindent=0.5in
\vspace{.5in}
\caption{Fit to the resistivity of  La$_{1-x}$Ce$_x$Cu$_{2.2}$Si$_2$ by 
Kim and Cox [1997].  Theory (open symbols) 
is from an NCA calculation in the $M=2$ channel 
Kondo limit using the NCA as discussed in Sec. 5.1.6; small dots are 
experimental points.  }
\label{fig8p20}
\end{figure}

The electrical resistivity appears to show linear in $T$ behavior down
to 1.4K,  but in
view of the results of Sec. 5 from the NCA (Cox and Makivic [1994], Kim
and Cox
[1995b]) a $T^{1/2}$ law should not be cleanly seen until about
0.05$T_K$, which
sits below 0.6K.  Indeed the theory curve of Fig.~\ref{fig8p20} is a fit of the
NCA calculations
of Kim and Cox [1995b] to the $x=0.1$ data, and it is credible given 
the limited range of the experimental data (about a decade).
However, a clear break is seen in the data which takes it below 
the theory curve. 

A separate qualitative confirmation of the validity of the
$M=2,S_I=1/2$ model is
that when magnetic field is applied, the specific heat coefficient
shows a significant
increase.  In quantitative detail, it is not as large as for the
$M=2,S_I=1/2$ model
for the given field strengths, but the behavior sharply contrasts with
the $M=1,S_I=1/2$
model where $C/T$ should simply drop as the Kondo resonance Zeeman
splits
(Andrei, Furuya, and Lowenstein [1993]).  We note that the presence of
a sizeable
background term in $C/T$ of Fig.~\ref{fig8p19} likely corresponds to the tail of
the lifetime
broadened Schottky anomaly of the excited $\gei$ quartet.  This should
drop in
magnetic field, which complicates the direct analysis of the data in
terms of the
$M=2,S_I=1/2$ model.

Clearly the material satisfies the requirement of cubic symmetry, and
the question
arises whether it satisfies the necessary requirement of greater
fluctuation weight to
$f^2$ than $f^0$ in the ground state.  As discussed in Secs.
2.2.2,2.2.3, and 5.3,
this is sampled by the thermopower which should be negative well above
$T_K$
(Cox [1993], Kim and Cox [1995,1997]).  Early measurements of the 
thermopower on La$_{1-x}$Ce$_x$Cu$_2$Si$_2$ by Aliev {\it et al.} [1984]
revisited more recently by Buschinger, Geibel, and Steglich [1996]
show that for low concentrations the thermopower goes positive.  A complication
is that the host system has a positive thermopower and so a proper way 
to isolate the Ce concentration must be worked out, which has not yet
been done.   For example, the Nordheim-Gorter correction has been used  
for the Kondo system  La$_{1-x}$Ce$_x$B$_6$ to isolate the Ce ion induced
thermopower (Winzer [1975]). According to the work of 
Buschinger, Geibel, and Steglich [1996], this correction 
actually enhances the positive peak in $S(T)$ for dilute samples. 

The thermopower has not been measured for the $x=0.1$
samples
of Andraka [1994].  However, it has been measured for $x=1$, and it is
found that
a sign change to negative thermopower occurs near $T=70$K, which is so
large
compared to $T_K$ that it is not attributable to lattice coherence
effects.  The
thermopower in this case reaches a maximum amplitude of about
-30$\mu$V/K,
which is too large to be attributed to anything but resonant
scattering, which the
Kondo effect provides quite naturally.   It is intriguing that a study
of Ce based
1-2-2 compounds shows that the thermopower at 20K has a strong
correlation with
the unit cell volume, undergoing an abrupt change from negative to
positive near
the CeCu$_2$Si$_2$ unit cell volume.  Based on the suggestions by Cox
[1993] and
Kim and Cox [1995a,b] that the $M=3,S_I=1/2$ Kondo effect can be
realized
with pressure, the crossing point is a promising region to look for
three channel
Kondo model candidate materials.

Andraka [1994] has argued that the quantitative discrepancies in the
specific heat,
and the apparent linear in $T$ resistivity argue against the
$M=2,S_I=1/2$ magnetic
Kondo effect.  However, as argued above, the resistivity actually is in
good agreement
with theory over the measured temperature range, and the specific heat
coefficient
increase is difficult to interpret given the large and possibly field
dependent background
term.   Andraka [1994] also argues that $x=0.1$ is a special
concentration, since
at $x=0.05,0.15$ the logarithmic anomalies are not seen.  Moreover, the
$x=0.15$
sample is argued to possess a `spin glass'  transition due to irreversibility
in the magnetization.  Caution
should be lent
to accepting these arguments uncritically, however, because the
CeCu$_2$Si$_2$
system is well known to be extremely sensitive to preparation
conditions.  After
nearly twenty years of study for the concentrated case, only recently
(Steglich {\it et
al.}, 1995) has the ternary phase diagram been carefully worked out.
It is found that
the concentrated system has an extreme sensitivity to small variations
(order percent)
in the Cu concentration.  The nominal 2.2 stoichiometry for Cu is based
on starting
composition during preparation and places the $x=1$ samples in the
slightly copper
rich stoichiometry which leads to the most cubic crystal field scheme
as well as
superconductivity and lattice non-Fermi liquid behavior.  Slight
deficiency of Cu from
this concentration leads to a still uncharacterized magnetic state.
Hence, for a
conclusive answer to the question of whether the $x=0.1$ physics is or
is not single
ion in origin, we must await more careful studies of samples with well
characterized
Cu concentration.

{\bf Th$_{1-x}$U$_x$Be$_{13}$}  Aliev {\it et al.} [1992,1993,1994]
have studied this system.
extensively.  At the value $x=0.9$ which is certainly far from
dilute, they find
$C/T\sim -\ln T$, $\chi(T) \sim 1-AT^{1/2}$, $\rho(T)\sim 1+ BT^{1/2}$ 
(with $B>0$), all of which
fit the two-channel cubic quadrupolar Kondo picture as discussed in
Secs. 2.2.1, and 5.2. 
A complication is that in this crystal structure no dopants
appear to leave
the volume unchanged which means the hybridization is strongly affected
by the
doping.  (Indeed, since the Th ions are larger they expand the lattice
and diminish the
hybridization which will lower $T_K$.  The data appear to suggest this
happens relative to the bulk $x=1$ system.)
An extensive study of M$_{1-x}$U$_x$Be$_{13}$ alloys by Kim {\it et al.}
[1990] revealed that while the specific heat could be
significantly altered
by doping, the magnitude of the low temperature magnetic susceptibility
was
hardly affected.  This suggests further that the origin of the specific
heat and
susceptibility are different, consistent with an interpretation in
terms of van Vleck
susceptibility which is important for the two-channel quadrupolar
Kondo effect in
cubic symmetry.  Clearly it is desireable to dilute further.   

A further consistency with the quadrupolar Kondo effect is the non-linear
susceptibility (Aliev {\it et al.}, [1995a,b]).  This was motivated 
in part by earlier measurements of Ramirez {\it et al.}[1994] on UBe$_{13}$ 
which shall be discussed below.  In theory, the 
non-linear susceptibility $\chi^{(3)}(T)$ defined from the magnetization
via 
$$ \chi^{(3)}(T) = 6 [M(H,T)-\chi(T)H]/H^3 ~~.\leqno(8.2.1)$$
For a magnetic doublet ground state, $\chi^{(3)}$ is expected to 
be large and negative, as is easily seen from straightforwardly 
expanding the Brillouin function magnetization to 
obtain $\chi^{(3)}\sim 1/T^3$ for localized moments.  
This would be modified, at low temperature, to $\sim 1/T_0^3$ for
a Kondo system.  
In a more general situation, $\chi^{(3)}$ 
depends upon the orientation of $H$.  For a
purely localized quadrupolar moment system with a cubic non-Kramers
$\gth$ ground doublet, Morin and Schmitt [1981]
have shown that for a field along a principle axis, $\chi^{(3)}$ will
display a {\it positive} 
Curie law divergence, while for a field along a body diagonal
$\chi^{(3)}$ will be of van Vleck character at low temperature and 
{\it negative}.
This result is easily understood in terms of the 
magnetoelastic coupling of the $\gth$ ground state--principle axis
fields induce tetragonal distortions which are quadratic in $H$ and
split the doublet.  There is no linear 
coupling however to strains along the
body diagonal (matrix elements do exist to excited states).   Hence,
the non-linear susceptibility for a 
principle axis field is essentially a measure of the quadrupolar 
susceptibility.  While the quadrupolar Kondo effect would modify 
this from a $1/T$ divergence to $-\ln T$, the divergence would still
be present, and the characteristic anisotropy provides an excellent 
test of the applicability of the quadrupolar Kondo  model
(Chandra {\it et al.},  [1993], Ramirez {\it et al.}, 
[1994]).  

\begin{figure}
\parindent=2.in
\indent{
\epsfxsize=7.in
\epsffile{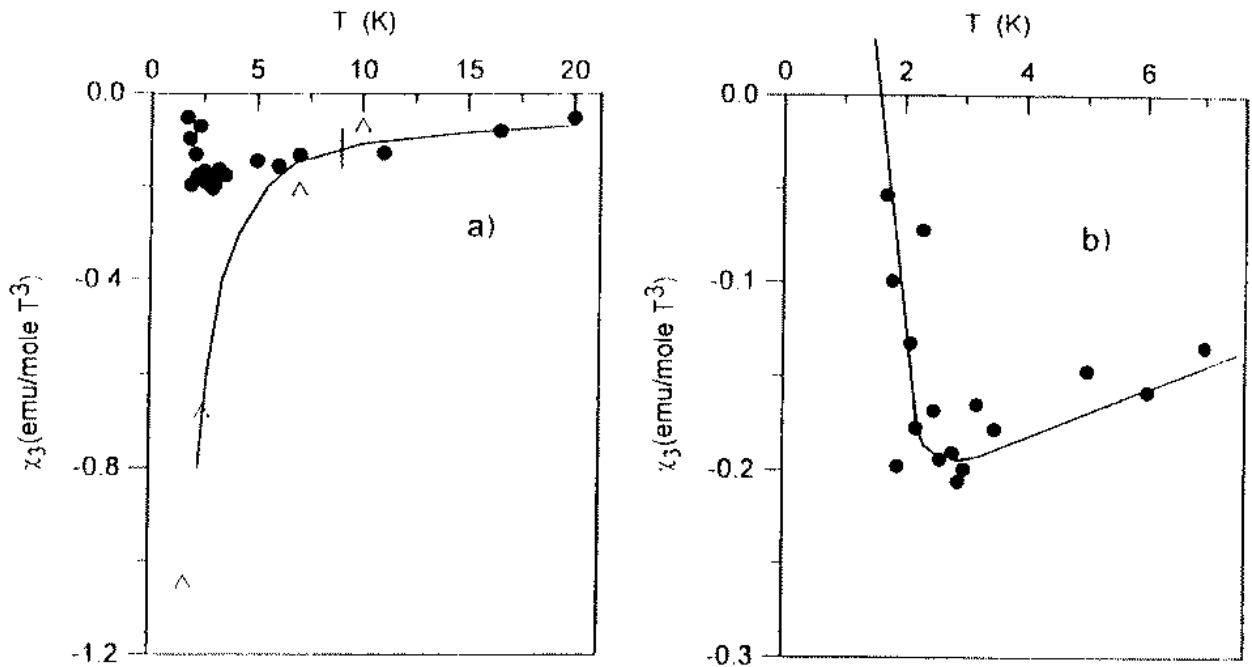}}
\parindent=0.5in
\vspace{.3in}
\caption{Non-linear susceptibility measurements on UBe$_{13}$ and 
U$_{0.9}$Th$_{0.1}$Be$_{13}$ from Aliev {\it et al.} [1995a,b].  
Triangles are measurements on a polycrystal of UBe$_{13}$, circles 
are measurements on a polycrystal of U$_{0.9}$Th$_{0.1}$Be$_{13}$. Line
is taken from the data of Ramirez {\it et al.} [1994]. Right hand figure
is a blow up of the low temperature region for
U$_{0.9}$Th$_{0.1}$Be$_{13}$ indicating the tendency to a sign change
expected for a quadrupolar Kondo material.}
\label{fig8p21}
\end{figure}

Aliev {\it et al.} [1995a,b] performed measurements only on polycrystalline
samples.  As shown in Fig.~\ref{fig8p21}, 
for $x=0.1$ they found that the powder averaged $\chi^{(3)}$ 
is predominantly negative at high temperatures but finds a minimum and 
changes sign as the temperature is lowered, which is in accord with the
expectations of the previous paragraph.  In contrast, when pure UBe$_{13}$
is measured for similarly prepared polycrystalline samples, $\chi^{(3)}$ 
is relatively large, negative, and decreases with decreasing temperature,
qualitatively in agreement with a magnetic ground state.  Indeed, the
polycrystalline data agrees excellently with Ramirez {\it et al.}'s [1994] 
single crystal data, which excludes the possibility of large moment
paramagnetic impurities giving rise to the $x=0$ $\chi^{(3)}$ results.  

\begin{figure}
\parindent=2.in
\indent{
\epsfxsize=5.in
\epsffile{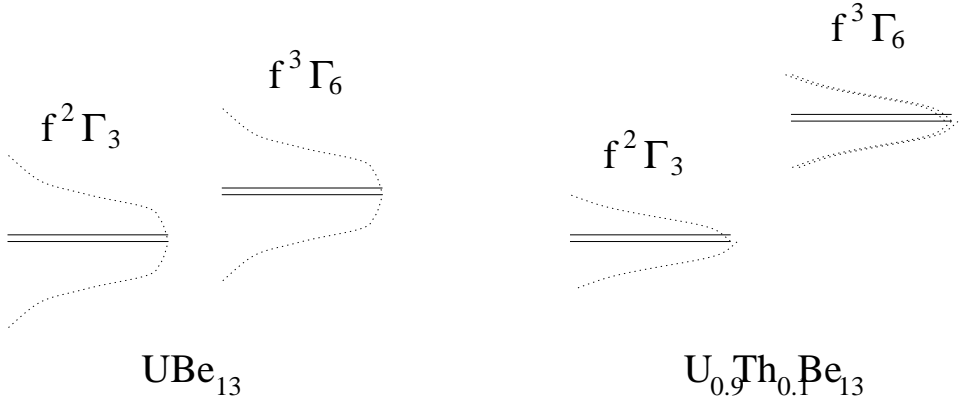}}
\parindent=0.5in
\caption{Mixed valence scenario for U$_{1-x}$Th$_x$Be$_{13}$ proposed by 
Aliev {\it et al.} [1995a,b].  For $x=0$, the picture is one of almost 
complete mixed valence between $f^2$ and $f^3$ configurations.  It is 
conjectured that the Th doping favors the $f^2$ energetically and diminishes 
the degree of valence mixing.  This favoring could arise from a reduction 
in the hybridization broadening of the configuration energy levels.  The
broadening is indicated schematically by the dotted curves about each 
doublet level.  
}
\label{fig8p22}
\end{figure}

Based upon this work and the {\it positive} coefficient of $\sqrt{T}$ in
the $x=0.1$ resistivity, Aliev {\it et al.} [1995a,b] put forward an 
interesting set of hypotheses: first, the $x=0.1$ samples are in the 
strong coupling regime.  That is, the coupling strength exceeds that 
of the non-trivial fixed point.  This can explain the positive coefficient
of the resistivity.  Second, the ionic ground state changes as a function
of Th doping, being an $f^3\gsi$ doublet for $x=0$, and an $f^2 \gth$ 
doublet for $x=0.1$. This would require the U ions to be strongly mixed
valent between $3+$ and $4+$, which is not implausible.  This scenario 
is illustrated in Fig.~\ref{fig8p22}.  This second hypothesis checks
with the first hypothesis because it is precisely in the mixed valence 
regime where strong coupling might plausibly occur (the dimensionless 
Schrieffer-Wolff exchange can grow to order unity).  The hypotheses
are very interesting because the different symmetry ground states would
seem to imply a novel quantum critical point at the precise point in $x$
where the levels cross.   

There are three major concerns  with the hypotheses:\\
1) Taking the Th ions as tetravalent, the substitution would add electrons,
and this would actually drive the uranium ions more towards trivalence. \\
2) While the non-linear 
susceptibility for pure UBe$_{13}$ is strongly  temperature dependent, 
the susceptibility is not, and hence this interpretation is problematic.  \\
3) It is not clear that a sufficiently small energy scale can be generated in
the mixed valent regime for the uranium ions, though some variational 
treatments of the ion with full spherical symmetry suggest this is 
possible (Read {\it et al.}, [1986]; Nunes, Rasul and Gehring [1986]).  \\

Nevertheless, the hypotheses of Aliev {\it et al.} are very intriguing
and deserve further exploration.

{\bf Th$_{1-x}$U$_x$Pd$_2$Al$_3$ and Th$_{1-x}$U$_x$Ni$_2$Al$_3$}
These
two hexagonal systems reveal $-\ln T$ specific heat coefficients at low
temperatures and
low concentrations ($x\simeq 0.1$) (Maple {\it et al.} [1994] (Pd), Kim
{\it et al.} [1993] (Ni)).
The susceptibility  in the Pd based system for a polycrystalline sample
can be fit
to either $-\ln T$ or $1-A\sqrt{T}$ behavior at low temperatures.  In
each case the
resistivity apparently saturates with a linear in $T$ law.   The U ions
in these systems are likely tetravalent so that they
are candidates for the quadrupolar Kondo effect in hexagonal symmetry
arising from
a non-Kramers doublet. The low concentration data clearly shows 
single ion scaling.   Kim {\it et al.} [1993] argue that in
the Ni case there
is a proximity to a spin glass ordering. Nevertheless, given the same 
crystal structure and the single ion scaling observed in 
Th$_{1-x}$U$_x$Pd$_2$Al$_3$, it is clear that these systems
deserve further
careful study as quadrupolar Kondo candidates.

{\bf Concentrated non-Fermi liquid Compounds:  UBe$_{13}$,
CeCu$_2$Si$_2$, and PrInAg$_2$}

{\it UBe$_{13}$}\\

UBe$_{13}$ is certainly the most anomalous of the heavy fermion
superconductors.
We identify the following set:\\
(1) The specific heat is very weakly dependent upon magnetic field and
highly sensitive
to pressure.  The low temperature value is of order 1000 mJ/mole-K$^2$,
corresponding
to an effective mass of several hundred free electron masses. \\
(2) The specific heat still rises with decreasing temperature on entering the
superconducting state
and with suppression of $T_c$ by magnetic field continues to rise down
to 0.3K (where
measurements stop).  This is clear evidence that the superconducting
instability does
not occur in a Fermi liquid normal phase.\\
(3) The magnetic susceptibility is weakly pressure dependent in
comparison to the
specific heat and under pressure 
has a completely different temperature dependence
(McElfresh {\it et al.}
[1993]).\\
(4)  Doping on the U sublattice which drives away the specific
heat anomaly leaves
the low temperature susceptibility essentially unchanged.  \\
(5) The magnetization is linear in field up to 20T. \\
(6) The dynamic magnetic susceptibility reveals no significant
structure on the scale of
1 meV as is evidenced in $C/T$, and instead shows a broad
`quasielastic' response on the
scale of 15 meV as evidenced both in neutron scattering and Raman
spectra.  Concommitant
with the peak in $\chi''$ is a Schottky anomaly in the specific heat,
suggesting that the
15 meV peak represents highly damped crystal field levels for which
further evidence appears
in nuclear magnetic relaxation of the $^9$Be sites.  This dynamic
susceptibility
peak integrates to give 80\% of the static susceptibilty up to the
experimental cutoff.  This
places a stringent bound on any hypothetical moment carrying state in
the low frequency
region:  given a 10K Kondo scale, to explain the residual susceptibility
the effective squared moment must be less than 0.25$\mu_B^2$ 
which would appear to
rule out an interpretation in terms of a 5$f^3\gsi$ ground state. \\
(7) The electrical resistivity at the superconducting transition is
reproducibly large (order
80-100$\mu-\Omega$-cm, and is reversibly suppressed by applied pressure
and magnetic
field (Aronson {\it et al.} [1989], Batlogg {it et al.} [1987], Andraka
and Stewart [1994]).
There is no region in ambient pressure or zero field in which a $T^2$
coefficient is visible
in the resistivity (rather, a linear in $T$ term is present).  In
applied magnetic field, there
is some evidence that the resistivity obeys a scaling law
$\rho(T,H)/\rho(T,0) \approx
f(H/|T-T_0|^{\beta})$ where $\beta =1,T_0=0$ (Batlogg {\it et al.}
[1987]) and
$\beta=0.7,T_0=0.5K$ (Andraka and Stewart [1993]) have been fit to the
data.  \\
(8) The muon Knight shift (Luke  [1995]), $^9$Be nuclear
Knight shift (MacLaughlin
{\it et al.} [1983]), and neutron scattering form factor (Stassis {\it
et al.} [1985]) show
no change upon entering the superconducting state.\\
(9) The upper critical field has a large low temperature limit which
has been estimated above
the Pauli paramagnetic value in at least one study (though not in
others).\\

Different interpretations of these experimental data abound, most
notably that the onset
to superconductivity occurs above the coherence temperature which marks
the beginning of well
defined Fermi liquid behavior.  We note that even if this is true, the
superconductivity occurs
in the absence of a well defined Fermi liquid state so that it is
meaningless to use a quasiparticle
picture for the description of the normal phase.  We take points
(1)-(9) to represent the
following picture: heavy fermions arise in UBe$_{13}$ in a non-Fermi
liquid state (evidenced
by points (2) and (6) particularly) which is likely not due to a magnetic
lattice
Kondo effect (evidenced by points (1),(3-5),(7,8)) but may potentially
arise from a non-magnetic
Kondo effect.  The logical candidate is the quadrupolar Kondo effect
(Cox [1987b], Cox [1988a,b],
Cox [1993], E. Kim {\it et al.} [1996];  see also secs. 2.2.1, 2.2.3,
and 5.2).  

In this body of
theory, it has been established that the quadrupolar Kondo effect can
plausibly explain the
thermodynamic data and dynamic magnetic response measured in NMR,
inelastic neutron
scattering, and Raman spectroscopy.
In particular, the susceptibility arises from vanVleck processes
(virtual
excitations to excited magnetic triplet levels) which have a different
physical origin and overall
energy scale than the specific heat coefficient.  The low temperature
relaxation of the NMR
corresponds to the triplet to non-magnetic doublet peak shown in the
schematic vanVleck
response of Fig.~\ref{fig5p23}.  This peak has little overall spectral
intensity and so is not clearly
resolved in neutron scattering data (though the more recent data of
Shapiro {\it et al.} [1993]
may be showing some evidence for this behavior).

An interpretation of the unusual
resistivity in terms of an infinite dimensional two-channel Kondo
lattice
picture is given in Cox [1996] (see also Sec. 9.3.2).  In particular,
it is argued that the finite
resistivity is an intrinsic feature of the two-channel lattice in the
absence of an ordering transition
due to the `spin-disorder' scattering off the degenerate many body
clouds surrounding each
Kondo site (see secs. 3.4.5, 5.1,6.1.4 for further discussion of this
spin-disorder
scattering).    It is noted that ordinary Kondo
lattice behavior cannot give the observed negative magnetoresistance at
low temperature
in this reference
and in McElfresh {\it et al.} [1993] (where an interpretation of the
pressure dependent $\chi(T)$
data in terms of the quadrupolar Kondo effect is given).  It is
possible that positive
magnetoresistance expected for the ordinary Kondo lattice is obscured
by the superconducting
transition, but the large extrappolated residual resistivity goes  
against this interpretation
(essentially the resistivity is already very close to its $T=0$ value
by any reasonable extrappolation
scheme).   The detailed scaling function predicted by the theory of Cox
[1996] does not agree
with the experimental data, but this may well be due to the
oversimplifications induced by the
infinite dimensional limit.

At this point, one main piece of experimental data is held up as an
objection to this point of
view:  non-linear susceptibility measurements (Chandra {\it et al.} [1993],
Ramirez {\it et al.}
[1994]).  
response can be expected, and
thus a low temperature {\it positive} susceptibility is anticipated
which is non-divergent.
The data of Ramirez {\it et al.} [1994] show apparently divergent
$\chi^{(3)}$ curves for
both body diagonal and principal axis fields.  This would appear to
rule out the quadrupolar
Kondo effect.  Three concerns may be raised about the data:\\
(1) $\chi^{(3)}$ measurements on Th doped samples by Aliev {\it et al.}
[1995a,b] show the
anticipated anisotropy and roughly correct temperature dependence.\\
(2) There is anisotropy in $\chi^{(3)}$ which goes the correct
direction: if the body diagonal
curve is subtracted from the principal axis curve, a weakly divergent
response may remain.\\
(3) Points (1,2) suggest that an extrinsic origin to large 
negative non-linear
susceptibility is possible.  
Contamination by large moment paramagnetic impurities such as
Ho$^{3+}$ ions
can weakly affect the magnetic susceptibility (which scales as the
effective moment squared
times the concentration of impurities) while dominating the non-linear
susceptibility (which
would scale as the concentration times the fourth power of the
effective moment).  However, this scenario would produce significant
sample dependence to $\chi^{(3)}$, while the polycrystalline data of 
Aliev {\it et al.} [1995a,b] fall right on top of that of Ramirez 
{\it et al.} [1994].  This would appear to argue against extrinsic
origin to $\chi^{(3)}$.\\

{\bf CeCu$_2$Si$_2$}\\

Careful recent studies of CeCu$_2$Si$_2$ coupled with attention to the
ternary phase diagram
show the following (Steglich {\it et al.} [1995]):\\
(1) In the slightly copper rich region which has only a superconducting
phase at low temperature
and magnetic field, the specific heat coefficient which has a large
slope at the superconducting transition continues to rise on initial
suppression of $T_c$ with magnetic field.  This, as for
UBe$_{13}$, argues that the normal state from which the
superconductivity occurs is not
a Fermi liquid state. \\
(2) The normal state has a dominant linear in
 $T$ resistivity with large (order 30-40 $\mu-\Omega$) residual value.
 The linear term goes
to zero in applied magnetic field while a $T^2$ term grows, and an
appreciable region over
which $T^2$ behavior is seen in the resistivity opens up at precisely
the region where some
kind of field induced magnetic order occurs.  The boundary of the $T^2$
region tracks the
boundary of the magnetic order.   The linear term in $\rho(T)$ vanishes
at precisely the
field where the magnetic  order first arises.  \\

Given the data for La$_{1-x}$Ce$_x$Cu$_{2.2}$Si$_2$ discussed earlier
in this subsection,
it is reasonable to propose CeCu$_2$Si$_2$ as a candidate magnetic
two-channel Kondo
lattice system.  In terms of the infinite dimensional resistivity
theory used to interpret
UBe$_{13}$ (Cox [1995]; see also Sec. 9.3.2), the low field behavior is
compatible with
the non-Fermi liquid behavior expected for the lattice.  In higher
fields, the combination of
magnetic order with applied field induces a crossover to Fermi liquid
physics associated
with the same crossover in the impurity problem (c.f, Secs. 4.2,
5.1,6.1.2,7.2).    The
crossover scale in the infinite dimensional theory would track
$(H_{mol}+H_{ext})^2/T_K$,
where $H_{mol}$ is the molecular field associated with the magnetic
orderand $H_{ext}$
is the applied field.  This gives a rough interpretation of the high
field $T^2$ region in
the resistivity.  \\

{\bf PrInAg$_2$}\\

This is the first compound studied in a promising program to look for 
quadrupolar Kondo physics in lanthanide intermetallics (Yatskar {\it et al.},
[1996]). In this material, 
neutron scattering confirms that the Pr ions are in a $\gth$ ground state.
The material shows anomalous properties, though not so unusual as UBe$_{13}$.
The specific heat at low temperatures is large ($C/T$ tends to a low temperature
value of around 6 J/mole-K$^2$!) which is strongly indicative of Kondo effect
physics. There is a pronounced region of $-\ln T$ behavior in the specific 
heat prior to saturation.  However, the residual resistivity is clearly
finite, although the low temperature behavior is not quadratic in the 
temperature.  There is no evidence for long range quadrupolar or 
superconducting order.  

The discovery of the first unambiguous quadrupolar Kondo lattice candidate
presents reasons for theorists like the present authors to feel both 
excited and challenged.  In particular, this result stands in stark 
contrast to the $d=\infty$ calculations discussed briefly above in 
relation to UBe$_{13}$ and more in Sec. 9.3.2, where a residual resistivity
is present for the lattice.  

A tentative reconciliation of existing
theory, data for UBe$_{13}$, and data for PrInAg$_2$ may rest upon 
``banding'' effects which are excluded rigorously in infinite dimensions.
Namely, for the quadrupolar(magnetic) two-channel lattice, inclusion of
realistic inter-orbital hybridizations leads to a $k$-dependent splitting
of the spin(channel) states in momentum space, except at special points
in the Brillouin zone (only the $\Gamma$ point and $X$ point [zone-corner])
for the cubic lattice have degeneracy of the spin(channel) labels).  
As argued later in Sec. 9.3.2, these effects manifest in the $k$-dependence
of the self energy which enters as a $1/\sqrt{d}$ correction.  This 
splitting is of no concern in the conventional single-channel Kondo 
lattice where only one band plays a role in the physics.  However, in 
this case, a new route to Fermi liquid physics is offered: if the self-consistent
band splitting is small compared to the lattice Kondo scale, the system 
will pass close to the non-Fermi liquid fixed point prior to reaching 
a ground state which removes the residual degeneracy of the two-channel 
Kondo screening clouds. However, if the renormalized splitting exceeds 
the two-channel scale, a novel metallic state will be formed in which 
the finite paraquadrupolar susceptibility stabilizes the system against
quadrupolar/collective-Jahn-Teller ordering.  Of course, if the intersite
coupling or strain-quadrupole coupling exceeds the Kondo scale and banding
energy, this will separately cut off the non-Fermi liquid physics.  If 
the Kondo scale exceeds the intersite strain-quadrupole coupling and 
banding energy, we create the situation most favorable for the 
formation of a heavy fermion superconducting state through odd-frequency
staggered pairs. 

Regardless, this is an exciting experimental development and offers a
number of directions.  For example, PrPb$_3$ is also a cubic system with
a $\gth$ ground state, but in this case collective Jahn Teller order
does set in at low temperature (Miksch {\it et al.} [1982]).  However,
only about $R/2\ln2$ entropy is removed per Pr at below the ordering
transition.  Moreover, dilution studies might reveal a pure quadrupolar
Kondo ground state in the impurity regime.  The study of such Pr based 
compounds will be a very interesting
new thrust to watch in the coming years.  

{\bf Summary}  Six heavy fermion alloys show non-Fermi liquid behavior
for which leading
candidates models are the two-channel quadrupolar Kondo effect for the
five U based alloys,
and the two-channel magnetic Kondo effect for the Ce alloy.  This
picture explains most
data but runs up against difficulties in explaining resistivity data
for most of these systems.
This picture naturally explains the {\bf U}biquity of uranium based
alloys, since the quadrupolar
Kondo model is far easier to obtain than the two-channel magnetic model
(c.f. Secs. 2.2,
3.4.4, 5.3).    An intriguing experimental correlation is that five of
these systems become
heavy fermion superconductors when the U or Ce sublattice is fully
occupied.  This covers
the known heavy fermion superconductors excluding UPt$_3$ for which no
Th based
reference compound exists (with  the same crystal structure).  Two of
the heavy fermion
superconductors (UBe$_{13}$ and CeCu$_2$Si$_2$) have superconducting
transitions
which occur relative to normal phases which are not describable in
terms of well defined
Fermi liquid quasiparticle states.  There are promising signs that
UBe$_{13}$ is describable
as a two-channel Quadrupolar Kondo lattice, and CeCu$_2$Si$_2$ as a
two-channel
magnetic Kondo lattice.  On the basis of the above discussion, all of
these systems
warrant further study to clarify the origin of the non-Fermi liquid
behavior and whether its
origin is in impurity or lattice versions of the two-channel Kondo
model.  \\


\section{Related Theoretical Developments} 

\subsection{Related Models} 

\subsubsection{Connection to Coulomb Blockade Physics} 

The basic idea of the mapping of the coulomb blockade to a multichannel
Kondo model hinges on the ability to view the charge variables
as a pseudospin.
The dynamical
charge fluctuations in a small metallic particle can be described by
a pseudospin if the number of electrons is fluctuating primarily between
two integer values $N,N+1$
with all the other charging states neglected.  It
is precisely this situation which  is called the ``coulomb blockade''.
The origin of the restriction to two charge states arises from a proper
setting of applied voltage to the metallic particle and a sufficiently
small capacitance so as to create a large charging energy that effectively
takes  charge states $..,N-1,N+2,...$ off to very large energies.

\begin{figure}
\parindent=2.in
\indent{
\epsfxsize=3.in
\epsffile{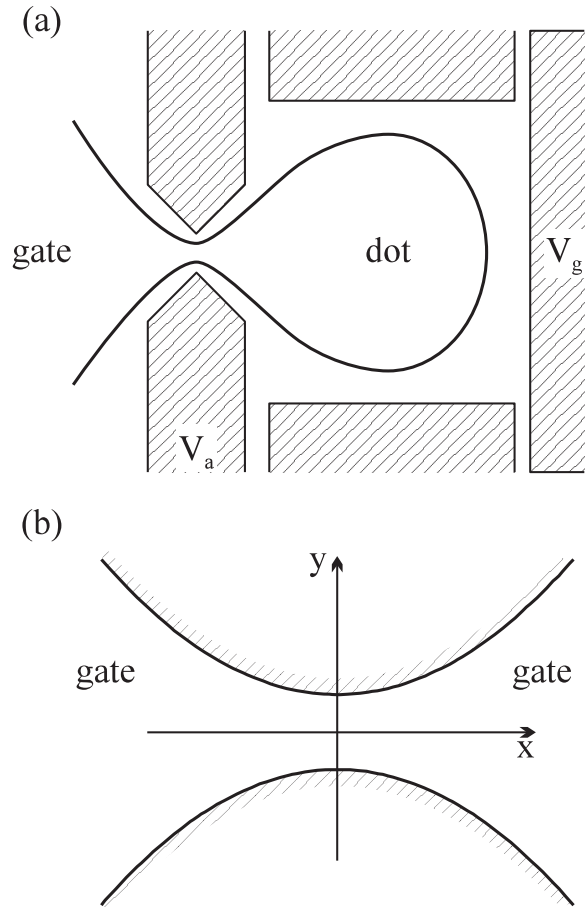}}
\parindent=.5in
\caption{Quantum dot device (for coulomb blockade).  (a) The voltage
applied to the shaded areas localizes the electron states in the area of
the dot and the electrode which are connected by the gate.  The gate voltage
$V_g$ controls the charge in the dot, while the strength of the link between
the electrode and the dot can be adjusted by the auxiliary potential
$V_a$ which controls width and the potential height of the gate.  The
electronic wave functions are appreciably non-zero only within the solid
line. (b) The area of the gate is enlarged, and the wave function in this
region has the form given by Eq. (9.1.2).  }
\label{fig9p1}
\end{figure}

\begin{figure}
\parindent=2.in
\indent{
\epsfxsize=3.in
\epsffile{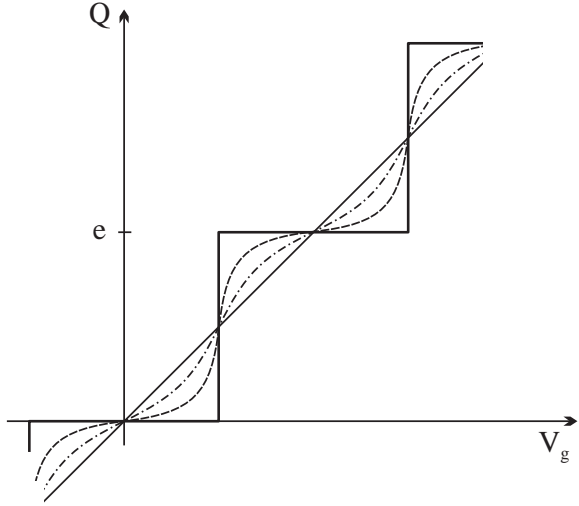}}
\parindent=.5in
\caption{Average quantum dot
charge as a function of bias number $N=CgV_g/e$ (c.f. Eq. (9.1.1)).
The solid line is in the limit of zero transmission coefficient, indicating
that in this limit where no charge leakage arises, the charge on the dot is
quantized.
If $N$ is half-integer, then the charge is not well defined in this limit.
For non-zero transmission amplitude, the number of electrons in the
dot can fluctuate, so that the time average of the charge $<Q>$ deviates
from an integer value (dotted line).  Further increase of the transmission
amplitude can lead to a complete washout of the charging steps (light solid
line).}
\label{fig9p2}
\end{figure}

Such systems can be fabricated using nanotechnology methods.  First,
a two-dimensional electron gas is produced in a semiconductor through
quantum well formation.  Then additional electrodes are added on the top
of the device to produce potential shifts in the quantum well and ``pinch''
the constriction through which electrons enter the blockade region, as
shown in Fig.~\ref{fig9p1} .   The Coulomb energy in the metallic particle or
quantum dot may be expressed in terms of the capacitance $C$, gate
voltage $V_g$, and dot charge $Q$ as
$$E_c = {Q^2 \over 2C} - gV_g Q = {(Q-CgV_g)^2 \over 2C} - {C g^2V_g^2\over 2}
\leqno(9.1.1)$$
where $g$ is a geometric factor. By changing $V_g$ the charge
corresponding to the minimal energy is fixed. This procedure is
unique except when $CgV_g = (N+1/2)e$, where $N$ is an integer.  In this
case the Coulomb energies of the $N,N+1$ dot charge configurations are
degenerate and this gives rise to strong charge fluctuations between
these two states (see Fig.~\ref{fig9p2} ).

The coupling of the quantum dot to the external electrode
is via the link of Fig.~\ref{fig9p1} which is characterized by the transmission
amplitude $T$.  The number of electrons can fluctuate on the dot, and
if the amplitude $T$ is gradually increased, the charging steps of
Fig.~\ref{fig9p2} can be washed out.  This has been seen experimentally
(van der Vaart {\it et al.} [1993]; Pasquier {\it et al.} [1993]).
More sophisticated arrangements have since been constructed by
Molenkamp {\it et al.} [1995], Waugh {\it et al.} [1995,1996], Livermore
{\it et al.} [1996], where two quantum dots connected by a weak are constructed in a similar
manner to Fig.~\ref{fig9p1}.  In this case, the charge fluctuation between the
two symmetrical dots can be studied in a very controllable way.

Now we turn to the connection of this system to the multi-channel Kondo
model as discussed by Matveev [1991,1995] and Matveev, Glazman, and Baranger  [1996a,b]
(see also G. Falci, G. Sch\"{o}n, and G. \zim [1995], where a path integral 
method is applied).
The key idea is again the mapping to the pseudospin as discussed above.
The charge change about the biased value of $(N+1/2)e$
is then analogous to the magnetic polarization of a spin
impurity, and $V_g$ is analogous to a magnetic field.
The tunneling of electrons into and out of the dot which changes the
charge states then plays the role of a transverse coupling in the spin
model, or of the assisted hopping in the TLS Kondo model of Sec. 2.  The
capacitance $C$ plays the role of the magnetic susceptibility.  As in
the TLS and quadrupolar Kondo effects, the real magnetic spin is a spectator
field and gives rise to the channel index.  In principle, by regulating
the auxiliary gate voltage $V_a$ of Fig.~\ref{fig9p1},
the number of channels can increase beyond two by allowing
occupancy of more than one transverse sub-band.  In this one assumes the
wave functions on either side (electrode or dot) are given by
$$\psi(x,y) = \phi_n(y) e^{ikx} \leqno(9.1.2)$$
with energy $E_n(k) = E_n + k^2/2m^*$.  However, in practice the fillings
of different sub-bands will not be identical, meaning that channel
degeneracy between occupied sub-bands will not be satisfied.
Hence, in contrast to Matveev [1991,1995] and Matveev, Glazman, and Baranger  [1996a,b]
one can expect that the condition $C_gV_g = (N+1/2)e$ is satisfied for 
charge fluctuations only in one sub-band, 
so that the only stable fix point must be a two-channel one.
See also Golden and Halperin [1996a,b].  

The main goal of the theory is to calculate the profile of the charging
steps shown in Fig.~\ref{fig9p2} as a function of the transmission coefficient
$T$.  Assuming just a single sub-band,
the effective model Hamiltonian has been formulated in Matveev [1991],
and is given by
$${\cal H} = {\cal H}_0 + {\cal H}_T \leqno(9.1.3)$$
where
$${\cal H}_0 = \sum_{k,\sigma} \epsilon_k c^{\dagger}_{k\sigma}c_{k\sigma}
+  \sum_{k,\sigma} \epsilon_p c^{\dagger}_{p\sigma}c_{p\sigma}
+ {Q^2\over 2C} + \phi Q \leqno(9.1.4)$$
and
$${\cal H}_T = \sum_{kp\sigma}[t_{kp}c^{\dagger}_{k\sigma}c_{p\sigma} + h.c.]
\leqno(9.1.5)$$
with $k$ indexing states on the left of the gate and $p$ states on the
right of the link.  The matrix element $t_{kp}$ describes the weak
transmission rate across the gate in a way similar the classic tunneling
Hamiltonian.  All energies are measures with respect to the Fermi energy.
$\phi$ in Eq. (9.1.4) is the energy shift due to the applied gate voltage.
$Q$ is the charge on the right hand side (in the dot)
measured with respect to the occupied states, {\it i.e.}
$$Q = e\sum_{p\sigma}[c^{\dagger}_{p\sigma}c_{p\sigma} -
\theta(-\epsilon_p)]
\leqno(9.1.6)]$$
such that $<Q(\phi=0)>=0$.

To see the presence of Kondo like logarithms in this problem, it is
sufficient to carry out perturbation theory in $t_{kp}$ to second order.
In that case one obtains a logarithmic singularity in $<Q>$ of the form
$$<Q^{(2)}> \approx N_e(0)N_d(0)t^2
\ln [{|e/2C - \phi|\over |e/2C+\phi|}] \leqno(9.1.7)$$
valid for $-e/2C<\phi<e/2C$, where $t^2$ is a suitably energy averaged
matrix element, and $N_{e,d}(0)$ is the Fermi level density of states for
the electrode ($e$) or dot ($d$).  This result
clearly demonstrates that the perturbation
theory breaks down as the steps of Fig.~\ref{fig9p2} are approached.

Next,
Matveev [1991] looked at the region $\phi = -e/2C+u$, with $u<<e/C$.  In
this case higher energy charged states can be removed through the use of
the projection operators $P_0,P_1$ which restrict to charge $Q=0,1$ (it
is clear that an equivalent procedure can be followed for $N,N+1$, with
$N$ arbitrary).  Then the Hamiltonian becomes
$${\cal H}_{01} = [\sum_{k,\sigma} \epsilon_k c^{\dagger}_{k\sigma}c_{k\sigma}
+  \sum_{k,\sigma} \epsilon_p c^{\dagger}_{p\sigma}c_{p\sigma}][P_0+P_1]
+ euP_1$$
$$~~~~~~~ + \sum_{kp}[t_{kp}c^{\dagger}_{k\sigma}c_{p\sigma}P_1
+ t^*_{kp}c^{\dagger}_{p\sigma}c_{k\sigma}P_0] ~~.\leqno(9.1.8)$$
We distinguish two cases here:\\
{(i) Point Contact} In this case the transverse ($y$) dimension of the
gate is comparable to the Fermi wave length $2\pi/k_F$  so that the
transverse part of the wave function $\phi_n$ is always the same,
e.g. $\phi_0$.  In this case a suitable dimensionless coupling is
$g=N_e(0)N_d(0)|t|^2$, where $t$ is the Fermi level value of $t_{kp}$. \\
{(ii) Wide Junction Limit}.  In this case the dimension is large compared
to the Fermi wavelength and a large number of transverse sub-bands
can contribute (a number of order $M_{trans} \approx Ak_F^2$ where
$A$ is the transverse gate area).
As mentioned above, however, the equivalence in this limit to a multi-
channel Kondo model with $M=2M_{trans}$ is questionable because the different
sub-bands have very different transmission amplitudes and occupancies.

The Hamiltonian of Eq. (9.1.8) actually has the two-channel Kondo form,
assuming just a single relevant sub-band, as we shall now demonstrate.
Replace the indices $k,p$ now by $k,\alpha$ $\alpha =\pm 1/2$ corresponding
to the electrode momenta ($+$) or dot momenta ($-$).  By restricting to
just the $Q=0,1$ subspace the projection operators can be written in
terms of $2\times 2$ matrices, {\it viz.}
$${\cal H}_{01} = \sum_{k\alpha\sigma} \epsilon_k  c^{\dagger}_{k\alpha\sigma}
c_{k\alpha\sigma} \left(\begin{array}{cc}
					 1 & 0\\
					0  &1
			   \end{array} \right)
+ eu \left(\begin{array}{cc}
				 0 & 0\\
				0  &1
	   \end{array} \right)$$
$$+ t\sum_{kk'\sigma} [c^{\dagger}_{k,+,\sigma} c_{k',-,\sigma}
						   \left(\begin{array}{cc}
							 0 & 0\\
							1  &0
						   \end{array} \right)
+c^{\dagger}_{k,-,\sigma} c_{k',+,\sigma}
											   \left(\begin{array}{cc}
													 0 & 1\\
													0  & 0
												   \end{array} \right)]
\leqno(9.1.9)$$
where we have replaced $t_{kp}$ by the constant value near the Fermi
energy.  Eq. (9.1.9)
can be rewritten in terms of spin 1/2 or Pauli matrices in the
conduction pseudospin space (electrode or dot index $\alpha$, denoted
by Pauli matrices $\sigma^i$) and
the ``impurity pseudospin'' (charge $Q$ denoted by spin 1/2 matrices
$S^i$)
to give the form
$${\cal H}_{01} = \sum_{k\alpha\sigma} \epsilon_k  c^{\dagger}_{k\alpha\sigma}
c_{k\alpha\sigma} + eu(1/2 - S^z) + {t\over 2}\sum_{kk'\alpha\alpha'\sigma}
[c^{\dagger}_{k\alpha\sigma}\sigma^+_{\alpha\alpha'} c_{k'\alpha'\sigma}S^-
+ c^{\dagger}_{k\alpha\sigma}\sigma^-_{\alpha\alpha'} c_{k'\alpha'\sigma}S^+]
\leqno(9.1.10)$$
which has the form of a purely transverse two-channel Kondo bare coupling
in a magnetic field $h=eu$ and transverse coupling $J_{\perp} = t$.
The channel index is the real magnetic spin index, as in the quadrupolar
and TLS Kondo models.
It is worth noting an oddity about this Hamiltonian, which is that the
``impurity spin'' enters solely through the Hilbert space restriction
to the $Q=0,1$ subspace.  This means that the ``impurity'' spin is
composed from the same electrons which form the dot portion of the
continuum states.  This rather odd situation is certainly to be distinguished
from TLS, magnetic, or quadrupolar Kondo effects where the continuum states
and impurity pseudospin states form disjoint Hilbert spaces.

Starting with the planar $2M_{trans}$ Kondo model and using either
multiplicative renormalization group methods (Matveev [1991]) or
bosonization methods (Matveev [1995]), for $u\to 0$
one approaches an intermediate coupling
fixed point at temperatures small compared to the Kondo scale
$$k_BT_K = D_0(N(0)J_{\perp})^{M_{trans}} \exp[{-\pi \over 4N(0)J_{\perp}}]
\leqno(9.1.11)$$
where $N(0)=\sqrt{N_e(0)N_d(0)}$,$D=\sqrt{D_e D_d}$, $D_i$ the bandwidth
on the electron or dot side.  Using the results from two-channel Kondo
theory the average charge is
$$<Q>-e/2 \simeq e{ue\over k_BT_K}\ln[{k_BT_K\over eu}] \leqno(9.1.12)$$
and the effective dot capacitance then
$$C_{eff} = {-\partial ^2 F_{ed}\over \partial u^2} \sim
{1\over T_K}\ln[{T_K\over T}] \leqno(9.1.13)$$
for temperatures $T\le T_K$, where $F_{ed}$ is the free energy per electron
of the electrode-dot system.

In the latter paper (Matveev [1995]), the ``tunneling'' formulation is
avoided by the introduction of a narrow conducting gate (see Fig.~\ref{fig9p1}(b), where the 
electronic charge density
is determined by the energy of the electrode and the dot.  In this
case, the electron energy varies continuously going through the gate, and
therefore the bosonization technique can be applied.  In this case the
transmission coefficient can be large, with $g$ of order unity, in which case
the smeared step in Fig.~\ref{fig9p2} is replaced by a straight Ohmic line that is only
slightly modulated near the special integer and half integer charging values.
Considering the theory of the coupled dots studied by Waugh {\it et al.} [1995,1996] and
Livermore {\it et al.} [1996] we refer to the works Golden and Halperin [1996a,b] and
Matveev, Glazman, and Baranger [1996a,b].

\subsubsection{Hopping Models with Several Sites}

The concepts of the TLS Kondo model can be generalized to more than
two sites, either with the sites forming a lattice or a cluster.  In
the first case, a proton or muon hopping between interstitial sites is
a good candidate, while the second may be realized when in a
crystalline solid a larger atom is replaced by a smaller one such that
the  substituent atom sits in a cavity where several equivalent
potential minima are present.  We believe the Pb$_{1-x}$Ge$_x$Te
described in Sec. 8.1 presents one such example.  In amorphous metals,
the formation of three or four almost degenerate sites can be ruled
out by the stress due to the non-uniformity.  In contrast, e.g., in
Pb$_{1-x}$Ge$_x$Te, much less stress is expected because the
substituent atoms provide the disorder (note the recent papers on 
more than two sites by \zar [1996] and by Moustakas and Fisher [1997]).

The Hamiltonian for the lattice system is given by Eqs. (2.1.37,38)
and (2.1.39), where the heavy particle is created by
$h^{\dagger}_{\vec R}$ at lattice position $\vec R$, and the
conduction electrons form a band.  The single heavy particle hops
on a lattice, so that its dispersion in the absence of coupling to
the conduction electrons is described by a tight binding band with
width proportional to the one site hopping rate.  We use the notation
for the momentum space heavy particle operator $h^{\dagger}_{\vec Q}$
defined by
$$h^{\dagger}_{\vec Q} = {1\over \sqrt{N}}\sum_{\vec R} e^{-i\vec
Q\cdot \vec R} h^{\dagger}_{\vec R}~~. \leqno(9.1.14)$$
Electron assisted hopping (see Fig.~\ref{fig2p3}) connects neighbors separated
by displacement $a\vec \delta$ where $a$ is the lattice spacing and
$\vec \delta$ a unit vector assuming the heavy particle hops on a
simple cubic lattice.

With this notation, the interaction Hamiltonian described by Eqs.
(2.1.38) and (2.1.39) has the form (\zow [1987])
$${\cal H}_{int} \sim \sum_{\sigma, \vec Q'+\vec k'= \vec Q'' + \vec k''+\vec K}
 [V + {1\over 2} u \sum_{\vec \delta} \cos ({\vec Q'' + \vec Q'\over
2}\cdot \vec \delta a)]h^{\dagger}_{\vec Q''}h_{\vec Q'}
c^{\dagger}_{\vec k''\sigma}c_{\vec k'\sigma} ~~.\leqno(9.1.15)$$
The non-local nature of the assisted hopping shows up in the second
term of that Hamiltonian as a form factor.  The momentum is conserved
to within a reciprocal lattice vector.  The calculation of the two
second order diagrams for the scattering amplitude shown in Fig~\ref{fig1p2}.
can be performed straightforwardly, and the diagrams to order $u^2$
don't cancel one another because of the presence of the form factor
in Eq. (9.1.15).  Actually, new form factors can be generated.  The 2D
and 3D cases are very complicated because an infinite set of couplings
is produced in that way.  The situation simplifies in 1D because only
one new form factor is generated and the couplings correspond to
forward and backward scattering in the problem of the 1D electron gas.
The renormalization results in logarithmic terms where the low energy
cutoff is due to the spontaneous hopping of the heavy particle on the
lattice.
From that work it is learned that the form factor of the assisted
hopping makes the model non-commutative.  The cluster problem has been
discussed recently by \zar [1996].

In general, we expect these results to be most relevant to the finite
cluster limit because the sites can be close to each other (on the
scale of a lattice spacing) whereas in the lattice the overlap of the
heavy particle wave function from site to site is apt to be very
small.  It is important to note that the low energy cutoff here is due
do the spontaneous hopping, as mentioned above.

We now turn to the related problem of occupation dependent hopping.
These models are given by the Hamiltonian of Eqs. (2.1.40) and
(2.1.41) and by an additional heavy/light particle Coulomb interaction
acting between the light and heavy particle on the same atomic site.
The idea is that in the presence of a light particle, hybridization is
assisted between light and heavy particles.
The heavy particle is assumed to correspond to a weakly dispersed
electronic band in this case as opposed to a muon or proton.
The advantage of these models is that the heavy particle hybridizes with
the conduction electrons so that the heavy particle wave function
overlap does not appear and limit the physical relevance of the model.
The model becomes noncommutative, because the light particle assisted
hybridization has  a form factor $t^{\gamma}(\vec R,\vec \delta)$
which has a simple Fourier transform proportional to
$$t(\vec k_1,\vec k_2;\vec k_3) = te^{i(\vec k_1 +\vec k_2 - \vec
k_3)\cdot \vec \delta a} \leqno(9.1.16)$$
if the light particle occupation with spin $\sigma$ assisted a light
particle of opposite spin to hop to a neighboring heavy site.
In Eq. (9.1.16) the momenta
$\vec k_1$ and $\vec k_2$ refer to the  annihilated light particles
and $\vec k_3$ to the created light particle.

\begin{figure}
\parindent=2.in
\indent{
\epsfxsize=3.in
\epsffile{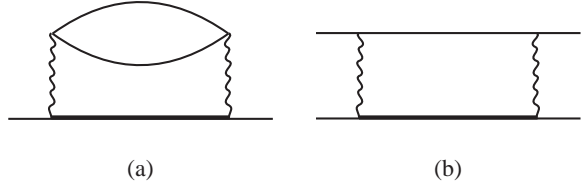}}
\parindent=.5in
\caption{Diagrams for Coulomb assisted hybridization of light
and heavy particles.  At left is the conduction electron self energy,
where the wavy line is the interaction, light lines are conduction
electron propagators, and the heavy line represents the heavy electron
propagator.  The diagram at right represents the induced
electron-electron interaction brought about by this coupling.}
\label{fig9p3}
\end{figure}

It is quite easy to show that the number of newly generated form factors
remains finite on renormalization, and that their form can be
generated from Eq. (9.1.16) by omitting one, two, or all three of the
momenta in that form factor.  (The situation is somewhat different if
the Coulomb interaction takes place on neighboring sites.)  The vertex
equation for this problem can be solved in the leading logarithm
approximation, and results in an enhancement of the bare assisted
hopping.  There is a low energy cutoff in this problem set by the
heavy particle energy $\epsilon_h$ (measured with respect to the fermi
energy). This parameter is crucial, as it can be shown that is not
renormalized to zero upon scaling of the bandwidth.  This therefore
cuts off the logarithmic divergence of the vertex function.  The
attractive feature of the model is that a mass enhancement of the
conduction electrons together with a strong anisotropic
pairing interaction are generated  from this relatively simple source,
namely, angular dependent form factors in the interaction matrix
elements.  Typical diagrams are shown in Fig.~\ref{fig9p3}.

An interesting competition arises between the mass enhancement and the
superconductivity.  The highest $T_c$ for superconductivity from this
mechanism can be obtained when $m^*\simeq 2m_e$.  Much higher mass
enhancements can be obtained, but in that case the superconductivity
is essentially suppressed to zero temperature as the wave fucntion renormalization 
$z<<1$ (Penc and \zow [1994]).
We note that these Coulomb assisted hopping models may be considered
in dilute or lattice limits, and may be of relevance for those heavy
fermion systems in which the $f$-level is close to the Fermi energy
(on the scale of $\epsilon_h/E_F \le 0.1$, say), or for those superconducting 
compounds in which atomic orbitals are near the Fermi energy and are weakly hybridized
with the conduction electrons so that they form a heavy band (\zow [1989b,c]).

Finally, we discuss the related impurity model proposed by
Giamarchi {\it et al.} [1993].  The model Hamiltonian is
$${\cal H} = \sum_{k,l} \epsilon_k c^{\dagger}_{kl}c_{kl} + {t\over
\sqrt{N_s}} (c^{\dagger}_{ko} h + h^{\dagger} c_{ko})
+ {1\over N_s}\sum_{kk'l} V_l(c^{\dagger}_{kl}c_{kl}-1/2)(h^{\dagger}h -
1/2) \leqno(9.1.17)$$
where the operator $c^{\dagger}_{kl}$ creates a spinless conduction
electron in angular momentum state $l$ about the impurity with radial
momentum $k$, and $h^{\dagger}$ creates a ``heavy'' particle at the
impurity site.  The hybridization is assumed to occur only in one of
the angular momentum channels ($l=0$), but the Coulomb screening takes
place between all the channels with the impurity.  Because of the
large number of channels, it turns out the an electron can be confined
to the impurity orbital as $T\to 0$, and this gives rise to non-Fermi
liquid behavior.  Clearly this model is much simpler than the previous
ones, because the interaction term has no form factors and is therefore
commutative.  The attractive feature of the non-Fermi liquid behavior
is, however, limited to the situation when the localized orbital sits
precisely at the Fermi energy.  As soon as one shifts it off by energy
$\epsilon_h$, the logarithms are cut off and Fermi liquid behavior is
recovered.  Unfortunately, just as in the previous model, the energy
$\epsilon_h$ is not renormalized upon scaling (\zow and \zim [1995]).
Physically it is hard to accept that accidentally small values of
$\epsilon_h/E_F$ are likely to occur.

In conclusion, it can be said that the impurity models of this type
are attractive only if the couplings are noncommutative so that they
can be enhanced by the logarithmic terms in perturbation theory.
However, due to the infrared cutoff imposed by $\epsilon_h$, they are
not good candidates to describe non-Fermi liquid systems.


\subsubsection{Application of Two-Channel Models to the Cuprate
Superconductors} 

Emery and Kivelson [1993,1994] have proposed a number of other 
models for realizing the two-channel Kondo effect in the context
of explaining marginal fermi liquid theory in the cuprate 
superconductors. We shall briefly
mention here two of them, both of which hinge critically upon 
the idea of phase separation in the cuprates.  

First, the idea came of a quantized hole running in a bound state 
around the 
boundary of a region of short ranged magnetic order.  The idea
is to think of the orbital moment of the hole as the longitudinal
component of the pseudo-spin.  A counter-clockwise traversal of
the boundary would count as an up-pseudo-spin, and a clockwise 
traversal would count as a down pseudo-spin. A mobile electron 
outside the domain can flip the orbital moment of the hole.  Since
the real spin of the electron is a spectator, this is argued to 
map to a two-channel Kondo problem.  

A second realization is that of a small region of segregated holes.  
If a motion of the region can be triggered by a passing carrier, 
there will be a net flip of an electrical dipole moment.  Because
the electric dipole moment is independent of the magnetic spin 
of the carrier, this again is expected to map to a two-channel 
Kondo problem.  

\subsection{Majorana Fermion Approach to the Two-Channel Model} 

Coleman and Schofield [1995] and Coleman, Ioffe, and Tsvelik [1995] have
developed an alternative approach to the two-channel model using Majorana
fermions to represent the conduction electrons.  The mapping to the
two-channel model rests upon introducing a single channel model in which
the impurity spin couples not only to the conduction electron spin but also
their isospin or axial charge.    This introduction of
Majorana fermions has some of the same flavor as the approach of Emery
and Kivelson [1992] and Sengupta and Georges [1994]; in particular, the
residual entropy, logarithmically divergent specific heat and susceptibility,
and marginal local dynamic susceptibility emerge from the presence of a
decoupled local Majorana degree of freedom.  However, in contrast to the
bosonization route, (i) no exchange anisotropy is introduced, (ii) the
Majorana fields emerge immediately at the initial stages of calculation,
and (iii) the non-trivial
fixed point, rather than being at intermediate coupling strength,
is shoved off to strong coupling. This has an advantage in that while
the intermediate coupling fixed point is not accessible to any finite order
perturbation theory schemes, an explicit perturbation theory may be set up
about the strong coupling fixed point in powers of the hopping over the
exchange integral.

The non-trivial physics then emerges from
the undispersed local Majorana fermion, which may be viewed as a bound state
of three other Majorana fermions.  This method has been used to
lend support to the idea that the channel anisotropic two-channel model
is in fact a non-Fermi liquid in contrast to conformal field theory and
numerical renormalization group results but in agreement with Bethe-Ansatz
calculations.  The particular representation of the model in terms of
spin and isospin does not generalize to the two-channel Kondo lattice;
however, a lattice extension does exist and we shall discuss that later
in this subsection.

The primary approach is outlined in the paper by Coleman, Ioffe, and
Tsvelik [1995] and we shall follow their arguments here.  The first
step is to write down the following Hamiltonian for a single channel
of conduction states, restricted to a one-dimensional chain assumed
to represent the radial quantization discussed for the NRG and CFT
approaches:
$${\cal H} = it\sum_{n,\sigma} [c^{\dagger}_{\sigma}(n+1) c_{\sigma}(n)
-h.c.] + {\cal J}\vec S_I\cdot [\vec \sigma_c(0) + \vec \tau_c(0)]
\leqno(9.2.1) $$
where $\sigma_c(0)$ is the conduction spin density at the origin
where the impurity sits, and $\tau_c(0)$ is the isospin density.
Because the low energy spin and isospin degrees of freedom decouple,
correlation functions of the two-channel model involving conduction
spins from each channel are completely equivalent to correlation
functions of the model specified by Eq. (9.2.1) where one of the
channel spins is replaced by spin, the other by isospin.  This
equivalence may be established formally, and is done so in an
appendix to Coleman, Ioffe, and Tsvelik [1995].  The authors call
this a ``compactified'' Hamiltonian in that uncoupled degrees of
freedom are removed; this jargon does not refer to the application
of compactification of dimensions used in string theory.

An unusual aspect of Eq. (9.2.1) is that the analogue to overcompensation
of the two-channel model is not possible at any finite coupling strength
because in view of the Pauli principle it is impossible to have both
maximal spin and isospin at a given site.  As a result, the nontrivial
coupling fixed point is pushed off to infinite coupling strength.
Practically, this corresponds to a change of the cutoff procedure; because
only the first two perturbative terms in the beta function are universal
in form, higher order terms can shift the fixed point coupling around.
The particular choice here shifts it to infinity.

The next step is to replace the conduction electrons by Majorana
fermion variables.  An explicit representation in terms of ``scalar''
and ``vector'' Majorana variables is
$$ (c^{\dagger}_{\uparrow}(n) c^{\dagger}_{\downarrow}(n))
= (0 i ) {1\over \sqrt{2}} (\Psi^{(0)}(n) - i \vec\psi(n)\cdot \vec\sigma)
\leqno(9.2.2)$$
where $\vec\sigma$ are Pauli matrices and the $\Psi^i ~(i=0,1,2,3)$ obey
the Majorana anticommutation relations
$$\{\Psi^{(i)},\Psi^{(j)}\} = 2\delta_{ij}~~. \leqno(9.2.3)$$
In terms of these Majorana variables, the spin plus isospin combination
of Eq. (9.2.2) may be written as
$$\vec \sigma_c(n) + \vec \tau_c(n) = -i \vec \Psi(n)\times\vec\Psi(n) ~~.
\leqno(9.2.5)$$
The $i=0$ Majorana component has the interpretation of a `charge' degree
of freedom, and the $i=1,2,3$ components make up a `spin' vector.

Since the coupling flows towards infinite strength, the fixed point
Hamiltonian at zero temperature is simply
$${\bf H}_{\infty} = -i{\cal J}\vec S_{imp}\times (\vec \Psi(0)\times
\vec \Psi(0)) \leqno(9.2.6)$$
which has a two fold degenerate ground state corresponding to either
a net spin singlet or isospin singlet (in this model the impurity has
the rather bizarre feature of either playing the role of spin or isospin).
The degeneracy is implied by the presence of the Majorana fermion operators
$\Psi^{(0)}$ and
$$\Phi = -2i\Psi^{(1)}(0)\Psi^{(2)}(0)\Psi^{(3)}(0)~~, \leqno(9.2.7)$$
both of which commute with the fixed point Hamiltonian.
Application of the complex fermion combination
$\zeta = (\Psi^{(0)}(0)-i\Phi)/\sqrt{2}$
will allow one to flip back and forth between the spin and isospin
singlet states.   The excited states of this Hamiltonian are spin and
isospin triplet states.

Adding the hopping back in to the nearest neighbor site on the chain,
one may admix singlet and triplet states of the strong coupling Hamiltonian.
If one eliminates the virtual triplet fluctuations through a canonical
transformation that projects to the singlet levels, the resulting effective
Hamiltonian is, for the `spin'  sector
$${\cal H}^* = it\sum_{n=1}^{\infty} \vec\Psi(n+1)\cdot \vec\Psi(n)
+ \alpha \Phi \Psi^{(1)}(1)\Psi^{(2)}(1)\Psi^{(3)}(1)
\leqno(9.2.8)$$
where $\alpha = 3t^3/4{\cal J}^2$, and the now strongly coupled state
at the origin is explicitly excluded.
The form of the interaction term in this  Hamiltonian is very similar
to that of Eq. (6.2.20) introduced by Sengupta and Georges [1994] within
the Emery and Kivelson [1992] bosonization approach. The $\Phi$ fermion
of Eq. (9.2.8) plays the same role as the $\hat a$ fermion of Sengupta
and Georges [1994], and as a result contributes a residual `half' degree
of freedom to the entropy.
 Indeed, perturbation
theory in $\alpha$ about the fixed point may be carried out, and
the diagrammatics are precisely analogous to the work of Sengupta and
Georges [1994], so that $\chi(T),C/T$ are second order in $\Phi$ and
diverge logarithmically with the temperature, and have a Landau-Wilson
ratio  of 8/3.
Moreover, (i) the mixed
susceptibility in which one line is a $\Phi$ propagator and the other
a $\Psi$ propagator is marginal in form (Varma {\it et al.} [1989]),
and the calculation is
completely analogous to that of Eq. (6.2.38);
(ii) the self energy of the
$\Psi$ fields is `marginal' (Varma {\it et al.} [1989])
in that $\Sigma_{\Psi}(\omega,T) \sim
-\omega\ln(\omega/\omega_c) + i\max{\omega,T}$
and the calculation is precisely analogous to that of the `spin' fermion
self energy of Emery and Kivelson [1992] (see Eq. (6.2.42)); (iii) the
self energy of the $\Phi$ fermion is regular, with an imaginary part
that vanishes as $T^2$.  It is this quasiparticle-like sharpness to the
$\Phi$ state which supports the marginal behavior at low temperatures.

Coleman, Ioffe, and Tsvelik [1995] then extend the model in two different
ways.  First, they notice that the original Hamiltonian enjoys an $O(3)$
symmetry.  A well regulated large $N$ expansion may be obtained when
this is extended to $O(N)$.  For even $N$, the $\Psi$ self energies are
analytic, while for odd $N$ they are non-analytic.  However, no divergent
thermodynamic properties arise apart from the physical case $N=3$.

Second, they extend the model specified by Eq. (9.2.8) to the lattice,
and note that the $\Phi$ fermions of each site will acquire a dispersion
in the lattice which will lift the residual entropy and induce a crossover
from the non-Fermi liquid state; the estimated crossover scale is
$\alpha^2/t$ which is of the order of the induced $\phi$ hopping through
the $\Psi$ fermions.  This lattice generalization is {\it not} the
same as the two-channel Kondo lattice model, but nonetheless offers a
potential theoretical playground for understanding non-Fermi liquid behavior
in a lattice model.

Coleman and Schofield [1995] have also considered the situation in
which the spin and isospin channels of Eq. (9.2.1) are not identical,
which then simulates the channel spin anisotropy of the original two-channel
Kondo model.  They find within this formalism that channel anisotropy 
may not drive the physics to a Fermi liquid fixed point. 
This possibility was anticipated by Andrei and Jerez [1995], who 
suggested that marginal operators could allow a flow to a line of 
Non-fermi liquid fixed points in the overcompensated model with 
exchange anisotropy, but called for further studies of asymptotic
correlation functions to support or refute 
this conjecture.   This consideration does not at all affect the 
thermodynamics presented by Andrei and Jerez [1995].  

The parent $O(3)$ symmetric Anderson model which maps to this 
compactified Kondo model in the limit of large Coulomb repulsion
has recently been studied with the numerical renormalization 
group by Bulla and Hewson [1997].  They find that the calculated
properties are indeed in agreement with those of the two-channel Kondo
model, but rather than a non-Fermi liquid fixed point in the presence 
of spin/isospin symmetry breaking (analogous to channel anisotropy for
the two-channel model) the crossover is to a Fermi liquid fixed point.  
These NRG results are independent of the Coulomb repulsion $U$ 
in the $O(3)$ Anderson model, indicating that the correspondence 
to the non-fermi liquid fixed point has greater validity than
anticipated by Coleman and Schofield [1995].  This has been further
confirmed by weak and strong coupling perturbation studies by Bulla, 
Hewson, and Zhang [1997].  

Finally, Schofield [1997] has shown that the Emery-Kivelson bosonization
(Emery and Kivelson [1992]) can be easily extended to yield a
description in terms of the compactified or $\sigma-\tau$ model, 
further cementing the equivalence of this formulation to the two-channel
model.

\subsection{Steps Toward the Lattice Problem} 

In this last subsection, we will briefly overview the steps made
to extend the theory of impurity models to the lattice.  These
steps consist of studies of two impurity single- and two-channel
Kondo models, which we discuss in  Sec. 9.3.1, and approaching the
problem from the $d=\infty$ limit ($d$ the spatial
dimensionality) which we discuss in Sec. 9.3.2. 
 
\subsubsection{Two Impurity Model} 

The two-impurity Kondo model has proven to be a source of non-Fermi 
liquid physics in both the one- and two-channel cases.  It is of 
interest to review the one-channel model first, both to set the 
tone of the discussion and to note a strong similarity between a 
non-trivial fixed point of that model with the two-channel one-impurity
model, first stressed by Gan [1995b].  Following that, we shall 
overview the more complex physics of the two-impurity two-channel 
Kondo fixed point.  

For both the one- and two-channel models, a competition results between the
Kondo effect and the intersite impurity coupling (RKKY interaction). 
For antiferromagnetic RKKY interaction, non-trivial non-Fermi-liquid
fixed points develop for particle-hole symmetric limits of the models.
For the one-channel model, a single (unstable) 
non-trivial fixed point emerges, 
while for the two-channel model an entire sheet of non-trivial fixed 
points emerge with continuously tuneable exponents.  However, in the  
case of the single-channel model, the non-trivial fixed point is 
removed with particle-hole symmetry breaking associated with asymmetry
of the density of states about the Fermi energy or the addition of 
potential scattering.  It is not yet clear
whether the manifold of fixed points is removed in the two-channel 
model, but the bias based upon the one-channel results is that it is
removed.  

{\it (1) One-Channel Two Impurity Model}\\
The Hamiltonian for this model is simply 
$${\cal H} = \sum_{\vec k\sigma} \epsilon_k 
c^{\dagger}_{\vec k\sigma}c_{\vec k,\sigma} + 
J\sum_{j=1,2}\vec S_I(\vec R_j)\cdot S_c(\vec R_j) \leqno(9.3.1)$$
where $j$ indexes the site of the two impurities.  
Although at second order in $J$, RKKY interactions between the impurities
will be generated, a ``bare'' interaction term $-I_0\vec S_I(\vec R_1)\cdot
\vec S_I(\vec R_2)$ is often added to Eq. (9.3.1) to allow for 
greater tuneability of the model parameters (with this convention, $I<0$ is antiferromagnetic).  
This model possesses a 
non-trivial fixed point for suitably defined particle hole symmetry
at antiferromagnetic RKKY coupling, as was first identified by 
Jones and Varma [1987,1989] and subsequently characterized by 
Jones, Varma, and Wilkins [1988].  

It is convenient to project the conduction electrons into local 
channels which are of even and odd parity about the midpoint of the 
line between the two impurities. It is only these states which  
couple to the impurity.  The projection allows for a reduction to 
an effective one-dimensional problem analogous to that of the 
two-channel model as outlined in Sec. 6.1.  
Assuming a symmetry conduction band
of width $2D$, the projected local annihilation operators are (Silva 
{\it et al.} [1996])
$$c_{0\sigma\pm} = A_{\pm} \int_{-D}^{D}d\epsilon
{1\over N_s} \sum_{\vec k}\delta(\epsilon-\epsilon_k)[e^{i\vec k\cdot\vec R_1} 
\pm e^{i\vec k\cdot\vec R_2}]c_{\vec k\sigma} \leqno(9.3.2)$$
$$~~~~~~=A_{\pm} \int_{-D}^{D}c_{\epsilon\sigma\pm} 
\sqrt{1\pm {\sin kR_{12}\over kR_{12}}}d\epsilon $$
where $A_{\pm}$ is a normalization constant, $R_{12}=|\vec R_1-\vec R_2|$ 
and +(-) indicates even(odd)
parity about the inversion center.  Note the correspondence to the operators for the TLS
defined in Eq. (A.2.5a,b) of App. II (also, the $\pm$ corresponds to the $e(o)$ 
labels of Moustakas and Fisher [1995,1996]). Defining the square-root factor in 
Eq. (9.3.2) by $N_{\pm}(E)/A_{\pm}$, it is possible to write the exchange interaction
term of Eq. (9.3.1) as (Affleck, Ludwig, and Jones [1995])
$${\cal H}_{int} = {J\over 4}
[\vec S_I(\vec R_1)+\vec S_I(\vec R_2)]
\cdot(\sum_{p=\pm,\mu\nu}\vec \sigma_{\mu\nu}) \int_{-D}^D d\epsilon\int_{-D}^D 
d\epsilon' N_p(\epsilon)N_p(\epsilon')
 c^{\dagger}_{\epsilon\mu p} 
c_{\epsilon'\nu p} \leqno(9.3.3)$$
$$~~~~+ {J\over 4}[\vec S_I(\vec R_1)-\vec S_I(\vec R_2)]\cdot(\sum_{p=\pm,\mu\nu}\vec \sigma_{\mu\nu})    
\int_{-D}^D d\epsilon\int_{-D}^D 
d\epsilon' N_p(\epsilon)N_{-p}(\epsilon')
c^{\dagger}_{\epsilon\mu p} 
c_{\epsilon'\nu -p} ~~.$$

As stressed by Affleck, Ludwig, and Jones [1995] and more recently, in 
the context of the TLS Kondo effect by Zawadowski {\it et al.} [1997], 
there are two kinds of particle-hole symmetry relevant to such a two-site
quantum impurity problem.  First (Type I), we require that 
$$N_p(E)=N_p(-E)~~and~~c_{\epsilon\mu p} \to (-1)^{1/2-\mu} 
c^{\dagger}_{-\epsilon\mu p} \leqno(9.3.4)$$
which preserves the parity index.  This simply says that the local
parity projected densities of states are invariant under inversion about 
the Fermi energy.  Another kind of particle-hole transformation
(Type II) corresponds to the mapping of electron minima to hole maxima
for a nearest neighbor tight binding model (under which $\vec k \to \vec k + \vec Q/2$, 
where $Q=\pi(111)/a$ in three-dimensions).  In this case, parity labels
get interchanged and the symmetry is specified by 
$$N_p(-E)=N_{-p}(E)~~and~~c_{\epsilon\mu p} \to (-1)^{1/2-\sigma}
c^{\dagger}_{-\epsilon\mu -p} ~~.\leqno(9.3.5)$$
It turns out that by considering, for example, a two-impurity Kondo 
model on a nearest neighbor lattice in one dimension, that if the
two sites differ by an odd number of lattice spacings, the model 
is invariant under type II symmetry, while for an even difference in
the number of sites, the model is invariant under type I symmetry
(Fye and Hirsch [1989]; Fye [1994]; Affleck, Ludwig, and Jones [1995]). 
In this particular model, the induced RKKY couplings will be antiferromagnetic
for odd separation and ferromagnetic for even separation.  
Affleck, Ludwig, and Jones [1995] further 
note that potential scattering will break 
Type I particle-hole symmetry but not type II particle-hole symmetry. 
In this sense, Type I particle-hole symmetry places a stronger constraint
on the model.  

\begin{figure}
\parindent=2.in
\indent{
\epsfxsize=6.in
\epsffile{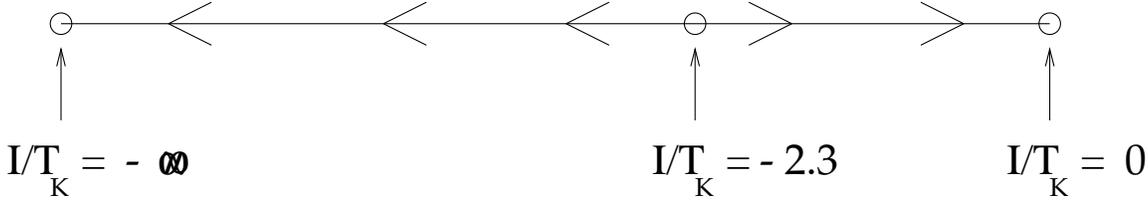}}
\parindent=.5in
\caption{Phase diagram of the two-impurity Kondo model with full particle-hole
symmetry.  For $I\to -\infty$, the system goes to the local singlet
fixed point, which is a Fermi liquid, in which the local moments just
lock out in a singlet.  For $I=0$, the local moments each are Kondo
compensated (this is actually extended to positive $I<\infty$).  
Most striking is the non-trivial unstable fixed point for 
antiferromagnetic RKKY 
interaction strength $I/T_K \approx -2.3$. 
After Jones and Varma [1987,1989].}
\label{fig9p4}
\end{figure}

In the presence of particle-hole symmetry of Type I, 
Millis, Kotliar, and Jones [1990] have given the following argument
to explain the existence of a non-trivial fixed point:  for zero 
total RKKY interaction strength (bare+induced), 
the ground state will be characterized
by the independent impurity fixed point (two isolated and screened 
Kondo impurities).  This is a stable fixed point (there are only irrelevant
operators about it), and the phase shift in each parity channel is $\pi/2$. 
On the other hand, for infinite antiferromagnetic RKKY coupling, the 
local moments are quenched into a singlet with no dynamics remaining, and
the phase shift for scattering off of the extended singlet is zero in 
each channel.  This fixed point is also stable for Type I
particle-hole symmetry (no relevant or marginal operators about the 
fixed point).  They also showed that these two possible phase shift
values, zero or $\pi/2$, are the only allowablve values in the presence
of Type I particle-hole symmetry.  
In consequence, as a parameter such as the intersite coupling strength 
tunes one between the independent impurity fixed point and the 
antiferromagnetic singlet fixed point while Type I symmetry is enforced, 
there must be a point arising in which the phase shift jumps
discontinuously between $0$ and $\pi/2$. This implies a critical point
separating the stability regimes of the two different fixed points along
the axis of the intersite coupling (normalized to the one impurity 
Kondo scale).  The renormalization group flow diagram for this
model is shown in Fig.~\ref{fig9p4} in terms of the total intersite coupling $I=I(J)+I_0$
where $I(J)$ is the induced RKKY coupling strength (measured in units 
of the Kondo scale).  (Note that for infinite 
ferromagnetic $I$ the odd channel drops from the problem, and the model 
maps to the single channel spin 1 Kondo impurity model, while in general 
for ferromagnetic $I<\infty$, a two-stage Kondo quenching occurs of the 
net impurity spin (first
even and then odd, or vice-versa), as
first envisioned by Jayaprakash, Krishna-murthy, and Wilkins [1980].  
The universality class of the successive quenching model is the same
as the two isolated impurities fixed point.)

The original NRG work of 
Jones and Varma [1987,1989] and Jones, Varma, and Wilkins [1988] together
with the CFT calculations of Affleck and Ludwig [1992] and Affleck, Ludwig,
and Jones [1995] confirms that the critical point is characterized by 
a second order transition, at which the staggered susceptibility, 
specific heat coefficient, and local pair field susceptibility in the
spin singlet sector diverge logarithmically.  Moreover, the residual   
entropy is $(R/2)\ln 2$ and the Fermi level scattering rate is 
at half the unitarity limit.  These results have a strong similarity
to the results for the two-channel one-impurity model.  It should be
noted that the original NRG work was based upon an ``energy-independent''
coupling constants approximation.  Namely, the coupling strengths in 
Eq. (9.3.3) were set to their Fermi level values and the energy dependence
ignored. This manifestly introduces full particle-hole symmetry and thus
to a critical point of some sort (as explained in the preceding
paragraph), although the non-trivial fixed point was an unexpected
result (instead a simple level crossing could have resulted).  

Gan [1995b] has proposed that this critical point is very similar to the
two-channel one-impurity non-trivial fixed point.  The idea is that the 
critical point is identified by a doubly degenerate ground state in each
case. For the two-channel fixed point, the ground state has a double 
degeneracy in spin.  For the two-impurity one-channel fixed point, the 
double degeneracy corresponds to a level crossing in a finite size 
spectrum.  The levels roughly are described by the isolated Kondo singlets
on the one hand, and the extended singlet on the other hand.  The level
degeneracy is lifted by the deviation of the RKKY coupling from the 
critical value, and in this sense the RKKY coupling plays exactly the same
role as a spin field in the two-channel Kondo model, so that for $I-I_c\ne 0$, 
the specific heat coefficient and staggered susceptibility diverge as
$(I-I_c)^{-2}$, similar to the $1/H_{spin}^2$ divergence for the two-channel
one impurity model.   

To understand why the mapping breaks down, it is helpful to mention a few
details of the CFT approach to the model as explained in detail in 
Affleck and Ludwig [1992], and Affleck, Ludwig, and Jones [1995].  
They first consider a free fermion theory.  For each parity channel of 
the free theory, when full particle-hole symmetry 
is maintained, there is a global conservation of $SU(2)$ axial charge or 
isospin (as defined in Secs. 4 and 6.1) and the $SU(2)$ spin symmetry. 
Hence, the total free field 
symmetry group may be taken to be $SU(2)_{iso,+}\otimes
SU(2)_{sp,+}\otimes SU(2)_{iso,-}\otimes SU(2)_{sp,-}$.  The conformal
charge of this theory provides a dimensionless measure of the numbers of
degrees of freedom present in the Hamiltonian (it is technically defined
in terms of the commutation relations of the Fourier transforms of the 
real space Hamiltonian density, which obey the so-called Virasoro algebra
when conformal invariance holds--a complete discussion is beyond the 
scope of this paper--we refer the reader to Affleck, Ludwig, and Jones 
and references therein).  Each of the $SU(2)$ spin and isospin currents
obey the level $k=1$ Kac-Moody algebra (c.f. the discussion of Sec. 1), and
each has a conformal charge of $c=1$, so that the total conformal charge
of the effective one-dimensional model is $c_{tot}=4$.  

Now, the presence of the impurities breaks the 
$SU(2)_{sp,+}\otimes SU(2)_{sp,-}$ spin symmetry into a global $SU(2)$ 
spin symmetry.  The impurities do nothing to the isospin symmetry, provided
we maintain full particle-hole symmetry.  
If the Kac-Moody commutation relations are computed for
these currents one finds that the level is $k=2$.  For a general 
$k$-level KM algebra, the conformal charge of the resulting Sugawara
Hamiltonian density quadratic in the KM currents is $c=3k/(k+2)$.  Thus
for example, for $k=1$ we recover $c=1$ as claimed above, while for $k=2$
we obtain $c=3/2$.  This means that
the sum of conformal charges for isospin and global spin currents is 
$c'=7/2$.  However, the representation in terms of the global spin 
current can be done for the free Hamiltonian which implies that we are 
not counting all the degrees of freedom (or else we would obtain 
$c'=c_{tot}=4$).  

The missing conformal charge is $c=1/2$. There is a unique, unitary conformal
theory with conformal charge $c=1/2$, and it is the Ising model.  Hence,
quite surprisingly, the remaining degrees of freedom obtained after 
coupling to the impurities and preserving maximal isospin and spin 
symmetry are that of the Ising model!  As an example, 
in terms of the primary $i=1/2$
field $h_p$ of the $k=1$ isospin algebra, the $j=1/2$ field $g_{\mu}$ 
of the $SU(2)$
$k=2$ spin algebra, and the order parameter field $\sigma$ of the 
Ising model, the one-dimensional fermionic operators are given by 
$$\psi_{\mu p}(x) \sim (h_p)_1 g_{\mu} \sigma \leqno(9.3.6)$$
where the $1$ on the $h_p$ field denotes the first component of the
spinor field and the $i$ is the species label.  The creation operator
would pick out the second component of the isospin spinor doublet.  
This operator has a scaling dimension of $1/2$ since $\Delta_h=1/4$,
$\Delta_g=3/16$, and $\Delta_{\sigma}=1/16$ add up to 1/2.  

Affleck and Ludwig [1992] first noticed that the non-trivial fixed point
of this model admits a very large $SO(7)$ symmetry.  Gan [1995a], proposed
that this $SO(7)$ symmetry can be understood from writing a Majorana 
fermion representation of the full model. The full free theory has an 
$SO(8)$ symmetry in this representation as discussed in Sec. 6.3 (Maldocena
and Ludwig [1996]).  In Gan's [1995a] approach, a single Majorana 
fermion is decoupled by the impurity leaving behind a global $SO(7)$ symmetry
to the remaining free Majorana fields.  It is apparent from the consideration
of these paragraphs, that a full mapping of the two-channel one-impurity
model to that of the two-impurity one-channel non-trivial fixed point 
(at particle-hole symmetry) cannot be established--the symmetries of the
problems are simply too different.  

It is unlikely that this non-trivial fixed point will be generically 
relevant to the understanding of experimental data for heavy fermion 
or TLS materials, although it is possible that for some systems a
crossover region may exist regulated by the properties of this 
fixed point.  As shown in detailed studies by Jones [1991], the
addition of simple potential scattering induces a line of Fermi-liquid
fixed points between the two isolated impurity fixed point and the 
antiferromagnetic fixed point, with continuously varying phase shifts.  
In extensive Quantum Monte Carlo studies, Fye [1994] found no evidence
for the non-trivial fixed point in nearest neighbor tight binding models
both in one- and three-dimensions.  In retrospect, this is related to 
the breaking of Type I particle-hole symmetry in his calculations.  
Finally, Silva {\it et al.} [1996] have performed an extensive series of 
NRG calculations in which no approximation is made for the energy 
dependence of the couplings.   They do not find the non-trivial fixed
point when they consider only the two impurity Kondo model (with $I_0=0$), 
but they do find that they can get close to the non-trivial fixed point
by shrinking the bandwidth parameter $\Delta=v_Fk_F$; this has the effect of 
dynamically restoring particle hole symmetry.  Further, they find that 
they can scale all of their data for specific heat coefficient and 
uniform susceptiblity onto common curves as a function of the bandwidth
parameter and fixed impurity separation.  The specific heat curve is 
sharply peaked as a function of RKKY interaction strength (tuned by $J$ and
$\Delta$) which indicates that at least for sufficiently small asymmetry, 
it is in principle possible to observe a specific heat peak due to the
proximity to the non-trivial fixed point.  \\

\begin{figure}
\parindent=2.in
\indent{
\epsfxsize=6.in
\epsffile{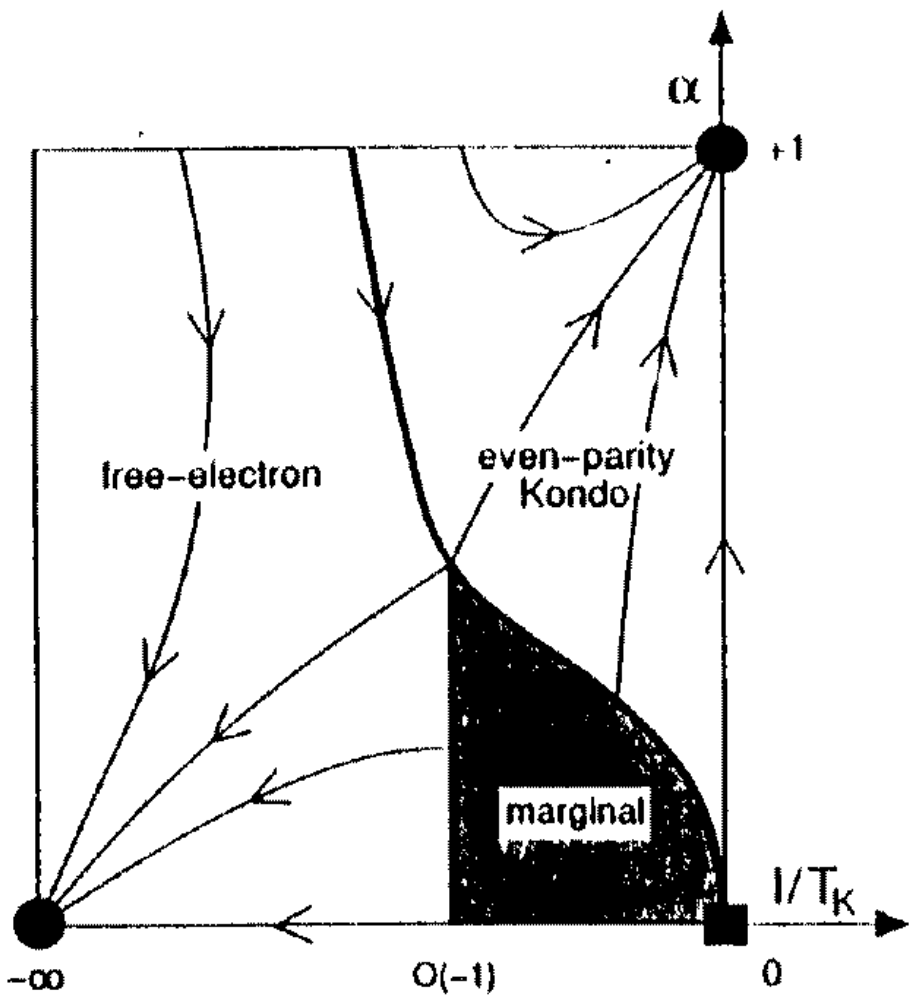}}
\parindent=.5in
\caption{Phase diagram of the two-channel, two-impurity Kondo model. Relevant parameters are
the RKKY coupling strength (measured in units of the Kondo scale) and the even-channel/odd-channel
exchange asymmetry (vertical axis).  Most striking is the marginal sheet for negative RKKY 
coupling.  From Ingersent and Jones [1994].}
\label{fig9p5}
\end{figure}

{\it (2) Two Impurity Two-Channel Model}.\\
To define the two-channel two-impurity model, we simply augment the 
conduction bands of Eqs. (9.3.1,9.3.2,9.3.3) by channel indices.  
The phase diagram for ground states of this model, obtained from NRG 
calculations under the assumption of energy independent coupling constants, 
is shown in Fig.~\ref{fig9p5}., taken from 
Ingersent, Jones, and Wilkins [1992], and Ingersent and Jones [1994] (see also Jones and Ingersent [1994]).  
The most salient features are arguably: \\
(i) an unstable non-Fermi liquid fixed point at the origin, corresponding to the
isolated impurities--we should not be surprised that the isolated impurity
fixed point is unstable, as the Kondo spin clouds around each two-channel
site must eventually feel each other for $T\to 0$ since the length scale
of the two-channel clouds is divergent (this is not the case for the one
channel model); \\
(ii) stable Fermi liquid fixed points for sufficiently 
large difference between even and odd exchange coupling strengths;\\
(iii) a complete manifold
of non-Fermi liquid fixed points for antiferromagnetic intersite spin 
coupling, with a marginally stable line marking the leftmost boundary of 
this manifold.

We shall now discuss points (i-iii) in order.  

The instability of the isolated two-channel one-impurity fixed point is often 
simplistically pointed to as evidence of the irrelevance of the single 
impurity model to any realistic description of heavy fermion or TLS materials. 
The difficulty with this simplistic argument is that it ignores the 
quantitative aspect of crossover:  namely, while the fixed point is unstable,
it may still regulate the physics over a large parameter range for sufficiently
small values of the intersite coupling strength and even-odd coupling difference. 
Indeed, as discussed in Sec. 8.2, the apparent single ion physics present for
Th$_{1-x}$U$_x$M$_2$Si$_2$ (M=Ru,Pd) suggests that over a wide temperature 
range the two-channel one impurity 
fixed point governs the low energy scale excitations until some interaction
physics enters (in this case, well below 1K for the concentrations studied). 
At issue then is the need for 
{\it quantitative} estimate of the crossover temperatures 
and exponents for this model. 

In addition to the intersite interaction strength, the two-impurity two-channel
model has an additional relevant parameter, which is the asymmetry between 
even and odd coupling strengths, with $J_p(0,0) = JN_p(0),~p=\pm$.  
The understanding of this is as follows, argued for the case of 
zero intersite spin interaction: if $J_e$ differs from $J_o$, 
then there are sufficient conduction degrees of freedom to fully
screen the spin for either channel, so whichever couples most strongly
will simply drive the system to the stable isolated impurity fixed point
of the single channel Kondo model, a Fermi liquid fixed point.  
We note that this crossover is related to the ``banding'' of the 
electrons, in that we cannot retain two-fold degenerate bands at all
points of the Brillouin zone for the full two-channel lattice.  Here we
are seeing that fact reflected by the generic difference of even and
odd coupling constants in the two-point Brillouin zone of the two-impurity
model.  As noted in Sec. 8.2, this banding effect may be relevant for 
understanding the unusual behavior of the quadrupolar Kondo lattice candidate
PrInAg$_2$ (Yatskar {\it et al.} [1996]).  

A fortunate point about the crossover to Fermi liquid behavior for 
non-zero $J_e-J_o$ is that the crossover exponent is extremely small.  
Specifically, the crossover temperature identified from the NRG (Ingersent
and Jones, [1994b]) is found from $|J_e-J_o|\sim T^{\Delta_{eo}}$
which gives a scale $T_{eo} \sim |J_e-J_o|^{1/\Delta_{eo}}$.  Numerically,
it is found $\Delta_{eo}\approx 0.1$!  This implies that unless $|J_e-J_o|/J_e
\approx 1$, the crossover is {\it extremely} slow.  

Finally, we turn to the manifold of fixed points.  Georges and Sengupta [1995]
have developed a complete conformal field theory (in concert with Abelian 
bosonization) for the particle-hole symmetric
model, which properly displays continuously tuneable scaling dimensions to
the primary field operators as the intersite coupling strength is 
tuned, indicating that the intersite interaction is a marginal parameter 
in the theory.  This theory is considerably more complex than that of the
one-channel two-impurity model, and we shall not 
go into details here.  Suffice to say, as for the two-impurity one-channel
model, it is a matter of concern whether the non-trivial 
fixed point manifold is robust to the lifting of particle-hole symmetry.  \\

\subsubsection{$d=\infty$ Limit} 

Given the very interesting data on UBe$_{13}$ and
CeCu$_2$Si$_2$, it is worth studying the properties of the
two-channel Kondo lattice.  One regime where this can be carried
out rigorously is in the limit of infinite spatial dimensions
where the lattice problem becomes a self-consistent impurity
problem.

{\it Key Ideas of the $d\to \infty$ limit}\\
The procedure for going to the infinite dimension limit is by
now well known and has been discussed extensively in the
original works of Metzner and Vollhardt [1989] and
M\"uller-Hartmann [1989] as well as two recent review articles
by Pruschke {\it et al.} [1995] and Georges {\it et al.} [1996].
There are two key ideas, which we illustrate for simplicity on a
hypercubic lattice assuming a nearest neighbor tight-binding
model for the conduction electrons.

First, the energy dispersion relation in $d$ dimensions is
$$\epsilon_{\vec k} = -2t\sum_{i=1}^d \cos k_ia \leqno(9.3.7)$$
where $t$ is the tight binding matrix element and $a$ is the
lattice constant.  This can be viewed as a sum of random
variables, each distributed on the interval $[-2t,2t]$.
Accordingly, the density of states must take a Gaussian form as
$d\to \infty$ by the central limit theorem.  The width of the
Gaussian is $\sqrt{2d} t=t^*$.  To have a sensible density of
states in the limit, we obviously should hold $t^*$ fixed, which
means $t\sim 1/\sqrt{d}$.

\begin{figure}
\parindent=2.in
\indent{
\epsfxsize=6.in
\epsffile{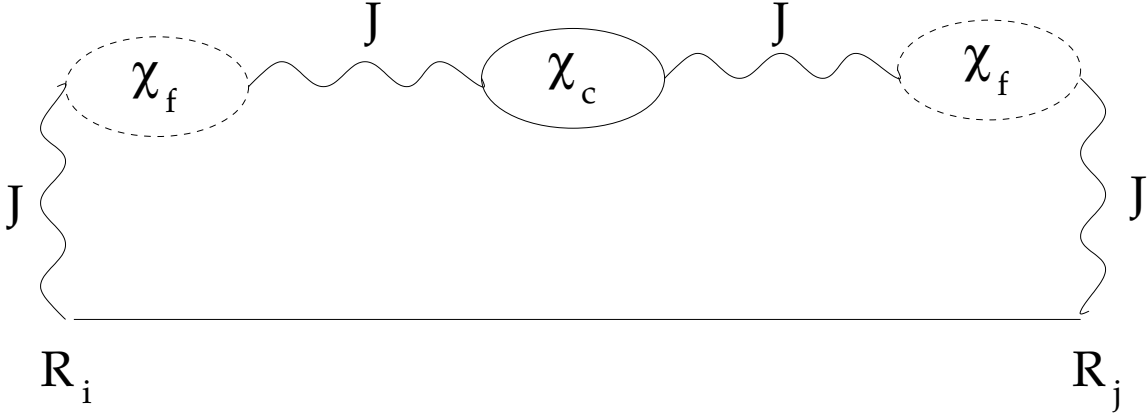}}
\parindent=.5in
\caption{Self energy diagrams for the two-channel Kondo lattice for $d=\infty$. 
A dashed bubble represents a local moment susceptibility, a solid bubble represents a 
conduction electron susceptibility; wavy lines represent the exchange interaction.  
For site index $i=j$, the diagram is of order $1/d^0$, while for $i\ne j$, the 
diagram is at least of order $1/d^{3/2}$ which holds for nearest neighbor sites. }
\label{fig9p6}
\end{figure}

This scaling immediately implies that the self energy becomes purely
local in this limit, as we illustrate in Fig.~\ref{fig9p6} for the two-channel
Kondo lattice model.  This illustrates a contribution to the conduction
electron energy in real space.  If the sites $i$ and $j$ are identical,
the diagram plus all higher order ones will be non-zero.  However,
suppose $i\ne j$.  Then the real space conduction propagator
$G(i,j,\omega)\sim t^{||i-j||}$, where $||i-j||$ measures the minimal
number of nearest neighbor hops required to connect sites $i,j$.
Because three intersite propagators appear in the diagram and $t\sim
1/\sqrt{d}$, then clearly this self energy contribution scales as
$d^{-3||i-j||/2}$ as $d\to\infty$. Note that for when $i,j$ are nearest
neighbors we get the largest contribution, and when Fourier transformed
this will give a momentum space contribution scaling as $1/\sqrt{d}$.

Because the self-energy becomes purely local, the problem is reduced to
an effective impurity problem.  Specifically, one picks a single site
(say $i$)
which is called the impurity.  One makes an initial guess for the
``medium'' propagator $\tilde G_c(i,i,\omega)$, i.e., the conduction propagator
for a lattice with this site plucked out.  One solves
then for the self energy of the conduction electrons by an appropriate
impurity method (e.g., quantum Monte Carlo or the NCA), and constructs an
estimate for the local propagator through
$$G_c(i,i,\omega) = {1\over N_s}\sum_{\vec k} {1\over \omega -
\epsilon_k + \mu - \Sigma(\omega)} \leqno(9.3.8)$$
where $N_s$ is the number of sites, and $\mu$ is the chemical potential. This equation
crucially illustrates the self consistency required to solve the
lattice--the same self energy which enters the momentum space propagator
enters the local propagator. The new
estimate for the medium propagator is given by
$$\tilde G_c(i,i,\omega) = (G_c(i,i,\omega)^{-1} +
\Sigma(\omega))^{-1}\leqno(9.3.9)$$
which is then used for the next iteration of the impurity problem.
Iterations continue until self consistency of Eq. (9.3.8) is reached, or
until the medium propagator, say, does not change from iteration to
iteration.  \\

{\it Examples:  Hubbard and Anderson Lattice Models}\\
The mapping to an effective impurity model affords considerable
qualitative insight into a number of problems. For example, the Hubbard
model then becomes a self-consistent Anderson impurity model.  This
implies that the density of states should possess satellite peaks
separated by the Coulomb interaction $U$ and
corresponding to transitions of the singly occupied state to empty
and doubly occupied orbitals.  In addition, in any metallic phase,
a quasiparticle ``Kondo resonance'' will appear in the vicinity of the
Fermi energy.  The energy scale of this quasiparticle resonance is
self-consistently determined and has no simple analytic form (Pruschke
{\it et al.} [1993,1995], Georges {\it et al.} [1996]).  For the
single-channel
Anderson lattice model, at particle-hole symmetry a band insulator forms
with an indirect gap determined by the ``coherently enhanced Kondo
scale''(Jarrell [1995]).  For sufficiently small hybridization (and 
therefore effective on-site exchange coupling) the band insulator 
may give way to an antiferromagnetic insulator, with the combination of
intersite RKKY and superexchange coupling driving the formation of the
antiferromagnetic state.  Finally, the energy scale may
in turn be coherently suppressed in the metallic phase
(Tahvildar-Zadeh {\it et al.} [1996]).  \\

\begin{figure}
\parindent=2.in
\indent{
\epsfxsize=4.5in
\epsffile{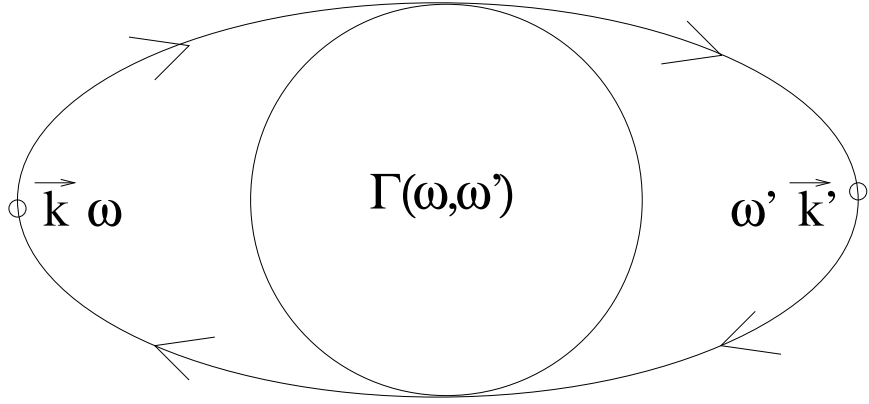}}
\parindent=.5in
\caption{Conductivity vertex correction in $d=\infty$.  The interaction vertex 
is momentum independent, and as a result, only the bubble diagram (with fully 
dressed electron propagators) need be retained.}
\label{fig9p7}
\end{figure}

{\it Two-particle Properties}\\
Because of the local character of the problem, great simplification also
results for two-particle properties (Pruschke {\it et al.} [1995],
Georges {\it et al.} [1996]).  Specifically, the irreducible
interaction functions for particle-hole and particle-particle
propagators become momentum independent.  This makes evaluation of the
magnetic susceptibility and $s-wave$ pairing susceptibilities
particularly straightforward.   Any quantity which involves off-site
vertices, such as the conductivity (current vertex) or a $d-wave$
pairing susceptibility will be formally of order $1/d$.  In the case of
the conductivity, the diagram can still be evaluated, but the local
character of the interaction implies that no vertex
corrections will arise to leading order in $1/d$.  
To see this, consider the lowest order vertex
correction, illustrated in Fig.~\ref{fig9p7}. Momenta which arise on each side of the diagram will be
independently summed, and the interaction function entering the vertex correction 
is independent of momentum, 
so that factors of $\sum_{\vec k}\vec k$ will be
present in the conductivity.  This will vanish.  Hence, only the bubble
diagram need be retained in a calculation of the conductivity.  

{\it Application to the Two-Channel Kondo Lattice Model}\\
The two-channel Kondo lattice model is amenable to such a treatment in
infinite dimensions.  In the simplest form of this model we simply
have two degenerate bands of electrons which have identical coupling to
the local moments $\vec S_I(\vec R)$
located on every site of the lattice $\vec R$, {\it
viz.}
$${\cal H} = \sum_{\vec k\sigma\alpha} \epsilon_{\vec k}
c^{\dagger}_{\vec k\sigma\alpha}c_{\vec k\sigma\alpha} + {\cal
J}\sum_{\vec R}\vec S_I(\vec R)\cdot \sum_{\vec k\vec k'\mu\nu\alpha}
e^{i(\vec k-\vec k')\cdot \vec R}\vec S_c|_{\mu\nu}c^{\dagger}_{\vec
k\mu\alpha}c_{\vec k'\nu\alpha} \leqno(9.3.10)$$
where $\vec S_c $ are spin 1/2 matrices in the conduction
space.  This model is unrealistic in the sense that it is
impossible for two spin 1/2 bands to be degenerate throughout
the Brillouin zone; however, we will see that this lack of
degeneracy is irrelevant in the infinite dimension limit.

In the two-channel Kondo model considerable attention has been paid to
the conductivity, motivated in large measure by the data for UBe$_{13}$
discussed in Sec. 8.2.  An initial effort with a Lorentzian density of
states for which self-consistency of Eq. (9.3.2) is trivial showed that
the paramagnetic state of this model should be an incoherent metal.
Namely,
the resistivity would be finite and the density of states finite if the
paramagnetic phase is extrapolated to $T=0$ (Cox [1996]).
Concommitant with this residual resistivity is a residual
resistivity which may be shown to be $R/2\ln2$ per site at half
filling of the conduction bands.  It was further argued that application
of a ``spin field'' (a field which couples linearly to the local moment
spin operators) or a ``channel field'' (which linearly couples to the 
conduction electron channel spin) would restore a phase shift description
at zero temperature and hence induce a crossover to Fermi liquid 
behavior.  Because the Lorentzian density of states is automatically 
self consistent, these crossovers would lead to the same arguments in 
scaling functions for physical properties that occur in the impurity 
limit.  Specifically, the crossover temperature is proportional to the
square of either applied spin or channel fields.  Non-trivial 
self consistency in infinite dimensions (due to, e.g., a starting Gaussian 
density of states) will not modify the fixed point structure of the
effective impurity problem but will generically lead to self-consistent
modification of the {\it next-leading} critical behavior.  Thus, for 
example, we would not expect the zero temperature amplitude of the $t$-
matrix to be modified in the particle-hole symmetric case, but we may well
expect the $T,H$ dependent corrections to be modified.  This appears 
to be the case as we argue below for the magnetoresistance.  

This unusual behavior may be
traced back directly to the behavior of the single particle
$t$-matrix at the Fermi energy, which is purely imaginary and
half the unitarity limit.  As discussed in Sec. 8.3, this
calculation indicates that: (i) the
application of a field which couples linearly to the
spin degrees of freedom or a field which couples linearly to the
channel degrees of freedom will induce a crossover to the Fermi
liquid state, (ii) the residual scattering may be understood as
corresponding to ``spin-disorder'' scattering off the degenerate
two-channel Kondo clouds.  In the absence of a phase transition
which either orders the moments or induces superconductivity,
there must be a residual resistivity analogous to that of Gd
metal above its Curie point (Cox and Jarrell [1996]).

Since the Lorentzian density of states has unphysically long
tails in energy and hence infinite moments in powers of the
frequency, this result understandably met some skepticism in the
community.  Accordingly, a study with a more physical density of
states was required to convincingly prove the result.

\begin{figure}
\parindent=2.in
\indent{
\epsfxsize=5.in
\epsffile{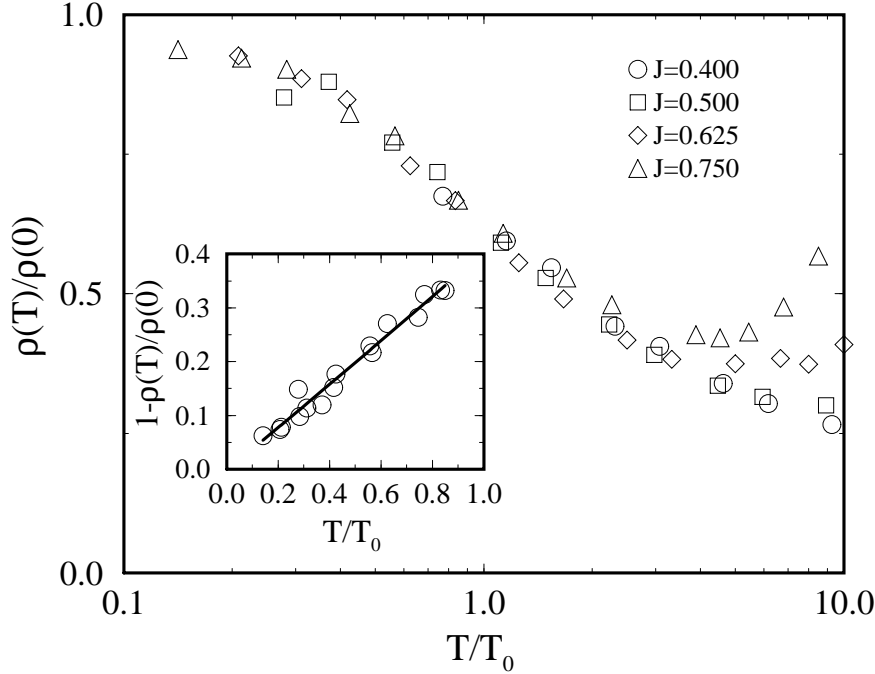}}
\parindent=.5in
\caption{Resistivity of the two-channel Kondo lattice in infinite
spatial dimensions at half filling.  The four different curves are 
four different $J$ values as indicated.  The best fit to the
lower temperature data is with a linear in $T$ behavior, as indicated
in the inset.  From Jarrell, Pang, Cox, and Luk [1996].}
\label{fig9p8}
\end{figure}

Quantum Monte Carlo (QMC) studies with a nearest neighbor tight
binding model (with a Gaussian DOS)
were first carried out by Jarrell {\it et al.} [1996a],
and these studies reached the same conclusion as for the
Lorentzian density of states: the resistivity at the Fermi
energy is finite, along with the density of states, which
develops a cusp at $T\to 0$. Resistivity calculations from this
work (at half-filling)
are displayed in Fig.~\ref{fig9p8} (note the scaling behavior for
curves computed with different exchange coupling values ${\cal
J}$; note also that $J$ of Jarrell {\it et al.} [1996a]
corresponds to ${\cal J}$ here). The QMC code was based upon
the Hirsch-Fye algorithm (Hirsch and Fye [1986]) as modified
by Fye [1986] for the Kondo model.  Calculation of real
frequency properties was carried out with the maximum entropy
method for analytically continuing imaginary time data
to real frequencies.  It should be noted that the
low temperature resistivity appears linear in $T$ to an
excellent approximation, which is in reasonable agreement with
experiments on a number of materials.   In the temperature range
covered by the QMC, the resistivity appears to monotonically
increase with decreasing temperature.  There are reasons to
believe that this may shift, at least at lower temperatures,
away from half-filling.  We discuss this point further below.

\begin{figure}
\parindent=2.in
\indent{
\epsfxsize=5.in
\epsffile{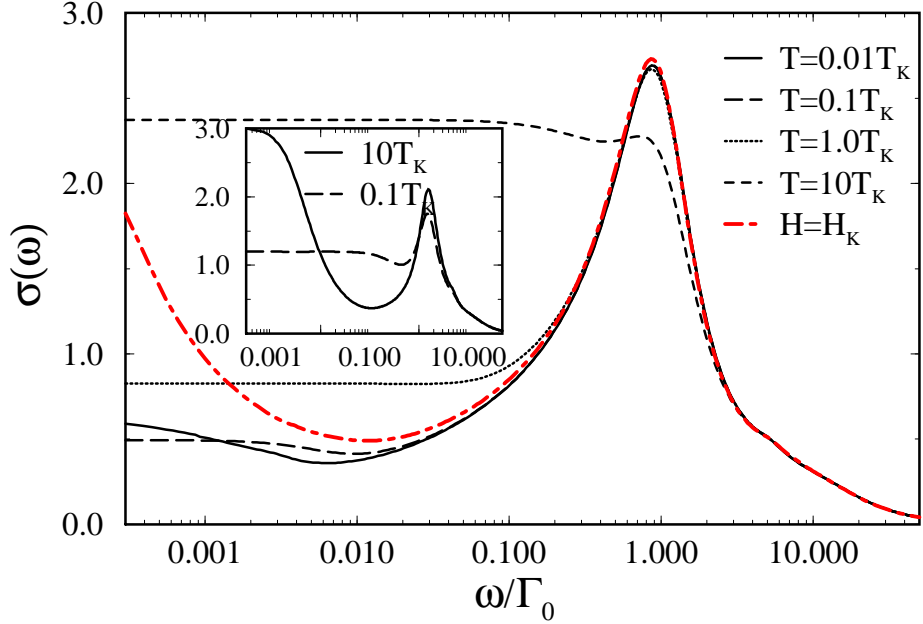}}
\parindent=.5in
\caption{Conductivity of the two-channel Anderson lattice in infinite
dimensions in zero and applied magnetic field.  The model parameters
assure a mapping of the effective impurity model to a Kondo model. 
Application of a magnetic field restores a Drude like peak. From 
Anders, Jarrell, and Cox [1997].}
\label{fig9p9}
\end{figure}

Corollary results include:\\
(i) A residual entropy (at half
filling) of $(R/2)\ln2$ per site. This implies
that the same ``spin-disorder'' scattering interpretation
of the preceding paragraphs holds in this more physical
case as well. \\
(ii) An absent Drude peak in
the optical conductivity at low temperatures
(there is a ``charge fluctuation'' peak at energies of the order of
${\cal J}$) (Fig.~\ref{fig9p9}).  These results for zero field 
appear to be in 
excellent agreement with the optical conductivity data for UBe$_{13}$
(Degeorgi [1997]).\\
(iii) The occupancy function $n(\epsilon_{\vec k},T)$ saturates
to a temperature independent {\it non-step function}
form at low $T$, with a width fixed
by the scattering rate of the conduction electrons.\\
(iv) a finite imaginary part to the conduction
electron self energy which immediately implies (through
Kramers'-Kronig analysis) that the real part has a {\it positive}
slope in the vicinity of the Fermi energy.
This in turn implies that the ``mass enhancement'' $1-\partial
Re\Sigma/\partial\omega|_0$ can be less than one and possibly
negative.  Taken together with (iii) above, we see that
Fermi liquid theory has broken down very severely for this
lattice system. (Note that this breakdown implies the usual
Fermi liquid relation between specific heat and mass enhancement
also breaks down, so heavy fermion behavior is not excluded.)

In addition to these interesting features, the lattice
calculations reveal a ``coherent enhancement'' of the Kondo scale
over the impurity limit.  Roughly, it is found
numerically that
$T_0^{latt} \approx E_F(T_0^{imp}/E_F)^{1/\sqrt{\pi}}$. The
QMC was unable to reach temperatures below $\simeq
0.1T_0^{latt}$ due to a combination of the familiar fermion sign
problem and the $L^3$ scaling of running time
with the number of time slices $L$. \\

{\it Magnetoresistance of the Two-Channel Kondo lattice}\\
Anders {\it et al.} [1996] have studied the magnetoresistance of
the two-channel Kondo lattice in infinite dimensions.  This is a
difficult problem to solve with the QMC approach,
particularly at low temperatures of order $\mu_B H/k_B$
where the magnetic field $H$ induces
pronounced effects.  Accordingly, Anders {\it et al.} [1996]
opted for the Non-Crossing Approximation (NCA) to solve the
effective impurity model.
As argued in Sec. 5, the NCA provides a
reliable method for calculating properties of the
overcompensated multichannel impurity models, in particular
giving exactly the right critical exponents for all
$SU(M)\otimes SU(N)$ models, and providing amplitudes correct to
within 7\% for the resistivity and residual entropy of the two-channel
spin 1/2 model ($N=M=2$).  The NCA is not well suited to a Fermi
liquid regime, but does correctly describe the crossover region
of the multichannel model in applied spin and channel fields, as
discussed in Sec. 5.

The NCA approach actually solves the two-channel Anderson
lattice model, in which a ground configuration ``spin''
doublet hybridizes through a quartet of conduction states (possessing
spin and channel indices) with an excited configuration
``channel'' doublet.  This corresponds to the lattice
generalization of the 3-7-8 model of Eq. (2.2.20).  The
restriction to two configurations assures that particle-hole
symmetry is broken, equivalent to moving away from half-filling
in the pure Kondo lattice case.

\begin{figure}
\parindent=2.in
\indent{
\epsfxsize=5.in
\epsffile{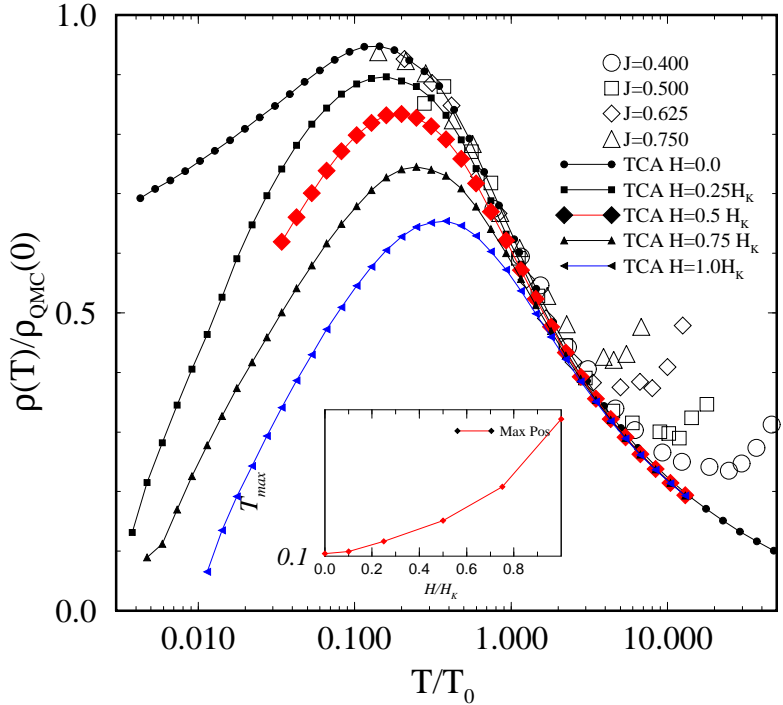}}
\parindent=.5in
\caption{Magnetoresistance of the two-channel Anderson lattice in 
infinite dimensions in the Kondo regime.  Application of a field reduces
the residual resistivity driving the system towards Fermi liquid
behavior.  The points are from the quantum Monte Carlo data of 
Jarrell {\it et al.} [1996] and Fig.~\ref{fig9p8}.  From Anders,
Jarrell, and Cox [1997].}
\label{fig9p10}
\end{figure}

The NCA results for the magnetoresistance are shown in Fig.~\ref{fig9p9}
First, focussing on the $H=0$ curve, it is clear that
it agrees well with the QMC results in the region above
$\simeq 0.1T_0$.  Below this an interesting dip in the
resistivity arises.  This is attributed by Anders {\it et al.}
[1996] to the breaking of particle-hole symmetry by the NCA
together with the self-consistent development of a downward
cusp.  The resistivity still extrapolates to a finite value as
$T\to 0$.

In applied field $H$, the resistivity drops dramatically, and in
fact, the extrapolated resistivity is negative.  This is taken
as a sign that a crossover to a Fermi liquid is being induced,
since the $T^2$ curvature of the Fermi liquid resistivity would
lead to a negative intercept for a linearly extrapolation from
the crossover regime.  The magnetoresistance obeys an
approximate scaling form, $[\rho(T,H)-\rho(T,0)]/\rho(T,0)
\approx f[H/(T+.006T_0)^{.39}]$, where $f(x)$ is a monotonic
decreasing function of $x$. Concommitant with this resistivity drop
is the inducement of a Drude like feature in the optical
conductivity near zero frequency and well below the charge
transfer peak. Hence, as argued for the Lorentzian case, a return to 
Fermi liquid behavior in applied spin field is strongly suggested.  
However, the approach to this Fermi liquid fixed point is significantly
different in {\it quantitative} detail from the analysis of the Lorentzian
case, due to the effects of lattice self consistency in infinite spatial
dimensions.


\subsection{Pairing Effects} 

In this subsection, we examine the possibility of pairing generated
by the local moment fluctuations of the two-channel Kondo model.  The
idea is intuitively clear as noted long ago by Cox [1990] (see also 
Cox and Jarrell [1996]).  Namely, if one examines the basic renormalization
group picture of Fig.~\ref{fig4p8}, it is clear that at each length scale, the
effective impurity at the core has a pair of electrons in a spin triplet,
channel spin singlet state.  In a first study using conformal theory, 
Ludwig and Affleck [1991] were unable to find evidence for such a triplet
pairing correlation and rather found an unusual singularity in the 
channel for spin singlet, channel-spin singlet, {\it odd-in-radial-parity}. 
Because the {\it odd-in-radial-parity} does not readily translate into 
a local pair correlation in a realistic model, it remained unclear whether
local pairing correlations could plausibly arise in the two-channel Kondo 
impurity model.  

The simplistic picture of triplet pairing is not totally wrong, 
as first clarified by Emery and Kivelson [1992] using their
Abelian bosonization method. The resolution  
lies in the understanding that 
the triplet pair is in {\it correlation} with the local moment spin at
the core of the RG picture.  This unusual pair field/spin correlation is 
then a net spin singlet (i.e., it transforms as a scalar under the $SU(2)$
rotations of the total spin of the system).  As such, it can mix with 
other spin-singlet, channel-spin singlet operators which also possess
charge $~~\pm 2$.  The complete spectrum of 
such pair operators with singular susceptiblities was worked out in detail by  
Ludwig and Affleck [1994] using conformal field theory.  The primary viable candidate for such an operator is 
an {\it odd-frequency} pairing state, which, put another way, has a node
in the relative time coordinate of the electron pair.   This possibility
of odd-frequency pairing has independently received considerable attention
as a candidate for novel superconductivity in real materials (Balatzky
and Abrahams [1992]; Balatzky {\it et al.} [1994]; Coleman, Miranda, and
Tsvelik [1993,1994]; Heid [1996];  
Heid {\it et al.} [1995,1996]; Jarrell, Pang, and Cox [1997]). 

Given a mechanism for generating strong local pairing correlations for a 
single impurity, one is tempted to view this result as a kind of Cooper
problem in real space for the lattice.  Namely, assuming the correct 
interpretation of the strong local pair correlations to be the formation
of a single pair about the impurity at $T=0$, it is tempting to speculate
that a lattice of two-channel Kondo sites may produce a superconductor 
provided the pairs are able to move coherently from site-to-site.  
Evidence for just such a superconductor has been found in the extreme 
limits of one-dimension (Zachar, Kivelson, and Emery [1996]) and infinite
dimensions (Jarrell, Pang, and Cox [1997]; Cox and Jarrell [1996]). 

Among the unusual properties of such an odd-frequency superconductor are
that:\\
(i) the anomalous Green's function has no equal time expectation value
and thus the gap function itself is {\it not} a suitable order parameter; \\
(ii) zone center (zero center-of-mass momentum) pairing 
is apparently at best {\it metastable}
(Coleman, Miranda, and Tsvelik [1993,1994]; Heid {\it et al.} [1995,1996]; Heid [1996])
though this point remains controversial (Balatzky {\it et al} [1994]); \\
(iii) zone boundary (finite center-of-mass momentum) pairing is apparently
stable (Coleman, Miranda, and Tsvelik [1993,1994]; Heid {\it et al.} [1995,1996];
Jarrell, Pang, and Cox [1997]); \\
(iv) the excitations of this system may be spectacularly unusual (one
treatment predicts a gapless branch of Majorana or ``real fermions'' in
the ordered state which has zero charge and spin at the Fermi energy but
contributes to the specific heat--see Coleman, Miranda, and Tsvelik 
[1993,1994]); \\
(v) the superconductivity is intrinsically intertwined with the local
non-Fermi liquid behavior in the lattice. 

The plan of this subsection is to first review the impurity results 
for a diverging local pair field susceptibility (Sec. 9.4.1). Then we'll  
review a Pairing symmetry analysis useful for constructing phenomenological
theories in the heavy fermion materials. This analysis not only gives
the local symmetry of the pair wave functions but with the assumption 
of ``negative pair hopping'' consistent with the staggered pairing of 
item (iii) in the preceding paragraph identifies which point(s) in the 
Brillouin zone are likely for the center-of-mass momentum.  In Sec.9.4.3
we will overview the status of understanding on odd-frequency pairing 
independent of the two-channel Kondo model.  Finally, in section
9.4.4 we will describe the microscopic evidence for pairing in the two-channel
Kondo lattice from studies in one- and infinite-dimensions.  \\

\subsubsection{Overview of Odd-in-frequency Pairing} 

In this subsection we shall overview the salient features of odd-in-frequency
pairing which are relevant to the subsequent discussion for the two-channel
Kondo lattice.  In order, we shall discuss: (1) Pauli Principle arguments
allowing for odd-in-frequency pairs; (2) phenomenology for a non-singular
pairing interaction; (3) equivalence of odd-in-frequency
pairs to appropriately defined even-frequency and composite pair fields; 
(4) the likely instability of odd-in-frequency pairs
with zero center-of-mass momentum; 
(5) the possible
connection of composite pairs to Majorana fermions and 3-body 
fermionic bound states; (6) Evidence for odd-in-frequency pairing in 
other models.  

{\it (1) Pauli Principle Arguments allowing Odd-in-frequency Pairs}.\\  
The odd-in-frequency idea was first introduced by Berezinskii [1974]
and later elucidated by Balatsky and Abrahams [1992].  Berezinskii was
interested in unusual pairing possibilities for superfluid $^3$He.  
The basic idea is simple.  If we for the moment suppress channel degrees
of freedom, a pair field operator at position $\vec r$ and imaginary 
time $\tau$ will, prior to any explicit symmetrization, have the form
$$P_{\mu\nu}(\vec r,\tau) = \psi_{\mu}(\vec r,\tau)\psi_{\nu}(0,0) \leqno(9.4.1)$$
where $\mu,\nu$ are spin indices. The conventional Pauli principle
analysis suggests that we can identify allowed pairing symmetries
by considering spatial parity $P_R$ and ``spin parity'' $P_S$ which is
the sign of the pair wave function under exchange of the spin labels.  
This clearly gives the standard results that for $P_R=+$, corresponding
to relative pair angular momentum $\ell=0,2,4...$ for the pair, we must have
$P_S=-$, or a spin singlet, 
and for $P_R=-$ corresponding to pair angular momentum $\ell=1,3,5...$
we must have $P_S=+$ or a spin triplet pair. 

Berezinskii [1974] noted that there is no reason to focus exclusively 
on spatial parity in considering the Pauli principle.  Rather, we can
also augment the conventional discussion with the notion of temporal 
parity $P_T$ (sign under the 
exchange of the imaginary or Euclidean time argument).  Note that
application of temporal parity in Euclidean time 
is {\it not} the same as time reversal, a point of easy confusion.  
This notion is particularly apt for the two-channel Kondo impurity model
given the conformal invariance at the fixed point:  the real radial 
coordinate and the Euclidean temporal coordinate are equivalent, so 
there should be no practical difference between spatial and temporal 
parity in this sense. Odd temporal parity immediately translates to  
odd-in-(Matsubara)frequency pairs in exactly the same way that odd
spatial parity translates to pair wave functions in momentum space
which are odd under $\vec k\to -\vec k$.  

With the new possibility of temporal parity, two new symmetry classes
are opened up for pairing of electrons with spin 1/2.  The $P_T$=+ case
was already covered.  For $P_T=-$, we may have spin singlet pairs ($P_S=-$) 
with {\it odd} angular momentum $\ell=1,3,5....$ ($P_R=-$), or 
spin triplet pairs ($P_S=+1$) with {\it even} angular momentum $\ell=0,2,4...$. 

The addition of the channel degree of freedom in the two-channel Kondo 
model generalizes this further.  We can then discuss the possibility 
of oddness under channel exchange or $P_{CH}$, the ``channel parity.''
The list of possible pair fields is 
as follows:\\

\noindent{$P_R=+$,$P_T$=+,$P_S=+$,$P_{CH}=-$ ($\ell$ even,spin
triplet,channel singlet);}\\
$P_R=+$,$P_T$=+,$P_S=-$,$P_{CH}=+$ ($\ell$ even,spin singlet,channel triplet);\\ 
$P_R=-$,$P_T$=+,$P_S=+$,$P_{CH}=+$ ($\ell$ odd,spin triplet,channel triplet);\\
$P_R=-$,$P_T$=+,$P_S=-$,$P_{CH}=-$ ($\ell$ odd,spin singlet,channel singlet);\\
$P_R=+$,$P_T$=-,$P_S=-$,$P_{CH}=-$ ($\ell$ even,spin singlet,channel singlet);\\
$P_R=+$,$P_T$=-,$P_S=+$,$P_{CH}=+$ ($\ell$ even,spin triplet,channel triplet);\\
$P_R=-$,$P_T$=-,$P_S=-$,$P_{CH}=+$ ($\ell$ odd,spin singlet,channel triplet);\\
$P_R=-$,$P_T$=-,$P_S=+$,$P_{CH}=-$ ($\ell$ odd,spin triplet,channel singlet).\\

The odd-in-frequency pair field  has an unusual property relative to 
an even-in-frequency pair field: it can have no equal time expectation 
value, and thus cannot by itself serve as an order parameter field. 
To see this, consider the case with no channel degrees of freedom 
which is a spin singlet, and thus we have the following anomalous 
Green's function in $\vec k,\omega$ space (repeated indices are summed):\\
$$F(\vec k,\omega) = i\sigma^{(2)}_{\mu\nu}\int d^3r\int_0^{\beta}d\tau 
e^{i\vec k\cdot\vec r - \omega\tau} <T_{\tau}\psi_{\mu}(\vec r,\tau)
\psi_{\nu}(0,0)> ~~.\leqno(9.4.1)$$
The Pauli matrix in front antisymmetrizes the pair amplitude. The 
numerator of this anomalous Green's function is, of course, just the 
``gap function''$\Delta(\vec k,\omega)$, specifically 
$$F(\vec k,\omega) =- {\Delta(\vec k,\omega) \over \omega^2 + \epsilon_k^2  
+ \Delta(\vec k,\omega)\Delta^*(\vec k,-\omega)} 
= {\Delta(\vec k,\omega) \over \omega^2 + \epsilon_k^2 
- |\Delta(\vec k,\omega)|^2} \leqno(9.4.2)$$
where we used the odd-in-frequency property to simplify the denominator and
neglected the normal self energy contribution for simplicity.
For an even in frequency superconductor, the squared gap term in the denominator
enters with a positive sign.  Since the denominator is even in frequency,
we can see that the numerator must be odd in frequency given the assumed
odd-in-frequency behavior of $F$.  Now, if we invert the Fourier transform
to find $\Delta(\vec k,0)$ we see that
$$\Delta(\vec k,0) = {1\over \beta} \sum_{omega} e^{i\omega\tau} 
\Delta(\vec k,\omega)|_{\tau=0} = 0\leqno(9.4.3)$$
which indicates that the gap function itself cannot serve as a suitable 
order parameter and rather some moment of the gap function with frequency
is an appropriate order parameter.  

For example, consider $d\Delta(\vec k,\tau)/d\tau$.  The time derivative
operation is explicitly odd-in-temporal parity, cancelling the odd-in-temporal
parity of the gap function.  The derivative explicitly pulls down a factor
of $\omega$ in the summand of  Eq. (9.4.3) which allows the sum to be 
non-vanishing for $\tau\to 0$.  This insertion of a time derivative is 
analogous to the insertion of a spatial form factor in the case of $p$-
or $d$-wave pairing for $P_T=+$.  

{\it (2) Phenomenology for a non-singular pairing interaction.}  \\
Balatsky and Abrahams [1992], Abrahams [1992], Abrahams {it et al.} [1993],
Balatsky {\it et al.} [1994], and Abrahams {\it et al.} [1995] have
considered extensively the phenomenology of an odd-in-frequency 
superconductor which has a non-singular pairing interaction.  Abrahams [1992]  
in particular considers a simple interaction form allowing for both 
$s$ and $p$-wave pairs which is separable
in incoming and outgoing electron energies and momenta
$$V_{\vec k,\vec k'}(\omega_n,\omega_n') = -2{\omega_n \omega_n'\over N(0)
\Omega_c^2}[\lambda_0 + 3\lambda_1(\vec k\cdot\vec k')] \leqno(9.4.4)$$
where $\Omega_c$ is a cutoff frequency for th mediating Boson (all  
interaction strength is assumed to die above the cutoff) and $\lambda_0,\lambda_1$
are assumed to be positive.  Note that
this interaction {\it can only be generated dynamically} so that there
is practically no simple way to derive odd-frequency pairing from a
mean field Hamiltonian in the spirit of the BCS approximation and 
Bogulubov theory.  

With this simple interaction form, Abrahams [1992] shows that the Eliashberg
equations can admit a solution with off-diagonal long range order in  
the $p$-wave channel 
between two temperatures $T_c^+>T_c^-$.  The condition for achieving
superconductivity depends upon the relative strengths of the couplings
$\lambda_0$ (which enters the normal self energy and therebye determines
mass renormalization of the couplings and dynamical pairbreaking) 
and $\lambda_1$.  Specifically, to 
develop $p$-wave spin singlet pairing requires $\lambda_1>\lambda_0+1/4$,
while to have a re-entrant normal phase ($T_c^->0$) requires $\lambda_1
<1+\lambda_0$.   Perhaps the most important aspect of these requirements
is that the renormalization of the interaction strength by the normal self energy 
requires a critical coupling strength of the net attractive interaction
to produce the odd-frequency pairing.   In contrast, for a net attractive
interaction even-in-frequency pairing can occur for arbitrarily small 
interaction strength.  
This result for odd-frequency pairing 
appears to be generically true for models with non-singular 
interactions.  

Abrahams {\it et al.} [1993] consider the specific model examples of 
phonon mediated pairing and an effective interaction generated by the
random phase approximation (RPA) for the Hubbard model.   In the case of 
phonon mediated pairing, while an attractive 
odd-frequency interaction may appear, it cannot satisfy the constraints
required to produce a superconducting transition at a finite temperature. 
In contrast, the RPA model interaction for the Hubbard model can produce
an odd-frequency transition.  They further argue that because of the
pseudo-gap (away from the central coherent quasiparticle band) 
between the Mott-Hubbard
sidebands, that $\partial \Sigma(\omega_n)/\partial \omega_n >0$ over 
the majority of the energy range of the electronic fluctuations responsible
for pairing.  This allows for the renormalization factor $\lambda_0\approx 0$
which makes odd-in-frequency pairing more plausible.  

{\it (3) Connection to Even-Frequency and Composite Pair Fields}\\
One of the intriguing aspects of odd-frequency pairing is that the 
pair field can linearly mix with other operators containing even 
frequency pair fields of different symmetry.  This fact was first 
noticed by Emery and Kivelson [1992] for the two-channel Kondo model.  
To illustrate the idea, we will restrict our attentions to this model.

With the notation introduced above, consider the pair field 
$P^{R+,T-}_{0;0}(\vec R,\tau)$ given by 
$$P^{R+,T-}_{0;0}(\vec R,\tau) = \sigma^{(2)}_{\mu\nu}\sigma^{(2)}_{\alpha\beta}
\Psi_{\mu\alpha}(\vec R,\tau){\partial \over \partial \tau} 
\Psi_{\nu\beta} (\vec R,\tau) \leqno(9.4.4)$$
with Einstein summation convention implicit on the spin indices $\mu,\nu$
and channel indices $\alpha,\beta$. This operator creates a single 
odd-frequency pair at position $\vec R$ and Euclidean time $\tau$.  
The point made by Emery and Kivelson [1992] is that the time derivative
may be explicitly carried out through the commutator with the Hamiltonian
of the fermion pair field operator.  

For definiteness, let us model the kinetic energy of the electrons through
a nearest neighbor tight-binding Hamiltonian, so that for a two-channel 
Kondo hypercubic 
lattice model we have 
$${\cal H} = -t\sum_{\vec R,\vec \delta,\mu,\alpha} 
\Psi^{\dagger}_{\mu\alpha}(\vec R+\delta)\Psi_{\mu\nu}(\vec R)
+ {J\over 2}\sum_{\vec R\mu\nu\alpha} \vec S_I(\vec R) \cdot \vec \sigma_{\mu\nu}
\Psi^{\dagger}_{\mu\alpha}(\vec R)\Psi_{\nu\alpha}(\vec R) \leqno(9.4.5)$$
where we assume the nearest neighbor vectors $\vec \delta$ are only
positively directed. Given $(\partial \Psi/\partial \tau) = [\Psi,{\cal H}]$,
the explicit evaluation of the time derivative in Eq. (9.4.4) gives 
$$P^{R+,T-}_{0,0}(\vec R,\tau) = -2t P^{R-,T+}_{0,0}(\vec R,\tau) 
+ 4J\vec S_I(\vec R,\tau)\cdot \vec P^{R+,T+}_{1;0}(\vec R,\tau) \leqno(9.4.6)$$
where $P^{R-,T+}_{0;0}$ is an {\it odd-radial-parity, even-frequency} pair 
field given by 
$$P^{R-,T+}_{0;0}(\vec R,\tau) = \sum_{\vec \delta} 
[\Psi_{\uparrow,+}(\vec R,\tau)\Psi_{\downarrow,-}(\vec R+\vec \delta,\tau)-\Psi_{\downarrow,+})(\vec R,\tau)
\Psi_{\uparrow,-}(\vec R+\vec \delta,\tau)
\leqno(9.4.7)$$
$$~~~~~~~+\Psi_{\downarrow,-}(\vec R,\tau)\Psi_{\uparrow,+}(\vec R+\vec \delta,\tau)
-\Psi_{\uparrow,-}(\vec R,\tau)\Psi_{\downarrow,+}(\vec R+\vec \delta,\tau)]
~~,$$
and $\vec P^{R+,T+}_{1;0}$ is a {\it spin-triplet even-parity, even-frequency }
pair field with components 
$$P^{R+,T+}_{1,1;0}(\vec R,\tau) = \Psi_{\uparrow,+}(\vec R,\tau)
\Psi_{\uparrow,-}(\vec R,\tau) ~~,\leqno(9.4.8.a)$$
$$P^{R+,T+}_{1,-1;0}(\vec R,\tau) = \Psi_{\downarrow,+}(\vec R,\tau)
\Psi_{\downarrow,-}(\vec R,\tau) ~~,\leqno(9.4.8.b)$$
and
$$P^{R+,T+}_{1,0;0}(\vec R,\tau) = [\Psi_{\uparrow,+}(\vec R,\tau)
\Psi_{\downarrow,-}(\vec R,\tau) + \Psi_{\downarrow,+}(\vec R,\tau)
\Psi_{\uparrow,-}(\vec R,\tau)] ~~.\leqno(9.4.8.c)$$

Eq. (9.4.7) describes a very peculiar pairing field, first 
suggested by Ludwig and Affleck [1991] (see also Ludwig and Affleck 
[1994]) about which a few
words are in order.  If one applies the full spatial parity operator to
it, one discovers it is even in parity.  Indeed, the closest
construction to this field for the single channel Kondo model is the 
``extended $s$-wave pair field'' which has the form 
$$\sum_{\vec\delta} [\Psi_{\uparrow}(\vec R,\tau)\Psi_{\downarrow}(\vec
R+\vec\delta,\tau) - \Psi_{\downarrow}(\vec R,\tau)\Psi_{\uparrow}(\vec
R+\vec\delta,\tau)]~~.$$
This field is both even in parity and even under exchange of sites 
$\vec R,\vec R+\vec \delta$.  In fact, it is easy to demonstrate that
the two operations are not distinct.  However, for the field of Eq.
(9.4.7), spatial inversion {\it does not} correspond to the exchange of
sites, for which the field is manifestly odd.  What is true is the
following: if we understand that the combination of annihilation 
operators summed
upon $\vec \delta$ for a given spin and channel destroys an ``$s$-wave'' 
symmetric (full crystal symmetry) electron state in the first shell 
of atoms about the indexed site $\vec R$, the operation of site exchange
corresponds to the inversion operation about the midpoint of the one
sided chain formed by applying the Lanczos tri-diagonalization procedure
to the reference site.  Namely, consider the effective one-dimensional 
chain first discussed by Wilson [1975] ({\it viz.}, Sec. 4.1).  When
considering the two-channel model, the inversion operator applied to 
the pair field of (9.4.7) (with inversion measured about the midpoint
between the origin and the first Wilson ``site'' or ``onion-skin
shell'') is precisely the same operator as site exchange.  It is {\it
not} however the full 3D parity operator.  Indeed, discretizing the 
continuum form of Ludwig and Affleck [1991,1994] yields precisely the 
same result.

Overall, this connection of the odd-frequency pair field to the 
odd-radial-parity and composite pair fields 
is a rather remarkable result.  Even though we are dealing with the
bare Hamiltonian, and not that of the low temperature excitations about 
the fixed point, Eqs. (9.4.6-8) hint at the outcome of the conformal 
field theory.  Namely, the even-parity, 
odd-frequency pair field linearly mixes with
an even-frequency, odd-radial-parity pair field.  This is an explicit real space
reflection of the approximate conformal invariance of the free-fermion 
Hamiltonian for points on opposite sides of the Fermi surface.  
This follows since 
the tight binding Hamiltonian, linearized about the Fermi energy, 
will indeed resemble a light-cone Hamiltonian for points on opposite sides
of the Fermi energy, with the Fermi velocity equal to the speed of light. 
Hence, spatial and temporal parity are seen to be equivalent operations 
on the pair fields.  Also, the dot-product of the spin triplet field 
with the local moment indicates that the interactions with the local 
moments will produce composite pair fields as candidate order parameters
which are ``bound states'' of the spin with the triplet pairs.  
The mixing with the odd-frequency is symmetry allowed since the inner
product of the triplet field with the local moment operator produces 
an object that is a scalar under spin rotations.  This result for the
composite field 
of course has a beautiful correspondence to the NRG pictures of the two-channel
Kondo fixed point, where at each length scale a core spin is correlated
with a triplet conduction pair. Indeed, as argued in Sec. 6.1, the  
leading irrelevant operator about the two-channel fixed point 
continues to have the form of the interaction term in Eq. (9.4.3) with
$S_I$ replaced by the conformal primary spin field $\Phi_s$.  

We note that to the extent the effective electron-electron interactions
remain local in position space, 
it is perhaps unlikely that the odd-radial-parity pair field will be the
first to condense. 
The reason is that unlike the odd-frequency field which requires only a
local attractive interaction for condensation, 
coupling to this field will require an interaction
that is extended at least over a lattice constant--an extended
discussion will appear later in Sec. 9.4.2 where it is noted that 
the odd-radial-parity field is rigorously excluded from condensation on
the infinite dimensional two-channel (hypercubic) lattice.  
However, this point should be explored in greater
detail numerically.  

It appears overall, then, from the impurity model, that 
the likely candidate for an observable pairing correlation
in the two-channel Kondo model is an odd-frequency, even-parity, spin-singlet,
channel-singlet pair field, or equivalently, a composite field consisting
of a local moment bound to an even-frequency, even-parity, spin-triplet, 
channel singlet.  \\

{\it (4) Apparent Instability of Odd-Frequency Pairing for $\vec q$=0 Pairs}.\\
Two theoretical works point the the possibility that odd-frequency pairing
is actually unstable for pairs with zero center of mass momentum $q$, at least
to the extent that a the solutions reside within Eliashberg-Migdal theory,
i.e., the vertex corrections in the self energies and conductivity are 
negligible.   

The most straightforward analysis is put forward by Coleman, Miranda, and
Tsvelik [1993], who estimate the superfluid density close to $T_c$ for
an assumed $q=0$ odd-frequency transition.  The general form of the
superfluid density or inverse penetration depth for  
$q=0$ pairs close to $T_c$ {\it
in the absence of vertex corrections to the conductivity} is (suppressing 
momentum and other dependencies of the gap function)
$$\rho_s \sim {1\over \lambda^2} \sim \sum_{\omega_n} 
{\Delta(\omega) \Delta^*(-\omega) \over |\omega_n|^3 }\leqno(9.4.9)$$
where $\Delta$ is the gap function.  
For even-frequency pairing, this is positive definite.  For odd-frequency
pairing, it is negative definite.  The interpretation of this result is
straightforward:  $\rho_s$ is proportional to the curvature of the 
free energy with respect to the order parameter at $\vec q$=0 (or, in 
position space in the center of mass coordinates, to the coefficient of
the gradient terms in the free energy).  $\rho_s(q=0)<0$ simply implies
that the uniform phase solution of the free energy is not stable.  
It does not rule out odd-frequency pairing.  Another loophole in 
this conclusion is that the denominator is modified to $(\omega_n^2 - 
|\Delta(\omega_n)|^2)^{3/2}$ for finite $\Delta$.  If a discontinuous
transition to finite $\Delta$ occurred with $|\Delta(\omega_n)|>|omega_n|$
for at least a few values of $n$, then $\rho_s(q=0)>0$ would be possible
(Heid [1995]).  

Heid [1995] pointed out another difficulty with $q=0$ odd-frequency pairs
as treated within Eliashberg theory.  Employing minimal assumptions, he
was able to show that within Eliashberg theory you cannot simultaneously
lower the free energy and find a solution to the self consistency equations
for $q=0$ odd-frequency pairs. 
This means practically that the Eliashberg solutions are extrema but rather
{\it maxima} of the free energy and not minima.  Within a simple model, 
there appear to be no such restrictions on pairs with finite center of mass
momenta (Heid [1995]). 

The physical interpretation suggested by Coleman, Miranda, and Tsvelik [1993,
1994,1995] is that the phase of the odd-frequency pairs is staggered at
the atomic level.  This conclusion is model dependent, since strictly 
speaking the above mentioned instability only implies finite $q$ pairing
is favored, not necessarily zone boundary pairing. 

\begin{figure}
\parindent=2.in
\indent{
\epsfxsize=6.in
\epsffile{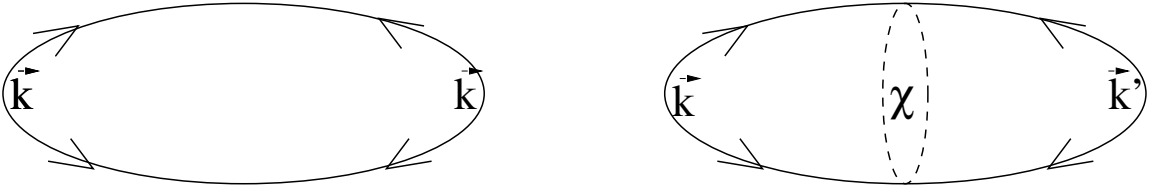}}
\parindent=.5in
\caption{Leading order diagrams for the Meissner ``stiffness'' for 
odd-frequency superconductors.  The 
Meissner stiffness, or ``superfluid density'' is proportional to this 
anomalous diagram of the electrical conductivity.  The
first diagram (at left) includes only renormalized anomalous propagators, 
and, as 
argued by Coleman, Miranda, and Tsvelik [1994] gives a negative
superfluid density or Meissner stiffness for an odd-frequency 
superconductor with $\vec q=0$ center of mass momentum.  The
second (at right) is the vertex correction discussed by Balatsky {\it et al.}
[1994] and Abrahams {\it et al.} [1995] which is argued to enforce a
positive sign to the Meissner stiffness for $\vec q=0$ center of 
mass momentum pairing in which the composite order
parameter representation is used.  The dashed bubble appearing 
as a vertex is a spin susceptibility.  Note that this diagram at right is 
ineffective (vanishes) for a purely local moment driven interaction,
because the different momentum sums integrate to zero. }
\label{fig9p11}
\end{figure}

Balatsky {\it et al.} [1994] and Abrahams {\it et al.} [1995] note that
in the composite operator picture, the leading contribution to the 
anomalous conductivity diagram does in fact have a vertex correction. 
The leading order diagram is displayed in Fig.~\ref{fig9p11}.  The interaction
is proportional to the local moment susceptibility.  This diagram 
contributes {\it positively} to the $q=0$ superfluid density.  Hence, 
the possibility remains that in moving beyond Eliashberg-Migdal theory
to include vertex corrections that $q=0$ pairs may be stabilized.  \\

{\it (5) Connection to Three-body Bound States and Majorana Fermions}.\\
Coleman, Miranda, and Tsvelik [1995] have noted on rather general grounds
that odd-frequency pairing can result from the formation of ``anomalous
three body'' bound states amongst fermions.  This intriguing possibility
rests on the potential to form Majorana fermion bound states.  Because 
the Majorana fermions have constant squared amplitudes, it becomes possible
to develop anomalous expectation values to three body Majorana bound states.

As a concrete example, Coleman, Miranda, and Tsvelik [1995] follow their
earlier work (Coleman, Miranda, and Tsvelik [1993,1994]) and focus on a 
mean field theory for the single channel Kondo lattice.  At a given site $\vec R$,
the interaction term in the Hamiltonian takes the form 
$${\cal H}_K(\vec R) = {J\over 2}\vec S_I(\vec R)\cdot \sigma_{\mu\nu}
\Psi^{\dagger}_{\mu}(\vec R)\Psi_{\nu}(\vec R) \leqno(9.4.10)$$
$$~~~~~= const.~~-J\xi^{\dagger}_{\nu}(\vec R)\xi_{\nu}(\vec R) $$
where the spinor field $\xi$ is defined by 
$$\xi_{\nu}(\vec R) = \vec S_I(\vec R)\cdot\sigma_{\mu\nu}\Psi_{\mu}(\vec R)
~~.\leqno(9.4.11)$$
If we represent the local moment by a pseudo-fermion, we can see that 
$\xi$ represents a composite field of three fermions.  To produce a 
mean field theory, Coleman, Miranda, and Tsvelik [1995] make the ansatz
that 
$$-J\xi_{\nu}(\vec R,\tau) = 2V_{\nu}(\vec R) \hat\phi(\vec R,\tau) + ~fluctuations
\leqno(9.4.12)$$
where $V_{\nu}$ is a c-number spinor field that serves as the mean field
three body amplitude, and $\hat\phi$ is a Majorana fermion field. 

The above transformation of the interaction term and mean field Ansatz
do not immediately permit an approximate model solution. However, it 
is easy to show that the operators $\eta^a(\vec R,\tau) = 
2S^a_I(\vec R,\tau) \hat\phi(\vec R,\tau)$ obey the canonical 
anticommutation relation $\{\eta^a(\vec R),\eta^b(\vec R')\}
=\delta_{ab}\delta_{RR'}$, and further, $(\eta^a)^2=1/2$, so the 
$\eta^a$ are Majorana fermions.  With the mean field Ansatz, this 
converts the interaction term on a given site into 
$${\cal H}_{K,MF}(\vec R) = [\Psi^{\dagger}_{\mu}(\vec R)
(\vec \sigma_{\mu\nu}\cdot\vec \eta(\vec R))V_{\nu}(\vec R) + h.c.] 
+ {2V^{\dagger}_{\nu}(\vec R)V_{\nu}(\vec R)\over J} ~~.\leqno(9.4.13)$$

This mean field Hamiltonian hybridizes the three branches of 
Majorana fermions $\eta^a$ with three of four Majorana fermion branches 
of the conduction electron fields.  The unhybridized Majorana fermion  
remains so in the superconducting state, leaving a linear spectrum
of excitations. Coleman, Miranda, and Tsvelik [1993,1994,1995] put
forward the intriguing speculation that this corresponds to the observed
linear specific heat coefficients below $T_c$ in some heavy fermion 
superconductors.  

Because the Majorana fermions
are admixtures of electron and hole operators, the full solution breaks
$U(1)$ gauge symmetry, i.e., it is a superconducting state.  The energy
is lowest for $V_{\nu}(\vec R) \sim \exp[i\vec Q\cdot\vec R/2]$, with
$\vec Q=[\pi,\pi,\pi]/a$ in three dimensions.  This is the phase of a 
single anomalous three body fermion field--this implies that the phase 
of an anomalous Majorana pair is staggered, i.e., alternates with a 
periodicity determined by $\vec Q$ itself.  

The presence of the Majorana field implies that because the Green's function
of the $\eta$ fields in the frequency domain goes as $\omega^{-1}$, the 
gap function $\Delta(\omega) \sim V^2/\omega$.  Namely, the gap function 
is {\it singular}.  Although Coleman, Miranda, and Tsvelik [1993,1994,1995]
do not construct an explicit realization of the effective fermionic 
interaction which can generate this anomalous gap function, one can infer
that is must be an interaction which is singular in the incoming and outgoing
relative frequencies of the pair.  

The singular pairing amplitude implies that the usual BCS coherence factors
have an unusual form in the superconducting state.  Specifically, the 
gapless Majorana excitations have vanishing charge and spin at the 
Fermi energy. This implies that the vanishing matrix element will 
induce power law behavior in nuclear relaxation and ultrasound as 
observed in the heavy fermion systems, but leave open the possibility
of non-vanishing linear specific heat coefficients and electronic Lorentz
numbers in the superconducting state.  \\

{\it (6) Evidence for Odd-Frequency Pairing Correlations in Other Models}\\
The mean field theory of Coleman, Miranda, and Tsvelik [1993,1994,1995]
suggests that a natural model in which to look for odd-frequency pairing states 
is the single channel Kondo lattice model, 
or the closely related single channel 
Anderson lattice model.  Unfortunately, searches in $d=\infty$ fail to 
produce any evidence for pairing in the Anderson lattice model 
except possibly at unphysically low fillings for the conduction band
(Jarrell [1995,1997]).  It must be emphasized that a possible reason the
correlations are missed in this case is that in the spin singlet sector, 
the odd-frequency pair fields must have odd-parity, and hence live only
at order $1/d$.  In the spin triplet sector, it is possible in principle 
to have an order $(1/d)^0$ contribution in principle, but in practice, 
the Pauli principle will tend to legislate against this, favoring at 
least $d$-wave pairs which are again of order $1/d$.  
Zachar, Emery, and Kivelson have identified a 
special parameter region of the one-dimensional Kondo lattice model 
in which a spin gap develops as an example of where staggered, 
odd-frequency, spin-singlet pairing correlations are enhanced. 
In this region,  
the physics maps to a special version of two-channel Kondo lattice model because of the
absence of back-scattering enforced by the spin gap.  
It remains to be seen whether these correlations can be made to survive in 
$d>1$.  

A possible understanding of this discrepancy between the $d=\infty$ results
and the Majorana mean field theory lies in the nature of the differing 
fixed point physics between the one-channel and two-channel Kondo models.  
The mean field theory depends crucially on a decoupling of the spin-dependent
interaction term of the single-channel model and assumes that this term
is dominant in the low energy scale physics.  However, for sufficiently
large $J$, which suppresses antiferromagnetism and enhances the Kondo 
physics, the single channel model is described by a collection of Kondo 
singlets which become mobile away from half filling.  Hence, the leading
irrelevant operator about the low energy fixed point are not likely to
be of the exchange interaction form required in the Majorana mean field
theory.  On the other hand, in the two-channel Kondo lattice the leading 
irrelevant operator about the paramagnetic fixed point (which is a 
collection of two-channel impurity fixed points), at least in $d=\infty$, 
is in fact of the exchange interaction form.  Hence, while the single
channel model on this basis does not look like a candidate for odd-frequency
pairing, the two-channel Kondo lattice looks rather promising.  As 
detailed in Sec. 9.4.4, in $d=\infty$ at least, this possibility of odd-frequency
pairing in the two-channel Kondo lattice is in fact realized.  

Bulut, Scalapino, and White [1994] have demonstrated that the pairing 
vertex of the Hubbard model in two-dimensions calculated from 
Quantum Monte Carlo has maximum eigenvalues  
for pair amplitudes 
corresponding to even-frequency $d$-wave singlet pairs, and odd-frequency,
$p$-wave singlet pairs.   However, there is no indication of either 
eigenvalue reaching unity for the temperature range studied, which would
indicate a superconducting transition.  \\

\subsubsection{Local Pair Field Susceptibility} 

The possibility of a truly divergent pair field susceptibility in the
two-channel Kondo model was first worked out by Ludwig and Affleck [1991,1994]
using conformal field theory.  The actual calculation of the pair field 
susceptibilities within the conformal theory is a {\it tour de force} in
the formalism.  We shall not discuss the details of these calculations here
but rather summarize and overview the resulting physical picture.  

To see the possibility of a diverging pair field susceptibility, 
it is helpful to 
recall our discussion of the scaling dimensions of 
possible operators discussed in Sec. 6.1.2.b.  For a pair field operator
with charge $Q=\pm 2$, spin $S=0$, and channel spin $S_{ch}=0$, the
scaling dimension is $Q^2/8 + S(S+1)/4 + S_{ch}(S_{ch}+1)/4$=1/2.  
This implies that the corresponding local pair field susceptibility 
must have a logarithmic divergence as $T\to 0$.  

However, the antisymmetry in spin and channel degrees of freedom 
requires that we antisymmetrize in some other variable to satisfy the
Pauli principle.  Let us denote the pair field operators by 
capital $P^{R\lambda,T\gamma}_{S;S_{ch}}$, where $\lambda=\pm,\gamma=\pm$
and the $R,T$ refer to spatial and temporal parity.  The Pauli principle
requires that the product $P_R P_T P_S P_{ch} = -1$ where $P_R$ is spatial
parity, $P_T$ is temporal parity, $P_S$ is spin parity (symmetry under 
spin exchange), and $P_{ch}$ is channel parity (symmetry under channel 
exchange).  The first guess implemented by Ludwig and Affleck [1991]
was to choose a pair field of the conduction electrons {\it odd} in the
one dimensional spatial coordinate about the origin.  Specifically, 
they chose to effect this by inserting a spatial derivative into 
the pair field operator (using the notation of Sec. 6.1.1) as given by
$$P^{R-,T+}_{0;0}(r,\tau) = \epsilon_{\mu\nu}\epsilon_{\alpha\beta}
c_{\mu\alpha}(r,\tau){\partial\over \partial r}c_{\nu\beta}(r,\tau) 
= 2ik_F \Psi_{L\mu\alpha}(r,\tau)\Psi_{R\nu\beta}(r,\tau) 
\epsilon_{\mu\nu}\epsilon_{\alpha\beta}\leqno(9.4.14)$$
which shows that the radial parity gets translated into antisymmetry 
under exchange of left and right fields.  Note that 
$\epsilon_{ij} = i\sigma^{(2)}_{ij}$ is the two-dimensional 
antisymmetric matrix, and that we have employed Einstein summation 
convention in the above equation.  Ludwig and Affleck [1991,1994]
then showed that this operator does indeed give a local susceptibility
which is log divergent.

\begin{figure}
\parindent=2.in
\indent{
\epsfxsize=4.5in
\epsffile{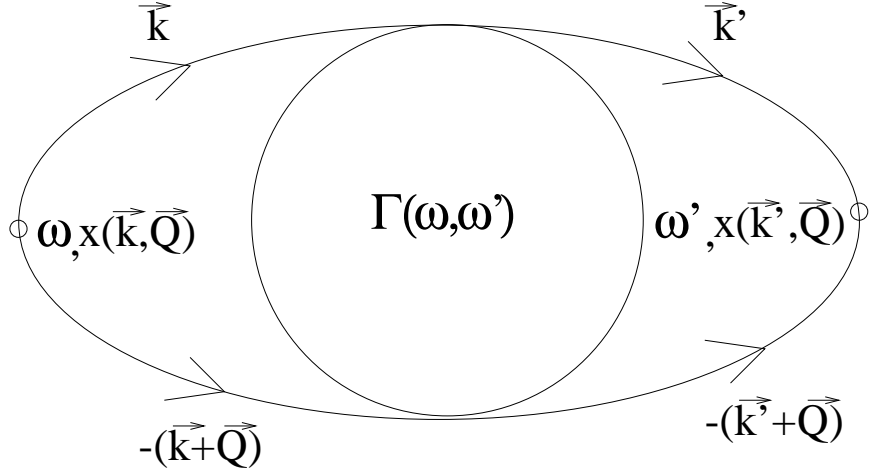}}
\parindent=.5in
\caption{Inefficacy of odd-radial-parity pairing for a local pairing 
interaction and presumed momentum independent hybridization/exchange
between conduction electrons and local electrons.  This Feynman diagram 
represents a contribution to either the full lattice
pairing susceptibility for pairs with center of mass momentum $\vec Q$ 
in the local approximation or $d=\infty$ limit,
or a term in the local pairing susceptibility for the impurity. This
local susceptibility is acquired by summing over $\vec Q$.  
Similar to the conductivity diagram of Fig.~\ref{fig9p6},
the locality of the interaction together with the $x(\vec k,\vec Q)=\sum_i
\cos[(k_i+Q_i)a]$ dependence 
of the vertices guarantees that this contribution must vanish. This term
wou }
\label{fig9p12}
\end{figure}

However, there is a physical problem with this particular 
operator when reconnected
to the full three dimensional physics.  When discretized, 
this operator corresponds the
amplitude of the spin-singlet, channel-singlet pair field operator 
described by Eq. (9.4.7).  At the local level, because the operator is
off-site, the on-site irreducible vertex will allow no coupling.  If
we consider the extension to the two-channel Kondo lattice, as
considered in Fig. \ref{fig9p12}, Fourier transformation gives a 
form factor of $\sum_{i=1}^d\cos[(k_i+Q_i)a]$ for pairs with CM momentum 
$\vec Q$.  Because the irreducible vertex function, which enters the 
bound part of the pair field susceptibility, is purely local in either 
the impurity limit or lattice limit, the momenta sums on either side 
of the lowest order ``bound'' correction are independent, and the
integral of the cosine sum vanishes.  We note that the argument hinges
in detail upon the assumption of momentum independent exchange or
hybridization vertices on-site--it is conceivable that non-zero
contributions may arise from inclusion of a more realistic momentum
dependence to the hybridization, but these would be likely smaller 
in magnitude than the purely on-site contribution of the odd-frequency
pair field.  More numerical work is in order to assess the importance of
this pair field in these more realistic circumstances.

One corollary which is, in retrospect, obvious is that one may equally
well insert a time derivative as a space derivative 
into the singlet pair field operator in 
view of the conformal invariance of the fixed point and the ensuing 
space-time equivalence.  This procedure completely circumvents the
problems with the radial derivative mentioned above.  This time
derivative pair field operator is 
$$P^{R+,T-}_{0;0} = \epsilon_{\mu\nu}\epsilon_{\alpha\beta}
c_{\mu\alpha}(r,\tau){\partial \over \partial \tau} c_{\nu\beta}(r,\tau) ~~.
\leqno(9.4.15)$$
Ludwig and Affleck [1994] considered this field as well as the 
composite operator which linearly mixes with this time derivative field
as discussed in the previous subsection and below.  

However, the first consideration of the equivalent composite operator 
field was by Emery and Kivelson [1992,1994] using the Abelian bosonization
approach, which we turn to next.  
Emery and Kivelson 
first considered the composite operator formed from the dot product
of the local moment spin with the three components of the spin triplet pair
field $P^{R+,T+}_{1,m;0}$ where $m=\pm 1,0$ indexes the 
spin component.  As discussed in the previous subsection and 
by Emery and Kivelson [1994], this is equivalent Because the starting model breaks the $SU(2)$
spin symmetry to a $U(1)$ symmetry about the $z$-axis, it is sufficient
to consider the composite operator (evaluated at the origin now where 
$\psi_{L\mu\alpha} = \psi_{R\mu\alpha}$)
$$\tilde P=\tau^z P^{R+,T+}_{1,0;0} = -i\hat a\hat b[\psi_{L,\uparrow,+}\psi_{L,\downarrow,-}
+\psi_{L,\downarrow,+}\psi_{L,\uparrow,-}] \leqno(9.4.16)$$
$$~~~~~~= -i\hat a \hat b \psi_c(\psi_{sf}+\psi_{sf}^{\dagger})$$
where we are using the notation of Sec. 6.2.1.  Recall that $\hat a$ is
the de-coupled local Majorana fermion of the impurity spin field, $\hat b$
is the local Majorana degree of freedom which couples to the conduction
electrons, $\psi_c$ is the charge fermion degree of freedom obtained   
from refermionization of the charge boson field, and $\psi_{sf}$ is the
spin-flavor fermion field which is obtained from refermionization of the
spin-flavor boson field.  

The calculation of the pair field susceptibility associated with
$\tilde P$ is given by the expression 
$$\chi_{\tilde P\tilde P}(T) = \int_0^{\beta} <\tilde P(\tau)\tilde P^{\dagger}(0)> ~~.
\leqno(9.4.17)$$
Emery and Kivelson argue that the leading singular term of the correlation
function in the above integral goes as $1\tau$ for large times.  The 
argument goes as follows:  first, for positive times, the Green's 
function of the $\hat a$ field $G_a(\tau)$ is just constant, i.e., we 
can just view $\hat a$ as a constant of the motion ({\it c.f.} 
Eq. (6.2.28)).  Second, because the
combination $\hat b(\psi_{sf}+\psi_{sf}^{\dagger})$ is precisely the
combination that enters the effective resonant level Hamiltonian of Eq. 
(6.2.19), we may view this too as a constant of the motion and extract
it from the expectation value.  A more precise way of putting this is
that when we perform the possible Wick contractions, because of the
special character of Majorana fermions this factor is an allowed 
contraction. A corresponding contraction for complex fermion fields  
would generally not be possible (except in a superconducting state). 
Contraction of $\hat b(\tau)(\psi_{sf}+\psi_{sf}^{\dagger})(\tau)$
with $\hat b(0)(\psi_{sf}+\psi_{sf}^{\dagger})(0)$ will generate 
subleading singular corrections to $\chi_{\tilde P\tilde P}$.  Finally,
the remaining Green's function is just the charge fermion Green's function
which behaves as $1/\tau$ for long times and so generates a log divergence.

We can evaluate $\chi_{\tilde P,\tilde P}$ more precisely using the 
Green's functions of Sec. 6.2.2 and App. IV, and it is illuminating to
do so.  First, denote $\psi_{sf}+\psi^{\dagger}_{sf}=\chi_{sf}$ and 
put $\tilde V= iJ_x/\sqrt{\pi a}$ in Eq. (6.2.19) (this is the effective
hybridization between $\hat b$ and $\chi_{sf}$.  Then the expectation
value 
$$<\hat b\chi_{sf}> = -<\chi_{sf}\hat b> = {\tilde V\over \beta N_s} \sum_k
\sum_n {1\over i\omega_n - v_Fk} {1\over i\omega_n + i\Gamma sgn\omega_n}
\leqno(9.4.18)$$
where $\omega_n=2\pi(n+1/2)/\beta$ is a Fermion Matusubara frequency.  
Recall that $\Gamma = \pi N(0)\tilde V^2/2$ is the resonant level width,
which is one kind of local Kondo scale in the approach.  
It is straightforward to evaluate this sum and obtain
$$<\hat b\chi_{sf}> \approx {[N(0)\tilde V]\over 2} \ln[{D\over \Gamma}] 
\leqno(9.4.19)$$
where $N(0)/2=1/(4\pi v_F)$ is the Fermi level DOS of the $\chi_{sf}$ 
fields, and $D=2v_F$ is the bandwidth (cutoff) of the linear dispersion
relation.  Putting this together and using 
$G_c(\tau)\approx G_s(\tau)\approx 1/[\beta D \sin(\pi \tau/\beta)]$
we obtain 
$$\chi_{\tilde P\tilde P}(T) \approx {1\over 2\pi D}[{N(0)|\tilde V|^2\over 4}]
\ln^2({D\over\Gamma}) \ln({D\over T}) \leqno(9.4.20)$$
$$~~~~~ \approx {1\over 2\pi D}{\Gamma \over D} 
\ln^2({D\over\Gamma}) \ln({D\over T})~~.$$

Given that $\Gamma$ measures the Kondo scale, the interesting point
to notice about this susceptibility is that while it is log divergent,
it is very small!  The effective energy denominator is not $T_K$ as
appears in the log divergent spin susceptibility but rather $D^2/T_K$
which is huge!  Hence, at the impurity level, this is a terrifically 
small effect, only presumably visible at very, very low temperatures. 
In fact, Quantum Monte Carlo studies on the two-channel impurity model
failed to reveal any hints of a logarithmic divergence (Pang and Jarrell
[1995]).   Nevertheless, as discussed in Sec. 9.4.4, a possibility for
odd-frequency pairing does emerge in the lattice.  

\begin{figure}
\parindent=2.in
\indent{
\epsfxsize=7.in
\epsffile{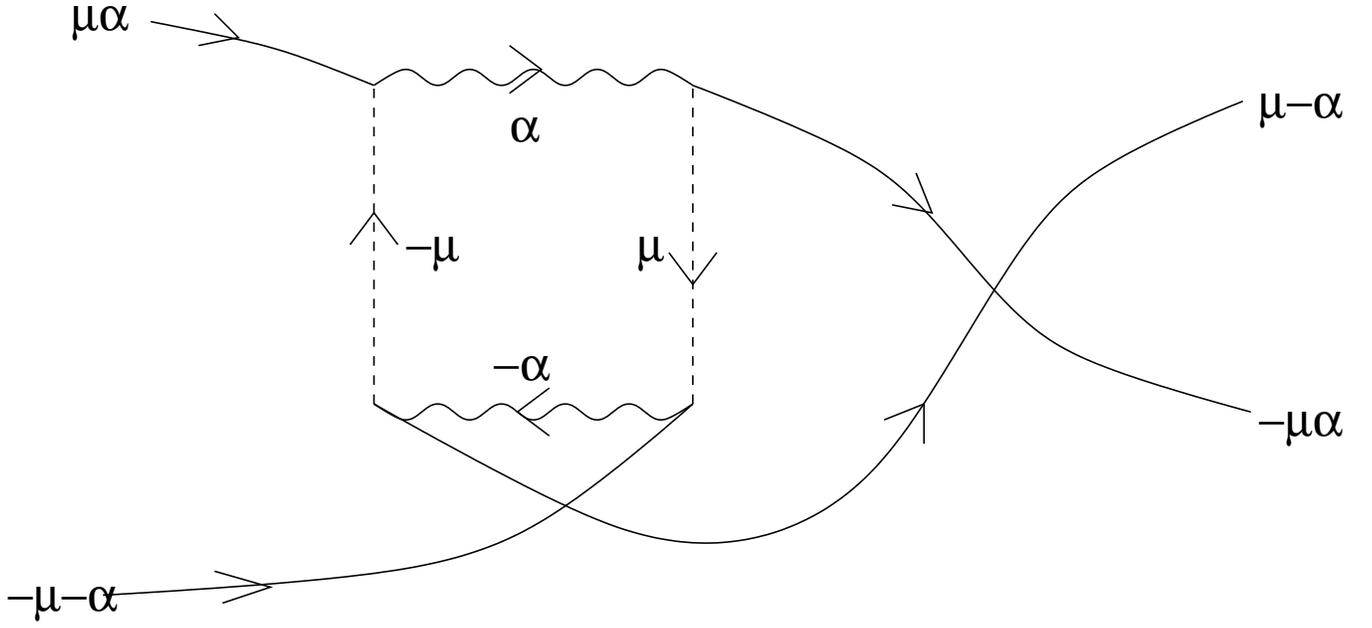}}
\parindent=.5in
\caption{On-site interaction vertex from the NCA for the two-channel 
Anderson model in the Kondo limit.  Each wavy line is an 
excited configuration propagator, carrying the channel index $\alpha$, 
each dashed line a ground
configuration propagator carrying the spin index $\mu$. 
Solid lines are external conduction 
legs, put in for illustrative purposes.}
\label{fig9p13}
\end{figure}

We close this subsection by noting that the local pairing vertex can 
be evaluated within the NCA discussed in Sec. 5.  This program has  
been carried out by Cox and Ruckenstein [1997]. The diagram is shown
in Fig.~\ref{fig9p13}.  Prior to a translation to the full physical subspace,
the interaction strength for pairs with zero center of mass energy is
manifestly antisymmetric in the exchange of either incoming or outgoing
frequencies at low frequencies.  The explicit asymptotic form valid for 
$|\omega|,|\omega'| <<T_0$ is 
$$ I(\omega,\omega') \approx - sgn(\omega)sgn(\omega')
{1\over 8N(0)^2} {1\over |\omega_>|} F({\pi/2},{|\omega_<|\over |\omega_>|}) \leqno(9.4.21)$$
where $F$ is the complete elliptic integral of the first kind, and $\omega_{>,<}$ are determined
from the $min,max$ of $|\omega|,|\omega'|$.  It is
clear that this interaction form is singular and non-separable in the incoming and outgoing frequencies.  
To translate back to three dimensions in the heavy fermion case 
requires multiplication on the left
by factors of $Y_{3m}(\hat k)Y^*_{3m'}(\hat k)$ where $\hat k$ is the
direction of incoming momentum and on the right by corresponding factors
in $\hat k'$,the direction of the outgoing momentum.  The momentum
dependence arising from this may allow for some small contribution to 
the odd-radial-parity pair field discussed by Ludwig and Affleck
[1991,1994], but a test of this requires explicit numerical calculations
for a more realistic model.

\subsubsection{Pairing Symmetry Analysis} 

By identifying the physical correspondence of the spin and channel indices
of the spin-singlet, channel-singlet pairs discussed in the previous 
subsection, one can identify the actual symmetry such pair correlations
would display in a real material assuming that the two-channel Kondo 
model serves as an appropriate starting point.  This information can 
then provide a useful input to phenomenological studies of odd-in-frequency
pairing.                                                                      

In the case of the Heavy Fermion materials, this symmetry analysis has been
carried out by Cox and Ludwig [1997] for the two-channel model in the allowed
symmetry groups as discussed in Sec. 2.2.4.  We shall only briefly describe
the method and results here, which are summarized in Table~\ref{tab9p1}. 

\begin{table}
\begin{center}
\begin{tabular}{|l|c|c|l|}\hline
{\bf Ion} & {\bf Group} & $P_{00}$ & $\nu$ \\\hline
\ufp&\cub & $\gtw\sim(x^2-y^2)(y^2-z^2)(z^2-x^2)$ & 2 \\\hline
\ctp & \cub & $\gtw\sim(x^2-y^2)(y^2-z^2)(z^2-x^2)$ & 1\\\hline
\ufp & \hex & $\gtw \sim(x^3 - 3xy^2)(y^3-3yx^2)$ & 2 \\\hline
\ctp & \hex &$\gtw \sim(x^3 - 3xy^2)(y^3-3yx^2)$ & 3\\\hline
\ufp & \tet &$\gtw \sim xy(x^2-y^2)$ & 3\\\hline
\end{tabular}
\end{center}
\caption{Symmetry of local spin-singlet, channel-singlet pair
fields ($P_{00}$)
and dimensionality $\nu$ of $P_{00}$ space due to
exchange anisotropy. From Cox and Ludwig [1997].}
\label{tab9p1}
\end{table}

The logic is that regardless of whether the local pseudo-spin is magnetic
or quadrupolar, one pair of conduction labels will be a magnetic pseudo-spin
determined by the vector space of a doublet irrep of the double point group 
at the impurity site,
and one pair of indices will be orbital/quadrupolar determined by a doublet
irrep of the ordinary point group at the impurity site.  On physical grounds, 
by analogy to the spin 1/2 problem, the magnetic pseudo-spin singlet should
always transform as $A_{1g}$, {\it i.e.}, have the full symmetry of the
underlying group.  This is indeed the case as explicitly verified from
the tables of Koster {\it et al.} [1963].  On the other hand,the orbital
singlet must change sign under any operations which interchange the orbitals,
and these are indeed effected by ordinary point group transformations.  
The only viable candidate is the pseudo-scalar irrep, present for 
any point group (apart from triclinic) and notated by $A_{2g}$.  As in 
the case of the magnetic pseudo-spin, it can be explicitly verified that 
this result for the orbital singlet holds for the cubic, hexagonal, and
tetragonal point groups of interest.  As a result, transformation 
properties of the overall singlet pair irrep under the local point group
operations must be $A_{1g}\otimes A_{2g}=A_{2g}$.  

The $A_{2g}$ irrep generically must be composed of high order angular 
momentum states about the local moment site.  For example, in cubic 
symmetry $A_{2g} \sim [x^2-y^2][y^2-z^2][z^2-x^2]$ which minimally 
derives from $l=6$ or $i$-wave pairs!  Similarly, in hexagonaly symmetry, 
the $A_{2g}$ state transforms as $(x^3 - 3x y^2)(y^3-3 yx^3)\sim \sin[6\theta]$
where $\theta = \tan^{-1}[y/x]$, which is also of minimal $i$-wave 
character.  Finally, in tetragonal symmetry, $A_{2g}\sim xy(x^2-y^2)$ which
is minimally of $g$ wave character.  We note that should such pair
correlations drive a superconducting instability in the lattice, it 
is unclear how the complex nodal structure of the pair wave functions 
{\it locally} will affect the excitation spectrum.  A complete theory 
of this must await a microscopic approach to the odd-in-frequency pairing
state found (at least in $d=\infty$) for the two-channel Kondo lattice.

We close this section by noting that inroads on Ginsburg-Landau 
phenomenology of odd-in-frequency staggered superconductors can be made
with the above result and an assumed ``negative pair hopping'' in 
real space consistent with the suggestion that the pairs prefer 
zone boundary momenta.  With the local pair symmetry and with the 
negative pair hopping assumption, one can uniquely identify likely 
zone boundary momenta as free energy minima, and construct unique
GL theories for all the heavy fermion superconductors.  This program
is in fact underway, initiated by the work of Heid {\it et al.} [1995,1996,1997]
and Martisovits and Cox [1997].  A review of the phenomenology is 
beyond the scope of this article, but perhaps the most interesting 
features to emerge are:\\
1) For the frustrated hcp lattice of UPt$_3$, a degenerate {\it space group}
irrep is favored which, in the presence of an ordered antiferromagnetic state,
can fully account for the complex $H- vs.- T$ phase diagram of this 
material (Heid {\it et al.}, [1995,1996,1997]).  In contrast, while a degenerage space group irrep is also 
favored for the 1-2-3 materials (UM$_2$Al$_3$, M=Pd,Ni) which have a
simple hexagonal structure, there can be no degeneracy lifting 
coupling of induced (by a 
magnetic moment) or applied strain to the superconducting order parameter
to all orders in the strain (Martisovits and Cox [1997]).  This is consistent with the simple single 
phase diagrams observed for these materials.  \\
2) In the two cases so far examined the presence of time reversal symmetry
breaking can induce microcurrents and thus additional staggered magnetic 
moments in the material (Heid {\it et al.} [1996,1997]; Martisovits and 
Cox [1997]).  For relative real phases between order parameter components,
an induced charge density wave is possible.  Observation of such induced
staggered order would provide a spectacular confirmation of the staggered
pairing hypothesis.  

Finally, we note that superconductivity induced by the TLS Kondo effect
has been suggested as a possibilty for A15 materials (T. Matsuura and 
K. Miyake [1986]; K. Miyake, T. Matsuura, and C.M. Varma [1989]; K. Miyake
[1996]) and the cuprate superconductors (N.M. Plakida, V.L. Aksenov, and 
S.L. Drechsler [1987]; J.R. Hardy and J.W. Flocken [1988]; A.M. Tsvelik [1989]).
Depending upon the nature of the anharmonic double well, a similar 
symmetry analysis should be possible for these models as well.  

\subsubsection{Superconductivity in the two-channel Kondo lattice}

{\bf $d=\infty$ Limit}\\

Given the residual entropy and correlated residual resistivity found 
in the two-channel Kondo lattice model in infinite dimensions by 
Jarrell {\it et al.} [1996,1997], it is natural to assume that in the
absence of external symmetry breaking fields (c.f. Anders, Jarrell, and
Cox [1997]) that the system will be driven to some spontaneous symmetry 
breaking to remove the residual entropy. Just such a possibility is found 
by Jarrell, Pang, and Cox [1997a,b], as we shall describe in this 
section.  

Following Jarrell, Pang, and Cox [1997a,b], we note that
in at least two regions of the 
phase diagram, there are very natural candidates for this symmetry breaking 
identified from the strong coupling limit ($J>>t$). 

\begin{figure}
\parindent=2.in
\indent{
\epsfxsize=5.in
\epsffile{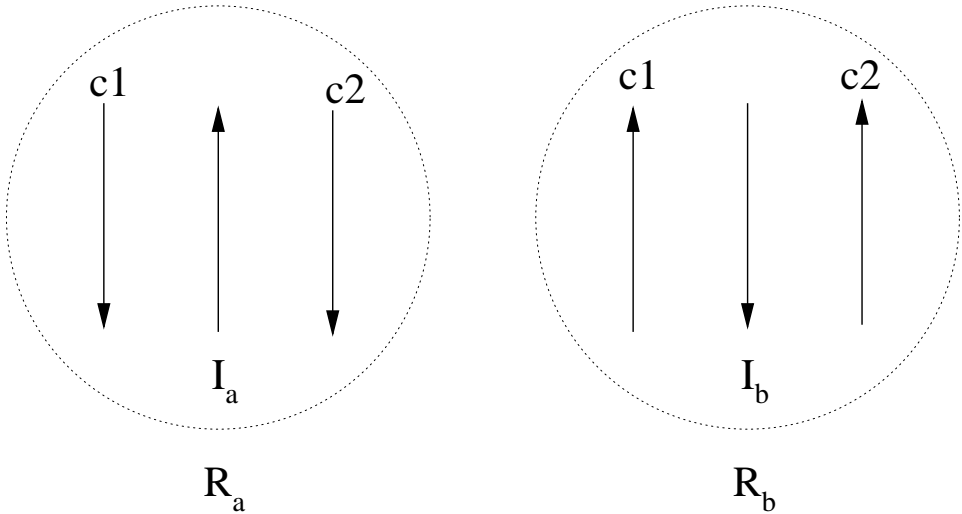}}
\parindent=.5in
\caption{Screening clouds about adjacent two-channel Kondo sites in the
lattice for strong coupling $J\ge E_F$, near half filling of each 
channel.  The overall spins are antialigned because the Pauli principle 
prevents hopping for conduction electrons of equal spin, which gives 
rise to a superexchange $\sim (t^*)^2/J$.  }
\label{fig9p14}
\end{figure}

First, near half filling (unit occupancy of each channel) antiferromagnetism
is certainly a favorable possibility.  To see this, examine Fig.~\ref{fig9p14}
which shows adjoining two-channel Kondo clouds in the strong coupling limit.
The overall spin of the clouds is anti-aligned because of superexchange--
by hopping an electron from channel one on site $a$ to channel one on site 
$b$ to form virtual singlets, followed by a subsequent hop of the original 
site b electron to site a, we see that the overall energy can be lowered 
by a factor $J_{eff} \simeq 4t^2/J>0$.  The positivity of this effective
exchange assures that the coupling is antiferromagnetic.  As usual, the Pauli
principle favors antiparallel spin coupling in such a superexchange process.

\begin{figure}
\parindent=2.in
\indent{
\epsfxsize=5.in
\epsffile{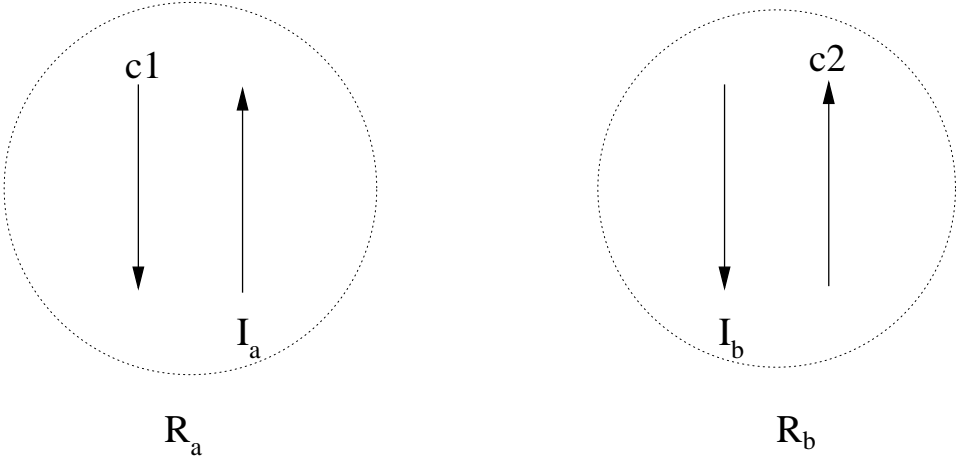}}
\parindent=.5in
\caption{Illustration of channel superexchange near quarter filling in
the two-channel Kondo lattice for strong coupling.  
At quarter filling, singlets on every 
site are energetically favored over doublets on every other site.  
However, hopping is blocked for spins on adjacent sites in the same 
channel, while it is not blocked for spins in opposite channels.  Hence,
a channel superexchange of order $(t^*)^2/J$ results.}
\label{fig9p15}
\end{figure}

Second, near quarter (half an electron per spin per channel) 
filling at strong coupling, in the absence of electron-
electron interactions, it is favorable to form Kondo spin singlets at each
site rather than two-channel spin doublets at every other site.  The former
gives an energy of $-3J/4$ per site, while the latter gives an energy of 
$-J/2$ per site.  The addition of on-site interactions between electrons
will further stabilize this state, while the addition of 
intersite density-density
electrons will presumably help destabilize the singlets on every site.  
Neglecting interactions, we see that because the spin singlets are of 
necessity channel doublets, that there is an intrinsic degeneracy in the
problem which the system would like to remove.  Fig.~\ref{fig9p15} illustrates
how this can occur via channel superexchange, in which a channel one 
electron at 
site a hops to site b to form a trivial local moment spin doublet at site a
and a two-channel screened spin doublet at site b.  The original channel
2 electron at b then hops to a.  The energy lowering in the process is
of order $J^{ch}_{eff} \simeq 4t^2/J$.  The positivity of this effective
exchange indicates that an antiferromagnetic pattern of channel spin is
favored near quarter filling.  

{\it A priori}, there is no reason to anticipate a substantial limitation 
of the boundaries of spin and channel ordering as a function of filling $n$,
though experience with other models (the Hubbard model and Anderson lattice
model, for example) suggest that the boundaries ought to be rather close
to half(quarter) filling for spin(channel) ordering.  If this is the case,
the third law of thermodynamics would again bias us to look for some other
kind of ordering in the intermediate fillings.  Based upon the outlook of
strong odd-frequency pairing correlations from the impurity model,it is 
natural to consider an instability to an odd-frequency superconducting 
phase.  

Most of these above expectations are born out in the calculations of 
the phase diagram of the $d=\infty$ two-channel Kondo lattice model 
as studied by Jarrell, Pang, and Cox [1997a,b] 
with the same Quantum Monte Carlo method discussed in 
Sec. 9.3.2 in the work of Jarrell {\it et al.} [1996,1997].  The  
exception is that no strong evidence for channel spin ordering is found
in the vicinity of quarter filling, though it may represent a possible 
secondary phase transition at very low temperatures.  

\begin{figure}
\parindent=2.in
\indent{
\epsfxsize=5.in
\epsffile{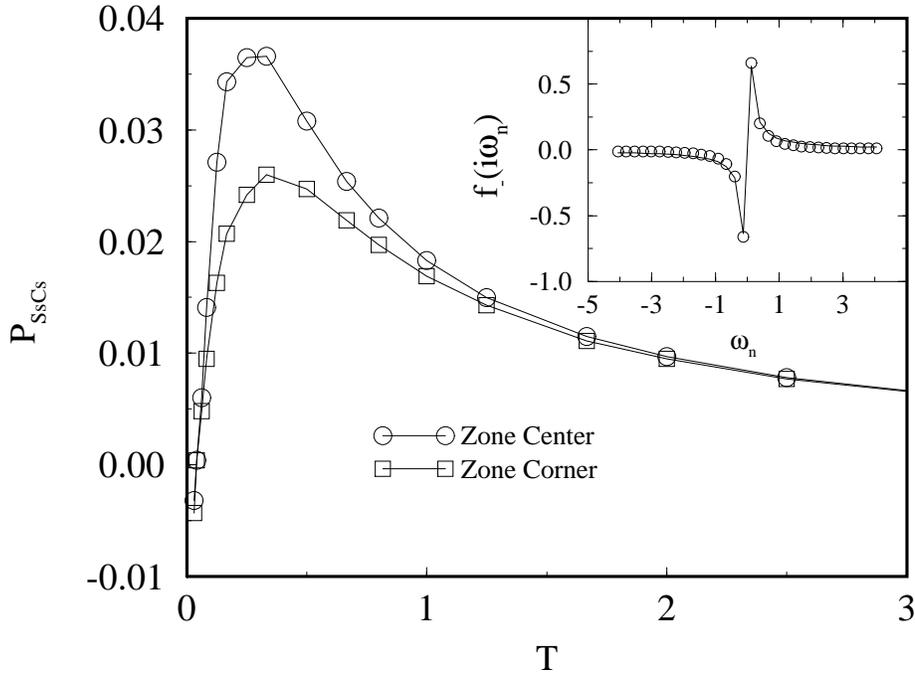}}
\parindent=.5in
\caption{Pair field susceptibilities for the spin singlet, channel
singlet, odd frequency (even parity) pairing channel in infinite 
dimensions.  Susceptibilities
for all momenta in the Brillouin zone go through zero at the same 
temperature, and, as argued in the text, this gives rise to a first
order transition.  From Jarrell, Pang, and Cox [1997].}
\label{fig9p16}
\end{figure}

\begin{figure}
\parindent=2.in
\indent{
\epsfxsize=5.in
\epsffile{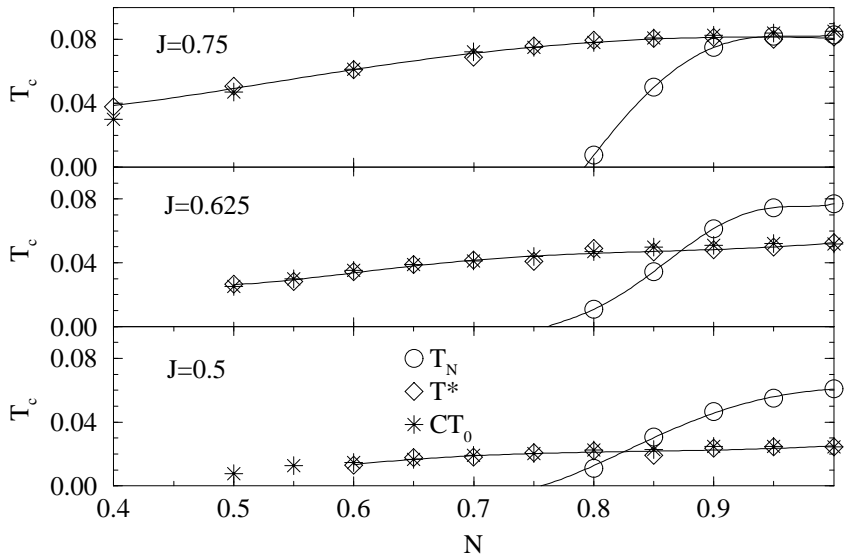}}
\parindent=.5in
\caption{Phase diagram of the two-channel Kondo lattice in infinite
dimensions.  From Jarrell, Pang, and Cox [1997].}
\label{fig9p17}
\end{figure}

To 
hunt for instabilities of the spin, channel, and pairing degrees of 
freedom, the corresponding susceptibilities were evaluated by Jarrell, 
Pang, and Cox [1996,1997].  These
are shown in Fig.~\ref{fig9p16}.  The resulting phase diagram is  
shown in Fig.~\ref{fig9p17}.  The remarkable features resulting from 
this study are: \\
1) The antiferromagnetism persists for a rather large range away
from half filling, and is commensurate except very close to the critical
value of the filling where it becomes incommensurate.  The antiferromagnetic 
transition temperature is a function of the bare coupling $J$, and for
arbitrary fillings and coupling strengths, a combination of RKKY and 
superexchange couplings determine the overall transition temperature.
It was not possible numerically to go to sufficiently large couplings 
to detect the $1/J$ behavior anticipated above.  \\
2) The antiferromagnetism is peculiarly weak on two counts: a) only 
a very small singularity is observed in the total bulk staggered 
magnetic susceptiblity (it is quite strong in the local-moment-only 
contribution to the bulk susceptibility, and so Jarrell, Pang, and 
Cox [1997a,b] employed this to accurately identify the antiferromagnetic
transition and critical behavior), and b) to within numerical accuracy, this 
singularity is characterized by a non-universal $J$ dependent critical 
exponent $\gamma <1$.  \\
3) The line of temperatures $T^*$ represents a {\it lower bound} on a
{\it first order} superconducting transition to an odd-in-frequency 
pairing state in which the conduction electrons pair into spin-singlets,
channel-singlets, and the pair wave function possesses even spatial parity.
The specific value of $T^*$ is determined by where the pair field 
susceptibility $P_S$ in this channel {\it changes sign} as a function of 
temperature.  As argued below, this must correspond to a lower bound for
a first order phase transition.  \\
4) The pair field susceptibility is computed for pairs with arbitrary 
center of mass momentum within the Brillouin zone.  To within numerical 
accuracy, the temperature $T^*$ at which $P_S(\vec q)$ changes sign is
identical for all $\vec q$ values.  As argued later in this subsection,
this is an artifact of infinite dimensions which will likely be removed 
at finite $d$ in favor of a zone-boundary center of mass momentum for 
the pairs.\\
5) Quite remarkably, the temperature $T^*$ goes as $C(J)T_0^{latt}$
where $T^{latt}_0$ is the lattice Kondo temperature determined from a 
fit of the local susceptibility in the local moment channel to the 
single impurity results for each value of $n,J$.  
The prefactor $C(J)\approx 0.5$ is a weakly
$J$ dependent number.  Since $T^{latt}_0$ signifies the energy scale
for formation of local {\it non-Fermi liquid} physics, this suggests
an interpretation of the superconducting instability in terms of pairs
best understood in real space and strongly associated to the non-Fermi 
liquid formation.  This point of view is sharpened by the observation for
at least $J<t$ that $T^*$ is close to the temperature at which 
$Z = 1 - \partial Re\Sigma(\omega)/\partial\omega|_0$ changes sign, 
where $\Sigma$ is
the conduction electron self energy. 
In a conventional Fermi liquid, $Z$ represents the dynamical
mass enhancement of the quasiparticles.  A negative value of $Z$ certainly
indicates a complete breakdown of the quasiparticle concept. \\
6) Though an examination of the staggered channel susceptibility with 
no coupling to pairing or spin degrees of freedom does indicate a narrow
region about quarter filling where ordering is possible, the transition 
temperature for such ordering is very small compared to $T^*$, and so
Jarrell, Pang, and Cox [1997a,b] focus instead upon the magnetic and 
superconducting order. \\
7) The frequency dependent form factor of the spin-singlet channel-singlet
pair field which was found from the maximum eigenvector of the pairing
kernel is singular in $\omega$--it goes as $1/\omega$, in precise agreement
with the result of Coleman, Miranda, and Tsvelik [1993,1994] obtained in 
a Majorana fermion mean field theory of the ordinary Kondo lattice.  
While this has a finite overlap with a pair field possessing a regular
form factor in frequency (e.g., going as $\omega$), the result suggests 
that it is necessary to have a singular interaction vertex to induce 
the pairing.  This further suggests that the physical order parameter,
while possessing an admixture of the time derivative pair field (which
produces a form factor linear in $\omega$ in the frequency domain),
is predominantly composed of more complicated non-local time operations 
in the time domain.  

To further understand the results and motivation behind the numerical study, it is 
convenient to consider the problem from the point of view of Ginsburg-Landau (GL)
mean field theory, though the numerical findings of  
Jarrell, Pang, and Cox [1997a,b] are more general than what is provided by this 
limited theoretical framework.  The following presentation follows
completely the more extensive work of Jarrell and Pruschke [1996], and 
Jarrell {\it et al.} [1997].  

\begin{figure}
\parindent=2.in
\indent{
\epsfxsize=5.in
\epsffile{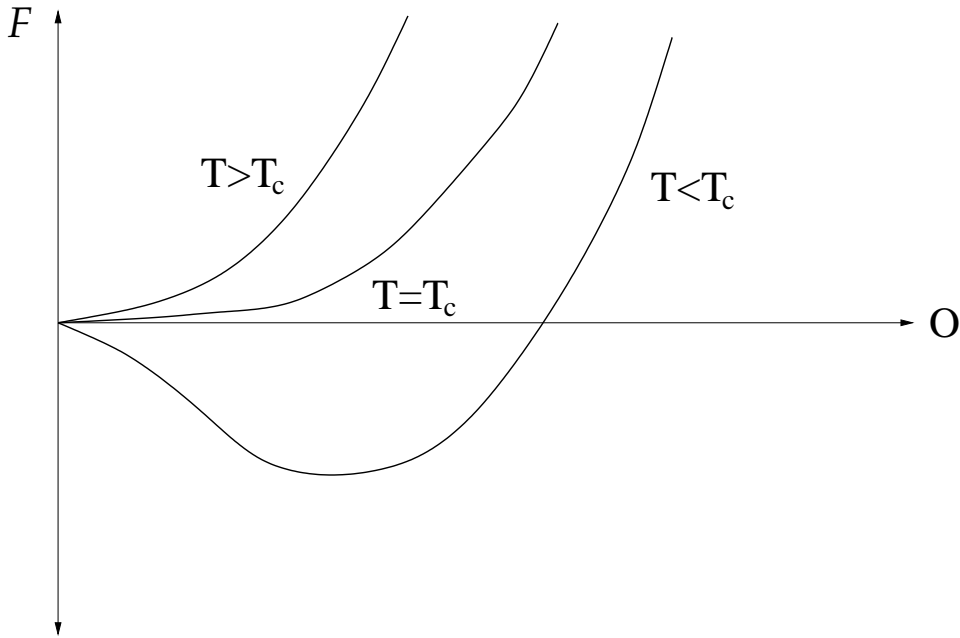}}
\parindent=.5in
\caption{Usual Landau scenario for a second order transition.  Sketched
is the free energy ${\cal F}$ as a function of the order parameter 
$O$.  At $T>T_c$, there is a quadratic curvature upwards.  For $T=T_c$,
the free energy is quartic in the order parameter.  For $T<T_c$, the
free energy has a quadratic curvature downwards. As a result, a stable
minimum at finite $O$ arises continuously from zero order parameter as the 
temperature is lowered. }
\label{fig9p18}
\end{figure}

 Consider the bulk free energy functional
for a general scalar order parameter $O(T)$.  It is given by 
$${\cal F}[O] \approx {O^2(T)\over 2\chi^{(1)}_O(T)} + b O^4(T) + c O^6(T) + .....
\leqno(9.4.22)$$
where $\chi_O^{(1)}$ is the static linear susceptibility corresponding to the 
order parameter $O$--for example, for antiferromagnetism it is the linear
response coefficient relating the induced staggered magnetization to an 
applied staggered field.  Of course, Eq. (9.4.22) is simplified in that
if the order parameter is not a scalar we must generally add more terms. 
Also, if the order parameter admits a cubic invariant (like volume) 
we need to include odd powers.  However, the above form is sufficient
for our purposes.   In the study for spin and channel ordering, Jarrell, Pang and
Cox [1997a,b] found it sufficient to hunt for divergences in the staggered 
spin and channel susceptibilities.  As discussed above, just such divergences
were found.   This common scenario is represented graphically in Fig.~\ref{fig9p18}.

\begin{figure}
\parindent=2.in
\indent{
\epsfxsize=5.in
\epsffile{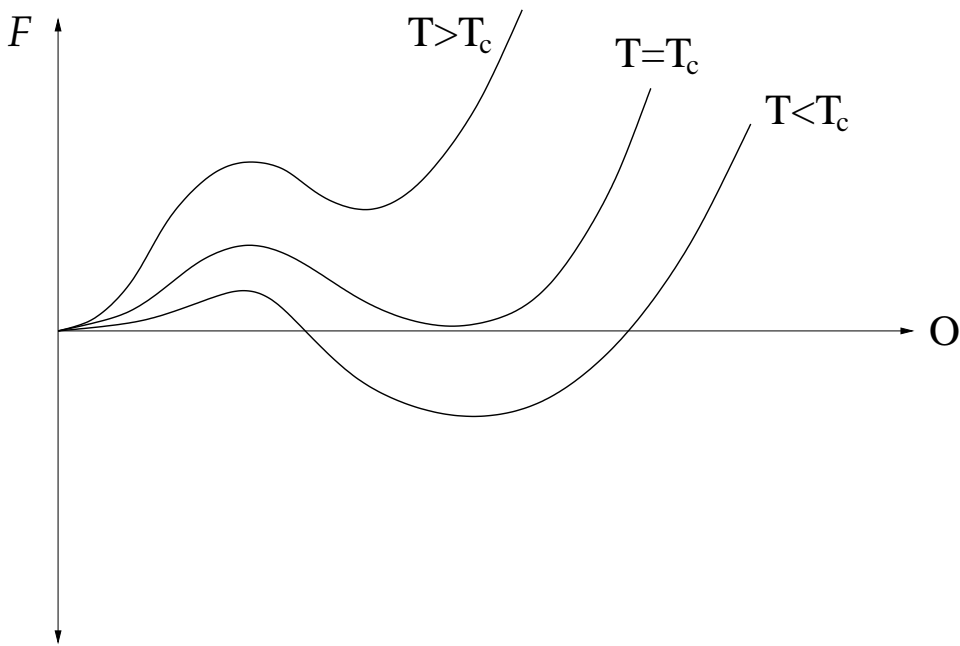}}
\parindent=.5in
\caption{Usual first order transition scenarios.  These are either
driven by the presence of a third order term (as occurs for liquid
crystals, quadrupolar, or structural transitions), or by a change of
sign of the fourth order term in the Landau free energy.  In either
case, the transition occurs when a metastable state is lowered through 
a degeneracy point with the zero order parameter minimum at a critical 
temperature $T_c$.  Below $T_c$ the metastable state becomes fully 
stable.  }
\label{fig9p19}
\end{figure}

Now, in terms of GL theory, the order parameter undergoes a second order
transition when the coefficient of the quadratic term vanishes. This 
clearly implies that $\chi^{(1)}_O(T)$ must diverge.  GL theory of course 
admits the possibility of first order transitions. The customary scenarios
require either: (i) that the order parameter admit a third order invariant
as happens for isostructural phase transitions or quadrupolar ordering, or
(ii) the fourth order term changes sign from positive to negative while 
the second order and sixth order terms remain positive.  This is the 
situation in which a mean field theory of tri-critical points becomes 
possible.  These scenarios are represented graphically in Fig.~\ref{fig9p19}.

\begin{figure}
\parindent=2.in
\indent{
\epsfxsize=5.in
\epsffile{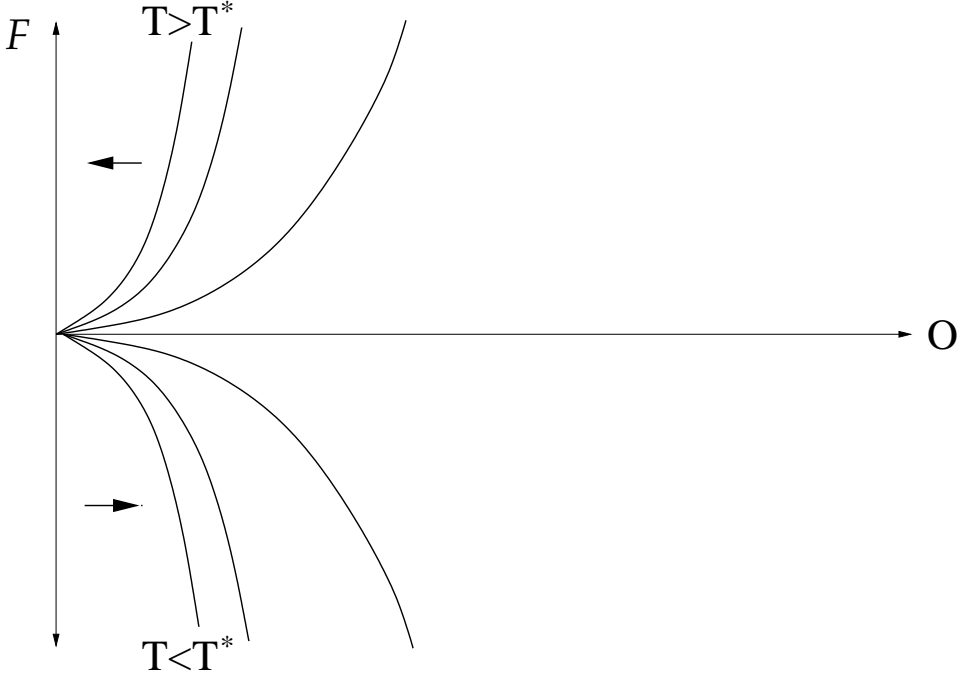}}
\parindent=.5in
\caption{Anomalous first order transition scenario for odd-frequency 
pairing in the two-channel Kondo lattice.  Given the finding of a sign
change to the pair field susceptibility at a temperature $T^*$, 
and noting that the order
parameter susceptibility is the inverse of the free energy curvature 
at the origin ($O=0$), we see that the low $O$ curves ``pile up'' on 
approach to $T^*$ at which point the free energy has infinite curvature 
at the origin.  Immediately below $T^*$, the free energy has {\it
negative} infinite curvature, and clearly stability requires a minimum
at a finite value of $O$.  The only consistent way to make this happen
is for there to be a minimum at finite $O$ above $T^*$, implying that 
$T^*$ is a lower bound for the first order transition temperature.  }
\label{fig9p20}
\end{figure}

Another rather unusual possibility signifies the presence of a first 
order transition (Jarrell and Pruschke [1996]; Jarrell {\it et al.}
[1997]): if $\chi^{(1)}_O(T)$ itself changes sign at a temperature 
$T^*$.  Clearly, such a sign change represents a thermodynamic instability.
This scenario is displayed in Fig.~\ref{fig9p20}.  As can be seen,
above $T^*$, a minimum at $O=0$ is clearly present, while below $T^*$,
a minimum at finite $O$ is present.  Because these cannot be smoothly
connected, we infer that the transition must be discontinuous.  
Of course, if $\chi_O^{(1)}$ passes through zero, the very assumption of a
Taylor's expansion in powers of $O$ for ${\cal F}[O]$ breaks down. 
One approach to remedy this problem is to apply a {\it gedanken} 
external field $h_O$ conjugate to $O$ and work in the Legendre transformed
free energy ${\cal A}[h_O] = {\cal F} - h_O O$.  
By using 
$${\cal A}[h_O] \approx -{\chi^{(1)}_O\over 2} h_O^2-{\chi^{(3)}_O\over 4!} O^4
- {\chi^{(5)} \over 6!} O^6 + .... \leqno(9.4.23)$$ 
where $\chi^{(i)}$is the $i$-th order non-linear susceptibility, 
for $h_O\to 0$, one can systematically invert
${\cal A}$ to obtain ${\cal F}$.  The point is that clearly ${\cal A}$ is
the sensible starting point free energy when $\chi_O^{(1)} \to 0$.  

Now, consider first the case $\chi^{(3)}_O<0$.  This gives the usual 
situation that the GL free energy is stabilized by the fourth order 
terms, as it is straightforward to show that for $O\to 0$, that 
$\beta \approx -\chi^{(3)}_O/[24(\chi^{(1)}_O)^4] >0$.  However, for
$\chi^{(1)}_O \to 0$, this expansion clearly breaks down.  Evaluation
right at $T^*$ where the linear susceptibility vanishes gives 
$${\cal F}[O] \approx -{\chi_O^{(3)}\over 4!} h_O^4 + h_O O +.... \leqno(9.4.24)$$
$$~~~~~= {\chi_O^{(3)}\over 8}({6 O\over |\chi_O^{(3)}|})^{4/3} +....$$
which is manifestly negative.  Stabilization must occur from the sixth
order and higher terms in the free energy.  Because this energy is already
negative at $T^*$, we infer as claimed that a transition to a finite value
of $O$ must have occurred at a temperature above $T^*$.  

For $\chi^{(3)}_O>0$, we cannot give quite so explicit a proof of this
fact.  However, we note that when the non-linear susceptibility is 
positive, the fourth order GL term in $O$ is negative, which indicates 
the presence of a free energy minimum at finite values of $O$.  Taken
together with the observation that the free energy must be negative 
for infinitesimal values of $T^*-T>0,O$, then we infer that again a 
transition to finite $O$ must have occurred above $T^*$.  

Finally, we return to the question of $1/d$ effects upon this phase
transition.  As mentioned in item (4) above in the discussion of the
results for the phase diagram, this transition in $d=\infty$ does not
distinguish between different center-of-mass momenta for the Cooper
pairs.  We expect this to change when $d<\infty$.  The argument is 
as follows.  The superfluid density or stiffness can be read off as
the coefficient of the diamagnetic term in the electrical conductivity.
This results from the anomalous diagram shown in Fig.~\ref{fig9p11}(a).
In a systematic $1/d$ expansion, as mentioned in Sec. 9.3.3, the 
conductivity itself is of order $1/d$.  The locality of the self energy
implies that correct to this order one can omit any vertex corrections
from the conductivity.  This implies that the conductivity has precisely
the bubble form assumed in the argument of Coleman, Miranda, and Tsvelik
[1993,1994] to infer that the superfluid density/stiffness at the zone
center is {\it negative} for an odd-frequency superconductor, while it 
is {\it positive} at the zone boundary.  Hence, it appears likely
that the restoration of a superfluid density at finite $d$ will lift 
the degeneracy between different $\vec q$ (center-of-mass momentum of 
Cooper pairs) and thus induce a staggered pairing state.  While it 
cannot be ruled out that vertex corrections will modify this result
(Balatsky {\it et al.} [1994], Abrahams {\it et al.} [1995]), to the 
extent the leading order (in $1/d$) term determines the sign of the
superfluid stiffness, it seems unlikely that the vertex corrections will
modify this conclusion.  

In closing this subsection, it is worth noting that this unusual pairing
must be described by physics which goes beyond Eliashberg-Migdal theory
in the local dynamics (in that the local spin fluctuations mediating the 
pairing must be self consistently adjusted) but not in the intersite dynamics.
Hence, it should be possible to develop a kind of real space Eliashberg
theory (Jarrell {\it et al.} [1997]) 
to understand the physics of this model in the superconducting 
state.  It will of course be important to develop a theory of a finite
order parameter to properly identify the location of the first order 
transition and to assess the impact of the antiferromagnetism on the 
superconducting transition. 

Jarrell, Pang, and Cox [1997a,b] have noted that the possibility of 
coexisting superconductivity with predominantly commensurate antiferromagnetism
along with the anomalous normal state properties makes the two-channel model
a leading candidate to explain the heavy fermion superconductors.  
It should be noted that the observed superconducting transition in these
materials appears to be second order.  Detailed numerical studies are
required to assess whether finite dimensionality effects induce a second
order transition or whether a sufficiently weak first order transition 
can be found to bring this fascinating state more into alignment with 
experiment.  \\

{\bf One-Dimensional Diluted Kondo Lattice}\\

Zachar, Kivelson, and Emery [1996] have studied a single channel 
Kondo lattice
in one dimension with Abelian bosonization techniques.  Dilution requires
that the spacing between local moments be significantly larger than 
the spacing between atoms yielding the conduction electrons.  The model
is soluble in the anisotropic limit analogous to that of Toulouse for
the single channel impurity model. Specifically, $J_z = \pi v_F$ must
be taken to obtain a soluble model.  

In this limit for incommensurate filling of the conduction band, there is
an antiferromagnetically ordered ground state in {\it transformed 
spin variables} and a gap to spin 
excitations of $\Delta_s \sim v_F[c J_{\perp}/v_F]^{2/3}$ where $c$ is
the local moment density. By transformed spin variables we mean  
$\tilde \tau^x(j) = U^{\dagger}(j)\tau^x (j) U(j)$ where 
$\tau^x(j)$ is the local moment spin ($x$-component) at site $j$ and 
$U(j)=\exp[-i\sqrt{2\pi}\sum_j\tau^z_j\theta_s(r_j)]$ is a stringlike
operator that effects a unitary rotation of the system. $\theta_s$ is 
a ``spinlike'' bosonic momentum variable introduced through the transformations
$$\Psi^{\dagger}_{\lambda,\sigma}(x) \sim \exp[i\sqrt{\pi}(\theta_{\sigma}(x)
+\lambda\phi_{\sigma}(x))+i\lambda k_F x] \leqno(9.4.25)$$
where $\lambda=\pm$ is the left moving (+) or right moving (-) index 
and $\sigma=\uparrow,\downarrow$ is the spin index.  The fields 
$\theta_{\sigma}$(``momentum'') and $\phi_{\sigma}$(``coordinate'') 
are canonically conjugate.   The field
$\theta_s = [\theta_{\uparrow}-\theta_{\downarrow}]/\sqrt{2}$.  
Note that the physical spin variables are disordered by virtue of the
excitation gap, and thus have an exponential decay in real space.  
Diluteness of the local moment array is self consistently determined  
by requiring that the spin correlation length $v_F/\Delta_s$ is large compared
to the inter-impurity spacing.  

The free fermi gas in one dimension would have spin, charge, and pair
correlations which decay as $x^{-2}$ for large distance $x$. If it is 
found that one has correlation functions decaying as $x^{-\alpha}$
for $\alpha<1$, one can have an indication from the one-dimensional model
that there is an enhancement of the ordering tendencies. Zachar, Kivelson,
and Emergy find just such enhancement for 
2$k_F$ ordering of the conduction charge and of the 
composite spin field $\Psi^{\dagger}_{L\downarrow}\Psi_{R\uparrow}\tau^x$
where $\Psi$ are conduction fields either of left ($L$) or right ($R$) 
moving character and $\tau^x$ is the local spin operator.  They also 
find enhanced ordering tendencies for the staggered singlet superconducting
field $(-1)^j\Psi^{\dagger}_{L\uparrow}(j)\Psi^{\dagger}_{R\downarrow}(j)$ 
and the staggered composite field 
$(-1)^j\Psi^{\dagger}_{L\downarrow}(j)\Psi^{\dagger}_{R\downarrow}(j)\tau^x(j)$, which, 
as argued earlier, should mix linearly with the staggered singlet
spin field.  Because $L,R$ act as channel indices here, the staggered 
singlet spin field is completely analogous to that of the two-channel  
model.  Satisfying the Pauli principle requires odd-in-frequency pairing.
This remarkable folding down of the one-channel model in one-dimension to
an effective two-channel model is possible because the spin gap stabilizes
backscattering terms to zero.  It is unclear from the study how generic
the spin gap is away from the soluble points, and this is obviously something
to be explored further. 

Subsequent to the work of Zachar, Kivelson, and Emery [1996], 
Coleman {\it et al.} [1997] have suggested that 
a novel two-channel pairing state can be bootstrapped out of a
superficially one-channel lattice when electrons of differing local
angular momentum (but the same parity) couple to a given local-moment site.

\section{Conclusions and Directions}

We have shown in this review that in order to arrive at exotic Kondo
states which display non-Fermi liquid fixed points the minimal
generalization of the single channel Kondo model is via the addition of 
channel or flavor quantum numbers to the conduction electrons.  The
two-channel spin 1/2 Kondo model obtained in this way has been shown to
describe a great variety of physical phenomena from disparate areas of
condensed matter physics, such as amorphous metals, nanometer scale
point contact devices, certain rare earth and transition metal ions in
metals (some of which become exotic superconductors at full
concentration of the ions), and quantum dots in the Coulomb blockade
regime.  In particular, it is apparent that {\it any} time a local 
pseudo-spin degree of freedom is non-magnetic, a mapping to a
two-channel Kondo model is possible in which the channel degree of
freedom is the magnetic spin or time reversal index of the conduction
electrons, guaranteed to be degenerate in the absence of magnetic field
by Kramers' theorem.  Given this broad variety of possibilities, we can
only anticipate that new models and experimental systems will be
discovered for which a mapping to a model with a non-Fermi liquid fixed
point will be on solid ground.  

We have stressed that the most likely {\it physical} 
route to a non-Fermi liquid Kondo fixed point is through the two-channel spin 
1/2 Kondo model.  However, other intriguing possibilities exist for
stable fixed points involving a large effective impurity spin (\zar
[1996], Moustakas and Fisher [1997]) or unstable fixed points with 
a large effective conduction
spin either for TLS Kondo sites or rare earth/actinide ions 
(Fabrizio and Gogolin [1994]; Fabrizio and \zar [1996]; Kim and Cox
[1996]; Sengupta and Kim [1996]; Kim, Cox, and Oliveira [1997]). 

On theoretical grounds, the two-channel Kondo model itself is very well
understood.  On the high temperature side, the multiplicative
renormalizatin group (Sec. 3) provides a solid indication of which
sectors of parameter space in the model are needed to describe various 
physical phenomena (valid for $T\ge T_K$, $T_K$ the Kondo temperature).  
Knowing the relevant sector of parameter space, one
is guided to the low temperature regime where a variety of powerful
methods are available to characterize the physics. Asymptotically, at
low temperatures, 
conformal field theory and Abelian bosonization may be brought to bear
(Sec. 6) (valid for $T<<T_K$).  
The numerical renormalization group (Sec. 4) and the
Bethe-Ansatz method (Sec. 7) may be employed to fully characterize the
crossover from low to high temperatures provided the model is
sufficiently simple (e.g., no excited crystal field states) and the
temperature range of interest is far from the band edges.  With these
methods it is, however, difficult (though not impossible) to obtain
dynamical properties.  The
non-crossing approximation provides an excellent way to study these
models in thermodynamic, transport, and spectral properties 
over all temperature ranges including a variety of interesting 
physical phenomena (especially realistic electronic structure inputs and
excited multiplet states) while being limited practically to the rare
earth and actinide ions.  

These methods all predict universality in physical properties and are
heavily intertwined, in that, for example, the conformal theory must
{\it assume} a fixed point, and have that assumption confirmed by
matching to finite size spectra from say the NRG or Bethe-Ansatz
approaches.  Once such a comparison is made, one may confidently employ
the predictions of the conformal theory for, say, a $T^{1/2}$ low
temperature behavior in the resistivity.   The non-crossing
approximation is strictly controlled only for large spin degeneracy $N$,
and its very successful application to the problem for $N=M=2$, where
$M$ is the channel degneracy, is justified {\it a posteriori} by a
comparison of the critical exponents, amplitudes such as the zero
temperature resistivity and entropy, and temperature dependent
thermodynamics to exact results such as arise from the Bethe-Ansatz and
CFT.  In this way, one has confidence in using the NCA to calculate the
full temperature dependence of quantities such as the thermopower,
electrical resistivity, and van Vleck susceptibility, which are accessed
either not at all or in limited ways by the other methods.  

Experimentally, there is growing evidence that the TLS interacting with
electrons in metals can be described by the two-channel Kondo model.
However, further studies are required to answer certain critical
questions, such as:\\
$\to$ How does the TLS form in metallic glasses, mesoscopic devices, and 
interfaces?\\
$\to$ Why are there so many TLS with small asymmetry, for example, in
the case of nanoscale point contact devices?\\
$\to$  What is the role of the complete set of excited states in the
TLS? \\
$\to$ What is the effect of static impurities in the metallic host on
the orbitals  and spherical waves of the conduction electrons and how
does that affect the orbital screening phenomenon about the TLS? 

Considering the physics of TLS is not only of interest for establishing
the existence and relevance of the two-channel Kondo problem, but also
for understanding the signals present in many mesoscopic samples,
devices, and interfaces, and thus is highly relevant to the ongoing
technological developments in nanoscale physics.  

Concerning the possibility of two-channel Kondo physics for heavy
fermion alloy systems, it must be said that an impressive body of
evidence argues for understanding of many systems in terms of the 
two-channel Kondo effect.  A complete verification of the behavior has
not proven possible for any one system however, usually due to: (i)
unusual behavior of transport coefficients such as the resistivity and
thermopower which do not appear reconcilable with the simple two-channel
Kondo model, (ii) unusual thermodynamic scaling in applied magnetic
field, or (iii) history dependent freezing effects associated with spin
glass physics.   However, the clear observation of single ion scaling
phenomena in Y$_{1-x}$(Th$_{1-y}$,U$_y$)$_x$Pd$_3$,
Th$_{1-x}$U$_x$Ru$_2$Si$_2$, Th$_{1-x}$U$_x$Pd$_2$Si$_2$,
Th$_{1-x}$U$_x$M$_2$Al$_3$, Th$_{1-x}$U$_x$Be$_{13}$, and, partially, in 
La$_{1-x}$Ce$_x$Cu$_2$Si$_2$ suggests that further experimental studies
on very well characterized crystals are worthwhile.  In the case of the
$U$ based systems, single crystal ultrasound and non-linear
susceptibility measurements are worthwhile to directly probe the
non-Kramers' ground state.  On the theory side, the effects of excited
crystal field levels and other physics beyond the simplest two-channel
model must be further investigated to assess whether the unusual
behavior in transport coefficients may be reconciled with the
observations.  Also on the experimental side, the recent discovery of 
anomalous Pr based systems for possibly 
studying the quadrupolar Kondo effect (Yatskar {\it et al.} [1996])
worthwhile to examine further, particularly by doping on the Pr
sublattice with La to search for dilute quadrupolar Kondo systems where
a known non-Kramers' ground state on the ion of interest is assured.   
Another extremely interesting theoretical question inspired by
experiment (Aliev {\it et al.} [1995a,b]) is to study the situation of
extreme mixed valence between two degenerate configurations
(c.f.,~\ref{fig8p22}).  

Perhaps the most wide open and exciting opportunidies for experimental and
theoretical research on the rare earth and actinide based multi-channel
Kondo candidates are in the periodic compounds such as UBe$_{13}$,
CeCu$_2$Si$_2$, and PrInAg$_2$.  Two of these are exotic
superconductors.  Investigations of the two-channel Kondo lattice in
infinite spatial dimensions yield a non-Fermi liquid normal state and 
either superconducting or antiferromagnetic ground states (with possible
coexistence) (Jarrell {\it et al.} [1996]; Jarrell, Pang, and Cox
[1997]; Anders, Jarrell, and Cox [1997]).  The normal state properties
have a good correspondence to UBe$_{13}$, while the superconductivity
that is found is {\it odd} in frequency and proceeds via a very novel
first order transition.  It is important to further examine whether this
transition is robust upon passing to finite spatial dimension (where
indications are that finite center of mass Cooper pairs will be
favored), what the effects of intersite magnetic correlations are upon
the normal state non-Fermi liquid behavior, and what the effects are of 
the inexactness of channel degeneracy (for the magnetic two-channel lattice) or
spin degeneracy (for the quadrupolar two-channel lattice) of the
conduction states.  Moreover, complete studies of the ordered phases are
necessary to characterize the degree of coexistence vs. competition for
the magnetic and superconducting order.  
Studies for a one dimensional model of the single
channel Kondo lattice which effectively maps to the two-channel lattice
in certain limits also support enhanced staggered odd-frequency pair
correlations (Zachar, Emery, and Kivelson [1996]).  
Because this is the first model Hamiltonian for which
odd-frequency pairing is definitively observed, it is important to
further develop the phenomenological consequences of this unusual
theoretical result. For example, what are the low temperature
excitations and how do they manifest in thermodynamics and transport?  
What happens to Josephson tunneling and Andreev
reflection? Some of these questions have been partially answered (see
for example Coleman, Miranda, and Tsvelik [1994], Heid {\it et al.}
[1995]) but clearly much work remains. 

It is clear that this is a field which has blossomed extensively in
recent years, and for which there remains a promising future for 
theorists and experimentalists with an Edisonian spirit.  

If this does not satisfy the reader, then he/she must go to hell
experiencing hectic non-Fermi liquid behavior with non-vanishing entropy
in marked contrast to heaven where single channel physics presides with
peace ensured by 
a singlet ground state and Fermi liquid excitation spectrum.  Thanks to 
free will, the choice belongs to the reader.  

\section{Acknowledgements}  

This work would not have been possible without the assistance of many of
our colleagues.  We especially want to thank I. Affleck, F. Aliev, N. Andrei, 
P. Fazekas, T. Giamarchi, J. Kroha, A.W.W. Ludwig, J. von Delft, and G.
\zar for carefully reading portions (or all) of the manuscript.  We also
wish to thank R. Buhrman, V. Emery, B. Jones, S. Kivelson, and D.C.
Ralph, and 
J.W. Wilkins for many useful discussions.  D.L. Cox wishes to thank his many
collaborators and colleagues over the years for shaping his views on this field,
especially M.B. Maple, C.L. Seaman, J.W. Allen, M. Jarrell, F. Steglich,
H.R. Ott, H. Pang, F.  Anders, T.S. Kim, J. Han, E. Kim, A. Schiller, L.
Oliveira, V. Libero, H.R. Krishna-murthy, P. Coleman, A. Tsvelik, 
R. Heid, V. Martisovits, Y. Bazaliy, and M. Makivic.  We are especially indebted to F.
(King) Anders for his substantial help in preparing this document, and 
to K. Hartlein for preparing most of the figures which appear here.  
We acknowledge the hospitality of the Institute of Theoretical Physics
at the University of California, Santa Barbara, where a portion of this
work was carried out during the workshop on Non-Fermi Liquid Metals in
1996, and the group of Prof. P. Erd\"{o}s at the University of Lausanne,
Switzerland, where a portion of this paper was written in the spring of
1996.  
The research efforts of D.L. Cox have been supported by the US
Department of Energy, Office of Basic Energy Sciences, Division of
Materials Research (normal state properties) and by the US National
Science Foundation under DMR-9420920 (superconducting properties).  
A. Zawadowski has been supported by Grants OTKA T-021228/1996 and 
OTKA T-024005/1997 from the Government of Hungary.  Both authors
acknowledge the generous support of Grant No. 587 of the
American-Hungarian joint research fund.

\pagebreak

{\bf Appendix I:  Model Calculations of $\Delta^x,V^x$.}\\

\begin{figure}
\parindent=2.in
\indent{
\epsfxsize=5in
\epsffile{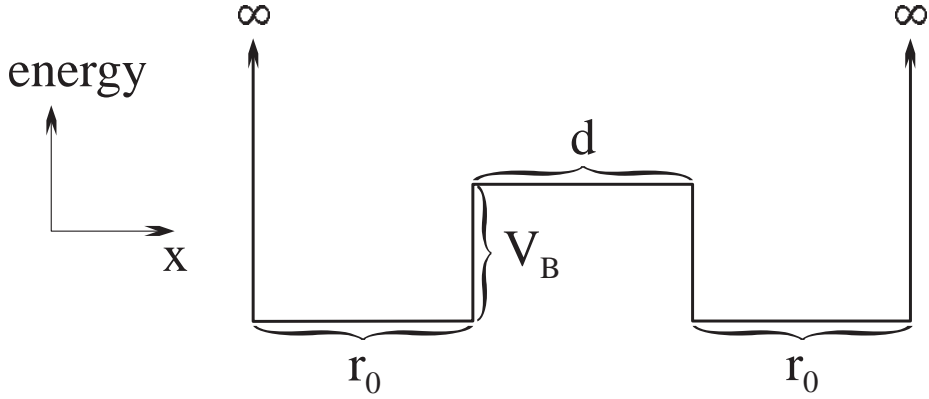}}
\parindent=0.5in
\caption{Double square well model potential 
for TLS Kondo effect. }
\label{figaip1}
\end{figure}

\begin{figure}
\parindent=2.in
\indent{
\epsfxsize=5in
\epsffile{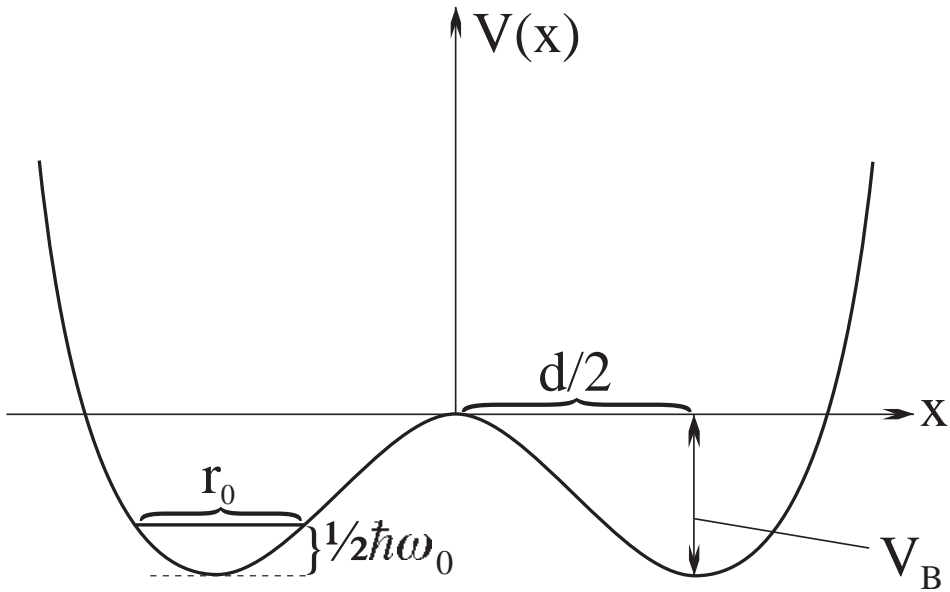}}
\parindent=0.5in
\caption{Quartic anharmonic well model for TLS Kondo effect.}
\label{figaip2}
\end{figure}

In this appendix, we will summarize two model calculations of
$\Delta^x,V^x$
which provide quantitative justification to the simple theory of Sec.
2.1.3.
The examples will follow from exact treatment of the square barrier
problem
displayed in Fig.~\ref{figaip1}, and the quartic anharmonic potential problem of
Fig.~\ref{figaip2}, 
where the non-orthogonal wave functions associated with each well
will be
taken to be approximately of harmonic oscillator (Gaussian) form.

The general prescription is as follows: We shall only treat the spatial
dependence along
the axis of the TLS, assuming for convenience a symmetric
model so that parity is a good eigenvalue.  Denote the lowest two
eigenfunctions
 of the double
well as $\phi_{\pm}(z)$ where the subscript is the parity level.  The
parity
wave functions are delocalized over the double well.  Define
orthogonalized
combinations of wave functions for each side (right $R$ or left $L$)
by
$$\phi_{R,L}(x) = {1\over \sqrt{2}}[\phi_+(x) \pm \phi_-(x)] ~~.
\leqno(A.1.1)$$
The spontaneous tunneling rate $\Delta^x$ is the splitting between the
two
states of definite parity, which is equal to the matrix element of the
{\it
total} Hamiltonian between the localized states.  That is,
$$\Delta^x=E_--E_+ = <-|H|->-<+|H|+> = -2<L|H|R> ~~.\leqno(A.1.2)$$
The assisted tunneling matrix element may be estimated by directly
computing
the matrix element of the electron-ion coupling taking the state $L\to
R$, i.e.,
$$V^x_{\vec k,\vec k'} = -2<\vec k,L|H_{el-ion}|\vec k',R>
\leqno(A.1.3)$$
$$~~~~ = -2U(\vec k'-\vec k) \int dx (\exp[i(\vec k-\vec k')\cdot \hat
x]
-\cos[(k'_x-k_x)(d/2)])
\phi_L(x)\phi_R(x) ~~.$$
The subtraction procedure is the same as discussed in the section on
the TLS
Hamiltonian.
We shall use  Eqs. (A.1.2,3) in estimating the parameters.

The above approach is
equivalent to the derivation based upon the Feynman-Hellman theorem
developed
by Ngai {\it et al} [1967] for estimating tunneling matrix elements.

{\it Square Barrier Model}.  Here the atomic wave functions feel an
infinite
barrier at $x=\pm(d/2+r_0)$, and a square barrier of height $V_B$ for
$-d/2<x<d/2$.  Following Ngai {\it et al.} [1967],
it is straightforward to find the wave functions $\phi_{\pm}(x)$ from
direct
solution of the Schr\"{o}dinger equation subject to the boundary
conditions
$$\phi_{\pm}(\pm(d/2+r_0))=0 ~~,\leqno(A.1.4.a)$$
$$\phi_{\pm}(\pm|(d/2)^+|)=\phi_{\pm}(\pm|(d/2)^-|)~~ \leqno(A.1.4.b)$$
where the superscript indicates whether the limit is taken from the
right (+)
or left (-), and
$${d \over dx}\phi_{\pm}(\pm |(d/2)^+|)={d\over
dx}\phi_{\pm}(\pm|(d/2)^-|)
~~.\leqno(A.1.4.c)$$
Given the eigenvalues $E_{\pm}$ and wavenumbers $k_{\pm} =
\sqrt{2ME_{\pm}}/\hbar$ and $q_{\pm} = \sqrt{2M(V_B-E_{\pm})}/\hbar$,
we take
the {\it Ansatz}
\[ (A.1.5) \begin{array}{ccc}
				   A_{\pm}\exp[ik_{\pm}x] \pm
				   exp[-ik_{\pm}x]
d/2 \le x \le d/2+r_0 \\
		 \phi_{\pm}(x) =  {B_{\pm} \over
2}(\exp[q_{\pm}x]\pm\exp[-q_{\pm}x])  -d/2\le x \le d/2 \\
 \pm A_{\pm}\exp[-ik_{\pm}x] + exp[ik_{\pm}x]
-(d/2+r_0) \le x \le- d/2 \end{array} ~~.\]
This yields the eigenvalue conditions,
$$ k_+r_0 + \tan^{-1}({k_+\over q_+} coth[q_+ d/2]) = \pi
\leqno(A.1.6.a)$$
and
$$ k_-r_0 + \tan^{-1}({k_-\over q_-} tanh[q_- d/2]) = \pi
\leqno(A.1.6.a)$$
the wave function coefficients
$$A_{\pm} = \mp \exp[-2ik_{\pm}(r_0+d/2)] \leqno(A.1.7)$$
and
$$B_{\pm} = \mp 2i({k_{\pm}\over
q_{\pm}}){\cos(k_{\pm}r_0)\exp[-ik_{\pm}(r_0+d/2)] \over
(\exp[q_{\pm}d/2]\mp
\exp[-q_{\pm}d/2])} ~~. \leqno(A.1.8)$$
We are interested in the limit of large barrier height, so that
$q_{\pm}
\approx \sqrt{2MV_B}/\hbar=q_0$, and to leading order $k_{\pm} \approx
k_0 =
\pi/r_0 $, and the overall normalization coefficient of the even and
odd parity
wave functions is $1/r_0$.  We denote the energy $E_0 =
\hbar^2k_0^2/2M$ by
$\hbar \omega_0$, and the Gamow factor $\lambda = q_0 d$.

Retaining terms only to leading exponential order, we find that the
tunneling rate $\Delta^x$ is given by
$$\Delta^x = E^- - E^+ \approx \hbar \omega_0 ({2d\over r_0})
{e^{-\lambda}\over \lambda} ~~.\leqno(A.1.9)$$
Using Eq. (A.1.3) and performing the expansion for
small $(k_x - k'_x) d = \Delta k_x d$,
we find that to leading exponential order the assisted
tunneling matrix element is given by
$$V^x_{\vec k,\vec k'} \approx {\hbar\omega_0 e^{-\lambda} \over
12}({d\over
r_0} ) (\Delta k_x d)^2 ({U(\vec k-\vec k') \over V_B})
\leqno(A.1.10)$$
$$~~~~~= \Delta^x[{\lambda \over 24} (\Delta k_x d)^2 ({U(\vec k'-\vec
k)\over V_B})] $$
which agrees with the heuristic derivation given in Sec.   2.1.

{\it Atom in a potential with quartic anharmonicity} We take the
potential
energy to have the form
$$V(x) = -\alpha x^2 + \beta x^4 \leqno(A.1.11)$$
with $\alpha,\beta>0$.  We identify the barrier width $d/2$ with the
potential
minima, {\it viz.}
$$({d\over 2})^2 = {\alpha \over 2\beta} ~~,\leqno(A.1.12)$$
the barrier height $V_B$ with the depth of the two wells
$$V_B = {\alpha^2 \over 4\beta} = {1\over 2} \alpha ({d\over 2})^2 ~~,
\leqno(A.1.13)$$
and the oscillator energy $\hbar\omega_0$ with the natural frequency of
harmonic oscillations in the well minima, so that
$$\hbar\omega_0 = \sqrt{2\alpha \over M} ~~.\leqno(A.1.15)$$

We assume the wave functions $\phi_{L,R}^0(x)$
 to orthogonalization have the Gaussian form
$$\phi_{L,R}^0(x) \approx {1\over \pi^{1/4}r_0^{1/2}} \exp(-{(x\pm
d/2)^2\over 2r_0^2}) \leqno(A.1.16)$$
where the oscillator length $r_0$ is given by
$$r_0^4 = {\hbar^2\over 2M\alpha} ~~.\leqno(A.1.17)$$
These wave functions have the overlap
$$S=\int dx \phi_L^0(x)\phi_R^0(x) = \exp(-{d^2\over 4r_0^2}) =
e^{-\lambda} \leqno(A.1.18)$$
which defines the Gamow factor $\lambda=d^2/4r_0^2$ in this model.
When we orthogonalize, we obtain the wave
functions
$$\phi_{L,R}(x) \approx \phi_{L,R}^0(x) -{S\over 2} \phi_{R,L}^0(x)
\leqno(A.1.19)$$
which is correct to leading order in $\exp(-\lambda)$.

With these approximately orthogonal orbitals, one can show exactly
that in the
expectation value specified by Eq. (A.1.2), only the quartic potential
term and
the kinetic energy of the orthogonalized pieces survives, so that
$$\Delta^x =-2<L|H|R> \approx \hbar\omega_0 e^{-\lambda}[1 + {\cal
O}({1\over
\lambda})] ~~.\leqno(A.1.20)$$
Similar considerations applied to Eq. (A.1.3)
yield the estimate for $V^x$ that
$$V^x \approx \Delta^x [{\lambda \over 16} {U(\vec k'-\vec k)\over V_B}
(\Delta
k_x d)^2][1 + {\cal O}({1\over \lambda})] ~~. \leqno(A.1.21)$$

The use of the orthogonalized orbitals also justifies the use of
$\delta\rho(\vec r) = \rho(\vec r) - (\rho(d/2)+\rho(-d/2))/2$.  The
argument
is simple:  by plugging $\rho(\vec r)$ into (A.1.3) and using the
orthogonalized orbitals given by Eq. (A.1.19), one finds that
$$<L|\rho(x)|R> \approx \int dx \phi_L^0(x)\phi_R^0(x)[\rho(x) -
{1\over
2}(<L|\rho|L>+<R|\rho|R>)] ~~.\leqno(A.1.22)$$
Assuming the oscillators to be well localized on the scale of density
variation
($\sim k_F^{-1}$) so that $<L,R|\rho|L,R>\approx \rho(\mp d/2)$,
 we see that the factor in braces is precisely $\delta
\rho(x)$.

Hence, in two cases we find results for $V^x$ estimated from the
formulation of
wave functions of orthogonal orbitals which agree to within factors of
order
unity with the Scalapino and Marcus formulation of the theory.\\

{\bf Appendix II: Local representation for electrons in the TLS problem}\\

In the most simple model for the TLS Kondo effect, the electrons are scattered
by the tunneling atoms in an s-wave channel assuming that the center
of the coordinate system is the center of the atom (for a discussion
of the more general  case, see \vld and \zow [1983a]).   In the case
of the TLS the scattering contributes to different angular momentum
channels if the mean position of the atom is chosen as the center of
the coordinate system.  In this case, the wave functions used earlier are 
not orthogonal.  Therefore it is useful to introduce states which are
even ($e$) and odd ($o$) about the center of the tunneling system as
has been done earlier in the two-impurity Kondo problem
(Krishna-murthy, Jayaprakash, and Wilkins [1980]) and in the TLS Kondo
problem by Moustakis and Fisher [1995].  We shall follow the latter 
description here.  

The non-interacting Hamiltonian used by Moustakis and Fisher [1995] is 
$${\cal H}_0 = \Delta_0 (d^{\dagger}_1d_2 + d^{\dagger}_2 d_1) + 
\sum_{\sigma} \int {d\vec k\over (2\pi)^3} 
\epsilon_k c^{\dagger}_{\vec k\sigma}c_{\vec
k\sigma} \leqno(A.2.1)$$
where $d^{\dagger}_i$ creates the tunneling atom at position $\vec R_i$,
$\Delta_0$ is the spontaneous hopping matrix element, and the last
term describes the conduction band.  The operators are assumed
to obey continuum commutation relations.  Turning on an interaction between
electrons and the TLS  which does not include assisted tunneling we
have the coupling 
$${\cal H}_1 = {V\over N_s}\sum_{\sigma,i} \int\int {d\vec k\over
(2\pi)^3}{d\vec k'\over (2\pi)^3} e^{-i(\vec
k-\vec k')\cdot \vec R_i} c^{\dagger}_{\vec k\sigma}c_{\vec k'\sigma}
d^{\dagger}_id_i \leqno(A.2.2)$$
where $V$ is the strength of the $s$-wave scattering off of the 
atom at either site. Measuring $\vec R_1,\vec R_2$ from the midpoint 
of the TLS, we see that $\vec R_1=\vec R/2$, $\vec R_2=-\vec R/2$,
where $\vec R$ is the distance between the impurities.  

We now introduce a new representation in which 
$$\int {d\vec k \over (2\pi)^3 } e^{\pm i\vec k\cdot \vec R/2} c_{\vec
k,\sigma}
= \int {dk\over 2\pi} c_{\pm,k,\sigma} \leqno(A.2.3)$$
which defines $\tilde c_{\pm,k,\sigma}$.  Note that the operators 
$\tilde c_{\pm,k,\sigma}$ are not orthogonal at this point.  
We can properly introduce now
odd and even electron operators which are correctly normalized as 
$$c_{ek\sigma} = {1\over \sqrt{N_e}}[c_{+,k,\sigma} + c_{-,k,\sigma}]~~,
\leqno(A.2.4.a)$$
$$c_{ok\sigma} = {1\over \sqrt{N_o}}[c_{+,k,\sigma} - c_{-,k,\sigma}]
\leqno(A.2.4.b)$$
with 
$$N_e(k) = {2k^2\over \pi^2}[1 + {\sin kR \over kR}] \leqno(A.2.5.a)$$
and
$$N_o(k) = {2k^2\over \pi^2}[1 - {\sin kR \over kR}]~~. \leqno(A.2.5.b)$$
With this definition, and $\alpha=e,o$, 
$$\{c_{\alpha,k,\sigma},c^{\dagger}_{\alpha',k',\sigma'}\} = 2\pi
\delta(k-k')\delta_{\alpha,\alpha'}\delta_{\sigma,\sigma'}~~. \leqno(A.2.6)$$
With this definition, we can introduce properly orthogonal 
left and right conduction operators as 
$$ c_{1k\sigma} = {1\over \sqrt{2}}[c_{e,k,\sigma} + c_{o,k,\sigma}]
\leqno(A.2.7.a)$$
and
$$ c_{2k\sigma} = {1\over \sqrt{2}}[c_{e,k,\sigma} - c_{o,k,\sigma}]~~.
\leqno(A.2.7.b)$$
Note that this definition of conduction basis is completely general
for a two-center problem and may thus be applied to the two-impurity
Kondo model as well.  

We now return to rewrite the interaction Hamiltonian of Eq. (A.2.2). 
In terms of the $e,o$ basis we have 
$${\cal H}_1 = \sum_{\sigma}\int\int {dk\over 2\pi}{dk'\over 2\pi} 
[V_1 (c^{\dagger}_{e,k,\sigma} c_{e,k',\sigma} +
c^{\dagger}_{o,k,\sigma} c_{o,k',\sigma}) + V_2
(c^{\dagger}_{e,k,\sigma} c_{e,k',\sigma} -
c^{\dagger}_{o,k,\sigma} c_{o,k',\sigma})$$
$$
+ V_3[(c^{\dagger}_{e,k,\sigma} c_{o,k',\sigma} +
c^{\dagger}_{o,k,\sigma} c_{e,k',\sigma})(d^{\dagger}_1d_1-d^{dagger}_2d_2)]
\leqno(A.2.8)$$
where we take $d^{\dagger}_1d_1+d^{dagger}_2d_2=1$ (appropriate to the
restricted Hilbert space) and 
$$V_1 = \pi N(0)V~~, \leqno(A.2.9)$$
$$V_2 = \pi N(0)V {\sin k_FR\over k_FR} ~~, \leqno(A.2.10)$$
and
$$V_3 = \pi N(0)V \sqrt{1 - ({\sin k_FR\over k_FR})^2} ~~.
\leqno(A.2.11)$$
This completes the derivation of the Moustakis and Fisher [1995] 
hamiltonian displayed in Eq. (3.5.12) (recall that $\sigma^z = 
d^{\dagger}_1d_1-d^{\dagger}_2d_2$).  

Finally, we mention that in Eqs. (A.2.9) and (A.2.10), an averaged 
atomic potential can be subtracted (i.e., a uniform background). \\

{\bf Appendix III: NCA Treatment of  Spin and Channel Magnetization in
the $SU(N)\otimes SU(M)$
Multi-channel Model}\\

We wish to pull out the singular low field dependence of the spin
magnetization when
$\gamma \ge 1$ and the channel spin magnetization when $\gamma \le 1$.
Our treatment follows the calculations of susceptibilities in Bickers
[1987], Sec. V.B.3. We shall
present a more detailed argument for the spin magnetization; the
details for the channel spin
magnetization will be similar.  Our notation follows Sec. 5.1.
The strategy is to develop and utilize an expression for the ground
state energy
as an integral over $d_b$ where the integrand contains only factors of
$g_f$ viewed as depending
on $d_b$ through inversion of the  constant of integration relation
(5.1.17).  In applied field, $g_f$
will carry all the explicit field dependence and an expression for the
magnetization may be
obtained.  The field is assumed to enter through the Zeeman energy $H_Z
= -\sum_{\sigma}
\mu H_{sp}\sigma$ where $\sigma$ runs between $-J=(N-1)/2$ to $J$.

First, we extend the NCA differential equations in the applied spin
field.  Now $g_f$ acquires a
$\sigma$ dependence so the equations are
$$ {dg_{f\sigma}\over d\omega} = -1 - {\gamma\tilde\Gamma\over \pi d_b}
\leqno(A.3.1.a)$$
and
$$ {dd_b\over d\omega} = -1 - {\tgam\over \pi N}\sum_{\sigma} {1\over
g_{f\sigma}} \leqno(A.3.1.b)$$
subject to the boundary conditions
$$d_b(-D) = D,~~g_{f\sigma}(-D) = D+\ef -\mu\sigma H_{sp}
~~.\leqno(A.3.2)$$
It is easy to see that we may write
$$g_{f\sigma} = g_{fJ} + (J-\sigma)\mu H_{sp} = g_{fJ} + m \tilde
H_{sp} \leqno(A.3.3)$$
where $m=J-\sigma$ and $\thsp = \mu H_{sp}$.
In consequence, the integration constant relation between $d_b,g_{fJ}$
is
$$g_{fJ} + {\tgam\over N\pi}\sum_{m} \ln({g_{fJ} + m\thsp \over D})+
{\cal C} =
d_b + {\gamma\tgam\over \pi }\ln({d_b\over D}) \leqno(A.3.4)$$
from which we infer ${\cal C}=-\tef + J\thsp$ by evaluation at $-D$.

In principle, Eq. (A.2.4) allows us to solve for $g_{fJ}$ as a function
of $d_b$, or vice versa.
This gives a route to integrate the differential equations of Eq.
(A.2.1.a,b).  We find
$$\omega = -d_b + \int_D^{d_b} dx {S_1(x)\over 1+S_1(x)}
\leqno(A.3.5)$$
with
$$S_j(x) = {\tgam\over N\pi}\sum_{m} {1\over (g_{fJ}(x) + m\thsp)^j}
~~.\leqno(A.3.6)$$
The $x$ in the integrand is expressing the dependence of $g_{fJ}$ on
$d_b$.
At the threshold which is also the ground state energy, $d_b$
vanishes.  Hence
$$E_0 = - \int_0^D dx {S_1(x)\over 1+S_1(x)} \leqno(A.3.7)$$
where the field dependence enters now only through $g_{fJ}$ in the
integrand.  Hence, we need
a knowledge of $\partial g_{fJ}/\partial \thsp$.
From Eq. (A.2.4), holding $d_b$ fixed and differentiating with respect
to $\thsp$, we see
that
$${\partial g_{fJ} \over \partial \thsp} = -{ J + S_1'\over 1+ S_1 }
\leqno(A.3.8)$$
where
$$S_j' = {\tgam\over N\pi} \sum_{m} {m\over (g_{fJ}+m\thsp)^j}
~~.\leqno(A.3.9)$$

Putting the above relations together, we obtain the spin magnetization
from
$$M_{sp} = -{\partial E_0\over \partial \thsp} = \int_0^D dx
{[S_2(x)(S_1'(x)+J)-S_2'(x)(1+S_1(x))]
\over (1+S_1(x))^3} ~~.\leqno(A.3.10)$$
Our goal is to simply pull out the low field dependence on $\thsp$.
Hence we may evaluate
the above equation assuming $g_{f}$ is above the crossover value, but
still in the
the low energy regime so that $\thsp \le g_f <<T_0$.
This imposes an infrared cutoff on the integral determined by the
relation between $g_f$ and
$d_b$ of $x_c \approx (\tgam/\pi)(\thsp/T_0)^{1/\gamma}$ in view of
Eq. (5.1.18).  The upper cutoff is
specified by the maximal value of $d_b$ in the low temperature regime
which is $\tgam/\pi$.
It is straightforward to show
that for small $\thsp$,
$$ [S_2(J+S_1')-S_2'(1+S_1)] \approx ({\tgam \over \pi})^2{1\over
g_f^4} {J(J+1)/3}\thsp
\leqno(A.3.11)$$
and since for $\thsp < g_f <<T_0$, $1+ S_1 \approx {\tgam/\pi g_f}$ we
see that
$$M_{sp} \approx {\pi J(J+1) \thsp \over 3\tgam}
\int_{x_c(\thsp)}^{\tgam/\pi} {dx\over g_f(x)}  \leqno(A.3.12)$$
$$~~~~ = {J(J+1)\over 3} ({\thsp\over T_0}) {1\over \gamma - 1}
[({\thsp\over T_0})^{1/\gamma-1} -1] ~~.$$
For $\gamma >1$, we see then that $M_{sp} \sim (\thsp/T_0)^{1/\gamma}$;
for $\gamma=1$,
$M_{sp} \sim (\thsp/T_0)\ln(T_0/\thsp)$.  For $\gamma<1$, $M_{sp}$ is
linear in $\thsp$
at low fields corresponding to the finite value of $\chi_{sp}(0)$.

We may follow analogous reasoning to obtain the channel spin
magnetization
$M_{ch}=-(\partial E_0 /\partial \tilde H_{ch})$ in applied channel
field which couples linearly to
the channel states.  The result is that for $\gamma <1$, $M_{ch} \sim
(\pi T_0/\tgam) (\pi \tilde H_{ch}/\tgam)^{\gamma}$, for
$\gamma=1$, $M_{ch} \sim (\pi T_0/\tgam) (\pi \tilde H_{ch}/\tgam)
\ln(\tgam/\pi\tilde H_{ch})$, and for $\gamma>1$, $M_{ch}$ simply
varies linearly in the applied
field corresponding to the finite value for $\chi_{ch}(0)$.    The
prefactor of $(\pi T_0/\tgam)=W_{ch}$
reflects the need to virtually excite to the channel configuration to
polarize the channel spin.\\

{\bf Appendix IV:  Green's Functions in the Abelian Bosonization
Approach to the Two-Channel
Kondo Model}\\

In this appendix we briefly review the derivation of the Green's functions
in the Abelian Bosonization approach to the two-channel Kondo lattice.  
Of interest are the Green's functions for the Majorana $\hat b$ field
and for the $\psi_{sf}$ field.  

Because the effective resonant level model of Eq. (6.2.19) only couples
the combination $\psi_{sf}+\psi_{sf}^{\dagger}$ to $\hat b$, it is 
convenient to decompose the $\psi_{sf}$ field into two real Majorana
fields, 
$$\hat\chi_{sf} = {1\over \sqrt{2}}(\psi_{sf}+\psi_{sf}^{\dagger}),~ 
\hat\lambda_{sf} = 
{1\over i\sqrt{2}}(\psi_{sf}^{\dagger}-\psi_{sf}) ~~.\leqno(A.4.1)$$
The $\lambda_{sf}$ field decouples from the problem.  The remaining 
resonant level model may be solved by equation of motion methods.  
Define the Green's functions 
$$G_b(\tau) = -<T_{\tau}\hat b(\tau)\hat b(0)>~~, \leqno(A.3.2.a)$$
$$G_{bk}(\tau) = -<T_{\tau}\hat b(\tau)\hat\chi_{sf,k}(0)> ~~,\leqno(A.4.2.b)$$
$$G_{kb}(\tau) = -<T_{\tau}\hat \chi_{sf,k}(\tau)\hat b(0)> ~~,\leqno(A.4.2.c)$$
and
$$G_{kk'}(\tau) = -<T_{\tau}\hat\chi_{sf,k}(\tau)\hat\chi_{sf,k'}(0)> ~~,\leqno(A.4.2.d)$$ 
where $\hat\chi_{sf,k}$ is the Fourier transformed Majorana field in 
momentum space.  

The equations of motion in Matusubara frequency space read
$$i\omega_n G_b = 1 + {\tilde V \over \sqrt{N_s}} \sum_k G_{kb}~~,\leqno(A.4.3.a)$$
$$(i\omega_n - \epsilon_k) G_{kb} = {\tilde V^*\over \sqrt{N_s}} G_{bb} ~~,\leqno(A.4.3.b)$$
$$i\omega_n  G_{bk'} = {\tilde V\over \sqrt{N_s}}\sum_{k''} G_{k''k'} ~~,\leqno(A.4.3.c)$$
and
$$(i\omega_n - \epsilon_k) G_{kk'} = \delta_{kk'} + {V^*\over \sqrt{N_s}}G_{bk'} ~~, 
\leqno(A.4.3.d)$$
where $\tilde V=i J\sqrt{2/\pi a}$ is the effective mixing matrix element
between $\hat b,\hat \chi_{sf}$, and $\epsilon_k$ is the energy of the
$\psi_{sf}$ fermions.   

It is now straightforward to solve these equations for the separate 
Green's functions, and we find 
$$G_b = {1 \over i\omega_n + i\Gamma sgn\omega_n} ~~,\leqno(A.4.4.a)$$
$$G_{kb} = {\tilde V^*\over \sqrt{N_s}}{1 \over (i\omega_n-\epsilon_k)
(i\omega_n+i\Gamma sgn\omega_n)} = {\tilde V^*\over \tilde V} G_{bk}~~,\leqno(A.4.4.b)$$
and
$$G_{kk'} = {\delta_{kk'}\over i\omega_n - \epsilon_k} + {|\tilde V|^2\over N_s}
{1\over (i\omega_n-\epsilon_k)(i\omega_n+i\Gamma sgn\omega_n) (i\omega_n -\epsilon_{k'})}
~~.\leqno(A.4.4.c)$$
This completes the derivation.

\pagebreak

\section{References}

Abrahams, E., 1992, J. Phys. Chem. Sol {\bf 53}, 1487.\\
Abrahams, E., A.V. Balatsky, J.R. Schrieffer, and P.B.
Allen,
1993, Phys. Rev. B{\bf 47}, 513.\\
Abrahams, E., A.V. Balatsky, J.R. Schrieffer, and D.J. Scalapino, 
1995, Phys. Rev. B{\bf 52}, 1271 (erratum, B{\bf 52}, 15649).\\
Abrikosov, A.A., 1965, Physics {\bf 2}, 5.\\
Abrikosov, A.A. and A.A. Migdal, 1970, J. Low Temp. Phys. {\bf 3},
519.\\
Affleck, I., 1990a, Nuc. Phys. B{\bf 336}, 517.\\
Affleck, I., 1990b, in {\it Fields, Strings, and Critical Phenomena},
eds. C.
Itzykson, H. Saleur, and J.-B. Zuber (North-Holland, Amsterdam), p.
563.\\
Affleck, I., 1993, Proceedings of 1993 Taniguchi 
Symposium (cond-mat/9311054).\\
Affleck, I., 1995, Lecture Notes from 1995 Polish Summer School on 
Theoretical Physics, 
Acta Physica Polonica B{\bf 26}, 1869.\\
Affleck, I., and A.W.W. Ludwig, 1991a, Nuc. Phys. B{\bf 352}, 849.\\
Affleck, I., and A.W.W. Ludwig, 1991b, Nuc. Phys. B{\bf 360}, 641.\\
Affleck, I., and A.W.W. Ludwig, 1991c, Phys. Rev. Lett. {\bf 67},
161.\\
Affleck, I., and A.W.W. Ludwig, 1992, Phys. Rev. Lett., {\bf 68},
1046.\\
Affleck, I., and A.W.W. Ludwig, 1993, Phys. Rev. B {\bf 48}, 7297.\\
Affleck, I., A.W.W. Ludwig, and B.A. Jones, 1995, Phys. Rev. B{\bf
52}, 9528.\\
Affleck, I., A.W.W. Ludwig, H.-B. Pang, and D.L. Cox, 1992, Phys. Rev.
B{\bf 45}, 7918. \\
Akimenko, A.I., and V.A. Gadimenko, 1993, Sol. St. Comm. {\bf 87},
925.  \\
Aliev, F.G., 1984, Sov. Phys. Sol. St. {\bf 26}, 682.\\
Aliev, F.G., S. Vieira, R. Villar, H.P. van der Menlen, K. Bakker, A.V.
Andreev, 1993, JETP Lett {\bf 58}, 762 (Sov. Phys.
ZhETP, Pisma {\bf 58}, 814).\\  
Aliev, F.G., H. El. Mfarrej, S. Vieira, P. Villar, 1994, Solid State
Comm.{\bf 91}, p.775.\\ 
Aliev, F.G., S Vieira, R. Villar,
J.L. Martinez, C.L. Seaman, 1995a, Physica B {\bf 206-207},
p.454.\\
Aliev,  F.G., H. El. Mfarrej, S. Vieiera, R. Villar, and J.L.
Martinez, 1995b, 
Europhysics Letters {\bf 32}, 765.\\
Aliev, F.G., H. El. Mfarrej, S. Vieira, and R. Villar, 1996a, 
Physica B{\bf 223\&224}, 464. \\
Aliev, F.G., S. Vieira, R. Villar, and J.L. Martinez, 1996b, 
Physica B{\bf 223\&224}, 475. \\
Aliev, F.G., H. El. Mfarrej, S. Vieira, and R. Villar, 1996c, 
Europhysics Lett. {\bf 34}, 605.\\
Allen, J.W., 1991,  Physica B{\bf 171}, 175.\\
Allen, J.W., L.Z. Liu, R.O. Anderson, C.L. Seaman, M.B. Maple, Y.
Dalichaouch, J.-S. Kang, M. Torakachvili, M.A. Lopez de la Torre,
1993, Physica B{\bf 186-188}, 307.\\
Altsch\"{u}ler, D., M. Bauer, and C. Itzykson, 1990, Comm. Math. Phys.
{\bf 132},
349. \\
Altshuler, B.L., and A.G. Aronov, 1985, in {\it Electron-Electron
Interactions in
Disordered Systems}, eds. A.L. Efros and M. Pollak (North-Holland, New
York), p. 1.\\
Amara, M, D. Finsterbusch, B. Luy, B. Luthi, and H.R. Ott, 1995, 
Phys. Rev. B{\bf 51}, 16407.\\
Amit, D.J., Y.Y. Goldshmidt, and G. Grinstein, 1980, J. Phys. A{\bf
13}, 585.\\
Amitsuka, H., T. Hidano, T. Honma, H. Mitamura, and T. Sakakibara,
1993,
Physica B{\bf 186-188 }, 337.\\
Amitsuka, H., and T. Sakakibara, 1994, J. Phys. Soc. Jap. {\bf 63},
736.\\
Amitsuka, H., T. Shimamoto, T. Honma, and T. Sakakibara, 1995, Physica B{\bf 206-207}, 
461.\\
Anders, F., 1995a, Physica B{\bf 206\& 207}, 177.\\ 
Anders, F., 1995b, J. Phys. Cond. Matt. {\bf 7}, 2801.\\
Anders, F.,  and N. Grewe, 1994, Europhysics Letters {\bf 26}, 551. \\
Anders, F., M. Jarrell, and D.L. Cox, 1997, Phys. Rev. Lett. {\bf 78},
2000.\\
Anderson, P.W., 1950, Phys. Rev. {\bf 79}, 350.\\
Anderson, P.W., 1961, Phys. Rev. {\bf 124}, 41.\\
Anderson, P.W., 1967, Phys. Rev. Lett. {\bf 18}, 1049.\\
Anderson, P.W., 1970, J. Phys. C. {\bf 3}, 2346.\\
Anderson, P.W., 1981, in {\it Valence Fluctuations in Solids}, eds.
L.M.
Falikov, W. Hanke, and M.B. Maple (North-Holland, Amsterdam), p. 451.\\
Anderson, P.W., B.I. Halperin, and C.M. Varma, 1971, Philos. Mag. {\bf
25}, 1.\\
Anderson, P.W., and C.C. Yu, Proc. Intl. School of Physics ``Enrico
Fermi'', eds. F. Bassani {\it et al.} (North-Holland, Amsterdam, 1985),
p. 767.\\
Anderson, P.W., G. Yuval, and D.R. Hamann, 1970, Phys. Rev. B{\bf 1},
4464.\\
Andraka, B., 1994a,  Phys. Rev. B {\bf 49}, 3589.\\
Andraka, B., 1994b, Phys. Rev. B{\bf 49}, 348.\\
Andraka, B. and A.M. Tsvelik, 1991, Phys. Rev. Lett. {\bf 67}, 2886.\\
Andraka, B., and G.R. Stewart, 1994, Phys. Rev. B{\bf 49}, 12359.\\
Andrei, N., 1980, Phys. Rev. Lett. {\bf 45}, 379.\\
Andrei, N., 1982, Phys. Letters {\bf 87A}, 299. \\
Andrei, N., 1995, unpublished.\\
Andrei, N. and C. Destri, 1984, Phys. Rev. Lett. {\bf 52}, 364.\\
Andrei, N., and A. Jerez, 1995, Phys. Rev. Lett. {\bf 74}, 4507
(1995).\\
Andrei, N., K. Furuya, and J.H. Lowenstein, 1983, Rev. Mod. Phys. {\bf
55}, 331.\\
Appelbaum, J.A., and L.Y.L. Shen, 1972, Phys. Rev. B{\bf 5}, 544.\\
Aronson, M.C., J.D. Thompson, J.L. Smith, Z. Fisk, and
M.W. McElfresh, 1989, Phys. Rev. Lett. {\bf 63}
2311.\\
Aronson, M.C., R. Osborn,
R.A. Robinson, J.W. Lynn, R. Chau, C.L. Seaman, and M.B. Maple, 1995,
Phys. Rev. Lett. {\bf 75},  725.\\
Auerbach, A. and K. Levin, 1986, Phys. Rev. Lett. {\bf 57}, 877.\\
Baeriswyl, D., P. Horsch, and K. Maki, 1988, Phys. Rev. Lett. {\bf 60},
70.\\
Balatsky, A.V., and E. Abrahams, 1992, Phys. Rev. B{\bf 45}, 13125.\\
Balatsky, A.V., E. Abrahams, D.J. Scalapino, and J.R. Schrieffer, 1994,
Physica B{\bf 199\&200},
363.\\
Balkashin, O.P., R.J.P. Keijsers, H. van Kempen, Yu. A. Koleshnichenko,
and O.I. Shklyarevskii, 1997, preprint.\\
Bander, M., 1976, Phys. Rev. D{\bf 13}, 1566.\\
Barnes, S.E., 1976, J. Phys. F{\bf 6}, 1375.\\
Barnes, S.E., 1988, Phys. Rev. B{\bf 37}, 3671.  \\
Berezinskii, V.L., 1974, JETP Lett. {\bf 20}, 287 
(ZhETP Pisma {\bf 20}, 628).\\
Bermon, S., and C.K. So, 1978, Phys. Rev. Lett. {\bf 40}, 53.\\
Bernal, O.O., D.E. MacLaughlin, H.G. Lukefahr, and B. Andraka, 
1995, Phys. Rev. Lett. 
{\bf 75}, 2023.\\
Bickers, N.E., 1987, Rev. Mod. Phys. {\bf 59}, 845.\\
Bickers, N.E., D.L. Cox, and J.W. Wilkins, 1987, Phys. Rev. B{\bf 36},
2036.\\
Black, J.L., 1981, in {\it Metallic Glasses}, eds. H.J. G\"{u}ntherodt
and H.
Beck (Springer, New York), p. 167.\\
Black, J.L., and B.L. Gy\"{o}rffy, 1978, Phys. Rev. Lett. {\bf 41},
1595.\\
Black, J.L., B.L. Gy\"{o}rffy, and J. J\"{a}ckle, 1979, Philos. Mag.
B{\bf
40}, 331.\\
Black, J.L. K. \vld, and A. Zawadowski, 1982, Phys. Rev. B{\bf
26}, 1559.\\
Broholm, C., H. Lin, P.T. Matthews, T.E. Mason, W.J.L. Buyers,
M.F. Collins, A.A. Menovsky, J.A. Mydosh, J.K. Kjems, 1992,
Phys. Rev. B{\bf 42}, 12809.\\
Bulla, R., and A.C. Hewson, 1997, preprint (cond-mat/9701152).\\
Bulla, R., A.C. Hewson, and G.-M. Zhang, 1997, preprint
(cond-mat/9704024).\\
Burin, A.L., and Yu. Kagan, 1996, JETP {\bf 82}, 159 
(Zh. Eksp. Teor. Fiz. {\bf 109}, 299).  \\
Buschinger, B., C. Geibel, and F. Steglich, 1996, unpublished.\\
Buyers, W.J.L., A.F. Murray, T.M. Holden, E.C. Svensson, P. de
V. Du Plessis, G.H. Lander, O. Vogt, 1980, Physica {\bf 102B},
291. \\
Caldeira, A.O. and A.J. Leggett, 1981, Phys. Rev. Lett. {\bf 46},
211.\\
Caldeira, A.O. and A.J. Leggett, 1983, Ann. Phys. (NY) {\bf 149},
374.\\
Cardy, J.L., 1984, Nuc. Phys. B{\bf 240}, 514.\\
Cardy, J.L., 1986a, Nuc. Phys. B{\bf 270}, 186.\\
Cardy, J.L., 1986b, Nuc. Phys. B{\bf 275}, 200.\\
Chandra, P., A.P. Ramirez, P. Coleman, E. Br\"{u}ch, A.A. Menovsky, Z.
Fisk,
and E. Bucher, 1994, Physica B{\bf 199\&200}, 426.\\
Chattopadhyay, A., and M. Jarrell, 1996, to be published in Phys. Rev. B
(Rap. Comm.) (cond-mat/9609238).\\
Chun, K. and N.O. Birge, 1993, Phys. Rev. B{\bf 48} 11500.\\
Clarke, D., T. Giamarchi, and B. Shraiman, 1993, Phys. Rev. B{\bf 48},
7070.\\
Cochrane, R.W., R. Harris, J.O. Strom-Olsen, and M.J. Zuckerman, 1975,
Phys.
Rev. Lett. {\bf 35}, 676.\\
Coleman, P., 1983, Phys. Rev. B{\bf 28}, 5255.\\
Coleman, P., 1984, Phys. Rev. B{\bf 29}, 3035.\\
Coleman, P., 1987, Phys. Rev. B{\bf 35}, 5072.\\
Coleman, P., and A.J. Schofield, 1995, Phys. Rev. Lett. {\bf 75}, 2184.\\
Coleman, P., L. Ioffe, and A.M. Tsvelik, 1995, Phys. Rev. B{\bf 52},
6611.\\
Coleman, P., E. Miranda, and A.M. Tsvelik, 1993, Phys. Rev. Lett. {\bf
70}, 2960.\\
Coleman, P., E. Miranda, and A.M. Tsvelik, 1994, Phys. Rev. B{\bf 49},
8955.\\
Coleman, P., E. Miranda, and A.M. Tsvelik, 1995, Phys. Rev. Lett. 
{\bf 74}, 1653.\\
Coleman, P., A. Georges, and A.M. Tsvelik, 1997, J. Phys. Cond. Matt.
{\bf 9}, 345.  \\
Coleman, P., A.M. Tsvelik, N. Andrei, and H.Y. Kee, 1997, preprint
(cond-matt/9707002). \\
Coqblin, B. and J.R. Schrieffer, 1969, Phys. Rev. {\bf 185}, 847.\\
Costi, T.A., P. Schmitteckert, J. Kroha, and P. W\"{o}lfle, 
1994, Phys. Rev. Lett. {\bf 73}, 1275.\\
Costi, T.A., J. Kroha, and P. Wolfle, 1996, Phys. Rev. B{\bf 53}, 1850.\\
Cox, D.L,  1987a, Phys. Rev. B{\bf 35}, 4561.\\
Cox, D.L., 1987b, Phys. Rev. Lett. {\bf 59}, 1240.\\
Cox, D.L., 1987c, Phys. Rev. Lett. {\bf 58}, 2730.\\
Cox, D.L., 1987d, in {\bf Proceedings of the 5th
International Conference on Valence Fluctuations},
eds. S. Malik and K.
Gupta (Plenum, New York) p. 553.\\
Cox, D.L., 1988a), Physica C{\bf 153}, 1642.\\
Cox, D.L., 1988b), J. Mag. Mag. Mat. {\bf 76/77}, 53.\\
Cox, D.L., 1990, preprint, unpublished. \\
Cox, D.L.,  1993, Physica B {\bf 186-188}, 312. \\
Cox, D.L., 1996, Physica B {\bf 223\&224}, 453.  \\
Cox, D.L., and A. Ruckenstein, 1993, Phys. Rev. Lett.{\bf 71}, 1613.\\
Cox, D.L., and M. Makivic, 1994, Physica B{\bf 199\&200}, 391.\\
Cox, D.L, and A.W.W. Ludwig, 1997, to be published.\\
Cox, D.L., and M.B. Maple, 1995, Physics Today {\bf 48}, 32.\\
Cox, D.L., K. Kim, and A.W.W. Ludwig, 1997, unpublished.\\
Cox, D.L., 1996, Physica B{\bf 223\&224}, 453.\\
Cox, D.L., and M. Jarrell, 1996, J. Phys. Cond. Matt. {\bf 8},
9825. \\
Cragg, D.M. and P.Lloyd, 1978, J. Phys. C {\bf 11}, L597.\\
Cragg, D.M., P. Lloyd, and P. \noz, 1980, J. Phys. C{\bf 13}, 803.\\
Cruz, L., P. Phillips, and A.H. Castro-Neto, 1995, Europhysics Lett.
{\bf 29}, 389. \\
Dai, P., H.A. Mook, C.L. Seaman, M.B. Maple, and J.P. Koster, 1995, Phys. 
Rev. Lett. {\bf 75}, 1202.\\
de Haas, W.J., J. de Boer, and G.J. van den Berg, 1933/34, 
Physica {\bf 1}, 1115.\\
DeGeorgi, L., and H.R. Ott, 1996, J. Phys. Cond. Matt. {\bf 8},
9901.\\
DeGeorgi, L., 1997, private communication.\\
Delft, J. von, D.C. Ralph, R.A. Buhrman, A.W.W. Ludwig, and V.
Ambegaokar, 1997a, submitted to Annals of Physics (cond-mat/9702048).\\
Delft, J. von, A.W.W. Ludwig, and V. Ambegaokar, 1997b, submitted to 
Annals of Physics (cond-mat/9702049).\\
Delft, J. von, and G. \zar, 1997, preprint. \\
Desgranges, H.-U., 1985, J. Phys. C{\bf 18}, 5481.\\
Dobrosavljevic, V., T. Kirkpatrick, G. Kotliar, 1992, Phys. Rev. Lett. 
{\bf 6}, 1113.\\
Emery, V.J., and S.A. Kivelson, 1992, Phys. Rev. B{\bf 46}, 10812.\\
Emery, V.J., and S.A. Kivelson, 1993, Physica {\bf 209C}, 597.\\
Emery, V.J. and S.A. Kivelson, 1993, Phys. Rev. Lett. {\bf 71}, 3701.\\
Emery, V.J., and S.A. Kivelson, 1994, in {\bf Fundamental Problems in
Statistical Mechanics, VII}, eds. H. van Beijeren and M.E. Ernst
(North-Holland-Elsevier, Amsterdam) pp. 1-26.\\
Endo, T., and T. Kino, 1988, J. Phys. F{\bf 18}, 2203.\\
Esquinazi, P., H.M. Ritter, H. Neckel, G. Weiss, and S. Hunklinger,
1986, Z. Phys. B{\bf 64},
81.\\
Esquinazi, P., R. K\"{o}nig, and F. Pobell, 1992, Z. Phys. B{\bf 87},
305.\\
Fabrizio, M. and A.O. Gogolin, 1994, Phys. Rev. B{\bf 50}, 17732.\\
Fabrizio, M. and G. \zar, 1996, Phys. Rev. B{\bf 54}, 10008.\\
Fabrizio, M., A.O. Gogolin, and P. \noz, 1995a, Phys. Rev. Lett. {\bf
74}, 4503. \\
Fabrizio, M., A.O. Gogolin, and P. \noz, 1995b, Phys. Rev. B{\bf
51}, 16088.\\ 
Falci, G., G. Sch\"{o}n, and G. \zim, 1995, Phys. Rev. Lett. {\bf 74},
3257. \\
Fendley, P., 1993, Phys. Rev. Lett. {\bf 71}, 2485.\\
Fowler, M. , and  A. \zow, 1971, Sol. St. Comm. {\bf 9}, 471.\\
Fradkin, E., 1991, {\bf Field Theories of Condensed Matter Systems},
(Addison Wesley,
Cambridge, MA)\\
Friedel, J., 1952, Philos. Mag. {\bf 43}, 153.\\
Fujimoto, S., N. Kawakami, and S.K. Yang, 1996, 
J. Phys. Korea {\bf 29}, 136.\\
Fulde, P., 1978, in {\bf Handbook on the Physics and Chemistry of the
Rare Earths, Vol. II}, eds. K.A. Gschneidner and L. Eyring
(Elsevier, Amsterdam), Ch. 17.\\
Fye, R.M., 1994, Phys. Rev. Lett.{\bf 72}, 916.\\
Fye, R.M., and J.E. Hirsch, 1989, Phys. Rev. B{\bf 40}, 4780.\\
Gammel, J.T. and D.K. Campbell, 1988, Phys. Rev. Lett. {\bf 60}, 71.\\
Gan, J., 1995a, Phys. Rev. B{\bf 51}, 8287.\\
Gan, J., 1995b, Phys. Rev. Lett. {\bf 74}, 2583.\\
Gan. J, N. Andrei, and P. Coleman, 1993, Phys. Rev. Lett. {\bf 70},
686.\\
Gaudin, M., 1967, Phys. Lett. A{\bf 24}, 55.\\
Georges, A., and A.M. Sengupta, 1995, Phys. Rev. Lett. {\bf 74}, 2808.\\
Georges, A., G. Kotliar, and Q. Si, 1992, Int. J. Mod. Phys. {\bf 6},
705 .\\
Georges, A., G. Kotliar, W.  Krauth, and M. Rozenberg, 1996,
Rev. Mod. Phys. {\bf{68}}, 13.\\
Giamarchi, T., and H.J. Schulz, 1988, Phys. Rev. B{\bf 37}, 325.\\988,
Phys. Rev. B{\bf 37}, 325.\\
Giamarchi, T., C.M. Varma, A.E. Ruckenstein, and P. \noz, 1993, Phys.
Rev. Lett. {\bf 70}, 3967.\\
Giordano, N. in {\bf Mesoscopic Phenomena in Solids}, 1991, eds. B.L.
Altshuler,
P.A. Lee, and R.A. Webb (Elsevier Science, Amsterdam), p. 131.\\
Gogolin, A.O., 1996, Phys. Rev. B{\bf 53}, R5990.\\
Golden, J.M., and B.I. Halperin, 1996a, Phys. Rev. B{\bf 53}, 3893.\\
Golden, J.M., and B.I. Halperin, 1996b, Phys. Rev. B{\bf 54}, 16757.\\
Golding, B., J.E. Graebner, A.B. Kane, and J.L. Black, 1978, Phys.
Rev.
Lett. {\bf 41},
1487.\\
Golding, B., N.M. Zimmerman, and S.N. Coppersmith, 1992, Phys. Rev.
Lett. {\bf
68}, 998.\\
Gorymychin,  E.A. and  R. Osborne, 1994, Phys. Rev.
 B{\bf 47}, 14280.\\
Graebner, J.E., B. Golding, R.J. Schutz, F.S.L. Chu, and H.S. Chen,
1978, Phys. Rev. Lett. {\bf 39}, 1480. \\
Gregory, S., 1992, Phys. Rev. Lett. {\bf 68}, 2070.\\
Grewe, N., 1983, Z. Physik B{\bf 53}, 271.\\
Grewe, N., 1984, Sol. St. Comm. {\bf 50}, 19.\\
Grewe, N. and H. Keiter, 1981, Phys. Rev. B{\bf 24}, 4420.\\
Grewe, N., and T. Pruschke, 1985, Z. Phys. B{\bf 60}, 311.\\
Grewe, N., and F. Steglich, 1991, in {\it Handbook of the
Physics and Chemistry of the Rare Earths, Vol. 14}, eds. K.A.
Gschneidner Jr.
and L.L. Eyring (Elsevier, Amsterdam), p. 343. \\
Gr\"{u}ner, G., 1974, Adv. in Phys. {\bf 23}, 941.\\
Gunnarsson, O. and K. Sch\"{o}nhammer, 1983, Phys. Rev. Lett. {\bf 50},
604.\\
Gunnarsson, O. and K. Sch\"{o}nhammer, 1984, Phys. Rev. B{\bf 28},
4315.\\
Haldane, F.D.M., 1981, J. Phys. C {\bf 14}, 2585.\\
Hamann, D.R., 1970, Phys. Rev. B{\bf 2}, 1373.\\
Hamermesh, M., 1962, {\it Group Theory in Physics} (Addison-Wesley,
Reading,
Mass.), pp. 161-181.\\
Han, J., 1994, private communication.\\
Heid, R., 1995, Z. Phys. B{\bf 99}, 15.\\
Heid, R., Ya.B. Bazaliy, V. Martisovits, and D.L. Cox, 1995,
Phys. Rev.  Lett. {\bf 74}, 2571.\\
Heid, R., Ya. B. Bazaliy, V. Martisovits, and D.L. Cox, 1996, 
Physica B{\bf 223\&224}, 33. \\
Herbst, J.F. and J.W. Wilkins, 1987, in {\bf Handbook
on the Chemistry and Physics of the Rare Earths, Vol. 10}, eds. K.
A. Gschneidner and L. Eyring (Elsevier, Amsterdam), p.321.\\
Hettler, M.H.,  J. Kroha, and S. Hershfield, 1994, Phys. Rev. Lett.
{\bf
73}, 1967. \\
Hewson, A.C., 1993, {\bf The Kondo Problem to Heavy Fermions}
(Cambridge Press,
Cambridge UK).\\
Hirsch, J., 1983, Phys. Rev. B{\bf 28}, 4059.\\
Hirsch, J., 1989, Phys. Lett. {\bf 138}A, 83.\\
Hirsch, J., and R.M. Fye, 1986, Phys. Rev. Lett. {\bf 56}, 2521. \\
Hirst, L.L, 1970, Phys. Kondens. Matt. {\bf 10}, 255.\\
Hirst, L.L., 1978, Adv. in Physics. {\bf 27}, 231.\\
Horn, S., E. Holland-Moritz, M. Loewenhaupt, F. Steglich, H. Scheuer,
A.
Benoit, and J. Flouquet, 1981, Phys. Rev. B{\bf 32}, 3260.\\
Hubbard, J., 1963, J. Proc. Roy. Soc. (London) {\bf 276}, 238.\\
Hunklinger, S., and W. Arnold, 1976, in {\it Physical Acoustics}, eds.
W.P.
Mason and R.N. Thurston (Academic, New York), v. 12, p. 1555.\\
Inagaki, S., 1979, Prog. Th. Phys. {\bf 62}, 1441.\\
Ingersent, K., B.A. Jones, and J.W. Wilkins, 1992, Phys. Rev. Lett.
{\bf 69}, 2594.\\
K. Ingersent, and B.A. Jones 1994, Physica B{\bf 199\&200}, 402 .\\
Ishiguro, T., H. Kaneko, Y. Nogami, H. Ishimoto, H. Nishiyama, J.
Tsukamoto,
A. Takahashi, M. Yamaura, T. Hagiwara, and K. Sato, 1992, Phys. Rev.
Lett. {\bf 69}, 660.\\
Jaccard, D., 1992, Phys. Lett. A{\bf 163}, 475.\\
Jansen, A.G.M., A.P. van Gelder, and P. Wyder, 1980, J. Phys. C{\bf
13}, 6073.\\
Jarrell, M., 1995, Phys. Rev. B{\bf 51}, 7429. \\
Jarrell, M., and T. Pruschke, 1996, private communication.\\ 
Jarrell, M., H.-B. Pang, D.L. Cox, and K.-H. Luk, 1996, Phys.
Rev. Lett. {\bf 77}, 1612.\\
Jarrell, M., {\it et al.}, 1997, unpublished.\\
Jarrell, M., H.-B. Pang, and D.L. Cox, 1997, Phys. 
Rev. Lett. {\bf 78}, 1996. \\
Jarrell, M., H.-B. Pang, D.L. Cox, F. Anders, and A.
Chattopadhyay, 1996c, Physica B {\bf 230-232}, 557.\\
Jerez, A., 1995, private communication (unpublished).\\
Jones, B.A., 1988, Ph.D. dissertation (Cornell Univ., unpublished).\\
Jones, B.A., and C.M. Varma, 1987, Phys. Rev. Lett. {\bf 58}, 843.\\
Jones, B.A., C.M. Varma, and J.W. Wilkins, 1988, Phys. Rev. Lett. {\bf
61}, 125.\\
Kac, V.G., and K. Peterson, 1984, Adv. Math {\bf 53}, 125.\\
Jones, B.A., and K. Ingersent, 1994, Physica B{\bf 199\&200}, 411.\\
Han, J., 1995, private communication.\\
Kagan, Yu. and N.V. Prokof'ev, 1986, Zh. Eksp. Teor. Fiz. {\bf 90},
2176 [Sov.
Phys. JETP {\bf 63}, 1276 (1987(a))].\\
Kagan, Yu. and N.V. Prokof'ev, 1987, Zh. Eksp. Teor. Fiz. {\bf 93}, 366
[Sov.
Phys. JETP {\bf 66}, 211 (1987(b))].\\
Kagan, Yu. and N.V. Prokof'ev, 1989, Zh. Eksp. Teor. Fiz. {\bf 96},
1473 [Sov.
Phys. JETP {\bf 69}, 836 (1989)].\\
Kagan, Yu. and N.V. Prokof'ev, 1992, in {\bf Quantum Tunneling in
Condensed
Matter},
eds. Yu. Kagan and A.J. Leggett (Elsevier, Amsterdam), p. 37.\\
K\"{a}stner, J., H.J. Schink, and E.F. Wasserman, 1981, Sol. St. Comm.
{\bf 33}, 527.\\
Katayama, S., and K. Murase, 1980, Sol. St. Comm. {\bf 36}, 707.\\
Katayama, S., S. Maekawa, H. Fukuyama, 1987, J. Phys. Soc. Jap. {\bf
56}, 697.\\
Keiter, H. and J.C. Kimball, 1971, Int. J. Magn. {\bf 1}, 233.\\
Keijsers, R.J., O.J. Shklyarevskii, and H. van Kempen, 1995, Phys.
Rev.
B{\bf 51}, 5628. \\
Keijsers, R.J., O.J. Shklyarevskii, and H. van Kempen, 1996, Phys.
Rev. Lett. {\bf 77}, 3411.\\
Kim, E.,  T.-S. Kim, J. Han, M. Makivic, and D.L. Cox, 1995, to be
published.\\
Kim, J.S., B. Andraka, C.S. Jee, S.B. Roy, and G.R. Stewart, 1990, 
Phys. Rev. B{\bf 41}, 11073.\\
Kim, J.S., B. Andraka, and G.R. Stewart, 1992, Phys. Rev. B{\bf 45},
12081.\\
Kim, T.-S., 1994, private communication.\\
Kim, T.-S., 1995, Ph.D. dissertation (The Ohio State University,
unpublished).\\
Kim, T.-S., and D.L. Cox, 1995, Phys. Rev. Lett. {\bf 75}, 1622.\\
Kim, T.-S., and  D.L. Cox, 1996, Phys. Rev. B{\bf 54}, 6494.\\
Kim, T.-S. and D.L. Cox, 1997, to be published in Phys. Rev. B, 
(cond-mat/9508129).\\
Kim, T.-S., L.N. Oliveira, and D.L. Cox, 1997,  to be published in 
Phys. Rev. B (cond-mat/9606095).\\
Kim, W.W.,J. S. Kim, B. Andraka, and G. R. Stewart, 
1993, Phys. Rev. B{\bf 47}, 12403.\\
Kivelson, S., W.P. Su, J.R. Schrieffer, and A.J. Heeger, 1987, Phys.
Rev. Lett.
{\bf 58}, 1899.\\
Kondo, J., 1964, Prog. Theor. Phys. {\bf 32}, 37.\\
Kondo, J., 1976a, Physica {\bf 84B}, 40.\\
Kondo, J., 1976b, Physica {\bf 84B}, 207.\\
Kondo, J., 1984a, Physica {\bf 123B}, 175.\\
Kondo, J., 1984b, Physica {\bf 126B}, 377.\\
Kondo, J., 1985, Physica {\bf 132B}, 303.\\
Kondo, J., 1986, Hyperfine Interactions {\bf 31}, 117 (1986).\\
Koster, G.F., J.O. Dimmock, R.G. Wheeler, and H. Statz, 1963, {\it
Properties of the
Thirty-Two Point Groups} (M.I.T. Press, Cambridge, Mass.). \\
Kotliar, G., E. Abrahams, A.E. Ruckenstein, C.M. Varma, P.B. Littlewood
and S. Schmitt-Rink, 1991,
Europhysics Lett. {\bf 15}, 655.\\
Kozub, V.I., 1995, private communication.\\
Kozub, V.I. and I.O. Kulik, 1986, Zh. Eksp. Teor. Fiz. {\bf 91}, 2243
(Sov. Phys. JETP {\bf 64}, 1332 (1986)).  \\
Krishna-murthy, H.R., J.W. Wilkins, and K.G. Wilson, 1980a, Phys. Rev.
B{\bf 21}, 1003.\\
Krishna-murthy, H.R., J.W. Wilkins, and K.G. Wilson, 1980b, Phys. Rev.
B{\bf 21}, 1044.\\
Kroha, J., P.J. Hirschfeld, K.A. Muttalib, and P. Wolfle, 1992, Sol. St. 
Comm. {\bf 83}, 1003.\\
Kroha, J., P. Wolfle, and T.A. Costi, 1997, Phys. Rev. Lett. {\bf 79},
261 (1997).\\ 
Kuramoto, Y., 1983, Z. Phys. B{\bf 53}, 37.\\
Kuramoto, Y. and H. Kojima, 1984, Z. Phys. B{\bf 57}, 95.\\
Kuramoto, Y., and E. M\"{u}ller-Hartmann, J. Mag. Magn. Mat. {\bf 52},
122.\\
Kusunose, H., and K. Miyake, 1996, J. Phys. Soc. Jap. {\bf 65}, 3032.\\
Lang, J.K., Y. Baer, and P.A. Cox, 1981, J. Phys. F{\bf 11}, 121.\\
Langreth, D.C., 1966, Phys. Rev. {\bf 150}, 516.\\
Lasjaunias, J.C., and A. Ravex, 1983, J. Phys. F{\bf 13}, L101.\\
Lee, P.A., and T.V. Ramakrishnan, 1985, Rev. Mod. Phys. {\bf 57},
289.\\
Lea, K.R., M.J.M. Leask, and W.P. Wolf, 1962, J. Phys. Chem. Solids
{\bf 23}, 1381.\\
Leggett, A.J., S. Chakravarty, A.T. Dorsey, M.P.A. Fisher, A. Garg, and
W.
Zwerger, 1987, Rev. Mod. Phys. {\bf 59}, 1.\\
Liu, L.Z., J.W. Allen, C.L. Seaman, M.B. Maple, Y. Dalichaouch, J.-S.
Kang,
M.S. Torikachvili, and M.A. Lopez de la Torre, 1992, Phys. Rev. Lett.
{\bf 68},
1034.\\
Livermore, C., C.H. Crouch, R.M. Westervelt, K.L. Campman, and A.C.
Gossard, 1996, Science {\bf 274}, 1332. \\
Lloyd, P., A.W. Mirtschin, and D.M. Cragg, 1981, J. Phys. A. {\bf 14},
659.\\
L\"{o}hneysen, M. Platte, W. Sander, H. J. Schink, G.V. Minigerode, and
K. Samwer, 1980, J. Physique {\bf 41}, C8-745.\\
Ludwig, A.W.W., 1994a, Int. J. Mod. Phys. B{\bf 8}, 347.  \\
Ludwig, A.W.W., 1994b, Physica B{\bf 199-200}, 406.\\
Ludwig, A.W.W. and I. Affleck, 1991, Phys. Rev. Lett. {\bf 67}, 3160.\\
Ludwig, A.W.W., and I. Affleck, 1994, Nucl. Phys. B{\bf 428}, 545.\\
Luk, K.-H., 1992, Ph.D. dissertation (The Ohio State University,
unpublished).\\
Luk, K.-H., M. Jarrell, and D.L. Cox, 1994, Phys. Rev. B{\bf 49},
16066.\\ 
Luke, G., 1995, private communication.\\
MacLaughlin, D.E., O.O. Bernal, and H.G. Lukefahr, 1996, J. Phys. Cond.
Matt. {\bf 8}, 9855.\\
Maekawa, S.,  S. Takahashi, and M. Tachiki, 1984, J. Phys. Soc. Jap.
{\bf 53}, 702.\\
Maekawa, S., S. Takahashi, S. Kashiba, and M. Tachiki, 1985, J. Phys.
Soc. Jpn.
{\bf 54}, 1955.\\
Mahan, G.D., 1990, {\bf Many Particle Physics (Second Edition)}
(Plenum, New York).\\
Maldacena, J.M., and A.W.W. Ludwig, 1996, to be published in Nuc. Phys. B.
(cond-mat/9502109).\\
Maple, M.B., C.L. Seaman, D.A. Gajewski, Y. Dalichaouch, V.B. Barbetta,
M.C. deAndrade,
H.A. Mook, H.G. Lukefahr, O.U. Bernal, and D.E. MacLaughlin, 1994, J.
Low. Temp. Phys.
{\bf 94}, 225.\\
Maple, M.B., R.P. Dickey, J. herrmann, M.C. de Andrade, E.J. Freeman,
D.A. Gajewski, and R. Chau, 1996, J. Phys. Cond. Matt. {\bf 8}, 9773.\\
Marumoto, K., T. Takeuchi, and Y. Miyako, 1996, Phys. Rev. B{\bf 54},
12194.\\
Matsuura, T., and K. Miyake, 1986a, J. Phys. Soc. Jap. {\bf 55}, 29.\\
Matsuura, T., and K. Miyake, 1986b, J. Phys. Soc. Jap. {\bf 55}, 610.\\
Matveev, K.A., 1991, Sov. Phys. JETP {\bf 72}, 892 [Zh. Eksp. Teor. Fiz. 
{\bf 99}, 1598].\\
Matveev, K.A., 1995, Phys. Rev. B{\bf 51}, 1743.\\
Matveev, K.A., L.I. Glazman, and H.U. Baranger, 1996a, Phys. Rev. B{\bf 53},
1034.\\
Matveev, K.A., L.I. Glazman, and H.U. Baranger, 1996b, Phys. Rev. B{\bf
54}, 5637.\\
McElfresh, M.W., M.B. Maple,
J.O. Willis, D. Schiferl, J.L. Smith, Z. Fisk, and
and D.L. Cox, 1993, Phys. Rev. B{\bf 48}, 10395. \\
McEwen, K., M.J. Bull, and R.S. Eccleston, 
1995, Physica B{\bf 206\&207}, 112. \\
Menge, B., and E. M\"{u}ller-Hartmann, 1988, Z. Phys. B{\bf 73}, 225.\\
Metzner, W.,  and D Vollhardt, 1989, Phys. Rev. Lett.
 {\bf 62}, 324.\\ 
Mezei, F. and A. \zow, 1971, Phys. Rev. B{\bf 3}, 3127.\\
Mih\'{a}ly, L. and A. \zow, 1978, J. Phys. Lett. {\bf 39}, L-483.\\
Miranda, E., V. Dobrosavljev\'{i}c, and G. Kotliar, 1996, J. Phys.
Cond. Matt. {\bf 8}, 9871.\\
Miranda, E., V. Dobrosavljev\'{i}c, and G. Kotliar, 1997, Phys. Rev.
Lett. {\bf 78}, 290.\\
Mirtschin, A.W., 1986, Ph.D. Dissertation, University of Monash
(unpublished).\\
Mirtschin, A.W., and P. Lloyd, 1984, J. Phys. C{\bf 17}, 5399.\\
Millis, A.J. and P.A. Lee, 1987, Phys. Rev. B{\bf 35}, 3394 (1987).\\
Molenkamp, L.W., K. Flensberg, and M. Kemerink, 1995, Phys. Rev. Lett. 
{\bf 75}, 23.\\
Mook, H., C.L. Seaman, M.B. Maple, M.A. Lopez de la Torre, D.L.
Cox, M. Makivic, 1993, Physica B{\bf 186-188}, 341.\\
Morin, P., and D. Schmitt, 1981, Phys. Rev. B{\bf 23}, 5936.\\
Moustakas, A. and D. Fisher, 1995, Phys. Rev. B{\bf 51}, 6908.\\
Moustakas, A., and D. Fisher, 1996, Phys. Rev. B{\bf 53}, 4300.\\
Moustakas, A., and D. Fisher, 1997, preprint (cond-mat/9607208).\\
M\"{u}ller-Hartmann, E., 1984, Z. Phys. B{\bf 57}, 281.\\
M\"{u}ller-Hartmann, E., 1989, Z. Phys. B {\bf 74}, 507.\\
Muramatsu, A. and F. Guinea, 1986, Phys. Rev. Lett. {\bf 57}, 2337.\\
Newns, D.M. and N. Read, 1987, Adv. in Phys. {\bf 36}, 799.\\
Ngai, K.S., J.A. Appelbaum, M.H. Cohen, and J.C. Phillips, 1967, Phys.
Rev. {\bf 163}, 352.\\
Niksch, M., W. Assmus, B. L\"{u}thi, H.R. Ott, and J.K. Kjems, 1982, Helv.
Phys. Acta {\bf 55}, 688.\\
\noz, P., 1969, unpublished.\\
\noz, P., 1974, J. Low Temp. Phys. {\bf 17}, 31.\\
Nozi\`{e}res, P. and A. Blandin, 1980, J. Phys. (Paris) {\bf 41},
193.\\
Nozi\`{e}res, P. and C.T. De Dominicis, 1969, Phys. Rev. {\bf 178},
1097.\\
\noz, P., J. Gavoret, and B. Roulet, 1968, Phys. Rev. {\bf 178},
1084.\\
Nunes, A.C., J.W. Rasul, and G.A. Gehring, 1986, J. Phys. C{\bf 19},
1017.\\
Ott, H.R., 1987,  Prog. Low. Temp. Phys.  {\bf 6}, 215.\\
Pang, H.-B., 1992, Ph.D. dissertation (The Ohio State University,
unpublished).\\
Pang, H.-B., 1994, Phys. Rev. Lett. {\bf 73}, 2736.\\
Pang, H.-B., and D.L. Cox, 1991, Phys. Rev. B{\bf 44}, 9454.\\
Parcollet, O., and A. Georges, 1997, preprint (cond-mat/9707337).\\
Pasquier, C., U. Meirav, F.I.B. Williams, D.C. Glattli, Y. Jin, and B. 
Etienne, 1993, Phys. Rev. Lett. {\bf 70}, 69.\\
Penc, K., and A. \zow, 1994, Phys. Rev. B{\bf 50}, 10578. \\
Phillips, W.A., 1972, J. Low Temp. Phys. {\bf 7}, 351.\\
Pruschke, Th., 1989, Ph.D. Dissertation, Technische Hochschule,
Darmstadt, Germany (unpublished).
Pruschke, Th., D.L. Cox, and M. Jarrell, 1993, Phys. Rev. B{\bf 47},
3553.\\
Pruschke, Th., M. Jarrell and J.K. Freericks, 1995, 
 Adv. in Phys. {\bf{44}}, 187.\\
Ralls, K.S.,  and R.A. Buhrman, 1988, Phys. Rev. Lett. {\bf 60},
2434.\\
Ralls, K.S., and R.A. Buhrman, 1991, Phys. Rev. B{\bf 44}, 5800.\\
Ralph, D.C., 1993, Ph.D. Dissertation (Cornell University,
unpublished).\\
Ralph, D.C.,  and R.A. Buhrman, 1992, Phys. Rev. Lett. {\bf 69},
2118.\\
Ralph, D.C., and R.A. Buhrman, 1995, Phys. Rev. B{\bf 51}, 3554.\\
Ralph, D.C., A.W.W. Ludwig, J. von Delft, and R.A. Buhrman, 1994,
Phys.
Rev. Lett. {\bf 72},
1064. \\
Ralph, D.C., A.W.W. Ludwig, J. von Delft, and R.A. Buhrman, 1995,
Phys. Rev. Lett. {\bf 75}, 770 (1995).\\
Ramakrishnan, T.V., 1981, in {\it Valence Fluctuations in Solids},
eds.
L.M.
Falikov, W. Hanke, and M.B. Maple (North-Holland, Amsterdam), p. 13.\\
Ramakrishnan, T.V., 1988, J. Mag. Mag. Mat. {\bf 76\&77},
657.\\
Ramakrishnan, T.V. and K. Sur, 1982, Phys. Rev. B{\bf 26}, 1798.\\
Ramirez, A.P., P. Chandra, P. Coleman, Z. Fisk, J.L.
Smith, and H.R. Ott, 1994, Phys. Rev. Lett. {\bf 73}, 3018.\\
Read, N., K. Dharamvir, J.W. Rasul, and D.M. Newns, 1986, J. Phys.
C{\bf 19}, 1597.\\
Read, N. and D.M. Newns, 1983, J. Phys. C{\bf 16}, 3273.\\
Roulet, B., J. Gavoret, and P. \noz, 1968, Phys. Rev. {\bf 178},
1072.\\
Rudin, S., 1983, Phys. Rev. B{\bf 28}, 4825.  \\
Sacramento, P.D. and P.Schlottmann, 1989, Phys. Lett. A{\bf 142},
245.\\
Sacramento, P.D. and P.Schlottmann, 1991, Phys. Rev. B{\bf 43},
13294.\\
Sakai, O., Y. Shimizu, N. Kaneko, 1993, Physica B{\bf 186-188},
323.\\
Scalapino, D.J. and P. Marcus, 1967, Phys. Rev. Lett. {\bf 18}, 459.\\
Schiller, A., and S. Hershfield, 1995, Phys. Rev. B{\bf 51}, 12896.\\
Schiller, A., and S. Hershfield, 1997, preprint.\\
Schlottmann, P., 1978, J. Phys. {\bf C6}, 1486.\\
Schlottmann, P. and P.D. Sacramento, 1993, Adv. Physics {\bf 42},
641.\\
Schofield, A.J., 1997, Phys. Rev. B{\bf 55}, 5627.\\
Schrieffer, J.R., 1967, J. Appl. Phys. {\bf 38}, 1143. \\
Schrieffer, J.R. and P.A. Wolff, 1966, Phys. Rev. {\bf 149}, 491.\\
Seaman, C.L., M.B. Maple, B.W. Lee, S. Ghamaty, M.S. Torikachvilli,
J.-S. Kang,
L.Z. Liu, J.W. Allen, and D.L. Cox, 1991, Phys. Rev. Lett. {\bf 67},
2882.\\
Seaman, C.L., M.B. Maple, B.W. Lee, S. Ghamaty, M.S. Torikachvilli,
J.-S. Kang,
L.Z. Liu, J.W. Allen, and D.L. Cox, 1992, J. Alloys and Compounds {\bf
181}, 327.\\
Seaman, C.L. and M.B. Maple, 1994, Physica B{\bf 199\&200}, 396.\\
Sengupta, A.M., and A. Georges, 1994, Phys. Rev. B{\bf 49}, 10020.\\
Sengupta, A.M., and Y.B. Kim, 1996, Phys. Rev. B{\bf 54}, 14918.\\
Sereni, J., G. Nieva, J.G. Huber, E. Braun, F. Oster, E. Bruch,
B. Roden, D. Wohlleben, 1986, J. Mag. Mag. Mat. {\bf 63\&64}, 597.\\
Sethna, J., 1981, Phys. Rev. B{\bf 24}, 698.   \\
Shiba, H., 1970, Prog. Theor. Phys. {\bf 43}, 601.\\
Silva, J.B., W.L.C. Lima, W.C. Oliveira, J.L.N. Mello, L.N. Oliveira, 
and J.W. Wilkins, 1996, Phys. Rev. Lett. {\bf 76}, 275.\\
Slichter, C.P., 1989, {\it Principles of Magnetic Resonance, 3rd
Edition} (Springer-Verlag, Berlin), pp. 486-497.\\
Smolyarenko, I.E., and N.S. Wingreen, 1997, private communication.\\
S\'{o}lyom, J., 1974, J. Phys. F. {\bf 4}, 2269.\\
S\'{o}lyom, J., 1979, Adv. in Phys. {\bf 28}, 201 (1979).\\
S\'{o}lyom, J. and A. Zawadowski, 1974, J. Phys. F. {\bf 4}, 2269.\\
Takano, S., Y. Kumashiro, and K. Tsuji, 1984, J. Phys. Soc. Japan {\bf
53}, 4309.\\
Steglich, F., 1996, private communication.\\
Steglich, F., P. Gegenwort, C. Geibel, R. Helfrich, P. Hellmann,
M. Lang, A. Link, R. Modler, G. Sparn, N. B\"{u}ttgen, and A.
Loidl, 1996, Physica B{\bf 223\&224}, 1.\\
Tahvildar-Zadeh, A.N., M. Jarrell, and J.K. Freericks, 1997, Phys. Rev
B{\bf 55}, 3332.\\
Tanabe, Y. and K. Ohtaka, 1986, Phys. Rev. B{\bf 34}, 3763.\\
Tinkham, M., 1964, {\it Group Theory and Quantum Mechanics},
(McGraw-Hill, New York).\\
Toulouse, G., 1970, Phys. Rev. B{\bf 2}, 270.\\
Trees, B., 1993, Ph.D. Dissertation (The Ohio State University,
unpublished).\\
Trees, B., 1995, Phys. Rev. B{\bf 51}, 470. \\
Trees, B., and D.L. Cox, 1994, Phys. Rev. B{\bf 49}, 16066.\\
Tsvelik, A.M., 1985, J. Phys. C. {\bf 18}, 159.\\
Tsvelik, A.M., 1988, Pis'ma Zh. Eksp. Teor. Fiz. {\bf 48}, 502 [Sov.
Phys. JETP
Tsvelik, A.M., 1990, J. Phys. Cond. Matt. {\bf 2}, 2833.\\
Tsvelik, A.M. and P.B. Wiegman, 1984, Z. Phys. B{\bf 54}, 201.\\
Tsvelik, A.M., and P.B. Wiegman, 1985, J. Stat. Phys. {\bf 38}, 125.\\
Upadhyay, S.K., R.N. Louie, and R. A. Buhrman, 1997, to be published.\\
van der Vaart, N.C., A.J. Johnson, L.P. Kouwenhoven, D.J. Maas, W. de Jong, 
M.P. de Ruyter van Steveninck, A. van der Enden, C.J.P.M. Harmans, 
and C.T. Foxon, 1993, Physica B{\bf 189}, 99.\\
Varma, C.M., P.B. Littlewood, S. Schmitt-Rink, E.
Abrahams, and A.E. Ruckenstein, 1989, Phys. Rev. Lett. {\bf 63},
1996.\\
Verlinde, E., 1988, Nuc. Phys. {\bf B300}, 360.\\
\vld~, K., 1991, Phys. Rev. B{\bf 44}, 1019.\\
\vld~, K., 1993, Prog. Th. Phys. {\bf 90}, 43.\\
\vld~, K. and A.  \zow, 1983a, Phys. Rev. B{\bf 28}, 1564.\\
\vld~, K. and A.  \zow, 1983b, Phys. Rev. B{\bf 28}, 1582.\\
\vld~, K. and A.  \zow, 1983c, Phys. Rev. B{\bf 28}, 1596.\\
\vld~, K., A.  \zow, and G.T. \zim, 1988a, Phys. Rev. B{\bf 37},
2001.\\
\vld~, K., A.  \zow, and G.T. \zim, 1988b, Phys. Rev. B{\bf 37},
2015.\\
\vld~, K. and G.T. \zim, 1985, J. Phys. C{\bf 18}, 3755.\\
Waugh, F.R., M.J. Berry, D.J. Mar, R.M. Westervelt, K.I. Campman, and
A.C. Gossard, 1995, Phys. Rev. Lett. {\bf 75}, 705.\\
Waugh, F.R., M.J. Berry, C.H. Crouch, D.J. Mar, 
R.M. Westervelt, K.I. Campman, and
A.C. Gossard, 1996, Phys. Rev. B{\bf 53}, 1413.\\
Wiegman, P.B., 1980, Zh. Eksp. Teor. Fiz. Pis'ma Red. {\bf 32}, 392
(JETP Lett. {\bf 1}, 364).\\
Wiegman, P.B. and A.M. Finkelshtein, 1978, Zh. Eksp. Teor. Fiz. {\bf
75}, 204 [Sov. Phys. JETP {\bf 48}, 102 (1979)].\\
Wiegman, P.B. and A.M. Tsvelik, 1983a, Pis'ma Zh. Exp. Teor. Fiz. {\bf
38}, 489
[JETP Lett. {\bf 38}, 591.]\\
Wiegman, P.B., and A.M. Tsvelik, 1983b, Advances in Physics {\bf 32},
453.\\
Wilson, K.G., 1973, in {\it Collective Properties of Physical Systems,
 Nobel Symposium 24} (Academic Press, New York), p. 68.\\
Wilson, K.G., 1975, Rev. Mod. Phys. {\bf 47}, 773.\\
Wingreen, N., B.S. Altshuler, and Y. Meir, 1995, Phys. Rev. Lett. {\bf
75}, 769. \\
Winzer, K., 1975, Sol. St. Comm. {\bf 16}, 521.\\
Wyatt, A.F.G., 1964, Phys. Rev. Lett. {\bf 13}, 401.\\
Yafet, Y., C.M. Varma, and B.A. Jones, 1985, Phys. Rev. B{\bf 23},
3171.\\
K. Yamada, Kei Yosida, and K. Hanzawa, 1984, Prog. Th. Phys. {\bf 71},
450.\\
Yang, C.N., 1967, Phys. Rev. Lett. {\bf 19}, 1312.\\
Yanson, I.K. and O.I.  Shklyarevskii, 1986, Fiz. Nizk. Temp. {\bf 12},
899 [Sov. J. Low. Temp.
Phys. {\bf 12}, 509 (1986).]\\
Yanson, I.K., V.V. Fisun, A.V. Khotkevich, R. Hesper, J.M. Kraus, J.A.
Mydosh,
and J.M. van Ruitenbeek, 1994, Soviet Phys. Low Temp. Phys. {\bf 20},
836
(Fiz. Nizk. Temp. {\bf 20}, 1062).  \\
Yanson, I.K., V.V. Fisun, R. Hesper, A.V. Khotkevich, J.M. Kraus, J.A.
Mydosh,
and J.M. van Ruitenbeck, 1995, Phys. Rev. Lett. {\bf 74}, 302. \\
Yatskar, A., W.P. Beyermann, R. Movshovich, and P.C. Canfield, 1996, 
Phys. Rev. Lett. {\bf 77}, 3637. \\
Ye, Jinwu, 1996a, Phys. Rev. Lett. {\bf 77}, 3224.\\
Ye, Jinwu, 1996b, submitted to Phys. Rev. B (cond-mat/9609058)\\
Ye, Jinwu, 1996c, preprint (cond-mat/9612029).\\
Ye, Jinwu, 1996d, preprint (cond-mat/9609057).\\
Ye, Jinwu, 1996e, preprint (cond-mat/9609076).\\
Yoshimori, A., 1976, Prog. Th. Phys. {\bf 55}, 67.\\
Yosida, K., and A. Yoshimori, 1973, in {\it Magnetism}, ed. H.Suhl
(Academic,
New York), Vol. V., p. 253.\\
Yu, C.C., and P.W. Anderson, 1984, Phys. Rev. B{\bf 29}, 6165.\\
Yu, C.C., and A. V. Granato, 1985, Phys. Rev. B{\bf 32}, 4793.\\
Yu, C.C., and A.J. Leggett, 1989, Comm. Cond. Mat. Phys. {\bf 14},
231.\\
Yuval, G. and P.W. Anderson, 1970, Phys. Rev. B{\bf 1}, 1522.\\
Zachar, O., S.A. Kivelson, and V.J. Emery, 1996, Phys. Rev. Lett. {\bf
77}, 1342.\\
\zar, G.,1993, Sol. St. Communications {\bf 86}, 413.\\
\zar, G., 1995, Phys. Rev. B{\bf 51}, 273.\\
\zar, G., 1996, Phys. Rev. Lett. {\bf 77}, 3609.\\
\zar, G., and L. Udvardi, 1996a, Physica B {\bf 218}, 68.\\
\zar, G., and L. Udvardi, 1996b, preprint (cond-mat/9605063).\\
\zar, G., and K. \vld, 1996, Phys. Rev. Lett. {\bf 76}, 2133.\\
\zar, G., and A. \zow, 1994a, Phys. Rev. Lett. {\bf  72}, 542. \\
\zar, G., and A. \zow, 1994b, Phys. Rev. B{\bf 50}, 392.\\
\zar, G., and A. \zow, 1995, Physica B {\bf 218}, 60.\\
\zar, G., J. van Delft, and A. \zow, 1997, preprint (submitted to 
Phys. Rev. Lett. [Comm.]).\\
\zim, G.T., K. \vld~, and A. \zow, 1987, Phys. Rev. B{\bf 36}, 3186.\\
\zow, A., 1970 (unpublished).\\
\zow, A., 1974, {\it Collective Properties of Physical Systems,
Nobel Symposium 24}, (Academic Press, New York), p. 76.\\
\zow, A., 1980, Phys. Rev. Lett. {\bf 45}, 211.\\
\zow, A. and K. \vld~, 1980, Sol. St. Comm. {\bf 35}, 217.\\
\zow, A. , 1987, Phys. Rev. Lett. {\bf 59}, 469.\\
\zow, A., 1989(a), Phys. Rev. B{\bf 39}, 4682.\\
\zow, A.,  1989(b), Sol. St. Comm. {\bf 70}, 439.\\
\zow, A., 1989(c), Phys. Scr. T{\bf 27}, 66.\\
\zow, A. and M. Fowler, 1970, in {\it Proc. 12th International Conf.
on Low Temp.
Phys.}, (Science Council of Japan), p. 324.\\
\zow, A. and P. Fazekas, 1970, J. Appl. Phys. {\bf 41}, 1155.\\
Zawadowski, A., K. Penc, and G.T. \zim, 1991, Prog. Th. Phys. Supp.
{\bf 106}, 11.\\
Zawadowski, A. and K. \vld, 1992, in ``Quantum
Tunneling in
Condensed Media,'' eds. Yu. Kagan and A.J. Leggett (Elsevier,
Amsterdam), p. 427.\\
\zow, A., and G. \zim, 1985, Phys. Rev. B{\bf 32}, 1373.\\
\zow, A., G. \zar, P. \noz, K. \vld, and G. \zim, 1997, preprint.\\
Zimmerman, N.M., B. Golding, and W.H. Haemmerle, 1992, Phys. Rev.
Lett.
{\bf 67}, 1322.\\
Zhang, F.C. and T.K. Lee, 1983, Phys. Rev. B{\bf 28}, 33.\\
Zhang, F.C. and T.K. Lee, 1984, Phys. Rev. B{\bf 30}, 1556.\\
Zhang, G., A.C. Hewson, and R. Bulla, 1997, preprint
(cond-mat/9705199).\\
Zhang, S. and P.M. Levy, 1989, Phys. Rev. Lett. {\bf 62}, 78.\\
Zhang, S.C., 1990, Phys. Rev. Lett. {\bf 65}, 120.\\
Zhang, X.Y., 1994, Physica B{\bf 199\&200}, 445.\\

\end{document}